\documentclass{aa}   

\usepackage{graphicx}
\usepackage[]{txfonts}
\usepackage{natbib}  
\usepackage[figuresright]{rotating}

\begin{document}

\title{CSI~2264: Simultaneous optical and X-ray variability in pre-Main Sequence stars}
\subtitle{I: Time resolved X-ray spectral analysis during optical dips and accretion bursts in stars with disks.}

\author{M. G. Guarcello\inst{1} \and E. Flaccomio\inst{1} \and G. Micela\inst{1} \and C. Argiroffi\inst{1,2} \and S. Sciortino\inst{1} \and L. Venuti\inst{1,2} \and J. Stauffer\inst{3} \and L. Rebull\inst{3} \and A. M. Cody\inst{4}
          }

\institute{INAF - Osservatorio Astronomico di Palermo, Piazza del Parlamento 1, I-90134, Palermo, Italy\\
              \email{mguarce@astropa.unipa.it}
          \and
              Dip. di Fisica e Chimica, Universit\`{a} di Palermo, Piazza del Parlamento 1, 90134 Palermo, Italy 
          \and
              Spitzer Science Center, California Institute of Technology, Pasadena, CA 91125, USA 
          \and
              NASA Ames Research Center, Kepler Science Office, Mountain View, CA 94035, USA
              }

\date{}

\abstract
{Pre-main sequence stars are variable sources. The main mechanisms
responsible for their variability are variable extinction, unsteady
accretion, and rotational modulation of both hot and dark photospheric
spots and X-ray active regions. In stars with disks, this variability
is related to the morphology of the inner circumstellar region
($\leq0.1\,$AU) and that of the photosphere and corona, all impossible to be
spatially resolved with present day techniques. This has been the
main motivation for the Coordinated Synoptic Investigation of
NGC~2264, a set of simultaneous observations of NGC~2264 with 15
different telescopes.}
{In this paper, we focus on the stars with disks. We analyze the X-ray spectral properties
extracted during optical bursts and dips in order to unveil the
nature of these phenomena. Stars without disks are studied in a
companion paper.}
{We analyze simultaneous CoRoT and {\em Chandra}/ACIS-I observations to
search for coherent optical and X-ray flux variability in stars with
disks. Then, stars are analyzed in two different samples. In stars
with variable extinction, we look for a simultaneous increase of optical extinction
and X-ray absorption during the optical dips; in stars
with accretion bursts, we search for soft X-ray emission and increasing
X-ray absorption during the bursts.}
{We find evidence for coherent optical and X-ray flux variability
among the stars with variable extinction. In 9/24 stars with optical dips, we observe a simultaneous increase of X-ray absorption and
optical extinction. In seven dips, it is possible to calculate the
N$_H$/A$_V$ ratio in order to infer the composition of the obscuring
material. In 5/20 stars with optical accretion bursts, we observe
increasing soft X-ray emission during the bursts that we associate to
the emission of accreting gas. It is not surprising that these
properties are not observed in all the stars with dips and bursts,
since favorable geometric configurations are required.} 
{The observed variable absorption during the dips is mainly due to
dust-free material in accretion streams. In stars with accretion
bursts, we observe on average a larger soft X-ray spectral component
not observed in non accreting stars. } 

\keywords{} 

\maketitle

\section{Introduction}
\label{intro}

Pre-main sequence (PMS) stars can be classified according to their
spectral energy distribution (SED) in the
infrared \citep{Lada1987}: The youngest PMS stars still surrounded by
both a contracting envelope and a circumstellar disk are classified as
Class~I objects. Class~II objects are PMS stars surrounded by
circumstellar disks, whose envelope is partially or completely
dissipated. Class~III sources are PMS stars whose disks have been
dissipated or at least evolved into debris disks. The intermediate
phases of stars with pre-transition disks (with an intermediate gap
separating inner and outer disks, \citealp{EspaillatCDH2007}) and
transition disks (with cleared inner regions,
\citealp{MuzerolleAMH2010}) have been more recently added to this
evolutionary scenario. The evolution and dissipation of circumstellar
disks involve several physical processes, such as gas accretion onto
the central star driven by viscosity and mediated by the magnetic
field \citep[e.g.][]{Koenigl1991}, photoevaporation
\citep[e.g.][]{StorzerHollenbach1999}, dust aggregation and settling
\citep[e.g. ][]{TestiBRA2014}, and environmental feedback
\citep[e.g.][]{GuarcelloDWA2016arXiv}, with timescales which roughly
range from few thousands to about $10^7$ years
\citep{HaischLL2001,HernandezHMG2007,Mamajek2009}. \par

  The disk inner region ($\leq0.1\,$AU) is very important for the
accretion process, the magnetic coupling between disk and central
star, and the evolution of the entire disk and the star itself. This
region is however very difficult to analyze, even in the systems close
to our Sun and even with infrared and radio interferometry. The
analysis of the SEDs of disk-bearing stars has achieved many successes
in this direction, such as the discovery of the inner disk wall and
pre-transitional gaps, but the physical parameters derived from SED
fitting are model-dependent and likely affected by the strong intrinsic
variability of these sources in the optical and infrared. An
unique insight on the very innermost region of circumstellar disks can be
provided by studying the simultaneous variability in optical, infrared
and even X-rays, which are intimately connected with the morphology and
properties of the circumstellar environment. \par 

Optical variability of PMS stars has been the subject of several
studies
\citep[e.g.][]{Joy1945,AlencarTGM2010,Morales-CalderonSHG2011,WolkRA2013,CodySBM2014AJ},
and originally it was one of the criteria used by \citet{Joy1945} to
identify the newly discovered class of T~Tauri stars.
\citet{HerbstHGW1994AJ} presented the first classification of
variability of young stars based on their light curves.
Type~I\footnote{The \emph{type} of the light curves must not be
confused with the \emph{class} used to classify YSOs} light
curves are periodic and often sinusoidal, resulting from the
rotational modulation of cold photospheric spots. Type~II light curves
are less periodic and interpreted as the result of variable veiling
continuum and rotational modulation of accretion hot spot. Type~III light curves
vary irregularly because of variable extinction. A more detailed
classification of light curves of disk-bearing stars has been recently
proposed by \citet{CodySBM2014AJ}, as part of the CSI~2264 project that
will be described below. They classify ``burster'' light curves as
those characterized by rapid (0.1-1 day) and symmetric increments of 
flux (the symmetric shape of the burst means that its rising part is
not impulsive and it is similar to the decaying phase, which distinguishes
them from flares);  ``dipper'' light curves showing transient fading
events (dips); ``periodic'' and ``quasi-periodic'' light curves
resulting from rotational modulation; ``stochastic'' light curves
which, even if not dominated by bursts or dips, are nevertheless
characterized by brightness changes over a variety of timescales;
``long-time variables'' with monotonic light variations (either
brightening of fading) over timescales of days and weeks. \par

        The different types of optical and infrared light curves of
stars with disks reflect the different morphology of the inner disks
and probe the geometry of the accretion process. Following the
classification of \citet{CodySBM2014AJ}, ``dipper'' stars are AA~Tau
analogs. AA~Tau is a well studied variable star with disk
characterized by recurrent occultation of the central star by warps
in the circumstellar disk
\citep{BouvierCAC1999,BouvierGAD2003,MenardBDM2003,BouvierABD2007,GrossoBMF2007}
located close to the co-rotation radius
\citep{RucinskiMKP2008,AlencarTGM2010,CodyHillenbrand2010}. These
warps in the inner disks are in general due to misalignment between
the rotation and magnetic axes, and they are located at the base of steady
accretion streams, which are stable over several stellar rotation periods
\citep{AlencarTGM2010}. \citet{AlencarTGM2010} have also shown that
AA~Tau like variability is common in stars with inner disks; in their
study of the optical and infrared variability of the stars in
NGC~2264, they have found that nearly 40\% of the stars with inner
disks are characterized by AA~Tau like variability. \par

Also, accretion contributes to variable optical and infrared extinction
\citep[e.g., ][]{McGinnisAGS2015}. The accretion streams are dust-free, but small amounts of dust can be trapped at the base of the
streams and survive until the temperature is higher than the
sublimation temperature. As suggested by \citet{StaufferCMR2015} the
dust particles trapped in the accretion streams can be responsible for
small dips in the optical light curves. Another way accretion may
contribute to optical variability is by the emergence of the optical
emission from accretion hot spots on stellar surface
\citep{StaufferCBA2014}. In fact, the accreting material funneled by
the magnetic field falls onto the stellar surface with a velocity of
several hundreds of km per second. This energy is released at the
accretion shock as soft X-ray, UV, and optical radiation. \par

Is it possible to observe variability in X-rays due to the accretion
process and variable circumstellar extinction? PMS stars are very
bright X-ray sources \citep{FeigelsonDecampli1981}, with their X-ray
emission exceeding that of main sequence stars with the same mass by
three or four orders of magnitude \citep[e.g.][]{Montmerle1996}. The
main component of this X-ray emission is the
quiescent\footnote{Or at least apparently quiescent, since the
``quiescent'' coronal emission can be actually the result of a
superposition of small flares, e.g. \citet{CaramazzaFMR2007AA}} emission
from a scaled up version of the Solar corona with plasma at
$10-30\,$MK, powered and confined by a dynamo-generated stellar
magnetic field \citep[e.g.][]{FeigelsonKriss1981}. However, the lack
of an evident main sequence-like rotation-activity relation in PMS
stars, together with the large and so far unexplained scatter of X-ray
brightness, indicate that the emission mechanism might be more
complicated than this. Intense flaring activity produced by magnetic
reconnection is observed in PMS stars
\citep[e.g.][]{FlaccomioDMS2003}, and they can be so powerful as to
require non solar geometry for the stellar magnetic field
\citep{JardineCDG2006}. Sometimes flares in PMS stars are modeled with
very large loops that may even reach the surface of the inner disks
\citep{FavataFRM2005ApJs}. Accretion also contributes to the emission of
soft X-rays \citep[e.g.][]{KastnerHSC2002}, produced in the accretion
shocks and observed in a few Class~II stars, such as TW~Hya and
BP~Tau, using detailed spectroscopic analysis
\citep{KastnerHSC2002,StelzerSchmitt2004,SchmittRNF2005,ArgiroffiFBD2011,CurranASO2011}.
However, this X-ray emission has been unambiguously identified only in
the nearest stars with disks, primarily because it is difficult to 
distinguish from the coronal soft X-ray emission.
Additionally, a significant part of this emission is likely absorbed by
accreting and circumstellar material itself
\citep{ArgiroffiFBD2011,BonitoOAM2014ApJ}. \par

    X-ray emission from PMS stars is strongly variable over a large
range of timescales and amplitudes. The most evident source of X-ray
variability is undoubtedly flares. The rise phase is
much shorter than the decline phase (which can last several hours),
and the peak flux can be $\sim$100 times the quiescent flux
\citep{FavataFRM2005ApJs}. There are, however, other sources of X-ray
variability. Stellar coronae are not homogeneous, and their X-ray
emission can be modulated by stellar rotation
\citep{FlaccomioMSF2005ApJ}. This has been observed in the {\em Chandra} Orion Ultradeep
Project \citep[COUP,][]{GetmanFBG2005}: a $\sim$13 day long
continuous {\em Chandra}/ACIS-I observation of the Orion Nebula Cluster.
Also accretion spots are not uniformly distributed over the
stellar surface, resulting in a rotational modulation of soft
X-ray emission. This, however, has only been observed in the T~Tauri
star V4046 Sgr \citep{ArgiroffiMMH2012ApJ}. Variable absorption of
the coronal emission by circumstellar and accreting material can be
another source of X-ray variability \citep[e.g.][]{FlaccomioMFA2010}.
\par

       Simultaneous optical, infrared, and X-ray variability in stars
with disks can be the consequence of unsteady accretion, variable
extinction, and rotational modulation \citep{FlaccomioMS2012}.
\citet{StassunBF2007ApJ} find no convincing evidence for coherent optical
and X-ray flux variability in the PMS stars in Orion. No
evidence of coherent X-ray and infrared variability in PMS stars is
found by \citet{FlahertyMWR2014ApJ} in their study of the PMS stars in
IC~348, concluding that X-rays are not an important source of heating
for the circumstellar material. A different result has been obtained
by \citet{FlaccomioMFA2010} in their study of the PMS stars in
NGC~2264. They find a significant correlation between optical and
X-ray flux variability using two $30\,$ksec {\em Chandra}/ACIS-I
observations (separated by 16 days) and simultaneous CoRoT data. This
correlation is observed only in Class~II sources, and it is not
observed in the hard X-ray band. This is interpreted as a
consequence of variable absorption of both photospheric and coronal
emission.  \par

	\begin{figure}[]
	\centering	
	\includegraphics[width=8.5cm]{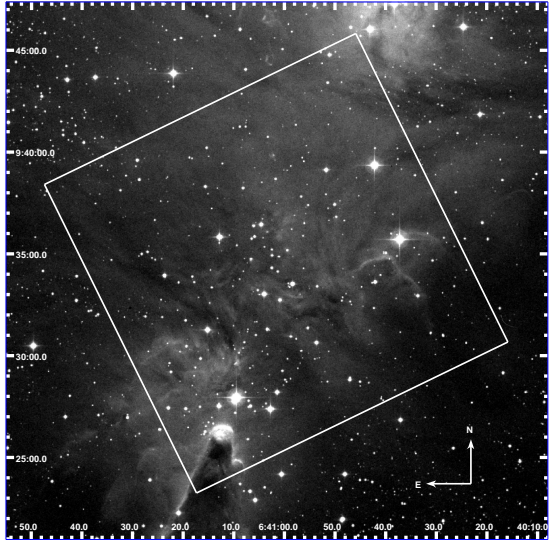}
	\caption{DSS-2 image of the central region of NGC~2264, with marked the field observed with {\em Chandra}/ACIS-I.}
	\label{field_img}
	\end{figure}

In this paper, we analyze new simultaneous X-ray and optical
observations of NGC~2264, obtained as a part of the CSI~2264 project,
to search for connections between optical and X-ray variability in PMS
stars with disks observed in the quiescent emission. We show the
effectiveness of time-resolved X-ray spectral analysis in stars with
disks using the optical light curves as template to isolate interesting
features such as accretion bursts and optical dips. NGC~2264, the
object of this study, and the CSI~2264 project are described in Sect.
\ref{2264_sec}. In Sect.~\ref{data_sec}, we describe the CoRoT and
{\em Chandra} data sets analyzed and the selection of the targets; in
Sect.~\ref{correl_II_sec}, we present evidence for coherent optical and X-ray flux
variability. In Sect.~\ref{II_var}, we present a
detailed analysis of the variability of the X-ray properties during
dips and bursts observed in the CoRoT light curves. Results are
summarized and discussed in Sect.~\ref{conclusions}.

\section{NGC~2264 and the CSI~2264 project}
\label{2264_sec}

  The study of the variability of young
stars offers the possibility of probing the very inner circumstellar region ($\leq0.1\,$AU) and
the morphology of stellar coronae and photospheres. This is one of the main
motivations of the Coordinated Synoptic Investigation of NGC~2264
\citep[\emph{CSI~2264},][]{CodySBM2014AJ,StaufferCBA2014}. This project
is a unique and unprecedented cooperative project involving
simultaneous observations of NGC~2264 with 15 ground and space
telescopes, covering the electromagnetic spectrum from X-rays to
mid-infrared. The entire list of the observations which are part of
the CSI~2264 project can be found in \citet{CodySBM2014AJ}. The main
optical photometric dataset is obtained from observations with the
Convection, Rotation and Planetary Transits satellite
\citep[\emph{CoRoT},][]{BaglinABD2006ESASP} from December 1$^{st}$ 2011 to
January 3$^{rd}$ 2012, using the second CCD designed for exoplanets
studies. CoRoT observed an area of $1.3 \times 1.3$ square degrees
centered on NGC~2264 with a cadence of 512$\,$sec, or 32$\,$sec for
the brighter sources. 

NGC~2264 is the only young cluster \citep[1-5
Myrs,][]{RebullMSH2002,Dahm2008} falling in one of the CoRoT eyes
(i.e., the two regions with a $10^{\circ}$ diameter close to the
galactic center and anti-center observed with CoRoT), making it a
unique target for monitoring the variability of young stars using this
telescope. This cluster is relatively nearby
\citep[$760\,$pc,][]{ParkSBK2000}, and part of the local spiral arm.
It is characterized by non-uniform extinction across the field, with
both richly populated low-extinction regions containing in total about
two thousand sources, and highly embedded regions where the star
formation process is still ongoing, as evidenced by the presence of
stars with thick disks, molecular outflows and Herbig-Haro objects
\citep{TeixeiraLML2012}. The median extinction of known cluster
members is quite low \citep[A$_V\sim$0.45$^m$,][]{RebullMSH2002}. The
cluster population is well defined, and it includes a few early type
stars, such as the O7V star S~Monocerotis \citep{SchwartzTOG1985} and
about a dozen B type stars. NGC~2264 is the only cluster within one
kpc of the Sun, besides the Orion Nebula Cluster, with such a
large mass spectrum. Fig.~\ref{field_img} shows a DSS-2 image of the
central region of NGC~2264. The field of view of the
{\em Chandra} observations analyzed here is indicated. The actual CoRoT field,
$1.3^{\circ} \times 1.3^{\circ}$ wide, is larger than the field shown
in Fig.~\ref{field_img}.

%%%%%%%%%%%%%%%%%%%%%%%%%%%%%%%%%%%%%%%%%%%%%%%%%%%%%%%%%%%%%%%%55
\section{Data analysis and targets selection}
\label{data_sec}

   \subsection{CoRoT light curves}
   \label{corot_sec}
   
  CoRoT photometry is available only for stars falling in pre-selected
pixel masks. For this reason, the list of targets has been carefully
pre-compiled before the observations, and it includes 1617 known
candidate members of NGC~2264 and 2129 field stars falling in the area
of the cluster \citep{CodySBM2014AJ}. The membership criteria used for
targets selection are described in \citet{CodySBM2014AJ} and they are
based on: 1) Optical photometry compatible with the cluster in
color-magnitude diagrams as defined by \citet{FlaccomioMS2006}; 2)
strong H$\alpha$ emission
\citep{RebullMSH2002,LammBMH2004,SungBCK2008}; 3) X-ray detection
\citep{RamirezRSH2004,FlaccomioMS2006}; 4) radial velocity compatible
with NGC~2264 \citep{FureszHSR2006}; 5) presence of a circumstellar
disk \citep{SungSB2009,CodySBM2014AJ}. Only targets with $11^m>R>17^m$
have been observed. \par

Light curves are produced after correction from gain and zero offset,
jitter, electromagnetic interference, and background subtraction,
following the standard CoRoT data reduction pipeline
\citep{SamadiFCD2006}. Data from hot pixels and outliers are flagged
and removed from the light curves. While most of the light curves
refer to the full CoRoT band, for a subset of stars three light curves
are provided, referring to  ``red'', ``green'', and ``blue'' bands.
However, since these bands are not calibrated, in this work we only use
the sum of the three bands, i.e. the white light curves. CoRoT light
curves are not in an absolute magnitude scale since the photometric
zero point of the CoRoT data varies between runs. A photometric zero
point of 26.74$^m$ specific for the CSI~2264 CoRoT observations has
been obtained by \citet{CodySBM2014AJ} after comparing the mean CoRoT
flux of selected stars with their available $R$-band photometry
\citep[from ][]{RebullMSH2002,LammBMH2004,SungBCK2008}. Flaccomio et
al. (in preparation) obtained a slightly smaller zero point (26.6$^m$), which is the value adopted in this paper.
\par

  A number of systematics are not corrected by the standard pipeline
and affect CoRoT light curves. The most important systematic,
affecting about 10\% of the observed light curves, consists of abrupt
jumps in flux which are due to rapid changes in detector temperature. In
this paper, we do not attempt any correction for this effect and we
reject when necessary CoRoT data that are affected by such jumps.
We use only CoRoT data which are not flagged as suspicious
data points.  \par

%%%%%%%%%%%%%%%%%%%%%%%%%%%%%%%%%%%%%%%%%%%%%%%%%%%%%%%%%%%%%%%%55
  \subsection{X-ray data}
  \label{xray_sec}

      	\begin{figure*}[]
	\centering	
	\includegraphics[width=19cm]{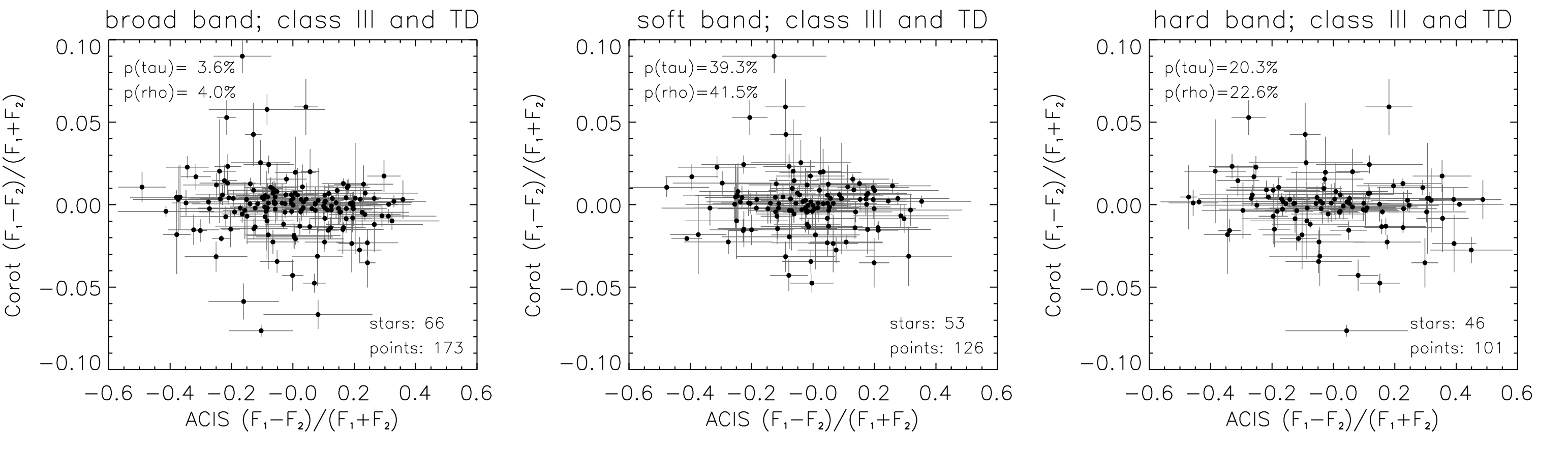}
	\includegraphics[width=19cm]{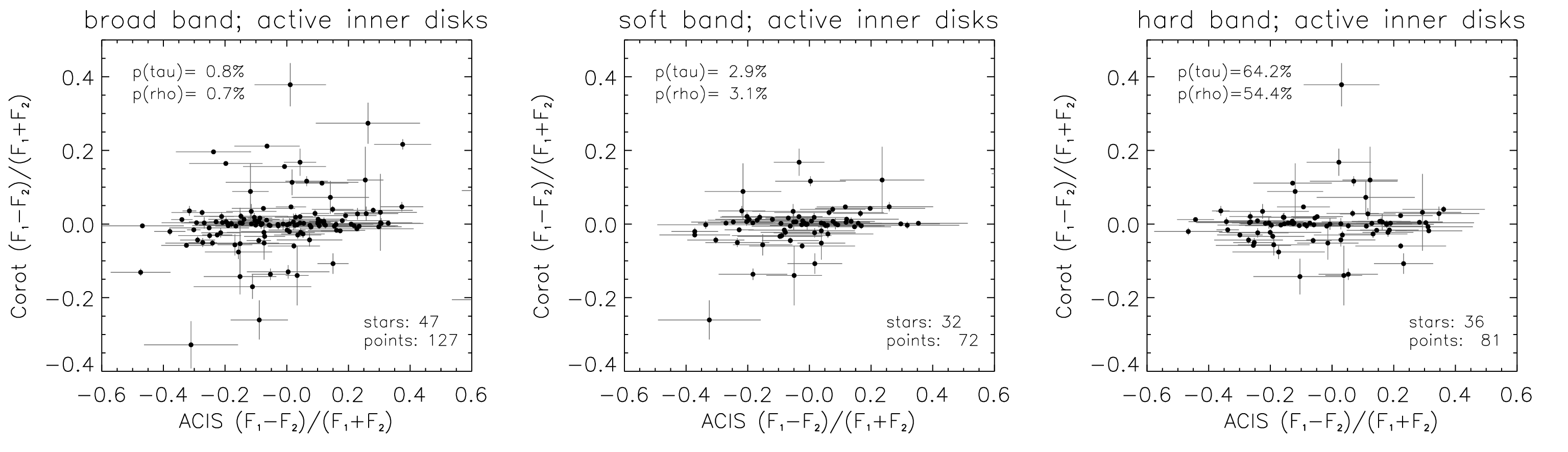}	
        \caption{Comparison between the optical and X-rays flux variability, calculated using in Eq. \ref{var_eq}, shown
separately for the stars with transition disks and class~III objects
(upper panels) and stars with inner disks, both passive and accreting
(lower panels). Note the different $y$-axis scaling for the top and
bottom rows. X-ray photons are selected in the broad band
($0.5-8\,$keV; left panels), soft band ($0.5-1.5\,$keV; central
panels), and hard band ($1.5-8\,$keV; right panels). In each panel, the
null-hypothesis probabilities of no correlation resulting from the
Spearman ($\rho$) and Kendall ($\tau$) rank correlation test are
shown. Each point corresponds to the variability observed between two consecutive
given {\em Chandra} frames for each source. The total number of stars and
points used in each plot is also indicated.}
	\label{fluxcompclass_plot}
	\end{figure*}
  
        In this paper, we use the data obtained from four {\em Chandra}/ACIS-I observations
taken from December 3$^{rd}$ to December 11$^{th}$ 2011 (P.I.\ G.\
Micela) during the CoRoT run. Table \ref{obsid_table} shows the log of
these observations. The total nominal exposure is $300\,$ksec, and all
the observations are pointed at
$\alpha_{J2000}$=06:40:58.70,$\,\delta_{J2000}$=+09:34:14, with almost
coincident roll angles. In order to make astrometry compatible with 2MASS, each image has been reprojected by matching bright 2MASS
sources detected in X-rays and correcting for the observed offsets both in RA and DEC. \par 

    \begin{table}
    \caption{{\em Chandra}/ACIS-I observations log.}
    \label{obsid_table}
    \centering                       
    \begin{tabular}{ccc}        
    \hline\hline                
    Obs.ID & Exposure (ksec) & Date \\
    \hline                        
    14368 & 74.44 & Dec. 3$^{rd}$ 2011\\ 
    13610 & 92.54 & Dec. 5$^{th}$ 2011\\
    13611 & 60.23 & Dec. 7$^{th}$ 2011\\
    14369 & 66.16 & Dec. 11$^{th}$ 2011\\
    \hline                                  
    \end{tabular}
    \end{table}

    A detailed analysis of the {\em Chandra}/ACIS-I
\citep{WeisskopfBCG2002,GarmireBFN2003} observations available for
NGC~2264, including source detection, photon extraction and spectral
fitting, is presented in Flaccomio et al.\ (in preparation). Briefly, all events have been fully reprocessed using the CIAO task
\emph{chandra-repro}. Sources were detected using the wavelet-based
algorithm PWDetect \citep{DamianiMMS1997}, adopting a significance
threshold of 4.4, roughly resulting in 10 expected spurious detections.
Event extraction, source repositioning, and validation were performed
with the IDL software $ACIS$ $Extract$ \citep[AE, ][]{BroosTFG2010}.
AE is capable of: i) defining the extraction region around each source,
accounting for crowding and the shape of the PSF at different off-axis
angles; ii) extracting both source and background events, the latter
in a suitable region around the source; iii) compiling photometry,
calculating light curves, and providing source statistics. After pruning
candidate spurious detections, a total of 694 X-ray sources were
validated. X-ray spectra were fitted using \emph{Xspec} v.12.8.1
\citep{Arnaud1996}. Observed spectra were rebinned in photon energy
in order to have a signal-to-noise ratio larger than one in each bin.

%%%%%%%%%%%%%%%%%%%%%%%%%%%%%%%%%%%%%%%%%%%%%%%%%%%%%%%%%%%%%%%%55
    \subsection{Targets selection}
    \label{sample_sec}

    The main objective of this paper is to analyze the simultaneous
optical and X-rays variability of stars with disks in NGC~2264.\par

We adopt the selection of stars with disks presented in
\citet{SungSB2009} and \citet{CodySBM2014AJ}, which is based on the slope of the SEDs in the
IRAC and MIPS $24.0\,\mu$m bands and suitable color-color diagrams
where the typical loci populated by disk-bearing stars can be defined.
Among the 95 candidate stars with disks observed with CoRoT and
falling in the ACIS-I field, 86 are detected in X-rays. We also
define a subsample of 79 stars with disks (75 detected in X-rays)
which are actively accreting. These stars are selected using two
criteria: H$\alpha$ equivalent width (EW) larger than 10$\,$\AA{}
\citep{RebullMSH2002} or using the $r^{\prime}-i^{\prime}$ vs.
$r^{\prime}-$H$\alpha$ color-color diagram from the INT (Isaac Newton
Telescope, 2.5$\,$m) Photometric H$\alpha$ Survey
\citep[IPHAS;][]{DrewGIS2008}. In this color-color diagram, in fact, it is
possible to select candidate accreting stars as those with red
$r^{\prime}-$H$\alpha$ color, and derive from this color an estimate
of the H$\alpha$ EW \citep{BarentsenVDG2011}. Disk-bearing sources
without signatures of accretion are identified as stars with passive disks
(15 stars, 10 detected in X-rays). We also selected 10 candidate stars
with transition disks (all detected in X-rays) as those showing
excesses only at $8.0\,\mu$m and 24$\,\mu$m. The infrared excesses in
each infrared band is calculated using the $Q_{VIJA}$ color indices
similar to those defined in \citet{GuarcelloMDP2009,GuarcelloDWD2013ApJ}.
These color indices compare the $V-I$ and $J-A$ colors, with $A$ being
$K$ or one of the Spitzer bands. Since these indices
increase as $J-A$ becomes more red, and they are independent from 
extinction, they can be used to separate the
extinguished stellar population from that with intrinsic red colors,
and to calculate the excess in each infrared band. We refer to
\citet{Damiani2006} and \citet{GuarcelloMDP2009} for a detailed
description of these color indices and their use. \par

 	\begin{figure}[]
	\centering	
	\includegraphics[width=9cm]{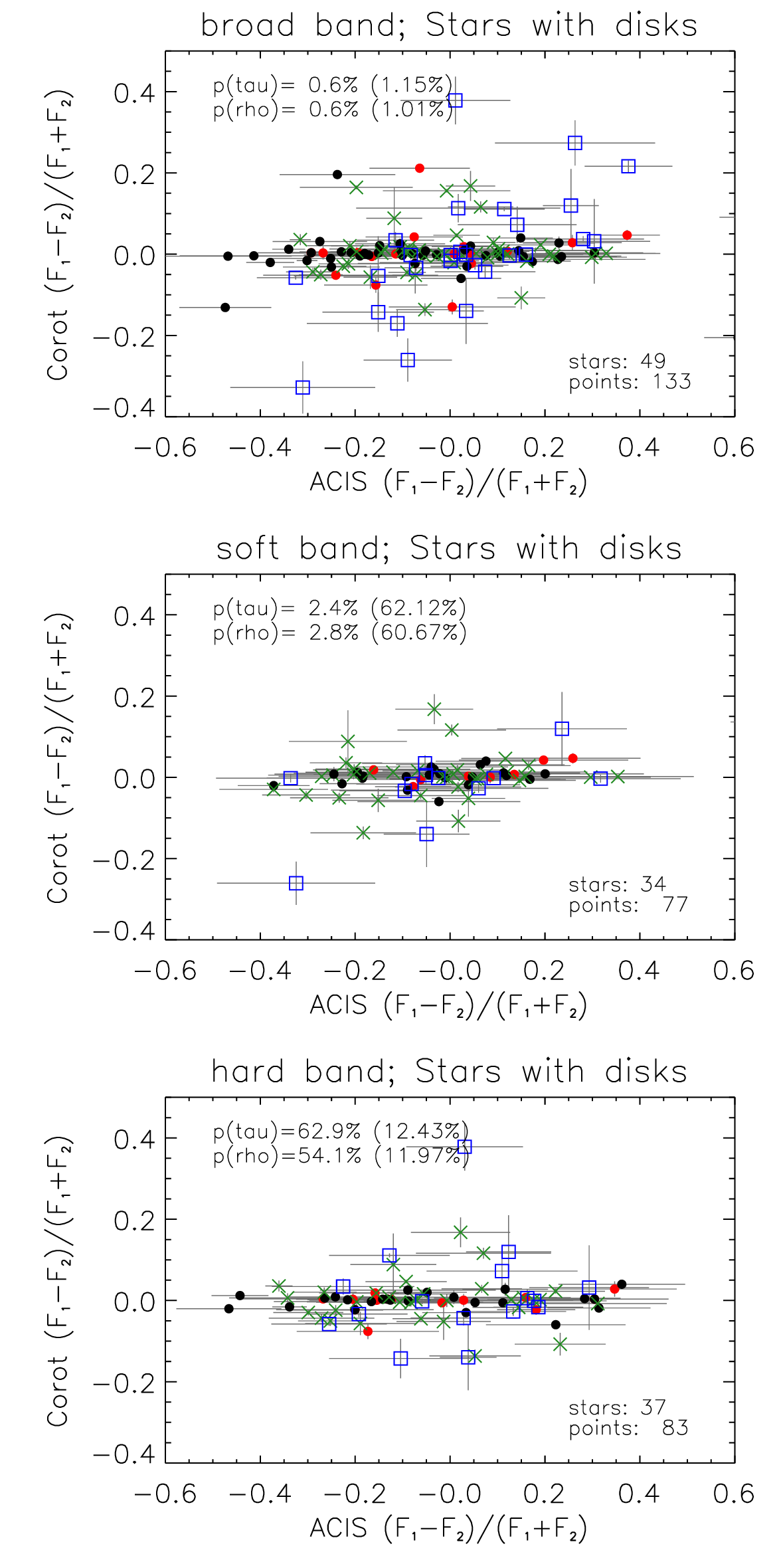}	
        \caption{Comparison between the flux variability in optical
and X-rays as in Fig.~\ref{fluxcompclass_plot}, for disk-bearing stars
whose optical variability has been classified by \citet{CodySBM2014AJ}.
``Burster'', ``unclassified'', ``not variable'', ``quasi-periodic'',
``long term variable'', and ``multi-periodic'' stars are marked with
filled dots (red for the ``bursters'' and black for the remainder),
while ``stochastic'' stars are marked with green $\times$ symbols and
``dipper'' stars with blue squares. In the upper left corner of each
panel we show the probabilities derived from the correlation tests
using both the entire sample and the dippers+stochastic stars (the
latter inside brackets).}
	\label{fluxcodyclass_plot}
	\end{figure}
 
  We also adopt the classification of the CoRoT light curves provided
by \citet{CodySBM2014AJ}: ``Bursters'' (13\% of the stars with disks in NGC~2264
observed with CoRoT); sources with variable extinction (``dippers'',
21.5\%, about half of which are periodic); stars with light curves
showing stochastic behavior (13\%, named ``stochastic''); non-variable 
stars (19\%); stars with periodic or quasi-periodic
variability (21\%); long-term variables (1\%); eclipsing binaries
(1\%); and sources with unclassified variability (11\%).   Since the
classification of the ``stochastic'' stars may vary according to different
time windows, we reviewed the behavior of these sources in the
period simultaneous with the four ACIS-I observations. Sometimes we
consider stars as ``dippers'' or ``bursters'' depending on the
dominant phenomenon occurring during the {\em Chandra} observations. \par 

%%%%%%%%%%%%%%%%%%%%%%%%%%%%%%%%%%%%%%%%%%%%%%%%%%%%%%%%%%%%55
\section{Coherent optical and X-ray flux variability}
\label{correl_II_sec}

In this section we analyze the simultaneous flux variability in
optical and X-rays. Coherent flux variability is expected to occur
when: i) The star is affected by variable extinction due to
circumstellar material simultaneously obscuring both stellar
photosphere and coronal active regions; or ii) photospheric
spots, accretion hot spots, and coronal active regions simultaneously emerging
during stellar rotation. \par

  \subsection{Existing studies}
  \label{back_sec}
  
  A search for coherent optical and X-ray flux variability in T~Tauri
stars has been attempted only in a very few cases, given the paucity of
existing simultaneous optical and X-ray observations of young
clusters. \citet{StassunBFF2006} studied $BVRI$ data of the Orion
Nebula taken with the WIYN 0.9$\,$m telescope at the Kitt Peak
National Observatory (KPNO) and the 1.5$\,$m Cassini telescope in
Loiano, Italy, simultaneous with the COUP observations. These optical
data have been taken with a cadence of one data point per hour, not
comparable to the excellent time resolution of the CoRoT data. These
authors find evidence of coherent optical and X-ray flux variability
in about 5\% of the analyzed T~Tauri stars. This
result has been interpreted in terms of rotational modulated emission
from accretion spots, and the lack of correlation in most of the
observed stars as evidence that X-ray emission arises mainly from
the stellar corona rather than from the accretion spots distributed
over the stellar surface. \par

\citet{FlaccomioMFA2010} search for correlations between optical and
X-ray flux variability in the young stars in NGC~2264 by comparing the
variation of the X-ray fluxes observed in two $\sim$30$\,$ksec
{\em Chandra} ACIS-I observations (Obs.~IDs: 9768 and 9769; P.I.\ G. Micela)
with that of the CoRoT ``white'' fluxes from simultaneous
observations. They find evidence of correlated flux variability in a sample of 24 low-mass T~Tauri stars with disks, and only in the
soft 0.5-1.5$\,$keV X-ray band. This is interpreted as evidence that
the correlation is a result of time-variable absorption by the surrounding
circumstellar material. These authors also suggest that the obscuring
material is dust-depleted, as expected from accretion streams covering
part of the stellar corona and photosphere. \par

\subsection{Coherent optical and X-ray flux variability in stars with disks of NGC~2264}  
\label{corr_class}
  
  The starting point for our search of coherent optical and X-ray
flux variability in the NGC~2264 T~Tauri stars is the result
obtained by \citet{FlaccomioMFA2010}. We first replicate their
approach to verify whether our new data confirm their finding. To this
aim, we select all the X-ray sources observed with CoRoT with more
than 10 counts detected in at least two {\em Chandra} observations,
discarding known massive and intermediate-mass stars (i.e., rejecting
stars more massive than 2$\,$M$_{\odot}$). Hereafter, we will call
{\em Chandra} {\em frames} the time intervals corresponding to the {\em Chandra}
observations listed in Table \ref{obsid_table}, i.e., where both CoRoT
and X-ray data are available. For each selected source, we calculate
the mean X-ray photon flux and CoRoT white-band flux observed in each
{\em Chandra} frame where no flares are detected\footnote{Flares are
automatically detected using the approach defined in
\citet{CaramazzaFMR2007AA}, i.e., dividing the X-ray light curve in
blocks of almost constant count rate (Maximum Likelihood Blocks) and
classifying these intervals according to the measured count-rates and
its time derivative. A detailed analysis of the flares observed in
NGC~2264 will be presented in Flaccomio et al.\ 2016 (in preparation)}.
We then calculate the flux variability among two {\em Chandra} frames $n$
and $m$ as:
  
  \begin{equation}
  \Delta_{flux}=\frac{F_n-F_m}{F_n+F_m}
  \label{var_eq}
  \end{equation}
  where $F_n$ is the optical or the X-ray photon flux detected during
the $n^{th}$ {\em Chandra} frame. \par

  The result of this approach is shown in Fig.\
\ref{fluxcompclass_plot}, where each point compares the value of
$\Delta_{flux}$ calculated for a given source in two consecutive {\em Chandra}
frames. Error bars are propagated from the uncertainties on X-ray and
optical fluxes, the former computed with Poissonian statistics, the
latter from the RMS of the optical light curve observed during the
given {\em Chandra} frames. We use X-ray data in different energy bands
(broad in the left panel, soft in the central, and hard in the right).
In Fig.~\ref{fluxcompclass_plot} we consider separately the sample of
stars with an inner disk, both passive or accreting, and that of stars
with a transition disk and class~III objects\footnote{The
classification of class~III objects and the analysis of their
variability is the subject of a companion paper.} (whose variability
is expected to be similar, \citealt{CodySBM2014AJ}).  \par

  The amplitude of the observed variability is different in these two samples of stars. 
We observe a smaller amplitude of optical variability in
stars without close circumstellar material than in those with inner
disks, with the difference of a factor between two and four.
Conversely, the range of variability in X-rays is similar in the two
cases. This is due to the fact that rotational
modulation of photospheric spots and active regions, typical of the
inner-disk-free sample, results in smaller amplitude modulation than, for
instance, variable extinction \citep{VenutiBIS2015AA}. 
In each panel, we also show the results of correlation tests, which in general
indicate that there is no obvious correlation between broad band X-ray and optical measures for stars with disks, though some of the extreme points (obtained from 17 stars) drive a statistically significant correlation in the X-ray broad energy band in the stars
with inner disks. This is not observed in
the stars without close circumstellar material. \par

Using the classification of optical light curves provided by
\citet{CodySBM2014AJ}, we can obtain deeper insight into the different
mechanisms responsible for coherent optical and X-ray flux
variability. In Fig.~\ref{fluxcodyclass_plot}, we investigate correlations between
optical and X-ray flux variability for those stars with
circumstellar disks whose light curves are classified by
\citet{CodySBM2014AJ}. The X-rays count-rates are indicated separately
for the the broad band (upper
panel), soft band (central panel), and hard band (bottom panel). It is evident that the stars with
large amplitude optical variability, dominating any possible
correlation, are ``dipper'' or ``stochastic'' stars. The results of
the correlation tests are shown for the entire sample and for the
``stochastic''+``dippers'' sample (values inside the brackets),
and they indicate that a correlation between optical and X-ray flux
variability is possible only in stars with disks and variable
extinction. This is reinforced by the fact that the ``stochastic''
stars with large variability in optical in Fig.\
\ref{fluxcodyclass_plot} have optical dips during the {\em Chandra}
frames. \par

%%%%%%%%%%%%%%%%%%%%%%%%%%%%%%%%%%%%%%%%%%%%%%%%%%%%%%%%%%%%%%%%%%%%%%%%%%%%%%%%%%%%
\section{Variability of the X-ray properties during optical dips and bursts}
\label{II_var}

  \subsection{Time resolved analysis of stellar X-ray properties}
  \label{xvar_sec}

The unique data set analyzed in this work allows us to study in detail
how the X-ray properties of disk-bearing stars vary during events
observed in optical, specifically dips due to variable extinction and bursts
due to accretion. Among the 86 stars with disks observed both with
CoRoT and Chandra, 51 are bright enough in X-rays to allow a
reliable analysis of their variability. Among them, 24 show
well-defined flux dips in the CoRoT data, and 20 show accretion burst
signatures, with some stars showing both properties.  \par

    To this aim, we divide each {\em Chandra} time frame into smaller {\em time
intervals}, defined in order to isolate interesting features in the
CoRoT light curves such as dips and bursts. For each time interval, we then calculate
the mean CoRoT flux and extract the X-ray photons detected during the time interval to calculate the corresponding X-ray photon flux F$_X$ (in
units of photons$\,$cm$^{-2}\,$s$^{-1}$), the hydrogen column density
N$_H$ (in units of $10^{22}\,$cm$^{-2}$), the temperature of the
emitting plasma kT (in keV), and the 10\%, 25\%, and 50\% photon
energy quantiles (E$_{10\%}$, E$_{25\%}$, E$_{50\%}$, respectively, in
keV). \citet{Flaccomioinprep1} demonstrate that E$_{10\%}$ and
E$_{25\%}$ are well-correlated with the hydrogen column density obtained
fitting the observed X-ray spectra of NGC~2264 low mass members with
1T or 2T thermal plasma models, so they are useful probes for the
X-ray absorption affecting the stars. Some of the time intervals are
narrow with few photons observed, but despite the small signal-to-noise 
ratio, such small time intervals are necessary to isolate those features that we want to
analyze. \par

The X-ray properties in each time interval are calculated fitting the
observed X-ray spectra with 1T and 2T APEC ionization-equilibrium
plasma isothermal model \citep{SmithBLR2001}, assuming the sub-solar
elements abundance defined by \citet{MaggioFFM2007}, and affected by
photoelectric absorption from both interstellar and circumstellar
material treated using the TBABS model \citep{WilmsAM2000}. Best-fit
models are chosen with the C-statistic and the quality of the fit is
tested using the Xspec tool \emph{goodness}. The limit for acceptable
fits in this paper is set to null-hypothesis probability of a good fit
(P$_{\%}$) equal to 5\%. The significance of the parameters obtained
with the spectral fit is tested by the analysis of the confidence
contours in the C-stat space with the Xspec tool \emph{steppar}. \par

  \subsection{Disks properties}
  \label{sed_sec}

  We analyzed the SEDs of some stars
using the on-line SED fitting tool presented by
\citet{RobitailleWIW2007}\footnote{Now available at
https://sedfitter.readthedocs.org/en/stable/}. With this tool it is
possible to compare the observed SEDs with a set of YSO
models covering an extensive parameter space, 20000 models, each at 10
different inclination angles, for a total of 200000 distinct SEDs. For
the fit, we constrain the source distances to that of NGC~2264
($760\,$pc) and explore a wide range of possible extinctions (from
$A_V$=$0.1^m$ to $A_V$=$100^m$). The best-fit models are those that
satisfy the condition: $\chi^2-\chi_{best}^2\leq 3$, where
$\chi_{best}^2$ is the reduced chi-square of the best-fit model (as
suggested by \citealp{RobitailleWIW2007}). Accretion rates and
H$\alpha$ EW of the stars discussed in this section are taken from \citet{DahmSimon2005} and
\citet{VenutiBFA2014AA}, and shown in Table \ref{accret_table}. \par  

	\begin{figure*}[]
	\centering	
	\includegraphics[width=10cm]{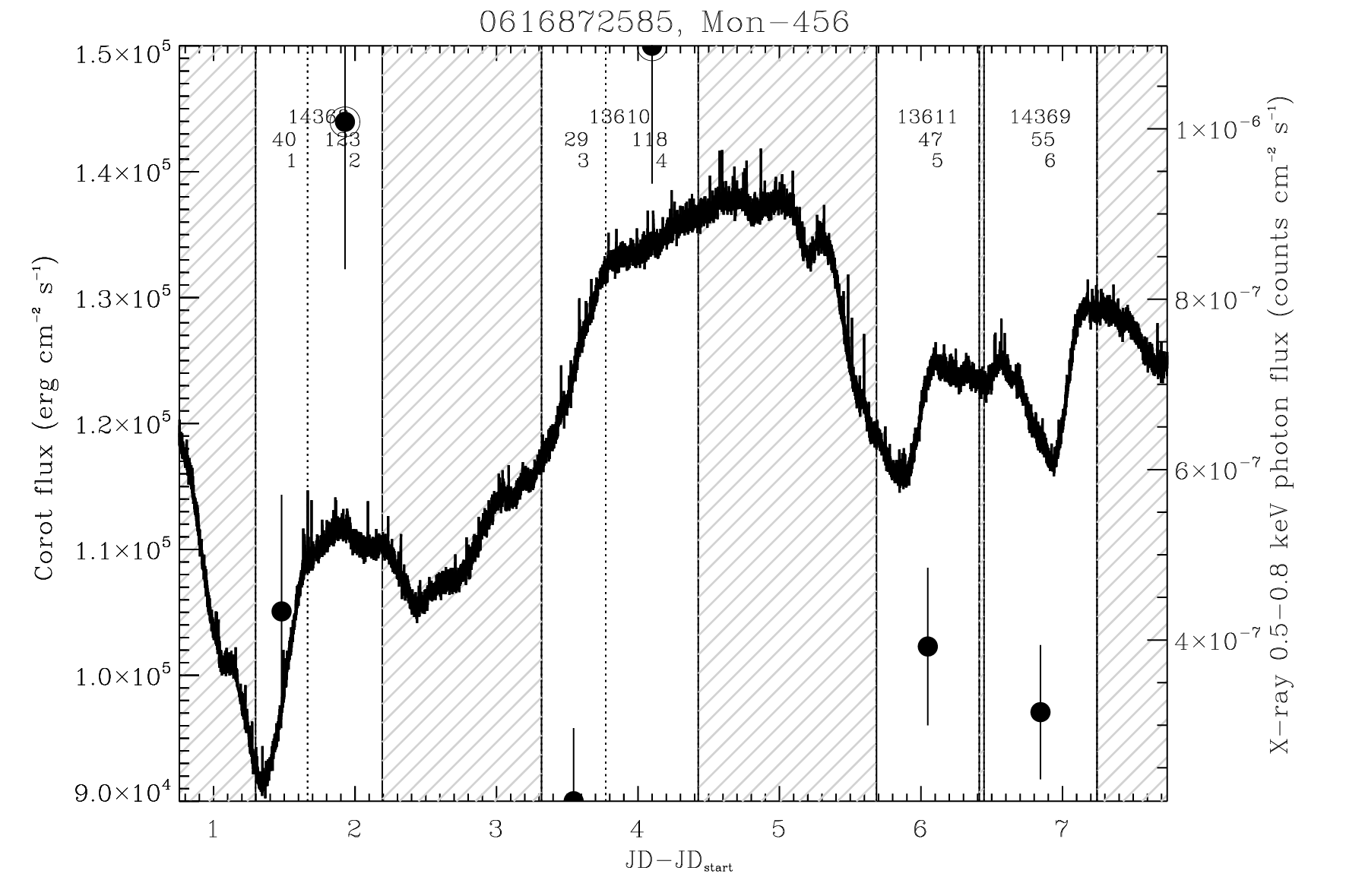}
	\includegraphics[width=8cm]{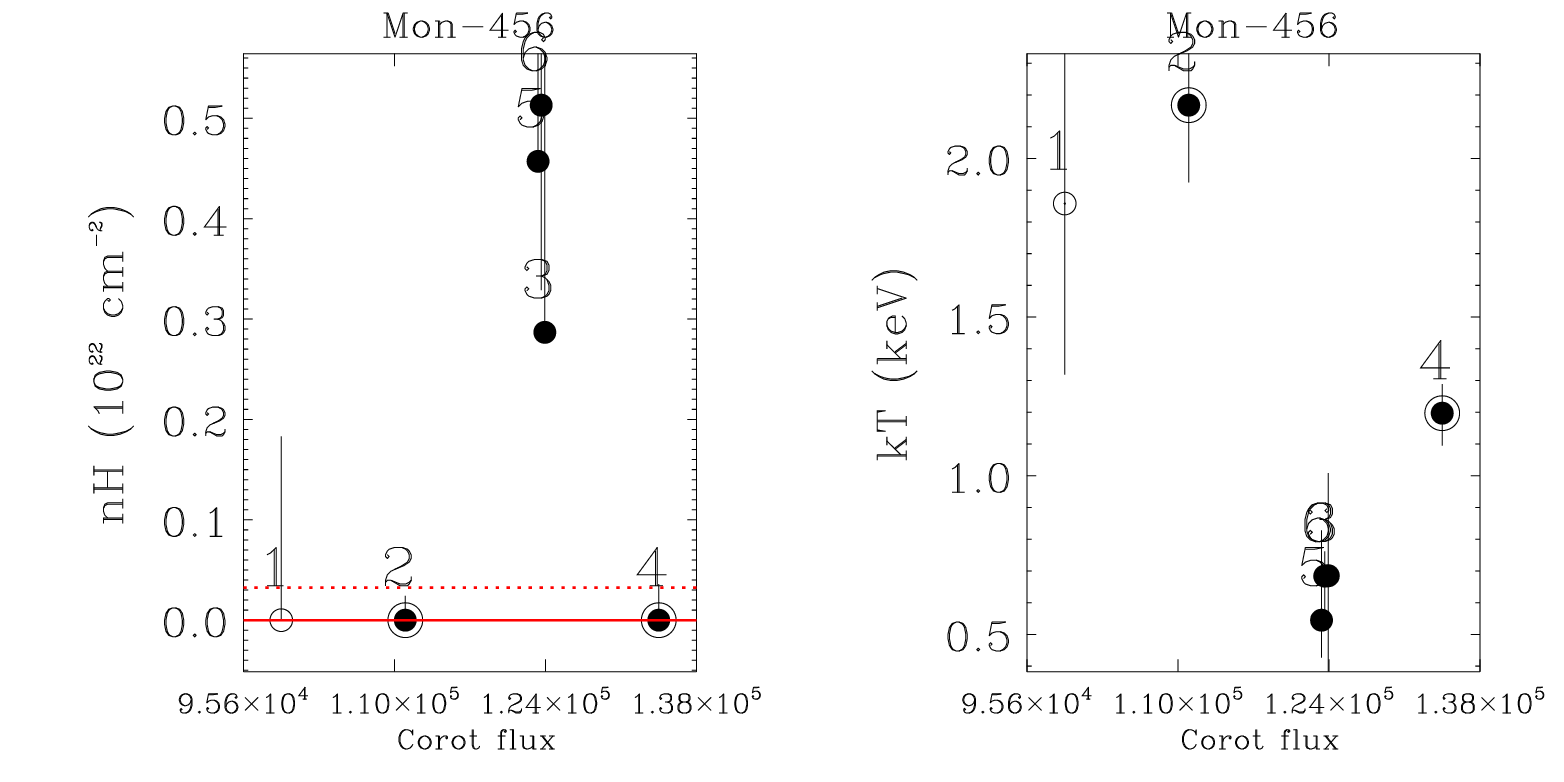}
	\includegraphics[width=7cm]{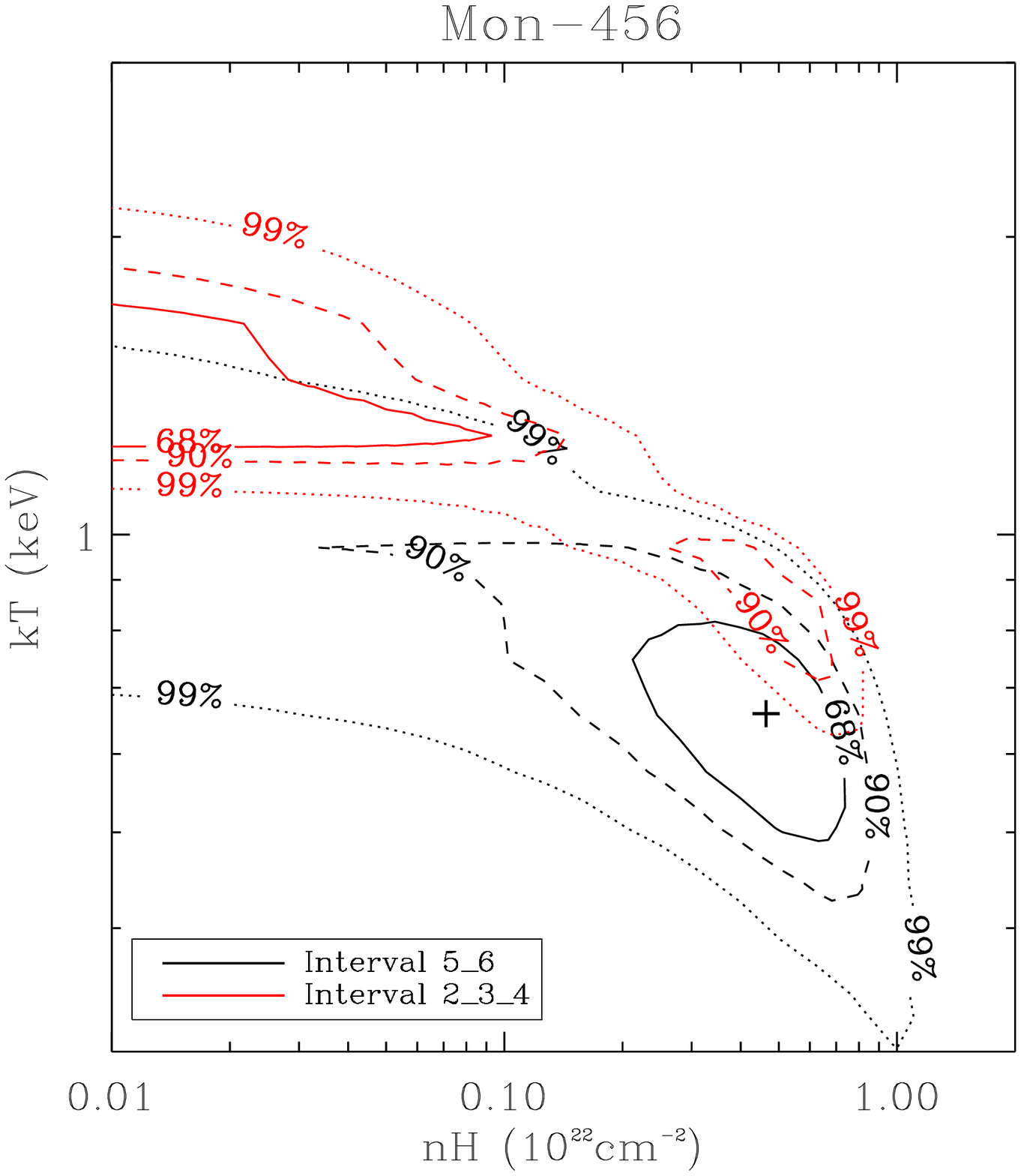}
	\caption{Optical and X-ray variability of Mon-456. The upper
        left panel shows part of the CoRoT light curve. The {\em Chandra}
frames are the unshaded intervals (the shaded intervals are
those not observed with {\em Chandra}). The dotted vertical lines
delimit the time intervals we defined; the black dots mark the
observed X-ray photon fluxes in the broad band (circled if, during the
given time interval, the source flared in X-rays). The numbers above the
light curve indicate the Chandra Obs.ID of the given {\em Chandra}
frame (top number), the number of detected X-ray photons (middle
number) and the label (bottom number) of the given time interval. The
upper right panels show the variability of selected X-ray properties (in
this case the hydrogen column density N$_H$ in units of
$10^{22}\,$cm$^{-2}$; and the plasma temperature kT, in keV) vs.\ the
median CoRoT fluxes observed in the time intervals indicated by the
labels. Empty circles mark intervals with not acceptable X-ray spectral fit.
The bottom panel shows contour levels in the C-stat space enclosing solutions within
68\% (solid contour), 90\% (dashed contour), and 99\% (dotted contour)
statistical confidence from the X-ray spectral fit of Mon-456 during
the intervals \#2+\#3+\#4 (red contours) and \#5+\#6 (black contours). 
The cross indicate the values obtained
from the best fit in the latter case.}
	\label{variab_mon456}
	\end{figure*}

	\begin{figure*}[]
	\centering	
	\includegraphics[width=10cm]{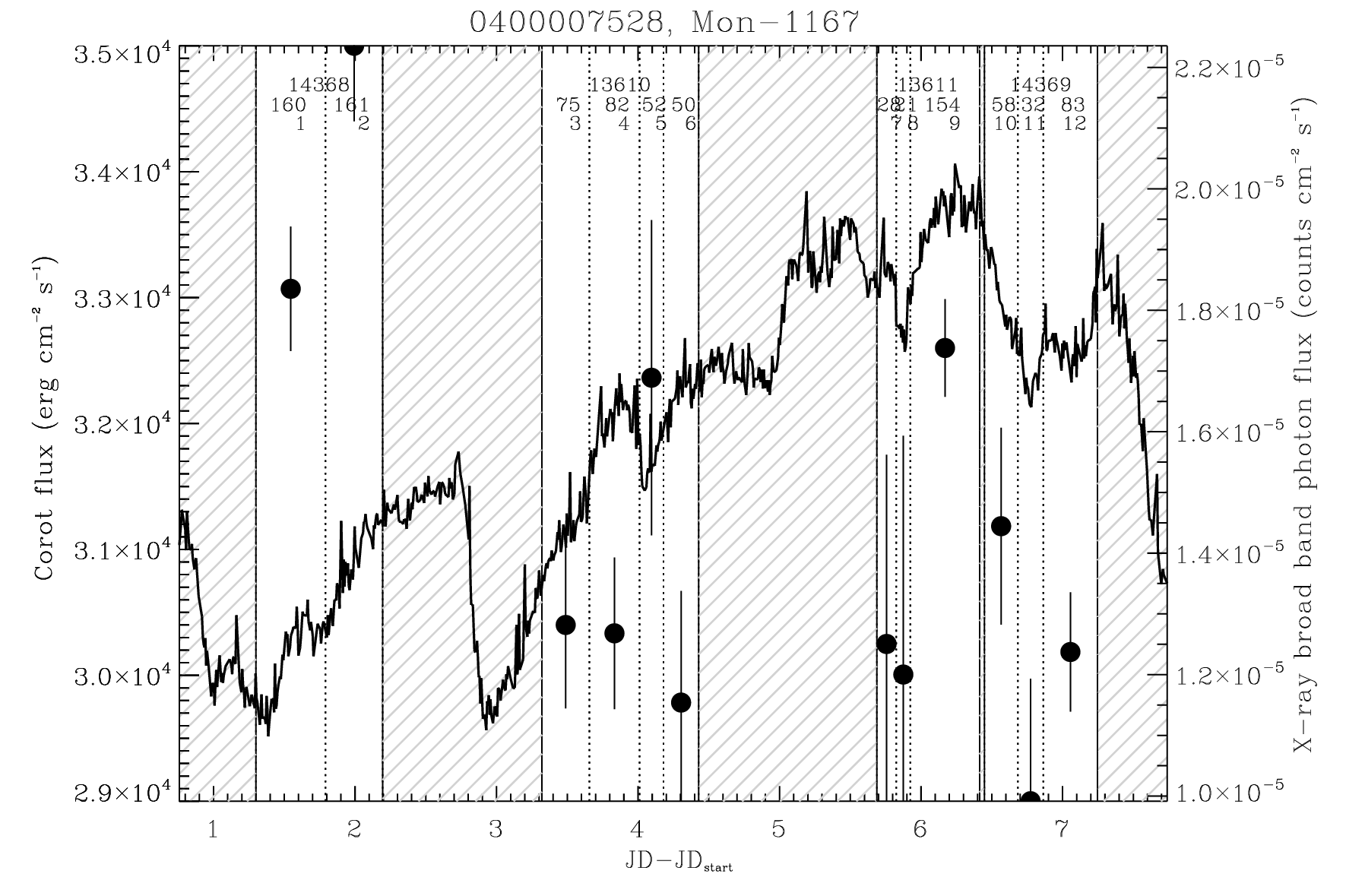}
	\includegraphics[width=8cm]{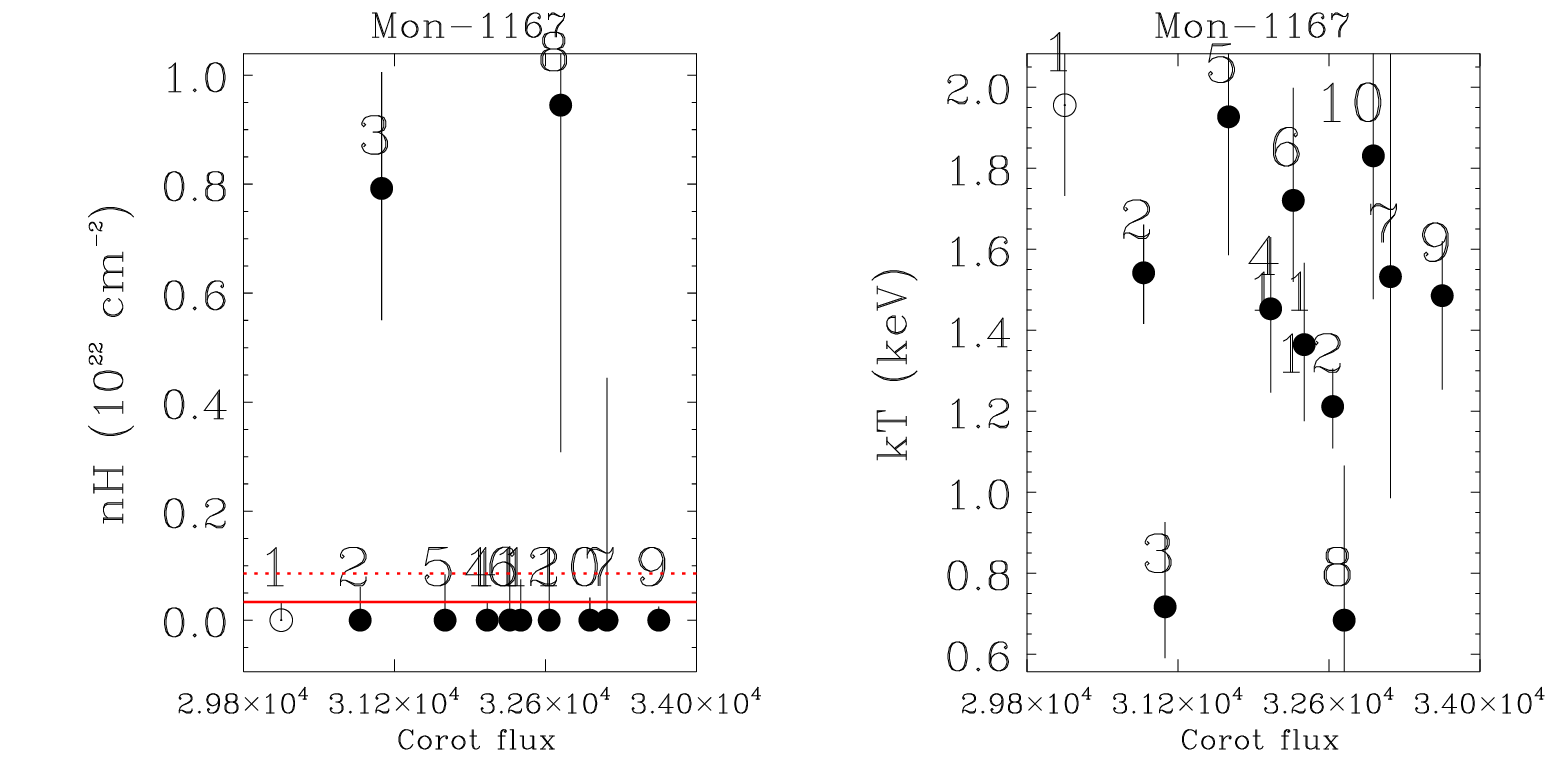}
	\includegraphics[width=7cm]{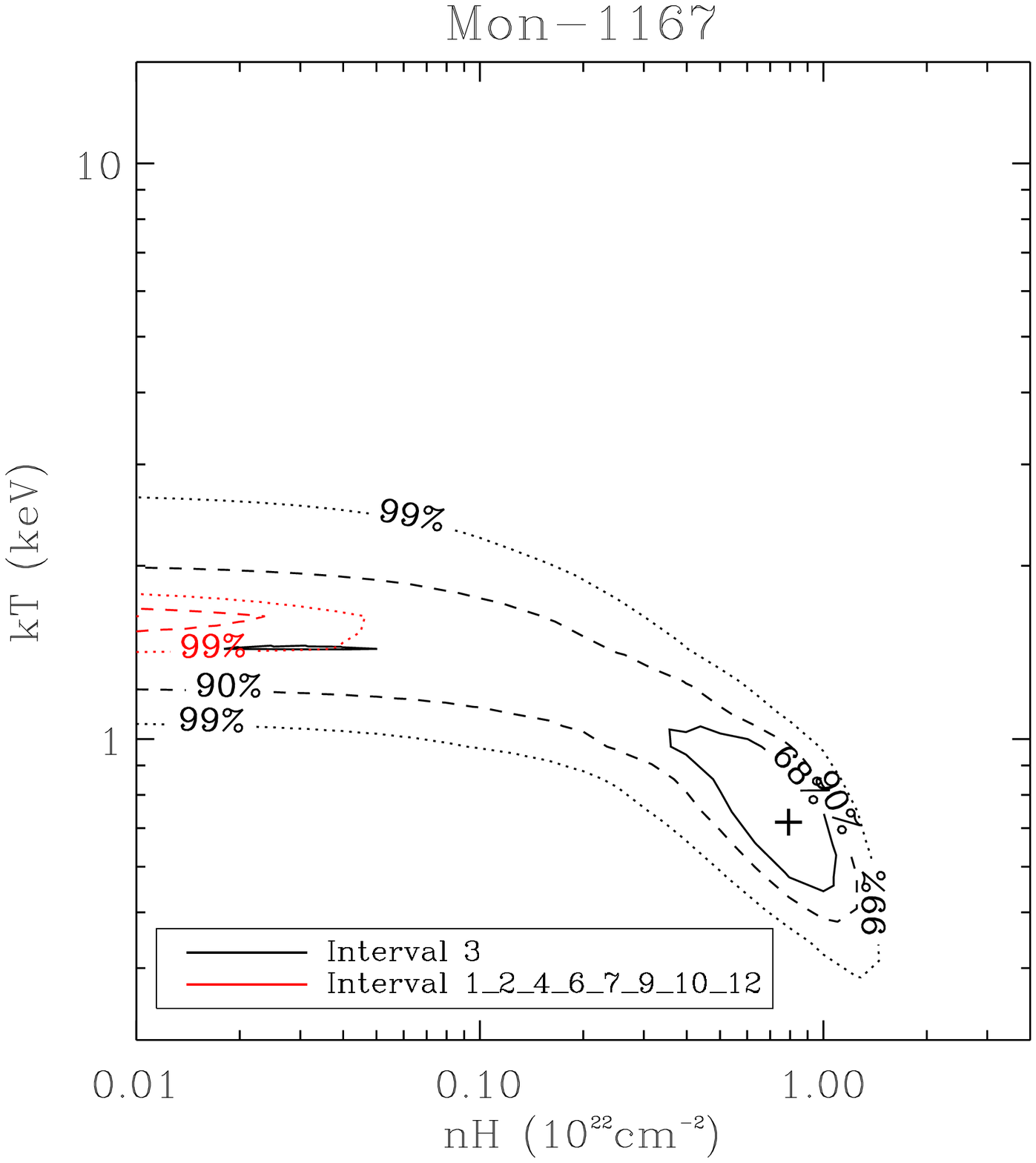}
	\includegraphics[width=7cm]{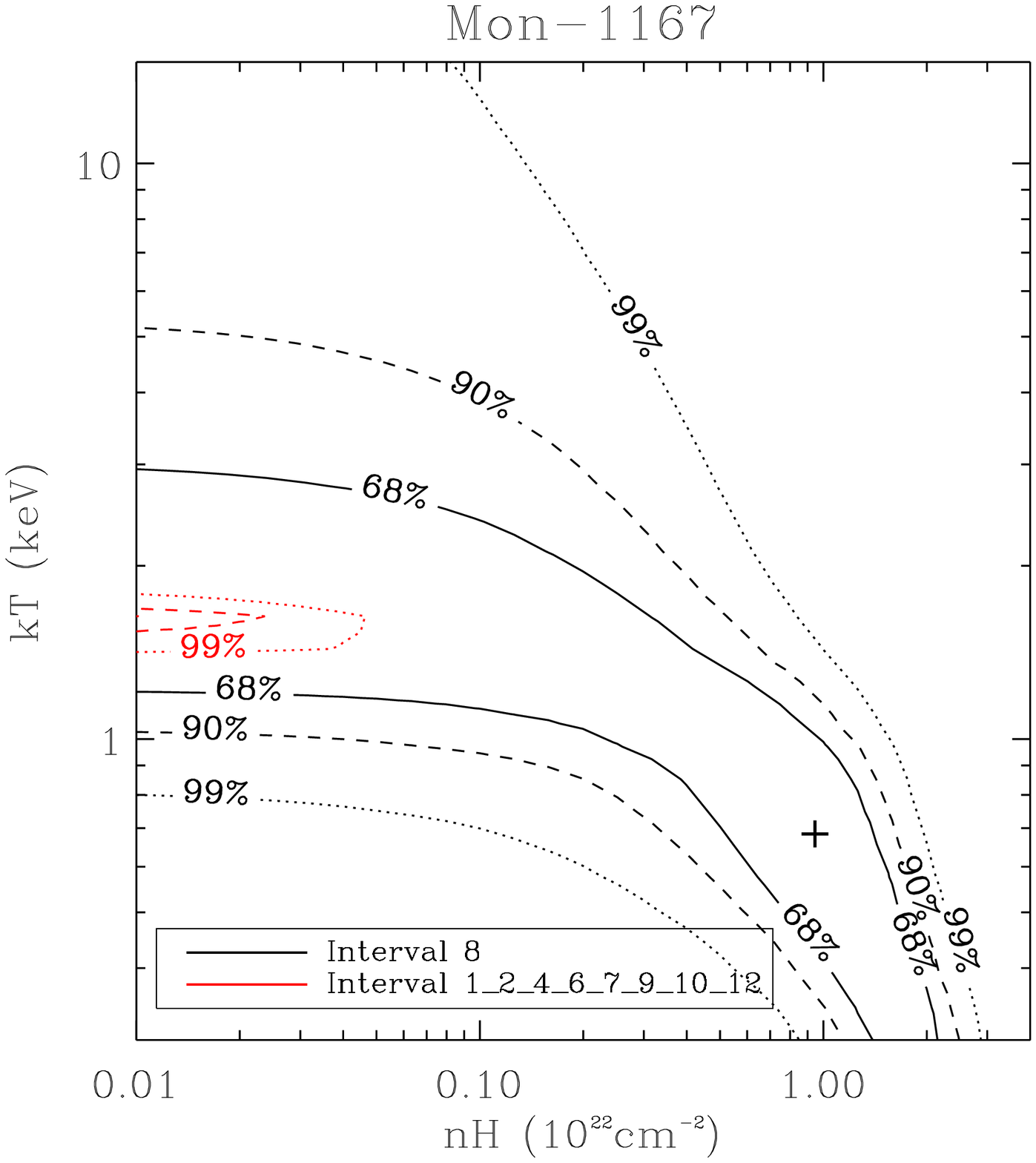}
        \caption{Optical and X-ray variability of Mon-1167. Panel
        format and content are as in Fig.~\ref{variab_mon456}. The
contours are shown for the X-ray spectral fit of the spectrum observed
during the intervals \#3 (black contours, bottom left panel) and \#8
(black contours, bottom right panel). In both panels, the contours from
the X-ray spectral fit in the intervals
\#1+\#2+\#4+\#6+\#7+\#9+\#10+\#12 are shown in red.}
	\label{variab_mon1167}
	\end{figure*}

	\begin{figure*}[]
	\centering	
	\includegraphics[width=10.0cm]{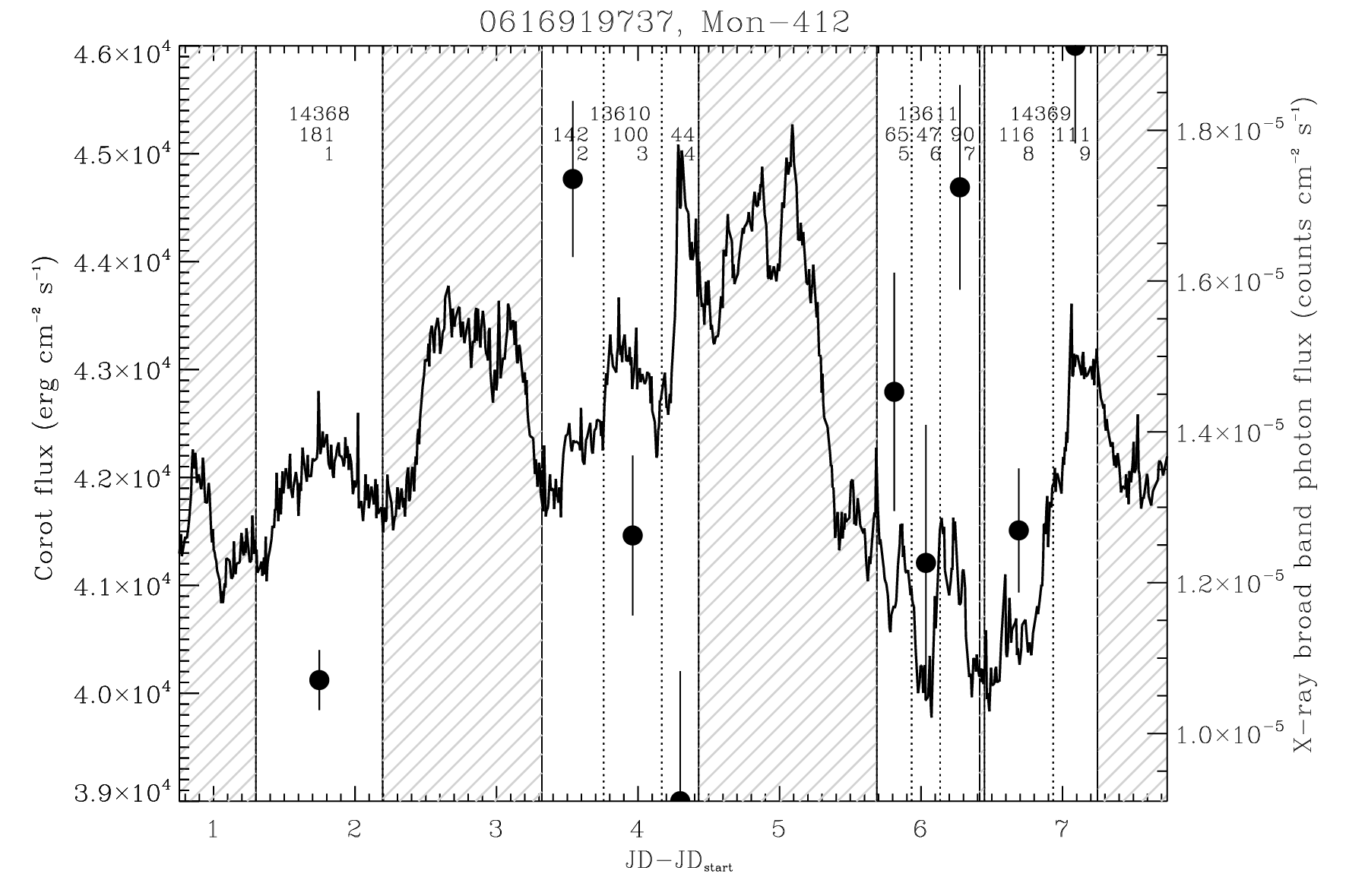}
	\includegraphics[width=8.0cm]{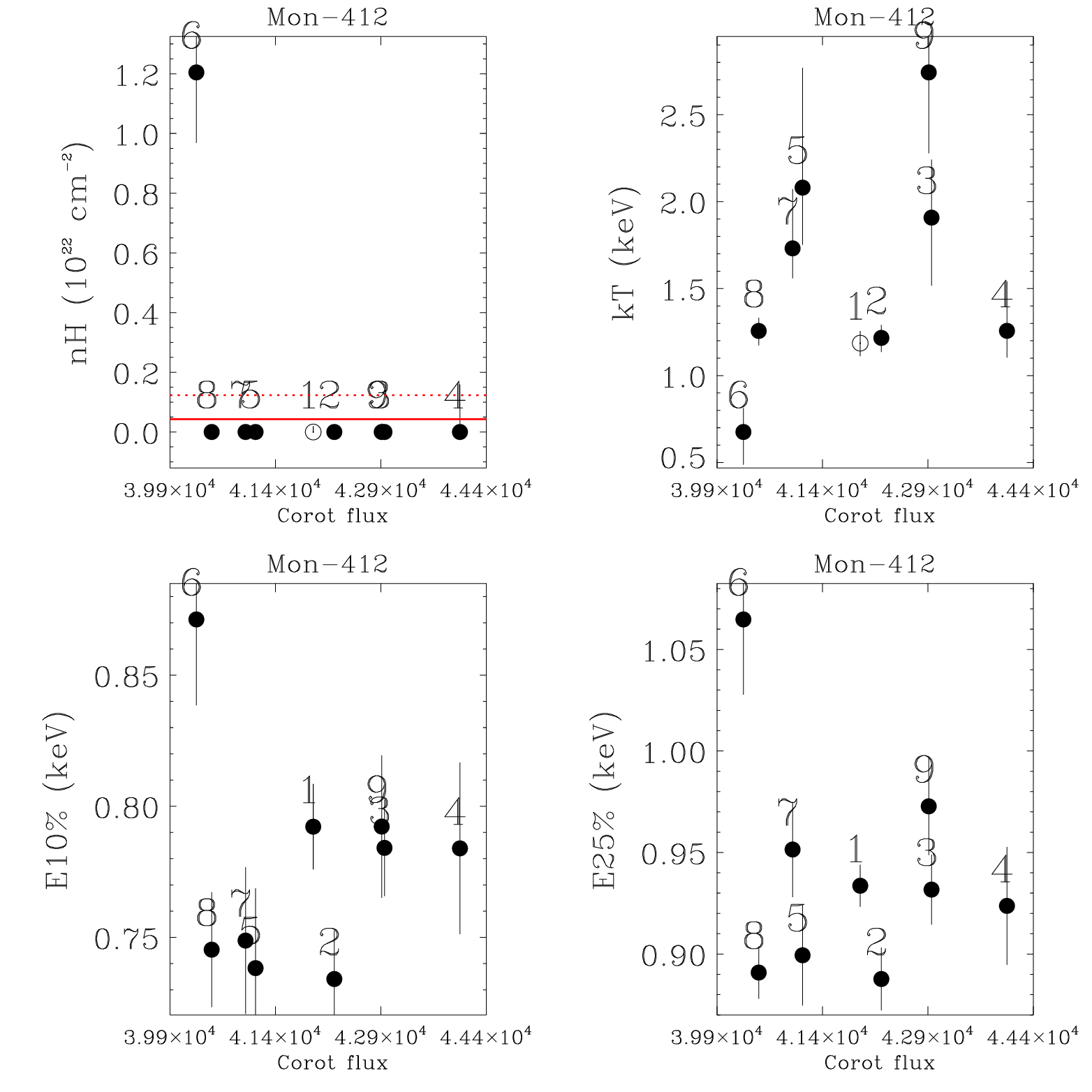}	
	\includegraphics[width=7cm]{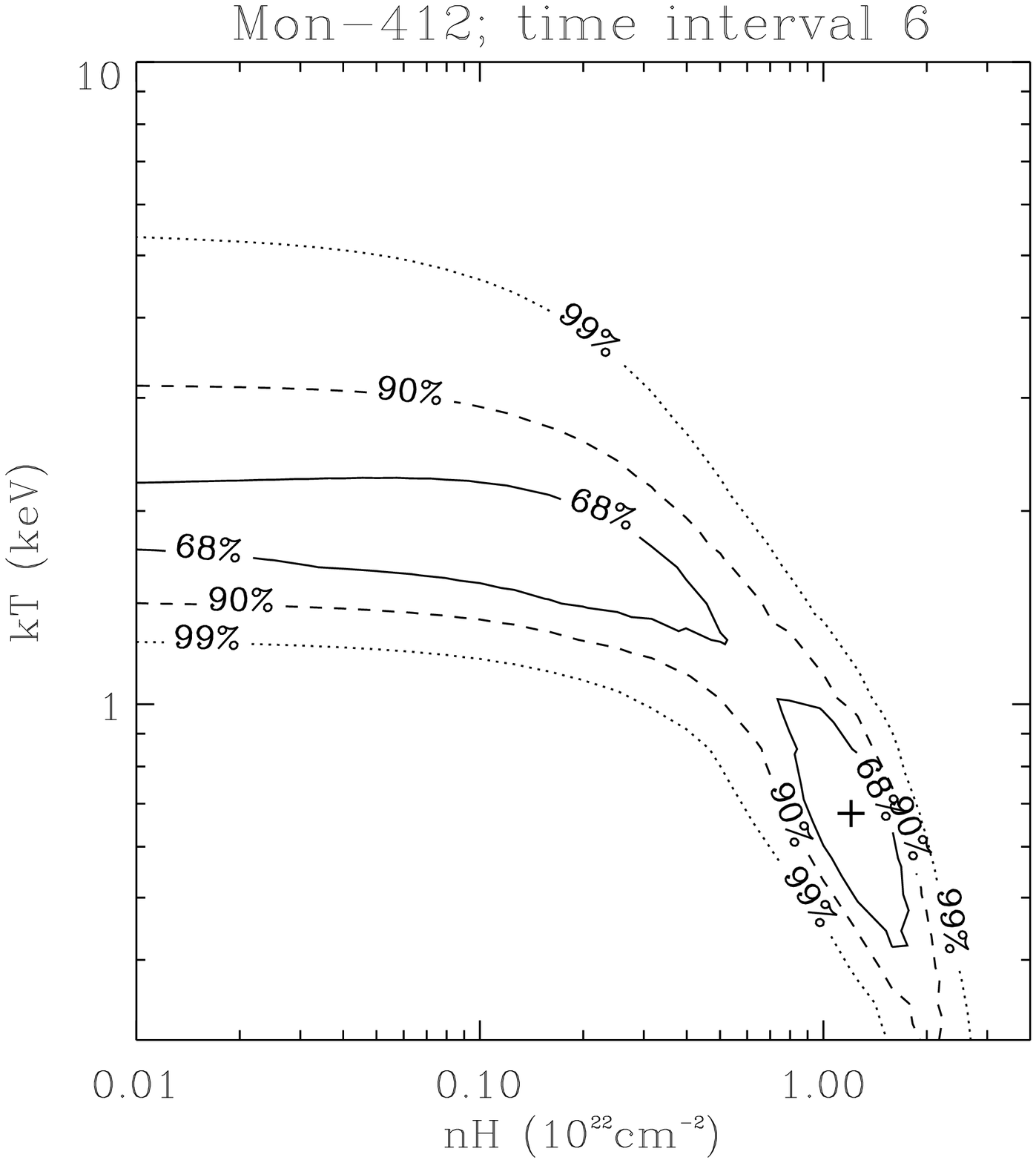}
        \caption{Optical and X-ray variability of Mon-412. Panel
        format and content generally follows Fig.~\ref{variab_mon456}. 
        In this case, are shown the hydrogen column density N$_H$ in units of
$10^{22}\,$cm$^{-2}$, the plasma temperature kT and the 10\% and 25\%
photon energy quantiles in keV, and contour levels in the C-stat
space. The contours are from the X-ray spectral fit
of the spectrum observed during the interval \#6.}
	\label{variab_mon412}
	\end{figure*}

	\begin{figure*}[]
	\centering	
	\includegraphics[width=10cm]{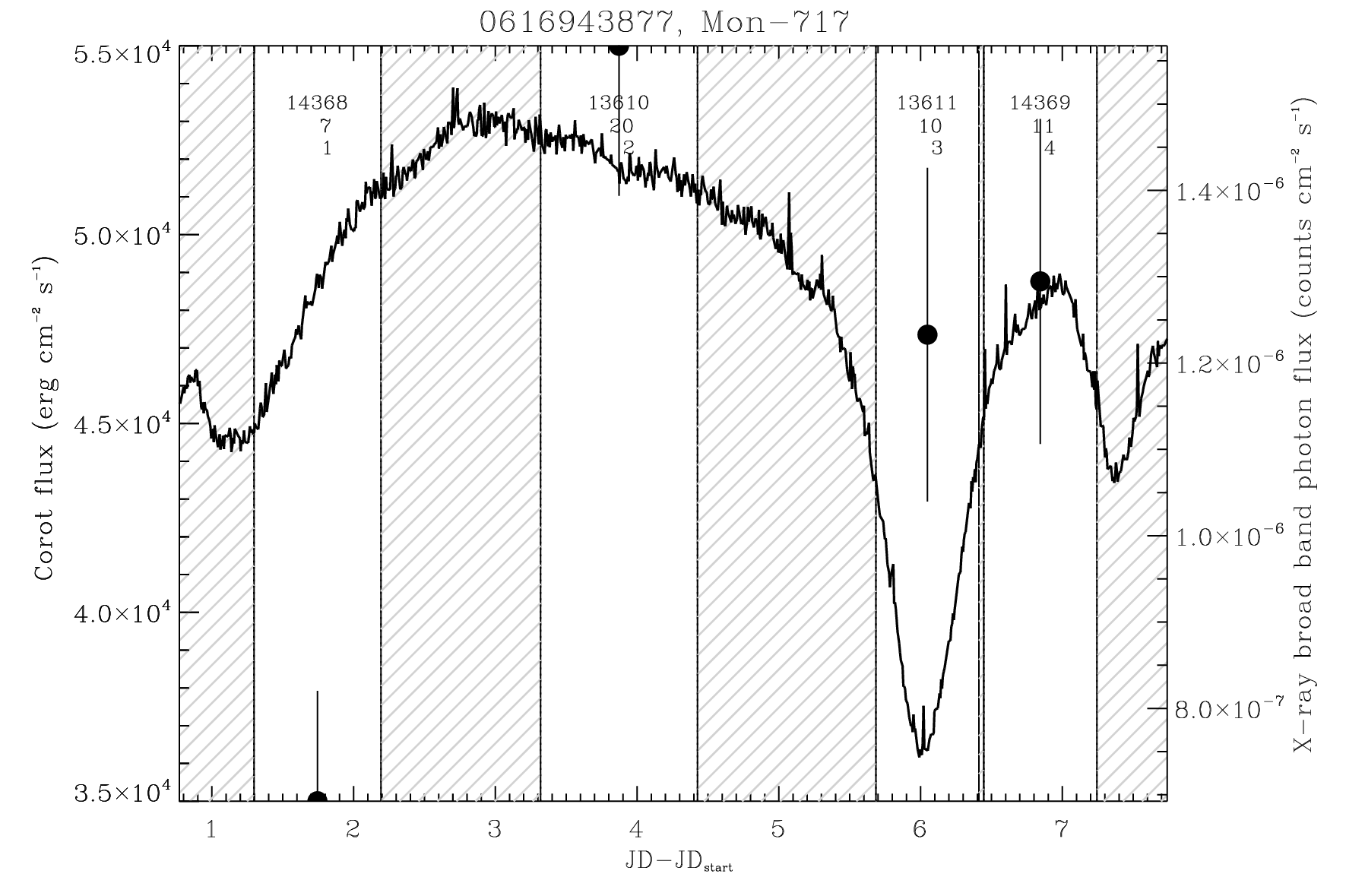}
	\includegraphics[width=8cm]{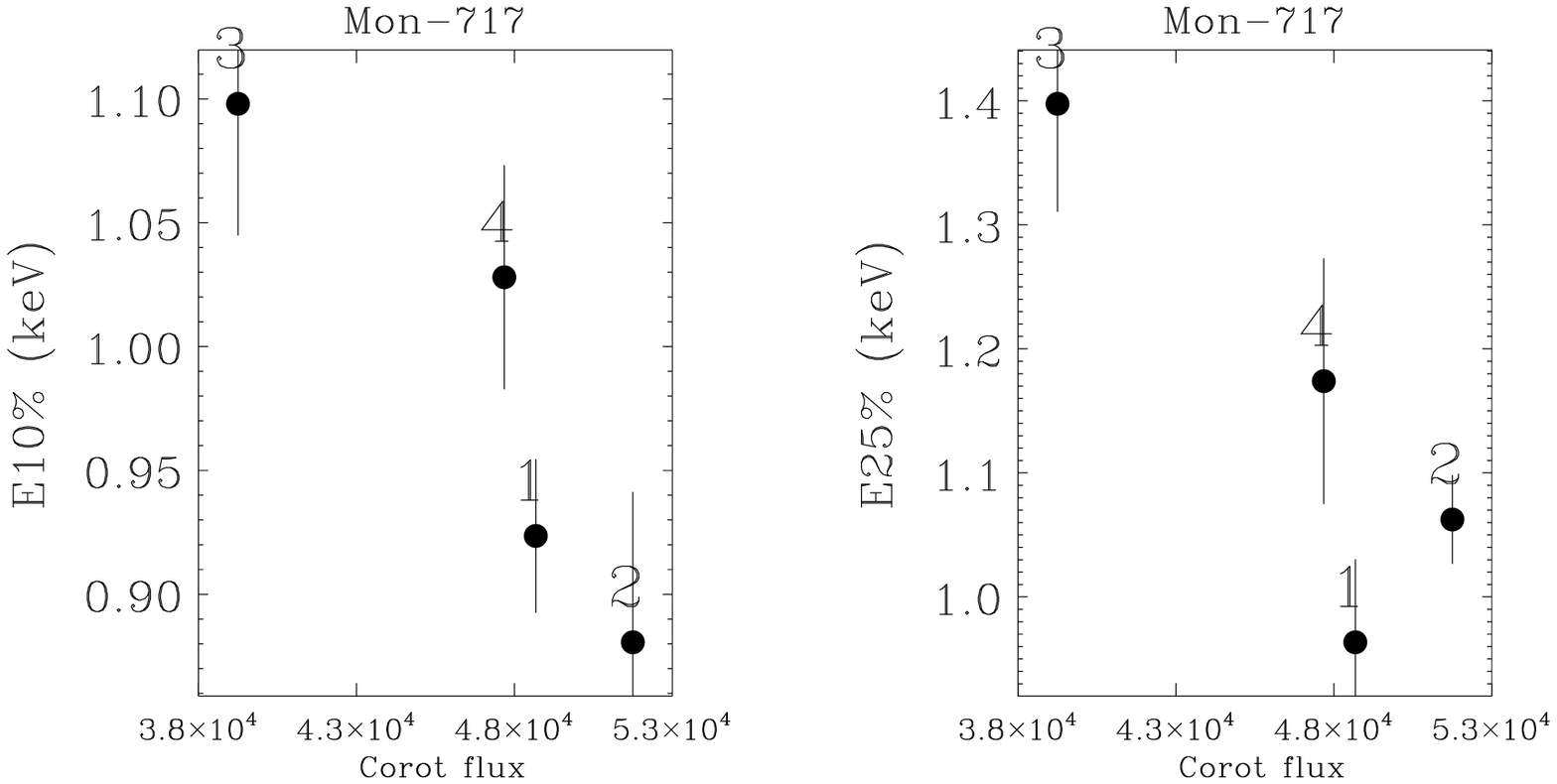}	
        \caption{Optical and X-ray variability of Mon-717.  Panel
        format and content generally follows Fig.~\ref{variab_mon456}.
        In this case, the 10\% and 25\% photon energy quantiles in
keV are shown.}
	\label{variab_mon717}
	\end{figure*}

	\begin{figure*}[]
	\centering	
	\includegraphics[width=10cm]{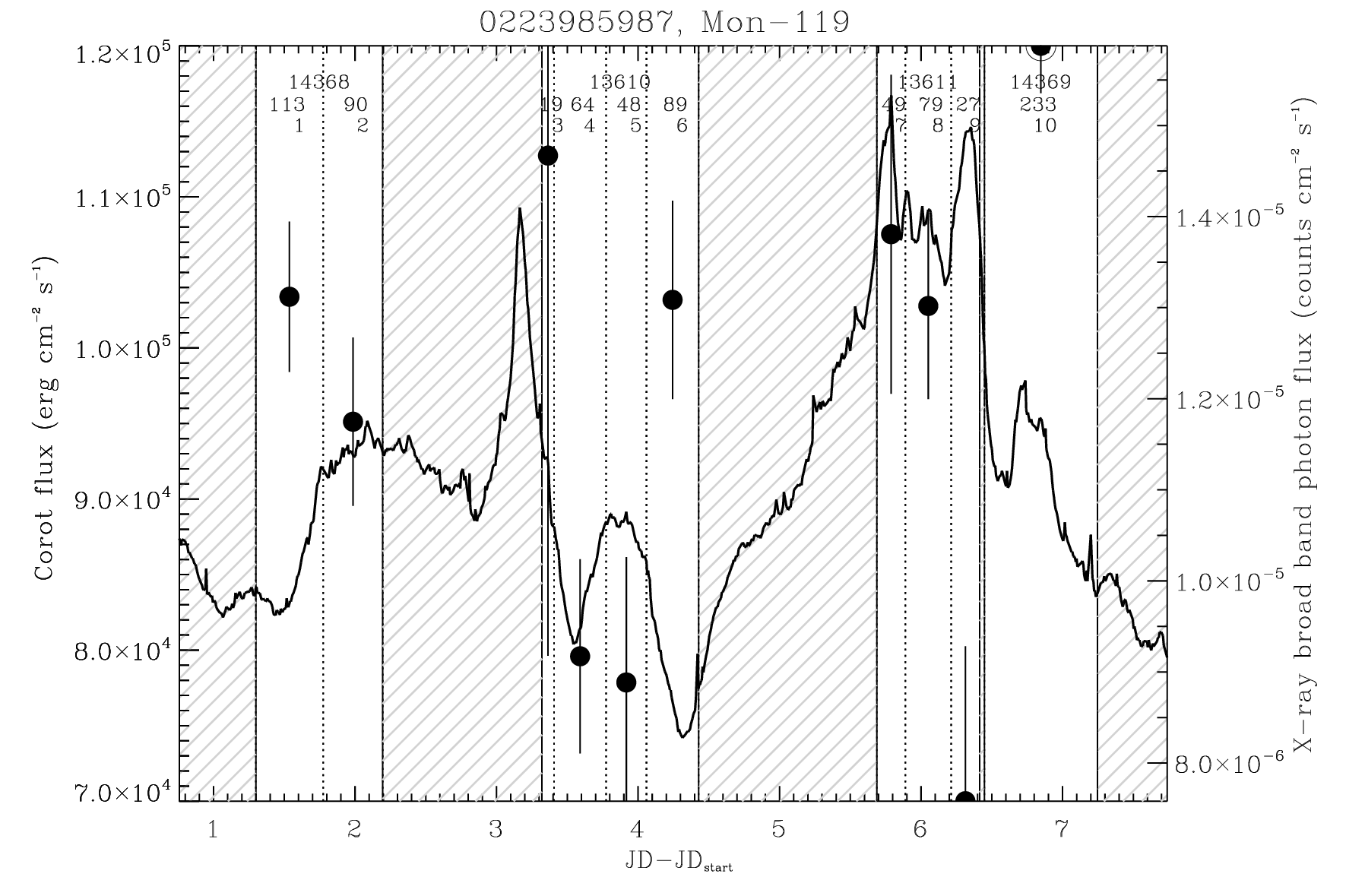} 
	\includegraphics[width=8cm]{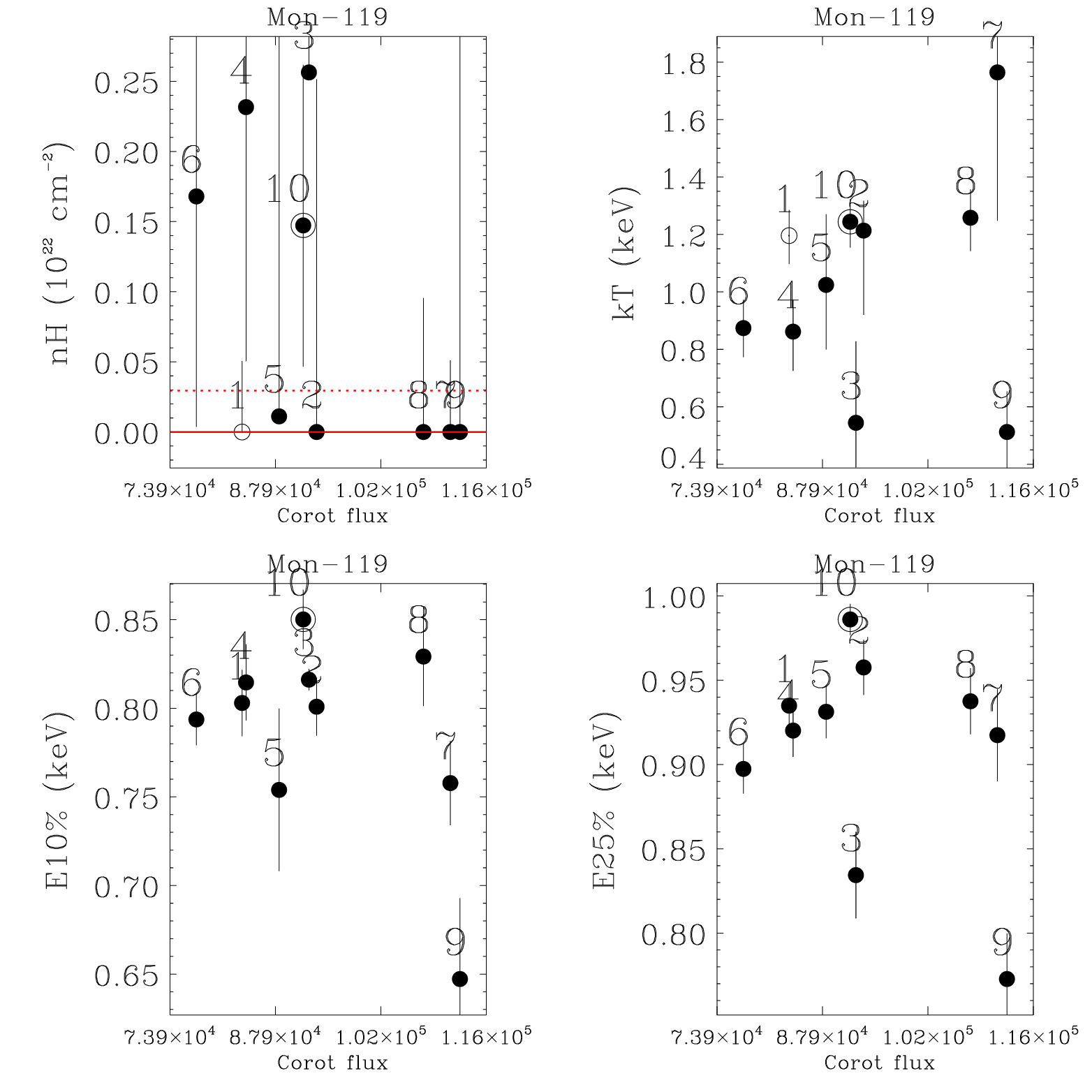}	
	\includegraphics[width=7cm]{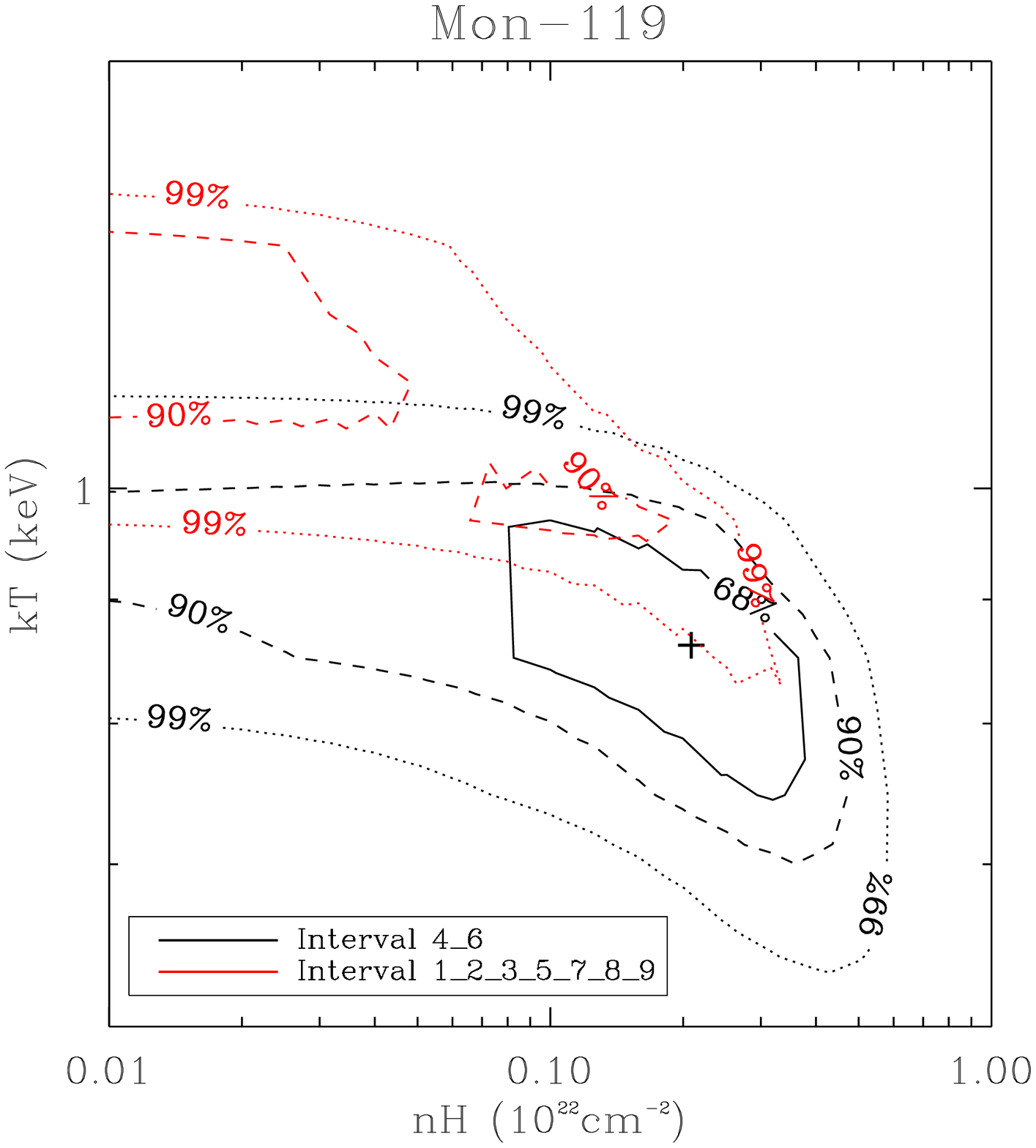}
        \caption{Optical and X-ray variability of Mon-119, with panel
        format and content generally following Fig.~\ref{variab_mon456}.
        The contours are from the X-ray spectral fit
of the spectrum observed during the interval \#4+\#6 (black) and
\#1+\#2+\#3+\#5+\#7+\#8+\#9 (red).}
	\label{variab_mon119}
	\end{figure*}

	\begin{figure*}[]
	\centering	
	\includegraphics[width=10cm]{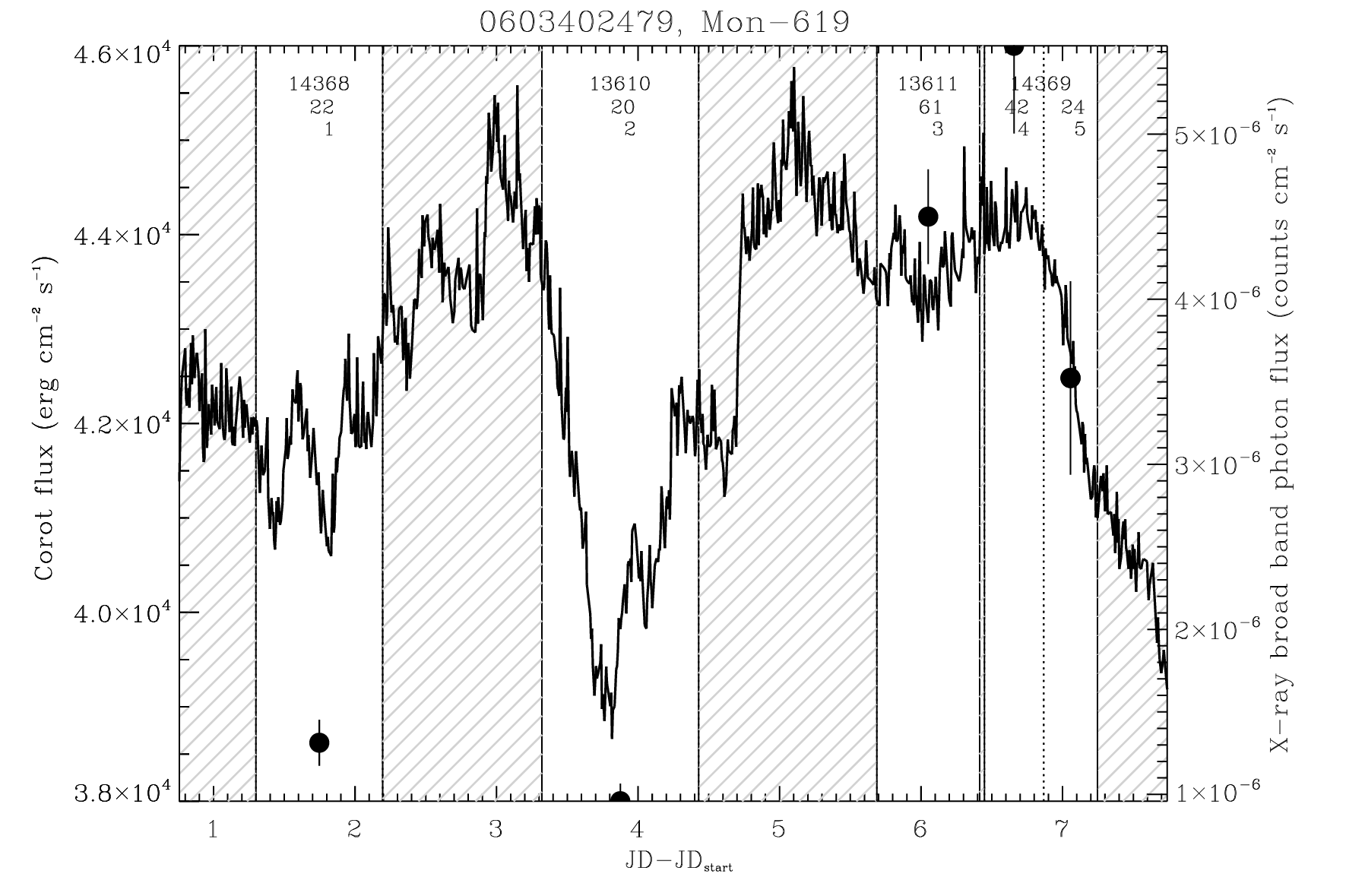}
	\includegraphics[width=8cm]{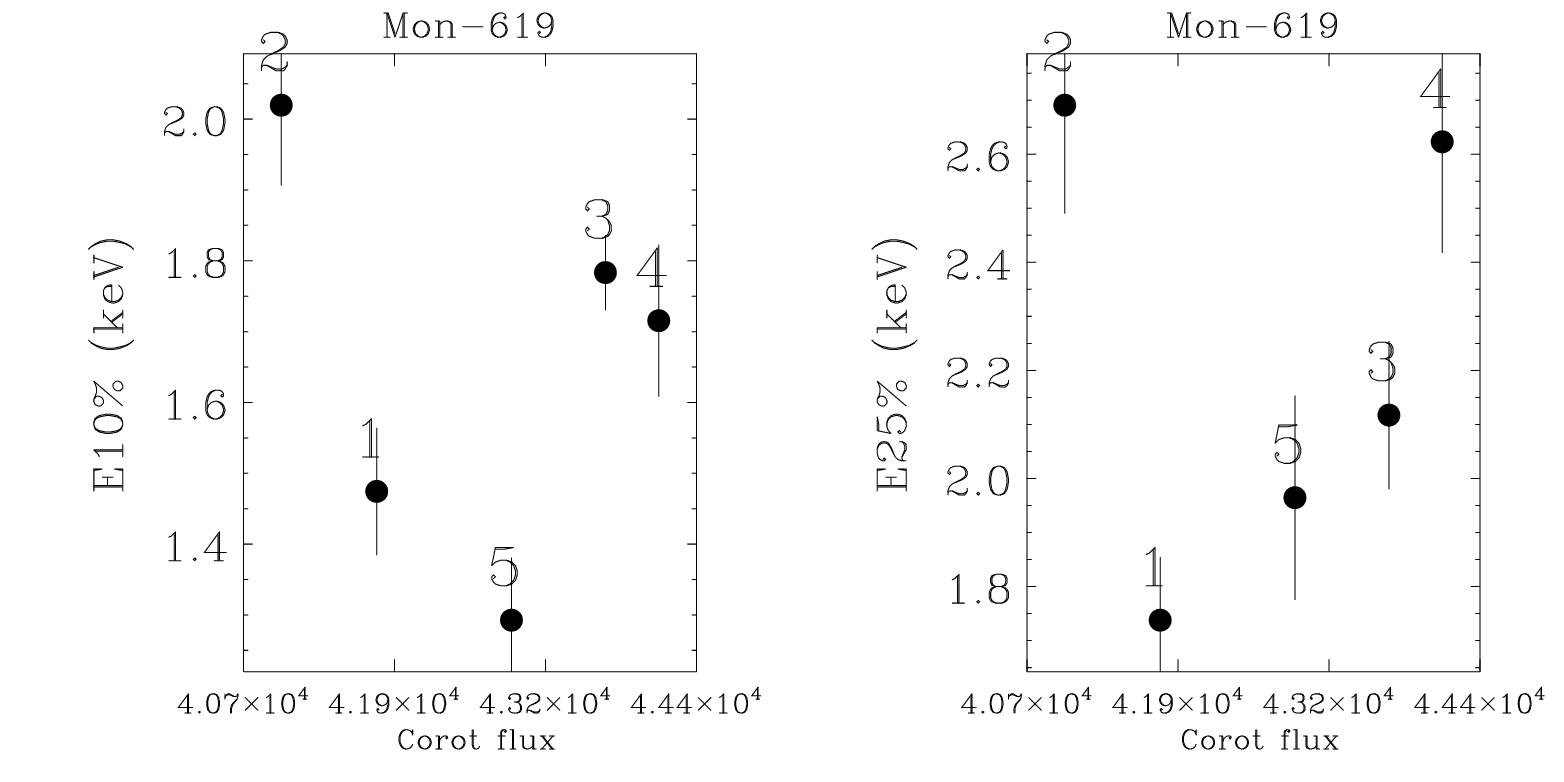}
        \caption{Optical and X-ray variability of Mon-619, with panel
        format and content generally following Fig.~\ref{variab_mon456}.}
	\label{variab_mon619}
	\end{figure*}

	\begin{figure*}[t]
	\centering	
	\includegraphics[width=9.5cm]{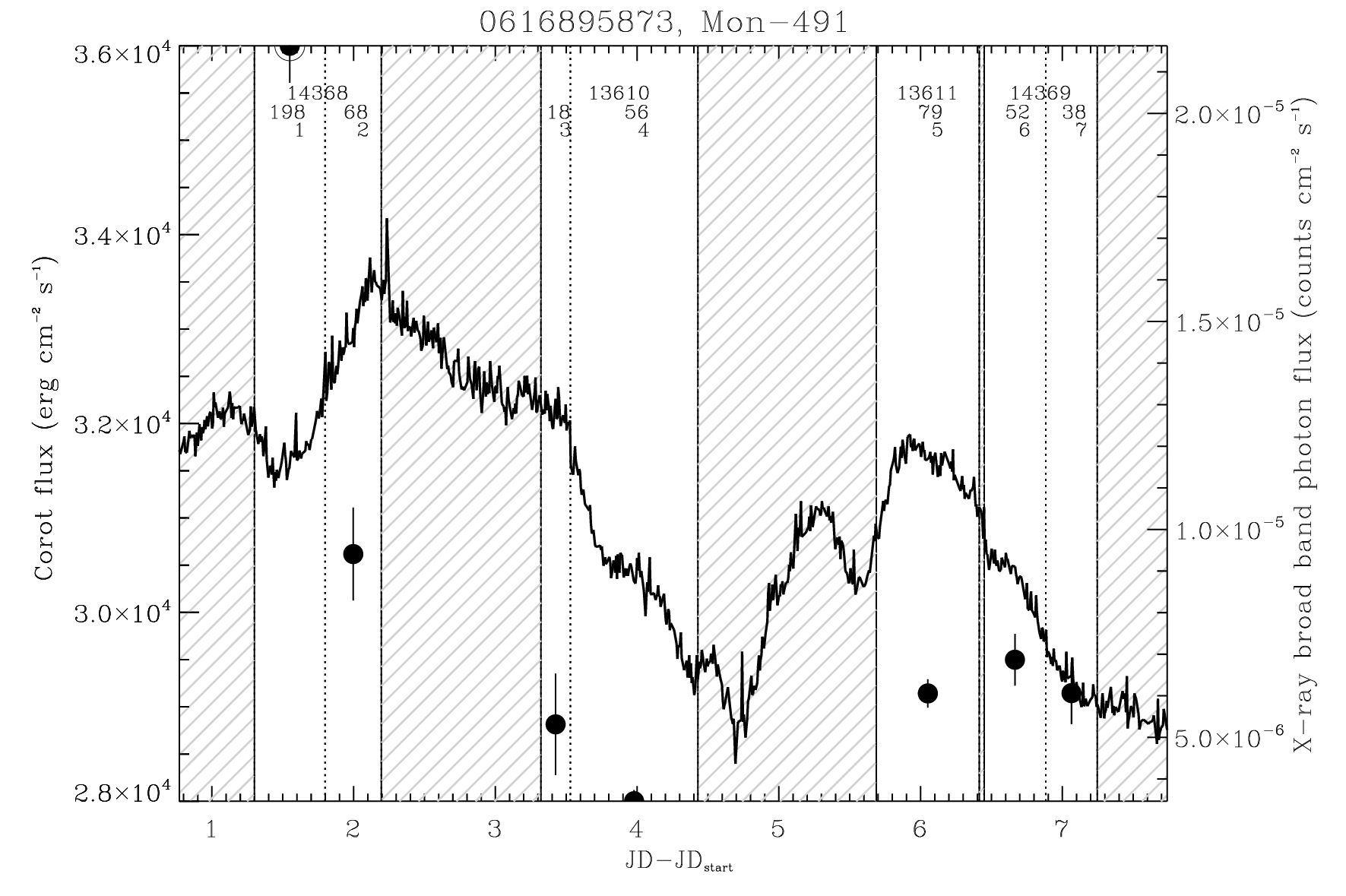}
	\includegraphics[width=7cm]{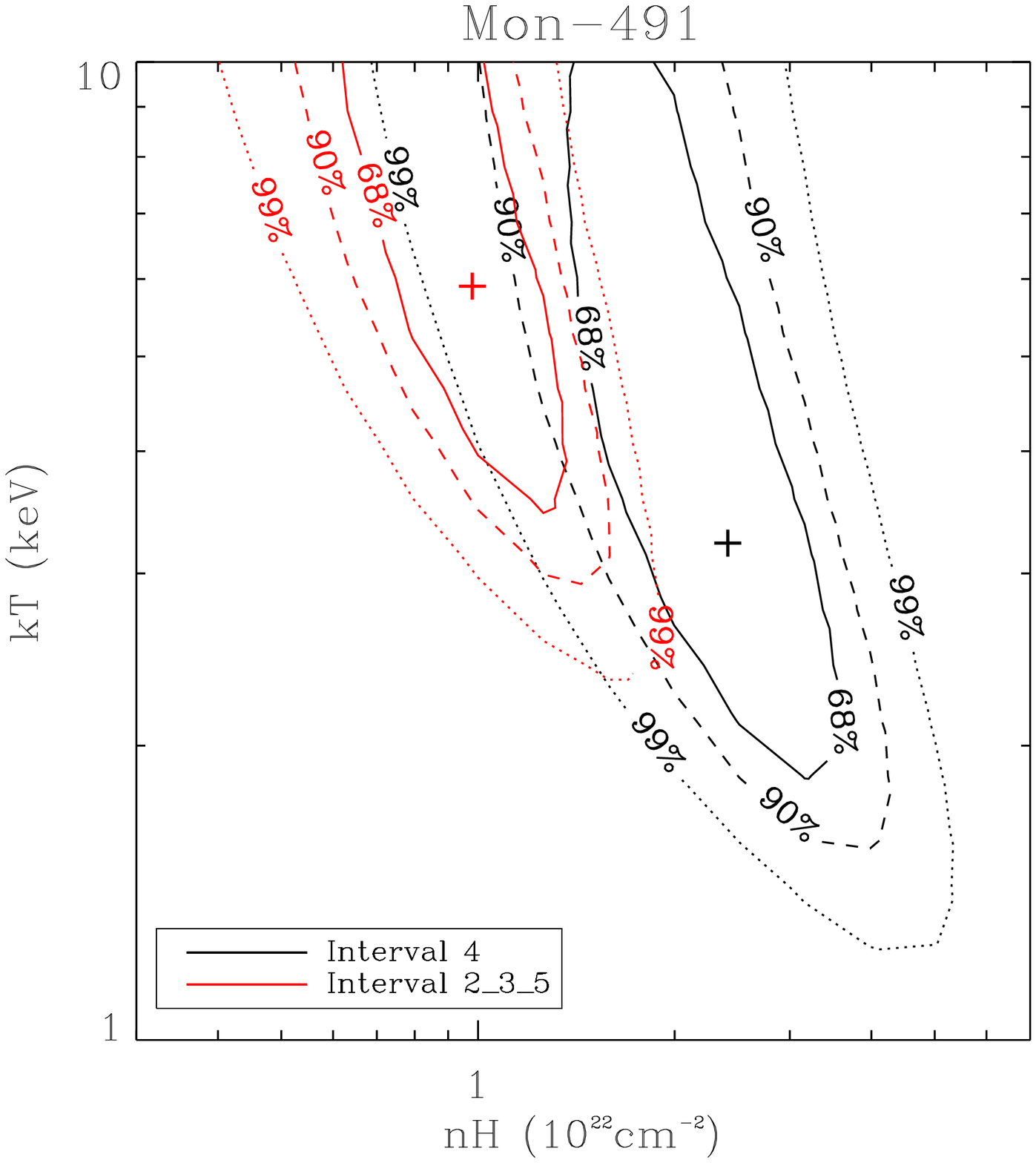}
        \caption{Optical and X-ray variability of Mon-491, with panel
        format and content generally following Fig.~\ref{variab_mon456}.
        The contours are from the
X-ray spectral fit of the spectrum observed during the interval \#4
(black) and \#2+\#3+\#5 (red).}
	\label{variab_mon491}
	\end{figure*}

	\begin{figure*}[]
	\centering	
	\includegraphics[width=9.5cm]{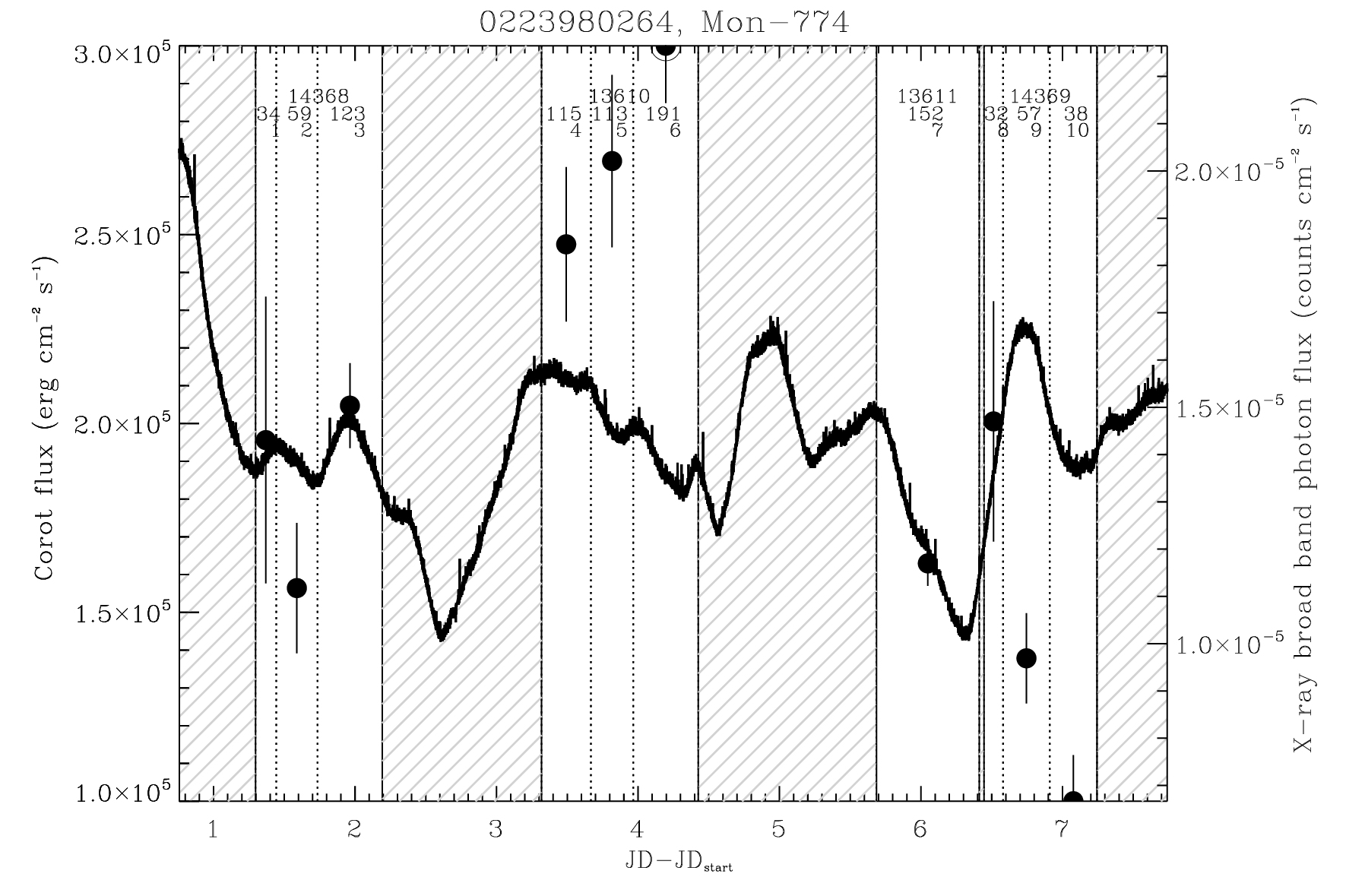}
	\includegraphics[width=8cm]{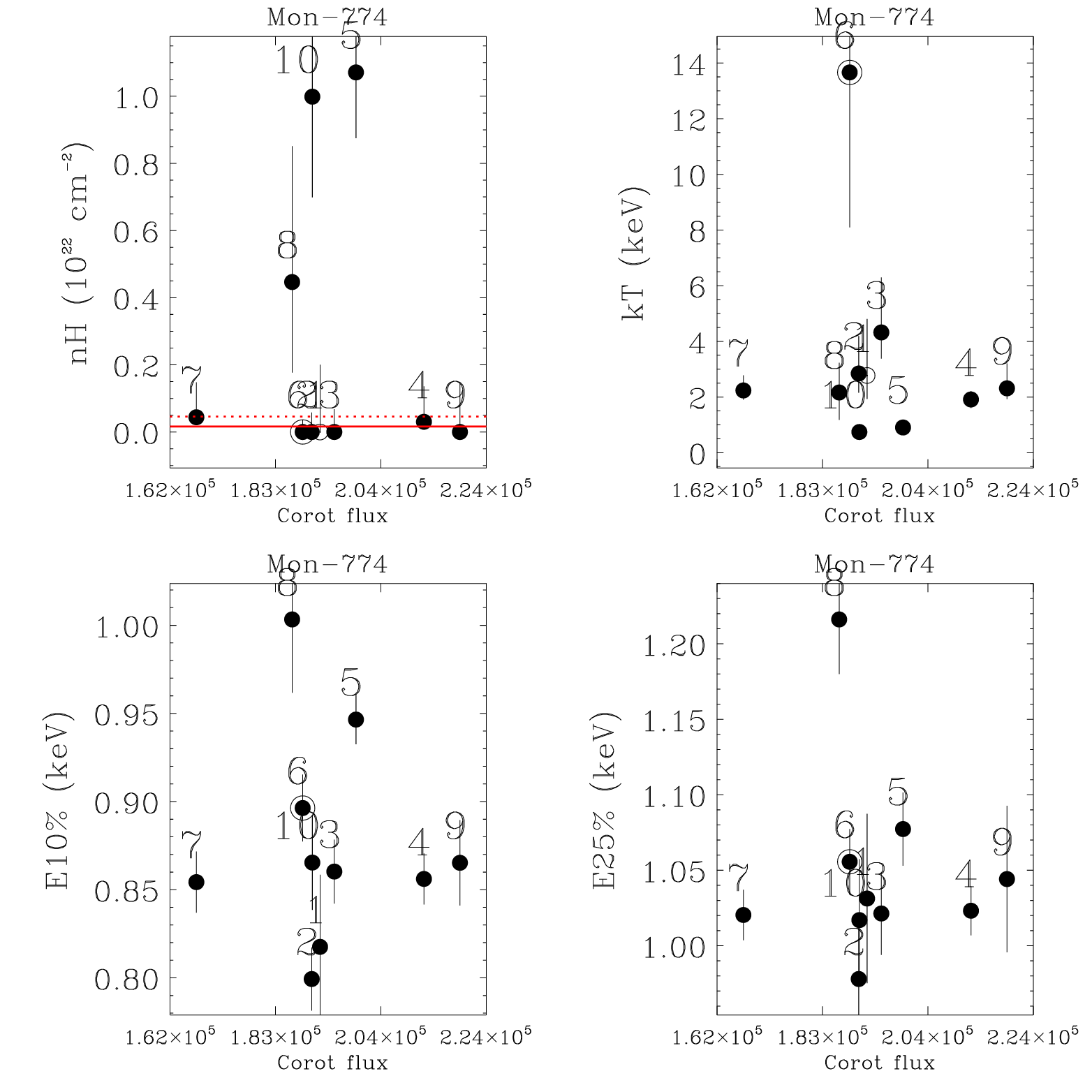}	
        \includegraphics[width=7cm]{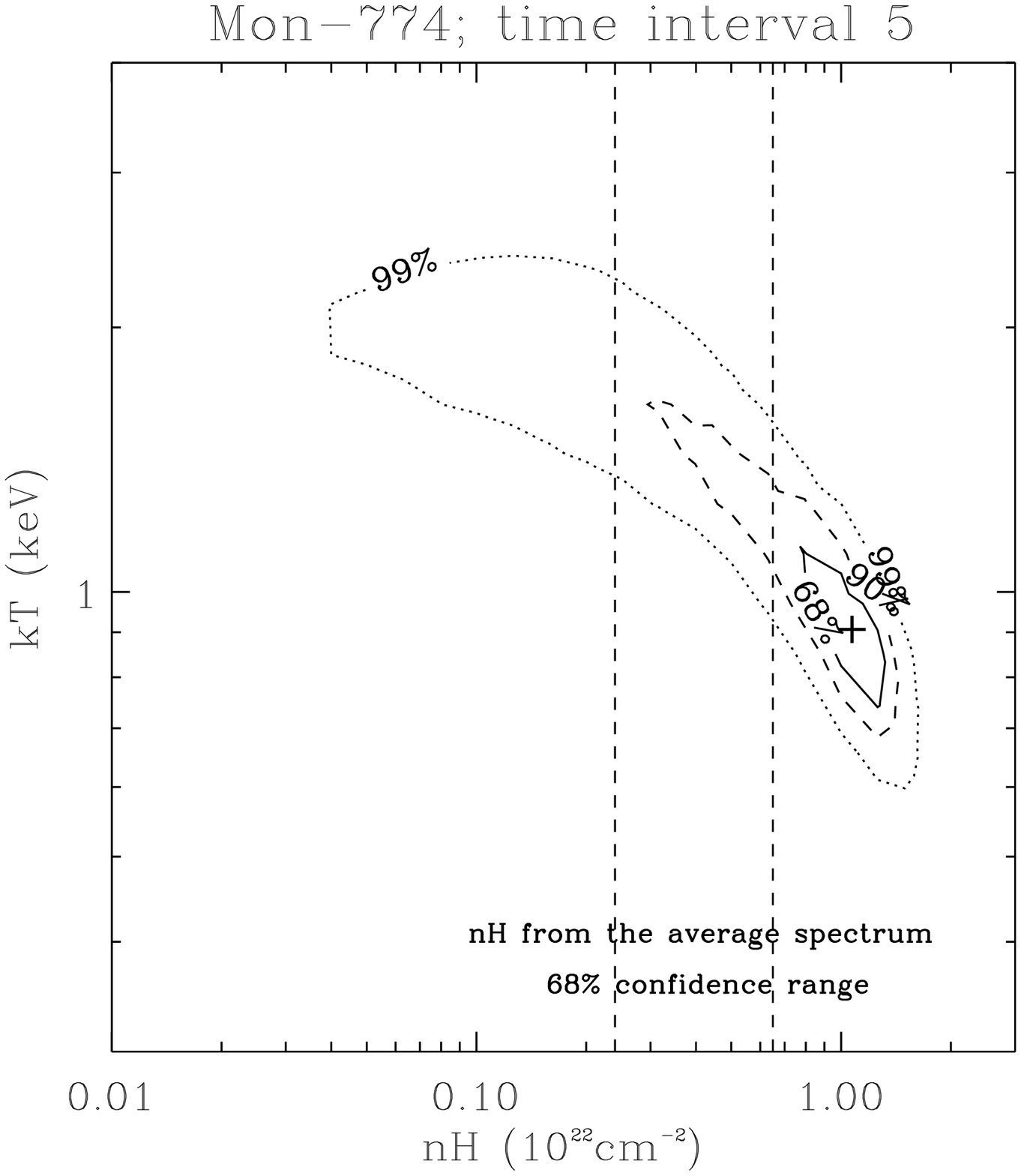}
	\includegraphics[width=7cm]{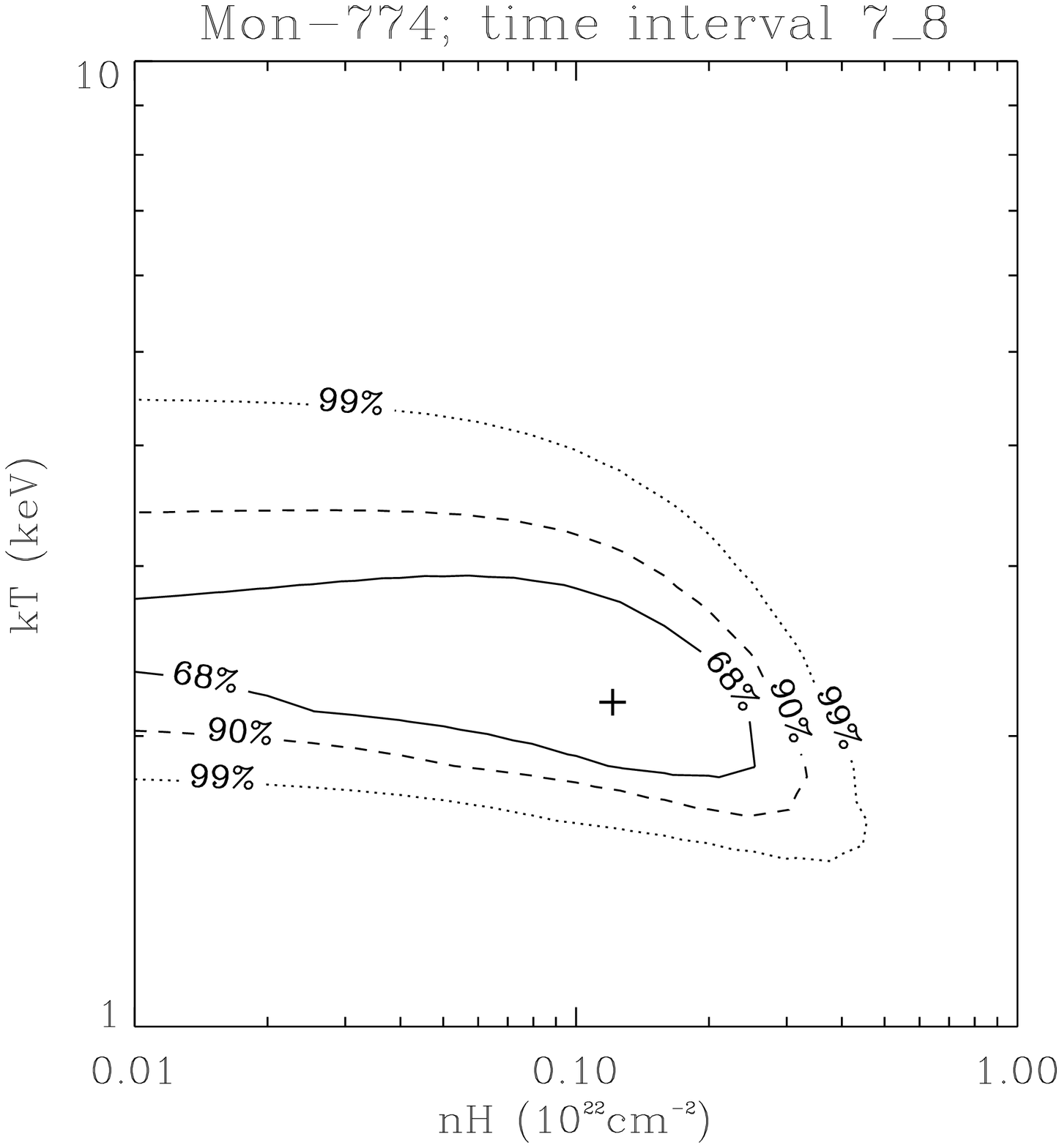}
        \caption{Optical and X-ray variability of Mon-774, with panel
        format and content generally following Fig.~\ref{variab_mon456}.
        The contours are from the X-ray spectral fit
of the spectrum observed during the interval \#5 (left bottom panel)
and \#7+\#8 (right bottom panel). The vertical dashed lines in the top
panel mark the 68\% confidence range for the N$_H$ obtained fitting
the average spectrum.}
	\label{variab_mon774}
	\end{figure*}

	\begin{figure*}[]
	\centering	
	\includegraphics[width=9.5cm]{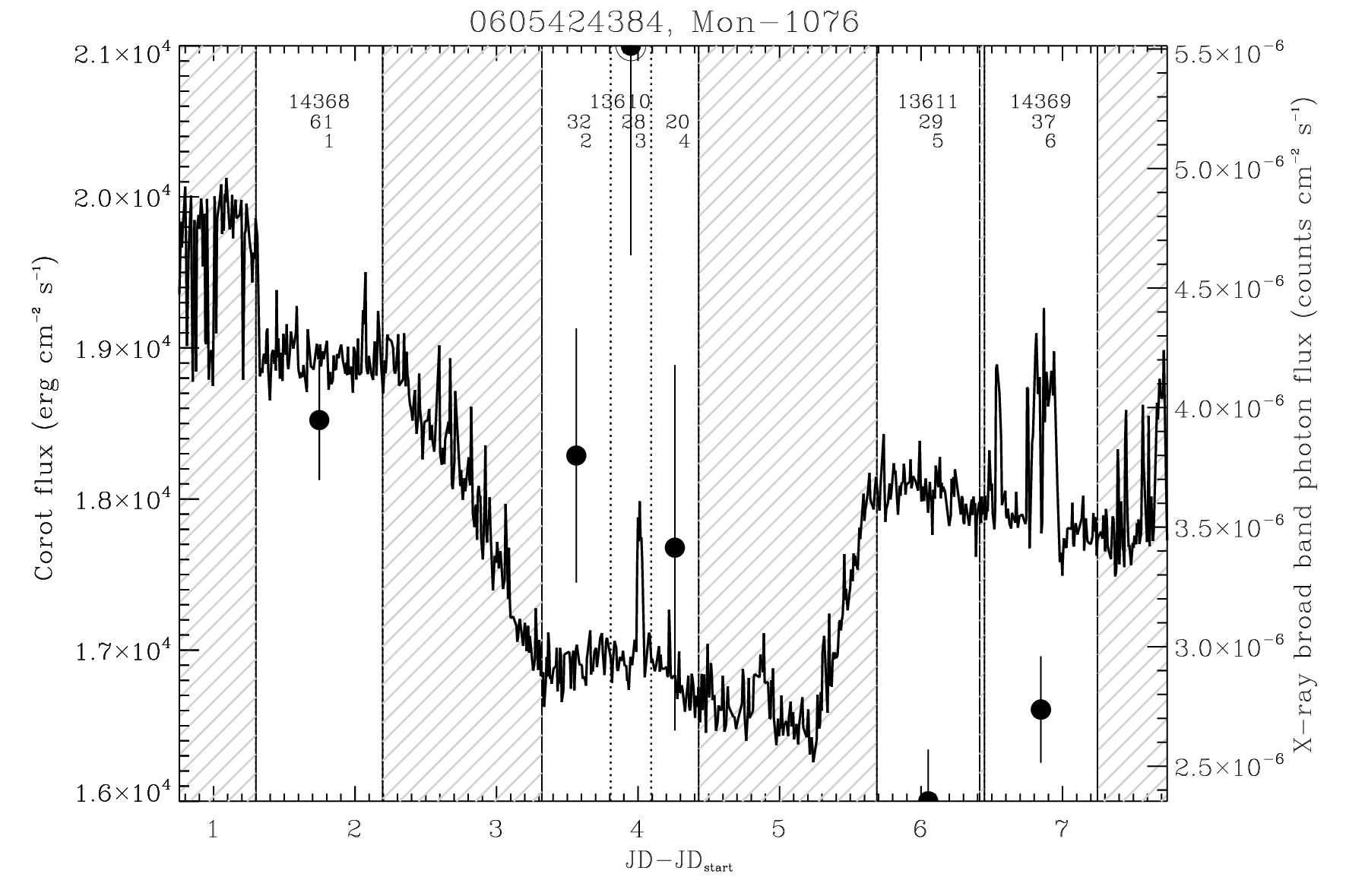}
	\includegraphics[width=8cm]{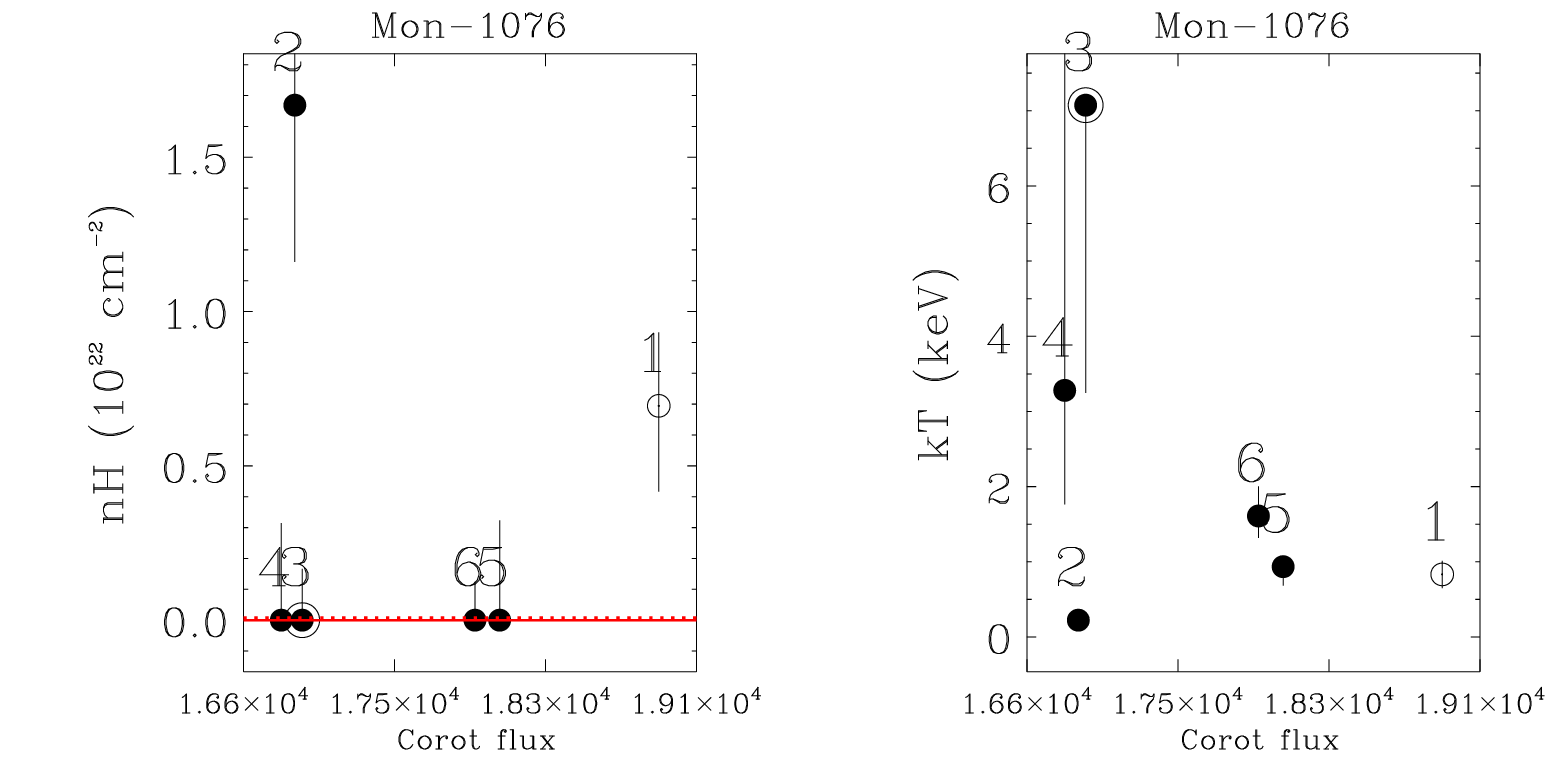}
	\includegraphics[width=7cm]{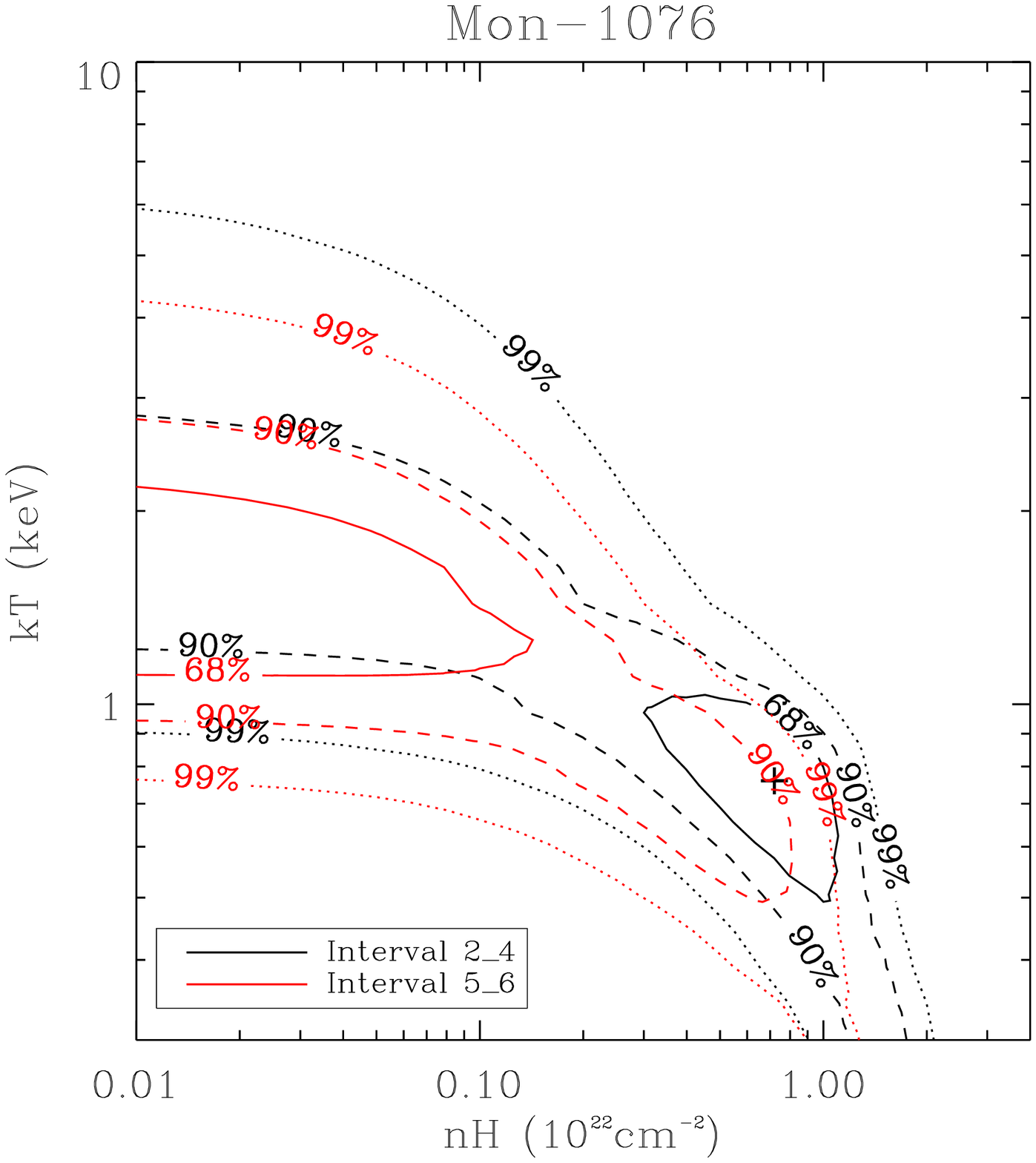}
        \caption{Optical and X-ray variability of Mon-1076, with panel
        format and content generally following Fig.~\ref{variab_mon456}.
        The
contours are from the X-ray spectral fit of the spectrum observed
during the interval \#2+\#4 (black contours) and \#5+\#6 (red
contours).}
	\label{variab_mon1076}
	\end{figure*}

	\begin{figure*}[!t]
	\centering	
	\includegraphics[width=10cm]{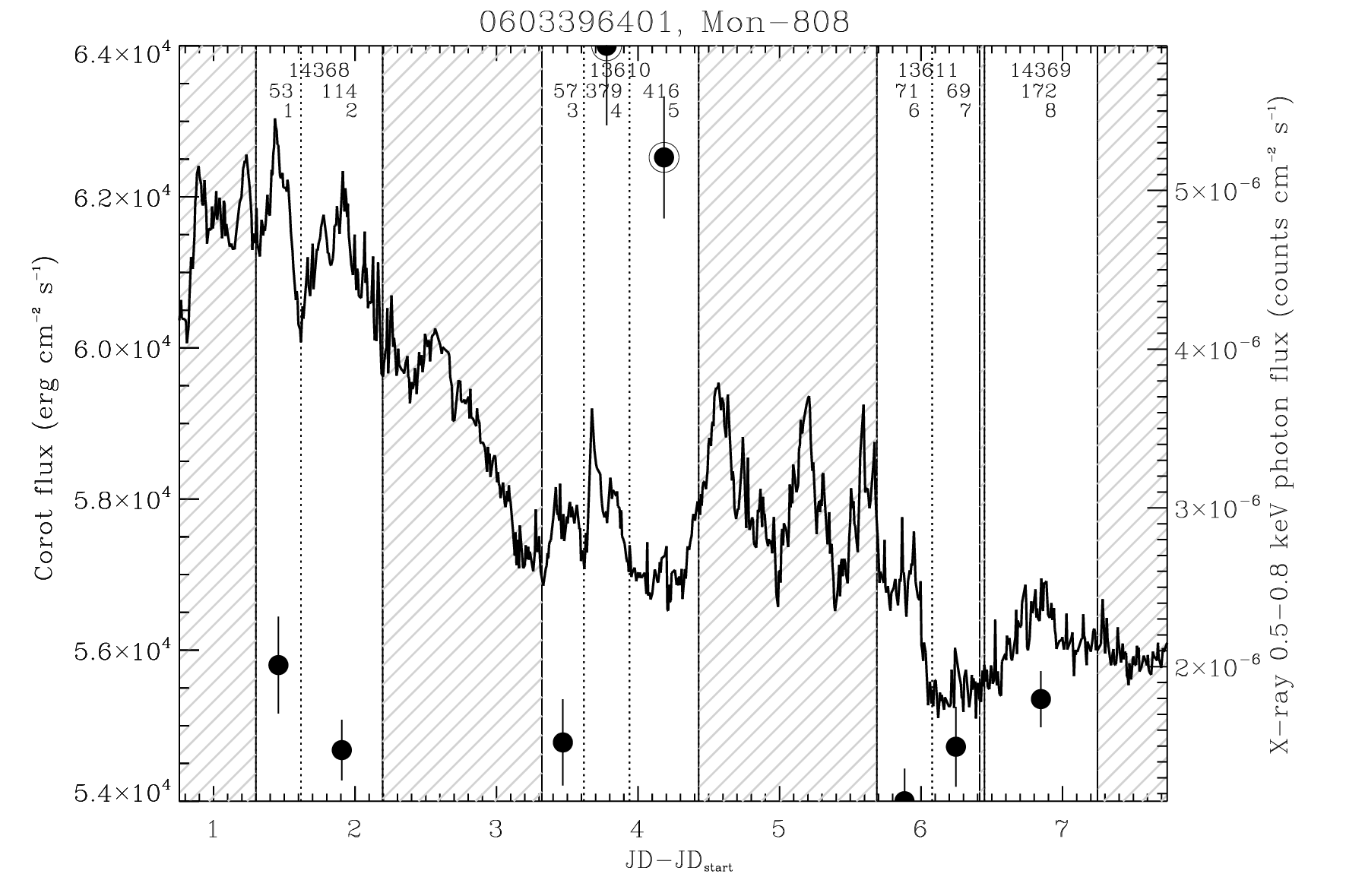}
	\includegraphics[width=8cm]{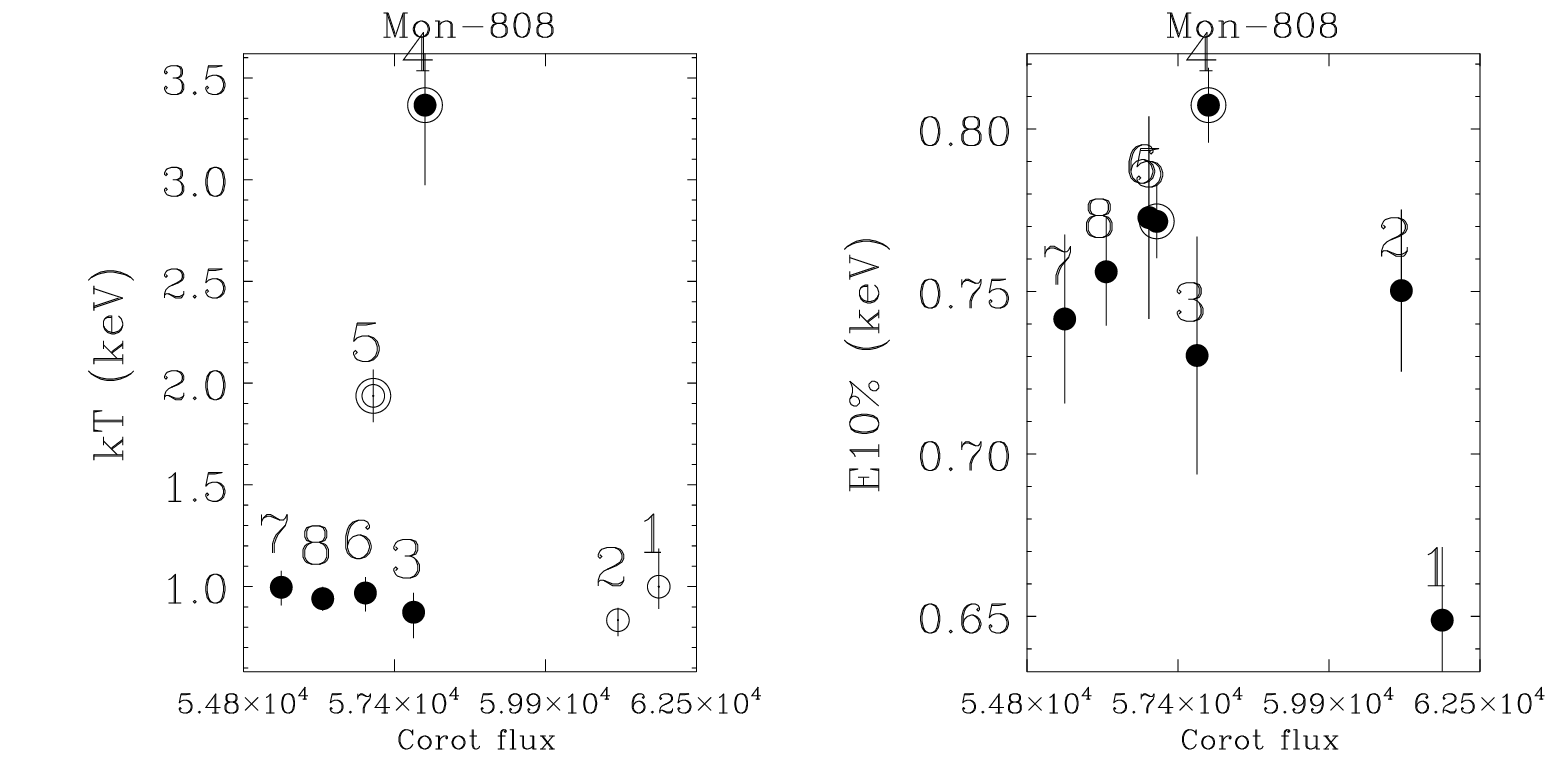}	
	\includegraphics[width=18cm]{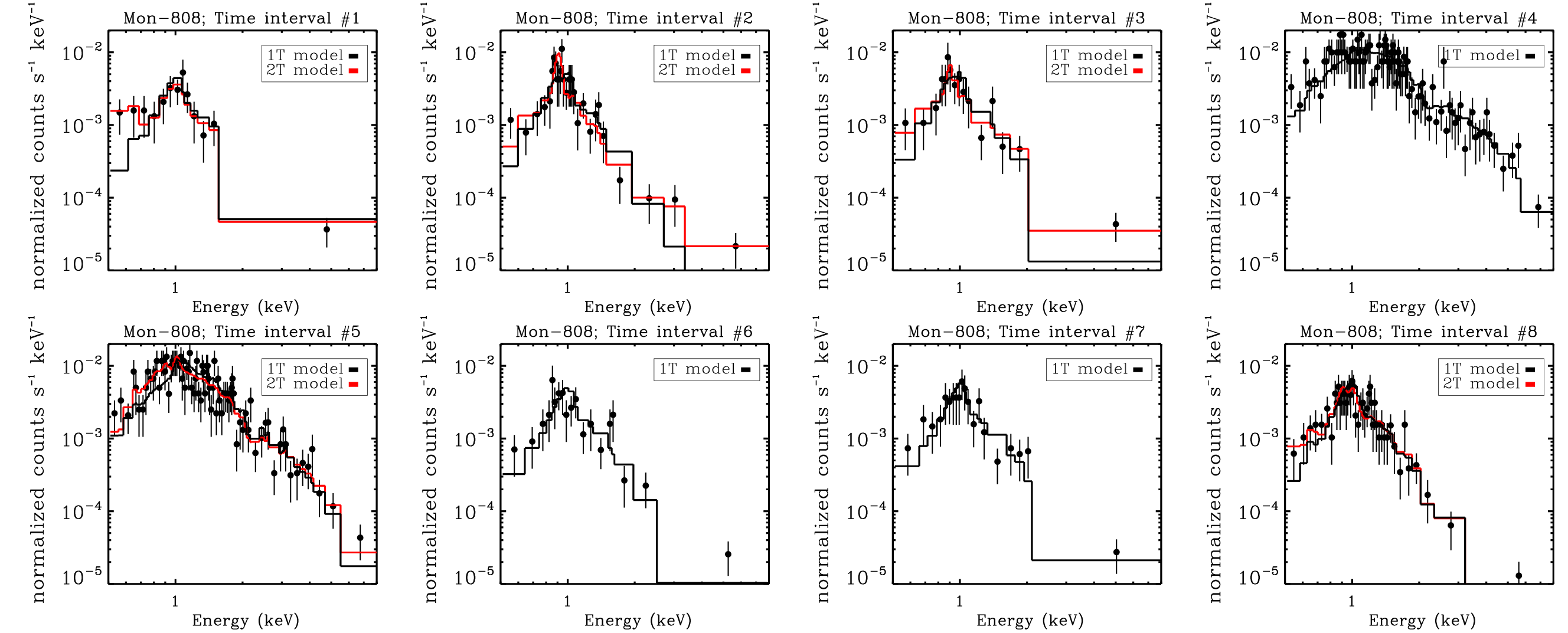}
	\includegraphics[width=8cm]{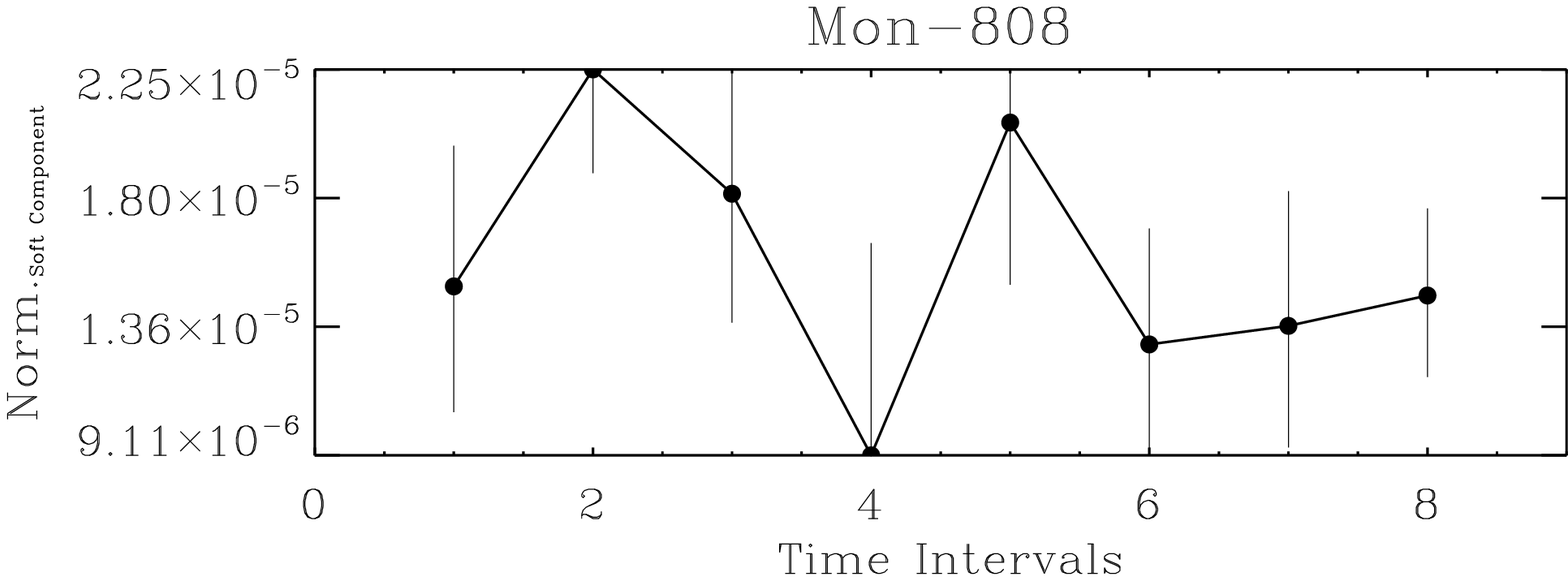}
	\includegraphics[width=7cm]{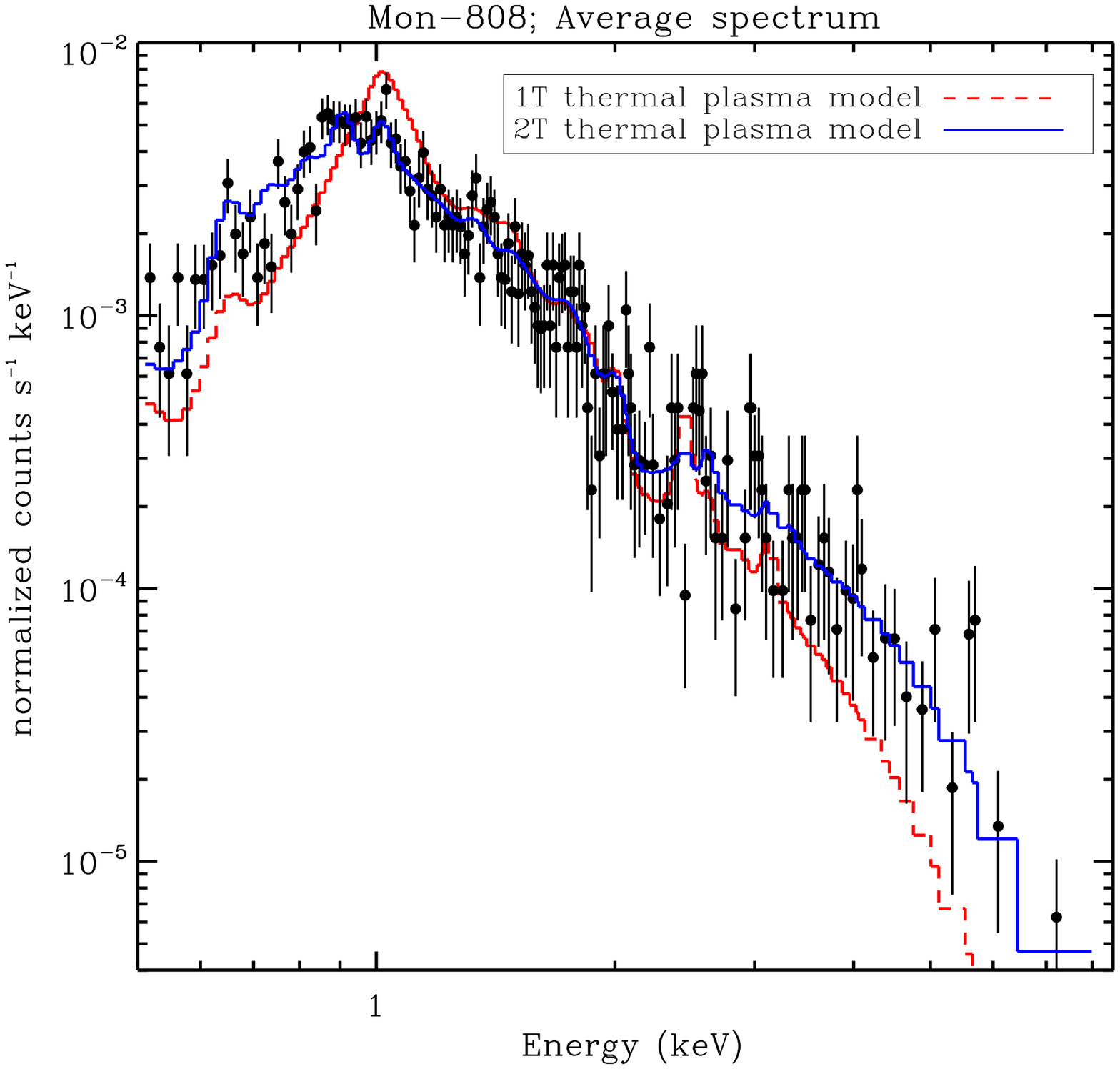}
        \caption{Optical and X-ray variability of Mon-808, with panel
        format and content generally following Fig.~\ref{variab_mon456}.
        There are two differences with respect to prior figures: i) the upper right
panels show the variability of kT and E$_{10\%}$ in units of keV, and ii)
the dots in the upper left panel show the X-ray photon flux in the
0.5-0.8$\,$keV energy band. The central panels show the X-ray spectra
observed in the time intervals (dots) with the best fit 1T thermal
plasma model (black line), together with the 2T model (red line) when
needed. The bottom left panel shows the normalization of the soft
component obtained fitting the X-ray spectrum observed in the time
intervals with 2T thermal plasma model, with plasma temperatures fixed
at kT$_{soft}$=0.3$\,$keV and kT$_{hard}$=1.6$\,$keV and N$_H$ fixed
at the value obtained from the optical extinction
(0.009$\times$10$^{22}\,$cm$^{-2}$). The bottom right panel shows the
average X-ray spectrum (dots). The dashed red line shows the best fit
1T thermal plasma model, while the blue line is the 2T model.}
	\label{variab_mon808}
	\end{figure*}

\begin{table}
\caption{H$\alpha$ EW from \citet{DahmSimon2005} and $\dot{M}_{acc}$ from \citet{VenutiBFA2014AA} of the stars discussed in Sect. \ref{N$_H$_vs_dips} and \ref{N$_H$_vs_burst}.}
\label{accret_table}
\centering                       
\begin{tabular}{cccc}
\hline
  \multicolumn{1}{c}{Mon-ID} &
  \multicolumn{1}{c}{H$\alpha$ EW} &
  \multicolumn{1}{c}{H$\alpha$ shape} &
  \multicolumn{1}{c}{$\dot{M}_{acc}$} \\
\hline
  \multicolumn{1}{c}{ } &
  \multicolumn{1}{c}{\AA{}} &
  \multicolumn{1}{c}{} &
  \multicolumn{1}{c}{M$_{\odot}$/yr} \\
\hline
456  &  13.1  &	 asymmetric  &                       \\
1167 &  23.4  &              & 4.6$\times$10$^{-9}$  \\
412  &  30.7  &  asymmetric  & $\sim$10$^{-7}$        \\
717  &  24.5  &              & 2.8$\times$10$^{-8}$   \\
619  &  94.3  &              & 4$\times$10$^{-8}$     \\
491  &  67.2  &              & 3$\times$10$^{-8}$     \\
774  &  14.3  &              & \\
1076 &  2.6   &              & \\
808  &  50.2  &  asymmetric  & 1.7$\times$10$^{-8}$   \\
370  &  113.2 &              & \\
326  &  27.9  &              & 1.7$\times$10$^{-9}$   \\
474  &  104.7 &              & \\
357  &	8     &              & \\
945  & 	66.3  &	 asymmetric  & \\
771  & 	28.9  &              & \\
765  &  18.2  &              & \\
103  &  6.4   &              & \\
378  &	8.5   &              & \\
\hline\end{tabular}   
\end{table}

%%%%%%%%%%%%%%%%%%%%%%%%%%%%%%%%%%%%%%%%%%%%%%%%%%%%%%%%%%%%%%%%%%%%%%%%%%%%%%%5
    \subsection{Optical dips with increasing X-ray absorption}
    \label{N$_H$_vs_dips}

    In this section, we analyze the variability of those stars with
disks with increasing X-ray absorption during the optical dips, i.e.,
where optical extinction and X-ray absorption may increase
simultaneously. For each of these stars, we show the CoRoT light
curve, with indicated the {\em Chandra} frames and the time intervals we
defined, together with the variability of some of the following X-ray
quantities: N$_H$ in units of $10^{22}\,$cm$^{-2}$, kT, E$_{10\%}$ and
E$_{25\%}$ in keV. The X-ray spectra observed during the time
intervals together with the best fit models are shown in
Appendix \ref{spectra_app}; Appendix \ref{all_LC} contains the
entire CoRoT light curves of all the stars analyzed in this paper.
Each star is labeled with both the CoRoT ID and the Mon- ID. \par

%%%%%%%%%%

$ $\par {\bf Mon-456:} Mon-456 (Fig.~\ref{variab_mon456}) is an
interesting case of simultaneous optical and X-ray variability. This
class~II K4 star, with a rotation period of 5.05 days, is moderately
accreting and it is classified as a ``quasi periodic dipper'' by
\citet{CodySBM2014AJ} and as a ``AA~Tau like'' star by
\citet{McGinnisAGS2015} according to both their 2008 and 2011 data.
The latter authors also observe increasing reddening of the CoRoT
light curve during the dips, typical of AA~Tau \citep{BouvierGAD2003},
and that several flux dips appear in the light curve during each
rotation period, suggesting the presence of various secondary
accretion streams. Large optical dips are observed during three {\em
Chandra} frames (left panel in Fig.~\ref{variab_mon456}): 

\begin{itemize}
\item A deep optical dip is observed in part during the time interval \#1.
\item The intervals \#2 and \#4 are dominated by X-ray flares.
\item The time intervals \#5 and \#6 are characterized by two optical dips
with similar X-ray properties (right panels in Fig.
\ref{variab_mon456}). 
\item Time interval \#3 has a rising optical light curve.
\end{itemize}

The time variability of N$_H$ (shown in Fig.~\ref{variab_mon456}) and
the time resolved X-ray spectra (Fig.~\ref{xspectra_mon456}) suggest a
larger N$_H$ during the dips in \#5 and \#6 than in the other time
intervals, but in each dip the X-ray absorption is not well
constrained. In order to better constrain the X-ray absorption during
the dips, we fit the average X-ray spectrum over the summed
time interval \#5+\#6 and compare the results with that obtained
from time intervals \#2+\#3+\#4. From the latter spectrum, we obtain
N$_H$=$0_0^{+0.11}\times 10^{22}\,$cm$^{-2}$ (P$_\%$=0.68), while for
the former N$_H$=$0.46_{-0.27}^{+0.32}\times 10^{22}\,$cm$^{-2}$
(P$_\%$=0.85), suggesting a larger hydrogen column density during the optical dips. The $>$2$\sigma$ significance of this
difference is confirmed by the contours in the C-stat space shown in
Fig.~\ref{variab_mon456}. \par

        There is no evidence for larger N$_H$ during the dip in \#1,
and the X-ray spectral fit of the spectrum observed in
\#1+\#5+\#6 results in a poorly constrained fit (P$_\%$=0.002).
However, some evidence for the X-ray spectrum getting harder during
this dip (i.e. increasing E$_{10\%}$) is presented in the Appendix \ref{e10lc_app}.
	
%%%%%%%%%%

$ $\par 	
{\bf Mon-1167:} Fig.~\ref{variab_mon1167} shows the optical and X-ray
variability of the M3 disk-bearing star Mon-1167, classified as a
``quasi-periodic stochastic'' by \citet{CodySBM2014AJ} and as an
aperiodic extinction dominated star during 2008 and an ``AA~Tau
analog'' during 2011 by \citet{McGinnisAGS2015}. This star is
moderately accreting and its light curve is characterized by:

\begin{itemize}
\item Small optical dips in the time intervals \#5, \#8, and \#11, together
with a large optical dip observed only partially during the time interval
\#3. 
\item The optical emission rises during \#1 and \#2, then steeply falls in \#10.
\item No peculiar features are observed in \#4, \#6, \#7, \#9, \#12.
\end{itemize}

During the dips, the CoRoT flux declines by the 5.5\% (\#3), 1.9\%
(\#5), 2.1\% (\#8), and 1.3\% (\#11). The obtained N$_H$ is significantly different from zero only during \#3 and
\#8, although in \#8 the error bar is large and marginally compatible
with the value observed in \#7 (right bottom panels in Fig.
\ref{variab_mon1167}). \par

The C-stat contours from the spectral fits in the time intervals \#3
and \#8 support the evidence of a larger N$_H$ during \#3, while
a possible solution with N$_H$=0 within 68\% confidence is possible
during the time interval \#8.

%%%%%%%%%%	
$ $\par
{\bf Mon-412:} Fig.~\ref{variab_mon412} shows the optical and X-ray
variability of the accreting M1 class~II object Mon-412, classified as
a ``burster'' by \citet{CodySBM2014AJ} and \citet{StaufferCBA2014}. CoRoT
light curve (upper left panel) is characterized by accretion bursts
and dips\footnote{It must be noted that
the CoRoT light curve of Mon-412 may have been contaminated by a
nearby disk-free cluster member more than one magnitude fainter in $I$
band and falling in the CoRoT mask.}:

\begin{itemize}
\item Small optical burst-like features are observed in \#1, \#2, \#3, \#8.
\item Prominent optical burst-like features are observed in part during \#4 and \#9.
\item In \#5, \#6, and \#7 the optical light curve shows a sequence of dip-like and/or burst-like features, with the CoRoT flux being 3.5\% lower in \#6 than in \#5 and \#7.  
\end{itemize}

It is not clear from the CoRoT light curve alone whether the
variability observed in the third {\em Chandra} frame is due to two
optical bursts in \#5 and \#7 or to an optical dip in \#6. Some hint
is provided by the variability of the X-ray properties (shown in the
right panels). During the time interval \#6, in fact, N$_H$ varies
from 0 to 1.2$^{0.38}_{0.24}\times10^{22}\,$cm$^{-2}$ together with E$_{10\%}$,
E$_{25\%}$, and E$_{50\%}$ (the latter not shown in Fig.
\ref{variab_mon412}) reaching the highest values observed in this
star. However, as shown in Fig.~\ref{variab_mon412}, the 68\%
statistical confidence region in the C-stat space for the interval \#6
also allows solutions at low N$_H$. We conclude then that there are
hints of a larger X-ray absorption during \#6, but this is not
significantly supported by the time resolved X-ray spectral analysis.

%%%%%%%%%%	
$ $\par
{\bf Mon-717:} Mon-717 shows a large optical dip (see Fig.
\ref{variab_mon717}), longer than the third {\em Chandra} frame and with
the CoRoT flux decaying by 25.3\% (corresponding to an increase\footnote{Ignoring the size of the obscuring feature
with respect to the stellar disk.} of
$A_V$ by 0.38$^m$) and a FWHM of 0.6 days. This star is a
Class~II M0.5 star that is moderately accreting. The source is faint in
X-rays, with only 10 photons detected during the dip and 20 in the
previous {\em Chandra} frame. The X-ray emission is, however, harder during
the dip than in the other frames, as suggested by the variability of
the photon energy quantiles (Fig.~\ref{variab_mon717}), suggesting
that not only the optical emission but also the X-ray emission is more
absorbed during the third {\em Chandra} frame.

%%%%%%%%%%	

$ $\par {\bf Mon-119:} The optical and X-ray variability of Mon-119
(spectral type K6, \citealp{DahmSimon2005}), classified as a
``stochastic'' star by \citet{CodySBM2014AJ} and as a star whose
variability is mostly driven by variable accretion by
\citet{StaufferCRH2016AJ}, is shown in Fig.~\ref{variab_mon119}. The
CoRoT light curve shows several features: 

\begin{itemize}
\item The optical emission rises by a factor 1.12 during \#1, followed by a more quiescent phase (\#2).
\item An intense optical burst occurred between the first and second {\em Chandra}
frames, whose final part is observed in the time interval \#3 (with
only 19 X-ray photons detected). 
\item Two optical dips are observed in the time intervals \#4 and \#6 separated by a quiescent phase (\#5).
\item The optical emission is higher during the third {\em Chandra} frame
with at least two evident bursts (intervals \#7 and \#9), separated by
a more quiescent phase (\#8).
\item The fourth {\em Chandra} frame (\#10) is dominated by an intense X-ray and optical flare. 
\end{itemize}

In the two dips (\#4 and \#6), the CoRoT flux decreases by 8.9\% and
15.7\% with respect to the optical emission observed during the time
interval \#5. As shown in the top right panels in Fig.
\ref{variab_mon119}, excluding the flare, N$_H$ is always compatible
with zero in all the time intervals except in the second {\em Chandra} frame
(\#4, and \#6). In order to verify whether a significantly larger
N$_H$ is observed during these two optical dips, we fit the 1T thermal
plasma model to the average X-ray spectrum summing the
time intervals \#4+\#6. The average value of N$_H$ suggested by the
best fit model is
N${_H\,\#4+\#6}$=$0.21^{+0.23}_{-0.16}\times10^{22}\,$cm$^{-2}$, which is
significantly larger than zero (and larger than the absorption observed in
the remainder intervals) within a 68\% confidence range. This is confirmed
by the contours in the C-stat space (Fig.~\ref{variab_mon119}). 

%%%%%%%%%%	

$ $\par
{\bf Mon-619:} Mon-619 is a K8.5V star actively accreting from its
disk, and both \citet{CodySBM2014AJ} and \citet{McGinnisAGS2015}
classified this star as an aperiodic extinction dominated star. The
CoRoT light curve of Mon-619 in Fig.~\ref{variab_mon619},
shows:

\begin{itemize}
\item A large optical dip that dominates the second {\em Chandra} frame (\#2).
\item Two smaller optical dips or an accretion burst occurring during a large dip during \#1. 
\item No evident features are observed in \#3 and \#4, even if there
are several spikes that may be the consequence of unsteady accretion.
\item A steep decline of the optical emission during \#5.
\end{itemize}

In the dip observed in the second {\em Chandra} frame the optical flux
decreases by 10.4\%, corresponding to the extinction increasing by
0.14$^m$.  We observe the largest values of
E$_{10\%}$ and E$_{25\%}$ during the large dip in the second {\em Chandra}
frame, suggesting a larger X-ray absorption during the optical dip.
Fig.~\ref{variab_mon619} also shows that the time variability of
both the CoRoT and X-ray flux is coherent and correlated, as expected
in AA~Tau like stars. The X-ray spectrum of Mon-619 is one of the hardest observed in our
sample (Fig.~\ref{xspectra_mon619}), with a median photon energy of
3.26$\,$keV, (the typical value in ours sample is 1.3$\,$keV). Given the
low X-ray counts, the spectral fits are not well constrained and
thus not discussed.

%%%%%%%%%%	

$ $\par	{\bf Mon-491:} Mon-491 is a K3V star actively accreting from
its disk, and it has been classified by \citet{CodySBM2014AJ} as a long
term variable and by \citet{StaufferCRH2016AJ} as a star with variable
accretion. The CoRoT light curve of Mon-491 (Fig.~\ref{variab_mon491})
shows:

\begin{itemize}
\item A rising phase in optical during \#1 and \#2.
\item A very steep optical decline with the CoRoT flux decreasing by about the
20\% during the second {\em Chandra} frame (\#4), after a phase with a
more constant optical flux (\#3), which may be due to variable
extinction.
\item The optical emission remains almost constant during \#5 and \#6 and then declines during \#7.
\end{itemize} 

Even if the light curve of Mon-491 does not show dips during the
{\em Chandra} frames, we want to verify whether during the decline of
optical emission observed in \#4, there is evidence for increasing
X-ray absorption. We compare the X-ray properties of Mon-491 during
the intervals where the optical light curve does not decline (i.e.,
\#2+\#3+\#5) with those observed during \#4. N$_H$ is smaller during
the former time intervals than during \#4 at $\sim$96\% confidence:
N$_H$$_{\#4}=2.14^{+1.07}_{-0.45}\times10^{22}\,$cm$^{-2}$ while
N$_H$$_{\#2+\#3+\#5}=0.98^{+0.46}_{-0.42}\times10^{22}\,$cm$^{-2}$.
The significance of this difference is confirmed by the contours in
the C-stat space shown in Fig.~\ref {variab_mon491}. \par

%%%%%%%%%%%%%%%%
$ $\par	{\bf Mon-774:} The CoRoT light curve of Mon-774 (spectral type
K2.5, classified as ``stochastic'' by \citealt{CodySBM2014AJ} with an
AA~Tau phase observed during 2008 and analyzed in detail by
\citealt{McGinnisAGS2015}), shown in Fig.~\ref{variab_mon774}, is
characterized by:

\begin{itemize}
\item A sequence of optical high and low phases in \#1, \#2, and \#3
\item Two large optical dips, one dominating the third {\em Chandra} frame (time
intervals \#7 and \#8), and one between the first and second {\em Chandra}
frames. 
\item An optical dip observed during \#10 after a higher phase in \#9.
\item A small optical dip observed in \#5 during the decline phase from \#4 to \#6.
\end{itemize}

The CoRoT flux variation during the dip in \#7 is about 16.8\%,
corresponding to an increase of optical extinction $\Delta A_V=0.2^m$,
while the dip observed between the first and second {\em Chandra} frames
is less deep, with a 6.4\% CoRoT flux variation ($\Delta A_V=0.09^m$).
The variability of the X-ray properties is shown in the right panels
of Fig.~\ref{variab_mon774}. N$_H$ is larger than zero in three time intervals. In the
interval \#10, the best fit predicts a N$_H$ larger than zero but the
C-stat contours, not shown here, admit solutions with N$_H$=0 within
68\% confidence level, so it will not be discussed further. \par

        In the interval \#8 N$_H$ is only slightly larger than zero at
68\% confidence. Interval \#8 is actually part of the large dip which
dominates \#7, so the two intervals must be considered together. However, the
X-ray spectral fit of the spectrum observed during \#7+\#8 does not
result in a well-constrained estimate of N$_H$, with solutions ranging
from 0 to about 0.3$\times10^{22}\,$cm$^{-2}$ within 68\% confidence.
\par

The small dip isolated in the time interval \#5 is more interesting in
this respect. The N$_H$ obtained using 2T
thermal plasma model\footnote{Using a 1T model, the fit is not
statistically acceptable}
(1.07$^{+0.19}_{-0.19}\times10^{22}\,$cm$^{-2}$) is significantly
different than zero and larger than the value
obtained from the average spectrum
(0.54$^{+0.25}_{-0.24}\times10^{22}\,$cm$^{-2}$) at 68\% confidence level, as proved by the C-stat contours shown in Fig.
\ref{variab_mon774}.  

%%%%%%%%%%

$ $\par	{\bf Mon-1076:} The CoRoT light curve of the non-accreting
star Mon-1076 (spectral type M1), listed as a star with periodic flux
dips in \citet{StaufferCMR2015}, is shown in Fig.\
\ref{variab_mon1076}: 

\begin{itemize}
\item During the second {\em Chandra} frame, the CoRoT light curve is
dominated by a large dip (time intervals \#2, \#3, and \#4), with a
10.6\% CoRoT flux variation (corresponding to an increase of optical
extinction by $\Delta$A$_V$=0.15$^m$, see Table \ref{dips_table}). 
\item An optical and X-ray flare is isolated in the interval \#3. 
\item Almost constant optical emission is observed during \#1, \#5, and \#6 (in the latter case probably with some variability).
\end{itemize}

The variability of the X-ray properties, shown in the right panels in
Fig.~\ref{variab_mon1076}, suggests a larger N$_H$ observed at the
beginning of the optical dip, compared with the remaining intervals.
The significance of this difference is confirmed in the bottom panel
in Fig.~\ref{variab_mon1076}, which compares the confidence regions in
the C-stat space obtained from the fit of the X-ray spectrum observed
in the intervals \#2+\#4 (during the dip) and \#5+\#6 (after the dip).
The X-ray absorption in the former interval is larger than that
observed in the latter within a 68\% confidence.

%%%%%%%%%%%%%%%%%%%%%%%%%%%%%%%%%%%%%%%%%%%%%%%%%%%%%%%%%%%%%%%%%%%%%%%%%%%%%%%%%%%%%%%%%%%%%%%%%%%%%
\subsection{Variability of the X-ray properties during the optical bursts}
\label{N$_H$_vs_burst}

In this section, we analyze the time-resolved X-ray properties during
the optical bursts.  We look for evidence of: i) A soft
component in the X-ray spectrum (below 1$\,$keV) which may result from
the emission of accreting gas in the accretion shock, and ii)
increasing X-ray absorption due to the gas in the accretion streams.
Hereafter, we classify the X-ray spectra with ``an intense soft X-ray
spectral component'' as those whose observed X-ray emission below
1$\,$keV is significantly larger than the flux predicted by the best
fit 1T thermal plasma model fitting the observed time resolved X-ray spectra as explained in Sect. \ref{xvar_sec}. With few exceptions, we restrict this definition to cases where the spectral fit using 1T thermal plasma
model is statistically unacceptable, accounting for the fact that
stellar coronae are not isothermal and 2T models may be always
required when there are enough X-ray photons detected. \par

    To analyze the simultaneous optical and X-ray variability of stars
with optical bursts, we use an approach similar to that adopted for
the stars with dips. For each star, we show a set of panels including the CoRoT light curve
observed during the {\em Chandra} frames, with superimposed the variability of the soft
X-ray flux; the observed time resolved X-ray spectra; and the variability of
the following X-ray properties: The plasma temperature; the E$_{10\%}$
and E$_{25\%}$ photon energy quantiles; and the median photon energy
E$_{50\%}$, for stars with soft X-ray emission during the optical
bursts, or the hydrogen column density N$_H$ for stars with
increasing X-ray absorption during the optical bursts. Together with
the time resolved X-ray spectra, we show the best
fit thermal plasma model, 1T or 2T if the fit using 1T thermal models
is statistically unacceptable. \par

The criterion adopted to search for a soft component (i.e., the
acceptance level for 1T models) depends on plasma
properties and on the number of photons collected. Since we aim to
monitor the presence of a soft component irrespective of the
photon statistics, for each star, we also analyze the variability of the normalization of the soft component over the defined time intervals. To attempt highlighting only the variation of the normalization of the soft component and thus the variability of the cold plasma emission measure, for this test we fit the time resolved X-ray spectra with 2T thermal plasma models with fixed N$_H$ (set equal to the value derived from the known optical extinction), kT$_{soft}$=0.3$\,$keV (typical of the soft emission from accretion spots), kT$_{hard}$=1.6$\,$keV (the typical coronal temperature of NGC~2264 members). We do not set the plasma temperatures and/or the absorption in our search of intense soft X-ray spectral component during the optical bursts to avoid that underestimating the X-ray absorption would affect our results. \par

%%%%%%%%%%%%%%%%%%%%%%%%%%%
        $ $\par {\bf Mon-808:} Mon-808 is an accreting disk bearing
star with spectral type K4 \citep{DahmSimon2005} classified as a
``burster'' by \citet{CodySBM2014AJ} and as a ``burst-dominated light
curve'' by \citet{StaufferCBA2014}. Its CoRoT light curve during the
{\em Chandra} frames (Fig.~\ref{variab_mon808}) shows:

\begin{itemize}
\item Large optical bursts in the time intervals \#1 and \#2.
\item Small optical bursts in the time intervals \#3 and \#8.
\item A bright optical and X-ray flare dominating the time intervals
\#4 and \#5 (the latter dominated by the decaying phase of the X-ray
flare). 
\item Micro bursts may be present in \#6 and \#7, with a
decline of the optical emission by about 3\% observed in the latter.
\end{itemize}

In some of the time resolved X-ray spectra shown in Fig.~\ref{variab_mon808}, the
soft X-ray emission below 0.8$\,$keV is larger than the
prediction of the 1T fitting model, specifically in \#1, \#2, and
\#3. The bottom panel shows how the normalization of the soft
component of the best fit 2T thermal plasma model (with fixed N$_H$
and kT) varies in the time intervals. The associated errors are too
large to allow us any meaningful comparison.  \par

Table \ref{mon808_2T_tab} shows the results of the time-resolved 
spectral analysis, specifically the predicted plasma temperatures, and the
null-hypothesis probabilities resulting from the best-fit 1T and 2T
APEC models. In particular, in the time intervals \#1 and \#2 the use
of 1T plasma models results in unacceptable fits, while the quality
of the fit improves significantly using 2T plasma models. In the time
intervals \#3 and \#8, the 1T fit is acceptable (in the former likely
because of the low X-ray counts), but the fitting thermal plasma
model does not reproduce the observed soft X-ray emission (below
0.7$\,$keV in the former and 0.9$\,$keV in the latter), which are
instead well-fitted with the 2T thermal plasma model. These are the
time intervals where we search for soft X-ray emission due to accretion. \par 

  Since we want to constrain the soft temperature as well as possible,
we fit the average X-ray spectrum
over the time intervals where an intense soft X-ray spectral
component is observed using 2T thermal plasma model. An acceptable fit is obtained from the average
spectrum observed in \#1+\#2+\#8, with a soft temperature
kT$_1=0.15_{-0.04}^{+0.06}\,$keV (P$_\%=91\%$), which corresponds to
$1.7_{-0.04}^{+0.07}\,$MK\footnote{We obtain the same soft component
temperature (kT$_1=0.14_{-0.05}^{+0.04}\,$keV) from the average
\#1+\#2 X-ray spectrum, but with a less constrained X-ray spectral fit
(P$_\%=8\%$)}. In the approximation of strong shock scenario, the
pre-shock velocity can be calculated from the post-shock plasma
temperature as:

  \begin{equation}
  v_{pre}^2=\frac{16 k T_{post}}{3 \mu m_H}
  \label{vel_eq}
  \end{equation}
  where $v_{pre}$ is the pre-shock gas velocity, $T_{post}$ the
post-shock temperature, and $\mu$ the mean molecular weight (0.61 in
our case). Adopting the computed soft temperature as $T_{post}$, this
calculation results in a pre-shock velocity of the accreting gas of
$352_{-48}^{+68}\,$km/s. Adopting a stellar mass of 1.2$\,$M$_{\odot}$
and radius of 1.7$\,$R$_{\odot}$ obtained interpolating the
\citet{SiessDF2000AAP} pre-main sequence isochrones with the values of
$L_{bol}$ and $T_{\rm eff}$ for Mon-808 (the former from a bolometric
correction on the dereddened $I$ magnitude and the latter from the
known spectral type), and an age of 4.5$\,$Myr, and with the
hypothesis of negligible energy loss during the accretion, this
pre-shock velocity corresponds to a free-fall from a distance of
$3.2_{-0.6}^{+1.7}\,$R$_{\odot}$, corresponding to $1.9_{-0.4}^{+1.0}$
stellar radii\footnote{It must be noted that these radii are
calculated from the center of the system.}. The fact that the pre-shock
velocity we obtain is significantly smaller than the free-fall
velocity from infinity (519$\,$km/s) suggests that the free-fall
radius (R$_{\rm ff}$) is sufficiently well constrained. Its lower limit is
comparable with the disk inner radius (1.4 stellar radii) which can be
predicted by SED analysis (see Sect.\ \ref{xvar_sec}). Since the light
curve of Mon-808 is not periodic, its rotation period is unknown, and
thus we can not calculate the co-rotation radius as $\left( GM_{star}
\right)^{1/3} \times \left( P_{star}/2\pi \right)^{2/3}$. However,
bearing in mind that this calculation is strongly uncertain, we note
that the free-fall radius is smaller than the typical
co-rotation radii of disk-bearing stars (5-10$\,$R$_{star}$,
\citealp{HartmannCGD1998,ShuNSL2000}).  \par %v dall'infinito: 519km/s
  %Calculation below
% 
The presence of a significant soft X-ray spectral component below
0.9$\,$keV in Mon-808 is also evident in the average spectrum, shown
in Fig.~\ref{variab_mon808}. Together with the observed spectrum (with
flares removed), we show the best fit 1T thermal plasma model, which
results in a poor fit (P$_\%$$\sim$0), and the best fit 2T model,
which results in an acceptable fit (P$_\%$=55$\%$) with the two plasma
temperatures kT$_1=0.37_{-0.04}^{+0.03}\,$keV and
kT$_2=2.6_{-0.2}^{+0.3}\,$keV.  \par

\begin{table}
\caption{Predicted plasma temperatures for Mon-808 from the time-resolved X-ray spectral fits. The results from the 2T model are shown when the best fit with the 1T model is poorly constrained.}
\label{mon808_2T_tab}
\centering                       
\begin{tabular}{|r|r|r|r|r|}
\hline
  \multicolumn{1}{|c|}{Interval} &
  \multicolumn{1}{|c|}{kT1} &
  \multicolumn{1}{|c|}{kT2} &
  \multicolumn{1}{|c|}{P(1T)} &
  \multicolumn{1}{|c|}{P(2T)} \\
\hline
  \multicolumn{1}{|c|}{\#} &
  \multicolumn{1}{|c|}{keV} &
  \multicolumn{1}{|c|}{keV} &
  \multicolumn{1}{|c|}{\%} &
  \multicolumn{1}{|c|}{\%} \\
\hline
1 & $0.16_{-0.12}^{+0.13}$ & $1.2_{-0.03}^{+0.4}$  & 2.1  & 98.8 \\
2 & $0.12_{-0.03}^{+0.08}$ & $4.6_{-3.50}$         & 0.3  & 50.0 \\
3 & $0.87_{-0.07}^{+0.10}$ &                       & 23.3 &      \\
4 & $3.37_{-0.40}^{+0.54}$ &                       & 54.8 &      \\
5 & $0.42_{-0.12}^{+0.11}$ & $2.6_{-0.51}^{+1.39}$ & 0.7  & 26.8 \\
6 & $0.97_{-0.09}^{+0.07}$ &                       & 22.2 &      \\
7 & $1.00_{-0.09}^{+0.08}$ &                       & 81.1 &      \\
8 & $0.94_{-0.16}^{+0.06}$ &                       & 55.4 &      \\
\hline\end{tabular}
\end{table}

	\begin{figure*}[]
	\centering	
	\includegraphics[width=10cm]{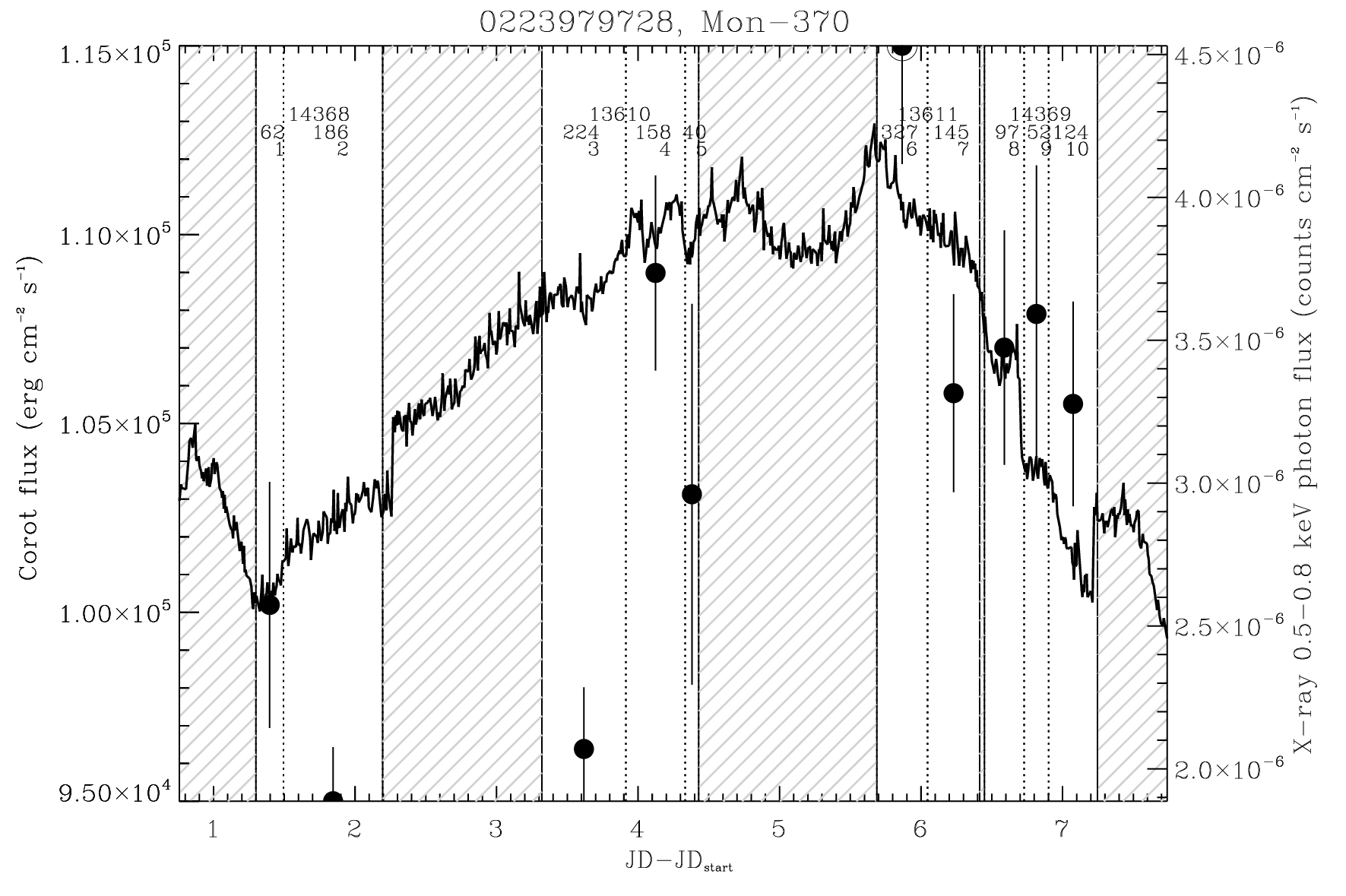}
	\includegraphics[width=8cm]{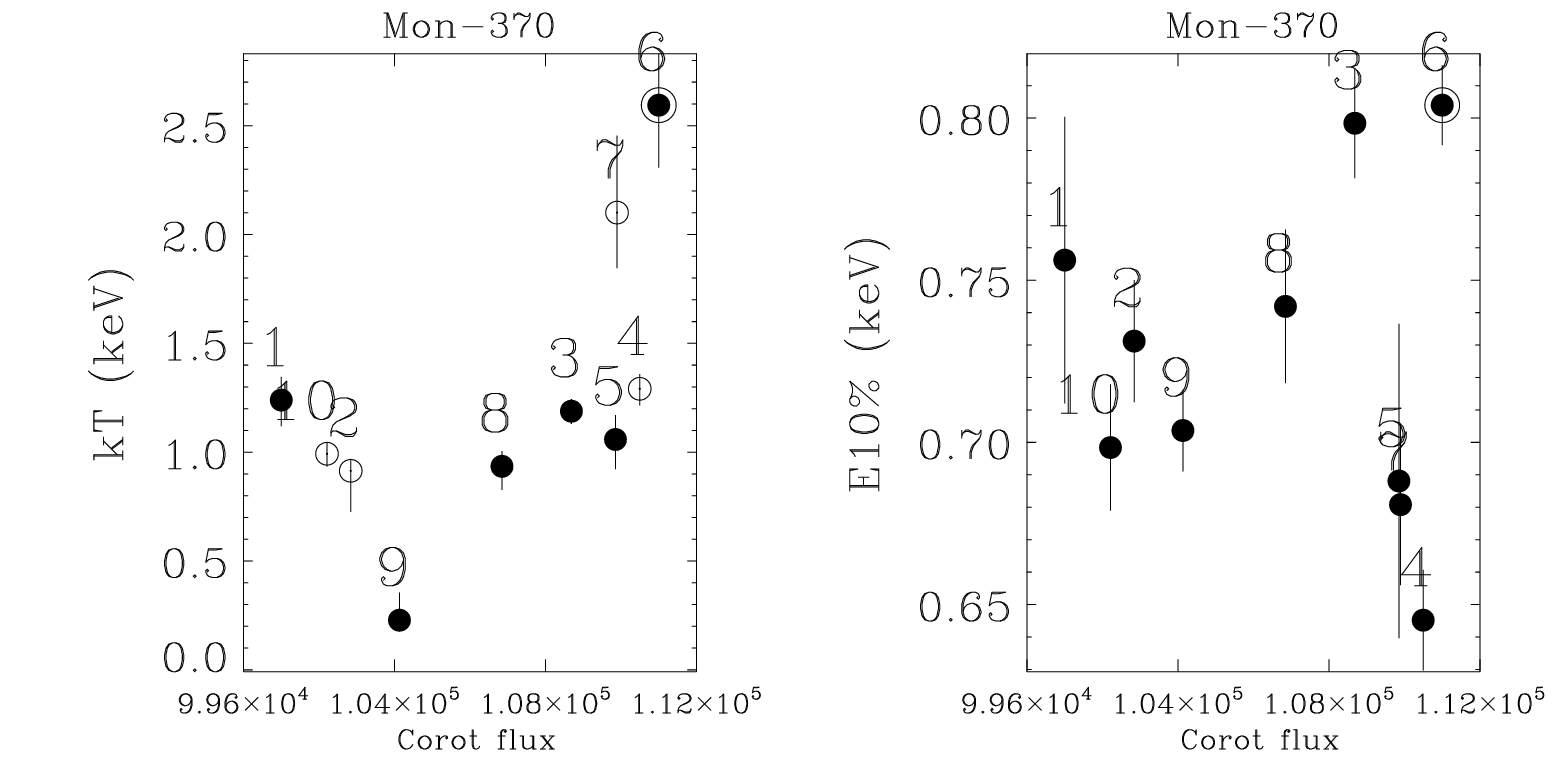}	
	\includegraphics[width=18cm]{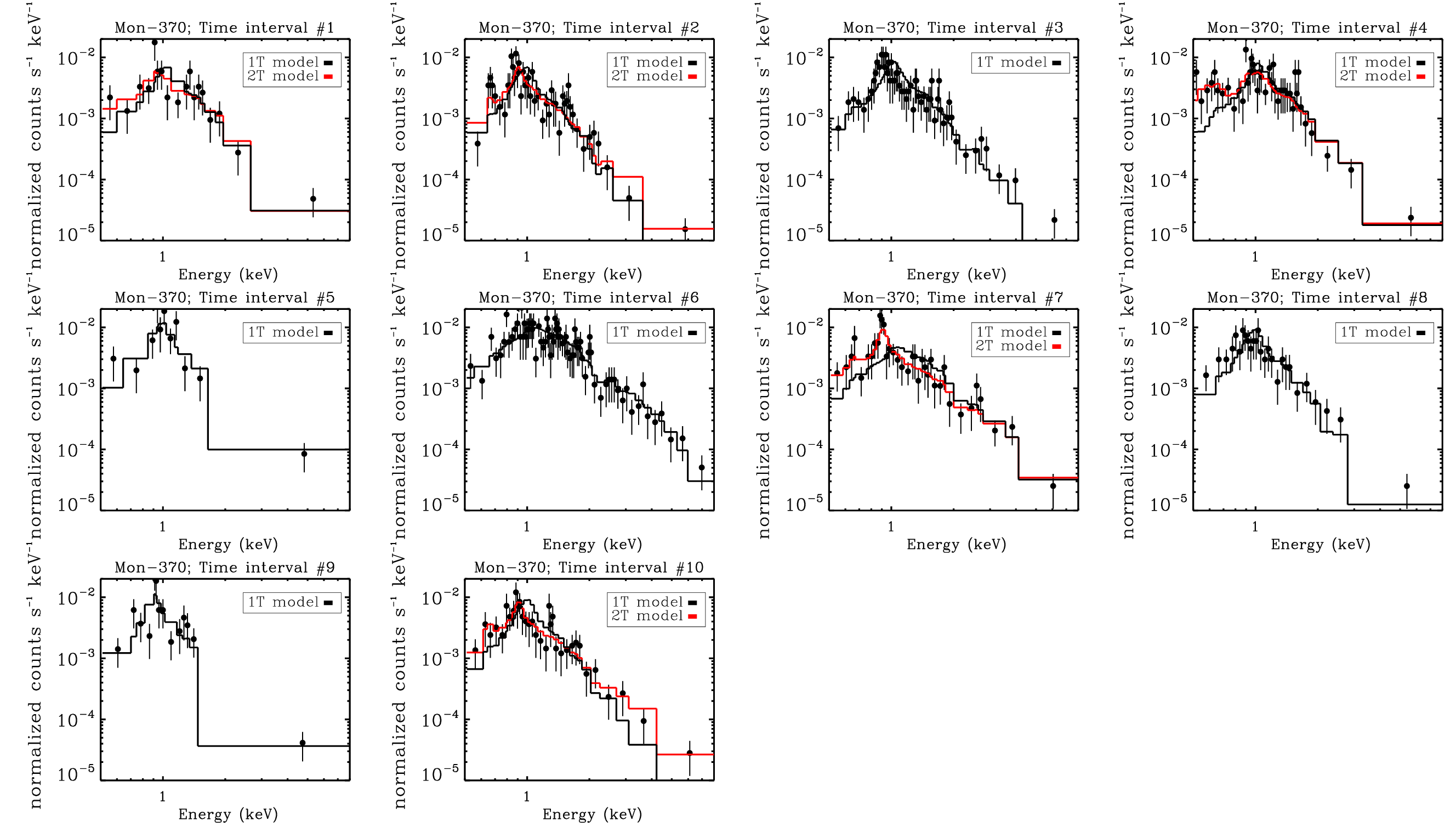}
	\includegraphics[width=8cm]{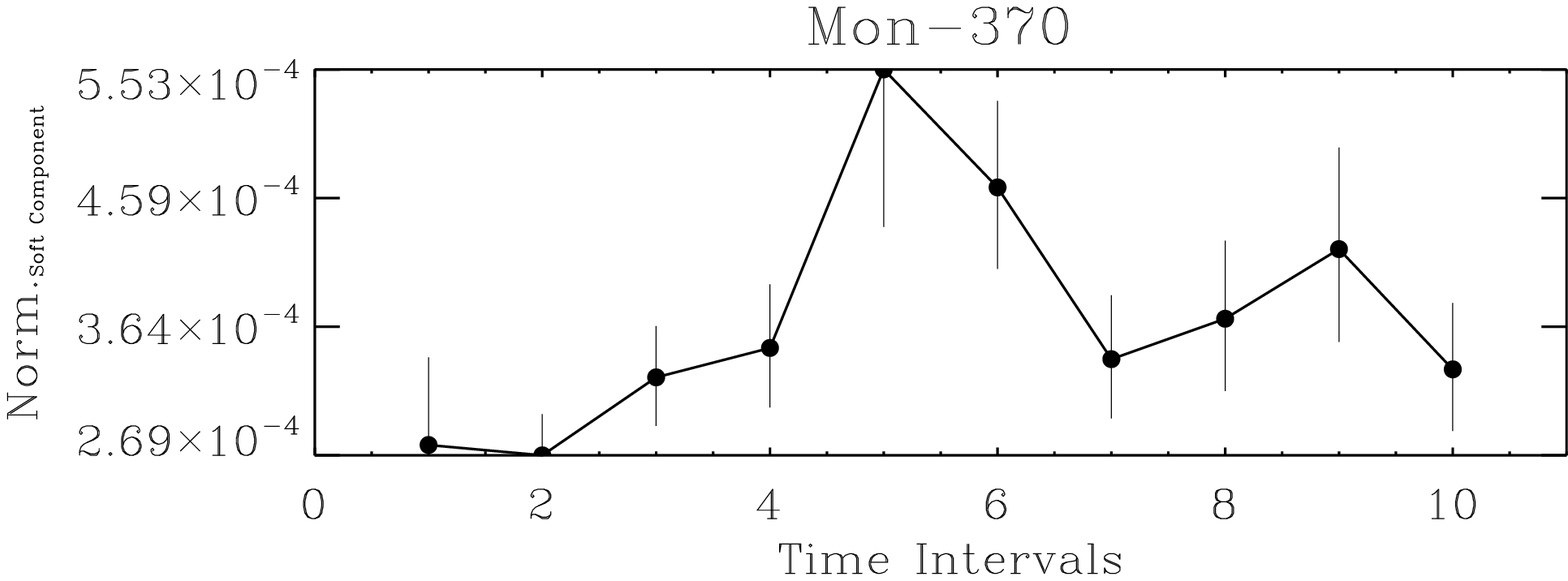}
	\includegraphics[width=7cm]{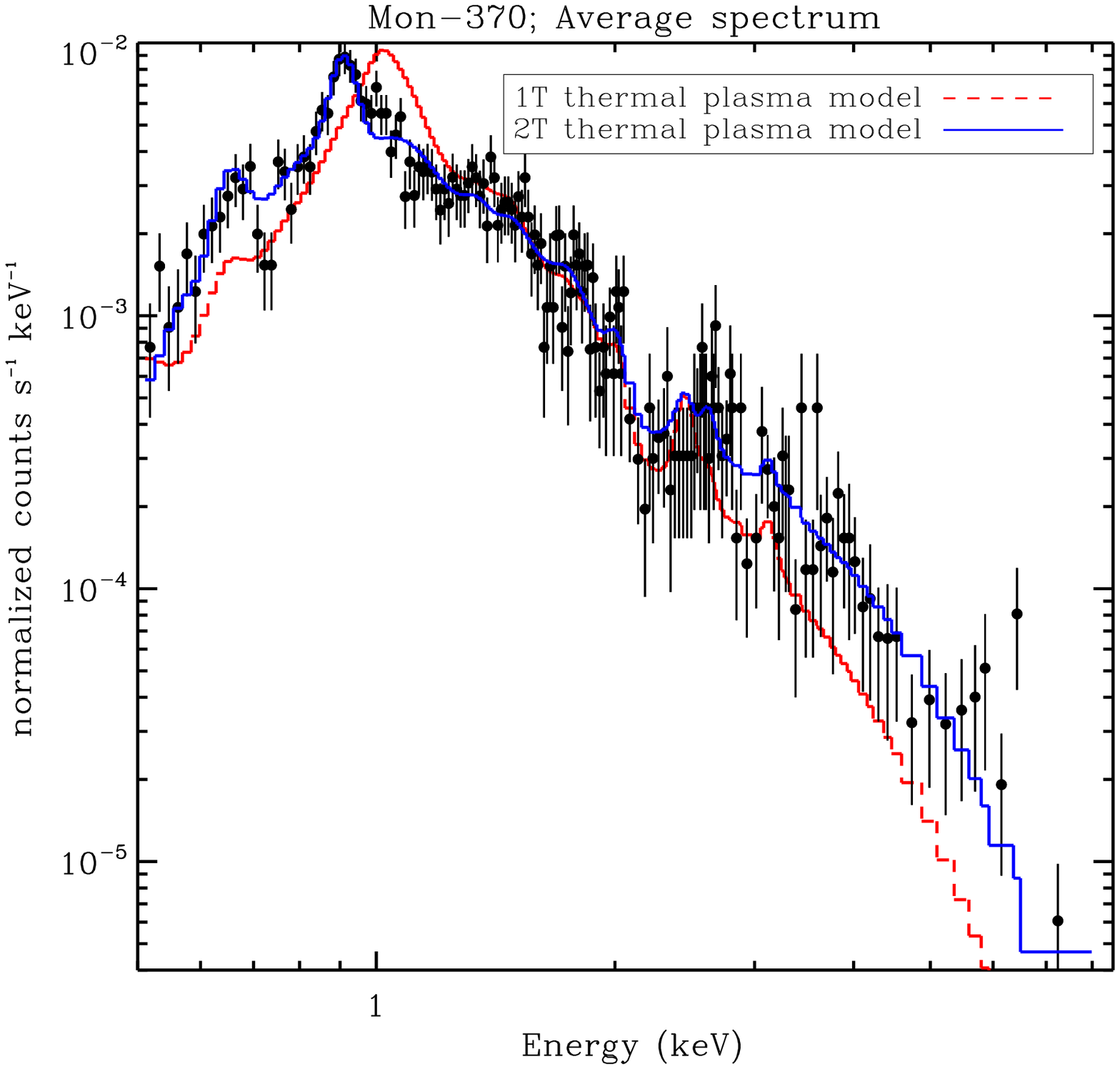}
	\caption{Time variability of the optical flux and X-ray properties of Mon-370 with the panel format and content as in Fig.~\ref{variab_mon808}.}
	\label{variab_mon370}
	\end{figure*}

	\begin{figure*}[]
	\centering	
	\includegraphics[width=9.5cm]{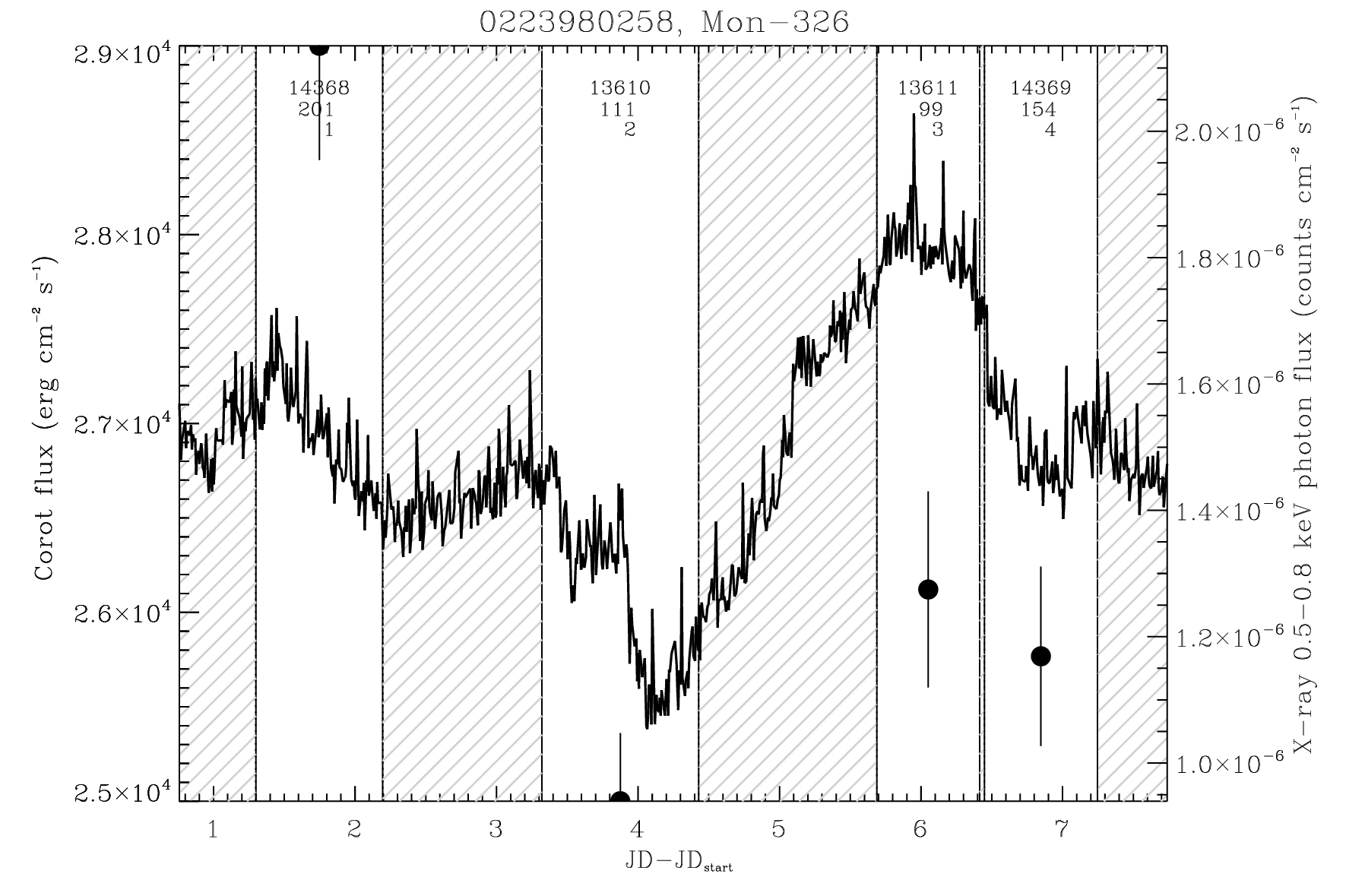}
	\includegraphics[width=8cm]{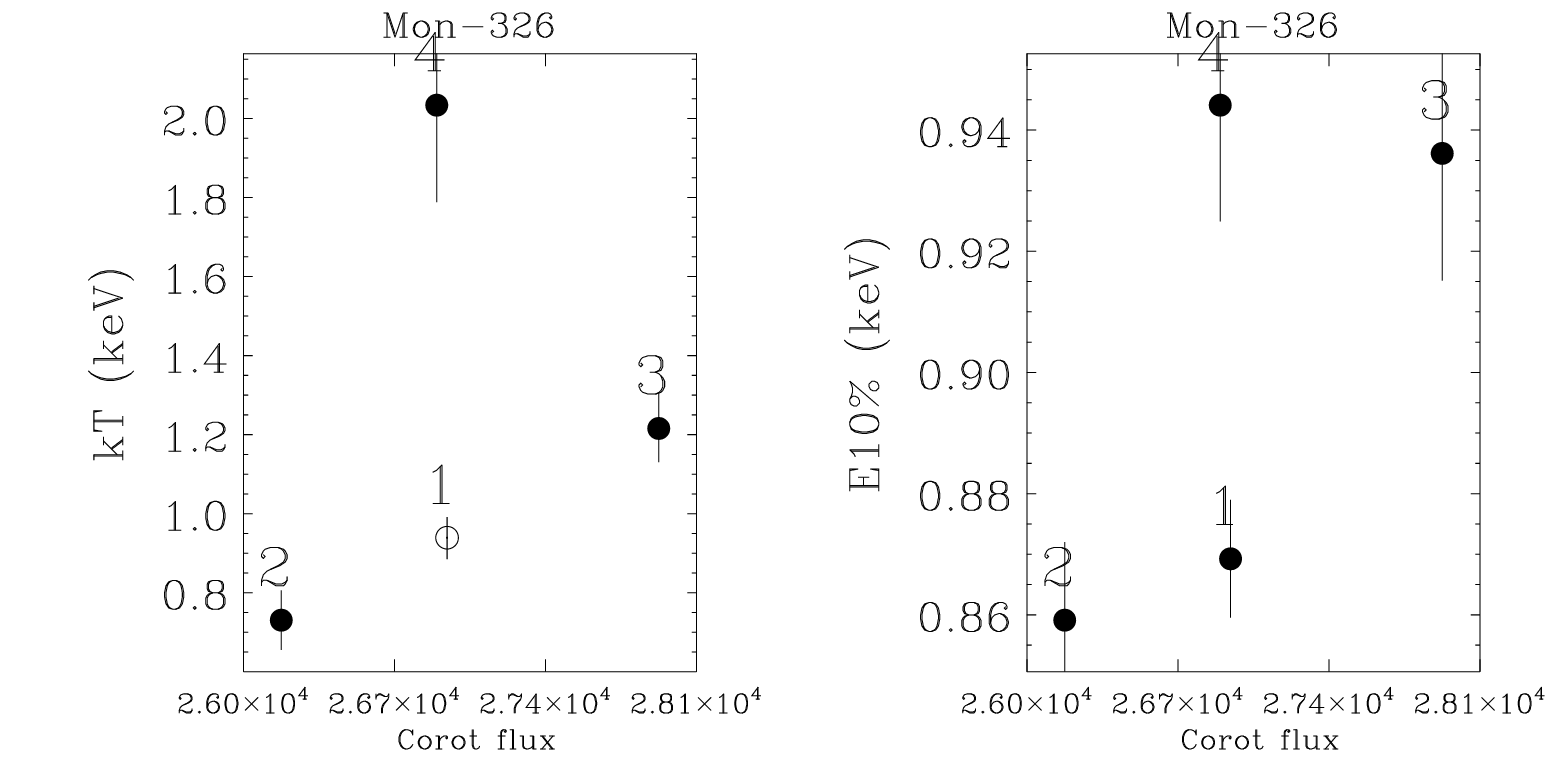}	
	\includegraphics[width=18cm]{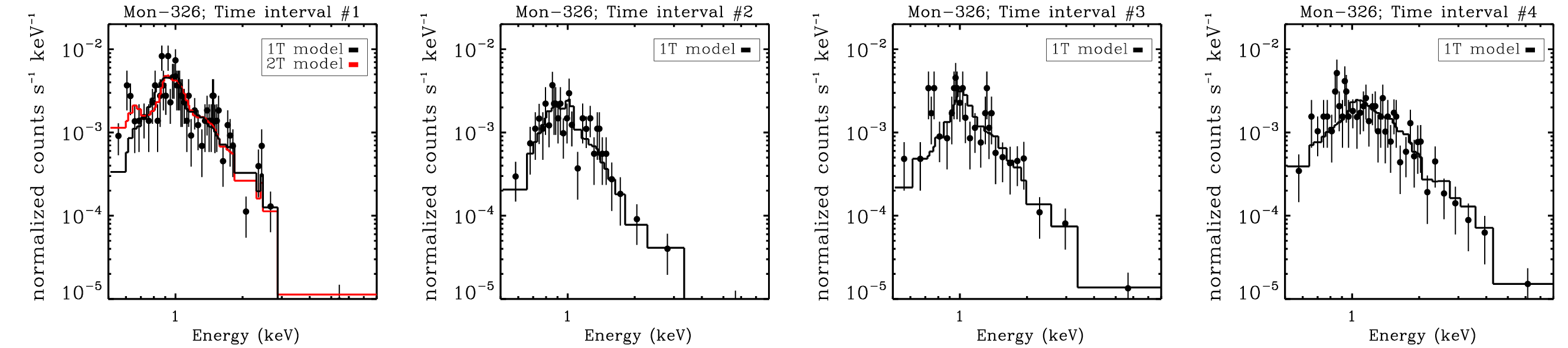}
	\includegraphics[width=8cm]{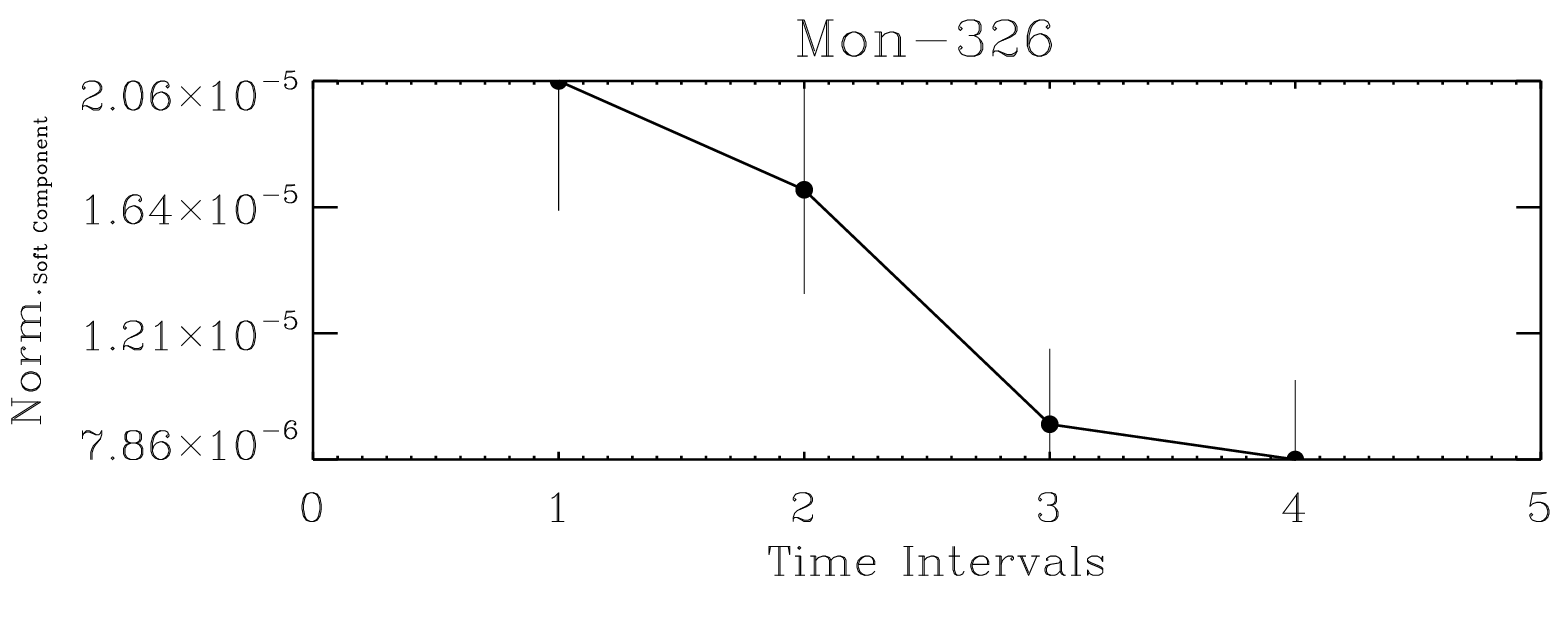}
	\caption{Time variability of the optical flux and X-ray properties of Mon-326 with the panel format and content as in Fig.~\ref{variab_mon808}.}
	\label{variab_mon326}
	\end{figure*}

	\begin{figure*}[]
	\centering	
	\includegraphics[width=9.5cm]{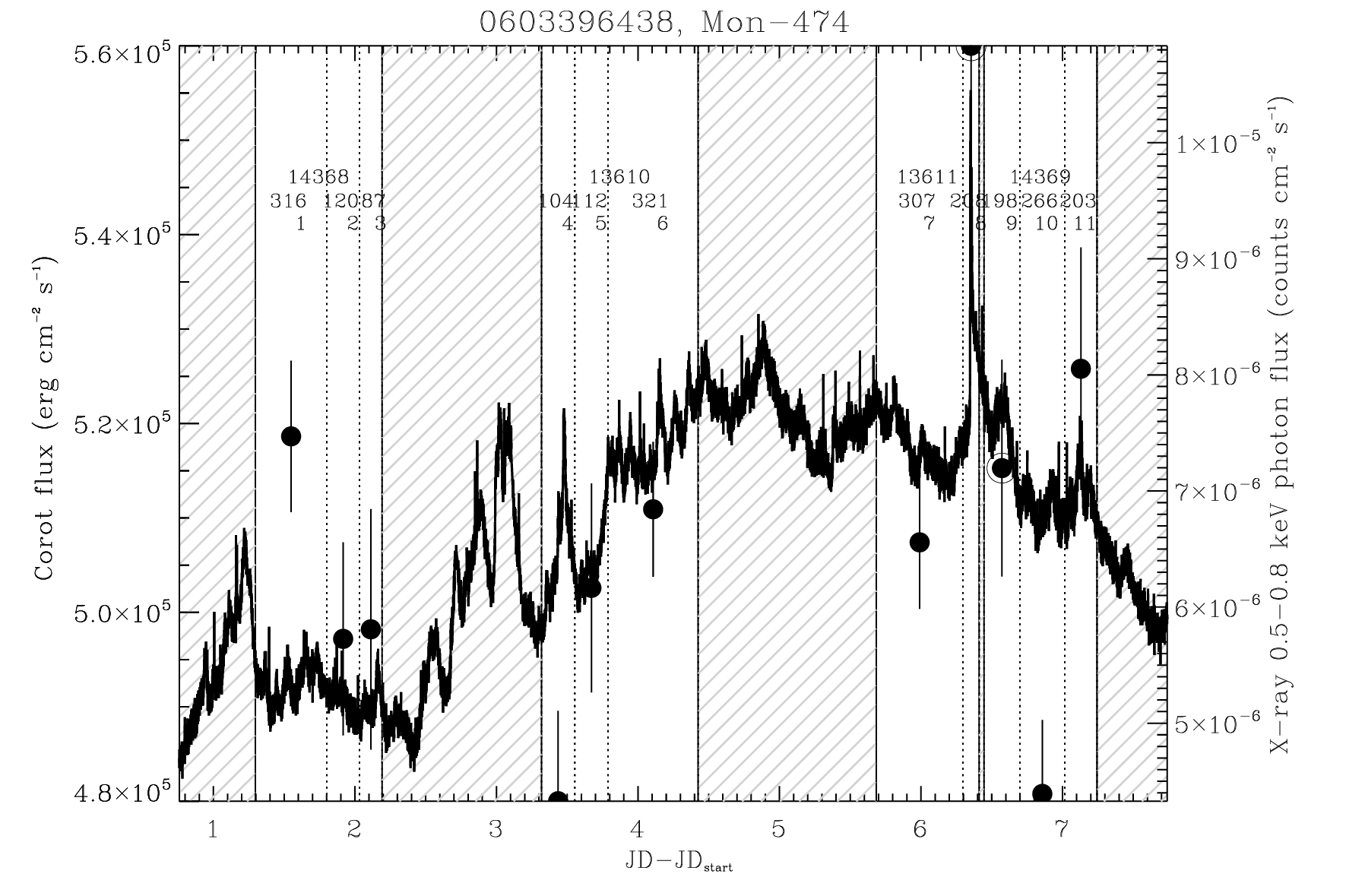}
	\includegraphics[width=8cm]{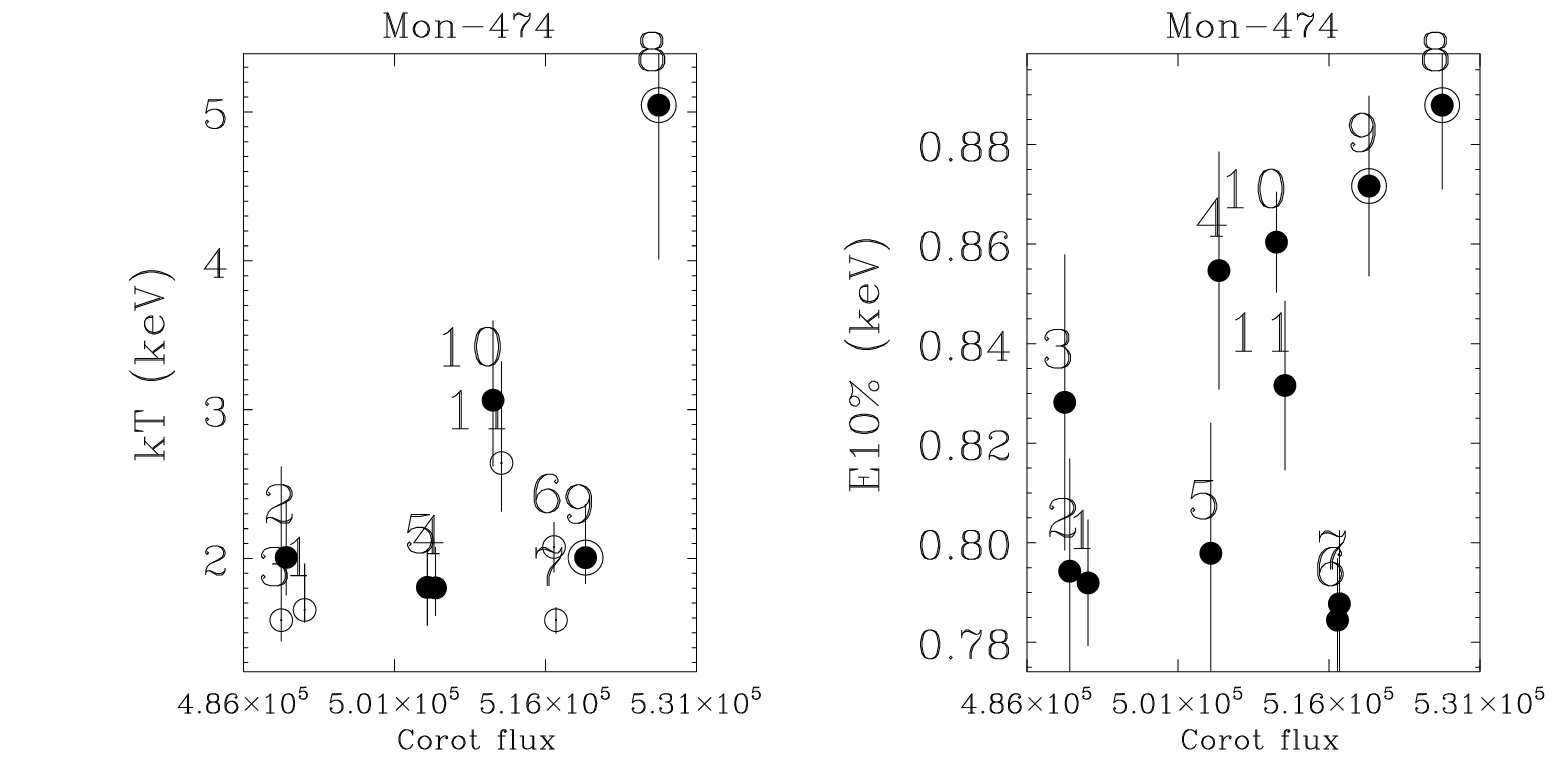}
	\includegraphics[width=18cm]{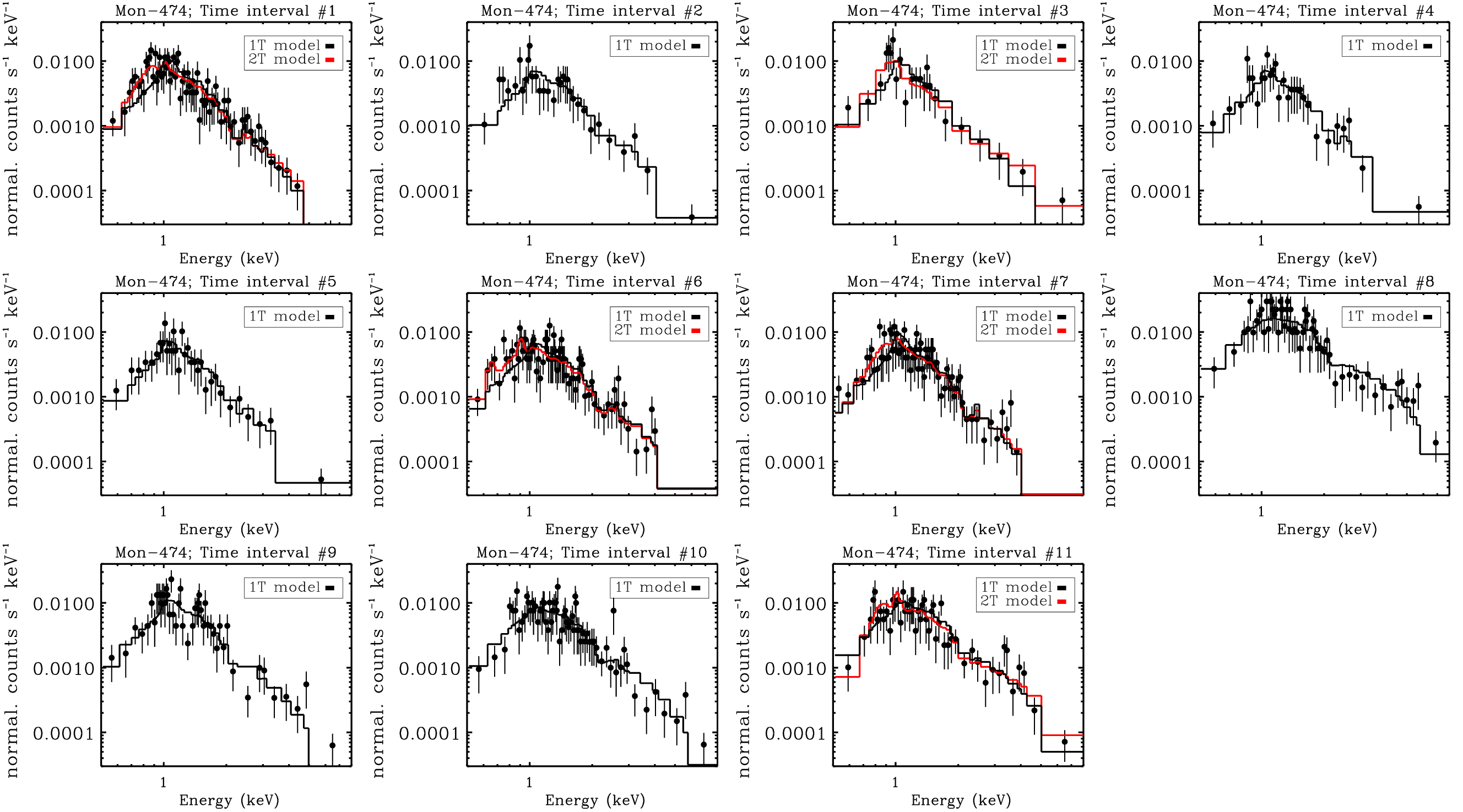}
	\includegraphics[width=8cm]{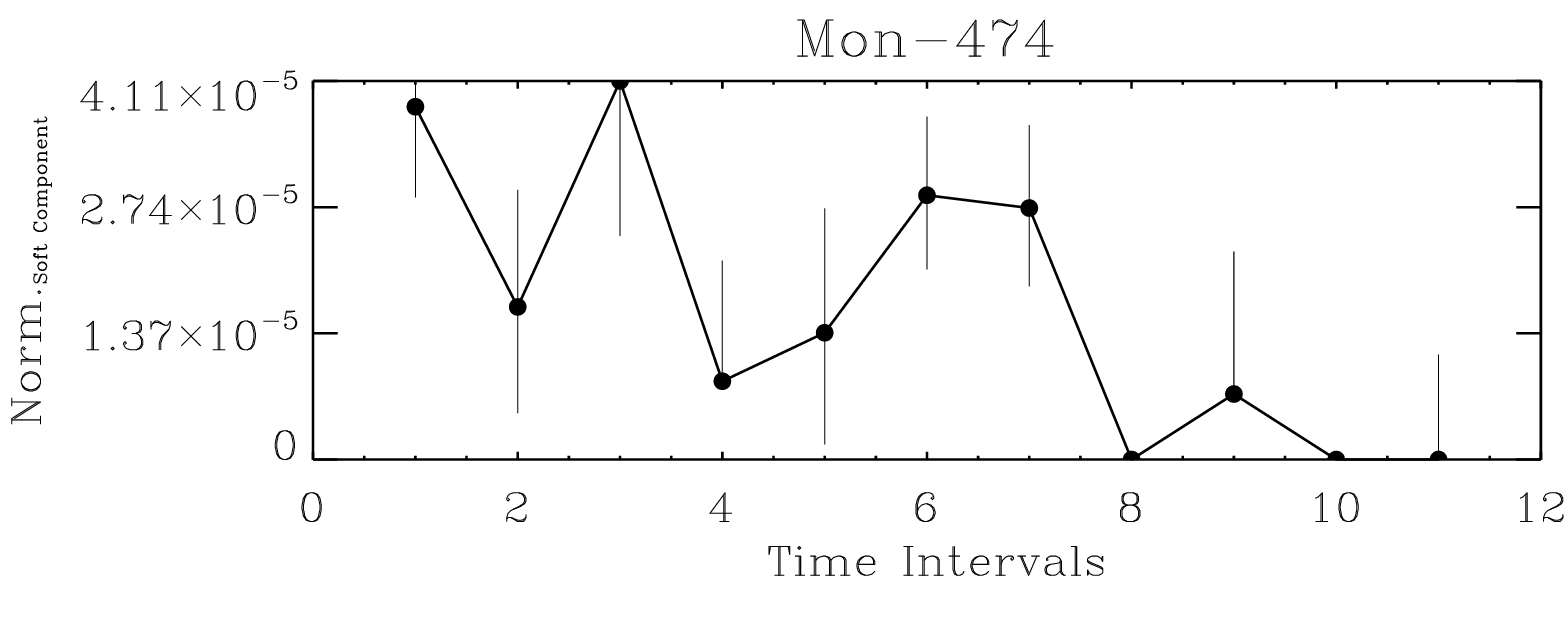}
	\caption{Time variability of the optical flux and X-ray properties of Mon-474 with the panel format and content as in Fig.~\ref{variab_mon808}.}
	\label{variab_mon474}
	\end{figure*}

%%%%%%%%%%%%%%%%5

        $ $ \par{\bf Mon-370:} The K5 star Mon-370 is among the 
strongest accretors in NGC~2264, with a H$\alpha$ EW roughly
corresponding to the 71\% quantile of the H$\alpha$ EW distribution of
all the accretors in our sample. The CoRoT light curve of this star
(Fig.~\ref{variab_mon370}) shows:

\begin{itemize}
\item A superposition of intense optical bursts in \#4.
\item In \#5, we isolate the begin of the rising phase of a burst occurring between the second and third {\em Chandra} frame.
\item An X-ray and optical flare during \#6.
\item A decline of optical emission by 9.1\% between \#9 and \#10.
\item Myriad small peaks during all the other intervals.
\end{itemize}

We can speculate that the light curve of this star is affected by 
myriad small bursts caused by an intense and chaotic accretion
process, with several short-lived accretion streams hitting the stellar
surface at various positions. The X-ray spectra observed in most of
the time intervals (shown in Fig.~\ref{variab_mon370}) confirm this
scenario, showing an intense soft X-ray spectral component (typically for
energies $\leq0.8\,$keV, sometimes $\leq1\,$keV). The most evident
soft X-ray spectral component is observed in the time interval \#4,
characterized by the most intense optical burst and the lowest
E$_{10\%}$ observed in this star. A smaller but evident soft X-ray
spectral component is also present in the time intervals \#7 and \#10,
all characterized by several small bursts. The X-ray spectral fit with
1T thermal plasma model is not well-constrained in most of the cases
(fourth column in Table \ref{mon370_2T_tab}): In the time intervals
\#1, \#2, \#4, \#7 and \#10 P$_\%$ is below the 5\% threshold. In some
cases, this is likely due to the fact that we have enough X-ray
photons to resolve different thermal components in the coronal
emission (e.g., \#2); in other cases by the presence of an evident soft
X-ray spectral component (e.g. \#4). An intense soft X-ray spectral
component is also evident in the average spectrum, as shown in the bottom
right panel in Fig.~\ref{variab_mon370}. \par

        The bottom left panel in Fig.~\ref{variab_mon370} shows the time
variability of the normalization of the soft component using 2T model
with fixed N$_H$ and kT. The time interval with the largest
normalization of the soft component is \#5, where we isolate the
rising part of a burst, followed by \#6 dominated by a X-ray flare.
The normalization observed in \#4 and \#7 are instead compatible with
those observed in the remainder intervals.  \par

\begin{table}
\caption{Predicted plasma temperatures for Mon-370 from the time-resolved X-ray spectral fits. The results from the 2T model are shown when the best fit with the 1T model is poorly constrained.}
\label{mon370_2T_tab}
\centering                       
\begin{tabular}{|r|r|r|r|r|r|r|}
\hline
  \multicolumn{1}{|c|}{\#} &
  \multicolumn{1}{|c|}{kT1} &
  \multicolumn{1}{|c|}{kT2} &
  \multicolumn{1}{|c|}{P(1T)} &
  \multicolumn{1}{|c|}{P(2T)} &
  \multicolumn{1}{|c|}{v$_{pre}$} &  
  \multicolumn{1}{|c|}{R$_{ff}^*$} \\
\hline
  \multicolumn{1}{|c|}{ } &
  \multicolumn{1}{|c|}{keV} &
  \multicolumn{1}{|c|}{keV} &
  \multicolumn{1}{|c|}{\%} &
  \multicolumn{1}{|c|}{\%} &
  \multicolumn{1}{|c|}{km/s} &  
  \multicolumn{1}{|c|}{R$_{star}$} \\
\hline
1 & $0.07_{-0.05}^{+0.17}$ & $0.90_{-0.22}^{+1.65}$ & 4.6      & 48.4 & $242_{-113}^{+207}$ & $4.5_{-3.2}^{+11.2}$\\  
2 & $0.22_{-0.15}^{+0.11}$ & $1.93_{-0.60}^{+1.81}$ & 0.0      & 7.8  & $429_{-187}^{+97 }$ & $1.4_{-0.5}^{+3.1}$ \\
3 & $1.19_{-0.06}^{+0.05}$ &                        & 8.7      &      &                     &                   \\
4 & $0.15_{-0.06}^{+0.06}$ & $1.42_{-0.44}^{+0.45}$ & 0.0      & 25.4 & $355_{-80 }^{+65 }$ & $2.1_{-0.6}^{+1.4}$ \\
5 & $1.06_{-0.14}^{+0.11}$ &                        & 12.7     &      &     &     \\
6 & $2.59_{-0.28}^{+0.41}$ &                        & 7.8      &      &     &     \\
7 & $0.14_{-0.06}^{+0.08}$ & $2.45_{-0.89}^{+2.29}$ & 0.0      & 92.4 & $342_{-83 }^{+87 }$ & $2.2_{-0.8}^{+1.7}$ \\
8 & $0.94_{-0.11}^{+0.06}$ &                        & 76.8     &      &     &      \\
9 & $0.23_{-0.05}^{+0.22}$ &                        & 16.2     &      &     &      \\
10& $0.20_{-0.11}^{+0.17}$ & $2.61_{-1.08}^{+3.74}$ & 0.13     & 97.8 & $409_{-134}^{+148}$ & $1.6_{-0.8}^{+1.9}$ \\
\hline
\multicolumn{7}{l}{$^*$ R$_{\rm ff}$ is calculated assuming M$_{star}=1.13\,$M$_{\odot}$, R$_{star}=1.64\,$R$_{\odot}$,} \\
\multicolumn{7}{l}{and age=4$\,$Myrs} \\  %v dall'infinito 512km/s
\end{tabular}
\end{table}

As for Mon-808, we fit the X-ray spectra of these time intervals with
two temperatures plasma models, which better reproduce the
observed spectra, primarily in the time intervals \#4 and \#7, but also
in \#2 and \#10 (see Fig.~\ref{variab_mon370} and Table
\ref{mon370_2T_tab}). The temperatures predicted by the best-fit models
are shown in Table \ref{mon370_2T_tab}, together with the pre-shock
velocity calculated using Eq.\ \ref{vel_eq} and the corresponding
free-fall launching distance from the central star, in units of
stellar radii. In the time intervals \#2, \#4, \#7, and \#10, the
pre-shock velocity ranges from 242$\,$km/s to 557$\,$km/s, sometimes
being compatible with the free-fall velocity from infinity
(512$\,$km/s). In order to better constrain these parameters, we fit
the average X-ray spectrum observed summing these four time intervals,
obtaining a good fit with a 2T thermal plasma model
(P$_\%\sim$100$\%$) with kT$_1$=$0.16_{-0.04}^{+0.08}\,$keV and
kT$_2$=$1.86_{-0.29}^{+0.55}\,$keV. The soft temperature corresponds
to a pre-shock velocity of $366_{-49}^{+83}\,$km/s from a free-fall
radius of $2.0^{+2.3}_{-0.4}\,$R$_{star}$. The X-ray spectral fit of
the average spectrum observed in the entire {\em Chandra} observation
results in similar temperatures (kT$_1=0.19_{-0.03}^{+0.02}\,$keV and
kT$_2=1.95_{-0.18}^{+0.18}\,$keV). Bearing in mind that this
calculation is approximate, the free-fall velocity is
smaller than the value from infinity, the upper limit of R$_{\rm ff}$ is
well below both the inner radius of the dusty disk predicted by the
SED analysis (8.9 stellar radii), and the co-rotation radius
calculated adopting the rotation period of 11.84 days (Venuti et al.\
in preparation), is equal to 13.8$\,$R$_{star}$.\par

%%%%%%%%%%%%%%%%%%%%

        $ $\par {\bf Mon-119:} Marginal evidence for an intense soft
X-ray spectral component during the optical bursts is also observed in
Mon-119 (see Fig.~\ref{variab_mon119}). This star has been discussed
in Sect. \ref{N$_H$_vs_dips}. We focus here on the optical bursts
during the time intervals \#3, \#7, and \#9. In particular, as shown
in the right panels of Fig.~\ref{variab_mon119}, during \#3 and \#9
both the plasma temperature and the 10\% and 25\% energy quantiles
suggest that the corresponding X-ray spectra are dominated by soft photons (see the spectra shown in Fig.~\ref{xspectra_mon119} and also in
the variability of E$_{10\%}$ shown in the Appendix \ref{e10lc_app}).
However, given the few X-ray photons detected, the two spectra are well
fitted by 1T thermal plasma models (P$_\%=84\%$ and 93\%,
respectively), as well as the average spectrum summing these two
time intervals (P$_{\%}$=58\%). \par

%%%%%%%%%%%%%%%%%%%%

        $ $\par {\bf Mon-326:} The M0 star Mon-326, whose variability
is shown in Fig.~\ref{variab_mon326}, is described as a star whose light
curve is dominated by accretion hot spots in \citet{StaufferCBA2014}.
It is not a strong accretor and in fact its CoRoT light curve does not
show many prominent bursts, except a 0.3 day long burst observed at
the beginning of the first {\em Chandra} frame (time interval \#1). In the first time interval, an
intense soft X-ray spectral component is evident (for
kT$\leq0.7\,$keV), and the normalization of the soft component of the best fit 2T model is
larger than in the other intervals (bottom panel). \par

        The fit with a 2T plasma model of the X-ray spectrum observed in
the first time interval (shown in Fig.~\ref{variab_mon326}) better
reproduces the observed soft X-ray emission below 0.7$\,$keV, with the
null-hypothesis probability increasing from almost zero to 34\%. In
\#1, the soft temperature obtained from the best fit 2T model is
kT$_{soft}=0.15^{+0.10}_{-0.08}\,$keV. Using Eq. \ref{vel_eq}, this
temperature corresponds to a pre-shock velocity of
$355^{+102}_{-113}\,$km/s, compatible with the free-fall velocity from
infinity (395$\,$km/s). \par

%%%%%%%%%%%%%%%%%%%%

        $ $ \par{\bf Mon-474:} Fig.~\ref{variab_mon474} shows the
variability of Mon-474, the only G type star \citep{DahmSimon2005}
analyzed in this section, which is actively accreting from its disk.
This star was listed as a star with accretion bursts in
\citet{StaufferCBA2014}. The CoRoT light curve during the {\em
Chandra} observations (but also in the remainder) is dominated by
several small bursts:

\begin{itemize}
\item \#4 is dominated by an intense optical burst with the optical emission increasing by 3.8\%.
\item In \#5, we isolate a transition from the large optical burst observed in \#4 and a sequence of small bursts.
\item The intervals \#1, \#2, \#3, \#6, \#7, \#10, \#11 are dominated by several small optical bursts.
\item \#8 and \#9 are dominated by a bright optical and X-ray flare.
\end{itemize}
	
The time resolved X-ray spectra (central panels of Fig.\
\ref{variab_mon474}) show an intense soft X-ray spectral component below
1$\,$keV in the time intervals \#1 and \#6. It may also be present even if less evident
during the interval \#7. No evident soft X-ray spectral
component is observed during \#3 and \#11. The X-ray spectra observed
in some of these intervals such as \#6 and \#7 are not well-fit
with 1T thermal plasma models, and they also show soft X-ray photon
energy quantiles. Table \ref{mon474_2T_tab} shows the plasma temperatures
predicted by the best fit 1T or 2T (when the former is poorly
constrained) plasma model and the associated null-hypothesis
probabilities. \par

        Repeating for Mon-474 the calculation made for Mon-808,
adopting a stellar mass of 1.9$\,$M$_{\odot}$ and a radius of
4.04$\,$R$_{\odot}$, we obtain a pre-shock velocity of the accreting
material of $744^{+85}_{-334}\,$km/s in the time interval \#1 and
$458^{+84}_{-92}\,$km/s in the time interval \#6, both compatible with
the free fall velocity from infinite distance from the star
(423$\,$km/s).  \par
	
\begin{table}
\caption{Predicted plasma temperatures for Mon-474 from the time resolved X-ray spectral fits. The results from the 2T model are shown when the best fit with the 1T model is poorly constrained.}
\label{mon474_2T_tab}
\centering                       
\begin{tabular}{|r|r|r|r|r|}
\hline
  \multicolumn{1}{|c|}{Interval} &
  \multicolumn{1}{|c|}{T1} &
  \multicolumn{1}{|c|}{T2} &
  \multicolumn{1}{|c|}{P(1T)} &
  \multicolumn{1}{|c|}{P(2T)} \\
\hline
  \multicolumn{1}{|c|}{ } &
  \multicolumn{1}{|c|}{keV} &
  \multicolumn{1}{|c|}{keV} &
  \multicolumn{1}{|c|}{\%} &
  \multicolumn{1}{|c|}{\%} \\
\hline
1  &	$0.66^{+0.16}_{-0.35}$  & $2.65^{+1.54}_{-0.65} $& 1.9  & 88.0 \\
2  &	$2.01^{+0.27}_{-0.26}$  &                        & 63.7 &      \\
3  &	$0.87^{+0.24}_{-0.19}$  & $23.1^{}_{-19.98}    $ & 1.7  & 79.4 \\
4  &	$1.80^{+0.28}_{-0.19}$  &                        & 76.5 &      \\
5  &	$1.80^{+0.26}_{-0.25}$  &                        & 98.7 &      \\
6  &	$0.25^{+0.10}_{-0.09}$  & $2.16^{+0.69}_{-0.45}$ & 0.5  & 26.9 \\
7  &	$0.78^{+0.21}_{-0.38}$  & $2.27^{+1.15}_{-0.45}$ & 3.5  & 64.2 \\
8  &	$5.05^{+1.68}_{-1.04}$  &                        & 83.6 &      \\
9  &	$2.01^{+0.25}_{-0.18}$  &                        & 11.0 &      \\
10 &	$3.06^{+0.54}_{-0.44}$  &                        & 13.2 &      \\
11 &	$0.54^{+0.09}_{-0.15}$  & $6.73^{}_{-3.76}    $  & 0.0  & 8.4  \\
\hline\end{tabular}
\end{table}

	\begin{figure*}[]
	\centering	
	\includegraphics[width=9.5cm]{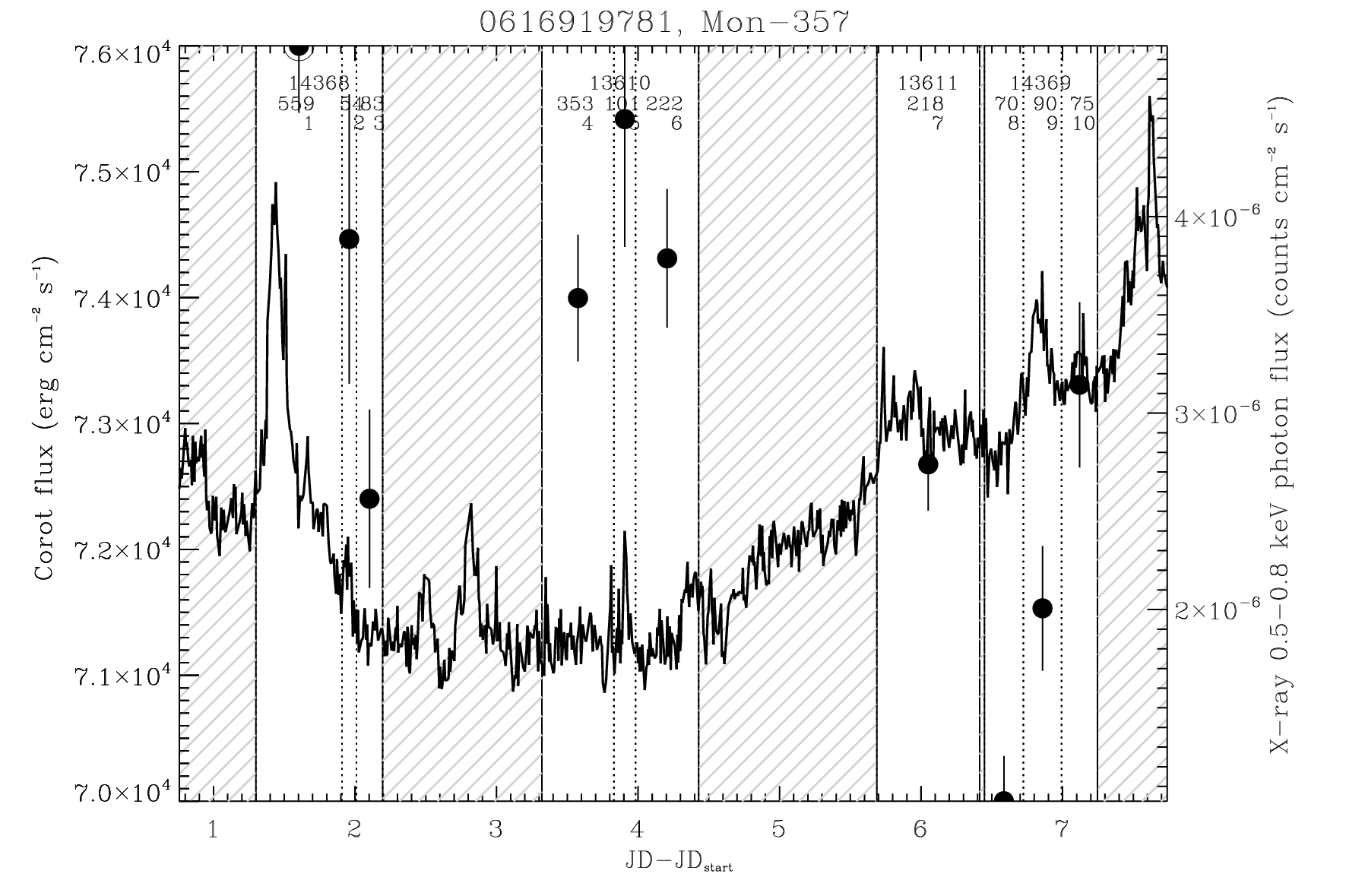}
	\includegraphics[width=8cm]{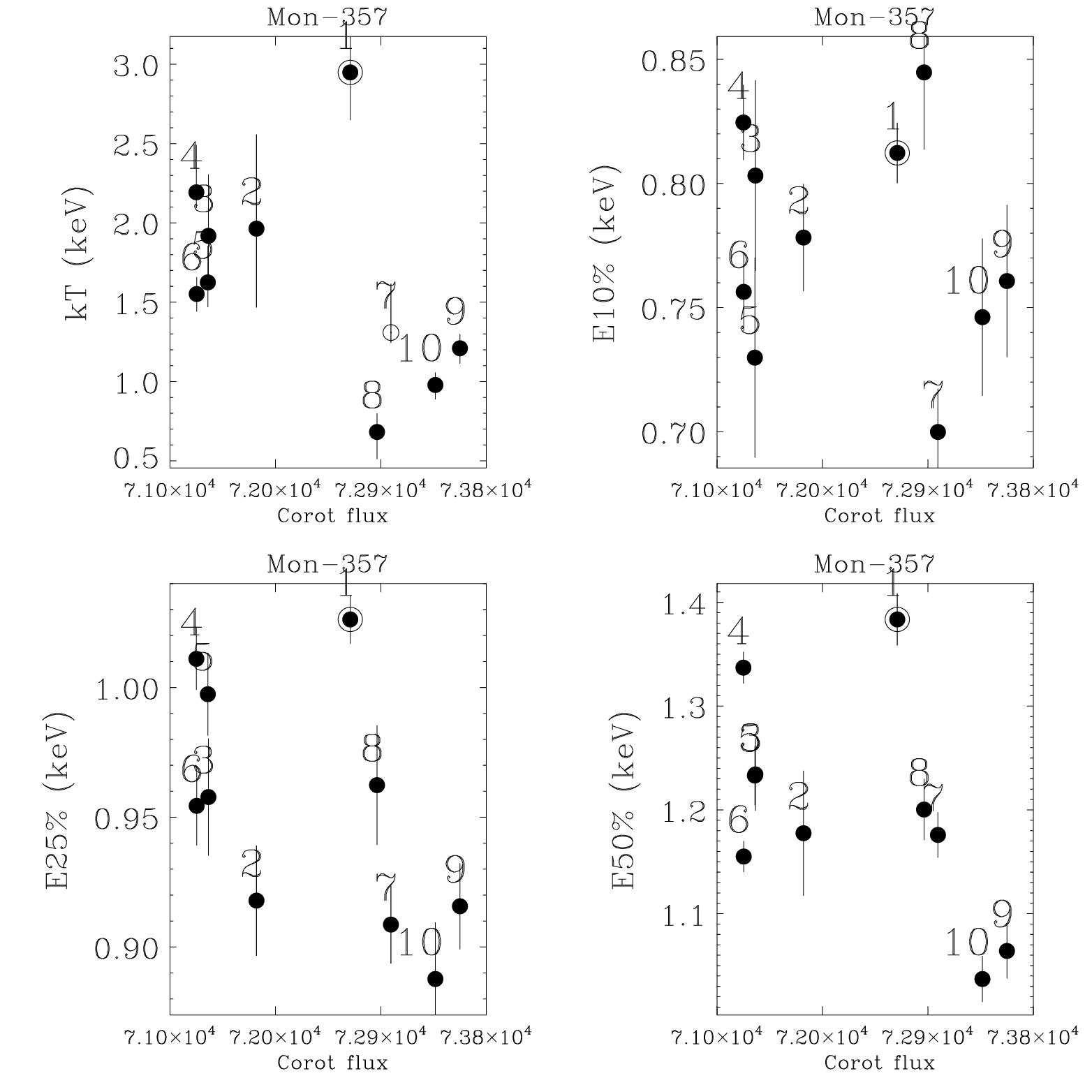} 
	\includegraphics[width=18cm]{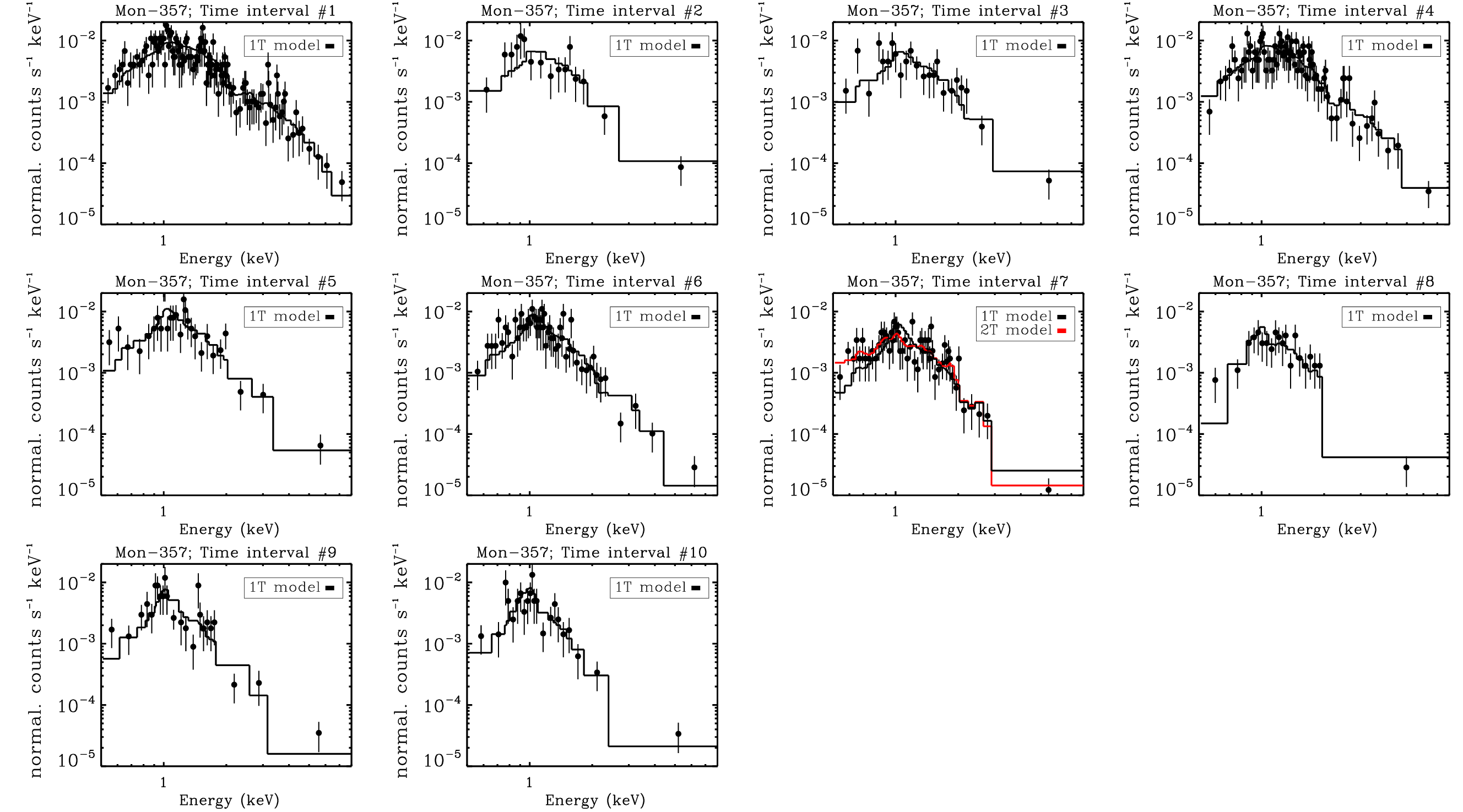}
	\includegraphics[width=7cm]{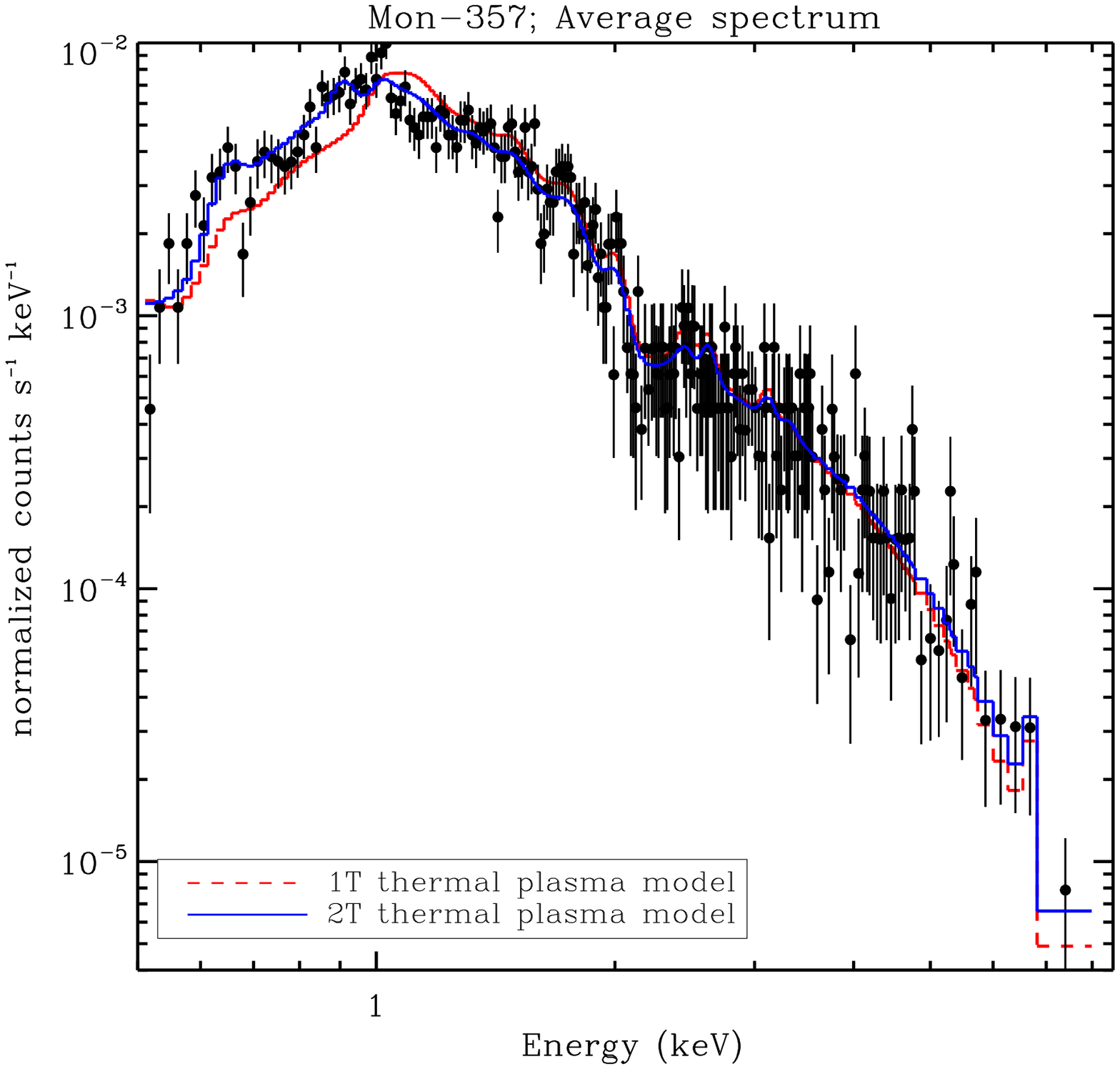}
	\caption{Time variability of the optical flux and X-ray properties of Mon-357 with the panels format and content as in Fig.~\ref{variab_mon808}.}
	\label{variab_mon357}
	\end{figure*}

  	\begin{figure*}[]
	\centering	
	\includegraphics[width=10cm]{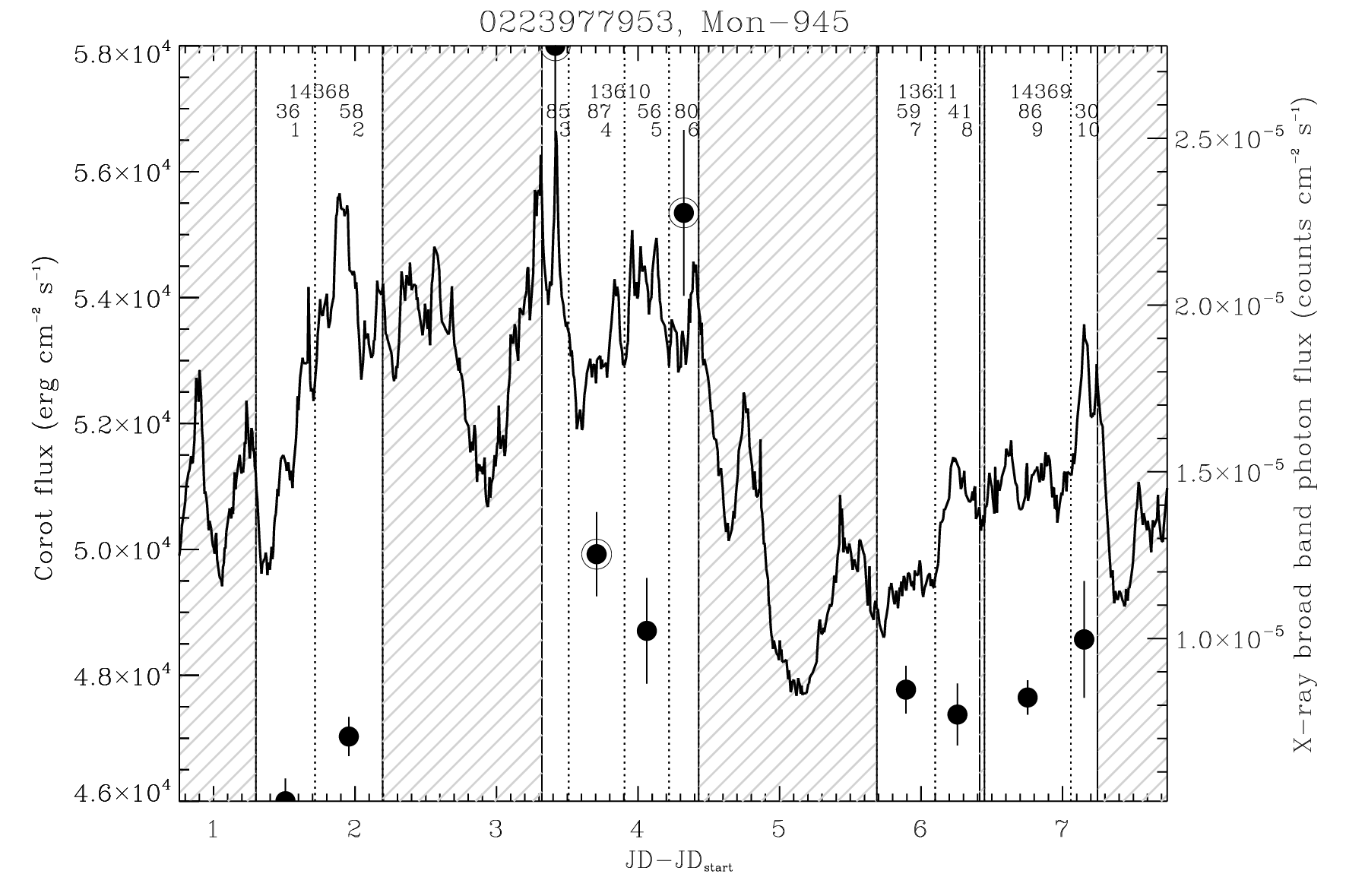} 
	\includegraphics[width=8cm]{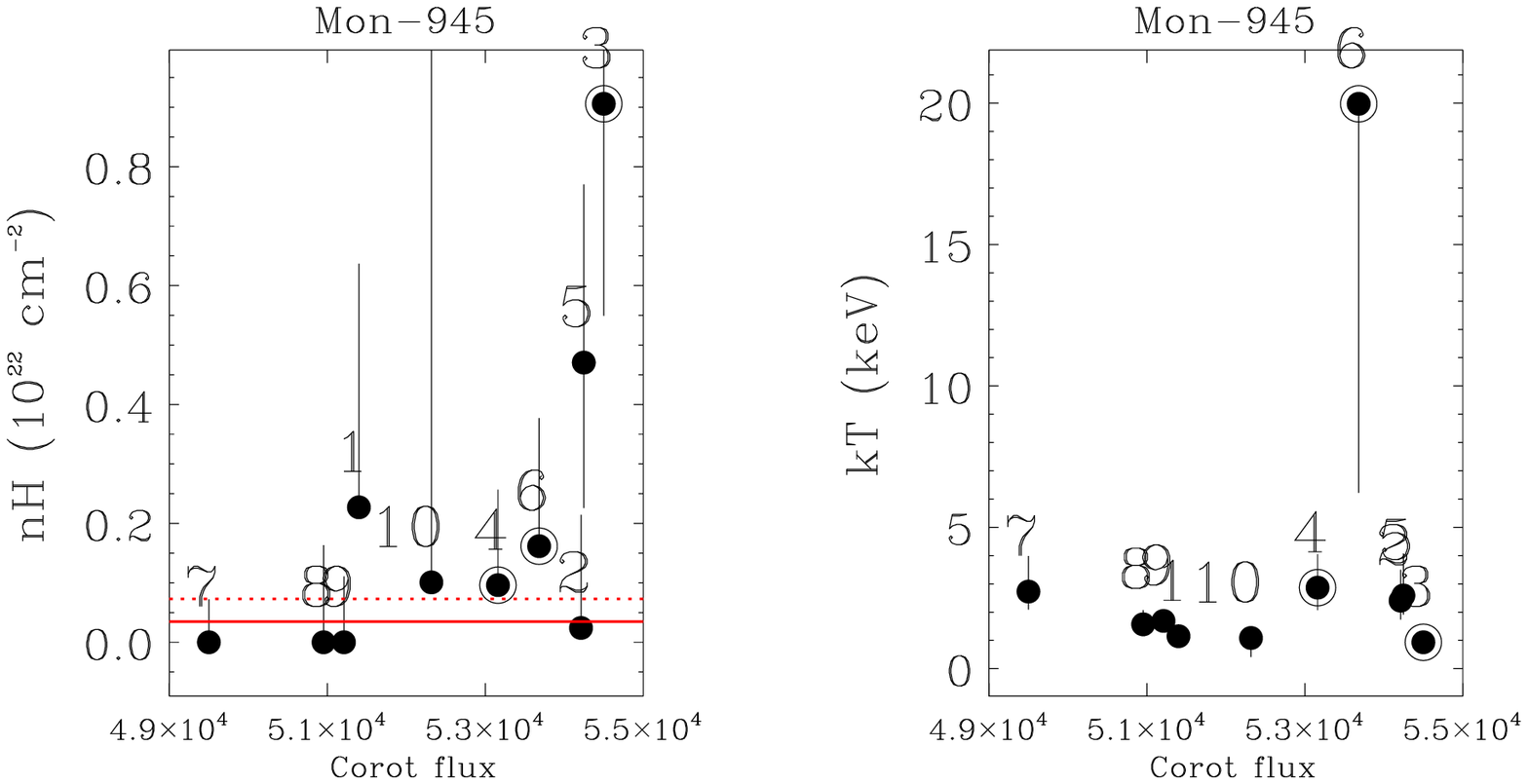}	
	\includegraphics[width=18cm]{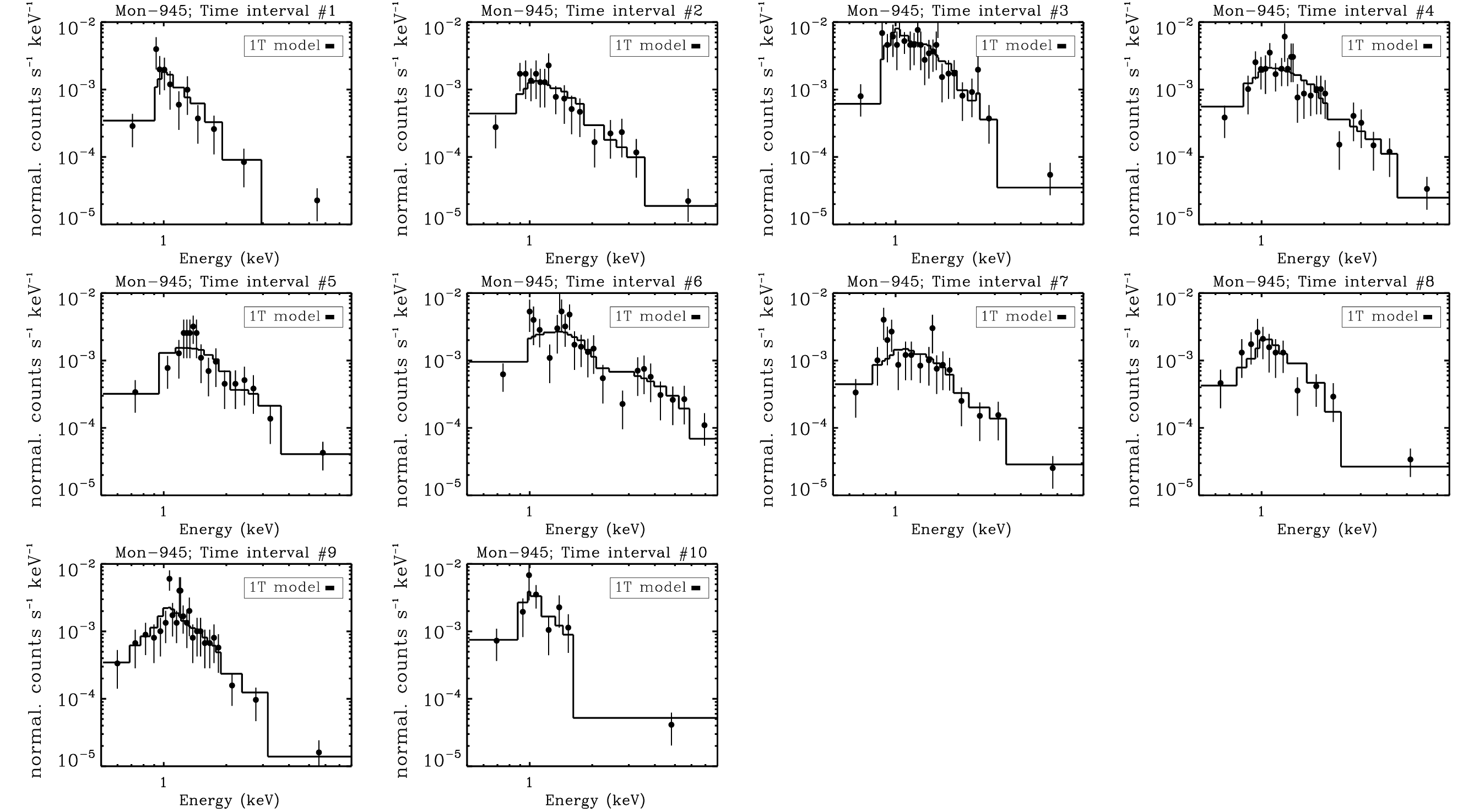}
	\includegraphics[width=5.5cm]{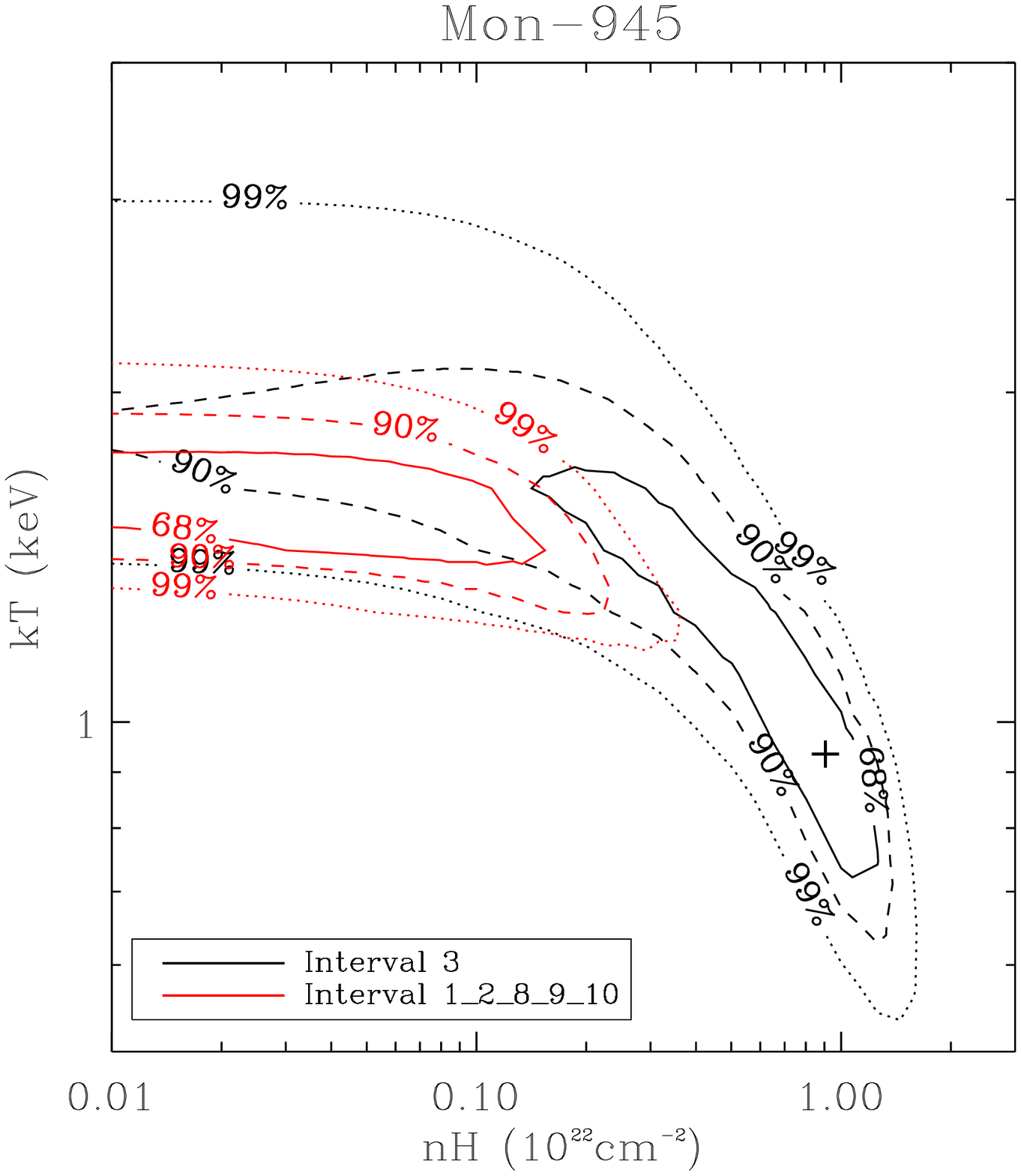}
	\includegraphics[width=5.5cm]{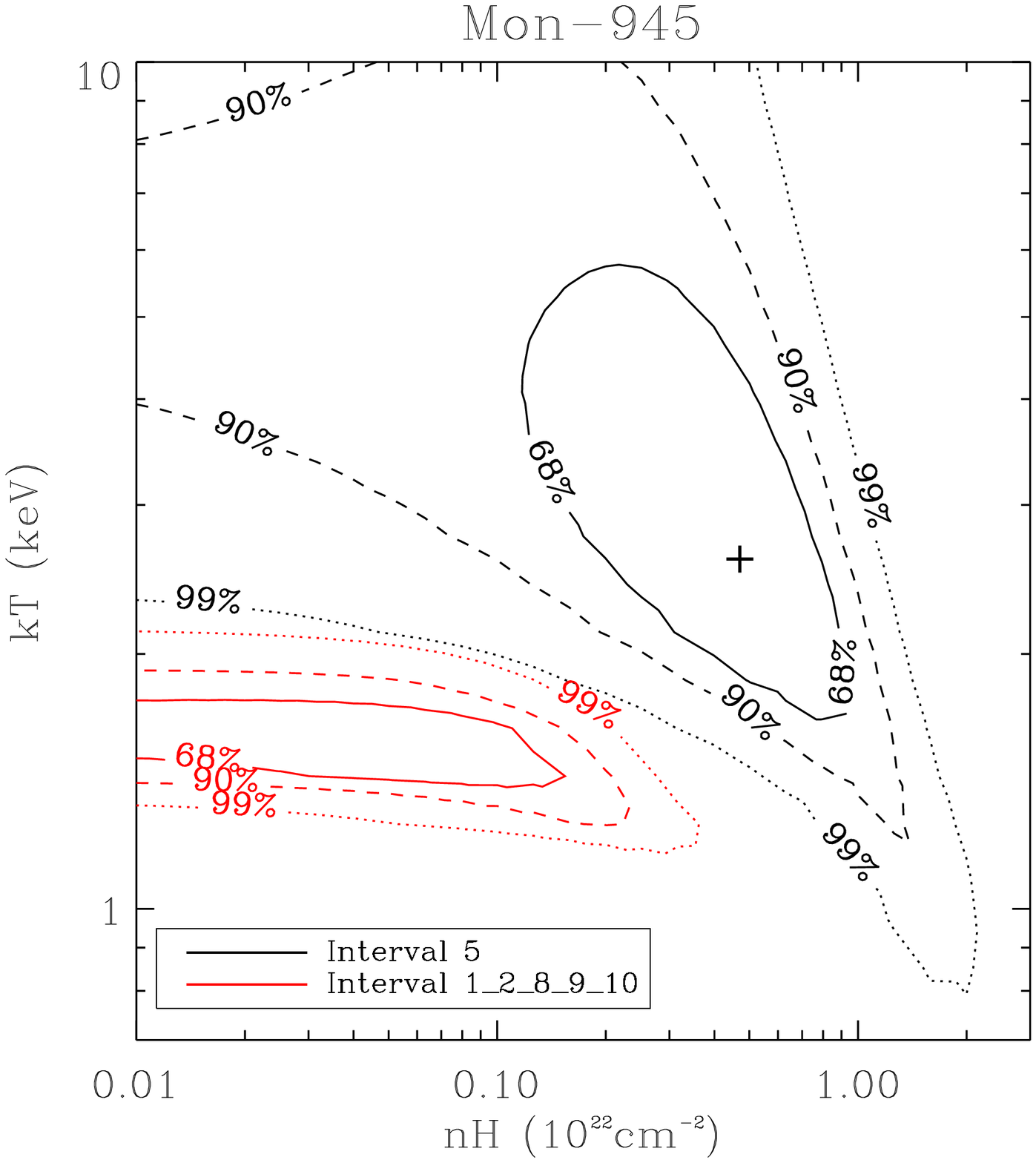}
        \caption{Time variability of the optical flux and X-ray
properties of Mon-945 with the panel format and content as in
Fig.~\ref{variab_mon456}. Contour levels in the C-stat space
corresponding to the 68\%, 90\%, and 99\% statistical confidence
regions from the fit of the X-ray spectrum of Mon-945 observed during
the intervals \#3 (bottom left panel, black contours) and \#5 (bottom
right panel, black contours) are also shown. In both panels the red
contours are obtained from the average spectrum observed in
\#1+\#2+\#8+\#9+\#10. The cross marks the values obtained from the
best fit.}
	\label{variab_mon945}
	\end{figure*}

	\begin{figure*}[!ht]
	\centering	
	\includegraphics[width=10cm]{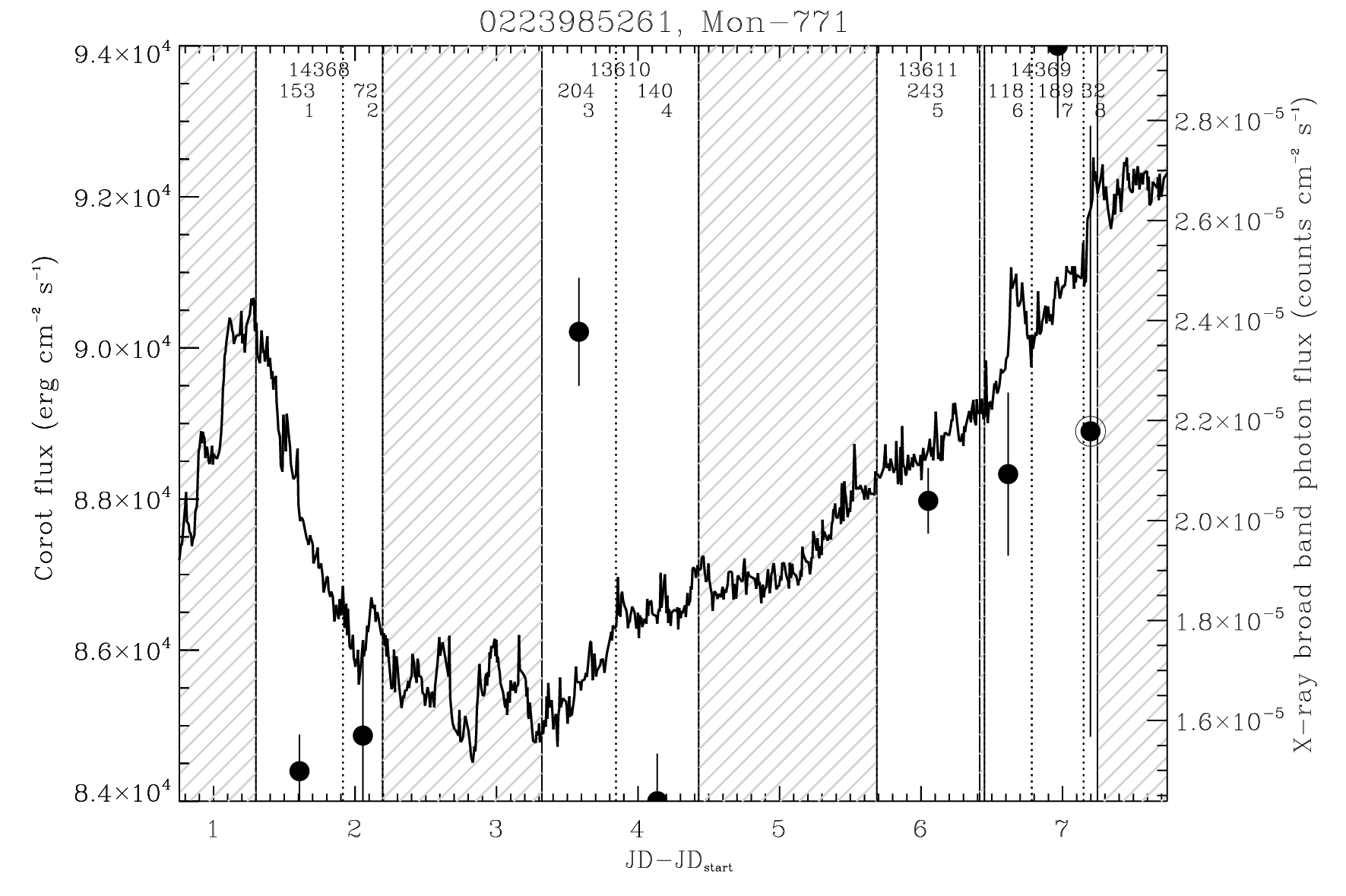} 
	\includegraphics[width=8cm]{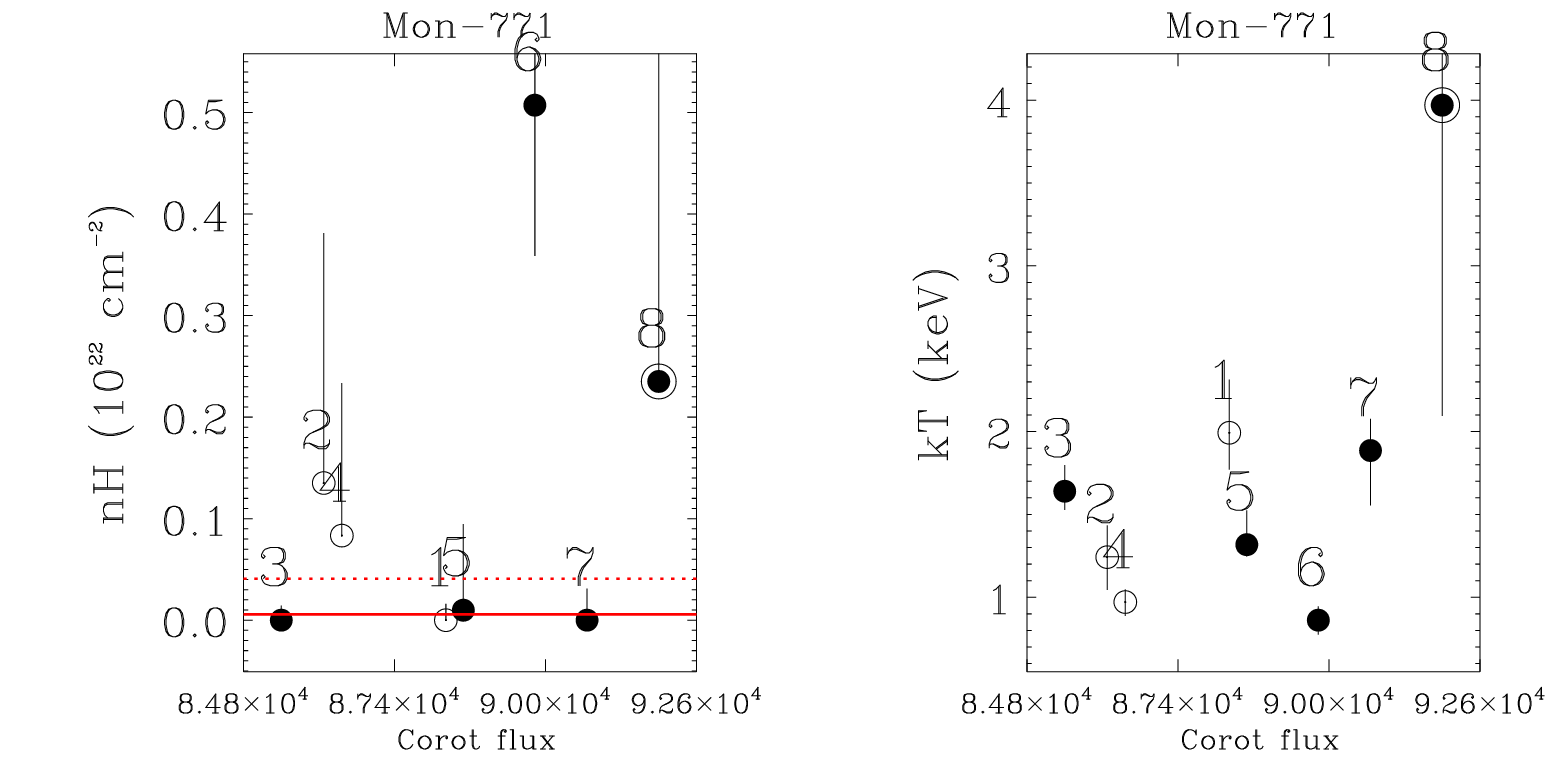}	
	\includegraphics[width=18cm]{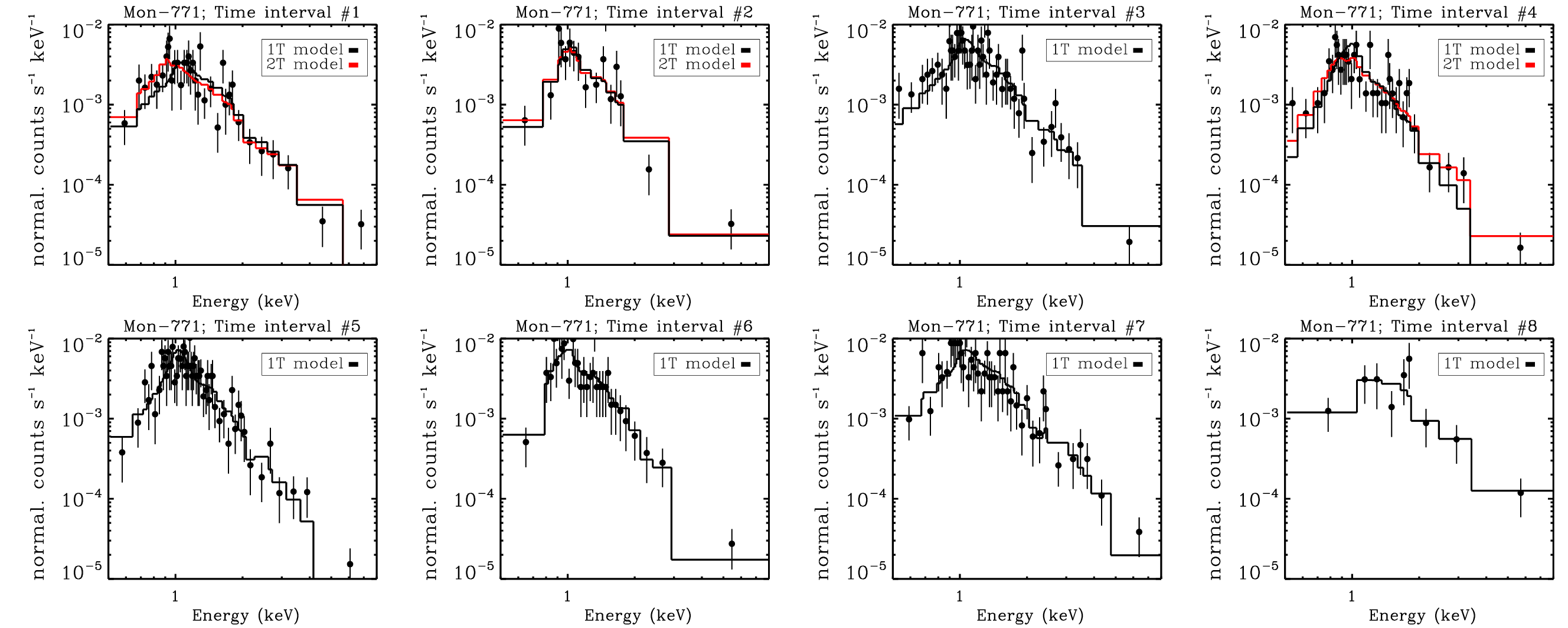}
	\includegraphics[width=6cm]{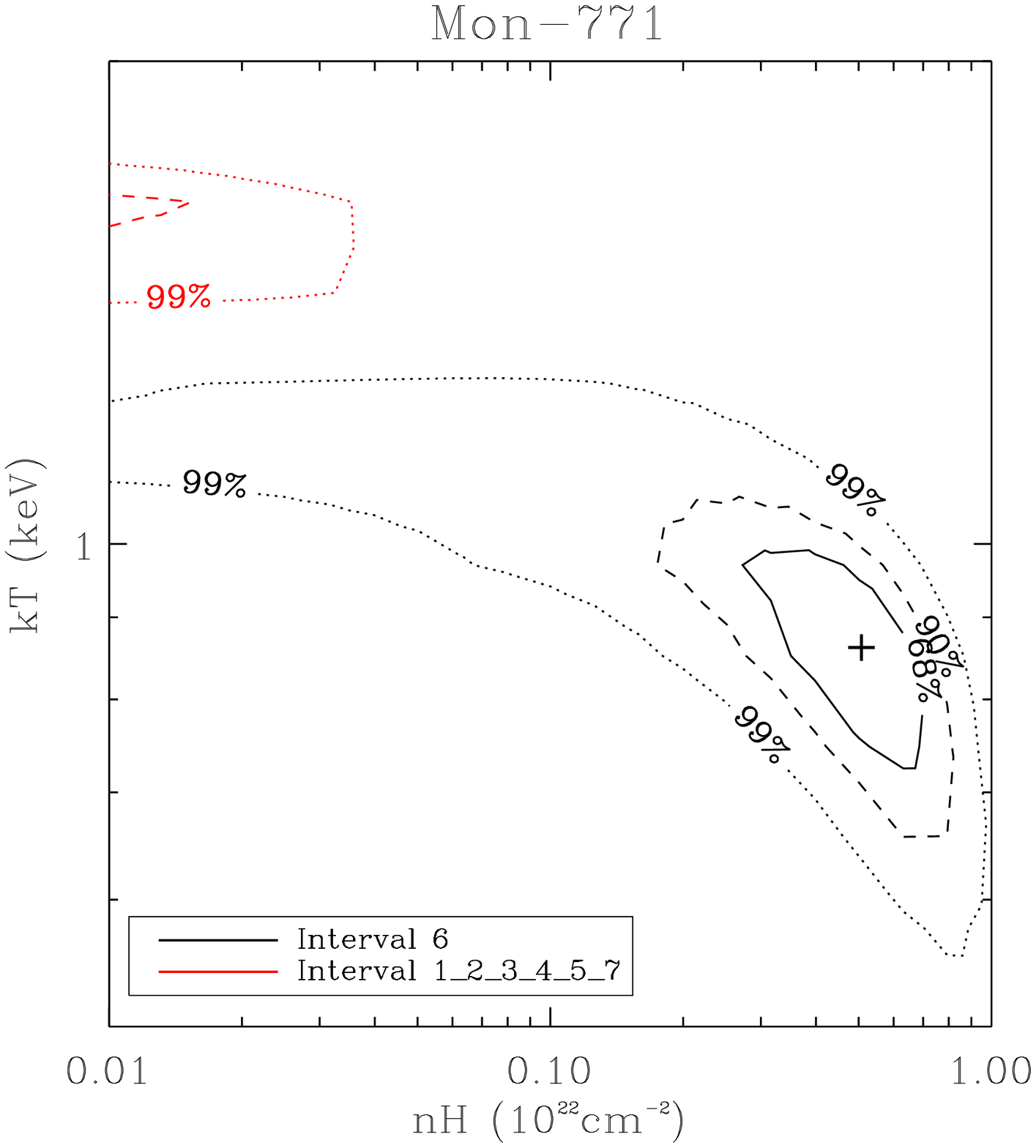}
	\includegraphics[width=7cm]{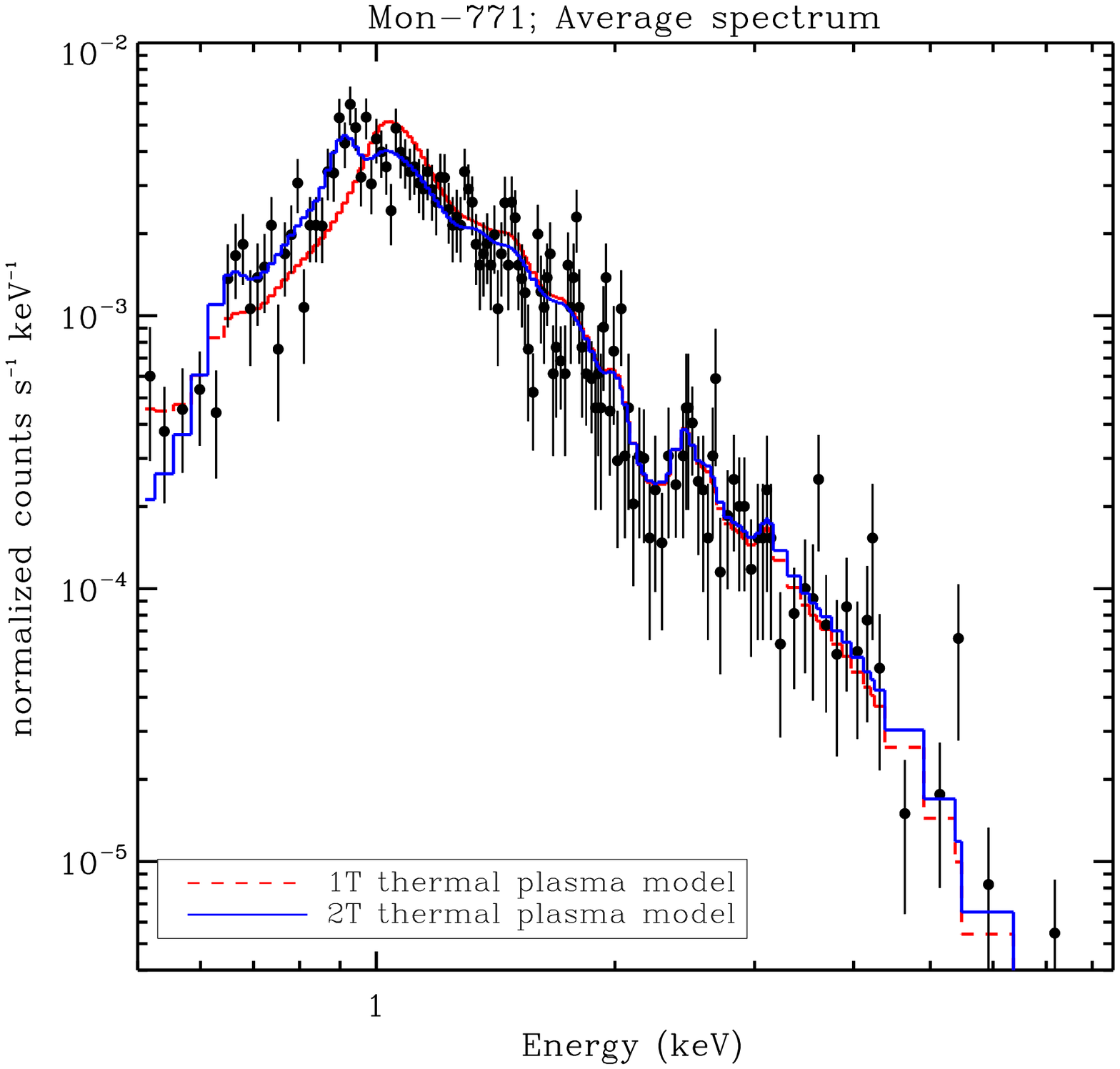}
        \caption{Time variability of the optical flux and X-ray
properties of Mon-771 with the panel format and content as in
Fig.~\ref{variab_mon456}. Contour levels in the C-stat space
corresponding to the 68\%, 90\%, and 99\% statistical confidence
regions from the fit of the X-ray spectrum of Mon-771 observed during
the time interval \#6 (black contours) are compared to that obtained
from the average spectrum observed during the remainder intervals
(excluding the flare, red contours). The cross marks the values
obtained from the best fit.}
	\label{variab_mon771}
	\end{figure*}

%%%%%%%%%%%%%%%%%%%%

        $ $ \par{\bf Mon-357:} The variability of the K5 star Mon-357
is shown in Fig.~\ref{variab_mon357}. Even though it has a small H$\alpha$ EW,
this star has a CoRoT light curve dominated by several bursts-like
features of various intensity:

\begin{itemize}
\item A bright optical and X-ray flare is observed during the time interval \#1. 
\item Large optical bursts are observed during \#5, \#6, \#7, and \#9.
\item Small optical bursts are observed during \#2 and \#10.
\item No interesting features are observed during \#3, \#4, and \#8.
\end{itemize}

The variability of the X-ray properties (Fig.~\ref{variab_mon357})
suggests that the X-ray spectrum becomes softer and the X-ray emitting
plasma colder with increasing CoRoT flux (ignoring the flare). The
Spearman's rank correlation test is significant for kT (correlation
coefficient $\rho=-0.72$ and the two-side significance of its
deviation from zero $P(\rho)=0.03$), E$_{25\%}$ ($\rho=-0.78$,
$P(\rho)=0.01$), and E$_{50\%}$ ($\rho=-0.70$, $P(\rho)=0.04$), while
it is not significant for E$_{10\%}$. This suggest that the observed variability 
is dominated by accretion. Intense soft X-ray spectral
components in the time-resolved X-ray spectra are observed in some
intervals. In \#7, when the CoRoT
light curve is dominated by a superposition of small bursts, it is
necessary to adopt a 2T thermal plasma model to obtain a good fit
(P$_{\%}$=3.3\% with 1T model, 93.2\% with 2T model). No useful
information is provided by the time variability of the normalization
of the soft component of the best fit 2T model (not shown) because of
the associated large errors. \par

The X-ray spectrum during \#7 shows a peak of soft X-ray emission at
about 0.7$\,$keV. The soft temperature predicted by the best fit 2T
thermal plasma model is well constrained, equal to
$0.06^{+0.01}_{-0.02}\,$keV. Since {\em Chandra}/ACIS-I is not sensitive
to such soft emission, we do not calculate the pre-shock velocity
from this plasma temperature. However, it must be noted that a similar
value for the temperature of the soft component is also obtained from
the average spectrum among those intervals with large bursts:
\#5+\#7+\#9+\#10 (P$_{\%}$=0.10), suggesting that the
soft part of the X-ray spectrum of Mon-357 is dominated by emission
from the accretion spots. \par

In Fig.~\ref{variab_mon357} we show the average X-ray spectrum of
Mon-357 with the best fit 1T (P$_{\%}=$0\%) and 2T (P$_{\%}=$88\%)
thermal models. The presence of an intense soft X-ray spectral
component is evident. 

%%%%%%%%%%%%%%%%%%%%

$ $ \par		
The search for increasing X-ray absorption during the optical bursts
has been less prolific. This is, however, not surprising given that
this effect can be observed only if the accretion streams obscure the coronal
active regions when the accretion hot spots are clearly visible, which is
strongly dependent on the geometry of the accretion and the
distribution of the active regions in the stellar corona. In only two
cases (Mon-945 and Mon-771), described below, there is evidence for
such a correlation.\par

$ $ \par{\bf Mon-945: }Mon-945 is a K4 accreting star, with evidence
for a larger veiling when the star is optically brighter, which
supports the classification as ``burster'' by \citet{CodySBM2014AJ} and
\citet{StaufferCBA2014}. The CoRoT light curve during the {\em Chandra}
observations is dominated by a large number of bursts (Fig.
\ref{variab_mon945}):

\begin{itemize}
\item Optical bursts are observed in \#2, \#5, \#8, \#9, and \#10.
\item \#1 and \#7 show dip-like optical features, more prominent in the former interval.
\item \#3, \#4, and \#6 are dominated by optical and X-ray flares.
\end{itemize}

 The bottom panels of Fig.~\ref{variab_mon945} show the X-ray spectra
observed during the time intervals, where there is no evidence of
intense soft X-ray spectral components. However, of particular
interest is the burst/flare observed during time interval \#3. The identification of this event by our
automatic routines as an X-ray flare is uncertain given that the rising part is not observed by
{\em Chandra}, as it lies between the first and second {\em Chandra} frames,
and we observe only the decaying phase at the beginning of the second
{\em Chandra} frame (the time interval \#3). Additionally, in \#3 we observe
the highest value of X-ray absorption
(N$_H$=$0.91_{-0.36}^{+0.14}\times10^{22}\,$cm$^{-2}$) and a low
plasma temperature (kT=$0.93^{+0.09}_{-0.16}\,$keV), similar to that
observed in other intervals. These properties are more compatible with
an accretion burst rather than a X-ray flare, with increasing X-ray
absorption due to the accreting gas falling into the line of sight.
This is the only time interval where the N$_H$ obtained from spectral
fit is significantly larger than zero, together with that observed in \#5
(N$_H$=$0.47^{+0.30}_{-0.14}\times10^{22}\,$cm$^{-2}$) during which
the CoRoT light curve shows several bursts. Fig.~\ref{variab_mon945}
also shows the contours in the C-stat space obtained for Mon-945 during the
time intervals \#3 and \#5, supporting the evidence for larger N$_H$.

%%%%%%%%%%%%%%%%%%%%
        $ $ \par {\bf Mon-771: }The moderately accreting K4 star
Mon-771 has not been classified as a ``burster'' by
\citet{CodySBM2014AJ}. However, its CoRoT light curve during the
{\em Chandra} observations (Fig.~\ref{variab_mon771}) is characterized by
several small bursts:

\begin{itemize}
\item During \#1, the optical emission declines by 5.5\%, with a subsequent slow rising phase longer than the whole {\em Chandra} observation.
\item An intense optical burst is observed during \#6.
\item Small optical bursts are observed during \#2 and \#4.
\item The small interval \#8 is dominated by the rising part of an optical and X-ray flare.
\item No interesting features are observed during \#3, \#5 and \#7.
\end{itemize}

 The time-resolved X-ray spectra are well fit by 1T thermal
plasma models except in \#1, \#2, and \#4. In \#1, an intense soft
X-ray spectral component between 0.7$\,$keV and $1\,$keV may be present.
The fit with a 2T thermal plasma model predicting
N$_H$=$0.21^{+0.59}_{-0.21}\times10^{22}\,$cm$^{-2}$,
kT$_1$=$0.31^{+0.67}_{-0.16}\,$keV, and
kT$_2$=$2.59^{+3.05}_{-0.89}\,$keV is acceptable (P$_{\%}$=$31.5\%$).
Repeating the calculation for the pre-shock velocity, we obtain a
value compatible with the free-fall velocity from infinity. Similar
results are obtained for \#4, where the spectral fit using 2T thermal
plasma model is good (P$_{\%}$=$66.9\%$) and it predicts
N$_H$=$0.0^{+0.22}_{-0.0}\times10^{22}\,$cm$^{-2}$,
kT$_1$=$0.68^{+0.20}_{-0.33}\,$keV, and
kT$_2$=$2.69^{+4.63}_{-1.34}\,$keV. The spectral fit for the time
interval \#2 is poorly constrained even using 2T thermal plasma models
(P$_{\%}=0.3\%$).\par  

Some evidence of intense soft X-ray spectral component is 
shown in the average spectrum (Fig.~\ref{variab_mon771}). The fit of
the X-ray spectrum with 1T thermal plasma model is poorly constrained
(P$_{\%}=0.0$), while an acceptable fit is obtained with 2T thermal
model (P$_{\%}=13\%$). This predicts a soft temperature of
kT$_2$=$0.86^{+0.09}_{-0.02}\,$keV, which is more likely a cool
coronal component.

 The variability of the X-ray properties during the defined time
intervals is shown in the upper right panels of Fig.
\ref{variab_mon771}. They suggest that the X-ray absorption is higher
in the time interval \#6, dominated by an intense burst. The C-stat
contour levels shown in Fig.~\ref{variab_mon771} confirm this
conclusion. The N$_H$ obtained during the time interval \#6 is the
only one different than zero at $>$90\% confidence level, and it is
significantly larger than the X-ray absorption obtained from the
average spectrum observed in the remanding intervals
(N$_H$=$0.0^{+0.02}_{-0.0}\times10^{22}\,$cm$^{-2}$, excluding \#8
dominated by a flare).  \par

	\begin{figure*}[]
	\centering	
	\includegraphics[width=10cm]{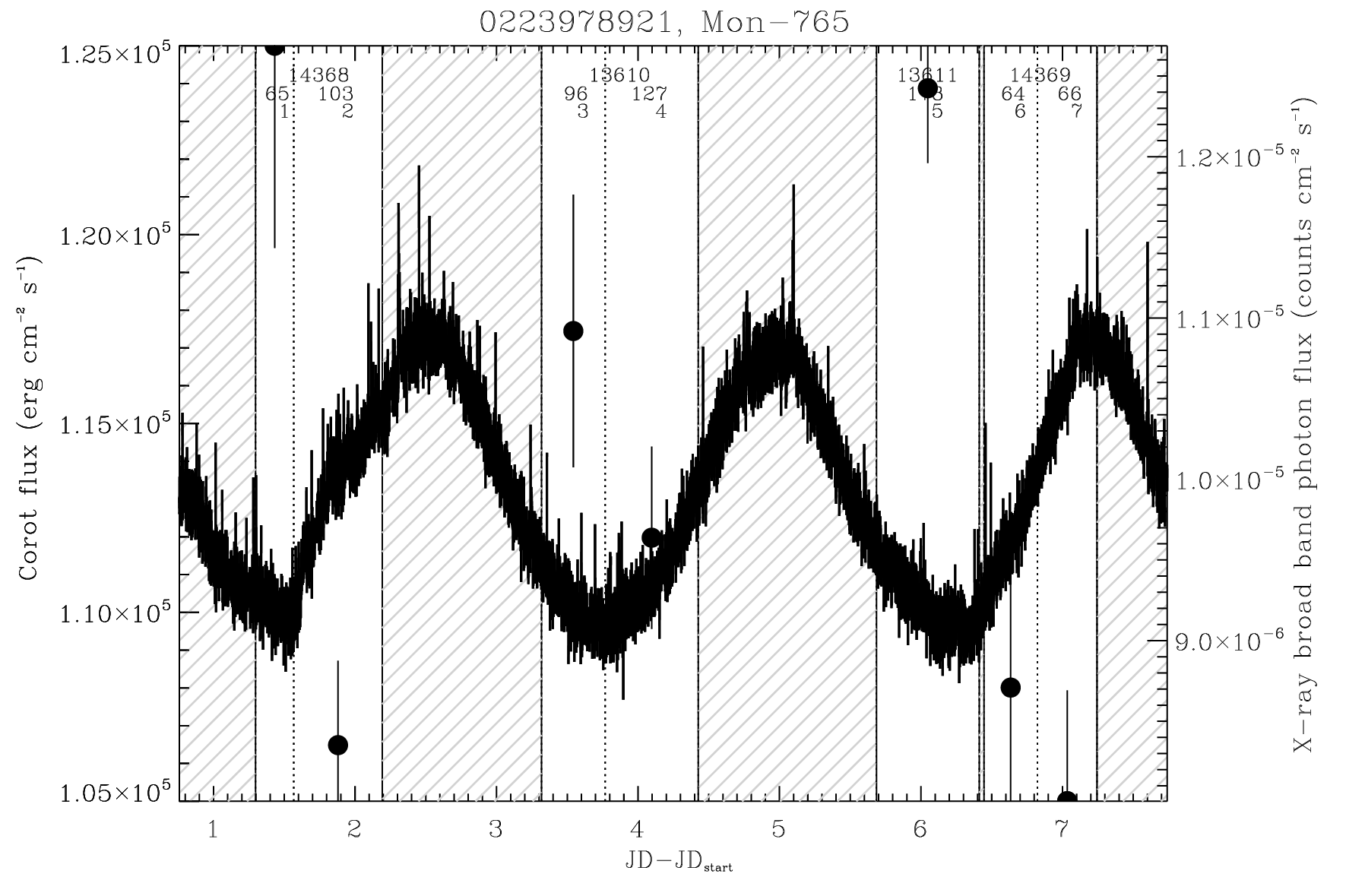}  
	\includegraphics[width=8cm]{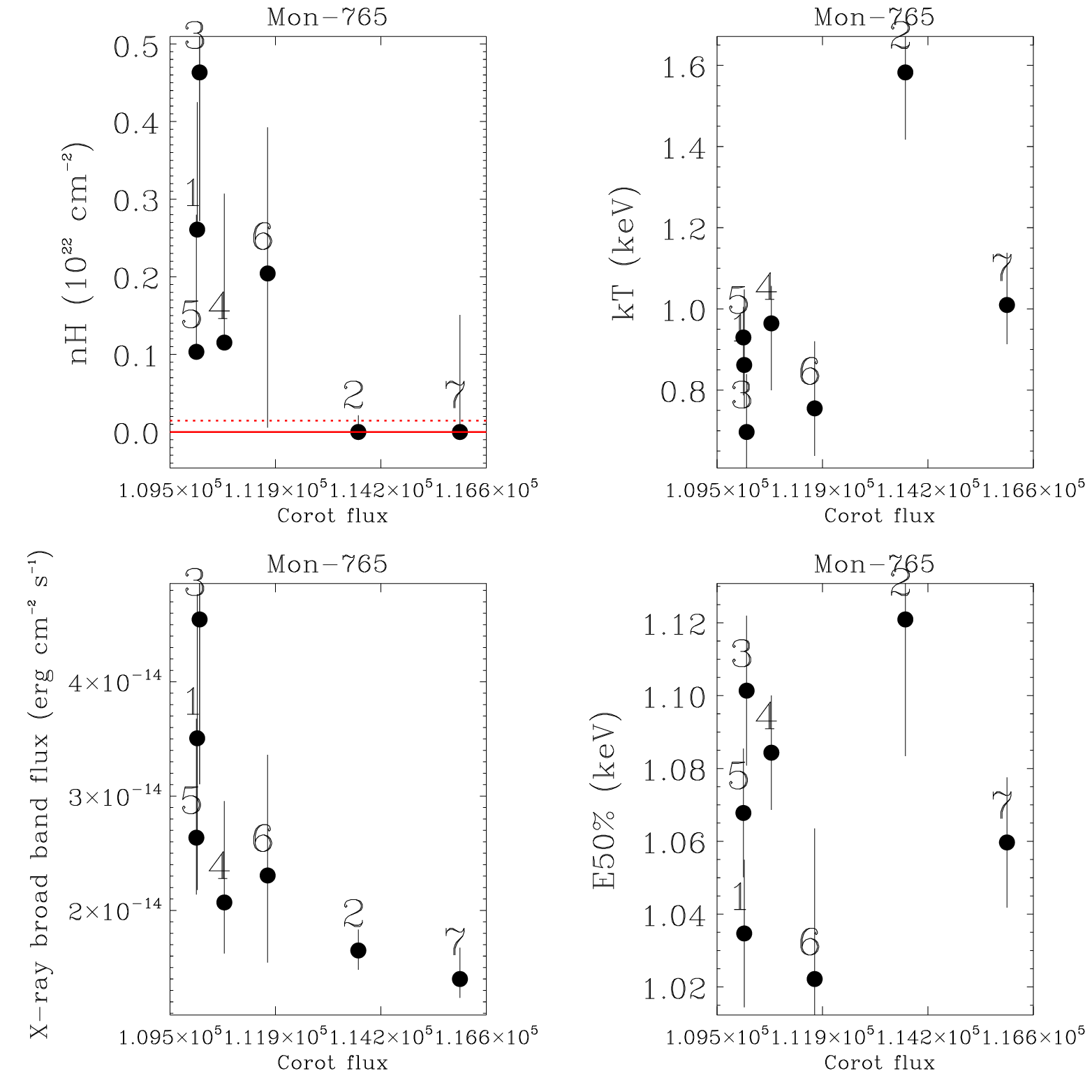}	
	\includegraphics[width=10cm]{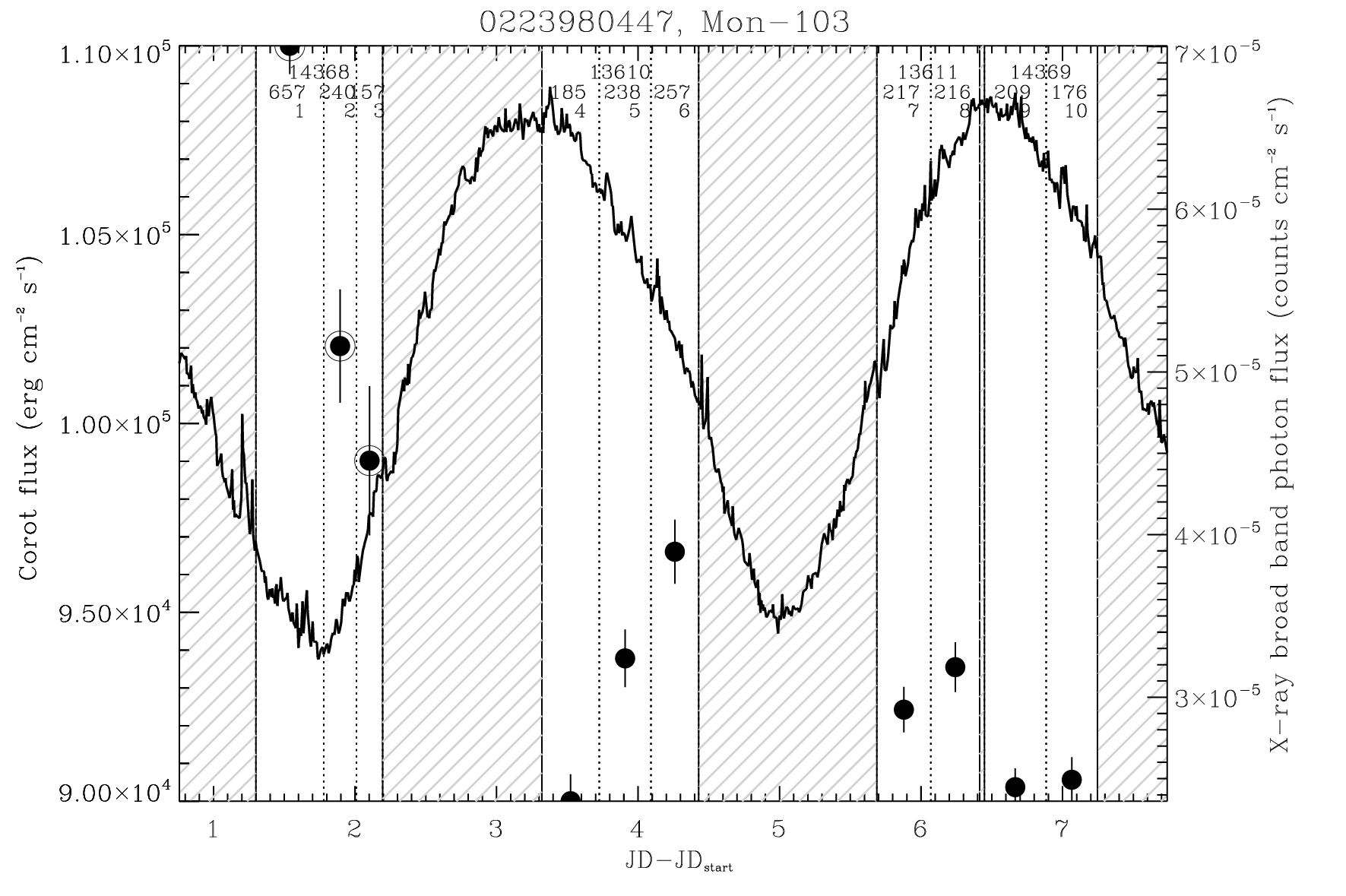} 
	\includegraphics[width=8cm]{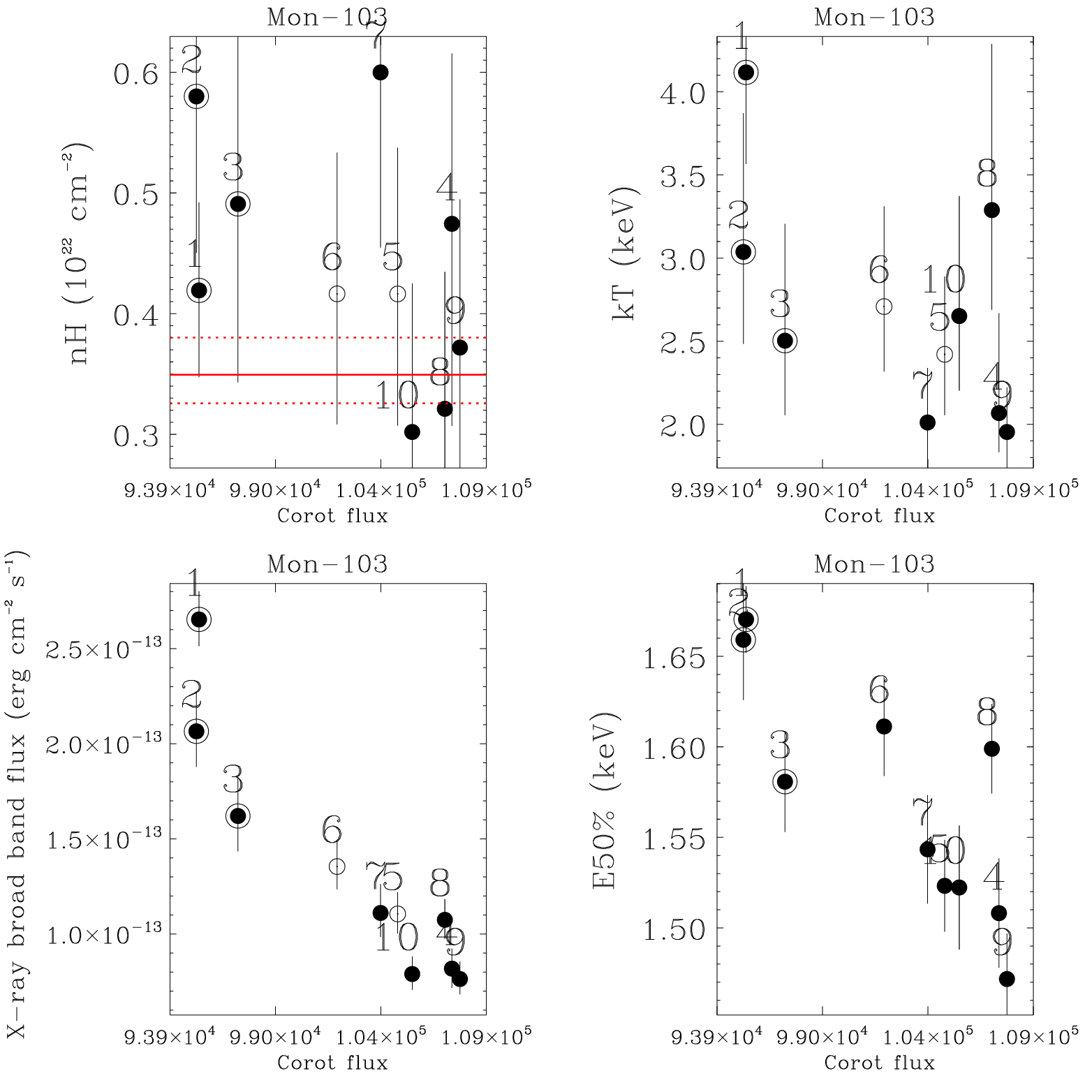}	
	\includegraphics[width=10cm]{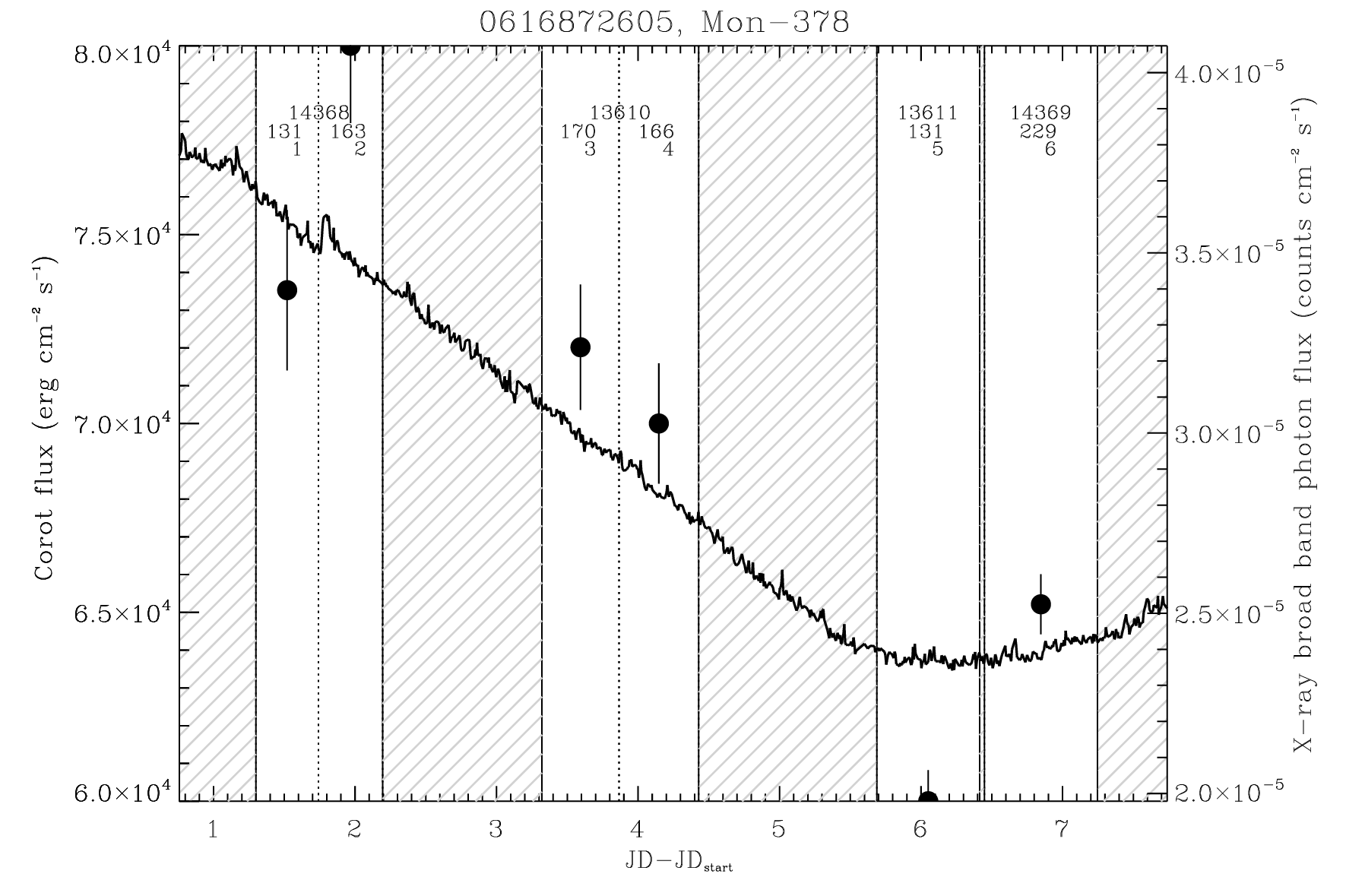} 
	\includegraphics[width=8cm]{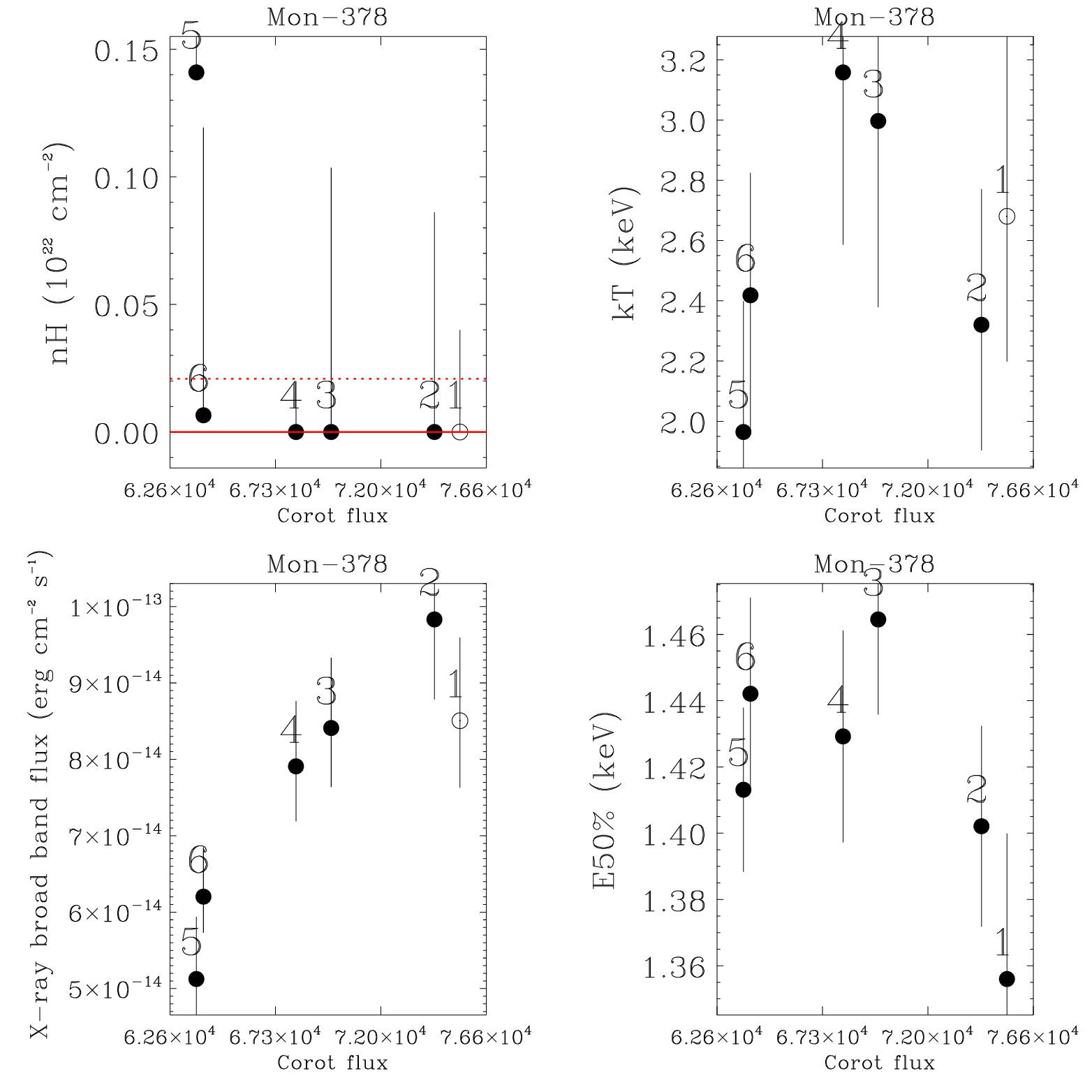}	
	\caption{Time variability of the optical flux and X-ray properties of stars with disks with periodic variability as shown in Fig.~\ref{variab_mon412}, with the difference that the X-ray quantities shown in the left panels are: N$_H$ (in units of $10^{22}\,$cm$^{-2}$), kT (in keV), F$_X$ (in erg$\,$cm$^{-2}\,$s$^{-1}$), and E$_{50\%}$ in keV.} 
	\label{periodic_disk}
	\end{figure*}

%%%%%%%%%%%%%%%%%%
	
\subsection{Stars with disks with periodic and quasi-periodic variability}
\label{disk_periodic}

Periodic optical and X-ray variability can be observed in stars with
disks. The emission from accretion hot spots, stellar occultation by
circumstellar material, optical darkening due to photospheric dark
spots, and enhanced X-ray emission due to coronal active regions can be
modulated by stellar rotation. In the sample of NGC~2264 stars with disks
observed with CoRoT and {\em Chandra}, six stars show 
periodic behavior in the CoRoT light curve, and three among them, discussed below, have coherent
optical vs.\ X-ray flux variability (Fig.~\ref{periodic_disk}). \par

$ $ \par {\bf Mon-765: } Mon-765 is a moderately accreting K1 star.
The CoRoT light curve (top left panel in Fig.~\ref{periodic_disk})
shows periodic variability with a period of $\sim$$2.7\,$days. The
variability of the X-ray properties is shown in the upper right
panels. The plasma temperature does not show significant variability
with the exception of the interval \#2. The variability of the X-ray
and CoRoT fluxes is anticorrelated, as confirmed by a Spearman rank
correlation test ($\rho$=-0.82, P($\rho$)=0.02). Rather than
rotational modulation of accretion hot spots or stellar occultation,
this behavior is more likely due to rotational modulation of spatially
coincident photospheric dark spots and coronal active regions. \par

$ $ \par {\bf Mon-103: } The hypothesis made for Mon-765 holds also
for Mon-103, a non-accreting K6 star whose variability are assumed to be
due to cold spots by \citet{StaufferCRH2016AJ}. This star has an evacuated
inner disk as suggested by its near-infrared colors. The CoRoT light curve of Mon-103 is very regular (central left panel in Fig.\
\ref{periodic_disk}) and a statistically significant anticorrelation
between the optical flux variability and both the X-ray flux and
median photon energy variability (respectively, $\rho$=-0.89,
P($\rho$)=0.006 and $\rho$=-0.78, P($\rho$)=0.04, in both cases
removing the first three time intervals dominated by flares). \par

$ $ \par {\bf Mon-378: }  Mon-378 is a K5.5 star with a different
behavior than Mon-765 and Mon-103. As reported by \citet{CodySBM2014AJ},
the light curve of this star is likely periodic with superimposed
fading episodes likely due to increasing circumstellar extinction.
This star has been listed as a star with periodic flux dips in
\citet{StaufferCMR2015}. The CoRoT light curve during the four
{\em Chandra} frames (bottom left panel of Fig.~\ref{periodic_disk}) is
observed in a decaying phase, which is part of its ``periodic''
behavior. We do not observe evidence of increasing X-ray absorption during the decline of optical emission, since during all the defined time intervals the X-ray spectral fit admit solutions at low N$_H$ within 68\% confidence. However the lower right panels show that the optical flux variability is correlated with the X-ray flux variability ($\rho$=0.94, P($\rho$)=0.004). This behavior is compatible with the scenario of recurrent occultation of the central star by circumstellar material. 

%%%%%%%%%%%%%%%%%%%%%%%%%%%%%%%%%%%%%%%%%%%%%%%%%%%%%%%%%%%%%%%%%%%%%%%%

\subsection{Summary of observed simultaneous events}
\label{suca}

The total of stars with disks with good detection with CoRoT and $Chandra$ analyzed in this paper (i.e. Sect. \ref{N$_H$_vs_dips}, \ref{N$_H$_vs_burst}, \ref{disk_periodic} and Appendix \ref{others_app}) is 51:

\begin{itemize}
\item A total of 24 stars are analyzed as ``dippers'', i.e. they show dips in their CoRoT light curves occurring during the $Chandra$ observations. Among them, in the 7 stars discussed in Sect. \ref{N$_H$_vs_dips} X-ray absorption increases during the optical dips. 
\item 20 stars are analyzed as ``bursters'', i.e. their CoRoT light curves show rapid bursts occurring during the $Chandra$ observations. Among them, 5 stars show an intense soft X-ray spectral component typically below 1$\,$keV during the optical bursts, while 2 stars show increasing X-ray absorption during the bursts (Sect. \ref{N$_H$_vs_burst}). It must be noted that 9 stars are analyzed both as ``bursters'' and ``dippers''.
\item 6 stars have periodic CoRoT light curves.
\item 8 stars are not analyzed because: i) they do not show any dominant phenomenon, or ii) X-ray detected photons are too few, or iii) their CoRoT mask is contaminated by nearby bright sources. 
\end{itemize}

It is important to understand whether the observed variations of N$_H$ in the time intervals can be due to statistical fluctuations rather that real variability. To this aim, we compare the occurrence of time intervals, including flares but excluding intervals with poor X-ray spectral fits, where N$_H$ is larger (within at least 1$\,\sigma$ significance) than the average absorption estimated by spectral fitting to the average spectrum of each star. In the 7 stars discussed in Sect. \ref{N$_H$_vs_dips} we define in total 60 time intervals, and the X-ray absorption is significantly larger than the individual average value in 13 intervals (21.6\% of the cases), and only in two cases (3.3\% of the cases) the CoRoT light curves do not decline or show dips during the intervals. In the other stars analyzed as ``dippers'' N$_H$ is larger than the individual average value in 4 time intervals over 103 defined (3.9\%), in 2 of which the optical emission declines. Considering together the stars analyzed as ``bursters'', those with periodic variability and those not analyzed, N$_H$ is larger than the individual average value in 17 time intervals over 145 defined (11.7\%), but only in 5 cases (3.4\%) the CoRoT light curves clearly do not decline or show dips. In these stars N$_H$ is observed to be variable compared to the individual average value in half of the time intervals with respect the stars analyzed in Sect. \ref{N$_H$_vs_dips}. Considering that stars with bursts and those not analyzed because of few detected X-ray photons can show intrinsically variable X-ray absorption, this strongly suggests that observed variability of X-ray absorption in these stars is not dominated by statistical fluctuations. \par 

    It is more easy to verify that the observed soft X-ray spectral components below 1$\,$keV in the 5 stars analyzed in Sect. \ref{N$_H$_vs_burst} is not due to statistical fluctuations. Such spectral component is in fact observed in 13 time intervals over 43 defined for these stars (occurrence of 30.2\%). Considering all the other stars analyzed in this paper, we observe 5 possible soft X-ray spectral components below 1$\,$keV (in three cases in stars analyzed as bursters) over 283 time intervals (1.7\% of the cases).  \par
%%%%%%%%%%%%%%%%%%%%%%%%%%%%%%%%%%%%%%%%%%%%555

\section{Discussion \& Conclusions}
\label{conclusions}

In this paper, we analyze the simultaneous variability in optical (from
CoRoT) and X-rays (from {\em Chandra}/ACIS-I) of stars with disks in
NGC~2264, focusing on two samples of stars, those with dips in their
CoRoT light curve due to variable extinction, and those with optical
bursts due to accretion. \par

%%%%%%%%%%%%%%%%%%%%%%%%%%%%%%%%%%%%%%%%%%%%%%%%%%%%%%%%%%%%%%%%%%%%%%%%%%%%%%%%%%%%%%%%%%%%%%%%%%%%%
  \subsection{The N$_H$/A$_V$ ratio during the optical dips}
    \label{N$_H$avratio_sec}

The hypothesis that stars with
disks can be affected by variable extinction due to circumstellar
material was first introduced by
\citet{Joy1945} and \citet{HerbstHGW1994AJ}. If circumstellar material is part of
large warps at or inside the co-rotation radius in the disk, then the
occultation occurs recurrently over several periods, and the dips are
deep. These systems are called ``AA~Tau like'' from the star which has
been the precursor of this class \citep{BouvierCAC1999}. The CSI~2264
project has also revealed that much shorter and irregular optical dips
can be observed in disk-bearing stars. \citet{StaufferCMR2015}
demonstrate that the most likely explanation for narrow and aperiodic
dips is the occultation of the central star by dust trapped in
unsteady accretion streams, while larger dips are more likely due to
disk warps. \par

	\begin{figure}[]
 	\centering	
 	\includegraphics[width=8.0cm]{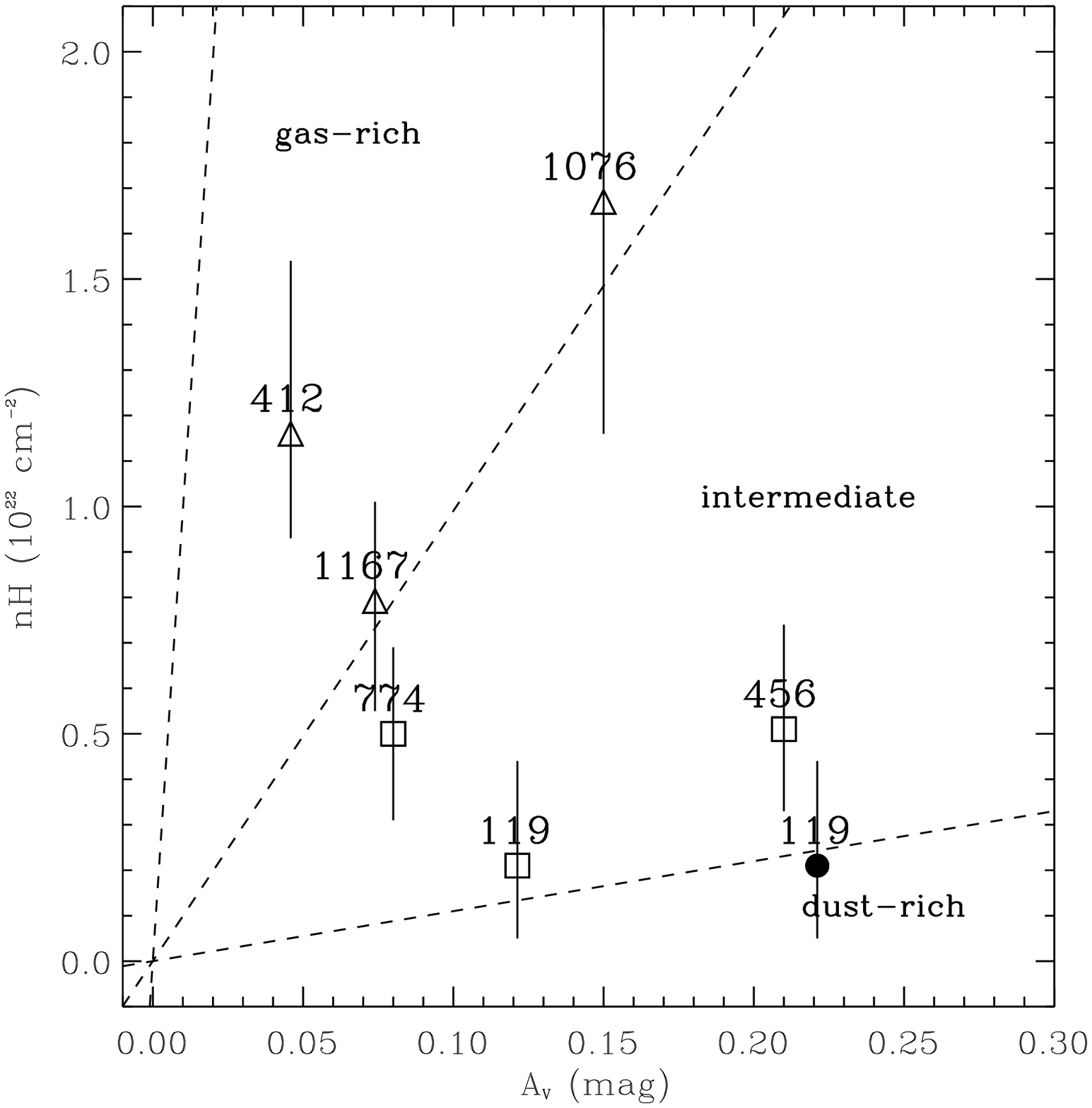}
 	\caption{N$_H$ vs. A$_V$ observed in the optical dips listed in Table \ref{dips_table}. The lines delimit the three loci populated by dips caused by dust-rich material, gas-rich, and with an intermediate gas to dust ratio. The labels indicate the stellar Mon-ID.}
 	\label{dips_class_fig}
 	\end{figure}

  However, in those cases where simultaneous increase of
optical extinction and X-ray absorption is observed, discussed in Sect. \ref{N$_H$_vs_dips}, we can attempt to
understand the nature of the obscuring material discriminating between
dust-free (likely associated with accretion columns) and dust-rich
(likely associated with disk warps) material. Two approaches can be
adopted. 

    One method developed by \citet{StaufferCMR2015} consists in
calculating the ratio between the FWHM of the observed dips and the stellar rotation
period. The distribution of this ratio is, in fact, observed to be
bimodal. Short dips (FWHM$_{dip}$/P$_{star} \leq 0.25$, where
FWHM$_{dip}$ is the FWHM of the observed dip and P$_{star}$ the
stellar rotation period) are due to occultation by accreting material.
Long dips (FWHM$_{dip}$/P$_{star} > 0.25$) are instead due to occultation by disk
warps. We can adopt the rotation periods of NGC~2264 members calculated by
Venuti et al.\ (in preparation) from CoRoT light curves, and calculate
the FWHM of the observed dips fitting the dip profile with a Gaussian
function. \par

It is also possible to infer the nature of the obscuring material by
calculating the N$_H$/A$_V$ ratio during the dip. Warps are in fact located
in the inner part of the circumstellar disk, near the co-rotation
radius. If the co-rotation radius is larger than the dust sublimation radius,
the N$_H$/A$_V$ ratio should be that typical of dust-rich
circumstellar disks, assumed to be $1.8-1.9\times
10^{21}$cm$^{-2}$mag$^{-1}$ \citep{BursteinHeiles1978,BohlinSD1978}.
The N$_H$/A$_V$ ratio is expected to be larger in dust-depleted
accreting material, since only the small $\sim$$\mu$m dust particles
can be dragged in the gas streams, sublimating as soon as the
temperature reaches $1000\,$K-1500$\,$K. \par

    	\begin{figure}[]
 	\centering	
 	\includegraphics[width=8.0cm]{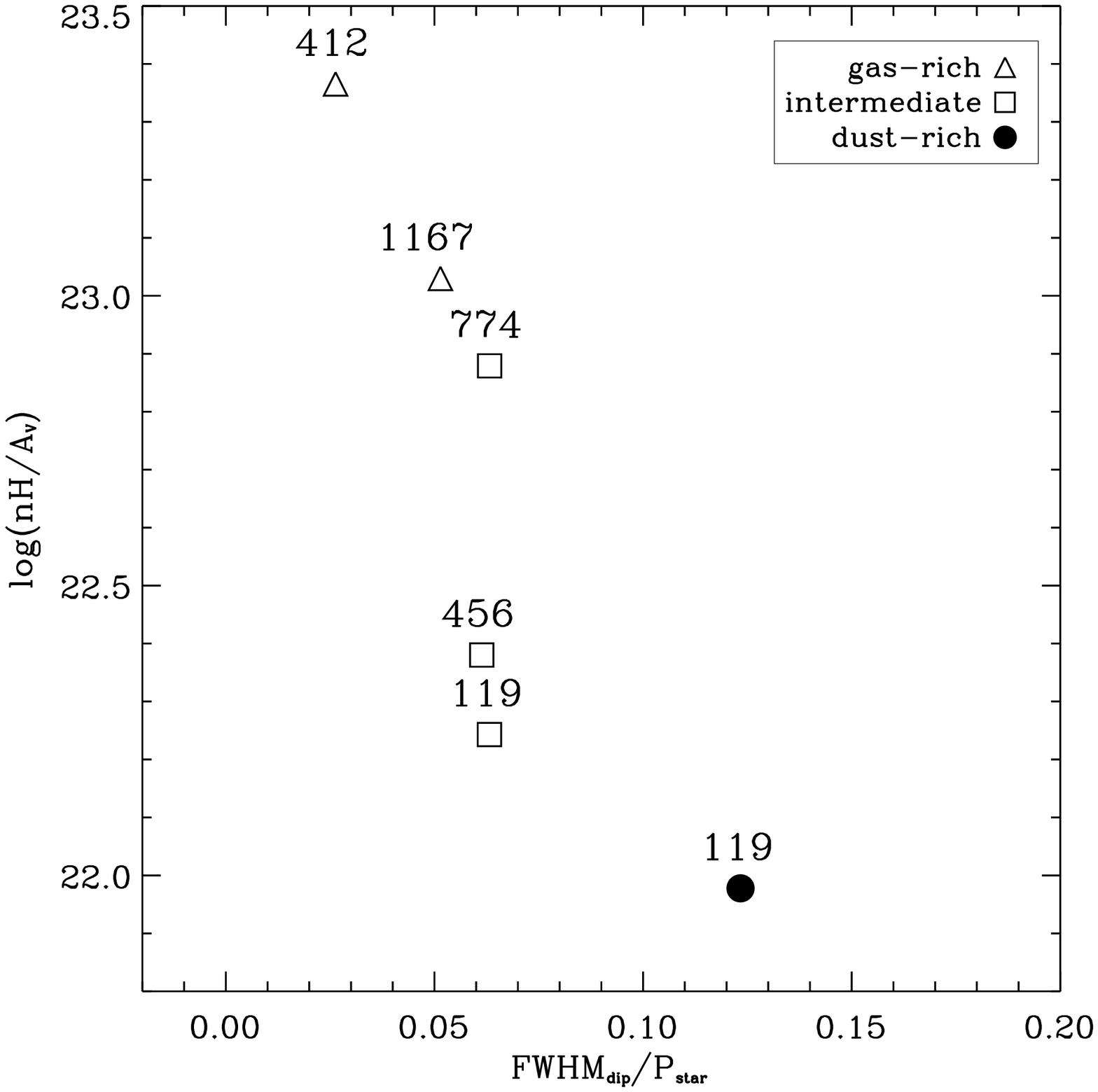}
 	\caption{log(N$_H$/A$_V$) vs.\ the ratio between the FWHM of the observed dips over the stellar rotation period. The labels indicate the stellar Mon-ID.}
 	\label{dips_fig}
 	\end{figure}

       	\begin{figure*}[]
	\centering	
	\includegraphics[width=6.0cm]{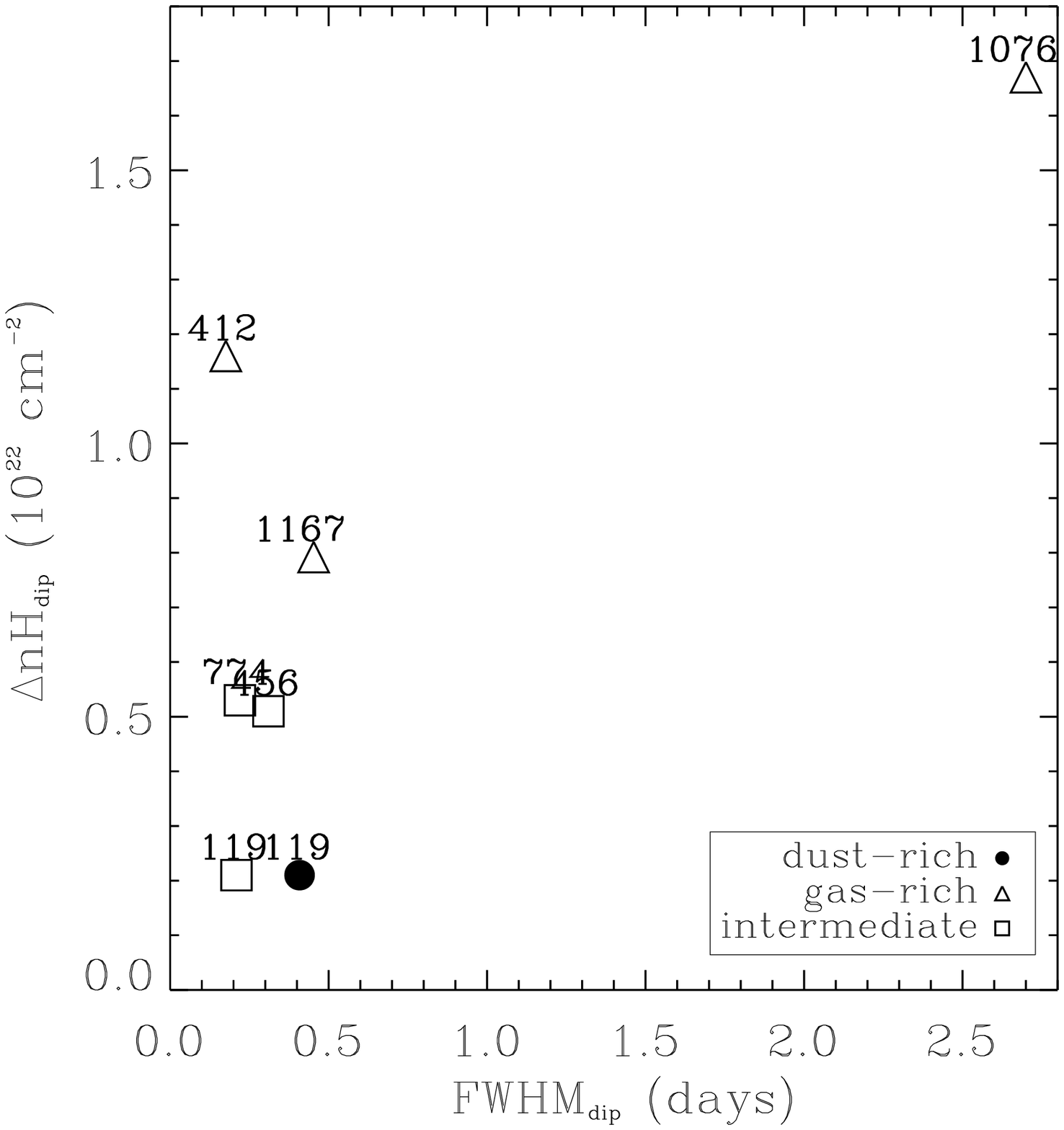}
	\includegraphics[width=6.0cm]{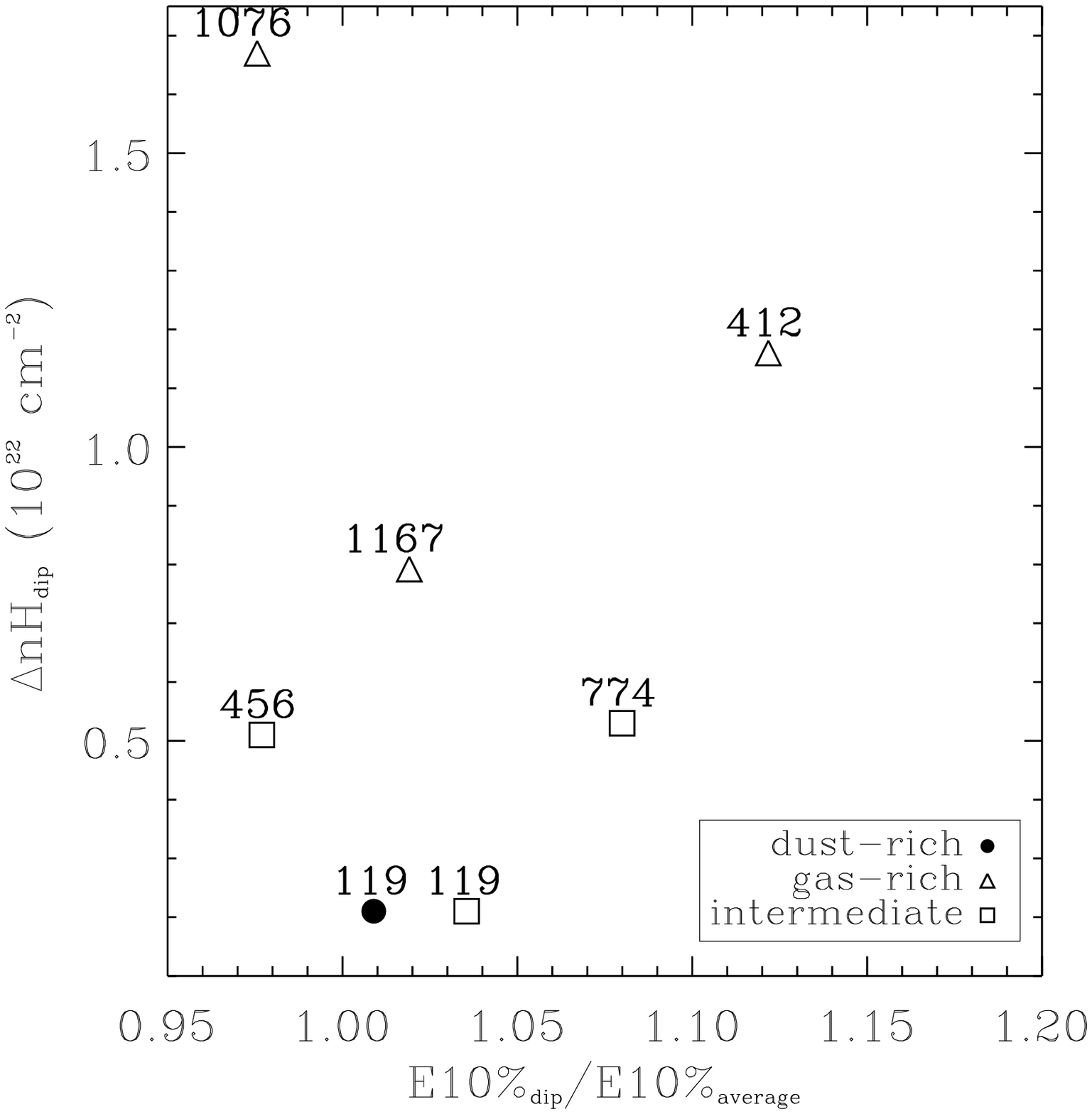}
	\includegraphics[width=6.0cm]{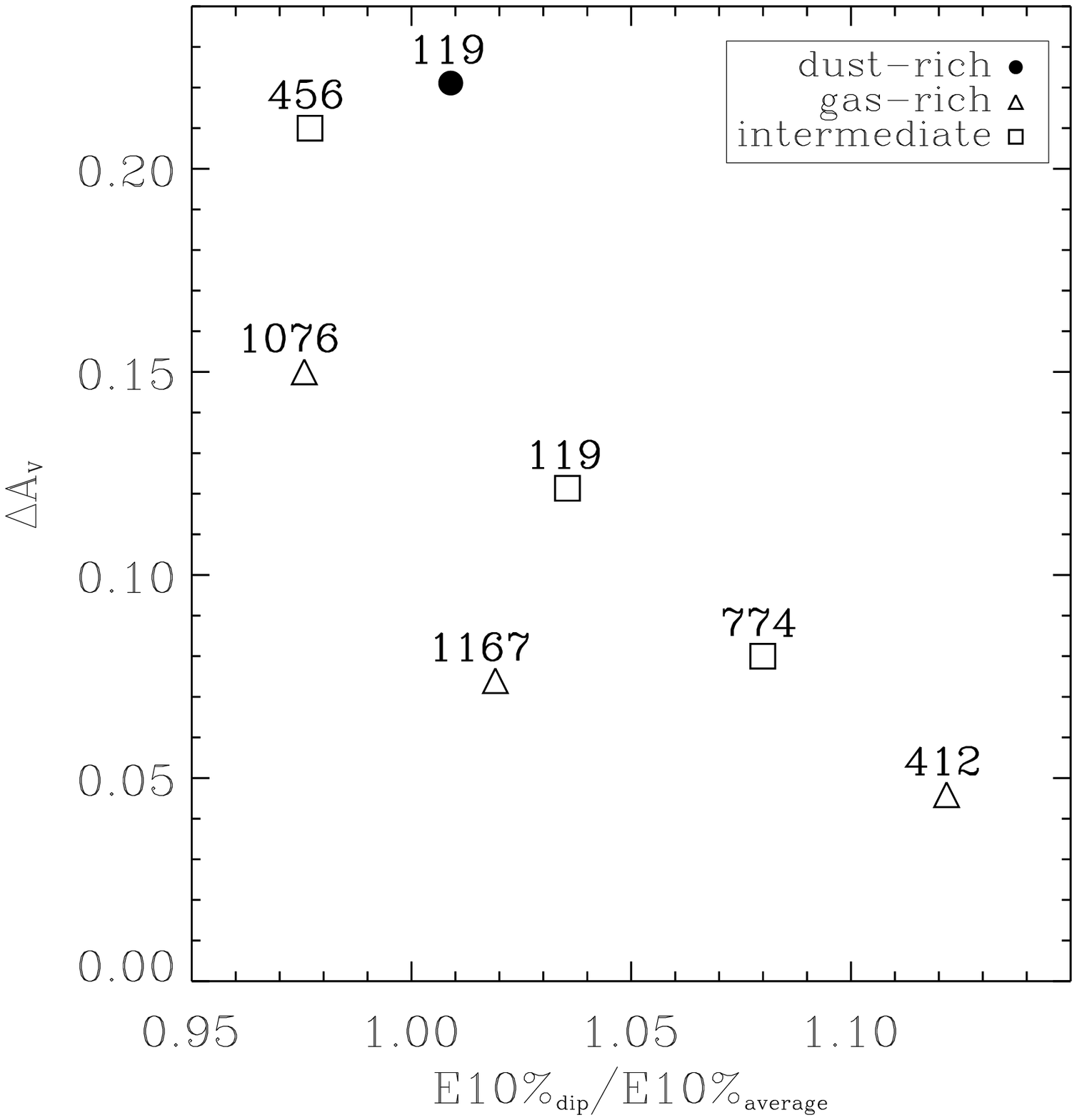}
        \caption{Optical and X-ray properties of the dips where we observe
an increasing X-ray absorption,
described in Table \ref{dips_table}. Left panel: N$_H$ vs.\ FWHM of
the dips. Central panel: $\Delta$N$_H$ due the dips vs.\ the ratio
between the 10\% photon energy quantile E$_{10\%}$ observed during the
dip and that in the average X-ray spectrum. Right panel: $\Delta$A$_V$
increment observed during the dip vs.\ the E$_{10\% \,dip}/$E$_{10\%
\, average}$ ratio. In all three panels, different symbols are used
to mark dips with different N$_H$/A$_{V}$ ratio and the labels show
the stellar Mon-IDs.}
	\label{dips_plot_fig}
	\end{figure*}

We can infer the A$_V$ necessary to reproduce the observed decline of
CoRoT flux from the optical flux absorbed during the dip, calculated
as the difference between the CoRoT flux observed at the top and at the bottom of the
dips, adopting a photometric zero point of
26.6$^m$ to convert the CoRoT fluxes into $R$ magnitudes, valid for
all members brighter than $R\leq14^m$ (Flaccomio et al.\ in
preparation), and the extinction law in $R$ band from
\citet{MunariCarraro1996}. We then compare it with the difference
$\Delta$N$_H$$_{dip}$ between the average X-ray spectrum (N$_H$$_{aver}$)
and N$_H$ observed during the dip, obtaining N$_H$/A$_V$. \par

    In this calculation, we make some assumptions. We assume that both
the photosphere and the coronal active regions are obscured by the
same structure, or structures with the same composition. We ignore
the size of the obscuring structure with respect to the star. We
assume that the disk's composition and column density is homogeneous.
We also ignore that the inner disk is irradiated by energetic UV and X-ray
radiation from the central star and it may photoevaporate
\citep[i.e.][]{ClarkeGS2001,PascucciSAA2011}. This process reduces the
amount of gas and likely also the amount of small dust particles in
the disk, affecting the N$_H$/Av ratio of the warps. \par

This calculation is performed over the seven optical dips discussed in
Sect.\ \ref{N$_H$_vs_dips} where optical extinction and X-ray
absorption are observed to increase simultaneously. The physical
properties of these dips are summarized in Table \ref{dips_table}. In
this table, each row corresponds to an observed dip, each star is
identified using the ``Mon-'' name defined by \citet{CodySBM2014AJ}, and
we also indicate the time interval containing the observed dip. The
CoRoT flux variability during the dip is calculated as
$(1-(F_{low}/F_{up}))\times100$, where $F_{low}$ and $F_{up}$ are the
CoRoT fluxes observed at the bottom and the top of the dip,
respectively. Similarly, the variation of the soft X-ray photon flux
$\Delta$F$_{X,soft}$ is calculated as
$(1-(F_{X,pre}/F_{X,dip}))\times100$, where $F_{X,pre}$ is the X-ray
soft photon flux observed during the time interval before the dip and
$F_{X,dip}$ during the dip. P$_{star}$, A$_V$, $\Delta$N$_H$$_{dip}$,
N$_H$$_{aver}$, and the FWHM$_{dip}$ are obtained as previously
explained. In Table \ref{dips_table}, the X-ray properties of the dips
observed in Mon-119 are the same since N$_H$ is obtained combining the
X-ray spectra observed in the two dips.

\begin{table*}
\caption{Properties of the analyzed optical dips.}
\centering                       
\begin{tabular}{|c|c|c|c|c|c|c|c|c|c|c|}
\hline
  \multicolumn{1}{|c|}{Star} &
  \multicolumn{1}{|c|}{Interval} &
  \multicolumn{1}{|c|}{$\Delta$F$_{CoRoT}$} &
  \multicolumn{1}{|c|}{$\Delta$Av} &
  \multicolumn{1}{|c|}{$\Delta$N$_H$$_{dip}$} &
  \multicolumn{1}{|c|}{N$_H$$_{aver}$} &
  \multicolumn{1}{|c|}{N$_H$/Av} &
  \multicolumn{1}{|c|}{FWHM$_{dip}$} &
  \multicolumn{1}{|c|}{P$_{star}$} &
  \multicolumn{1}{|c|}{FWHM$_{dip}$/P$_{star}$} &
  \multicolumn{1}{|c|}{$\Delta$F$_{X,soft}$} \\
\hline
  \multicolumn{1}{|c|}{MON-name} &
  \multicolumn{1}{|c|}{ } &
  \multicolumn{1}{|c|}{\%} &
  \multicolumn{1}{|c|}{mag} &
  \multicolumn{1}{|c|}{$10^{22}$cm$^{-2}$} &
  \multicolumn{1}{|c|}{$10^{22}$cm$^{-2}$} &
  \multicolumn{1}{|c|}{$10^{22}\,$cm$^{-2}\,$m$^{-1}$} &
  \multicolumn{1}{|c|}{days} &
  \multicolumn{1}{|c|}{days} &
  \multicolumn{1}{|c|}{ } &
  \multicolumn{1}{|c|}{\%} \\
\hline
119 & 4 & 8.9 & 0.12 & $0.21^{+0.23}_{-0.16}$ & $0^{+0.03}$           & 1.75 & 0.2 & 3.3 & 0.06 &  36.4\\  
119 & 6 & 15.7& 0.22 & $0.21^{+0.23}_{-0.16}$ & $0^{+0.03}$           & 0.95 & 0.4 & 3.3 & 0.12 & -45.2\\  
412 & 6 & 3.5 & 0.05 & $1.16^{+0.42}_{-0.19}$ & $0.04^{+0.08}$        & 23.2 & 0.2 & 6.8 & 0.03 &  13.7\\  
456 & 6 & 15.2& 0.21 & $0.51^{+0.23}_{-0.18}$ & $0^{+0.03}$           & 2.4  & 0.3 & 5.1 & 0.06 &   3.7\\  
774 & 5 & 5.7 & 0.08 & $0.53^{+0.22}_{-0.22}$ & $0.54^{+0.25}_{-0.24}$& 7.57 & 0.2 & 3.5 & 0.06 &   5.9\\
1076& 2 & 10.65&0.15 & $1.67^{+0.49}_{-0.51}$ & $0^{+0.01} $          & 11.4 & 2.7 &     &      &  -7.1\\  
1167& 3 & 5.5 & 0.07 & $0.79^{+0.22}_{-0.24}$ & $0^{+0.06}$           & 10.7 & 0.5 & 8.8 & 0.05 &  41.3\\  
\hline
\multicolumn{11}{l}{} \\
\end{tabular}
\label{dips_table}
\end{table*}

	\begin{figure}[]
	\centering	
	\includegraphics[width=9.0cm]{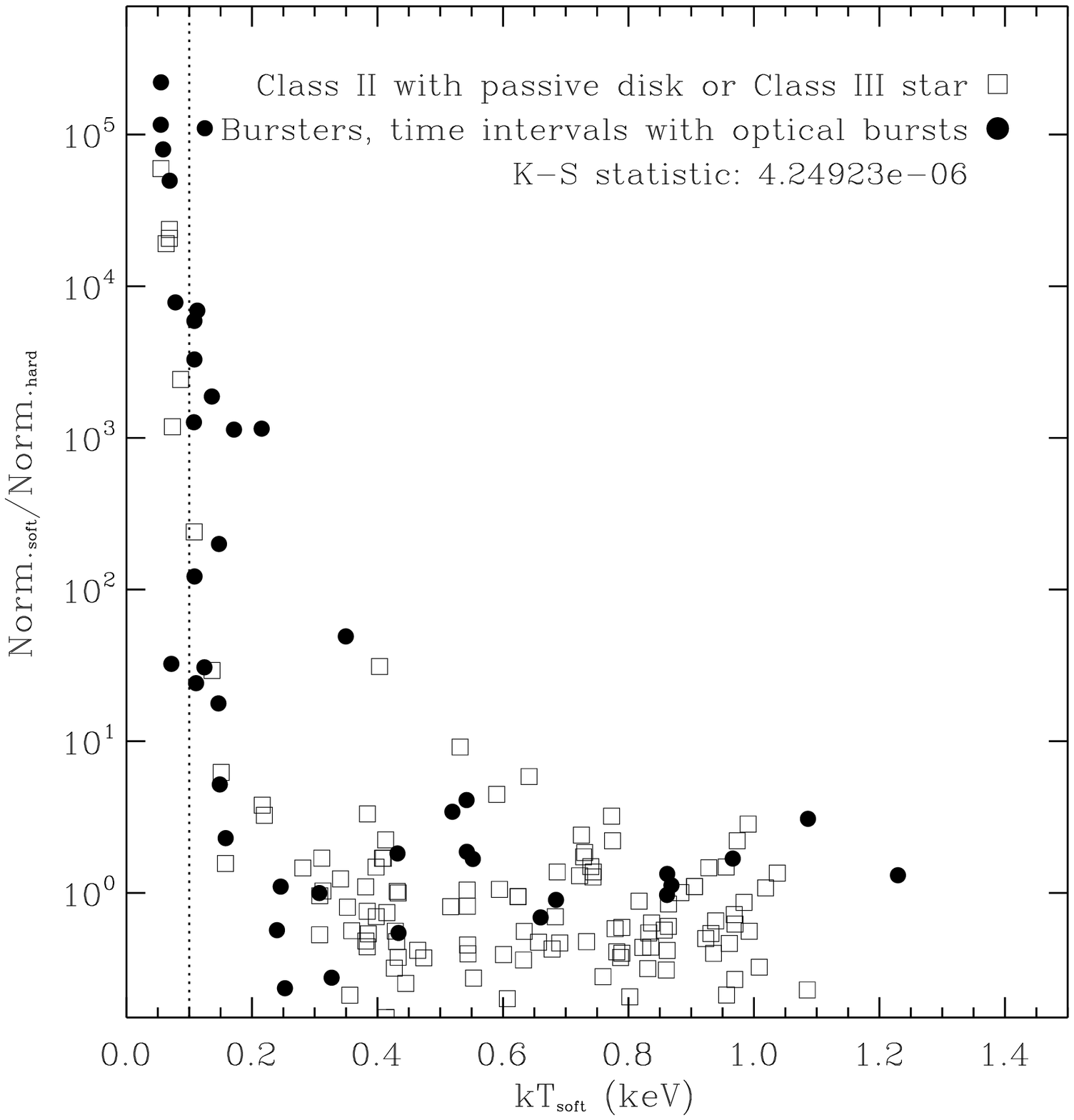}
	\includegraphics[width=9.0cm]{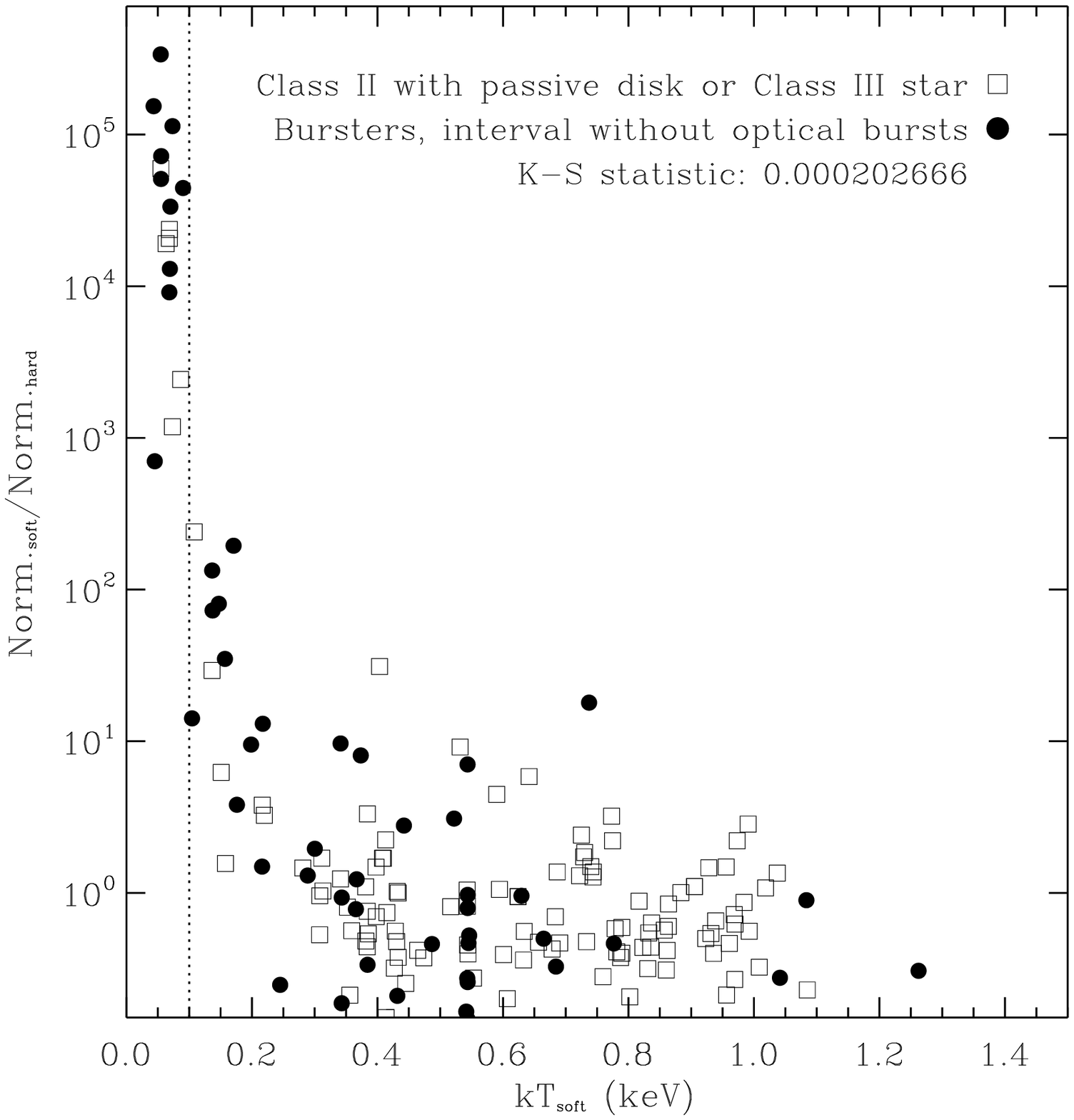}
        \caption{Ratio between the normalization of the soft and hard
components of the best fit 2T thermal plasma model vs.\ the correspondent soft temperature
of the X-ray spectra of the non accreting stars (class~III and class~II objects with passive disks)
and that observed in stars with optical bursts in the time intervals with bursts (upper panel) and in those 
with no bursts (bottom panel).}
	\label{em_vs_kt_plot}
	\end{figure}

A rough distinction between dips caused by dust-rich and gas-rich
material can be done selecting the dips with N$_H$/A$_{V}\leq
10^{21}\,$cm$^{-2}\,$mag$^{-1}$ and N$_H$/A$_{V}\geq
10^{23}\,$cm$^{-2}\,$mag$^{-1}$, respectively. This classification is
shown in Fig.~\ref{dips_class_fig}, where dips due to material with
different gas-to-dust ratios are marked with different symbols. The two points corresponding to Mon-119 show the values measured during the time intervals \#4 and \#6, containing two dips. \par

Only in one dip (Mon-119, interval \#6) is it the case that the absorbing material may be
dust-rich compared to the other dips (but gas-rich compared with the
typical interstellar material).
In 3/7 dips the absorbing material is likely dust-depleted, as expected
in accreting material, while the remaining 3/7 have intermediate
composition. This is confirmed by  the distribution of N$_H$/A$_V$ vs.
FWHM$_{dip}$/P$_{star}$, shown in Fig.~\ref{dips_fig}. Symbols mark
dips with different gas-to-dust ratios as in Fig.~\ref{dips_class_fig}.
All the dips studied in this paper have
FWHM$_{dip}$/P$_{star}\leq0.2$, as expected from occultation by narrow
accreting columns. In Fig.\
\ref{dips_fig}, the anticorrelation between N$_H$/A$_V$ and
FWHM$_{dip}$/P$_{star}$ is not statistically significant, according to a
Spearman rank correlation test. However, it is interesting to note that 
the dips due to the most and least
dusty obscuring material (i.e., the highest and lowest N$_H$/A$_V$
ratio), correspond to the largest and smallest FWHM$_{dip}$/P$_{star}$
ratio, respectively, in agreement with the hypothesis that dips with
larger FWHM$_{dip}$/P$_{star}$ are expected to be produced by
dust-rich obscuring material. \par

        The lack of dips due to dust-rich material is very likely a
selection effect, since we are studying dips occurring in timescales
comparable with the {\em Chandra} frames, missing those with larger
FWHM$_{dip}$ which are expected to be produced by disk warps. Additionally,
stars occulted by large disk warps are typically faint both in optical
and X-rays, so it is not possible to calculate N$_H$ (as in Mon-619
and Mon-717). We are also not very sensitive to small variations of
N$_H$. Taking all these effects together, we conclude that our results
are clearly biased toward narrow dips due to the accretion stream.
\par

%%%%%%%%%%%%%%%%%%%%%%%%%%%%%%%%%%%%%%%%%%%%%%%%%%%%%%
  \subsection{Global properties of the observed dips}
    \label{glob_dip_sec}

In the 7 dips analyzed in Sect.\ \ref{N$_H$avratio_sec} the fraction
of absorbed CoRoT flux ranges from 3.5\% (corresponding to an increase
of optical extinction by 0.05$^m$) to 15.7\% (A$_V$ increasing by
$0.22^m$), with a mean value of 9.3\% (the largest CoRoT flux
variation observed in the 33 stars with variable extinction is 58.1\%,
corresponding to A$_V$ increasing by $1.13^m$). The FWHM of these
seven dips ranges from 0.2 to 2.7 days; six dips are
very narrow, with a FWHM ranging from 0.2 days to 0.5 days. \par

More details on the connections between the optical and X-ray variability during these seven dips
are shown in Fig.~\ref{dips_plot_fig}. As shown in the
left panel, $\Delta$N$_{H\,dip}$ varies over a factor $\sim$6 while
the FWHM$_{dip}$ does not change significantly. The only
exception is the dip observed in Mon-1076 which has both the
largest FWHM$_{dip}$ and $\Delta$N$_{H\,dip}$ of the sample. This may
suggest that the FWHM$_{dip}$ is related to the extent of the
obscuring material with respect to the stellar disk, and a larger
extension does not necessarily imply a larger hydrogen column density.

As expected, there is a correlation between $\Delta$N$_H$ and 10\%
photon energy quantiles observed during the dip (not shown here). A
weak correlation is still observed when comparing the 10\% photon
energy quantile observed during the dip (E$_{10\% dip}$) with that
observed in the average spectrum (E$_{10\% average}$). A different
result is obtained comparing the extinction increment $\Delta$A$_V$
with the E$_{10\% dip}/$E$_{10\% average}$ ratio: The increment of
optical extinction is larger in those dips with small variations of the
10\% photon energy quantile with respect to the average X-ray
spectrum. The anticorrelation is significant based on a Spearman
rank correlation test.

  The direct observation of increasing X-ray absorption during the
optical dips is important for two reasons. First, X-rays have an
important role in regulating disks evolution. Disk photoevaporation,
which is one of the main mechanisms responsible for the dissipation of
circumstellar disks \citep{HollenbachYJ2000,AlexanderCP2006MNRAS} can be
induced by incident X-ray photons, which ionize the gas, raising the
disk temperature up to $\sim$$10^4\,$K within 1$\,$AU from the central
star \citep{ErcolanoDRC2008ApJ}. The resulting high thermal pressure
launches a photoevaporation wind, which results in significant
mass loss from the disk. Additionally, the more X-ray photons
absorbed by the circumstellar disk, the larger the ionization fraction
in the disk, and the more efficient the coupling between disk and stellar
magnetic field, enhancing magneto-rotational instabilities
\citep{BalbusHawley1991}, and thus enhancing the radial transport of gas across
the disk. In this context, it is of particular importance to observe
directly that stellar X-rays can be efficiently absorbed by the
circumstellar material in the inner disk. To date, the only evidence of 
interaction between energetic particles and protoplanetary disks is the observation
of fluorescent emission lines \citep{TsujimotoFGM2005ApJS}. Second, optical
and X-ray flux variability is observed to be correlated only in
stars with inner disks and/or actively accreting stars (Sect.\
\ref{corr_class} in this paper and \citealt{FlaccomioMFA2010}). This
correlation is interpreted as due to simultaneous occultation of
the stellar photosphere and corona by the circumstellar material. Our
study supports this hypothesis and provides further evidence of
increasing X-ray absorption during the occultation of the central star
by circumstellar material. \par

%%%%%%%%%%%%%%%%%%%%%%%%%%%%%%%%%%%%%%%%%%%%%%%%%%%%%%%%%%%%%%%%%%%%%%%%%%%%%%%%%%%%%%5
  \subsection{Accretion properties in the stars with optical bursts}
    \label{glob_burst_sec}

        In Sect. \ref{N$_H$_vs_burst}, we analyze the stars showing
evidence of increasing soft X-ray emission during the optical bursts.
In Fig.~\ref{em_vs_kt_plot}, we investigate whether such a correlation exists
during all the optical bursts and in the X-ray spectra of all the
accretors, even when optical bursts are not observed. To this aim, we fit the X-ray spectra of all the not accreting stars (class~III
objects and the class~II objects with passive disks, see Sect.\
\ref{sample_sec}), and the time-resolved X-ray spectra observed in
accreting stars during the optical bursts and in intervals not
containing bursts, using 2T thermal plasma models. We calculate, then, the
ratio between the normalization of the soft and the hard components
(Norm.$_{soft}$/Norm.$_{hard}$) for each spectrum. The result is shown in Fig.
\ref{em_vs_kt_plot}, where accreting and non-accreting stars are
marked with different symbols, and only the results from statistically
significant fits are shown. Even though in Fig.~\ref{em_vs_kt_plot} we
plot the soft temperatures down to $\sim$0$\,$keV, recall
that {\em Chandra} is not sensitive to very soft X-ray
emission, and thus values of kT$_{soft}$ below $\sim$0.1$\,$keV must
not be fully trusted. The first result is that in both panels,
Norm.$_{soft}$/Norm.$_{hard}$ is typically larger in accreting
stars than in non-accreting stars. 
A K-S test in both cases indicates that this difference
is significant. While in the bottom panel all the time resolved spectra
with Norm.$_{soft}$/Norm.$_{hard}$$\geq$500 have
kT$\leq$1$\,$keV, which can not be fully trusted, only
a small fraction of the time-revolved spectra during the bursts (top panel) have
kT$\leq$1$\,$keV.
This provides evidence for a larger emission measure of cold
plasma in accreting stars than in star with no accretion, and confirms the importance of the time-
resolved spectral analysis using the optical light curves as a template to isolate
the time intervals with optical bursts occurring. \par

  Our study also shed some light on the geometry of accretion in the analyzed stars. 
It is generally accepted that gas accretion from the inner region of
circumstellar disks is driven by stellar magnetic field.
Several magnetohydrodynamic models
\citep[e.g.][]{KulkarniRomanova2008,RomanovaUKL2013} show that a
stellar dipolar magnetic field inclined with respect to the rotation
axis produces two stable accretion streams from the inner region of
the disk, near the truncation radius, that impact the stellar surface
at near free-fall velocity. This is the ``stable accretion'' scenario,
in which the emission from the hot spots (from optical to soft X-rays)
is modulated by stellar rotation in a stable and periodic pattern
\citep{McKinneyTB2012,CemeljicSC2013}. \par 

Other existing models \citep[i.e.,
][]{RomanovaUKL2012,KurosawaRomanova2013} show that accretion can also
occur in an unstable regime, when Rayleigh-Taylor instabilities
occurring at the disk-magnetosphere boundary result in short-lived
accretion streams falling on the stellar surface \citep[see
also][]{ColomboOPA2016arXiv}. In this case, the optical light curve is
dominated by random, short accretion bursts such as those analyzed
in Sect.\ \ref{N$_H$_vs_burst} and by \citet{StaufferCBA2014}, whose
typical duration is of a few hours, smaller than the rotation periods
of stars and thus not compatible with stable accretion streams. \par

In some of the stars analyzed in Sect. \ref{N$_H$_vs_burst}, i.e. those with an intense soft X-ray spectral component observed during bursts, we can measure the
temperature of the plasma responsible for the emission of soft X-rays
(kT$_{soft}$) during the bursts, which we assume to be associated with
the hot accretion spots. Table \ref{bursts_tab} shows the values
of kT$_{soft}$ and the properties of the accretion streams obtained in
these bursts, the resulting values of the pre-shock velocity and
free-fall radius, together with the free-fall velocity from infinity
and the known H$\alpha$ equivalent width. The free-fall radius is
omitted in those stars where the pre-shock velocity is not constrained
(i.e., when it is compatible with the free-fall velocity from
infinity).

\begin{table}
\caption{Accretion properties of the stars with soft X-ray emission during the optical bursts.}
\label{bursts_tab}
\centering                       
\begin{tabular}{|r|r|r|r|r|r|r|}
\hline
  \multicolumn{1}{|c|}{Star} &
  \multicolumn{1}{|c|}{Interval} &
  \multicolumn{1}{|c|}{kT$_{soft}$} &
  \multicolumn{1}{|c|}{v$_{presh}$} &
  \multicolumn{1}{|c|}{R$_{FF}$} &
%  \multicolumn{1}{|c|}{R$_{cor}$} &
  \multicolumn{1}{|c|}{v$_{\infty}$} &
  \multicolumn{1}{|c|}{EW$_{H\alpha}$}\\
\hline
  \multicolumn{1}{|c|}{Mon-} &
  \multicolumn{1}{|c|}{\#} &
  \multicolumn{1}{|c|}{keV} &
  \multicolumn{1}{|c|}{km/s} &
  \multicolumn{1}{|c|}{R$_{star}$} &
%  \multicolumn{1}{|c|}{R$_{star}$} &
  \multicolumn{1}{|c|}{km/s} &
  \multicolumn{1}{|c|}{\AA} \\
\hline
326 & 1       & $0.15_{-0.08}^{+0.10}$  & $355_{-113}^{+103}$ &                        & 394 & 27.9 \\
357 & 7       & $0.06_{-0.02}^{+0.01}$  &                    &                         &     &      \\
370 &2,4,7,10 & $0.16_{-0.04}^{+0.08}$  & $366_{-49 }^{+83} $ &  $2.0_{-0.4}^{+2.3}$   & 512 & 113.2\\
474 & 1       & $0.66_{-0.35}^{+0.16}$  & $744_{-234}^{+85 }$ &                        & 423 & 104.7\\
474 & 6       & $0.25_{-0.09}^{+0.10}$  & $458_{-92 }^{+84 }$ &                        & 423 & 104.7\\
808 & 1,2,8   & $0.15_{-0.04}^{+0.06}$  & $351_{-47 }^{+69 }$ &  $1.9_{-0.4}^{+2.0}$   & 519 & 50.2 \\
\hline
\multicolumn{7}{l}{Typically the co-rotation radius R$_{cor}$ is 5-10$\,$R$_{star}$ in T~Tauri stars.} \\
\multicolumn{7}{l}{In Mon-326 R$_{cor}$ is 12.3$\,$R$_{star}$} \\
\end{tabular}
\end{table}

In three cases, the soft temperature is about 0.15$\,$keV (Mon-326,
Mon-370, and Mon-808), with the highest soft temperatures observed in
Mon-474 and the lowest in Mon-357. The calculated pre-shock velocities
vary over a wide range, although being well constrained only in two
stars, Mon-370 and Mon-808. In these two cases, v$_{presh}$ is quite
similar (about 350-360$\,$km/s). In Mon-370 and Mon-808, also the
corresponding free-fall radii are reasonably well constrained, being
smaller than the typical co-rotation radii of T~Tauri stars (usually
ranging between 5 and 10$\,$R$_{star}$,
\citealp{HartmannCGD1998,ShuNSL2000}). Considering that we can safely
ignore the energy loss during accretion, and thus that our
estimate of the free-fall radii is not seriously overestimated, this
is compatible with the hypothesis that the small accretion bursts
observed in these stars and the resulting soft X-ray emission are
likely due to unstable accretion rather than stable
accretion streams from the co-rotation radius. \par

\begin{acknowledgements}
We thank the referee for his/hers comments and suggestions that helped us improving our manuscript. M. G. G., E. F., and G. M. acknowledge the grant PRIN-INAF 2012 (P.I. E. Flaccomio). This research has made use of data from the Chandra X-ray Observatory and the CoRoT satellite. This research also made an extensive use of Xspec software and the NASA's Astrophysics Data System and Vizier databases, operated at CDS, Strasbourg, France.
\end{acknowledgements}

\newpage
\addcontentsline{toc}{section}{\bf Bibliografia}
\bibliographystyle{aa}
\bibliography{biblio}

%%%%%%%%%%%%%%%%%%%%%%%%%%%%%%%%%%%%%%%%%%%%%%%%%%%%%%%%%%%%
\begin{onecolumn}
\begin{appendix} %First onl

\section{Time resolved X-ray spectra of the stars with optical dips discussed in Sect. 5.3}
\label{spectra_app}

In this appendix we show the X-ray spectra observed in the time intervals defined for those stars with dips and with simultaneous increase of X-ray absorption, discussed in Sect. \ref{N$_H$_vs_dips}. 

	\begin{figure}[!hb]
	\centering	
	\includegraphics[width=16cm]{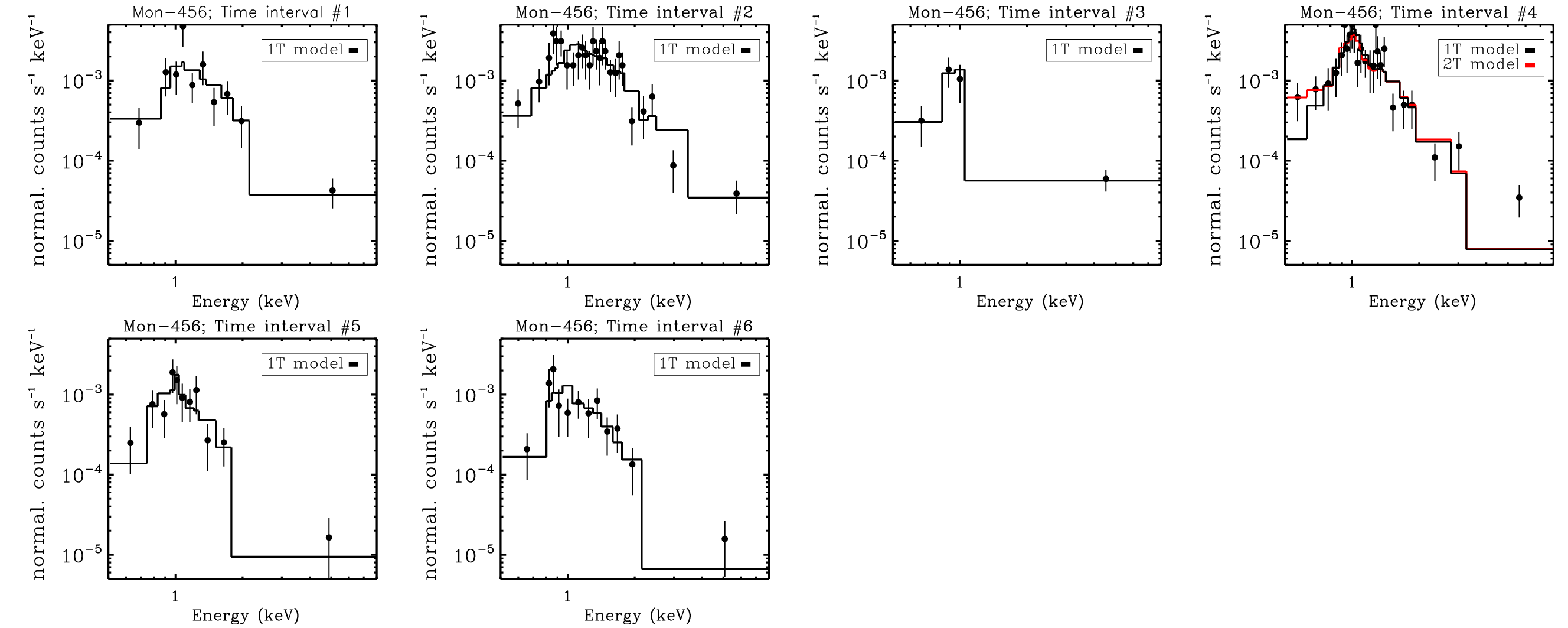}
	\caption{X-ray spectra observed in the time intervals defined for Mon-456. In each panel, the spectrum of the fitting model is marked with the solid line, while the black dots mark the observed normalized counts in the given energy bin. X-ray flares occur in \#2 and \#4.}
	\label{xspectra_mon456}
	\end{figure}

	\begin{figure}[!hb]
	\centering	
	\includegraphics[width=16cm]{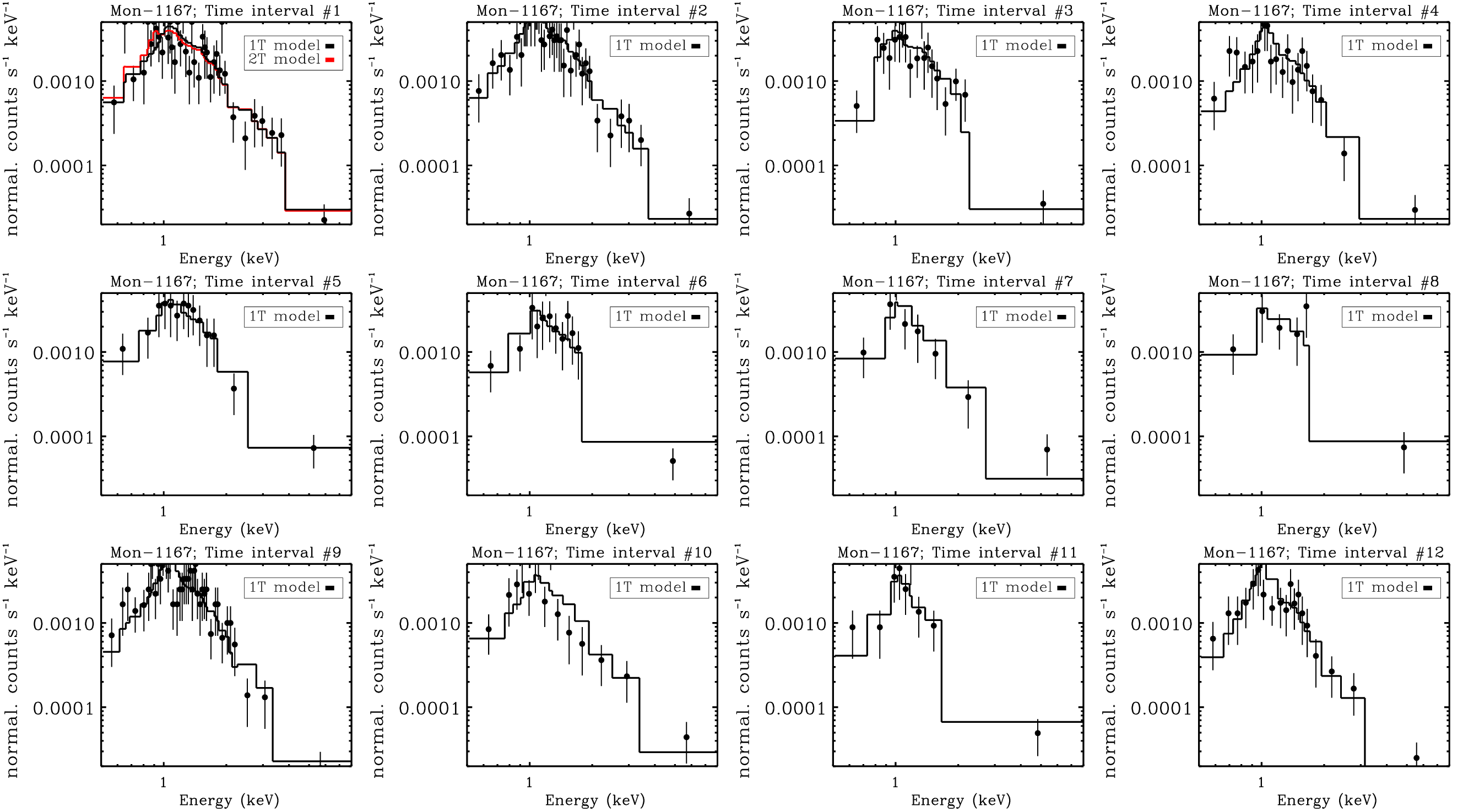}
	\caption{X-ray spectra observed in the time intervals defined for Mon-1167. In each panel, the spectrum of the fitting model is marked with the solid line, while the black dots mark the observed normalized counts in the given energy bin.}
	\label{xspectra_mon1167}
	\end{figure}

	\begin{figure}[]
	\centering	
	\includegraphics[width=15cm]{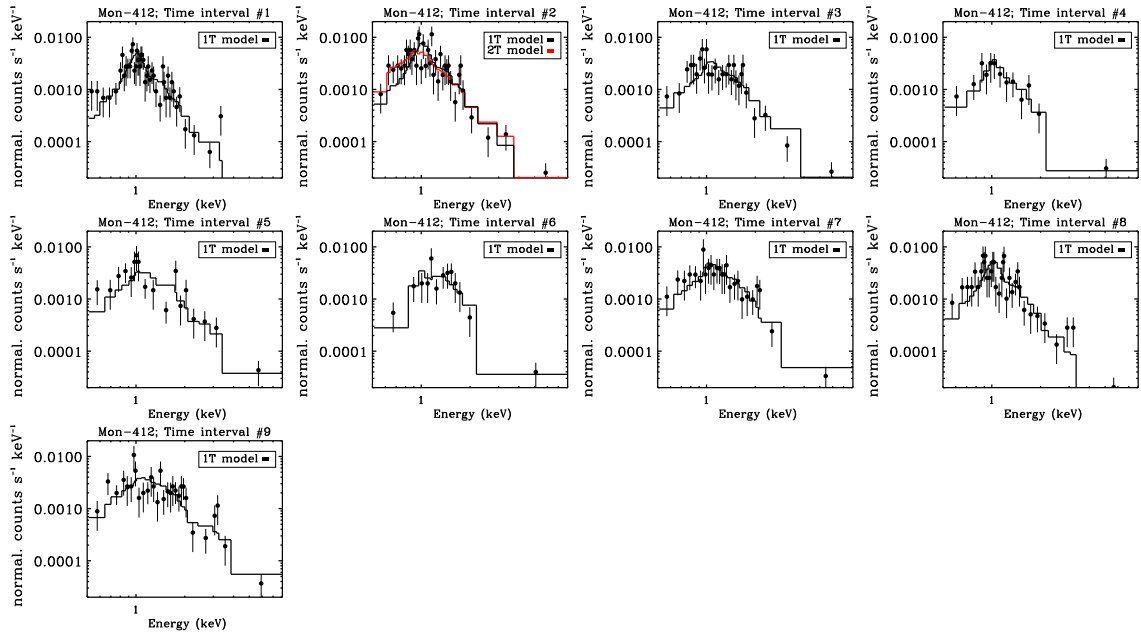}	
	\caption{X-ray spectra observed in the time intervals defined for Mon-412. In each panel, the spectrum of the fitting model is marked with the solid line, while the black dots mark the observed normalized counts in the given energy bin. This star is a burster and there may be soft X-ray emission related to the bursts, but not significant enough to be analyzed in details.}
	\label{xspectra_mon412}
	\end{figure}

	\begin{figure}[]
	\centering	
	\includegraphics[width=15cm]{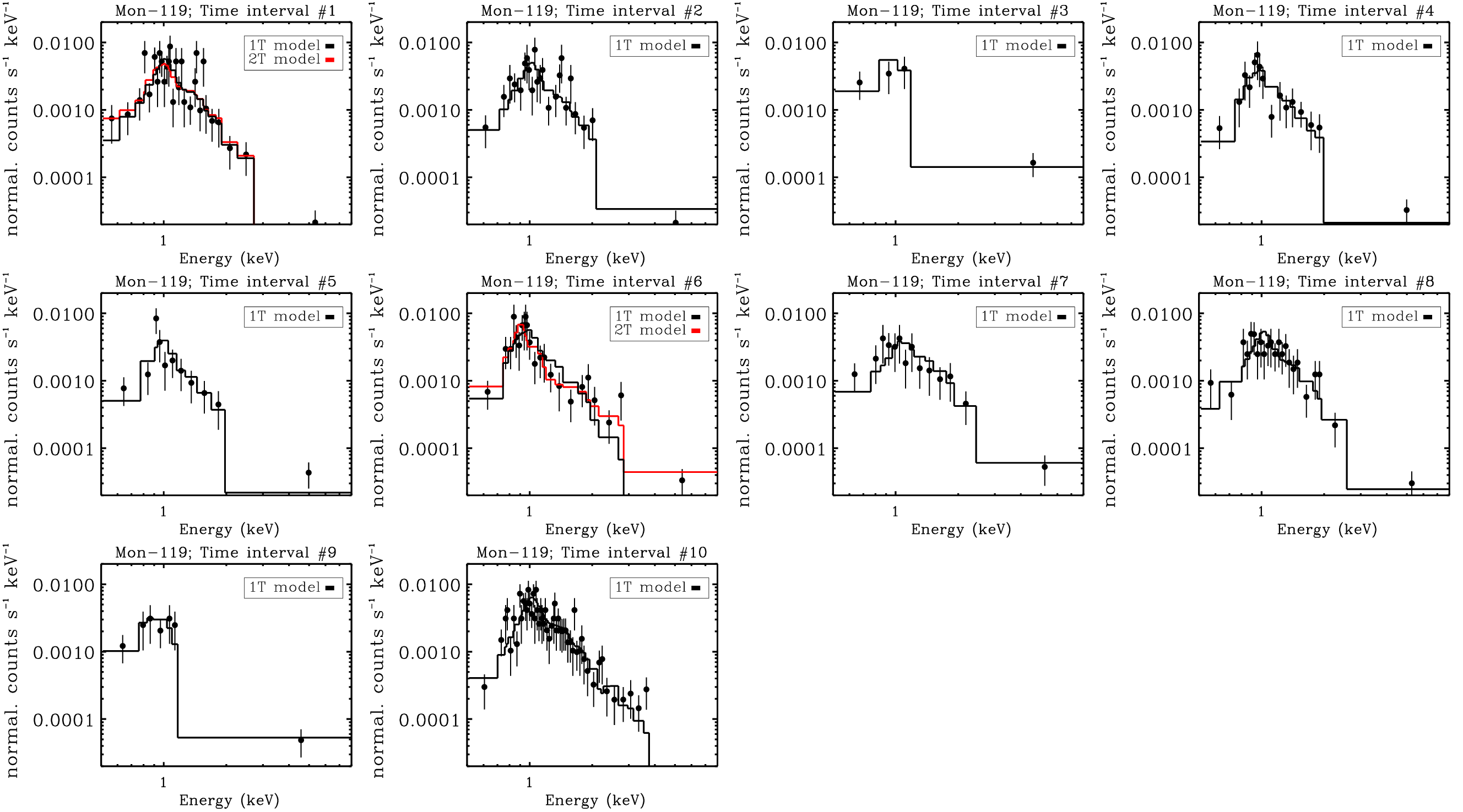}
	\caption{X-ray spectra observed in the time intervals defined for Mon-119. In each panel, the spectrum of the fitting model is marked with the solid line, while the black dots mark the observed normalized counts in the given energy bin. An optical and X-ray flare is occurs during \#10.}
	\label{xspectra_mon119}
	\end{figure}

	\begin{figure}[]
	\centering	
	\includegraphics[width=15cm]{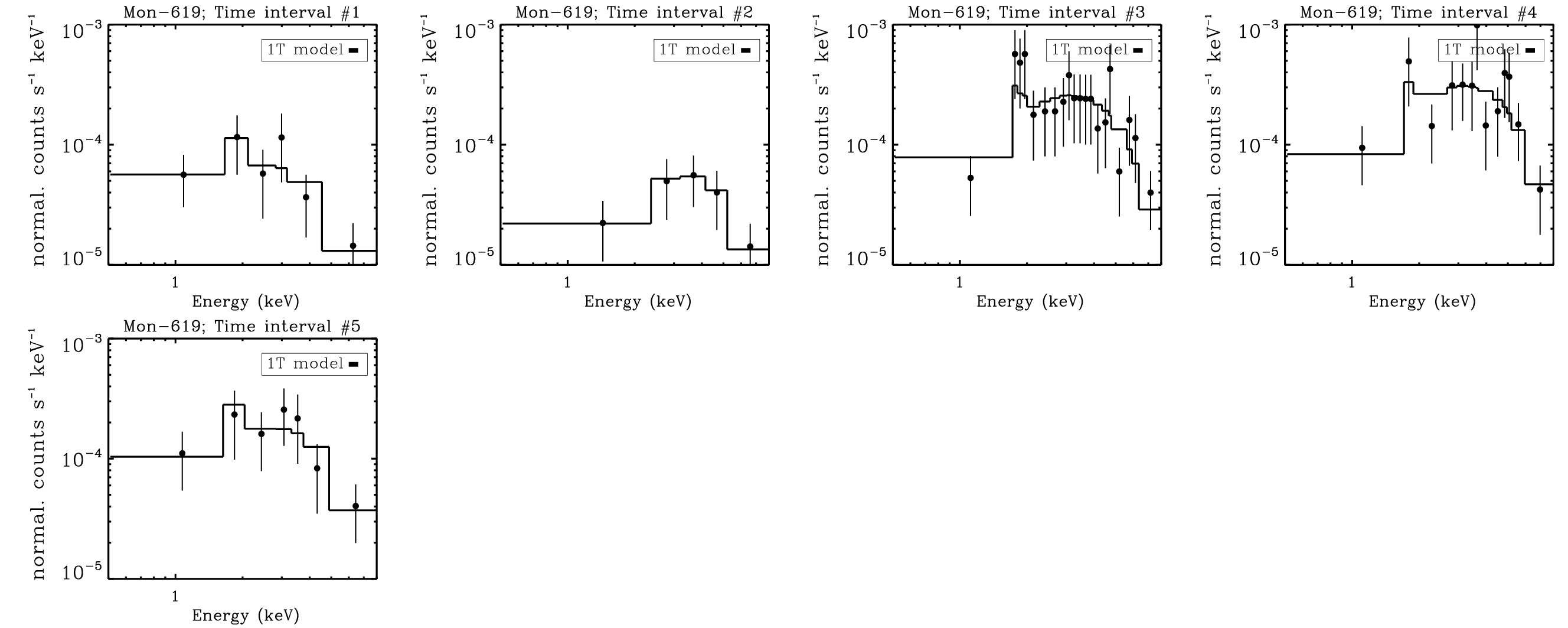}
	\caption{X-ray spectra observed in the time intervals defined for Mon-619. In each panel, the spectrum of the fitting model is marked with the solid line, while the black dots mark the observed normalized counts in the given energy bin.}
	\label{xspectra_mon619}
	\end{figure}

	\begin{figure}[]
	\centering	
	\includegraphics[width=15cm]{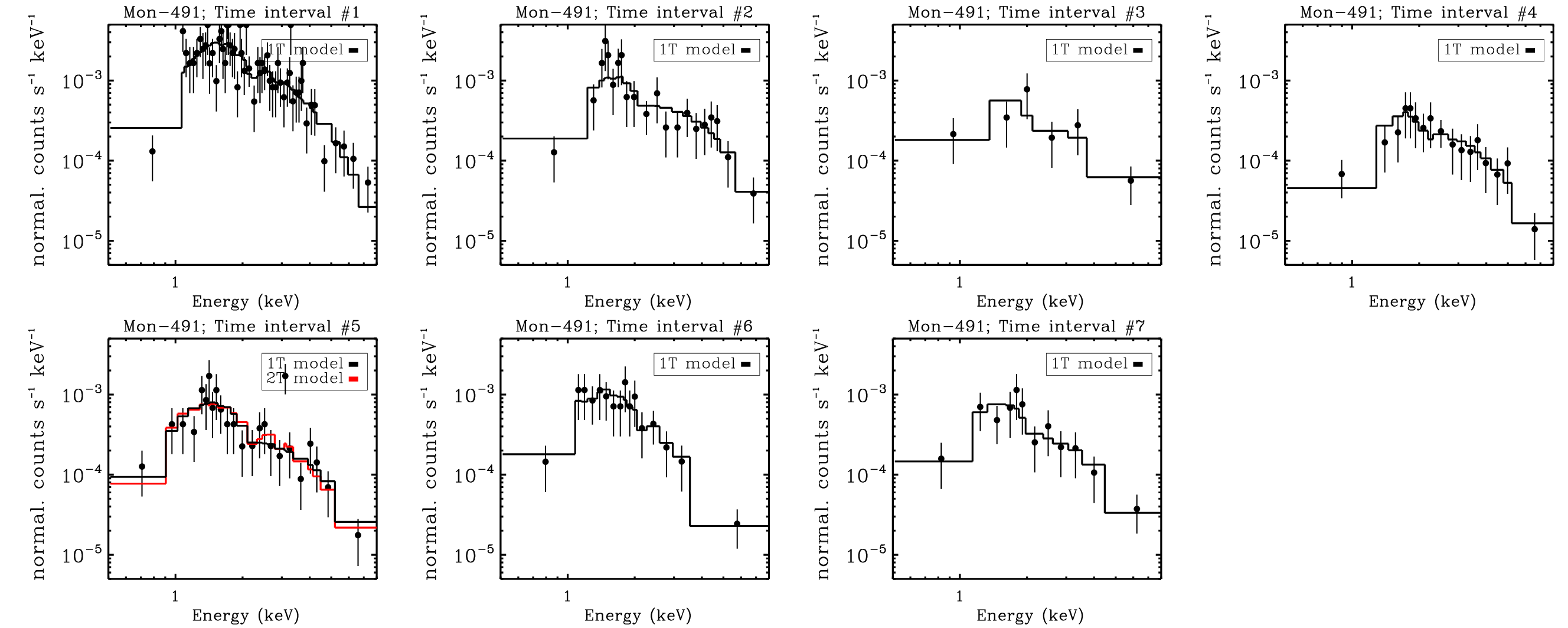}
	\caption{X-ray spectra observed in the time intervals defined for Mon-491. In each panel, the spectrum of the fitting model is marked with the solid line, while the black dots mark the observed normalized counts in the given energy bin. An X-ray flare is occurs during \#1.}
	\label{xspectra_mon491}
	\end{figure}

	\begin{figure}[]
	\centering	
	\includegraphics[width=15cm]{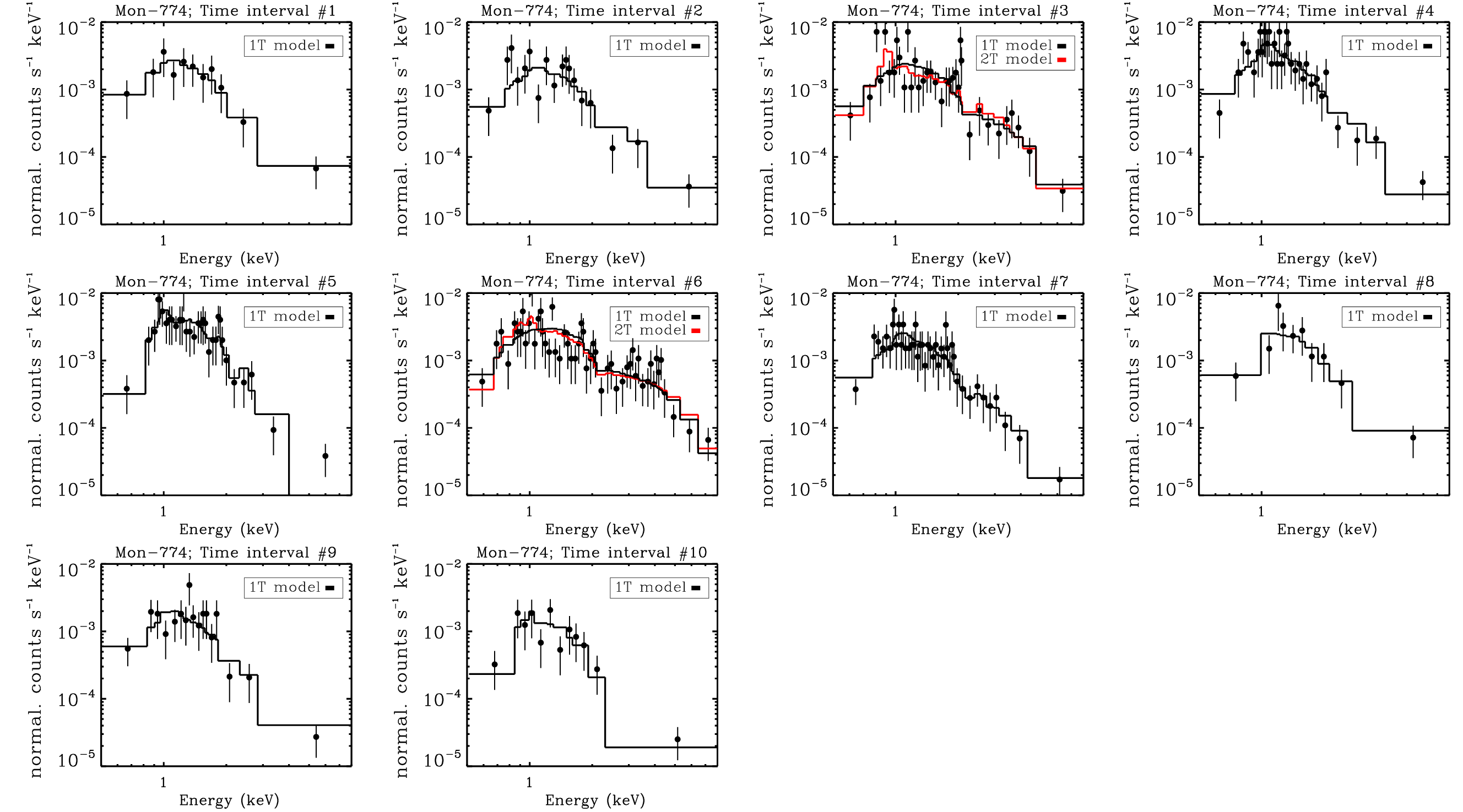}
	\caption{X-ray spectra observed in the time intervals defined for Mon-774. In each panel, the spectrum of the fitting model is marked with the solid line, while the black dots mark the observed normalized counts in the given energy bin. A X-ray flare is occurs during \#6.}
	\label{xspectra_mon774}
	\end{figure}

	\begin{figure}[]
	\centering	
	\includegraphics[width=15cm]{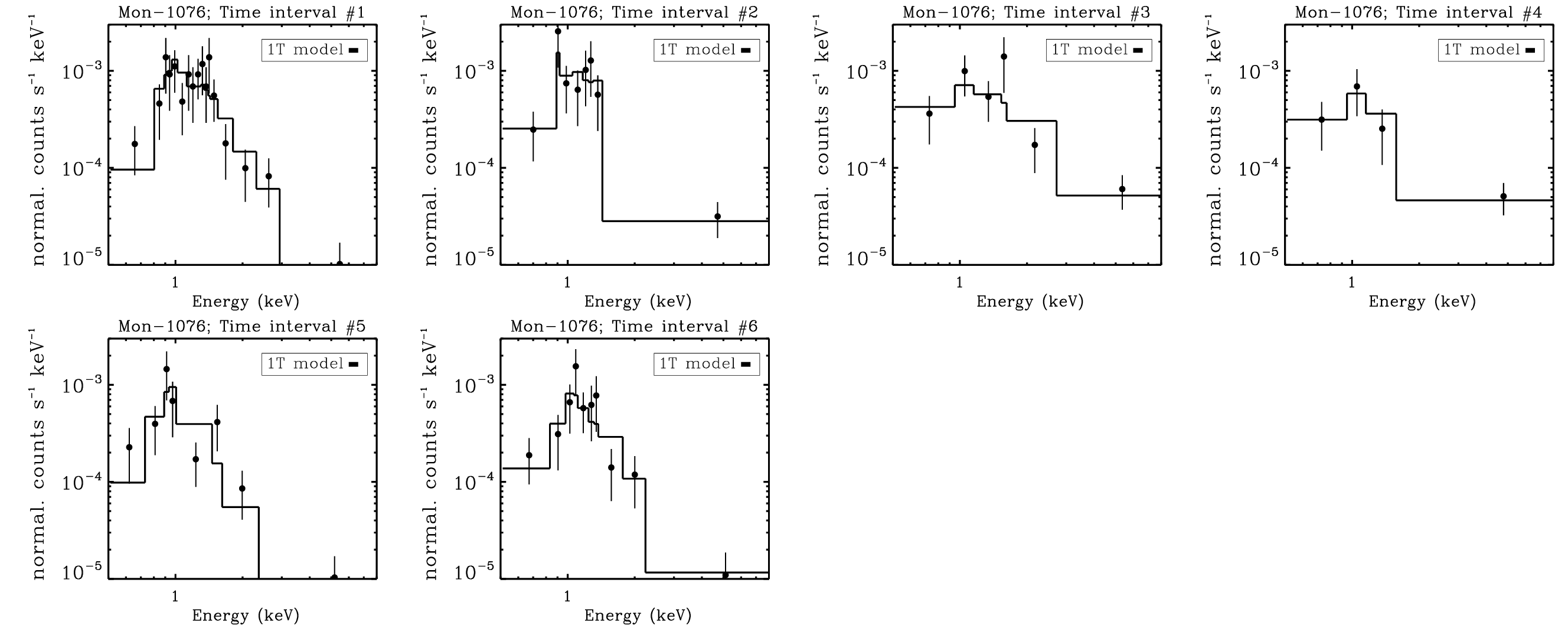}
	\caption{X-ray spectra observed in the time intervals defined for Mon-1076. In each panel, the spectrum of the fitting model is marked with the solid line, while the black dots mark the observed normalized counts in the given energy bin. An optical and X-ray flare is occurs during \#3.}
	\label{xspectra_mon1076}
	\end{figure}

\newpage
\clearpage

%%%%%%%%%%%%%%%%%%%%%%%%%%%%%%%%%%%%%%%%%%%%%%%%%%%%%%%%%%%%%

\section{Time variability of the 10\% photon energy quantile in stars discussed in the paper}
\label{e10lc_app}

In this appendix, we show how the 10\% photon energy quantile changes
with time in the stars discussed in the paper. For each star, the time
bin used to sample the variability of E$_{10\%}$ is the shortest one
resulting in a relative error $\sigma$(E$_{10\%}$)/E$_{10\%}$ smaller
than the threshold shown in the title of each panel. The values of
E$_{10\%}$ are marked with blue dots and the red lines indicate the
size of the time bin. In each panel, we also show the observed CoRoT
light curve, observed during the {\em Chandra} frames with the time
intervals dominated by X-ray flares covered in black. These figures
help us to understand how the soft X-ray spectrum of these stars changes over the time independently
from the time sampling based on the optical variability. (Note that the larger
N$_H$ the larger E$_{10\%}$, while the larger flux below 1$\,$keV the
lower E$_{10\%}$.)  \par

	\begin{figure}[!hb]
	\centering	
	\includegraphics[width=9.0cm]{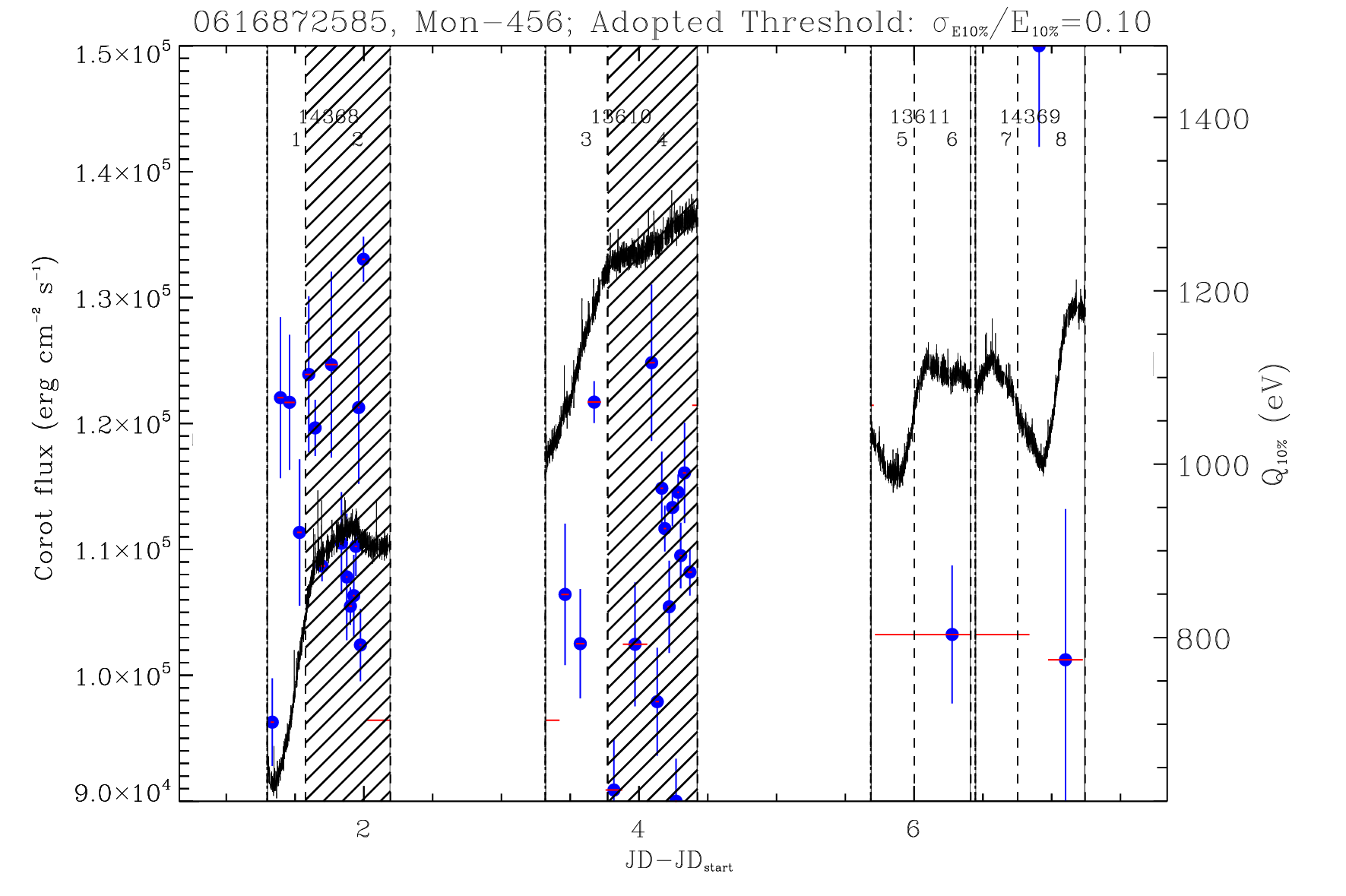}
	\includegraphics[width=9.0cm]{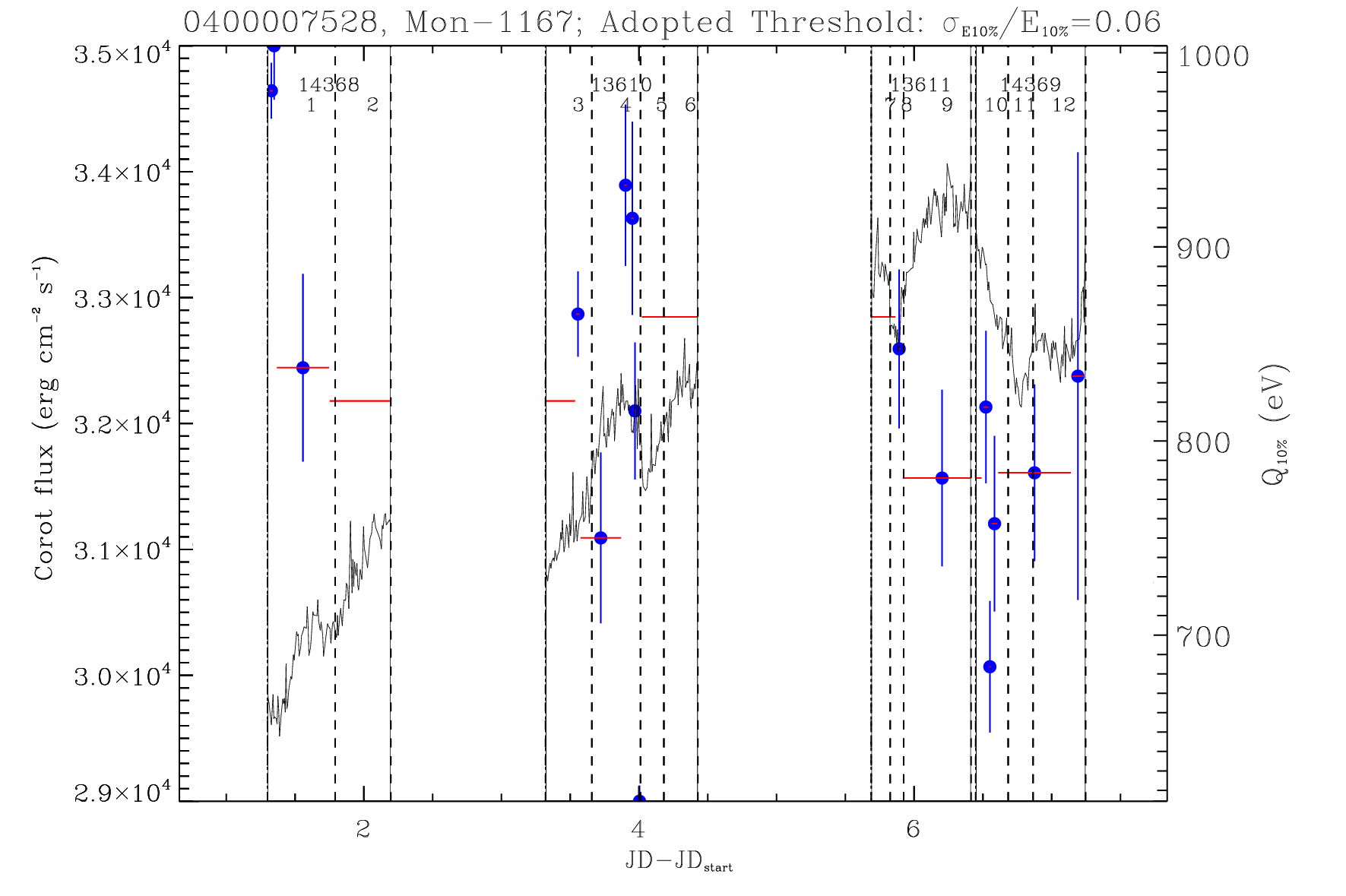}
	\includegraphics[width=9.0cm]{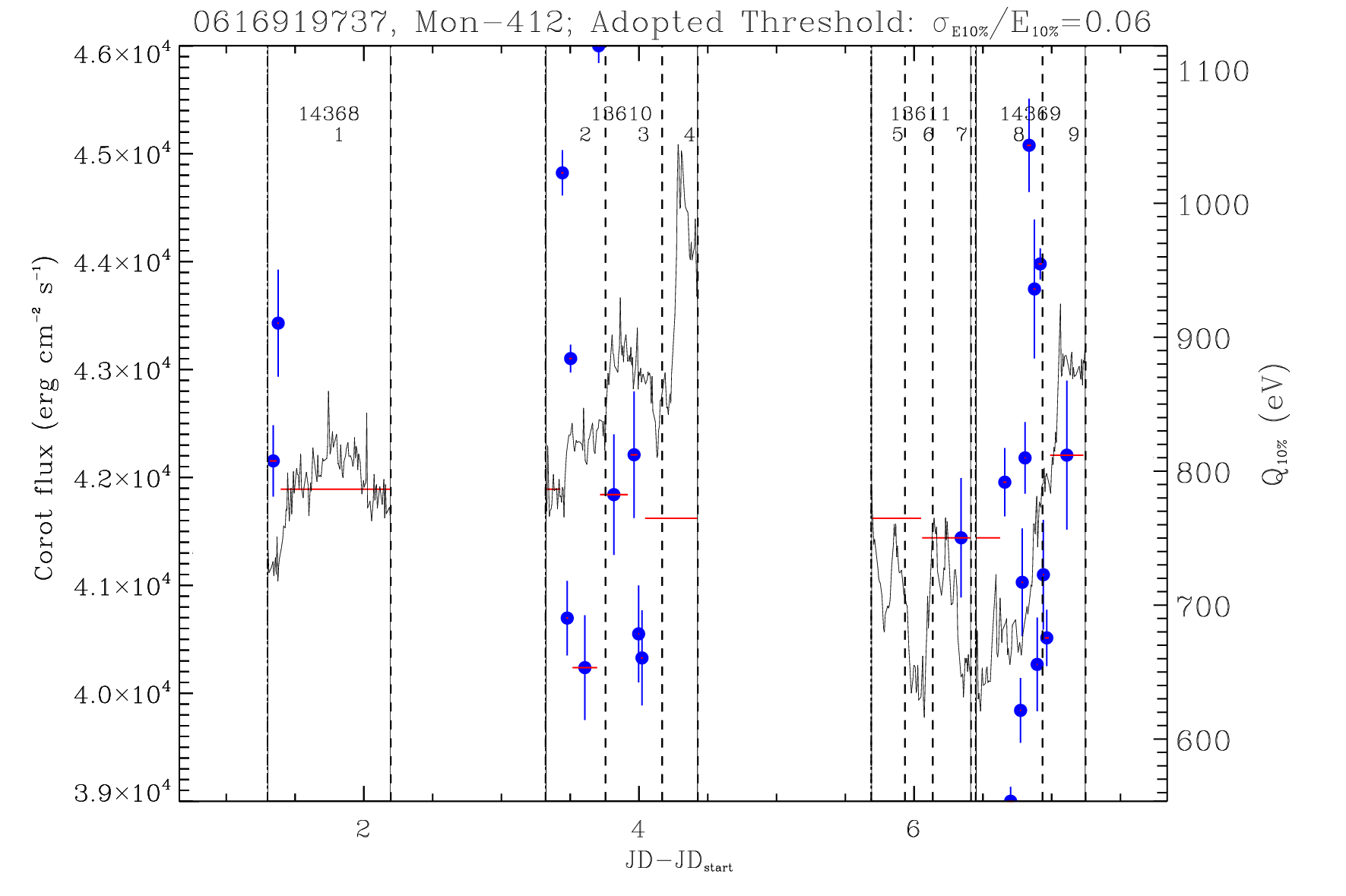}
	\includegraphics[width=9.0cm]{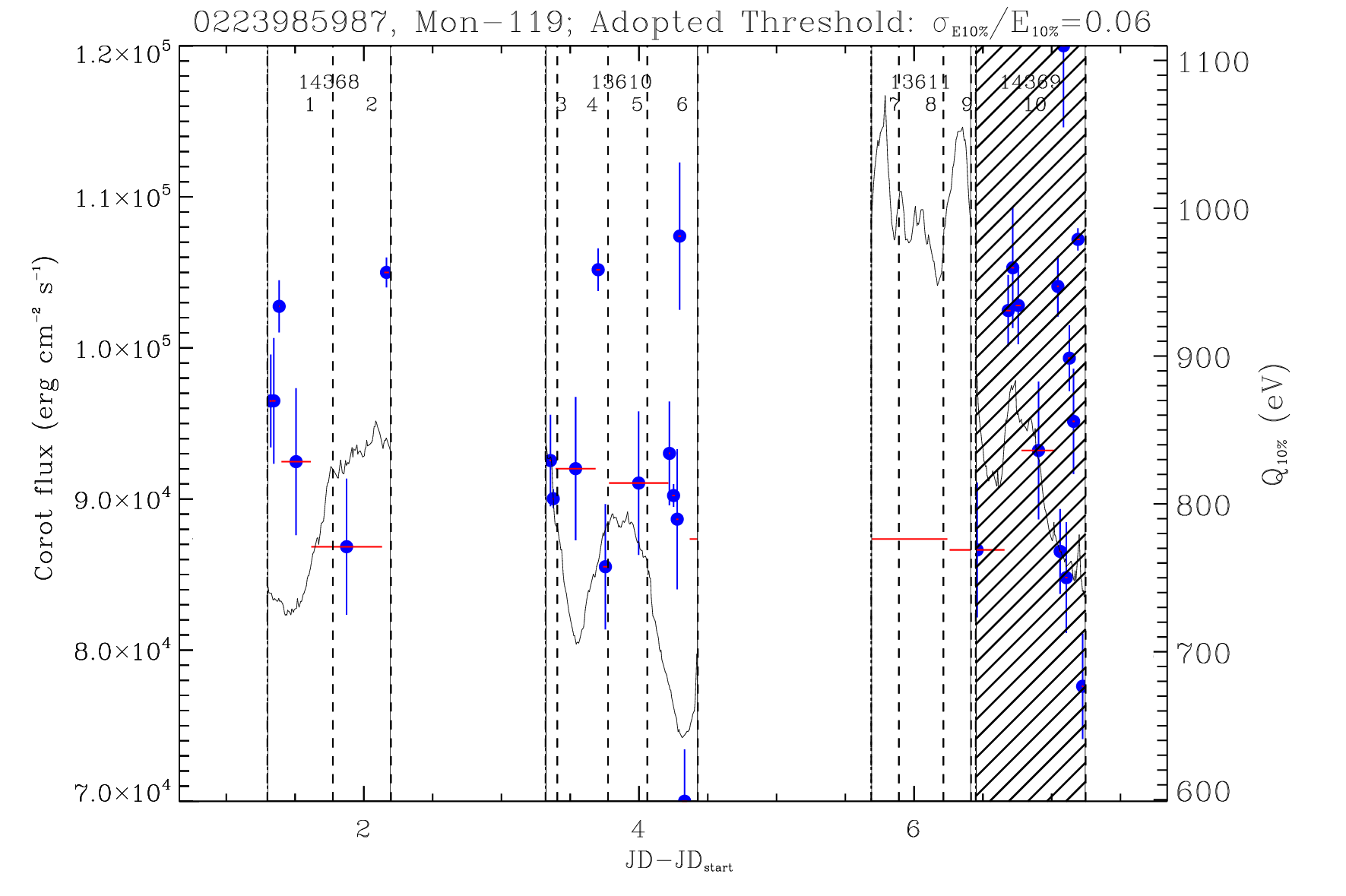}
	\includegraphics[width=9.0cm]{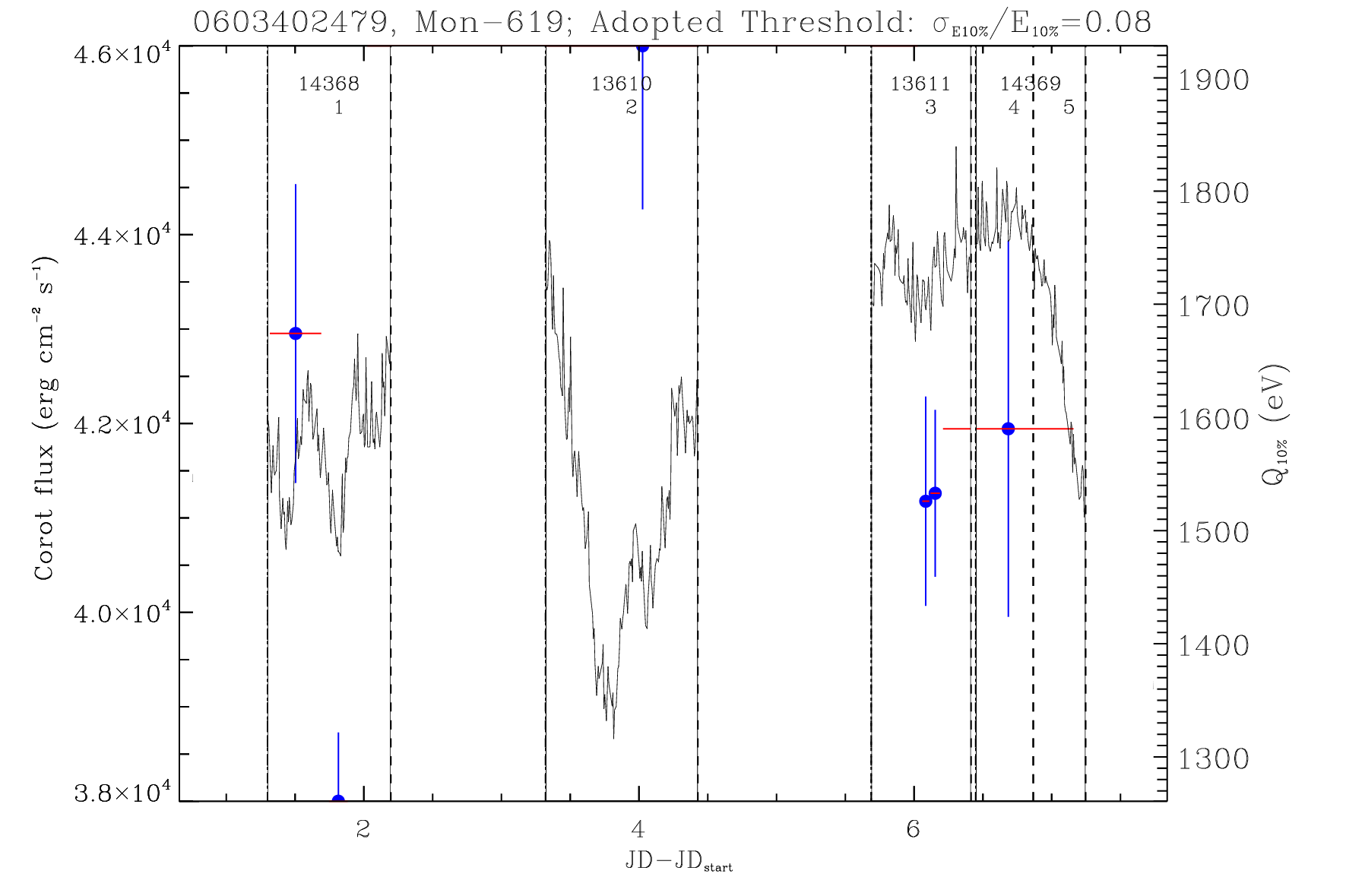}
	\includegraphics[width=9.0cm]{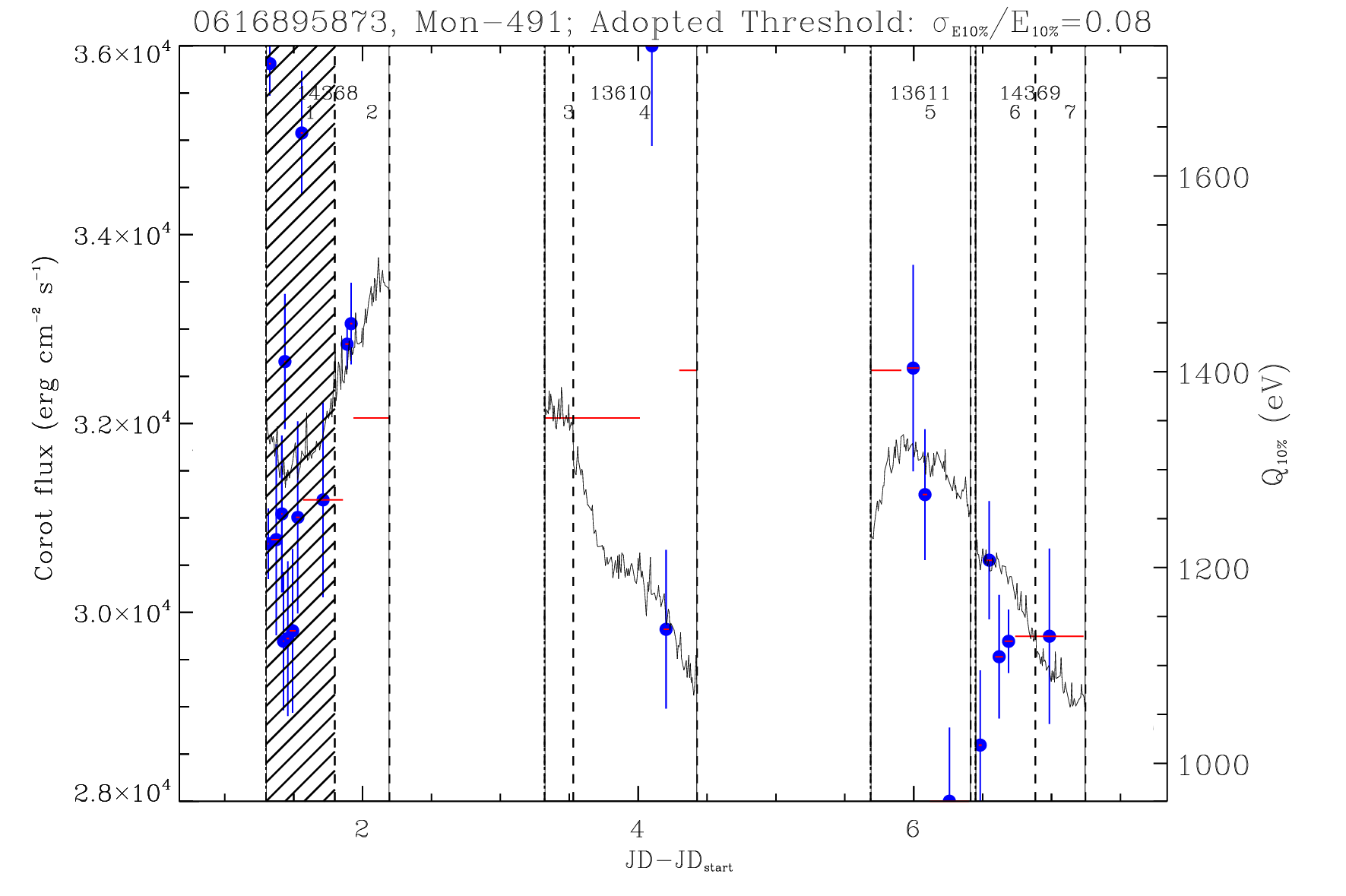}
	\includegraphics[width=9.0cm]{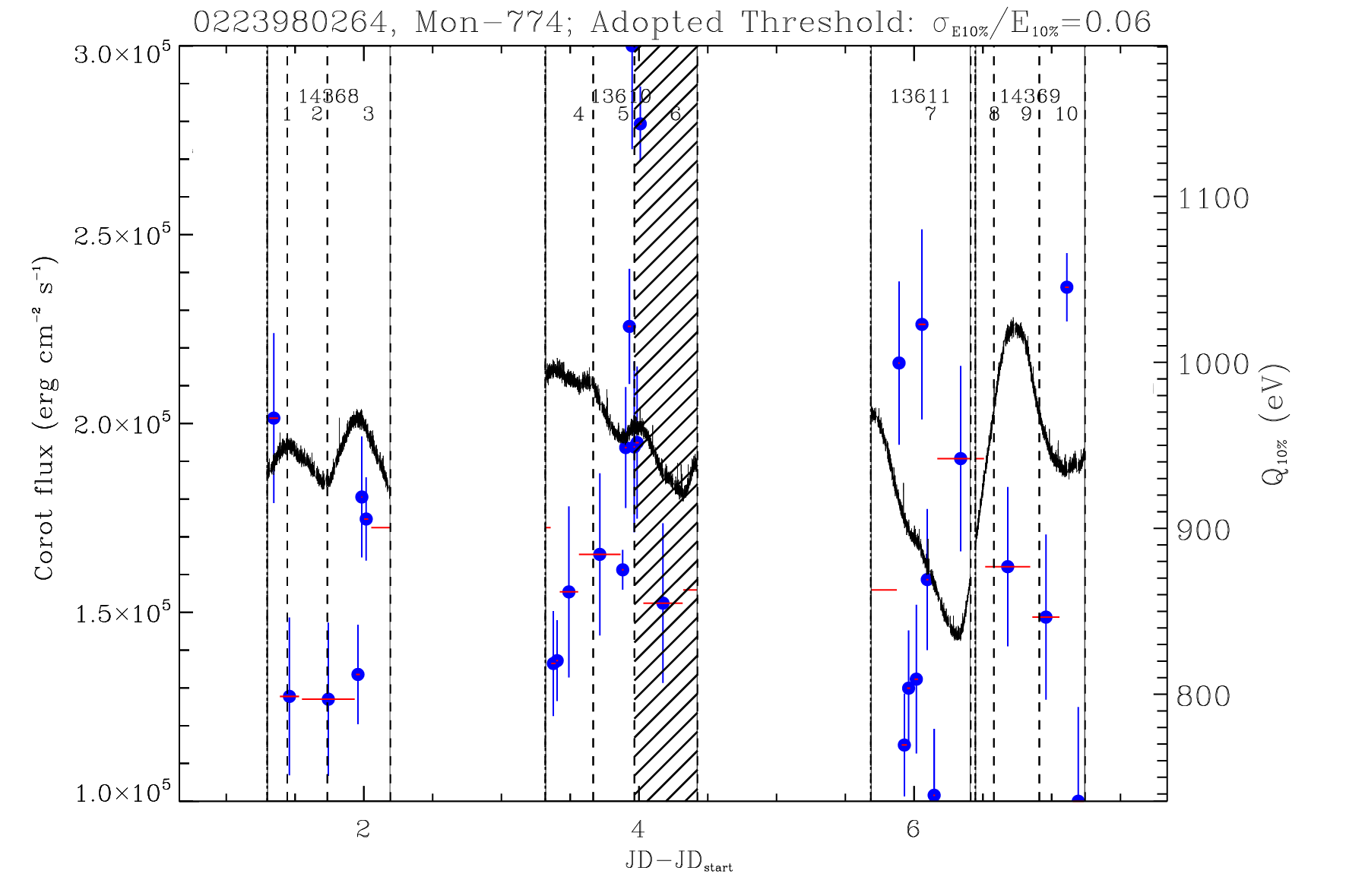}
	\includegraphics[width=9.0cm]{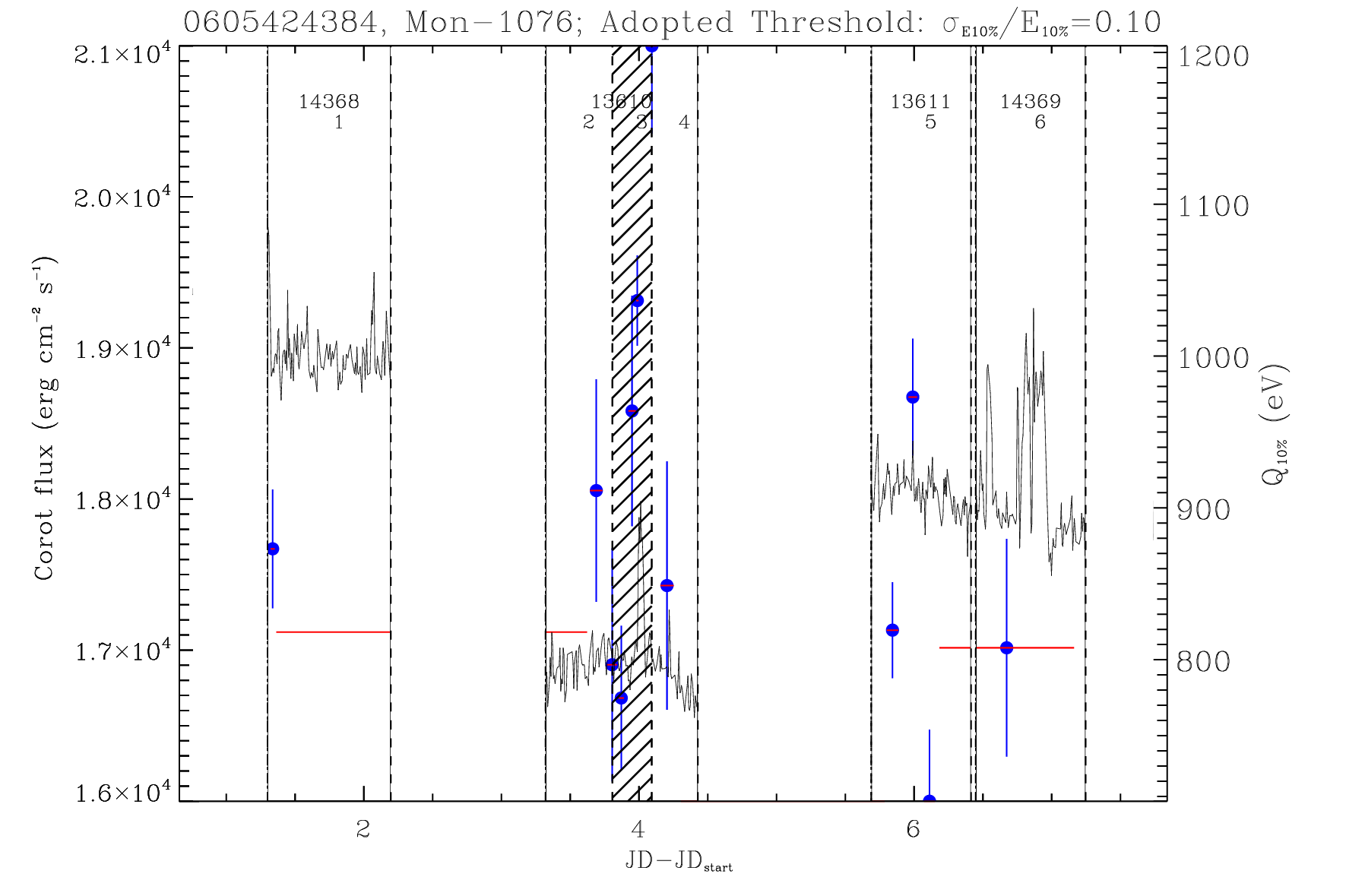}
	\label{runquant1}
	\end{figure}

	\begin{figure}[!hb]
	\centering	
	\includegraphics[width=9.0cm]{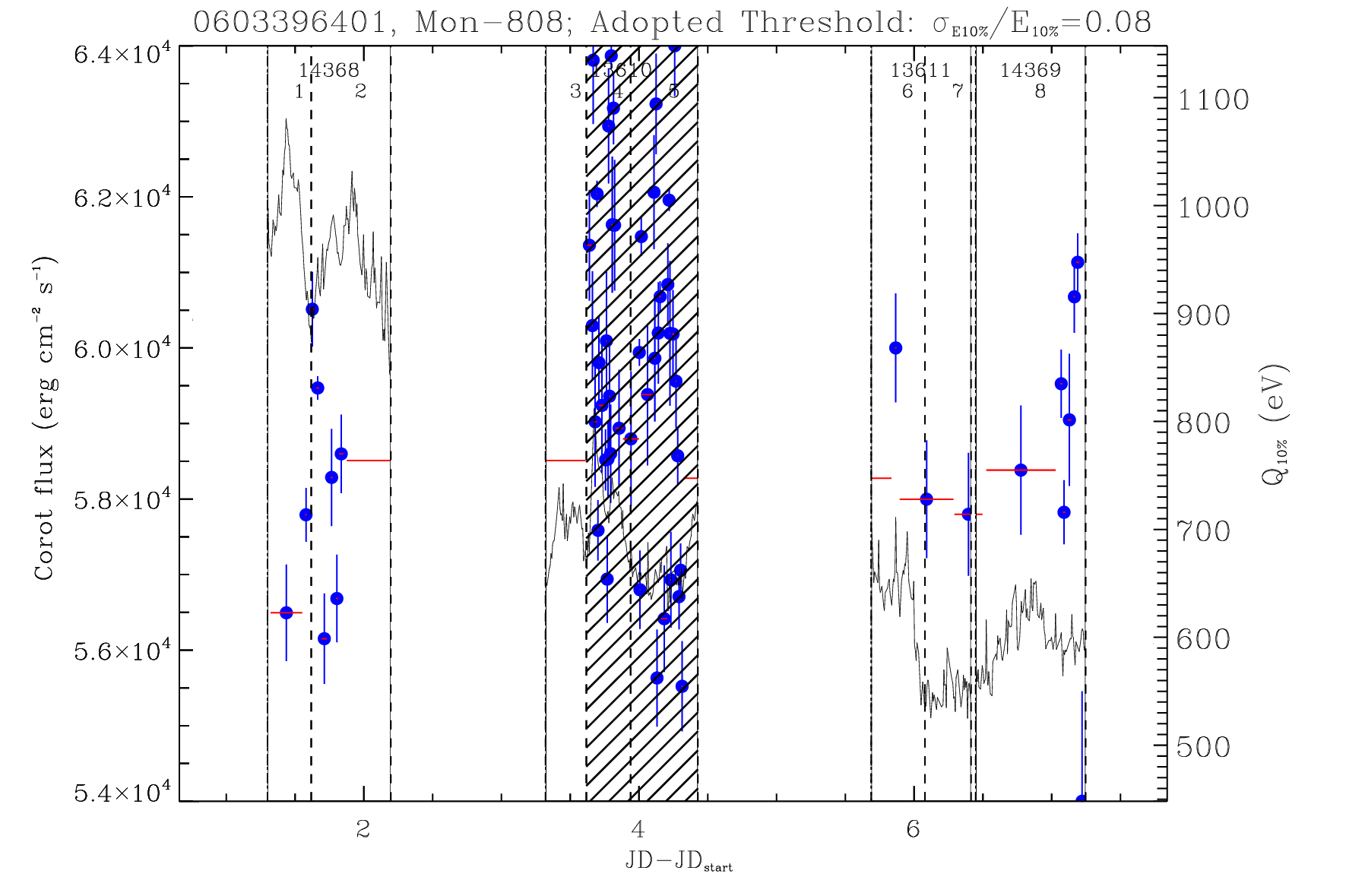}
	\includegraphics[width=9.0cm]{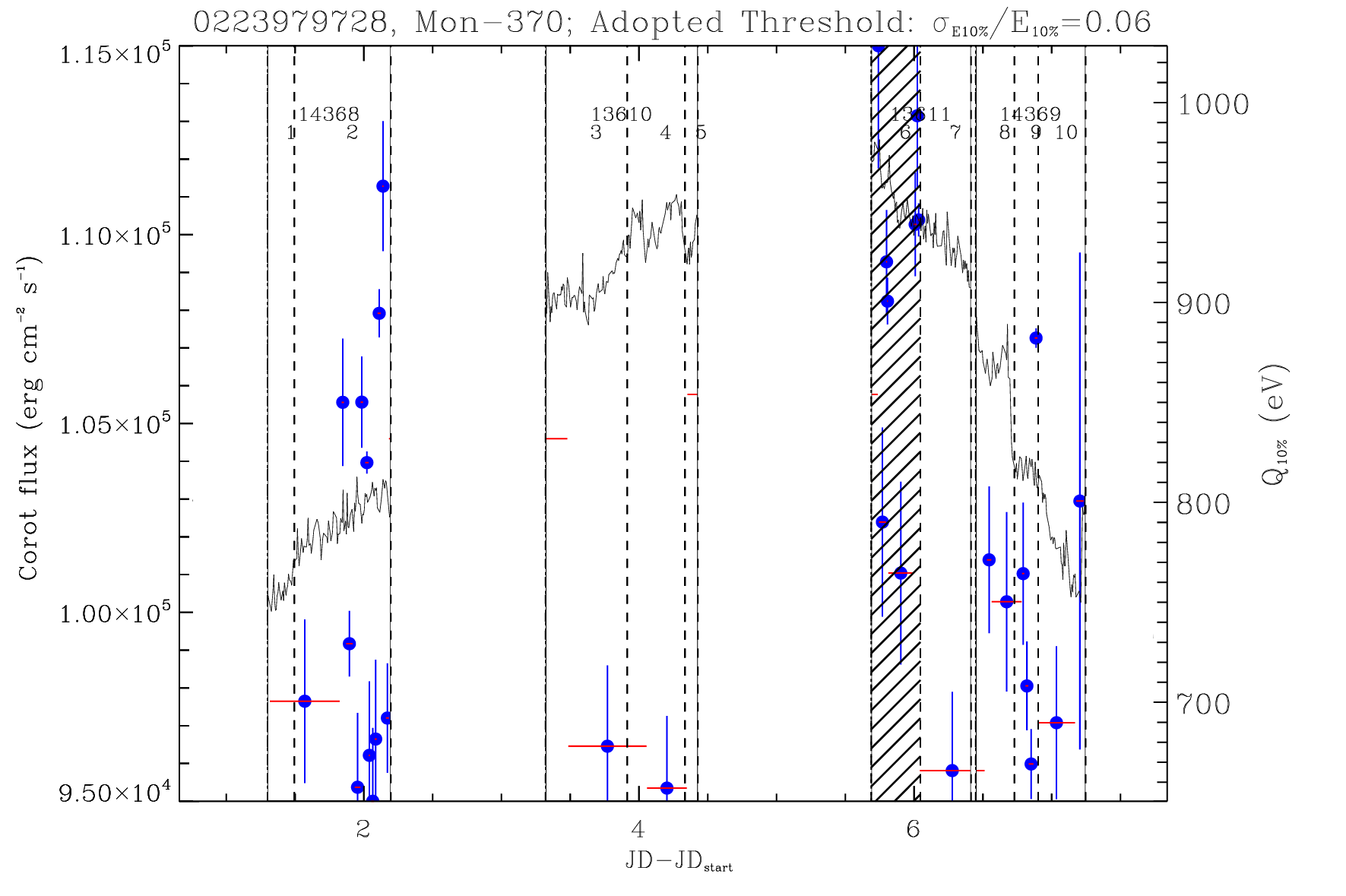}
	\includegraphics[width=9.0cm]{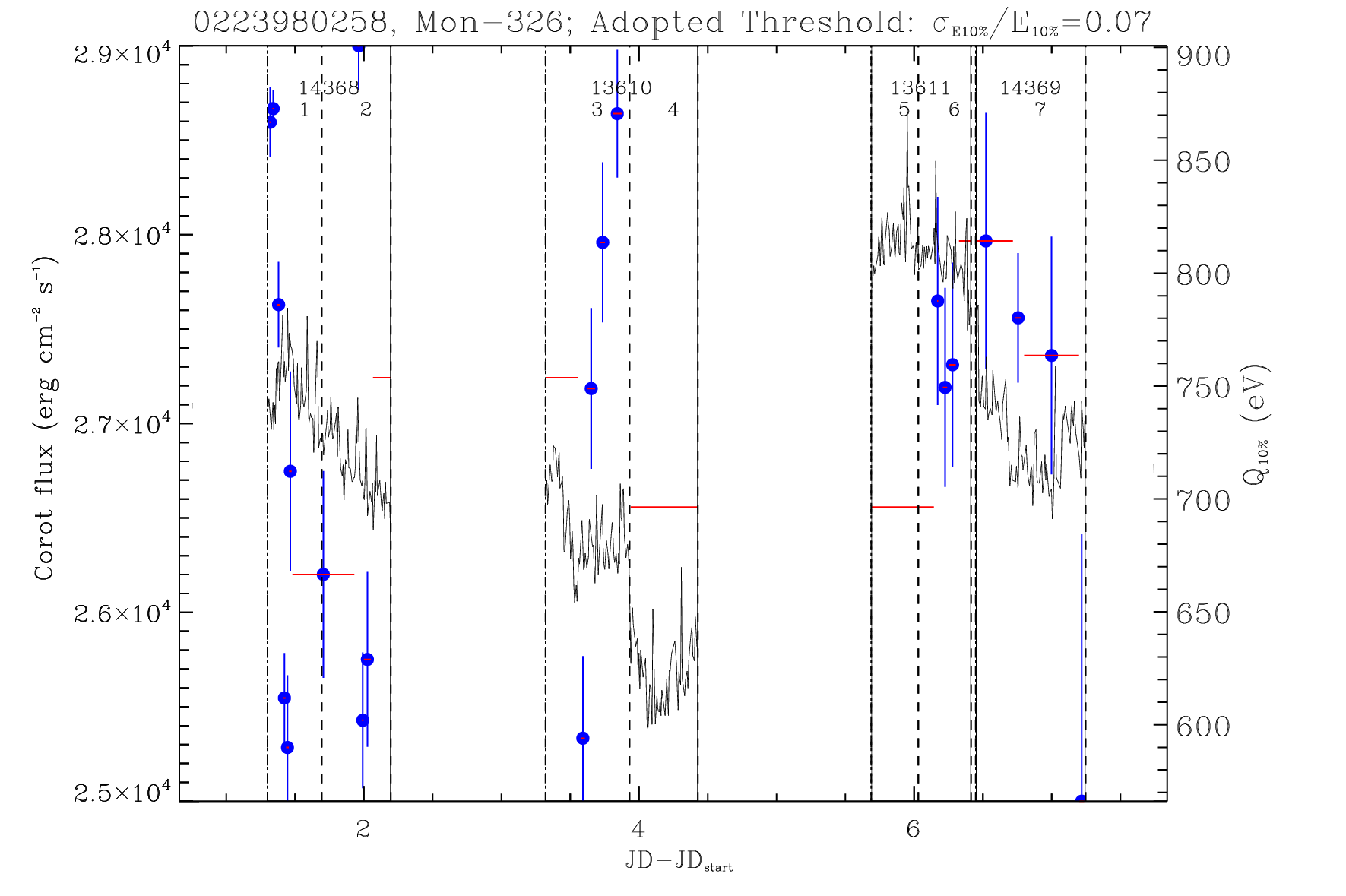}
	\includegraphics[width=9.0cm]{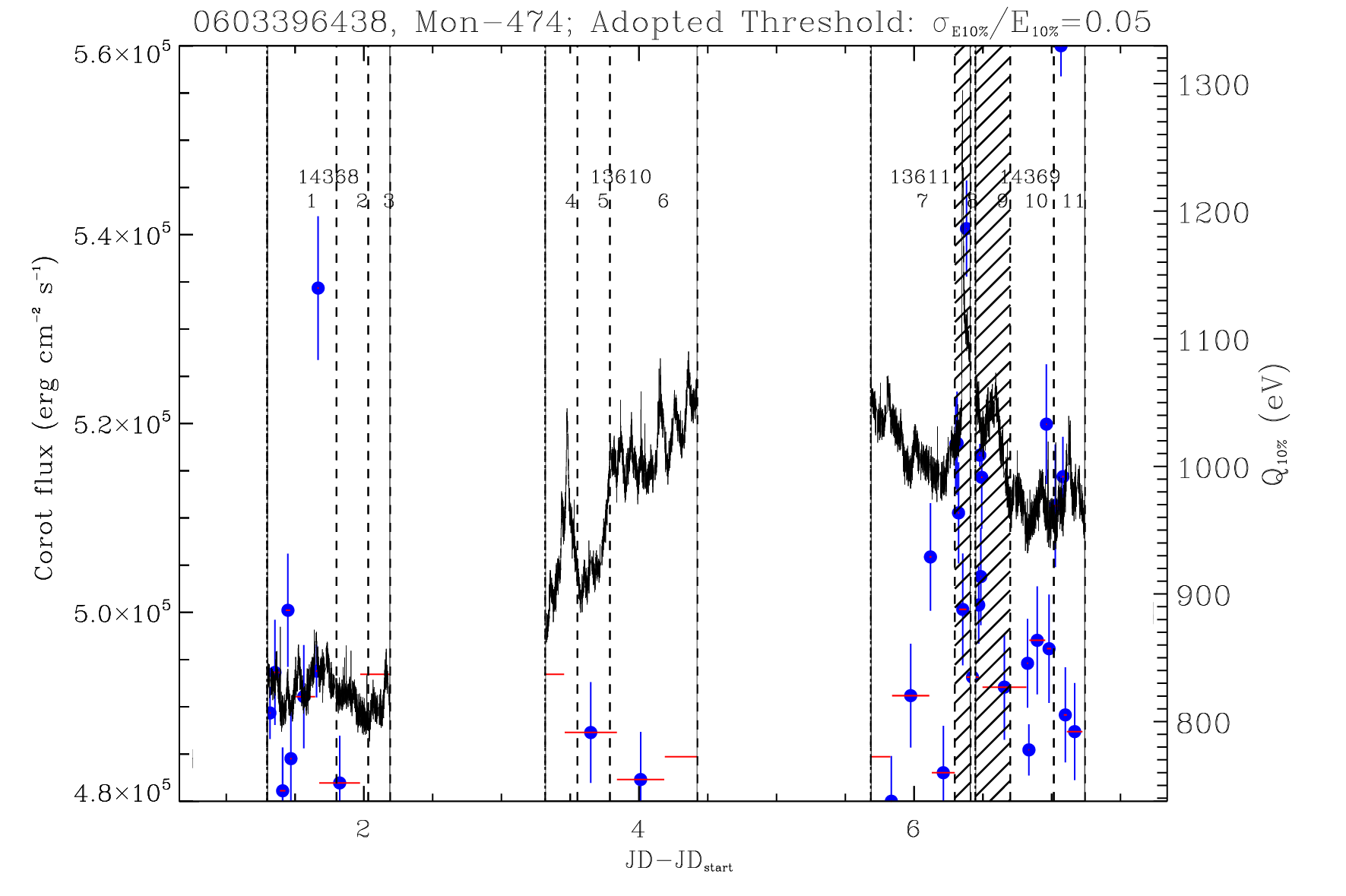}
	\includegraphics[width=9.0cm]{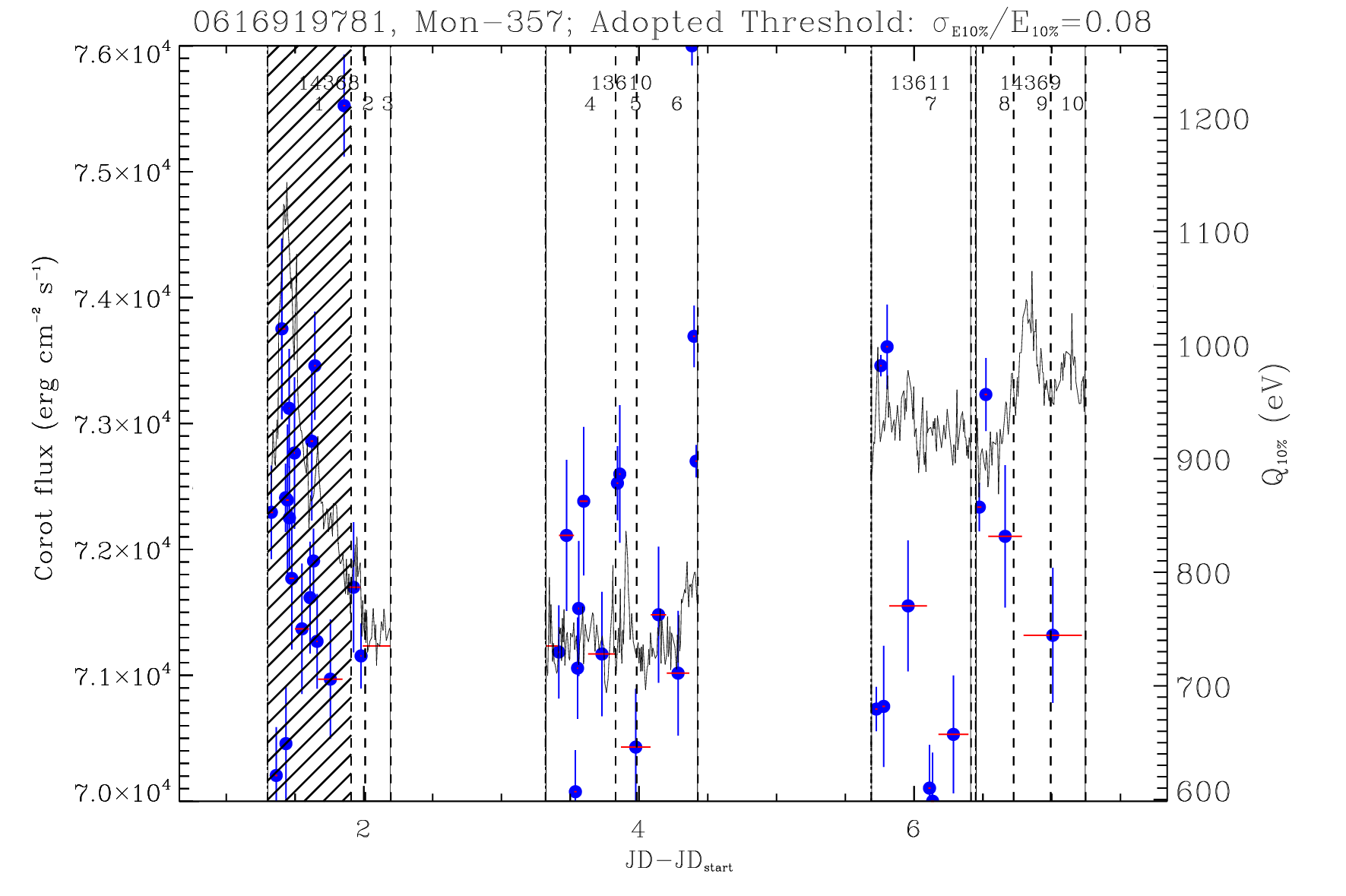}
	\includegraphics[width=9.0cm]{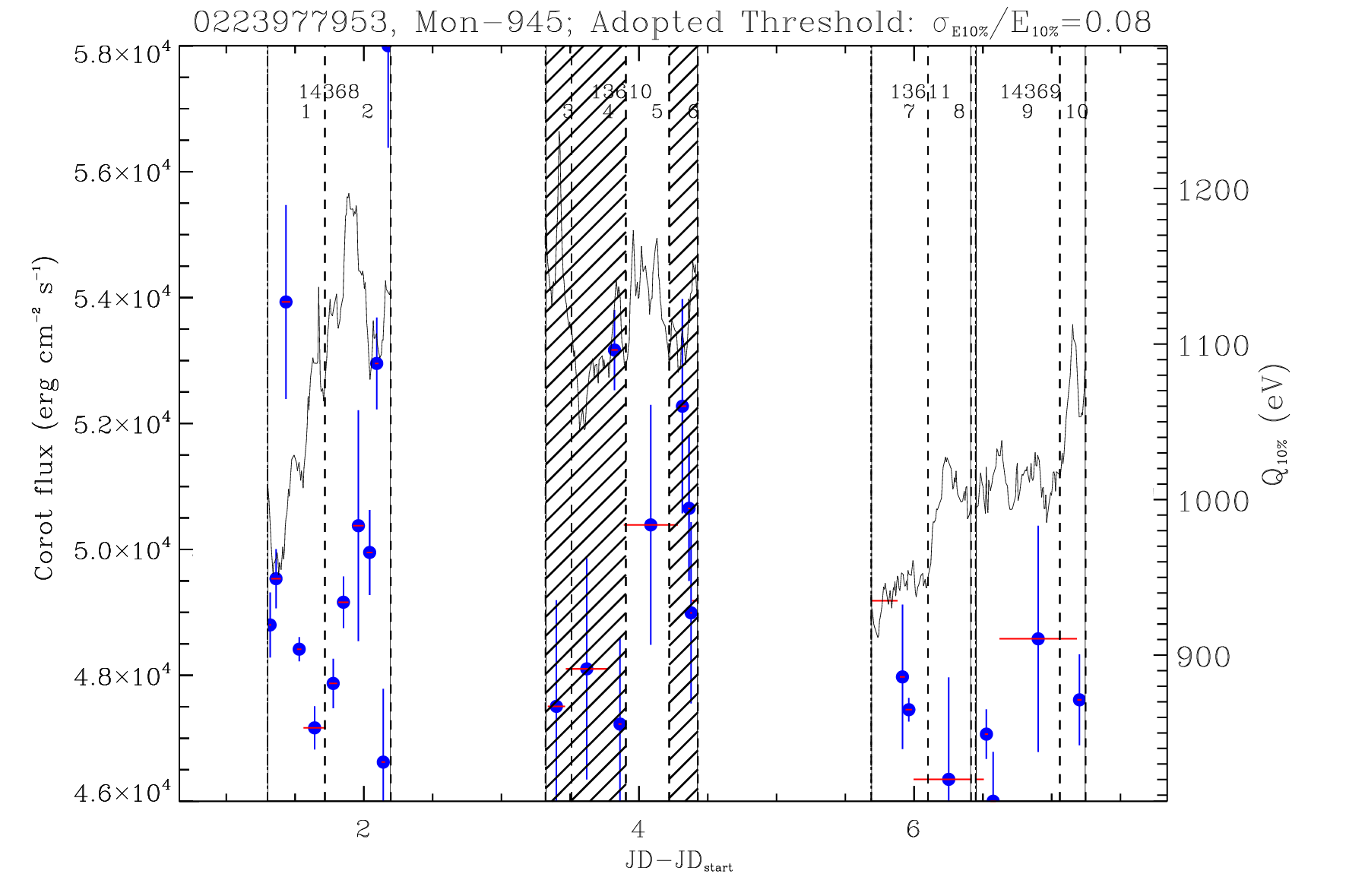}
	\includegraphics[width=9.0cm]{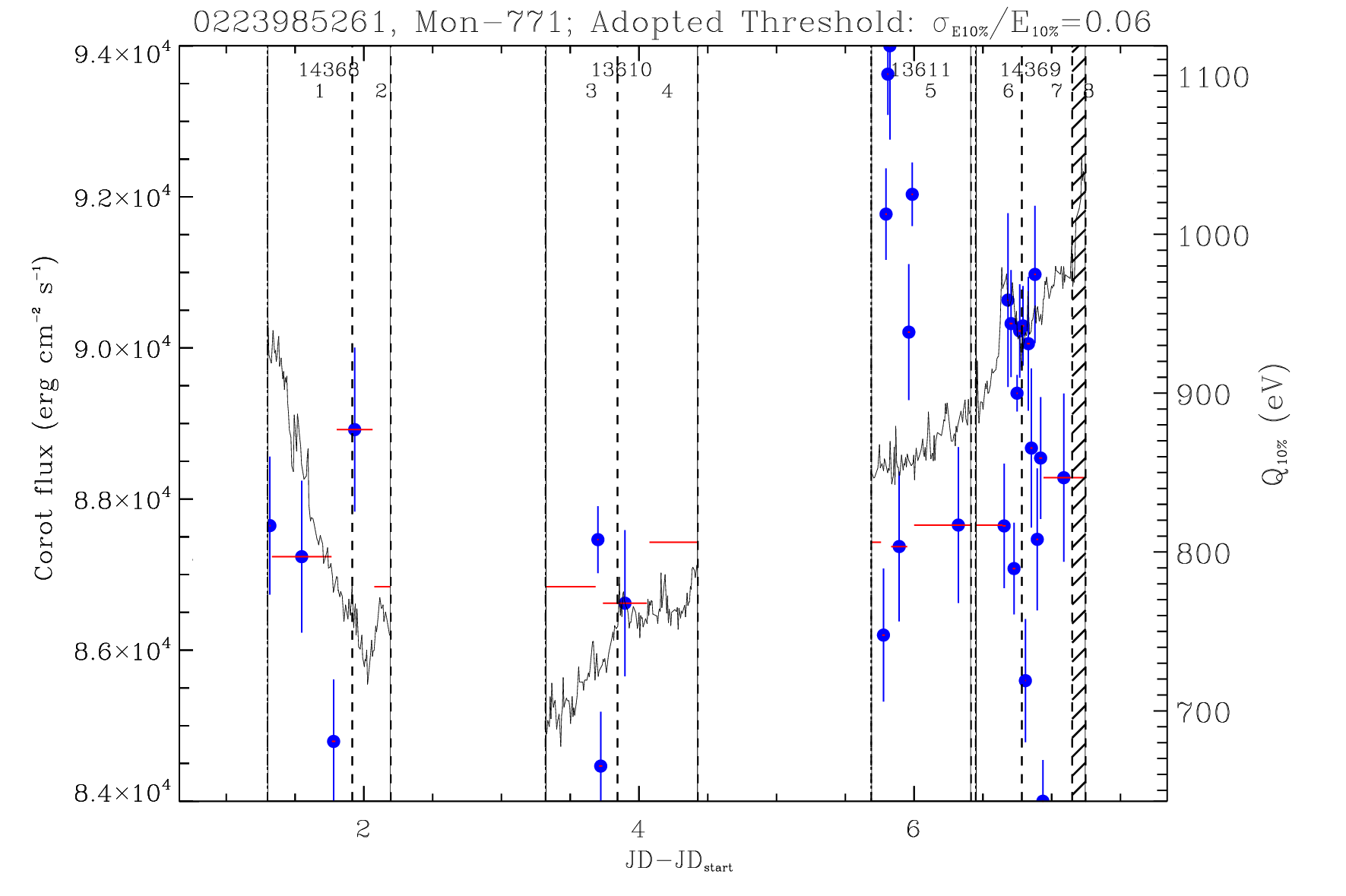}
	\label{runquant}
	\end{figure}

\newpage
\clearpage

%%%%%%%%%%%%%%%%%%%%%%%%%%%%%%%%%%%%%%%%%%%%%%%%%%%%%%%%%%%%%
\section{Optical and X-ray light variability of disk bearing stars not discussed in the paper}
\label{others_app}

In this appendix, we show the optical and X-ray variability and the
time resolved X-ray spectra of those stars with disks observed with
CoRoT and {\em Chandra} which have not been discussed in the paper for the
reasons explained in each caption. For each star, we show the CoRoT
light curve observed during the four {\em Chandra} frames; the X-ray
spectra observed in the time intervals; the variability of the
following X-ray properties: N$_H$ (in units of $10^{22}\,$cm$^{-2}$),
and kT (in keV), together with that of the 10\% and 25\% photon energy
quantiles (in keV) for the stars analyzed ``dippers''; kT and the
10\%, 25\%, and 50\% photon energy quantiles for the stars analyzed as
``bursters''; N$_H$, kT, F$_X$ (in erg$\,$cm$^{-2}\,$s$^{-1}$), and
the median photon energy for the stars analyzed as periodic or
quasi-periodic variable stars. We also show the variability of
E$_{10\%}$ as in the Appendix \ref{e10lc_app}. \par

	\begin{figure}[!hb]
	\centering	
	\includegraphics[width=9.5cm]{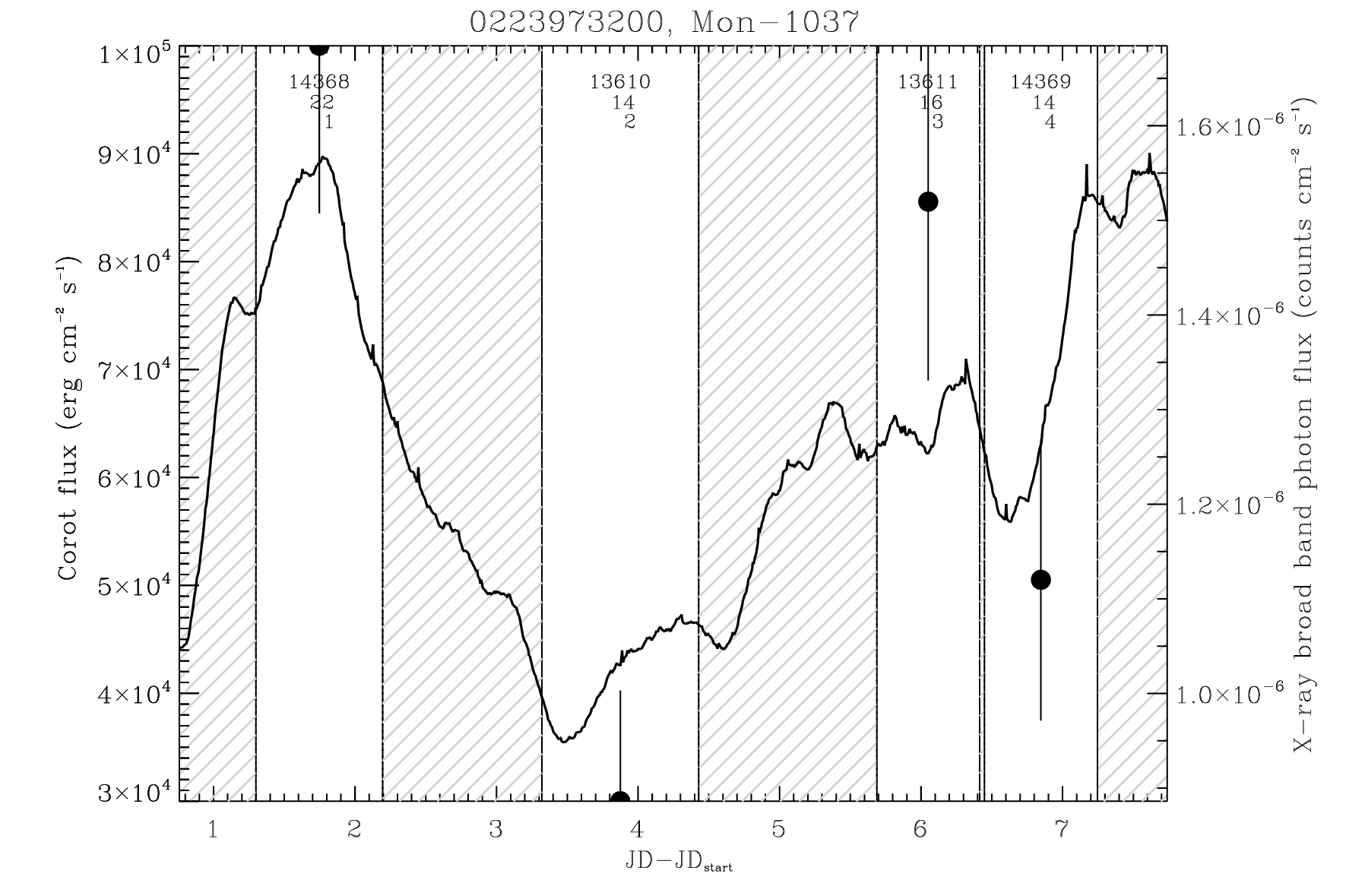}
	\includegraphics[width=9.5cm]{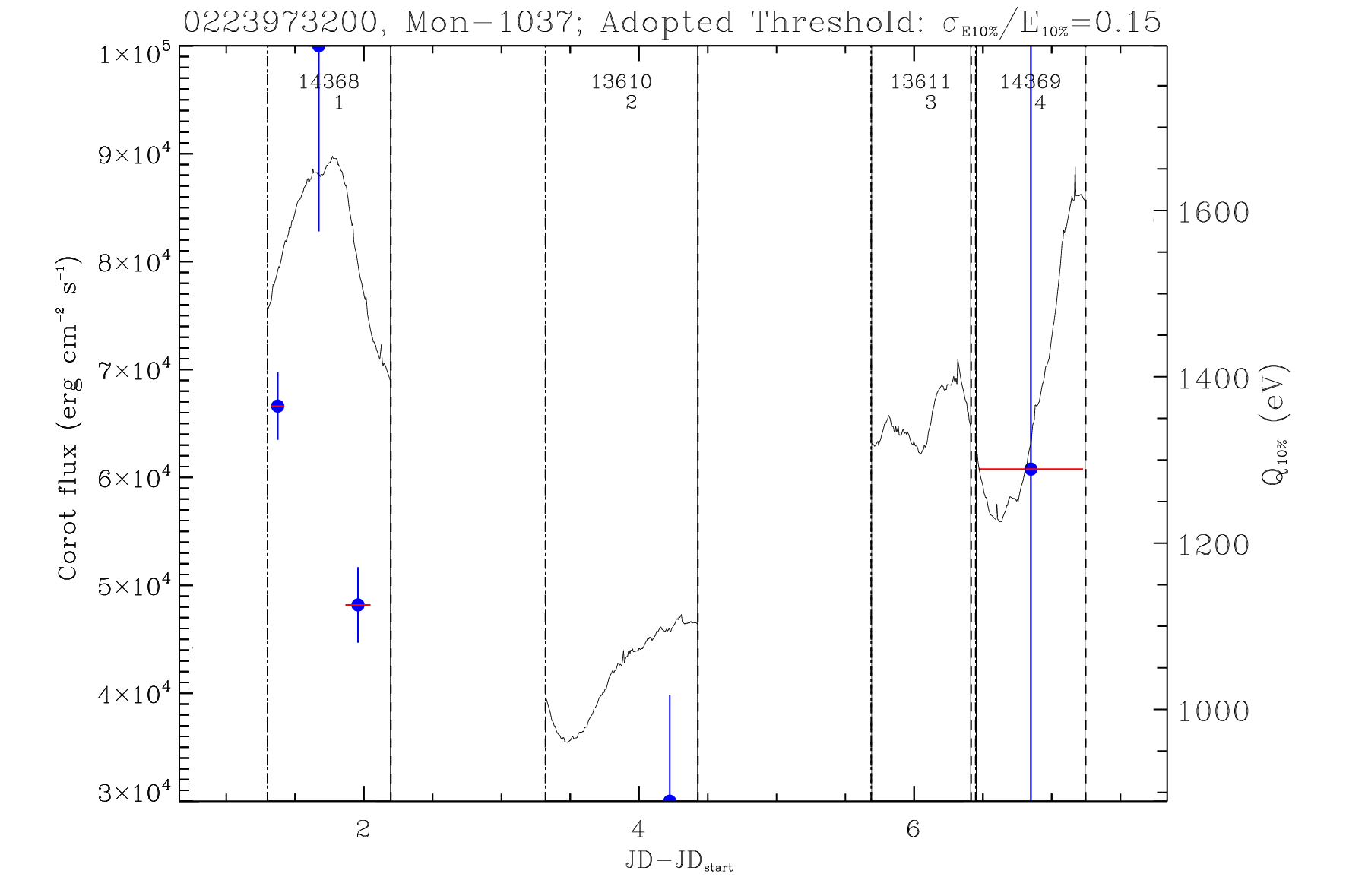}
	\includegraphics[width=8cm]{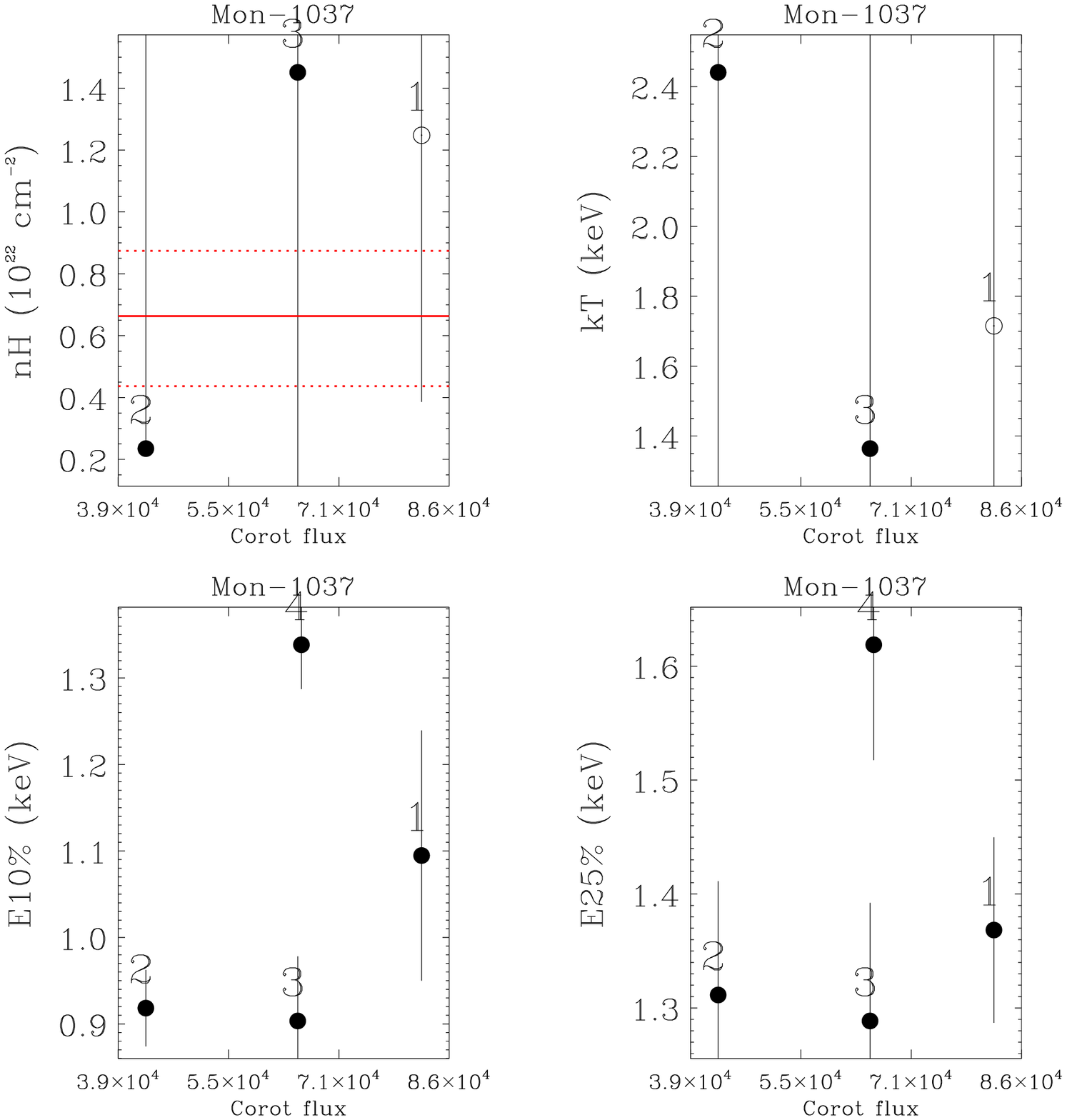}
	\includegraphics[width=18cm]{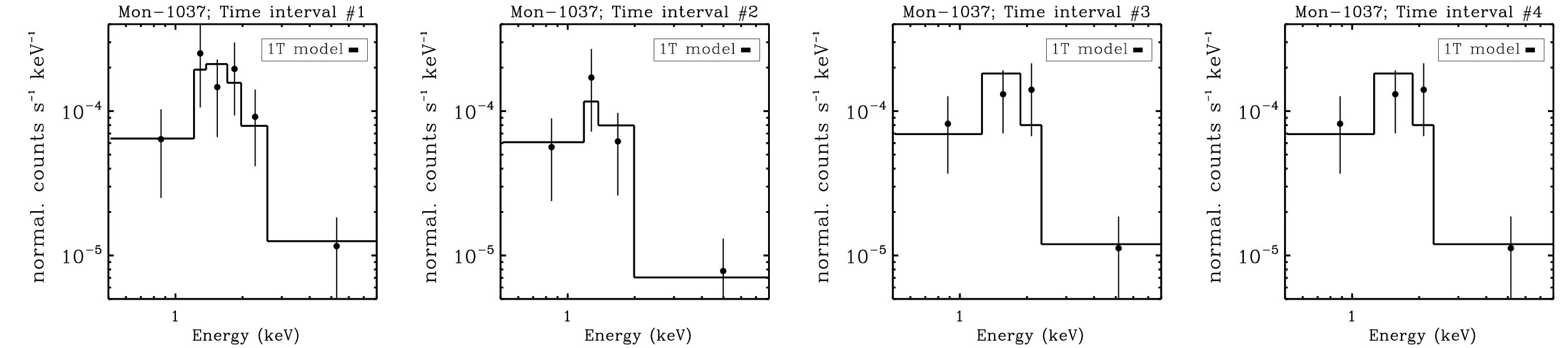}
	\caption{Variability and X-ray spectra of Mon-1037, analyzed as a dipper. E$_{10\%}$ is larger during the dip in \#4, but a significant variability of N$_H$ is not observed.}
	\label{variab_others_1}
	\end{figure}
  	
	\begin{figure}[!hb]
	\centering	
	\includegraphics[width=9.5cm]{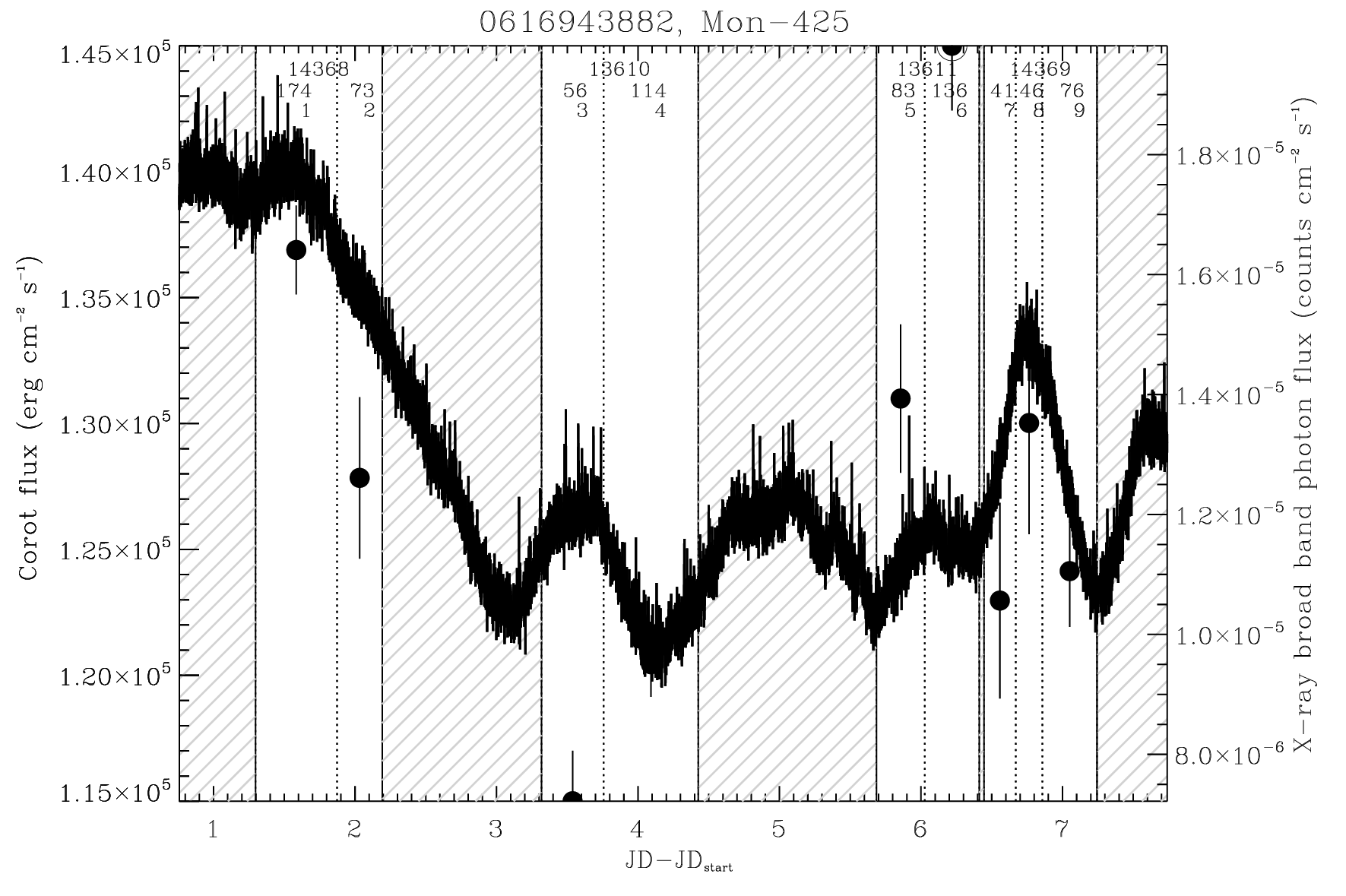}
	\includegraphics[width=9.5cm]{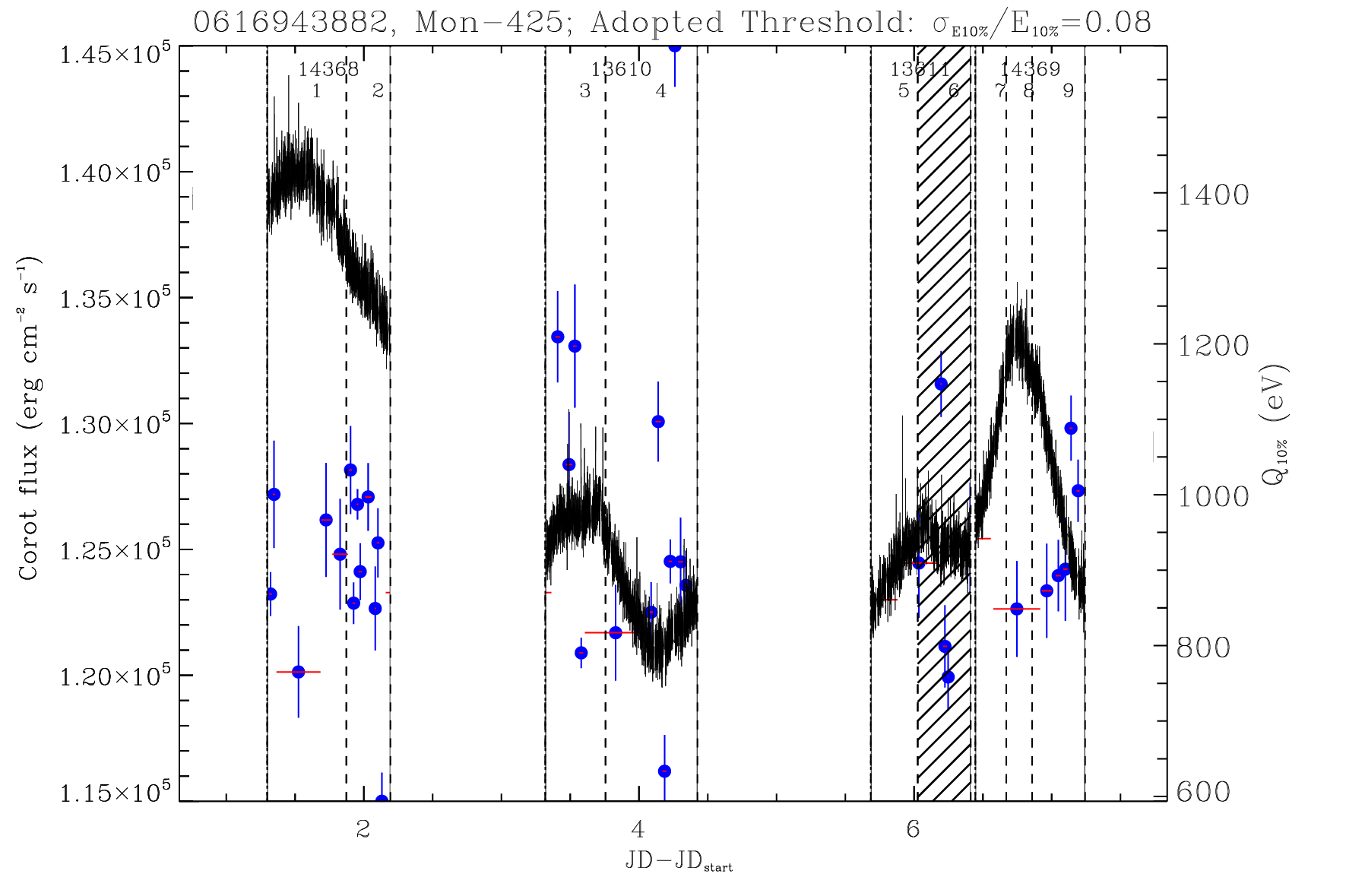}
	\includegraphics[width=8cm]{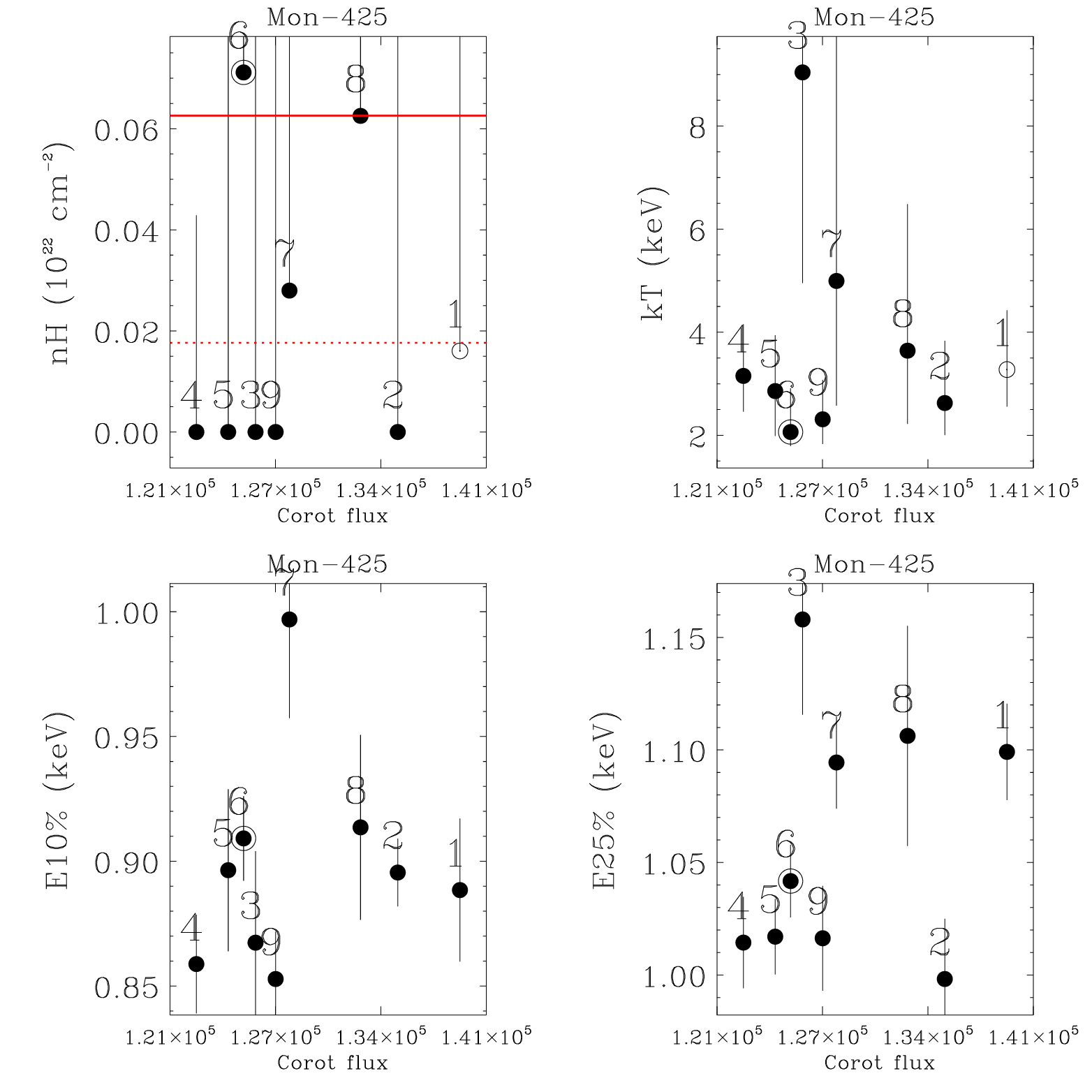}	
	\includegraphics[width=18cm]{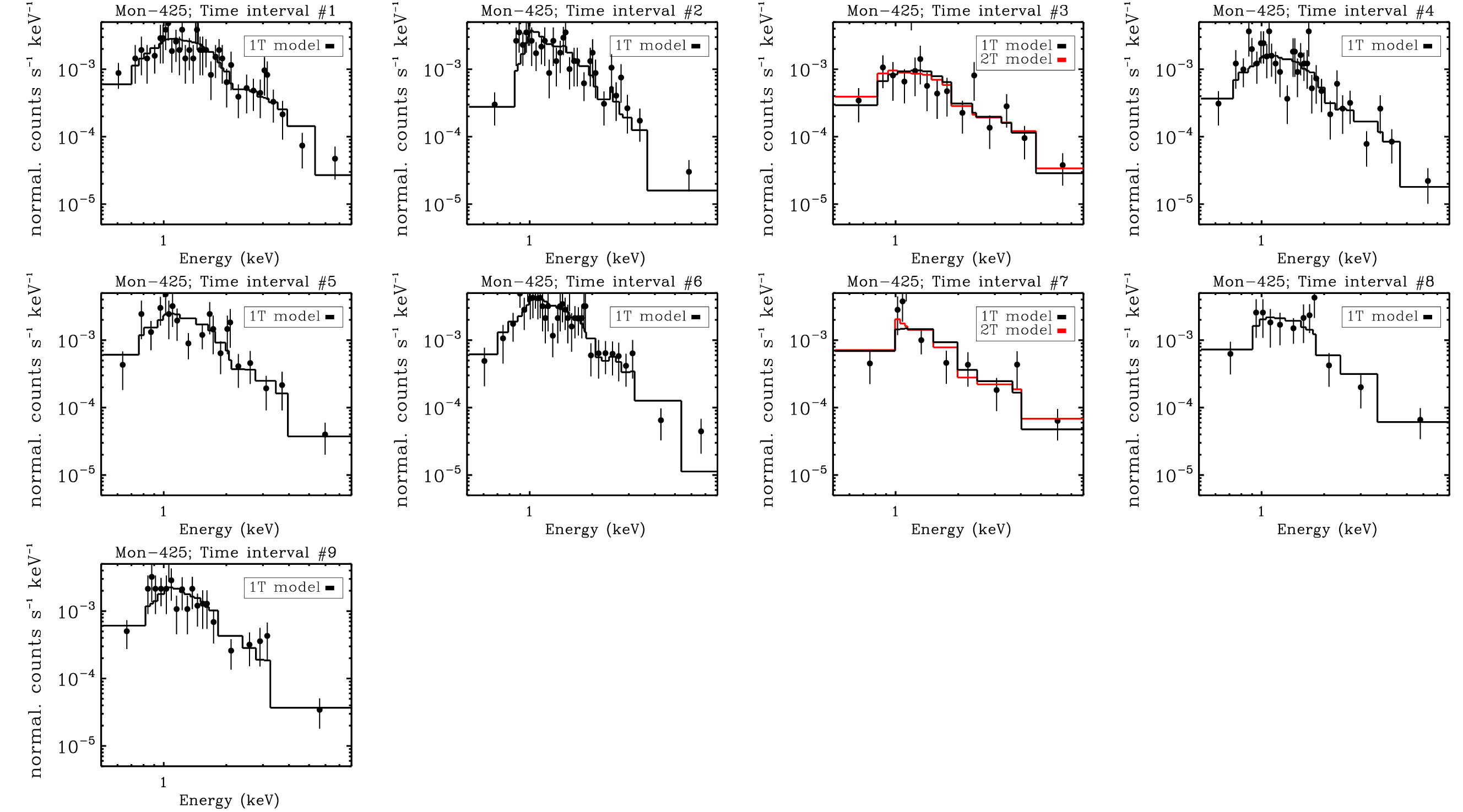}
	\caption{Variability and X-ray spectra of Mon-425, analyzed as a dipper. E$_{10\%}$ is larger during the dip in \#7, but there are no other significant correlations between the optical and X-ray variability.}
	\label{variab_others_2}
	\end{figure}

	\begin{figure}[]
	\centering	
	\includegraphics[width=9.5cm]{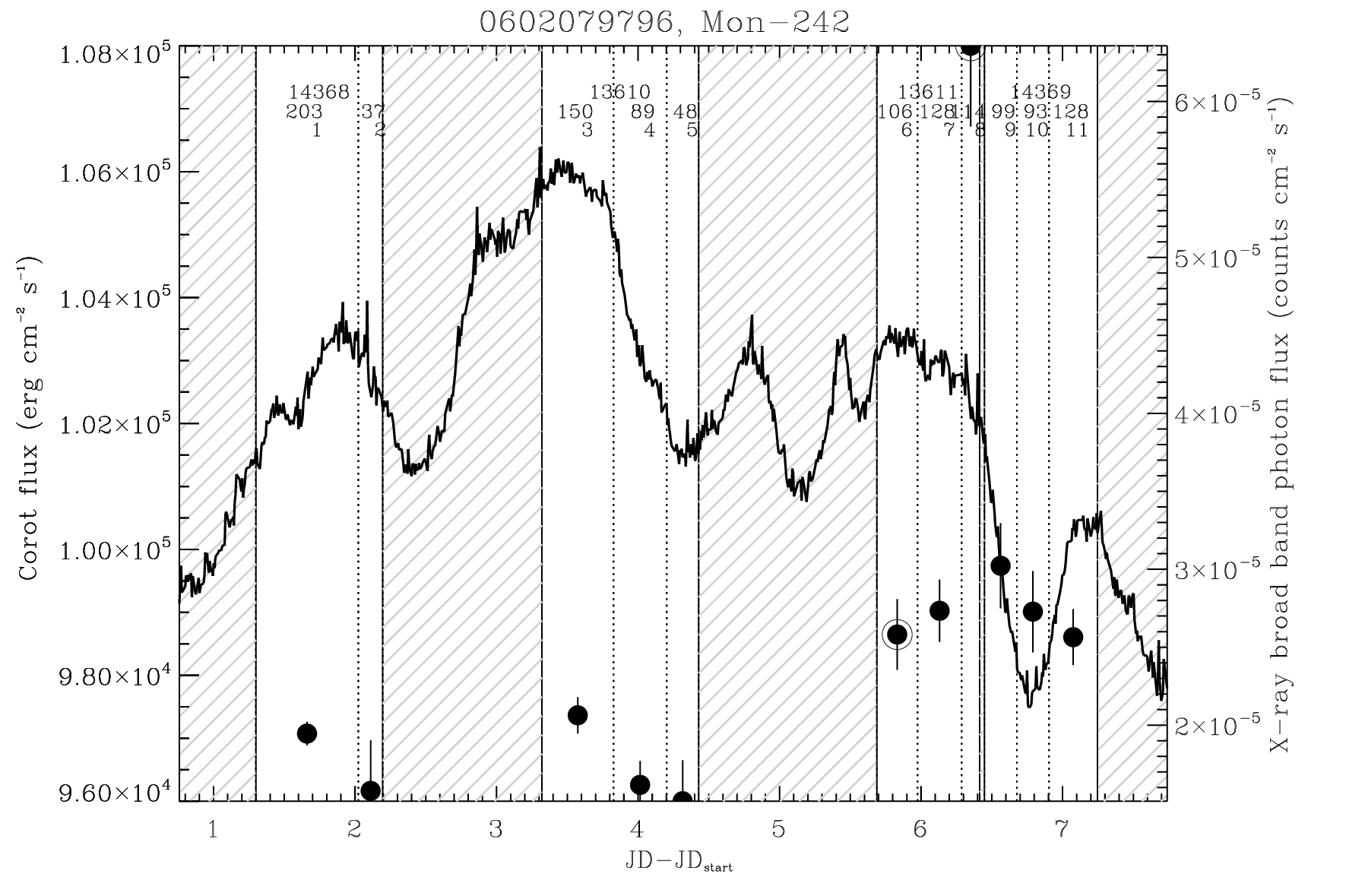}
	\includegraphics[width=9.5cm]{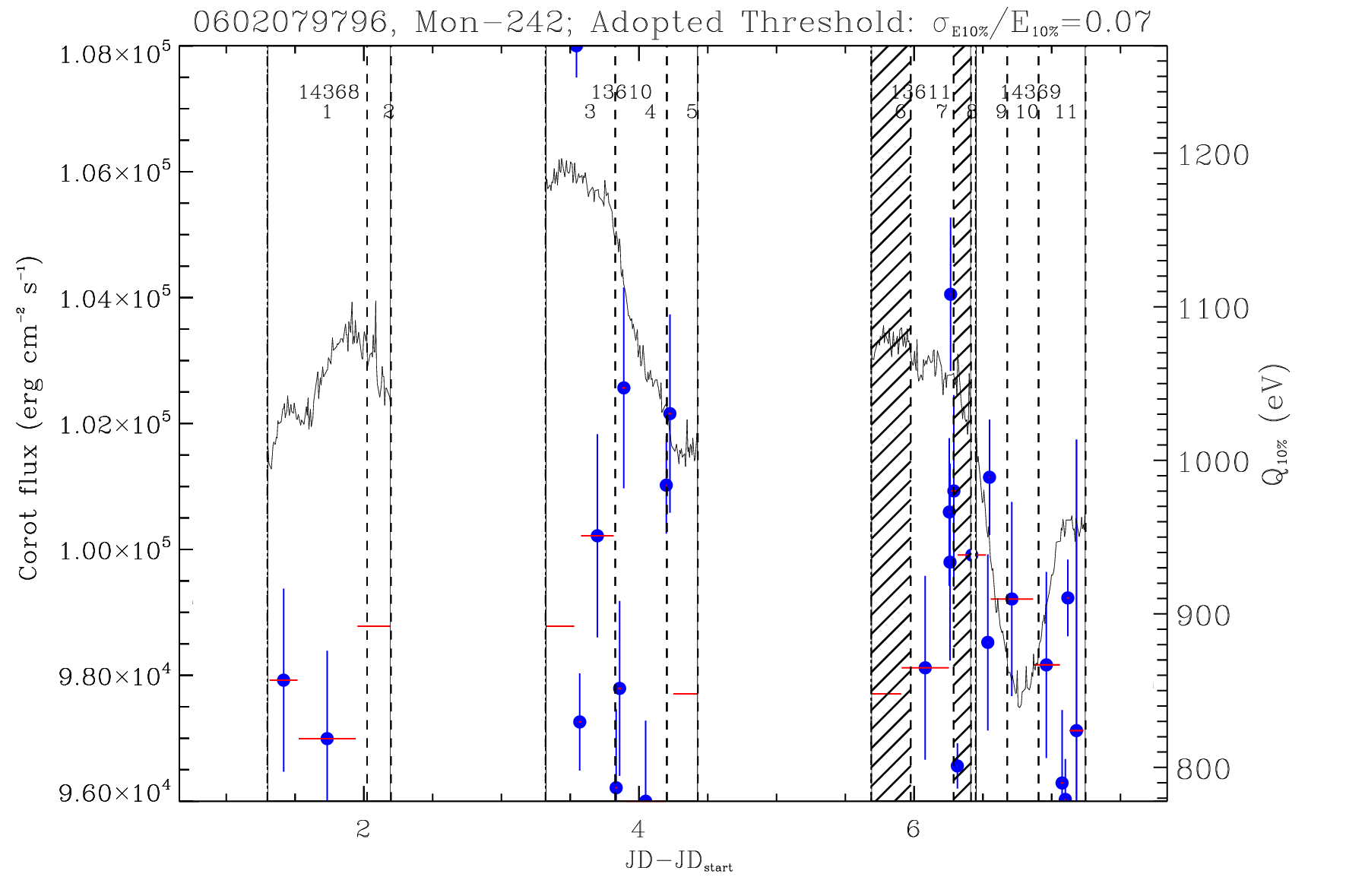}
	\includegraphics[width=8cm]{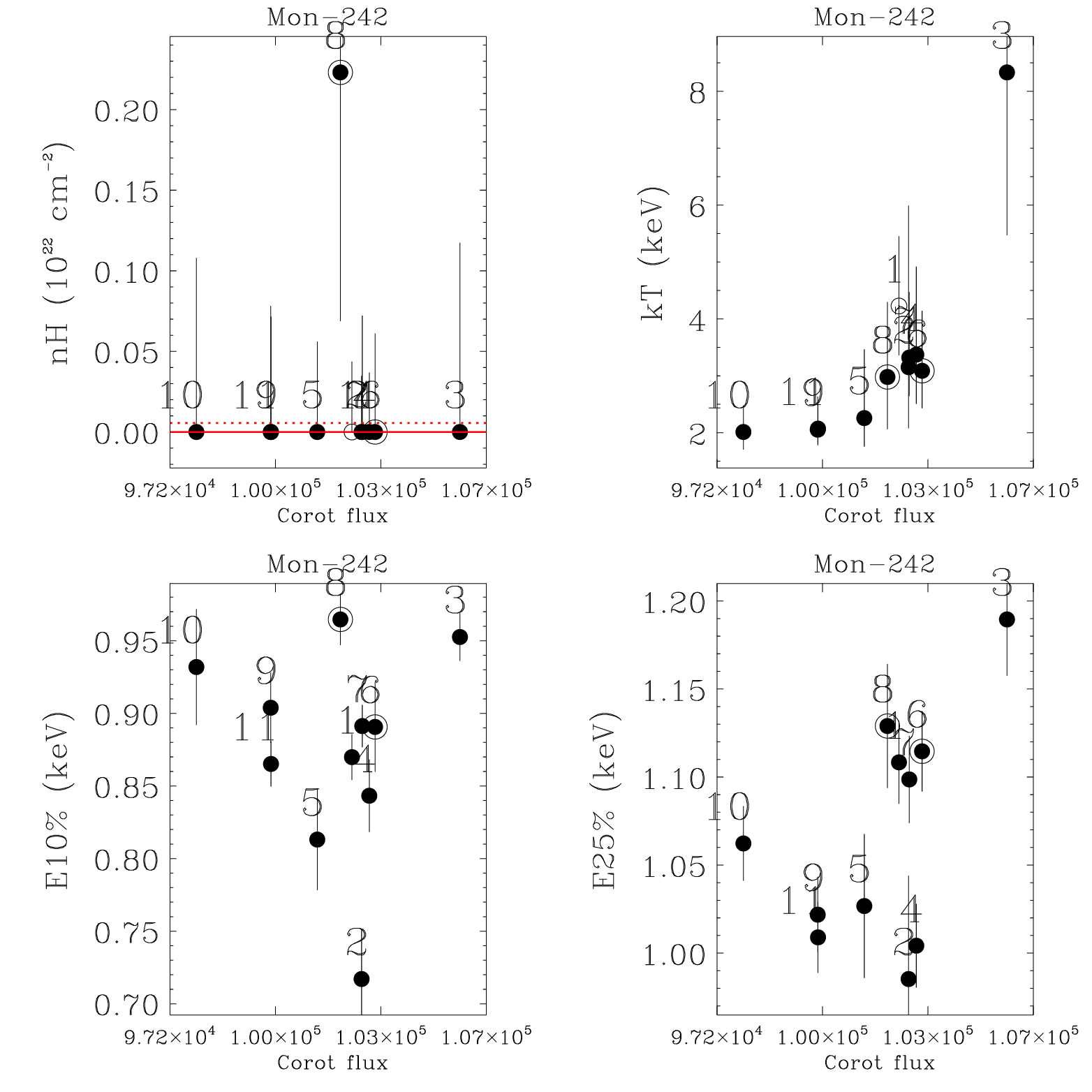}
	\includegraphics[width=18cm]{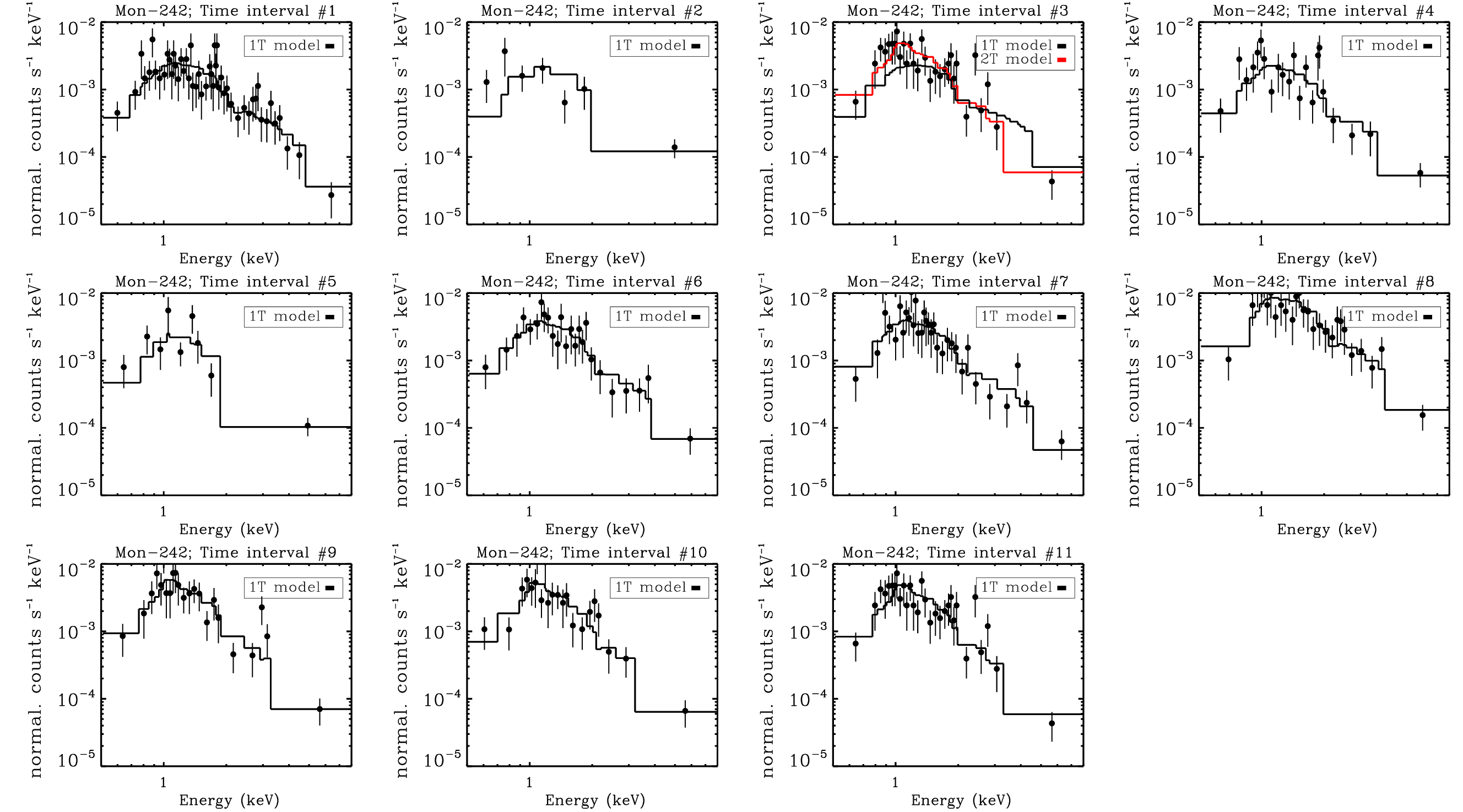}
	\caption{Variability and X-ray spectra of Mon-242, analyzed as a dipper. The CoRoT light curve shows several dips without evidence for increasing N$_H$. The spectrum in \#2 is very soft, but the low X-ray count rate does not allow us to verify the presence of a soft X-ray spectral component. The spectral fit with 2T thermal plasma model in \#3 is poorly constrained (P$_{\%}$=0.0)}
	\label{variab_others_3}
	\end{figure}

	\begin{figure}[]
	\centering	
	\includegraphics[width=9.5cm]{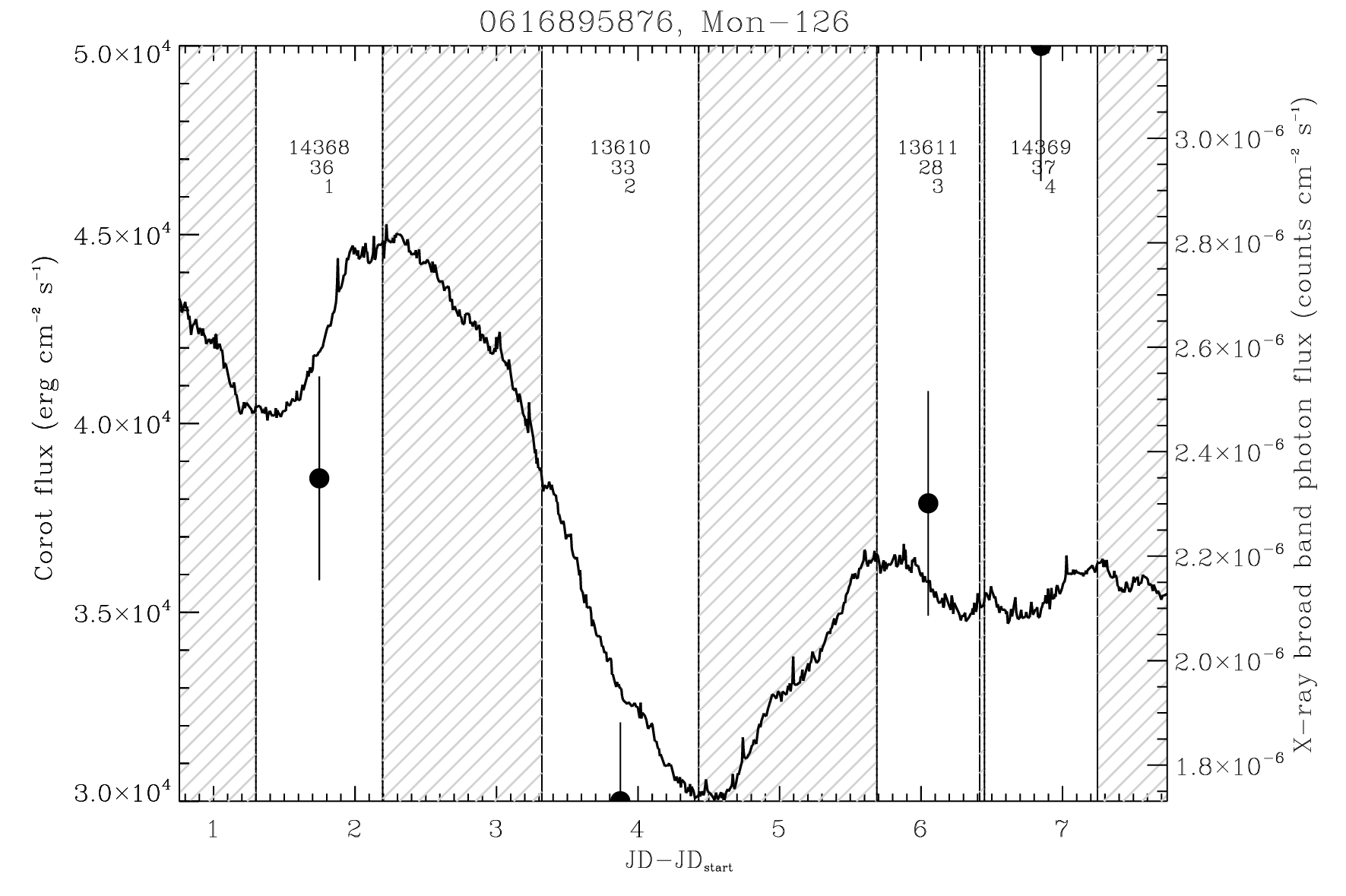}
	\includegraphics[width=9.5cm]{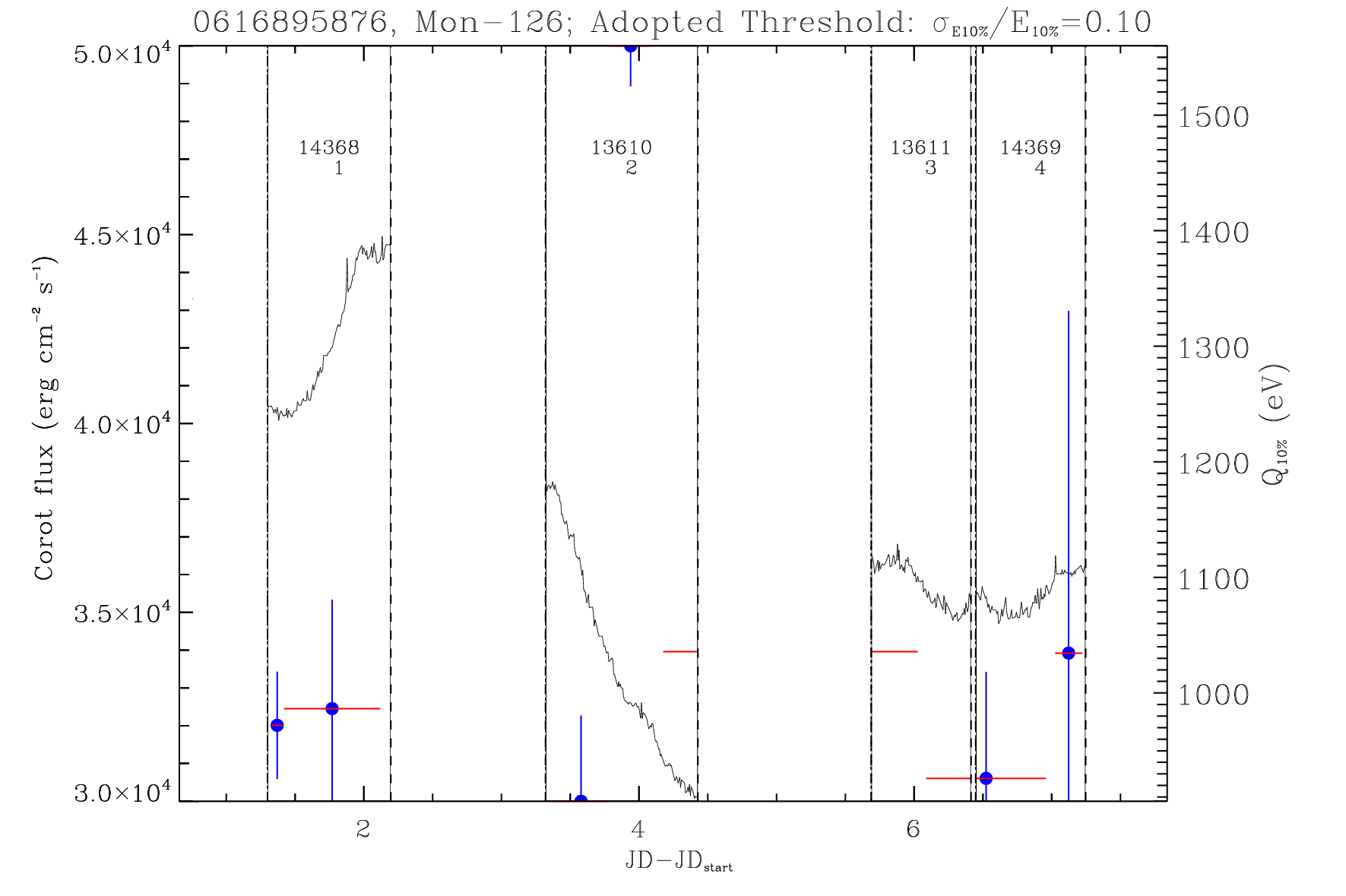}
	\includegraphics[width=8cm]{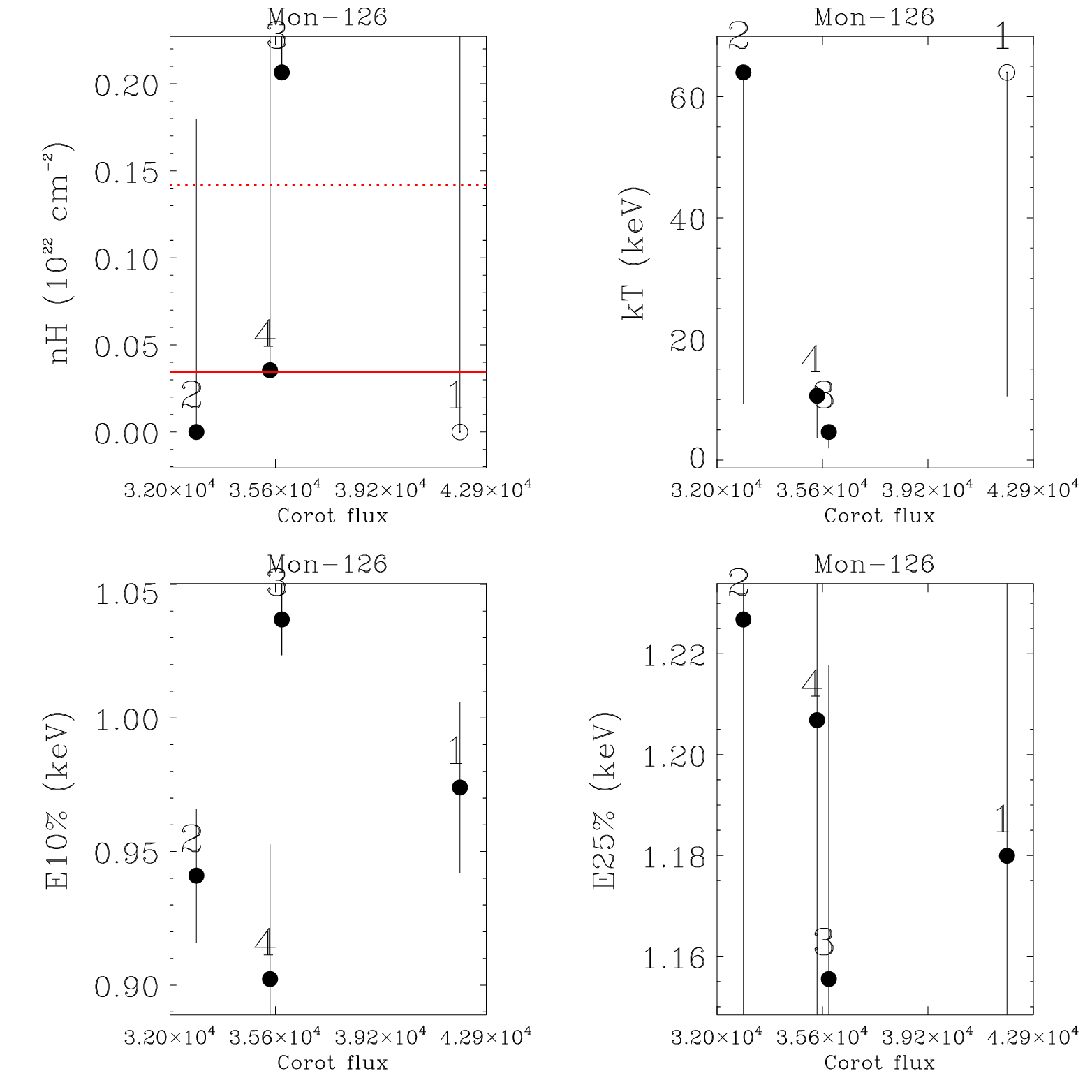}
	\includegraphics[width=18cm]{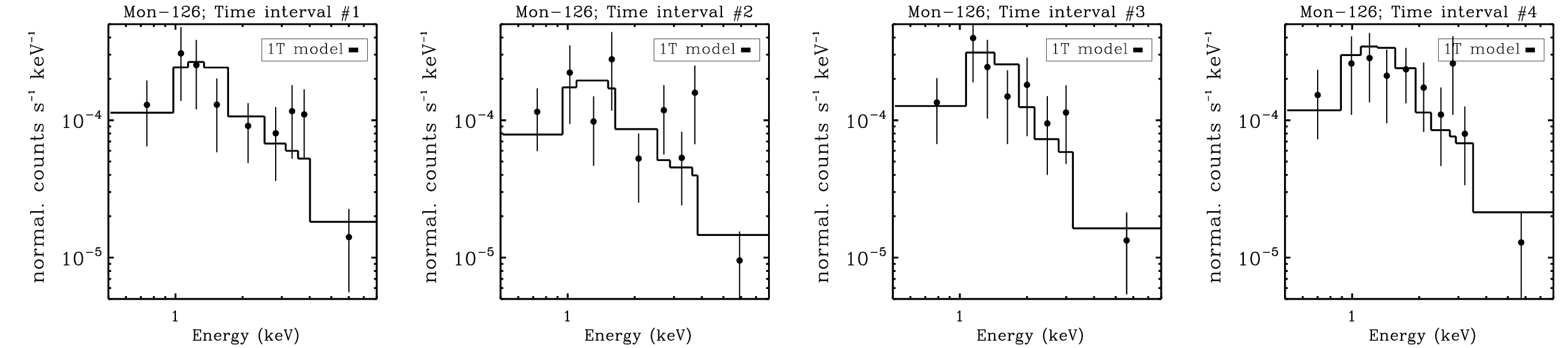}
	\caption{Variability and X-ray spectra of Mon-126, analyzed as a dipper. The CoRoT light curve shows dips without evidence of increasing N$_H$.}
	\label{variab_others_4}
	\end{figure}

	\begin{figure}[]
	\centering	
	\includegraphics[width=9.5cm]{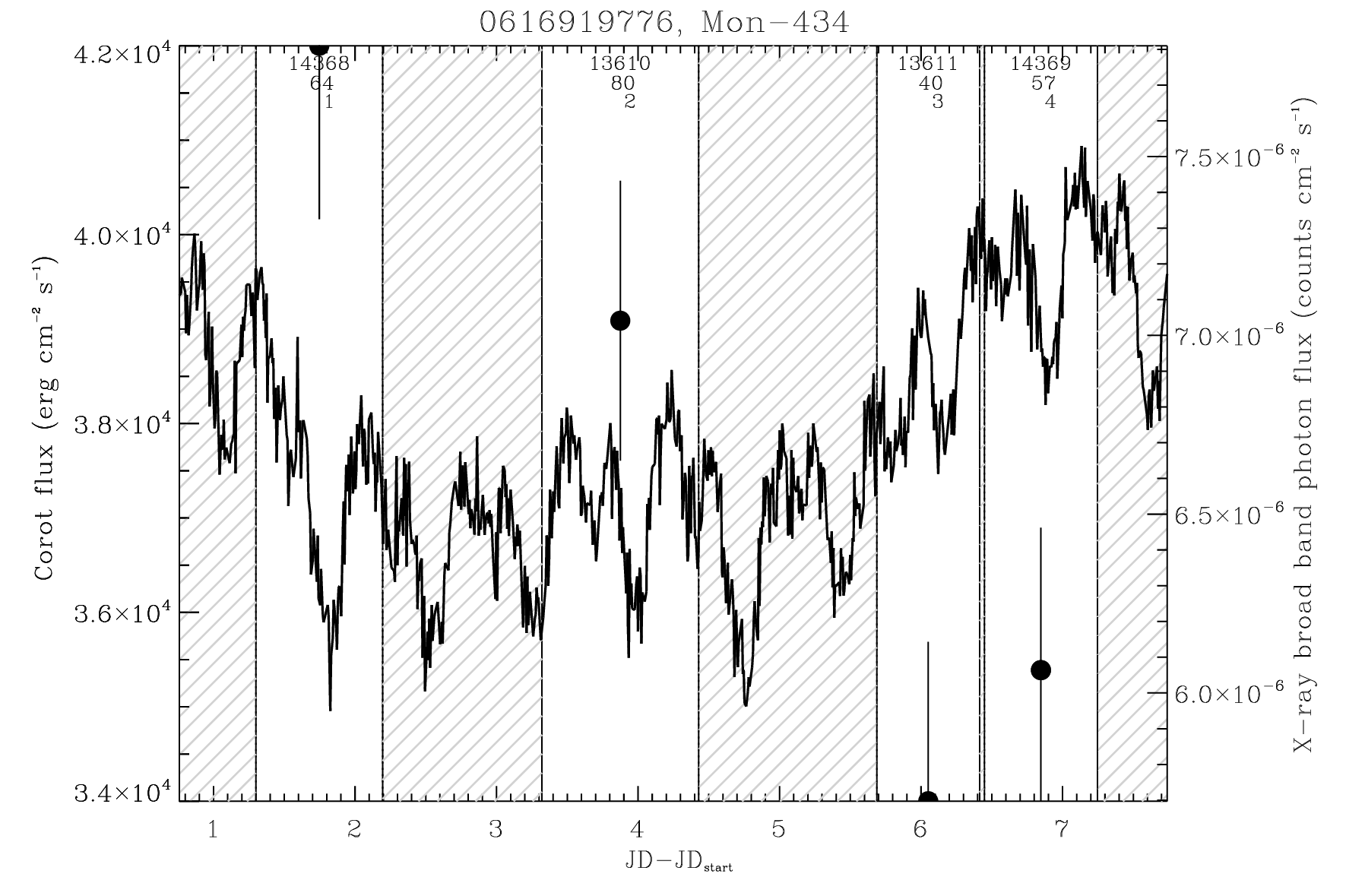}
	\includegraphics[width=8cm]{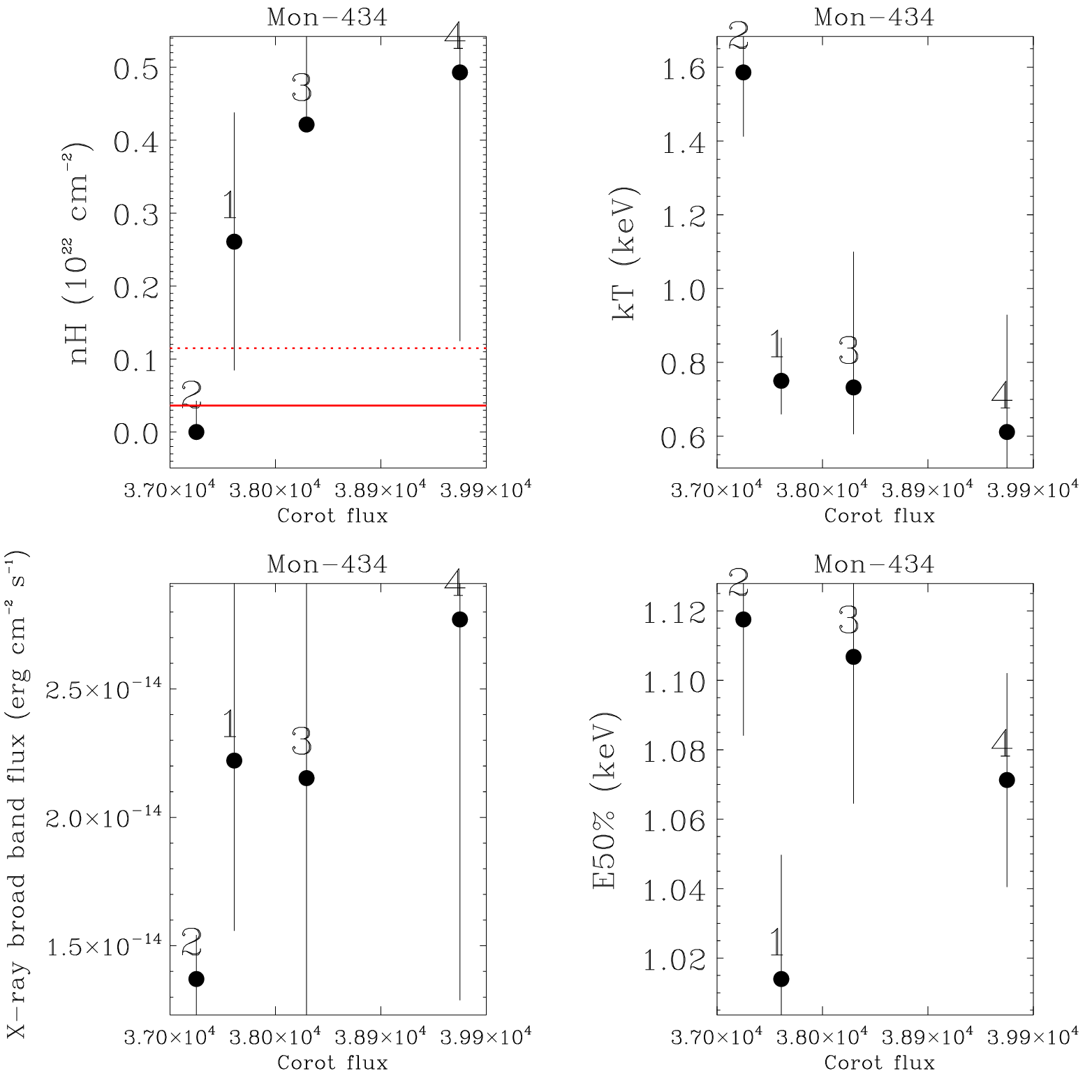}
	\includegraphics[width=18cm]{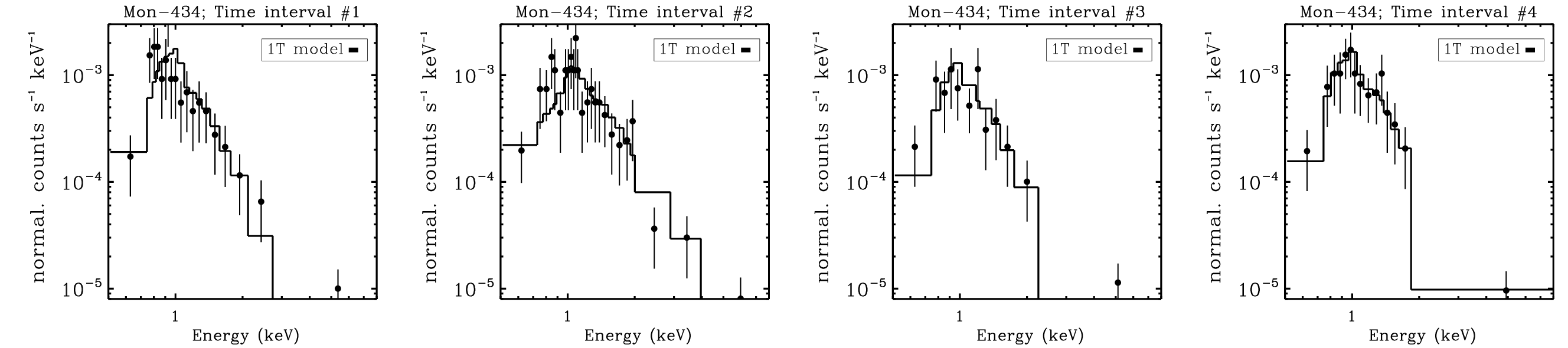}
	\caption{Variability and X-ray spectra of Mon-434. The CoRoT light curve is dominated by rotational modulation and likely pulsation, and it has been classified as a multi-periodic star by \citet{CodySBM2014AJ}. There may be evidence for correlated X-ray and CoRoT flux variability, but it is not significant.}
	\label{variab_others_5}
	\end{figure}

	\begin{figure}[]
	\centering
	\includegraphics[width=9.5cm]{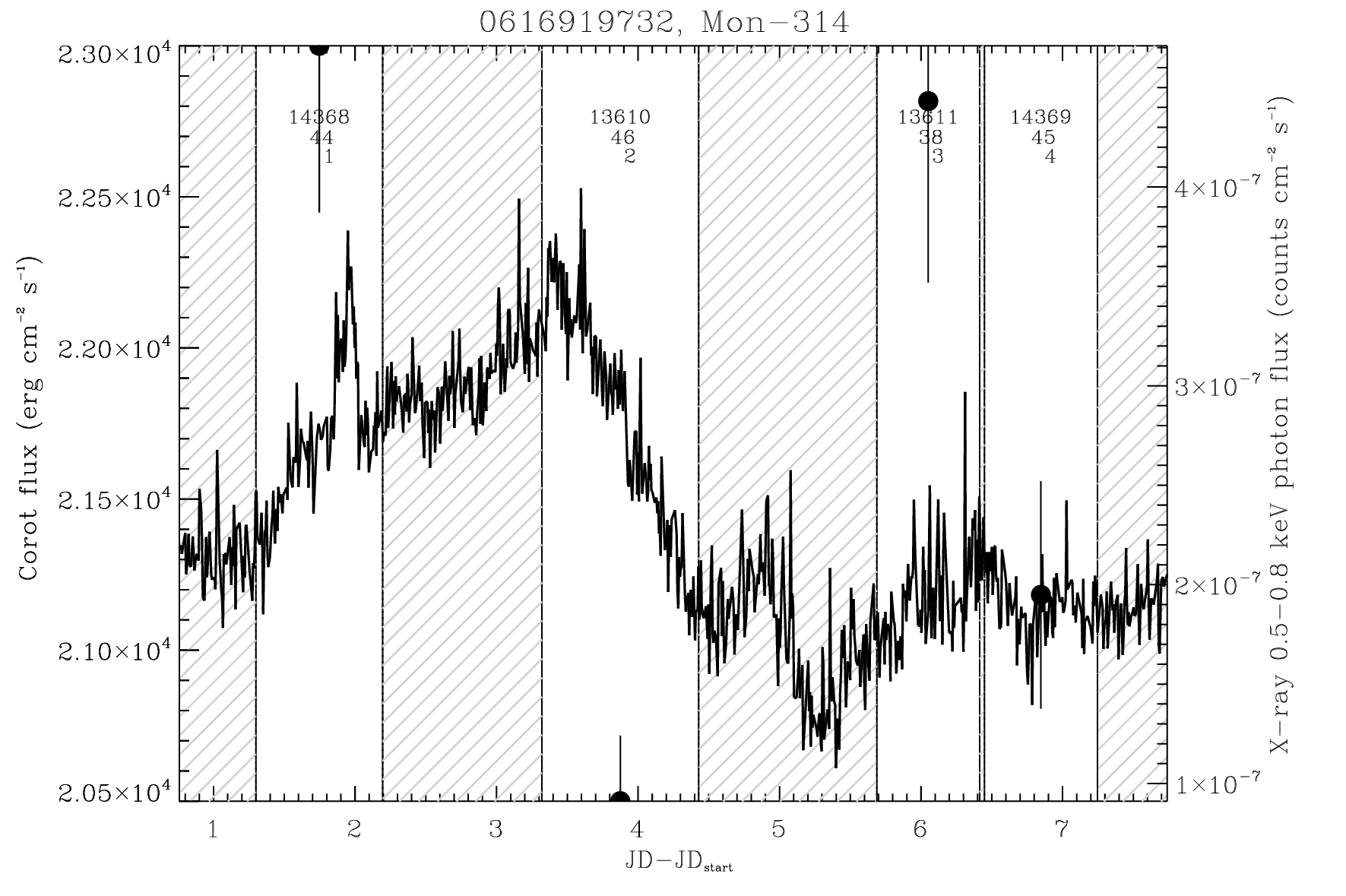}
	\includegraphics[width=9.5cm]{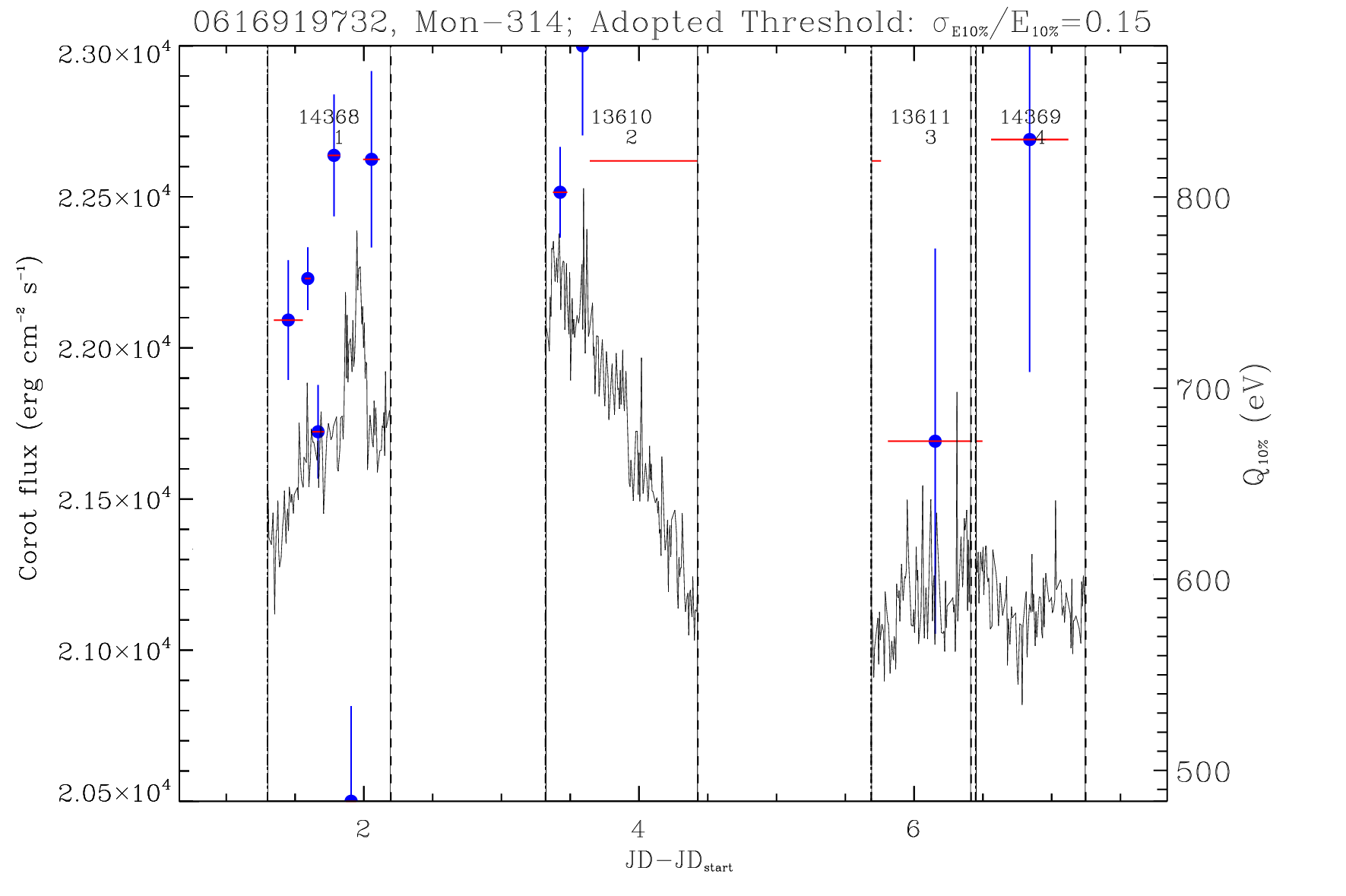}
	\includegraphics[width=8cm]{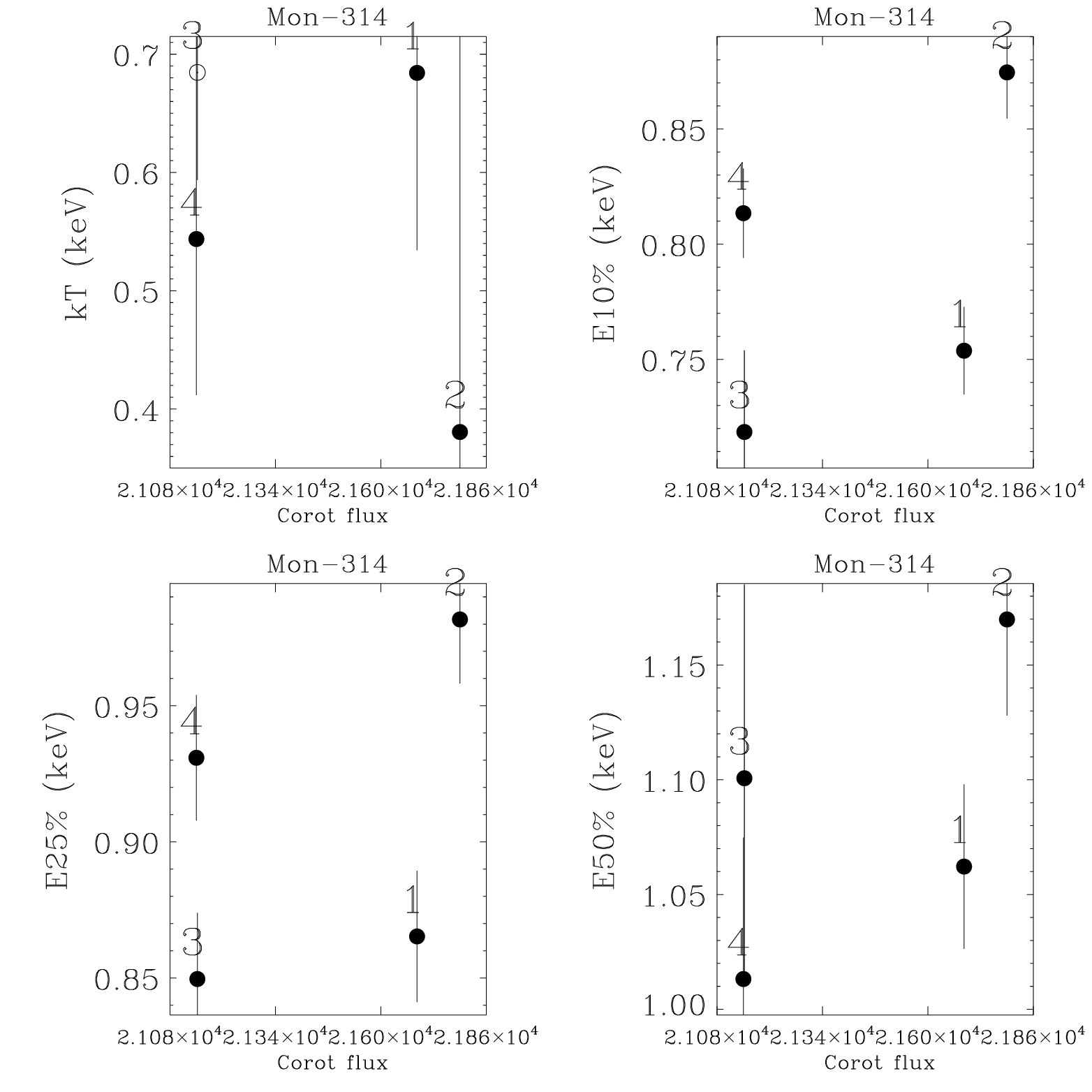}
	\includegraphics[width=18cm]{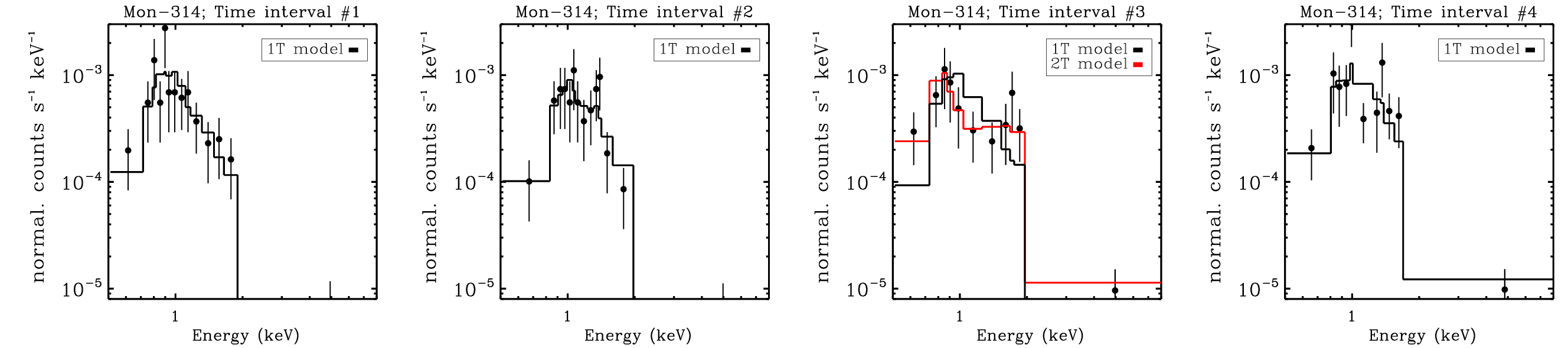}
	\caption{Variability and X-ray spectra of Mon-314, analyzed as a burster. The CoRoT light curve shows evident bursts and dips, but the poor X-ray count rate does not allow us to analyze the simultaneous X-ray variability. During \#1 the X-ray spectrum becomes very soft during the optical burst, as shown by the variability of E$_{10\%}$. The fit with 2T thermal plasma model in \#3 is not well constrained given the low X-ray count rate.}
	\label{variab_others_6}
	\end{figure}

	\begin{figure}[]
	\centering
	\includegraphics[width=9.5cm]{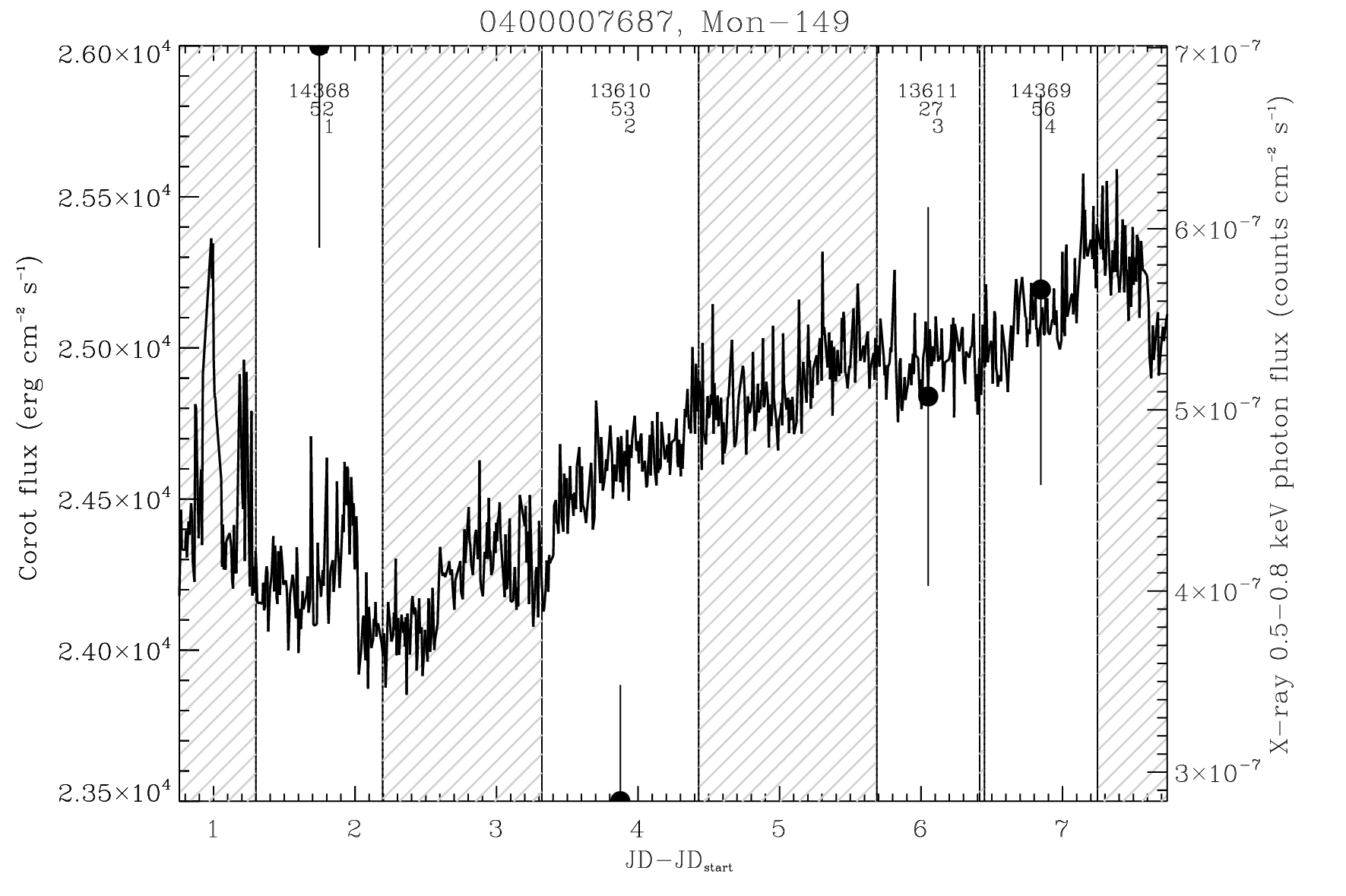}
	\includegraphics[width=9.5cm]{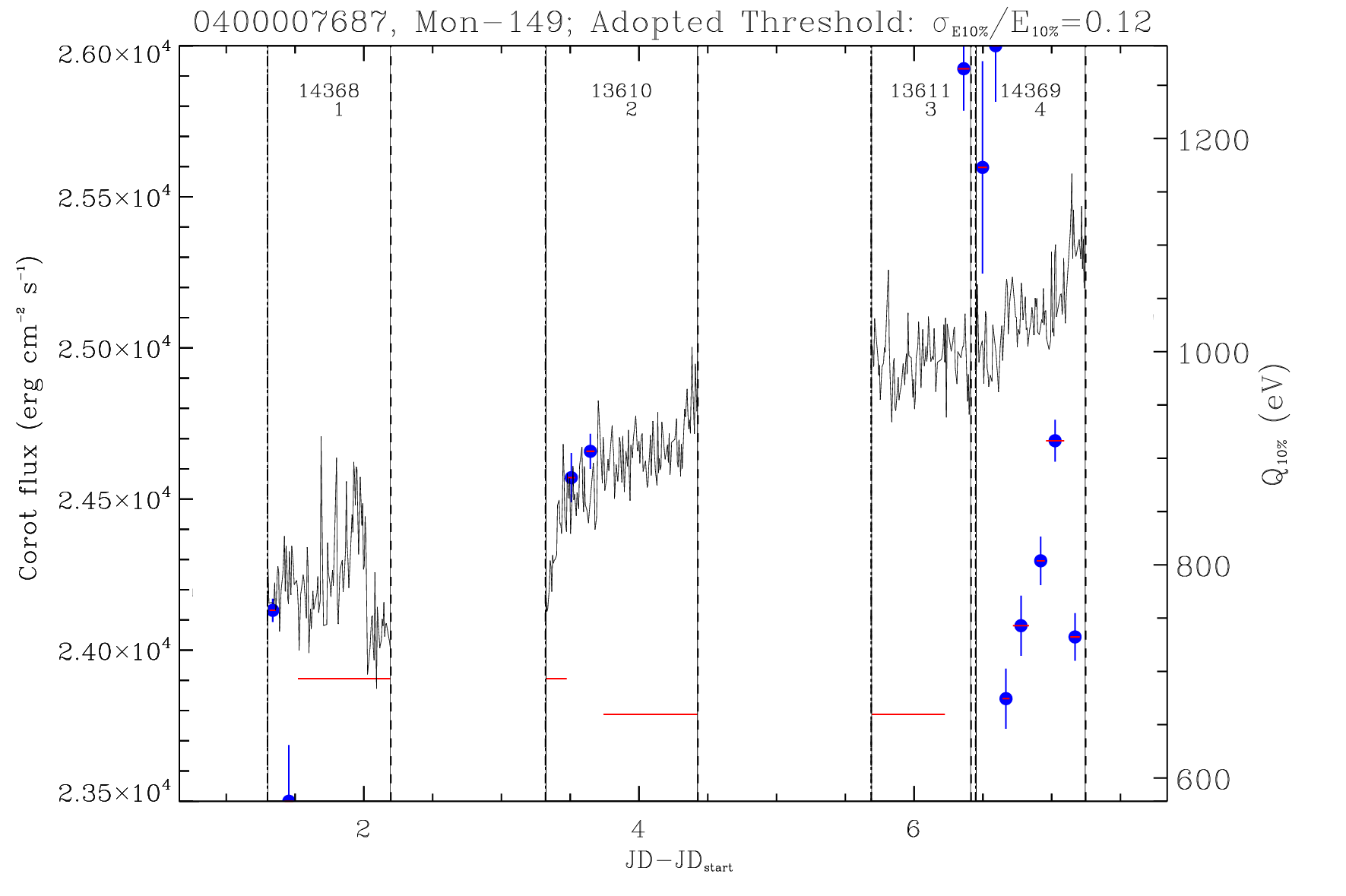}
	\includegraphics[width=8cm]{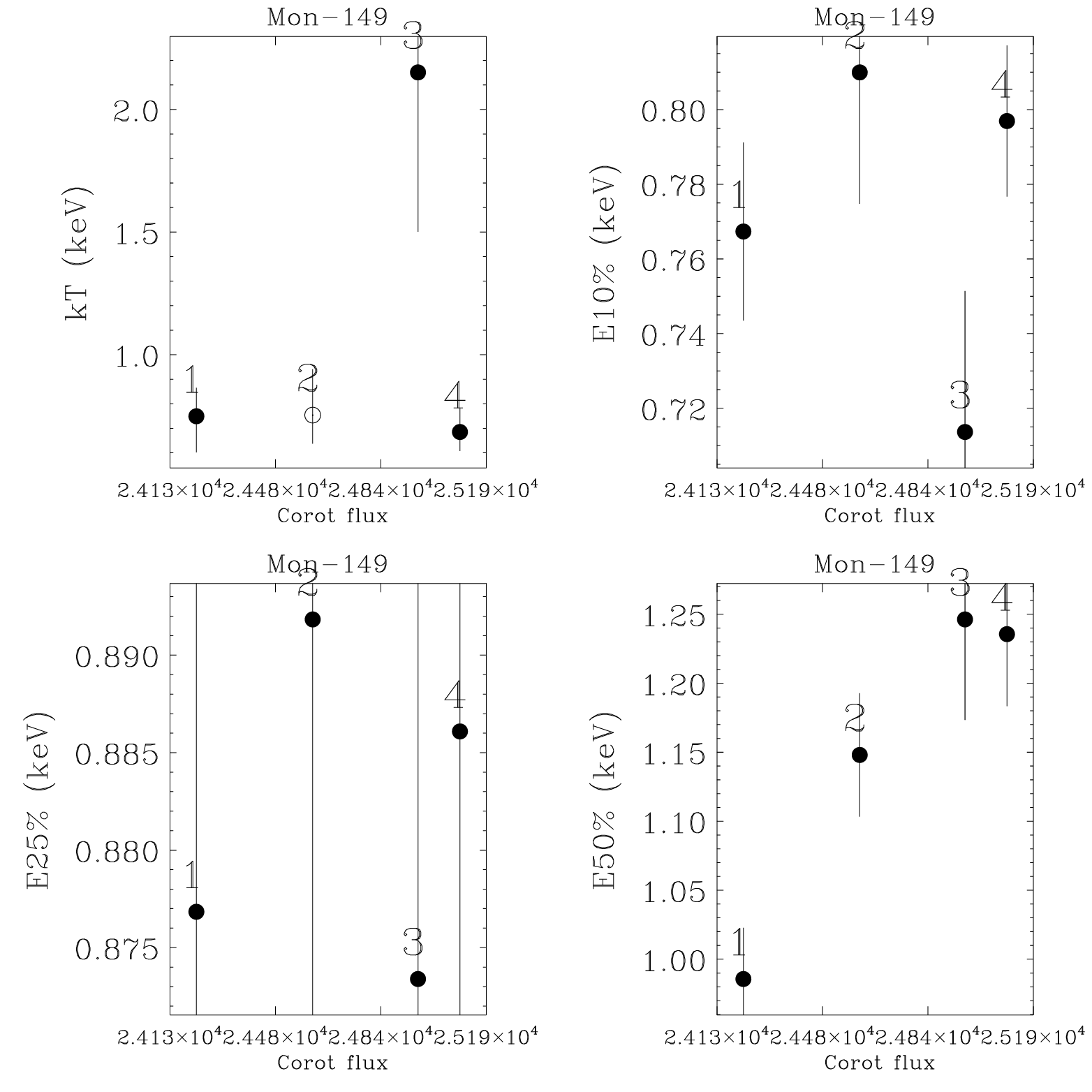}
	\includegraphics[width=18cm]{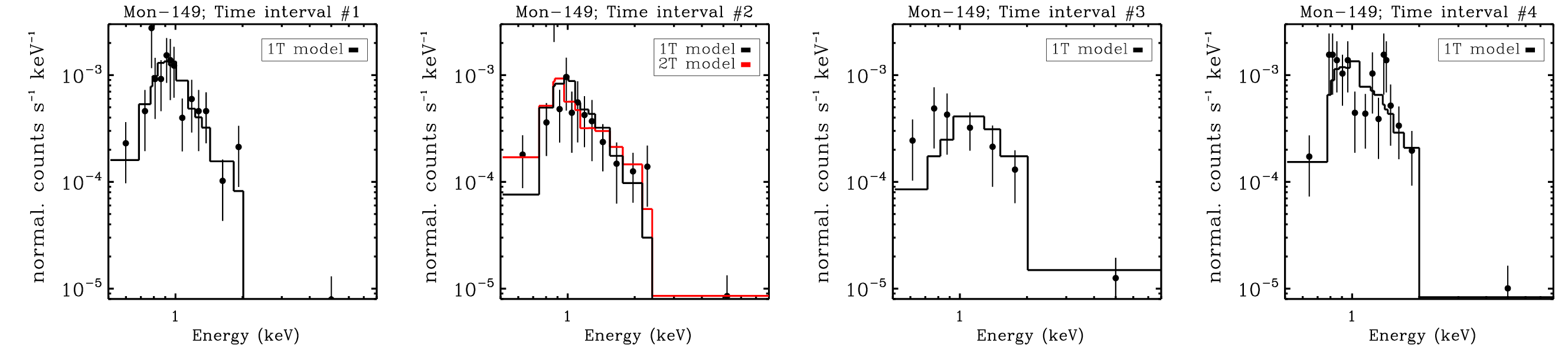}
	\caption{Variability and X-ray spectra of Mon-149, analyzed as a burster. The CoRoT light curve shows evident bursts and dips but few X-ray photons are detected. The variability of E$_{10\%}$ shows interesting patterns, such as during \#4 when it rises until it declines significantly when a large burst occurs at the end of the interval. The X-ray spectrum observed during \#3 is soft, but it is fitted with 1T thermal model. It is not clear whether a soft X-ray spectral component is observed during \#2.}
	\label{variab_others_7}
	\end{figure}

	\begin{figure}[]
	\centering
	\includegraphics[width=9.5cm]{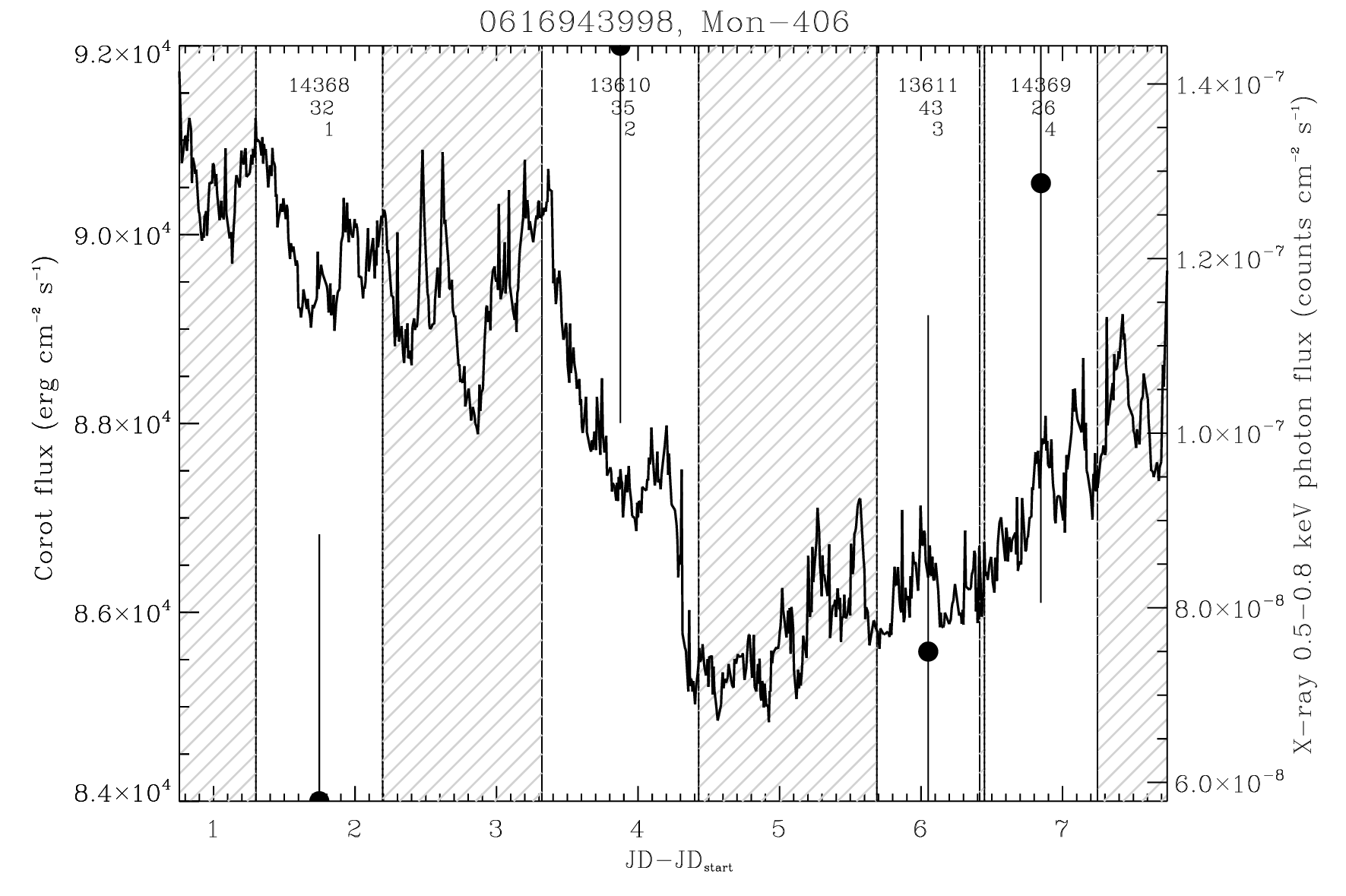}
	\includegraphics[width=9.5cm]{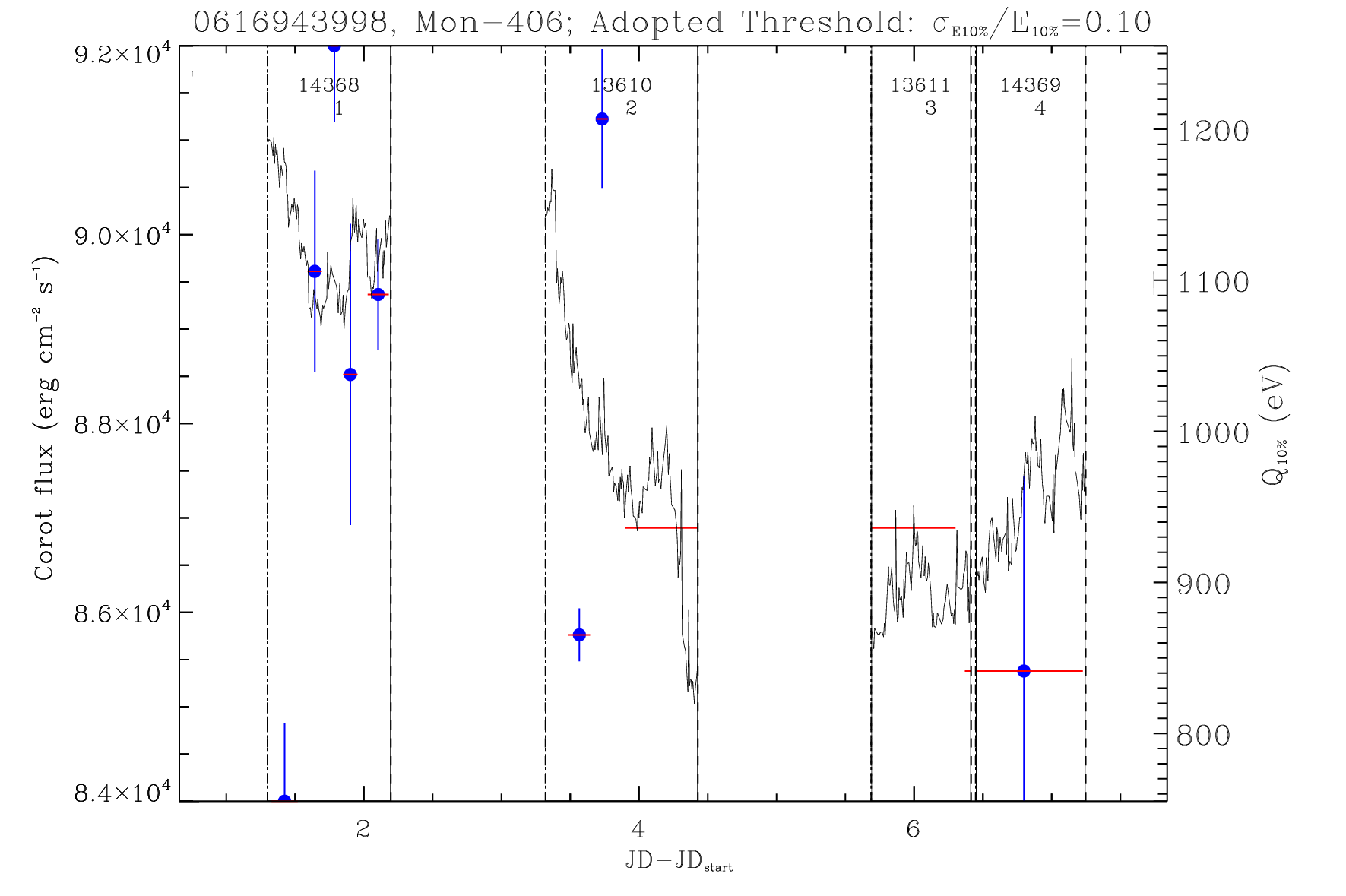}
	\includegraphics[width=8cm]{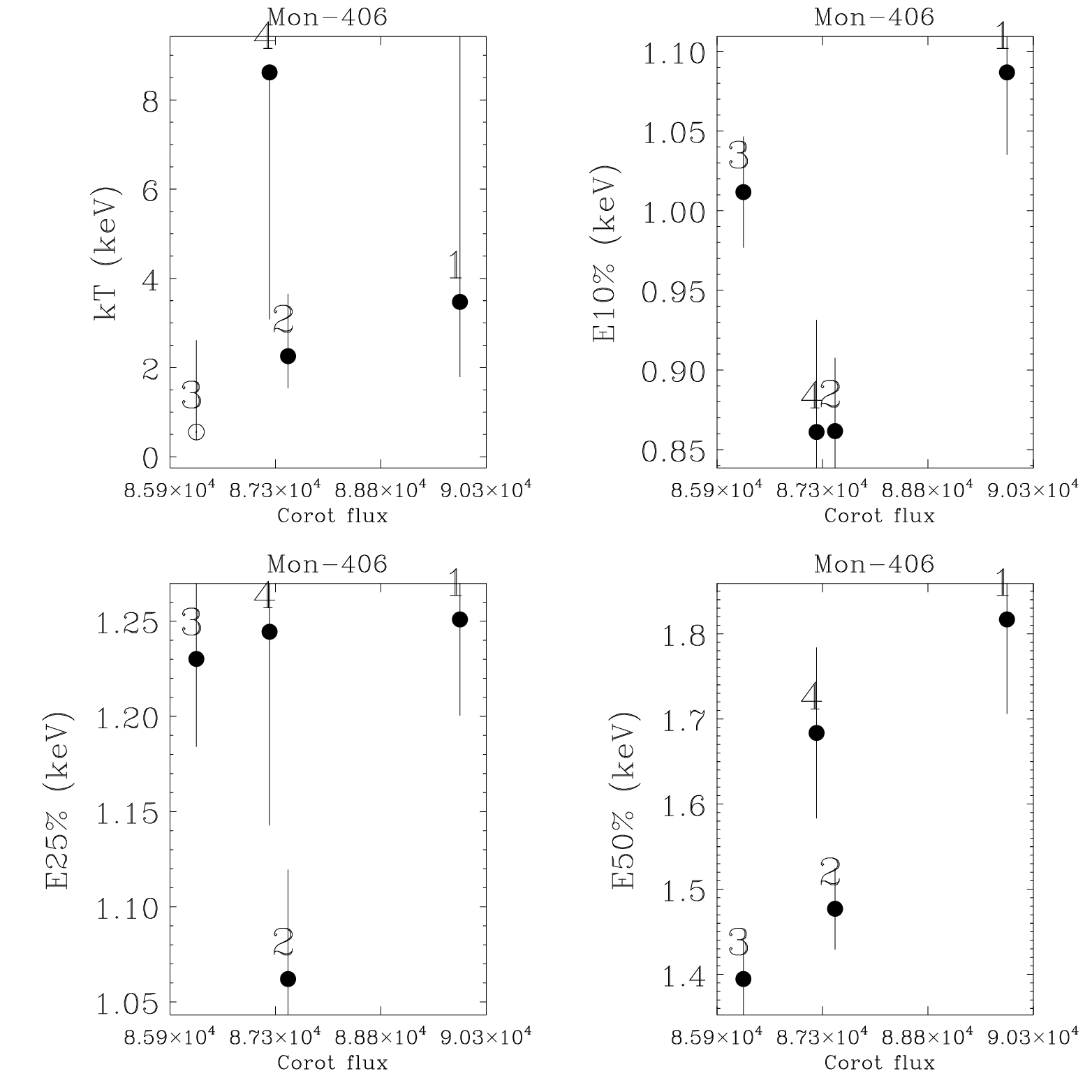}	 
	\includegraphics[width=18cm]{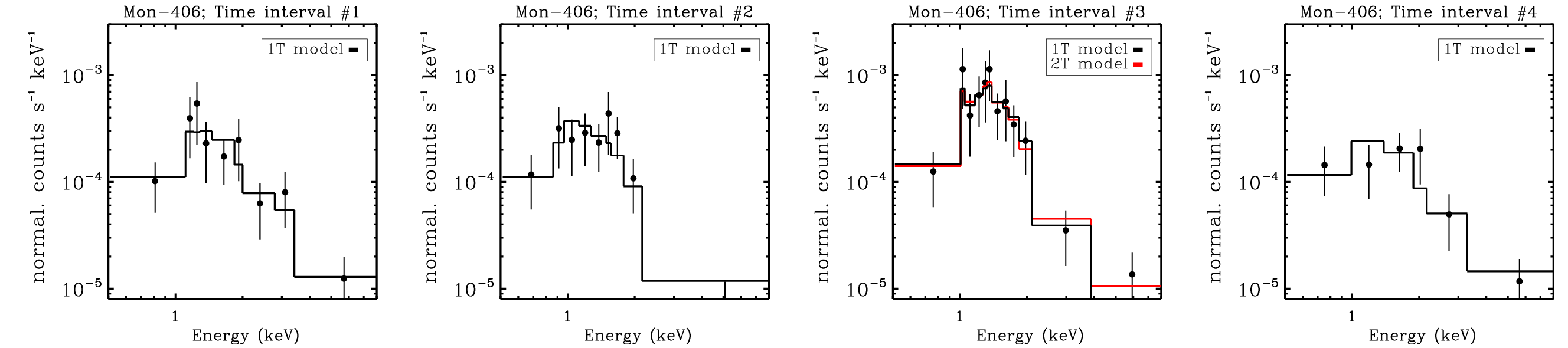}
	\caption{Variability of Mon-406, analyzed as a burster. The CoRoT light curve shows evident bursts and dips but few X-ray counts are detected.}
	\label{variab_others_8}
	\end{figure}
	
	\begin{figure}[]
	\centering
	\includegraphics[width=9.5cm]{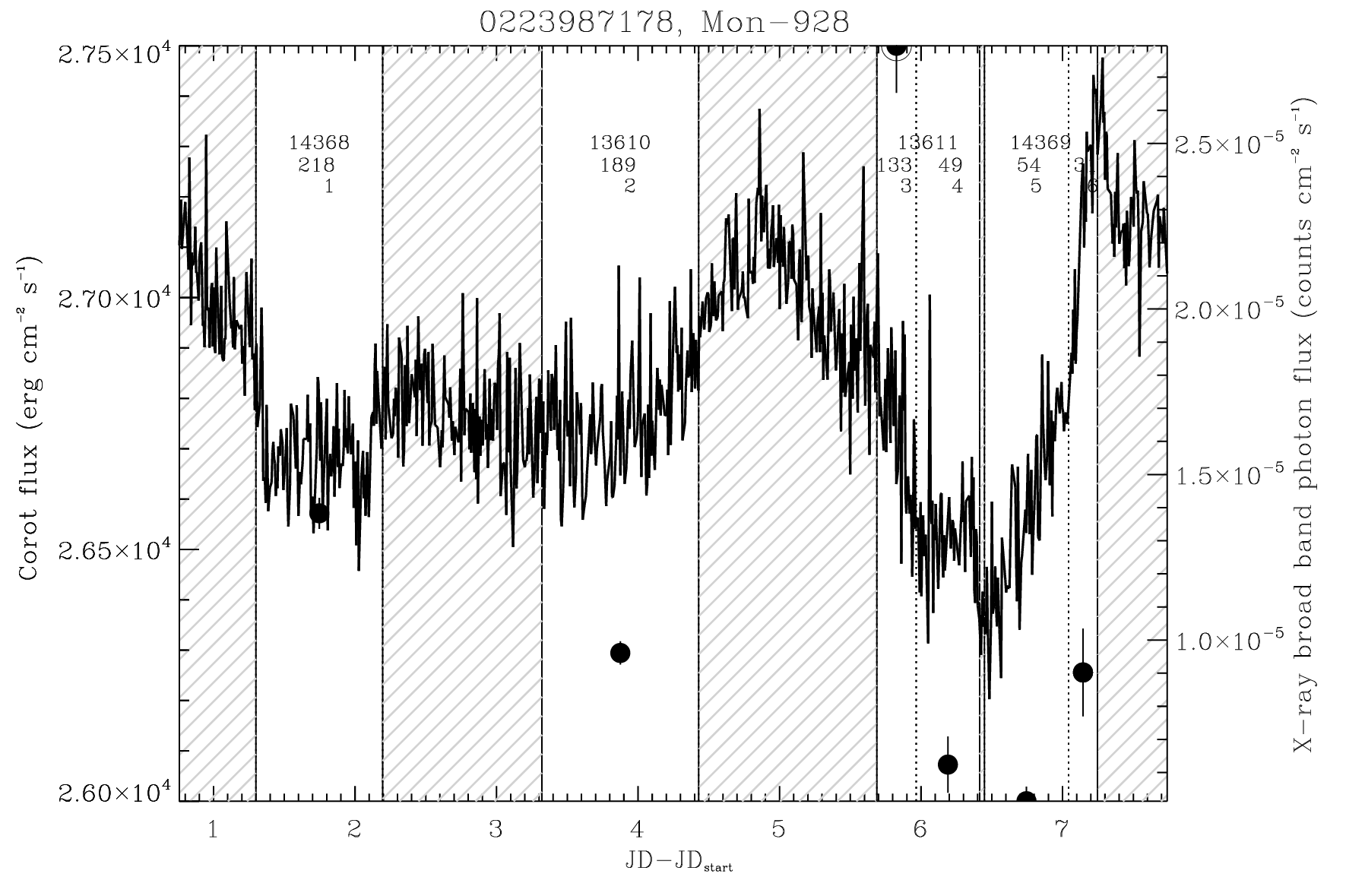}
	\includegraphics[width=9.5cm]{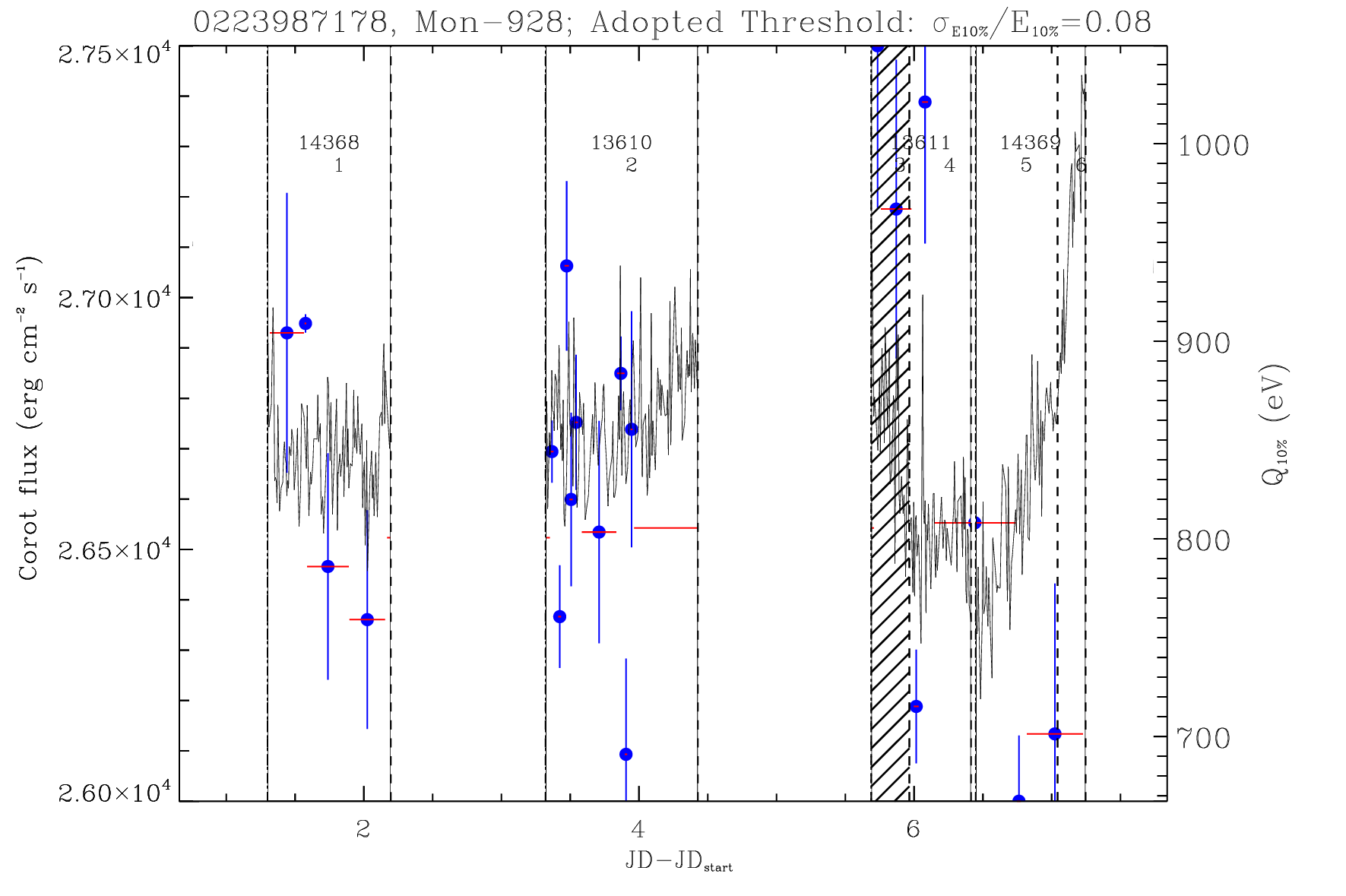}
	\includegraphics[width=8cm]{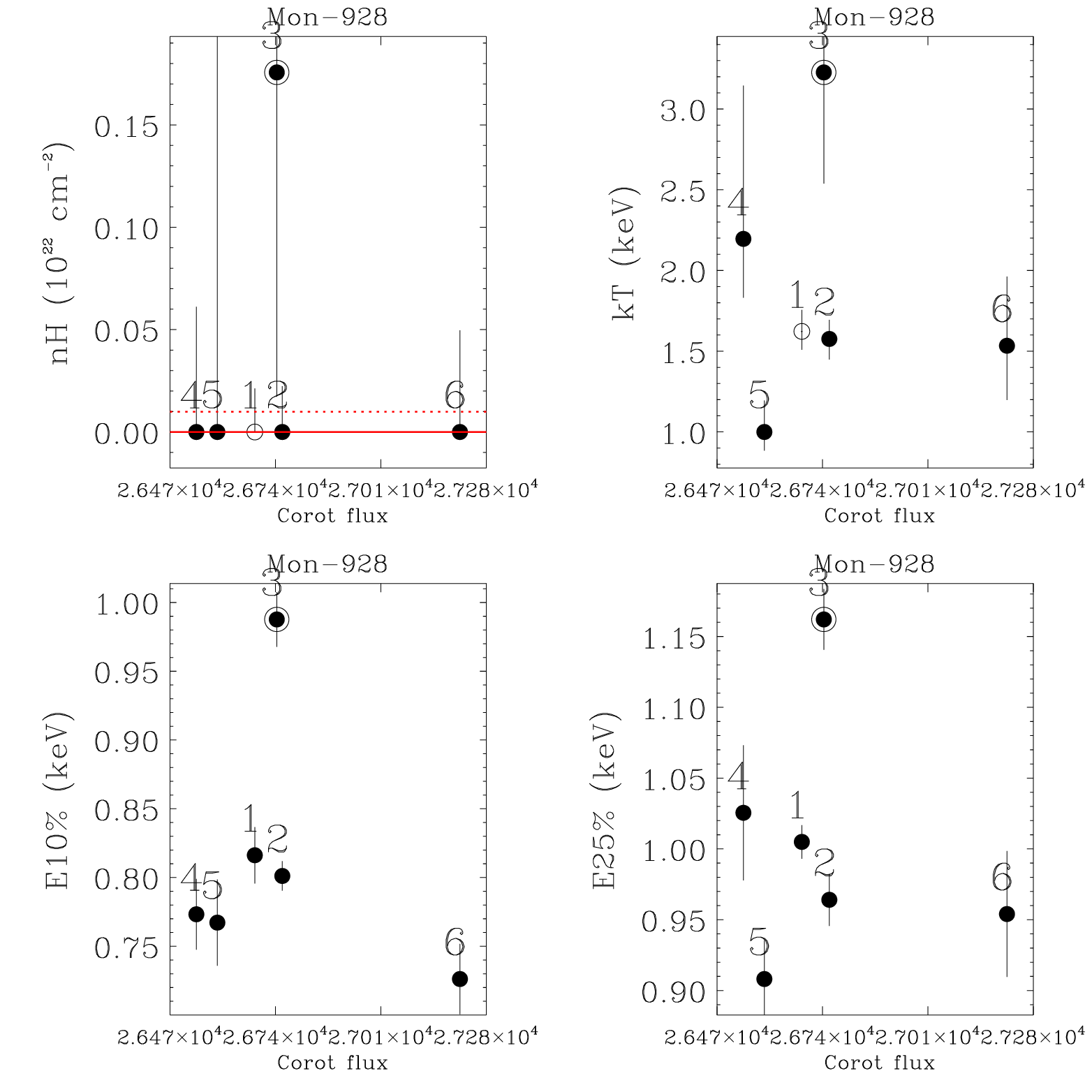}
	\includegraphics[width=18cm]{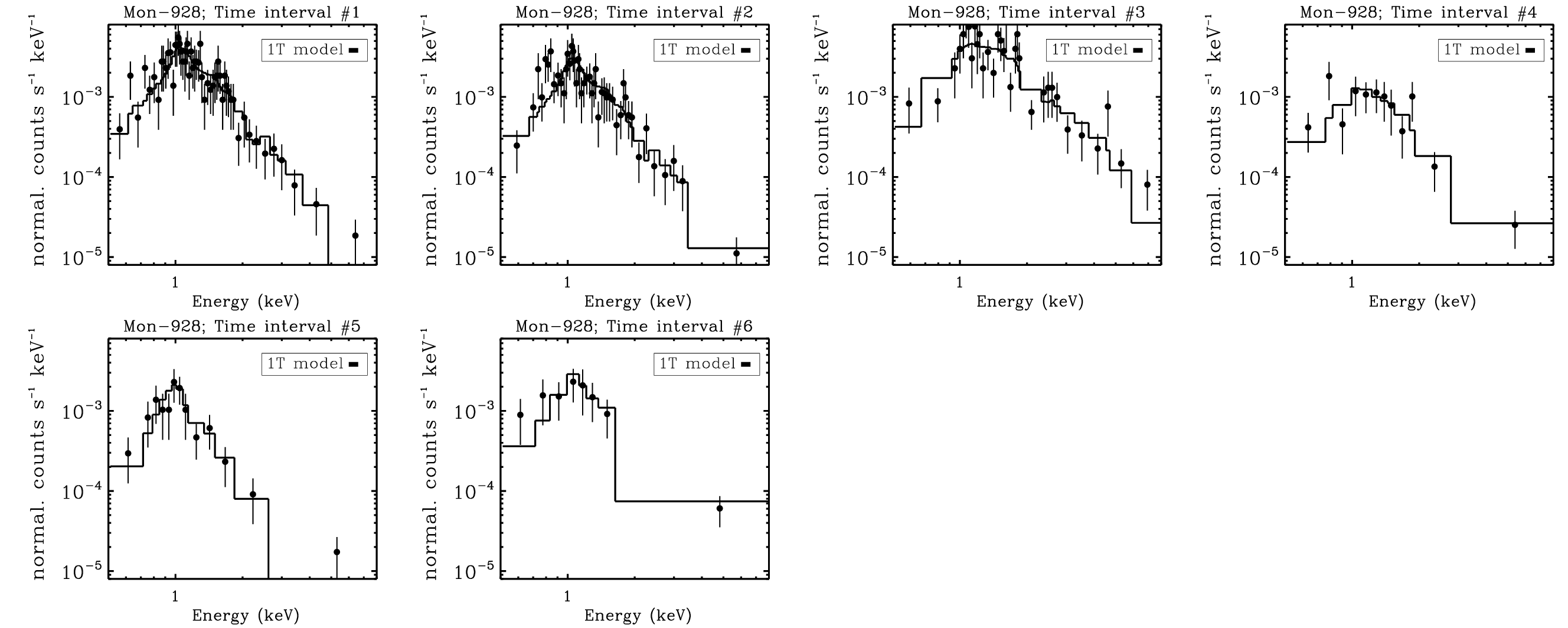}
	\caption{Variability and X-ray spectra of Mon-928 analyzed as a dipper. There is no evidence for increasing N$_H$ during the optical dip (\#4 and \#5) and it may have soft X-ray emission during the burst (\#6), but few photons have been detected.}
	\label{variab_others_9}
	\end{figure}

	\begin{figure}[]
	\centering
	\includegraphics[width=9.5cm]{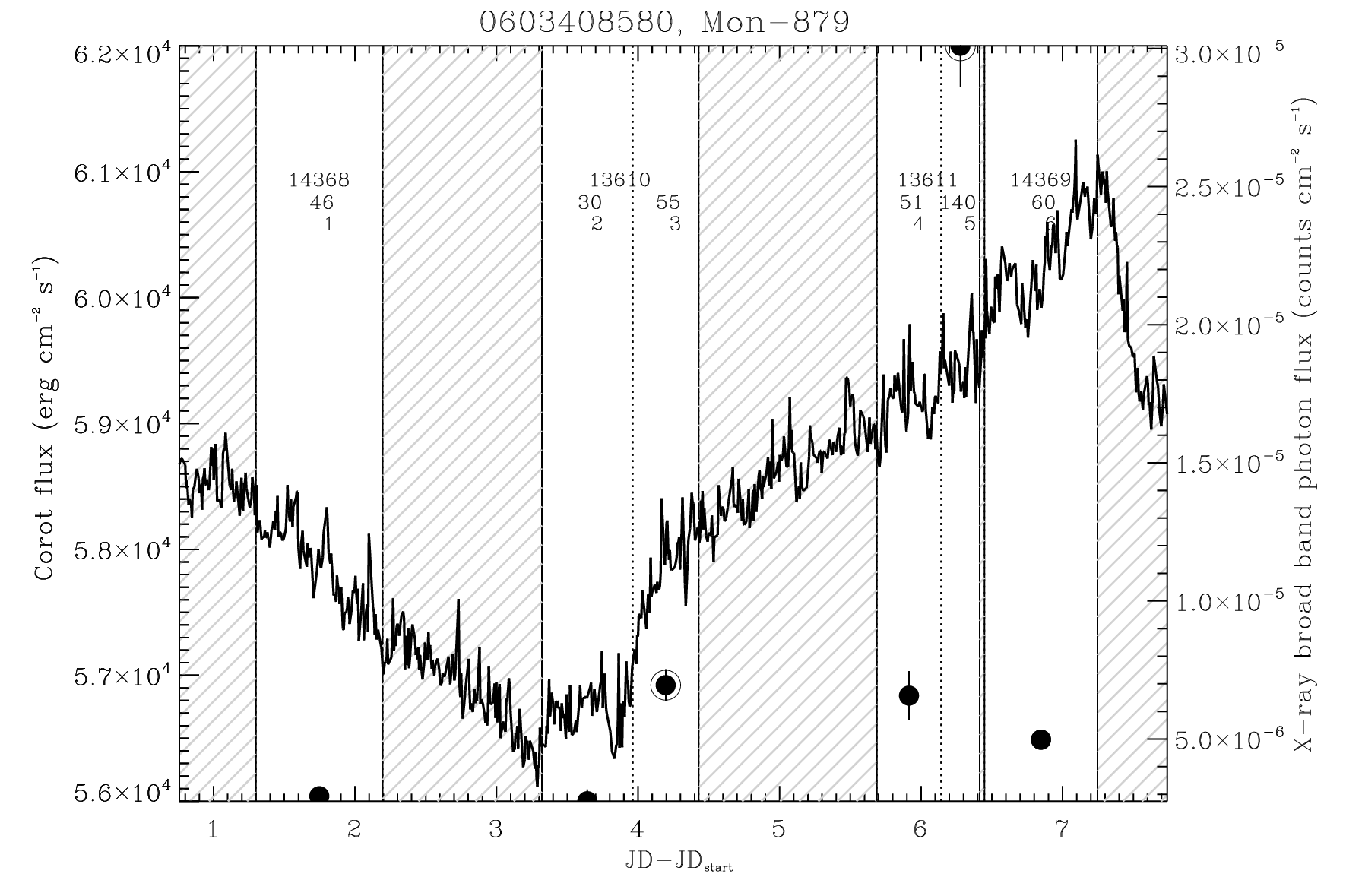}
	\includegraphics[width=9.5cm]{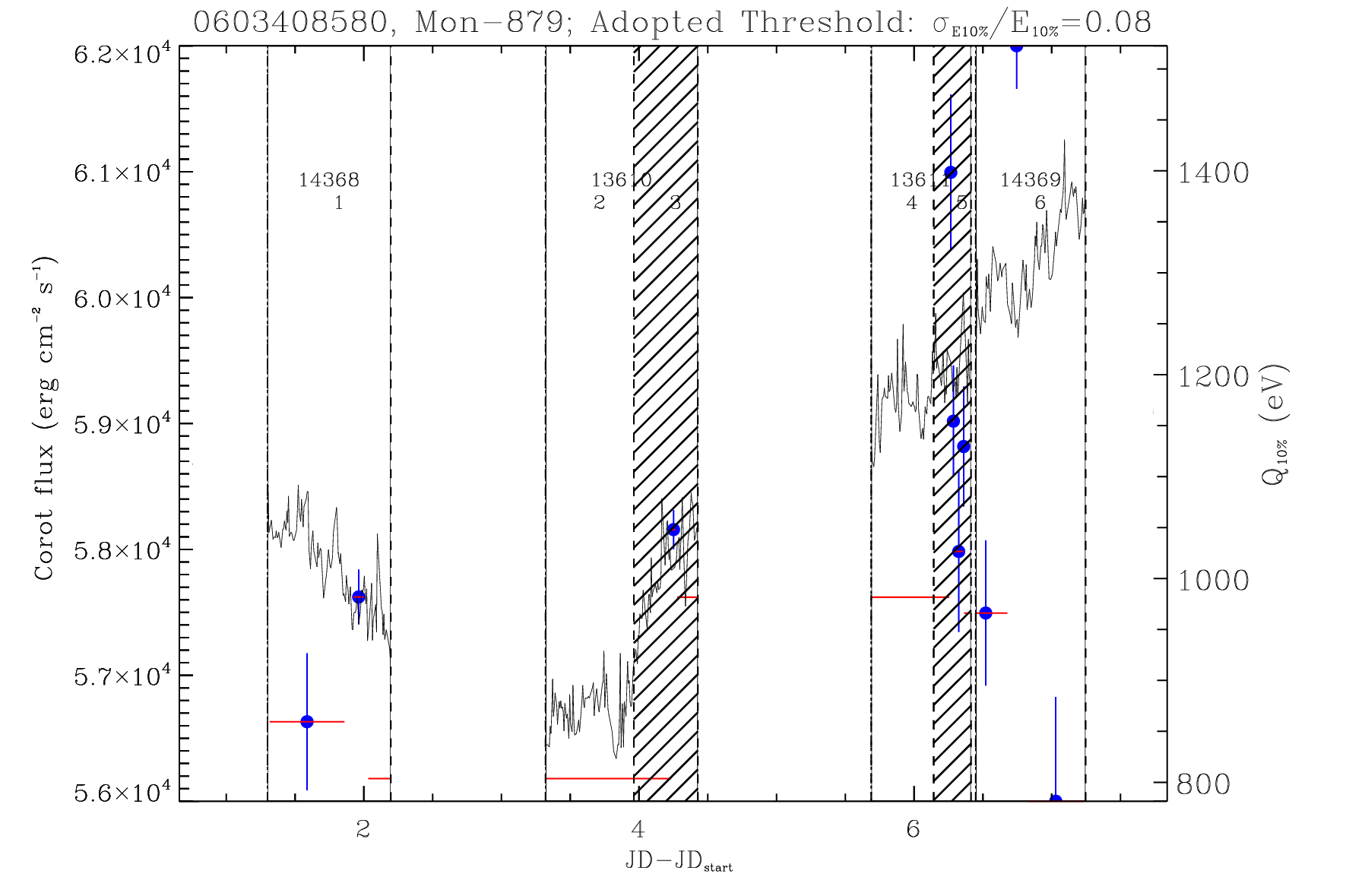}
	\includegraphics[width=8cm]{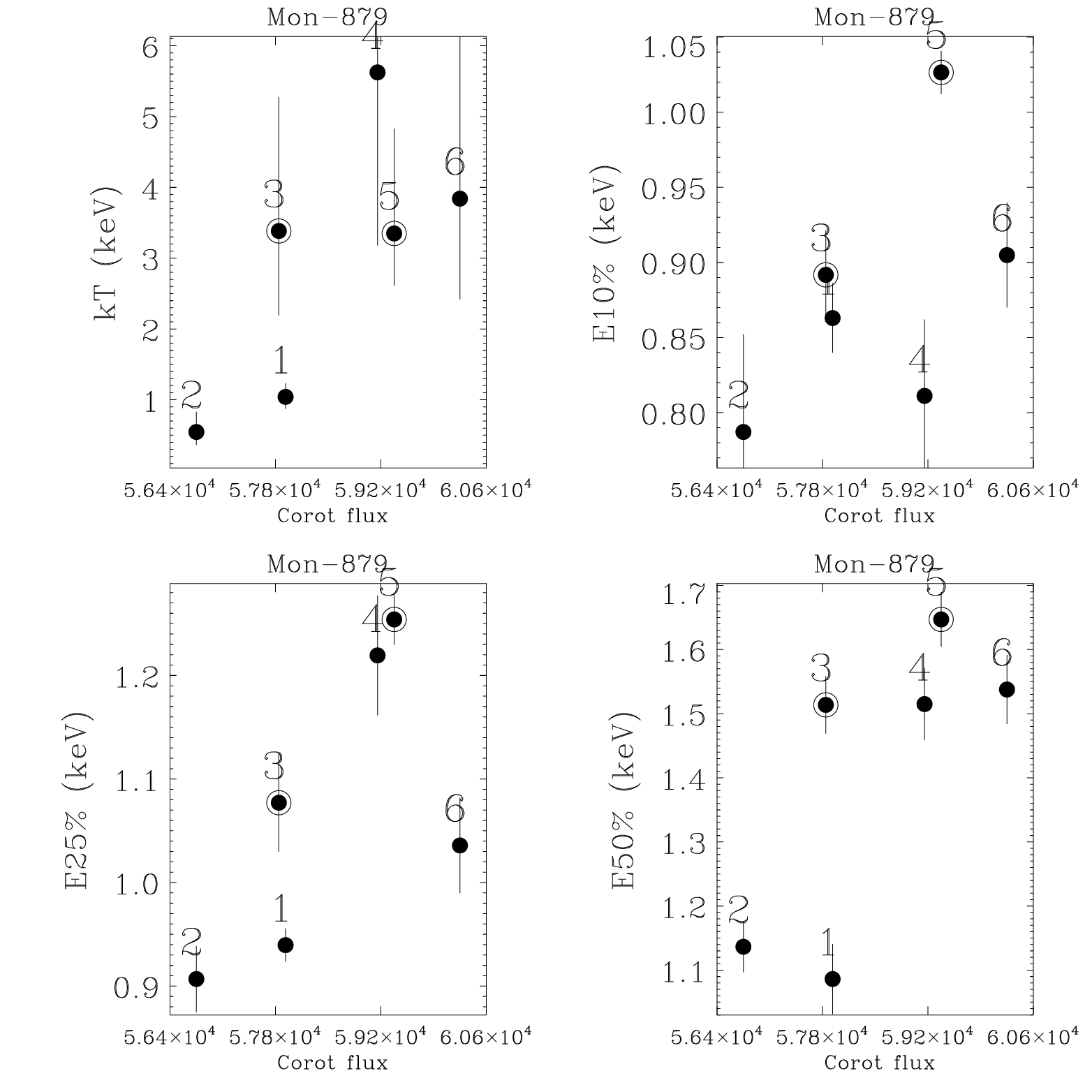}	
	\includegraphics[width=18cm]{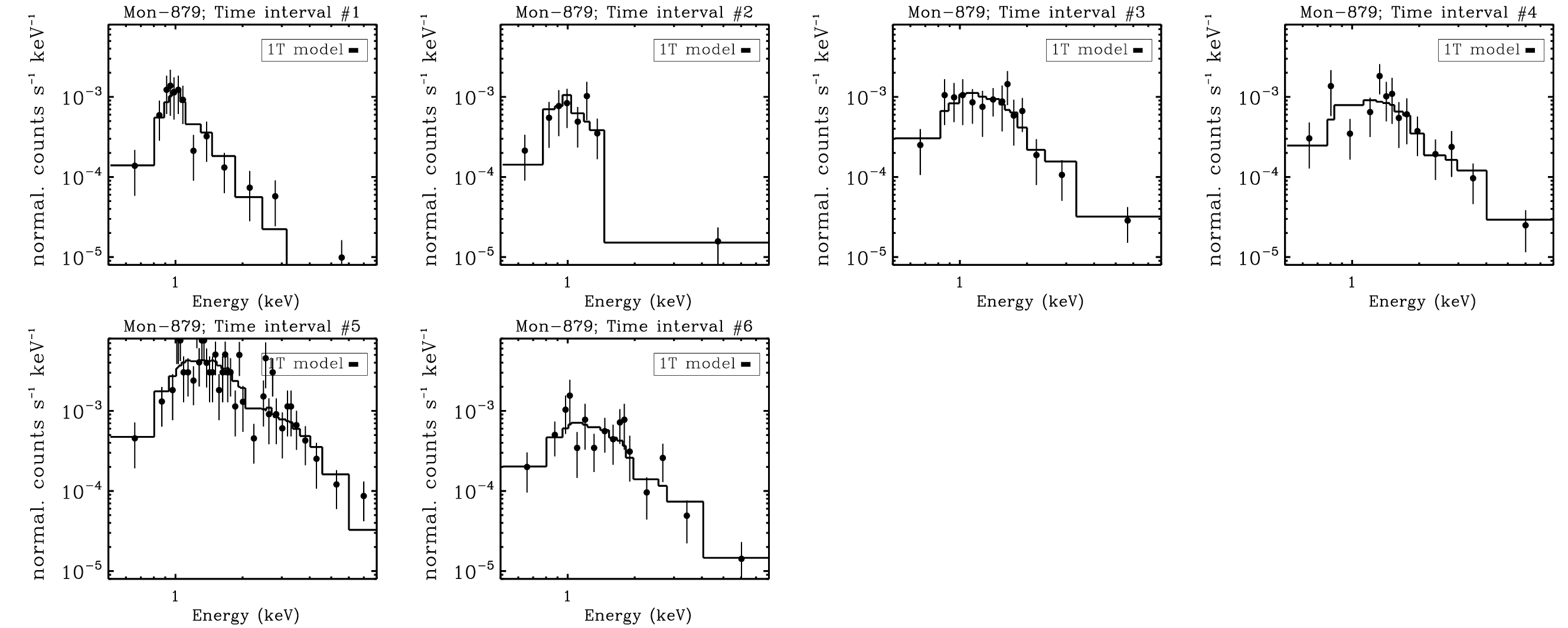}
	\caption{Variability and X-ray spectra of Mon-879, analyzed as a burster. There is no evidence for soft X-ray emission during the optical bursts, and the CoRoT mask may have been contaminated by a nearby bright optical source.}
	\label{variab_others_10}
	\end{figure}

	\begin{figure}[]
	\centering
	\includegraphics[width=9.5cm]{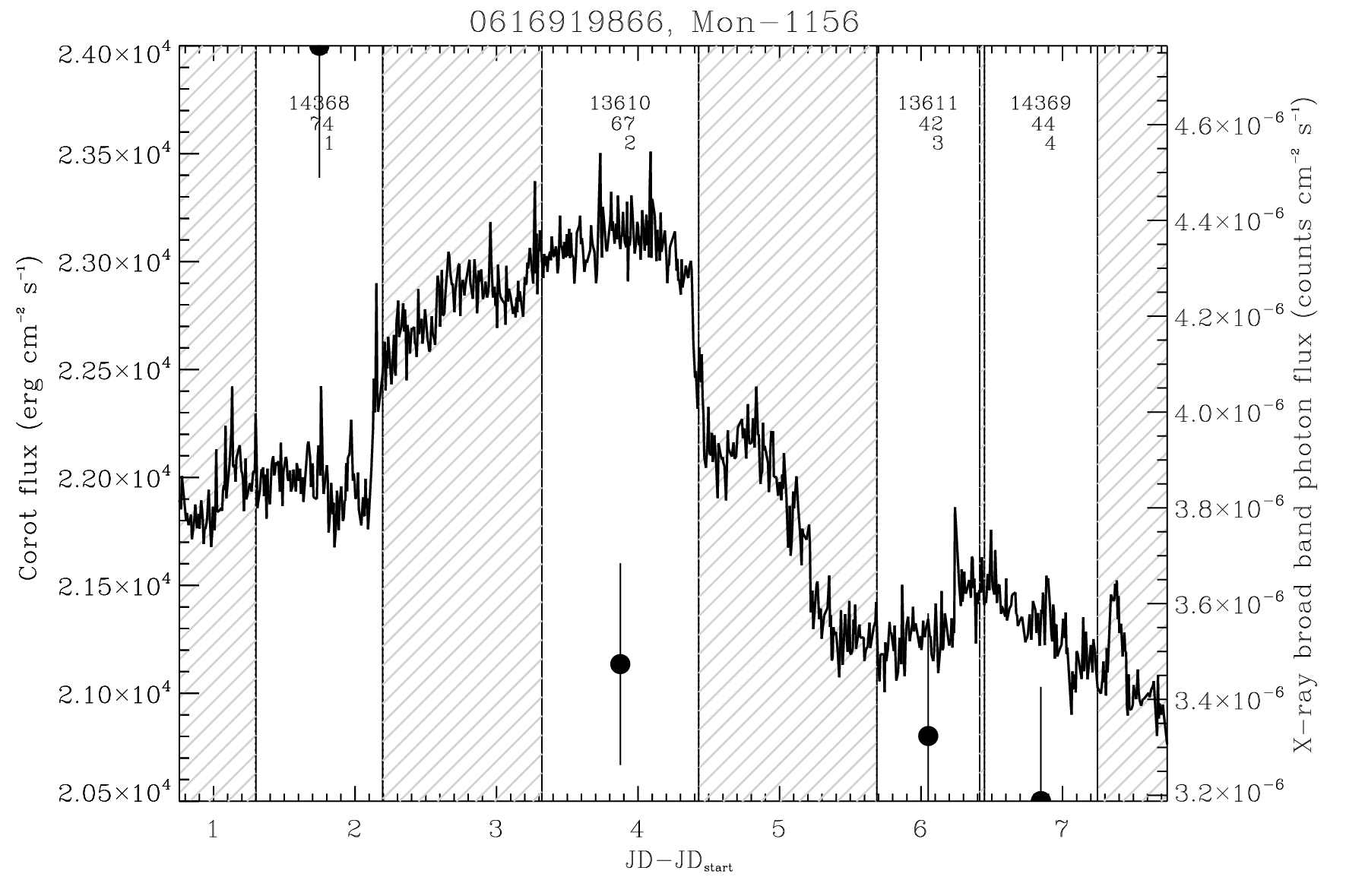}
	\includegraphics[width=9.5cm]{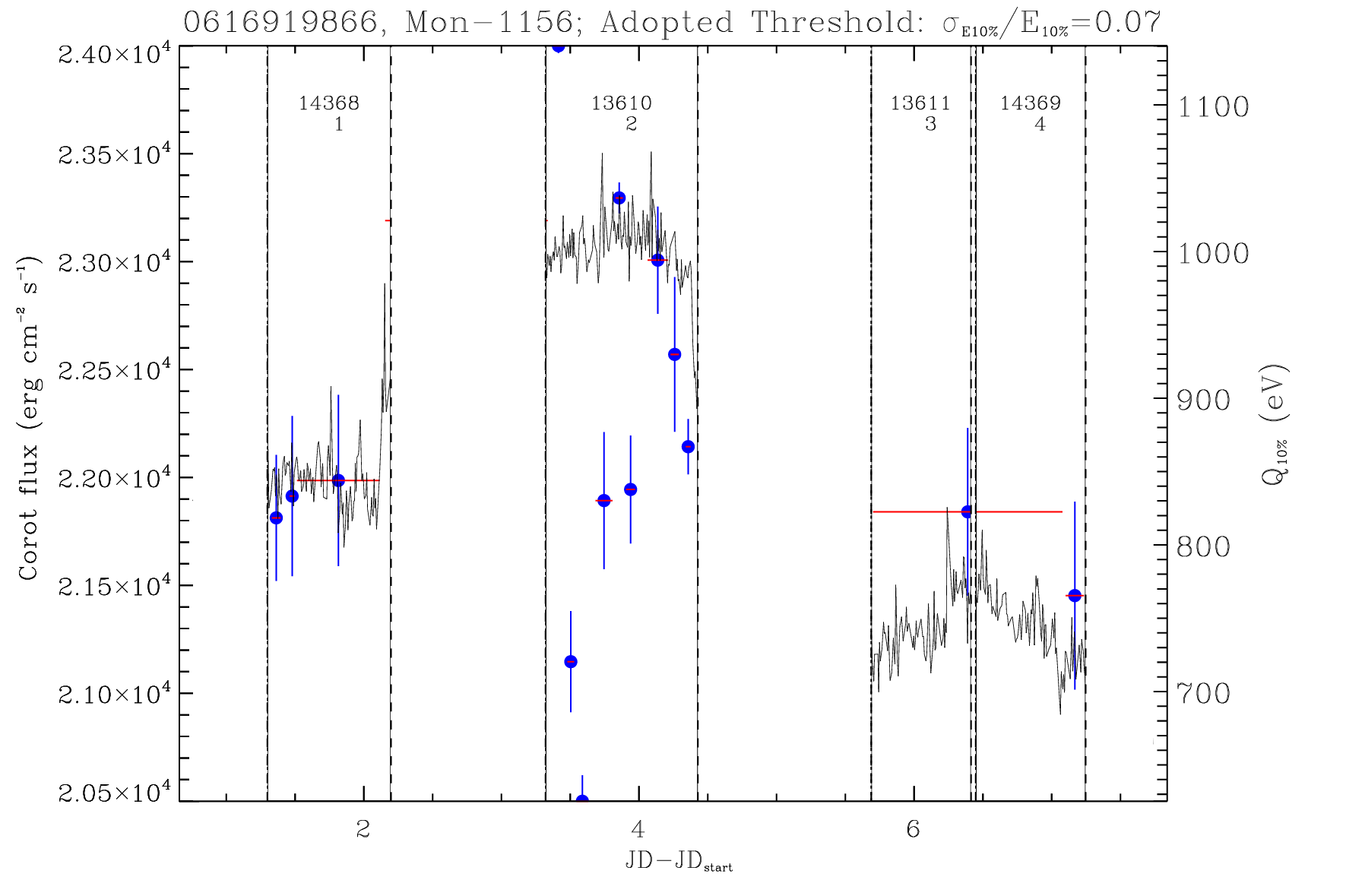}
	\includegraphics[width=8cm]{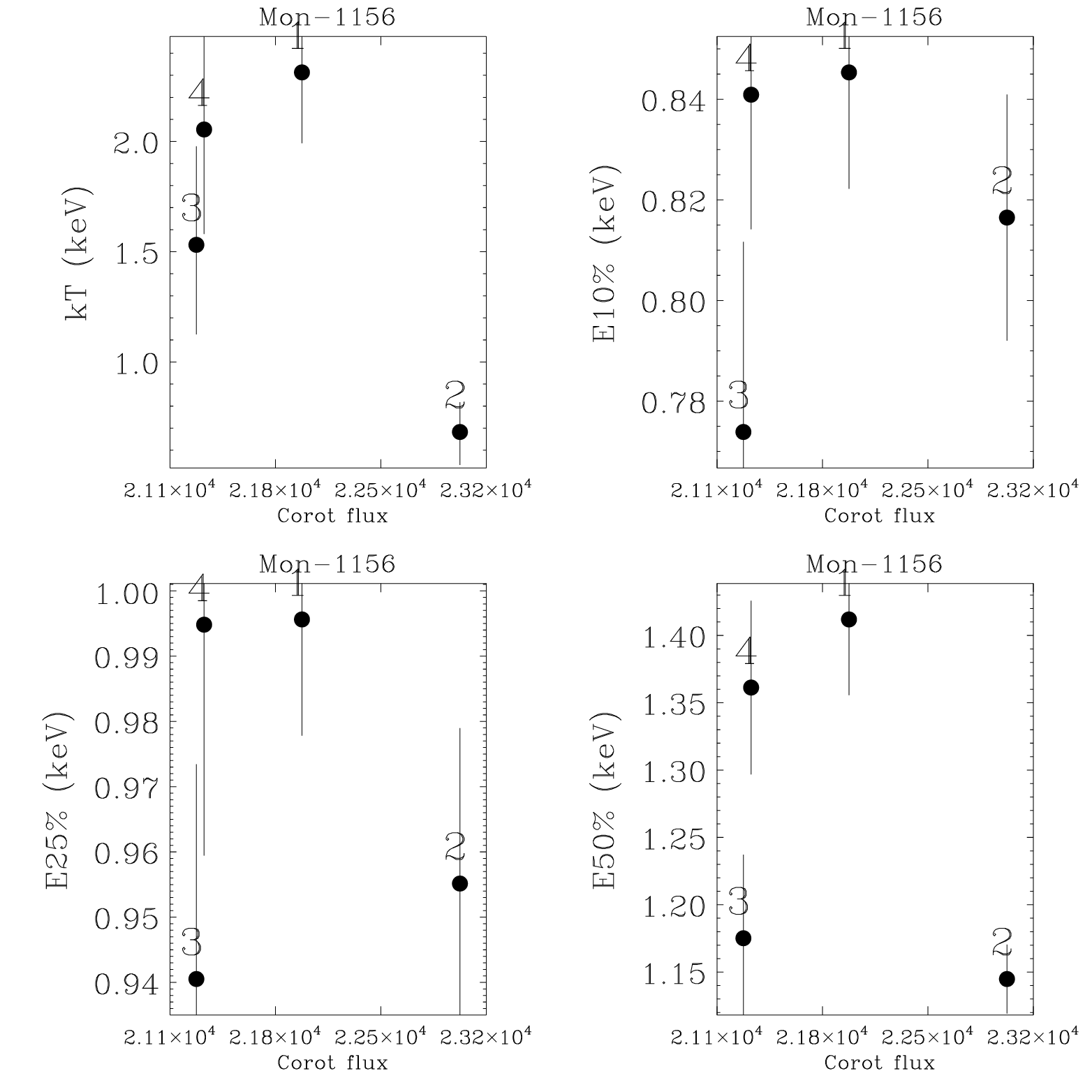}
	\includegraphics[width=18cm]{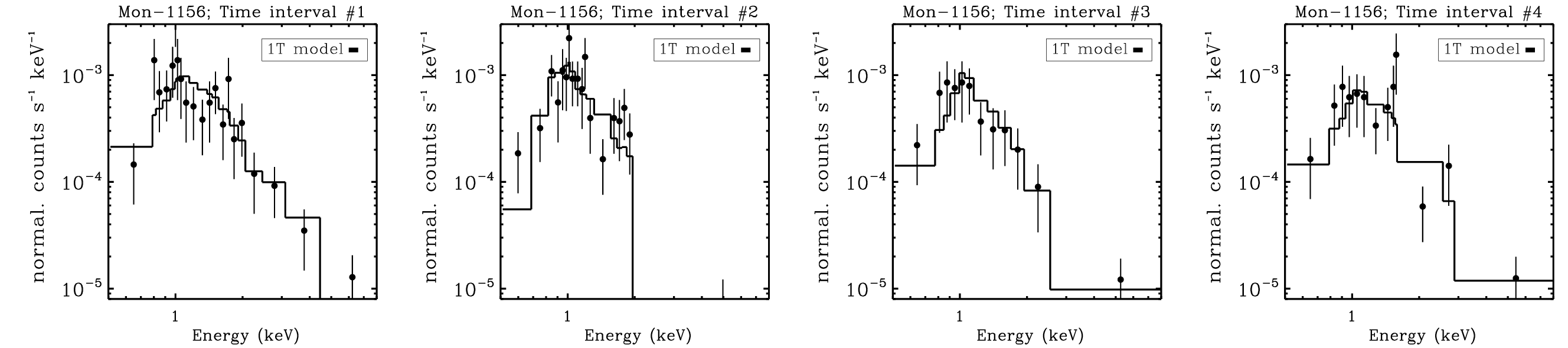}
	\caption{Variability and X-ray spectra of Mon-1156, analyzed as a burster. No significant X-ray variability is observed.}
	\label{variab_others_11}
	\end{figure}

	\begin{figure}[]
	\centering
	\includegraphics[width=9.5cm]{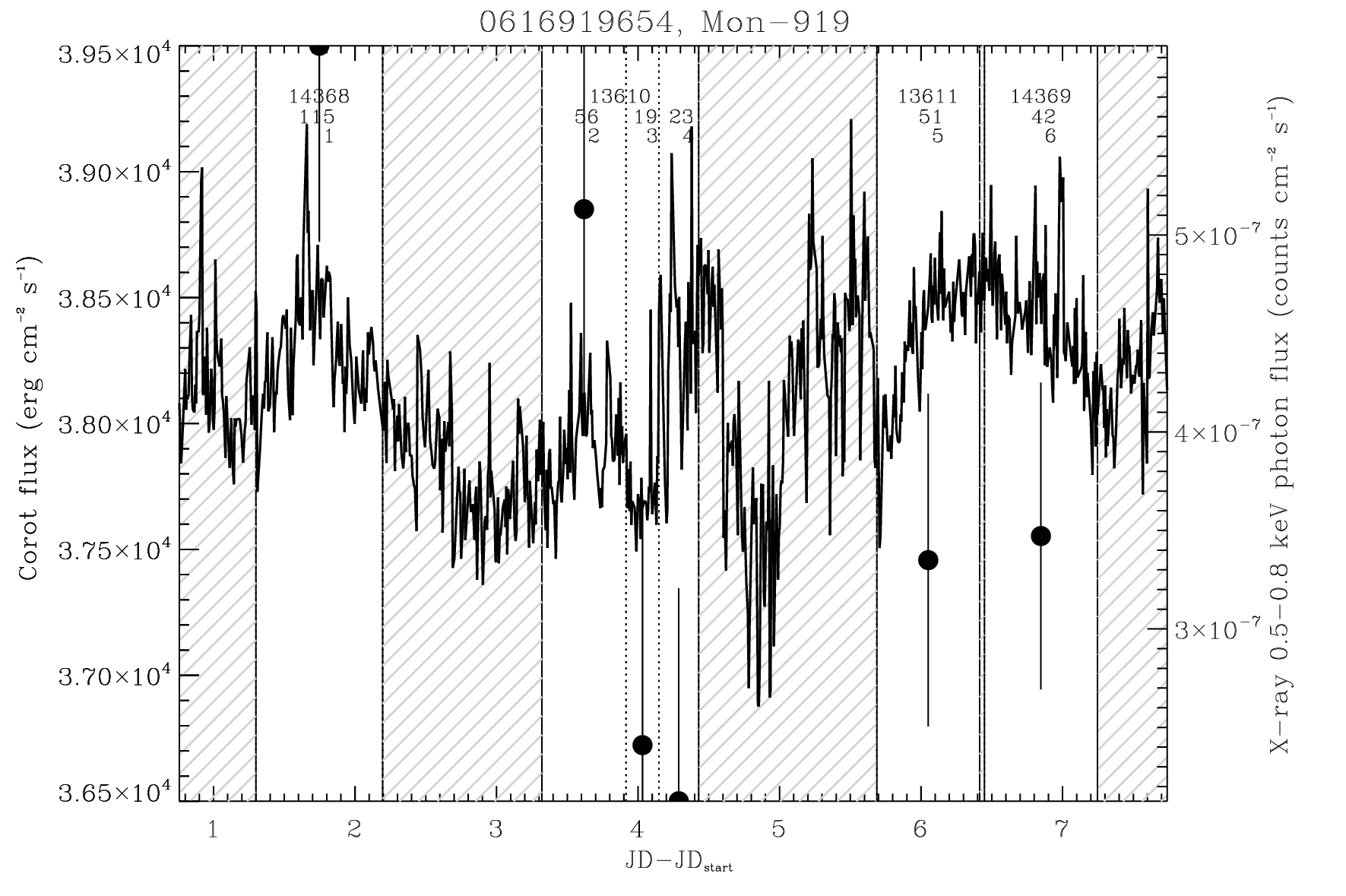}
	\includegraphics[width=9.5cm]{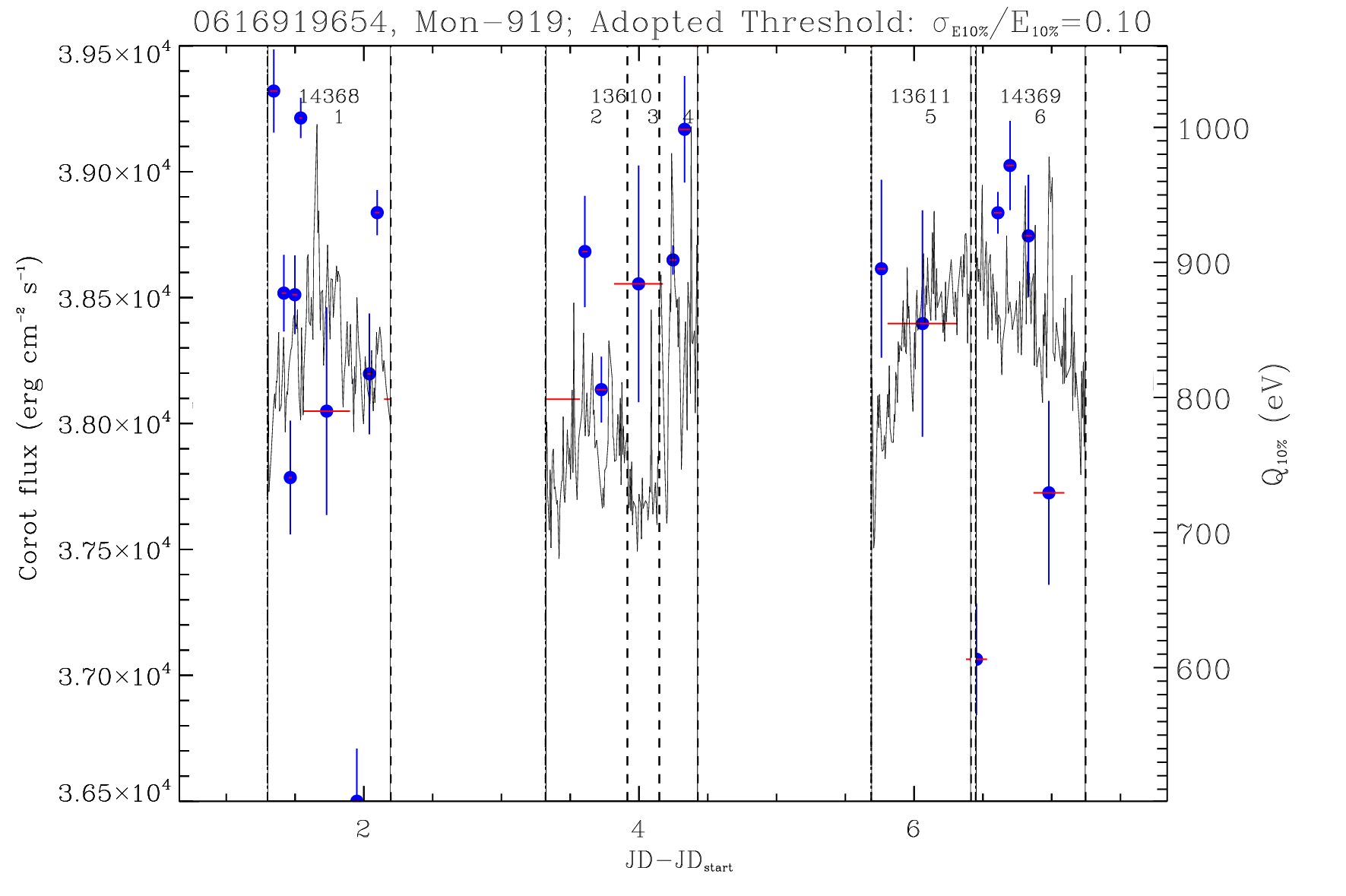}
	\includegraphics[width=8cm]{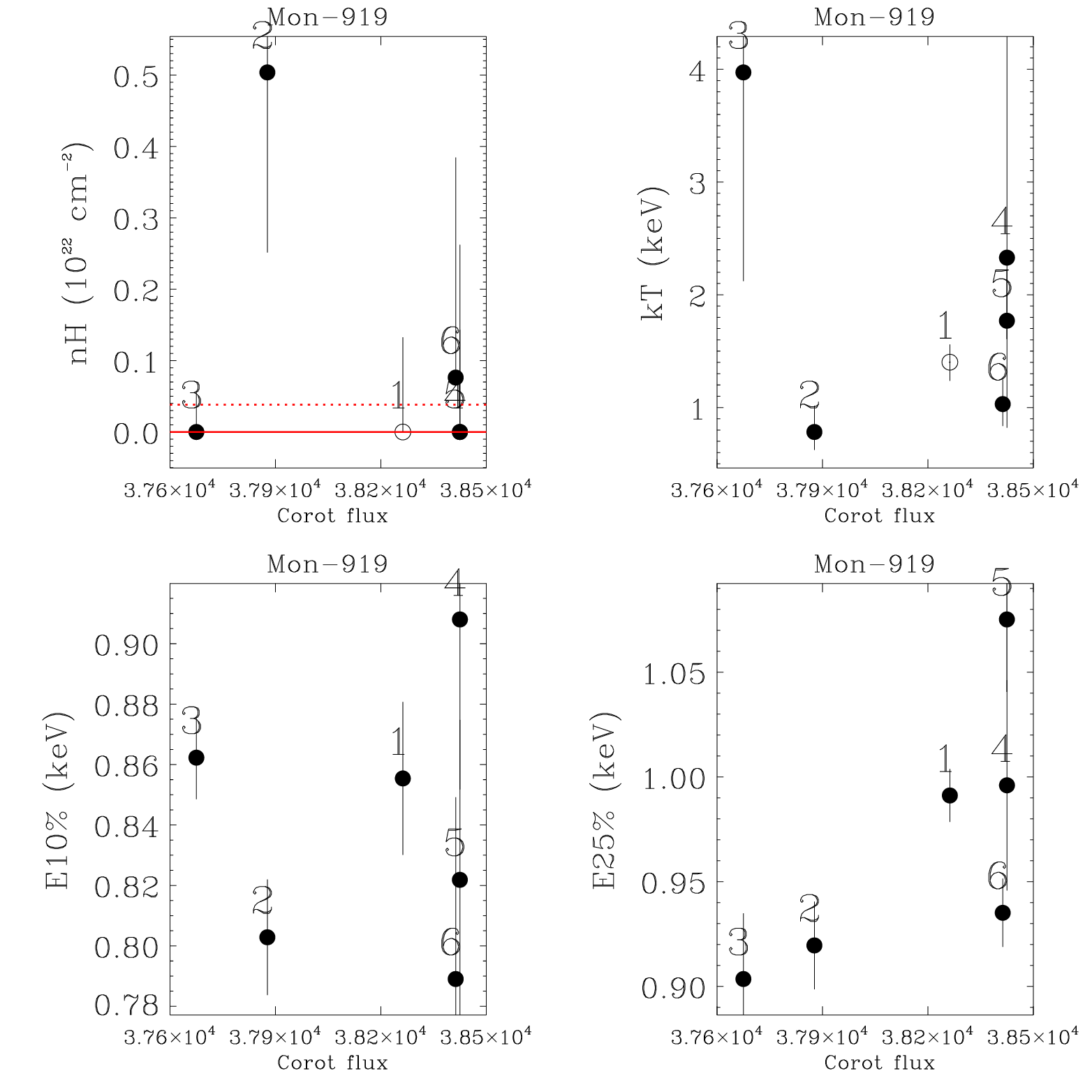}
	\includegraphics[width=18cm]{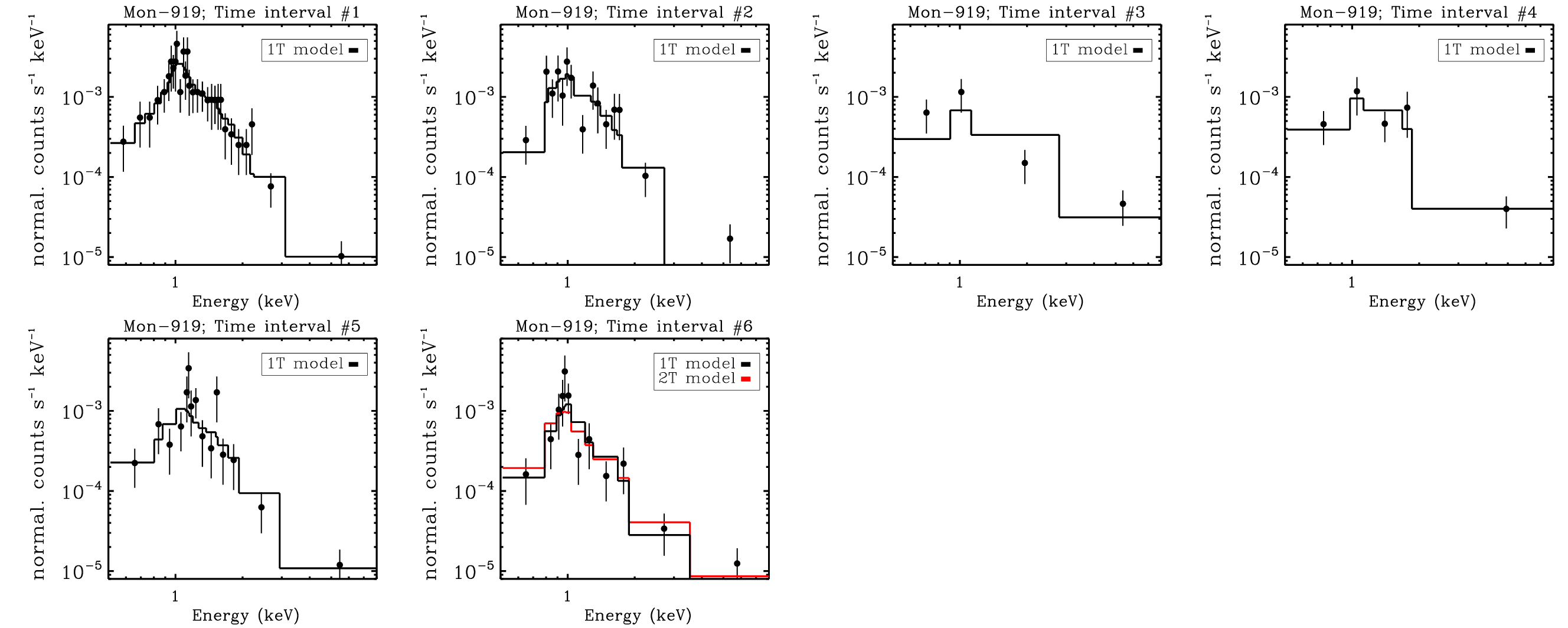}
	\caption{Variability and X-ray spectra of Mon-919, analyzed both as a dipper and burster, with no significant correlation between the optical and X-ray variability.}
	\label{variab_others_12}
	\end{figure}

	\begin{figure}[]
	\centering
	\includegraphics[width=9.5cm]{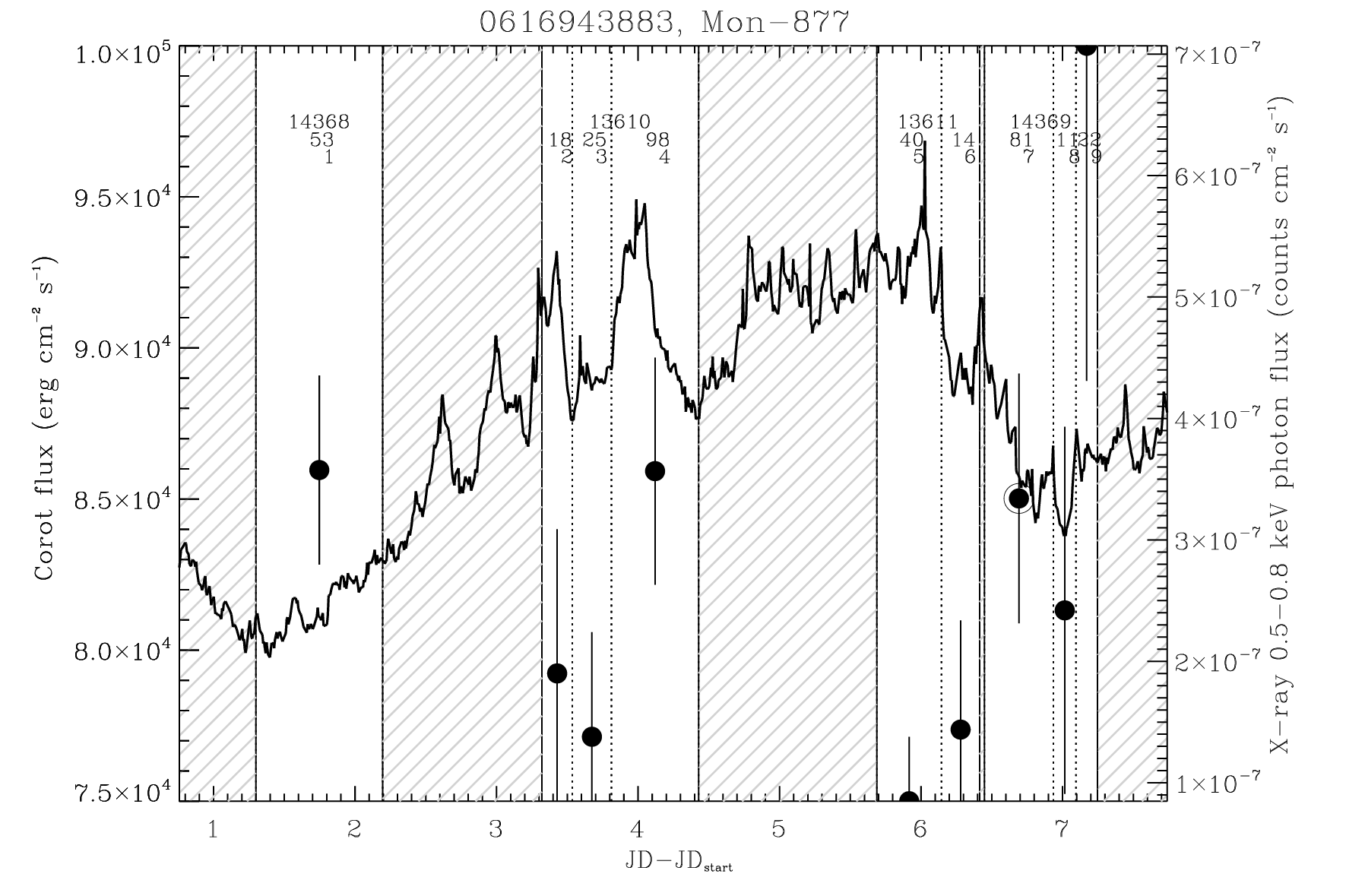}
	\includegraphics[width=9.5cm]{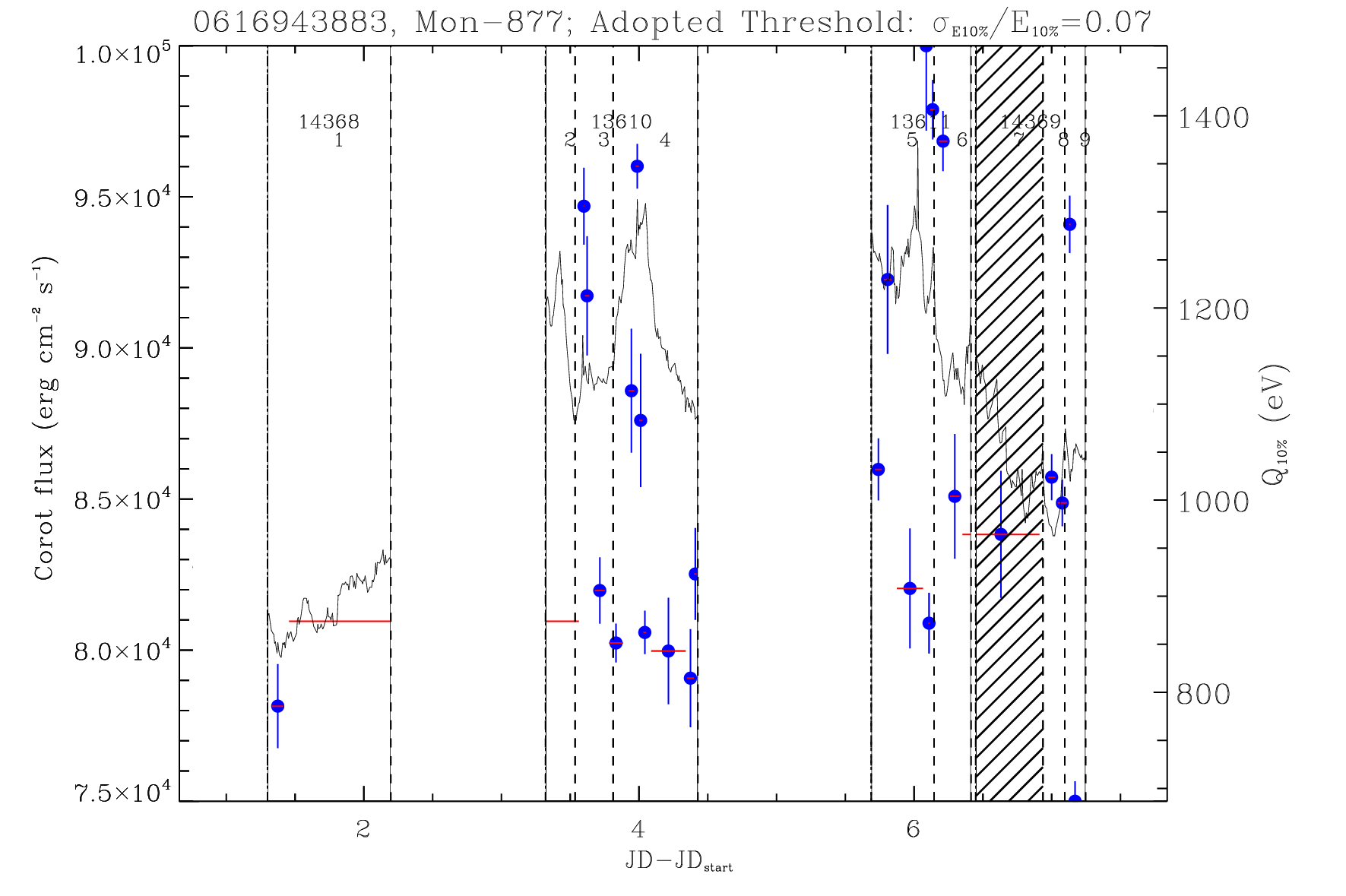}
	\includegraphics[width=8cm]{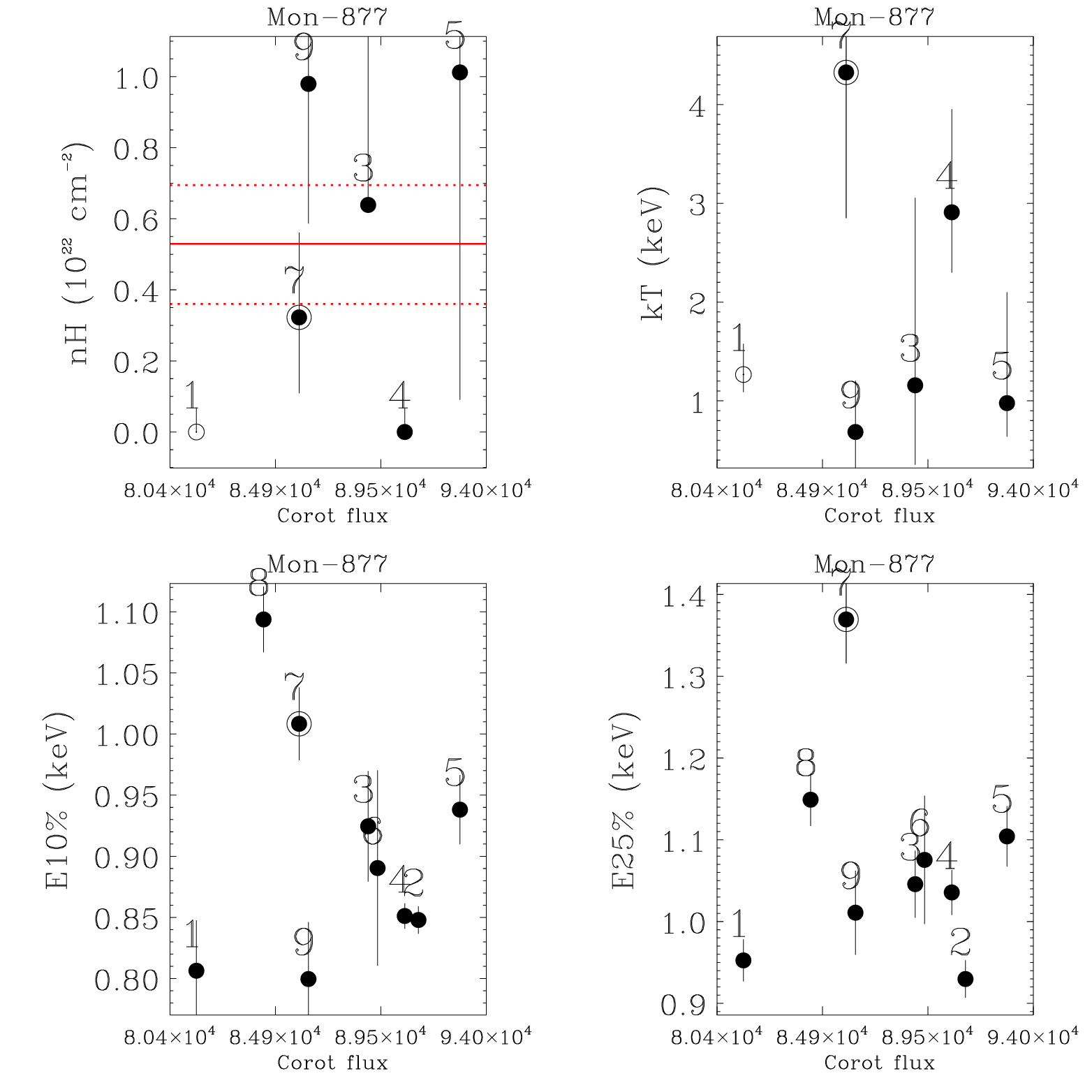}	
	\includegraphics[width=18cm]{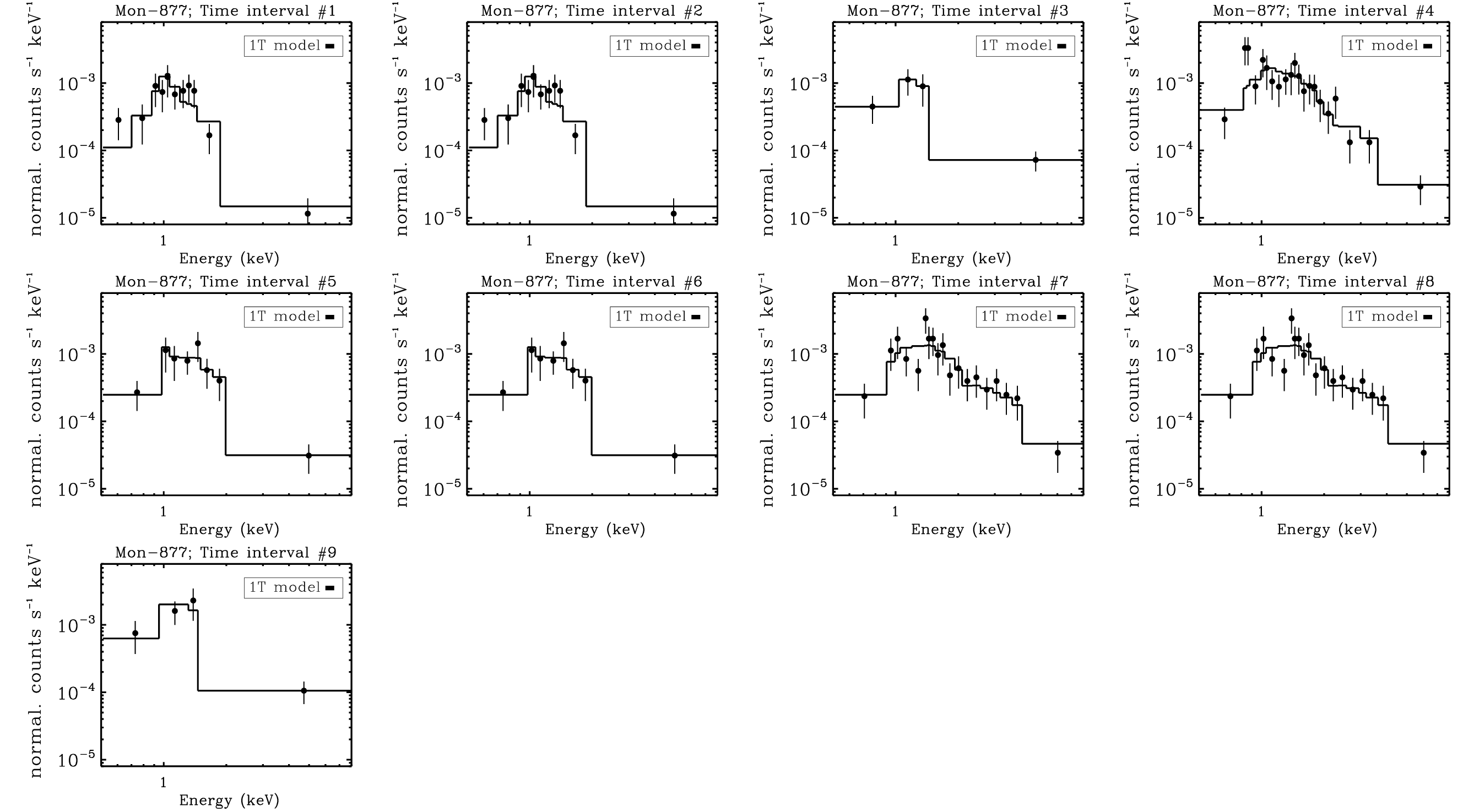}
	\caption{Variability and X-ray spectra of Mon-877, analyzed both as a dipper and burster, with no significant correlation between the optical and X-ray variability.}
	\label{variab_others_13}
	\end{figure}

	\begin{figure}[]
	\centering
	\includegraphics[width=9.5cm]{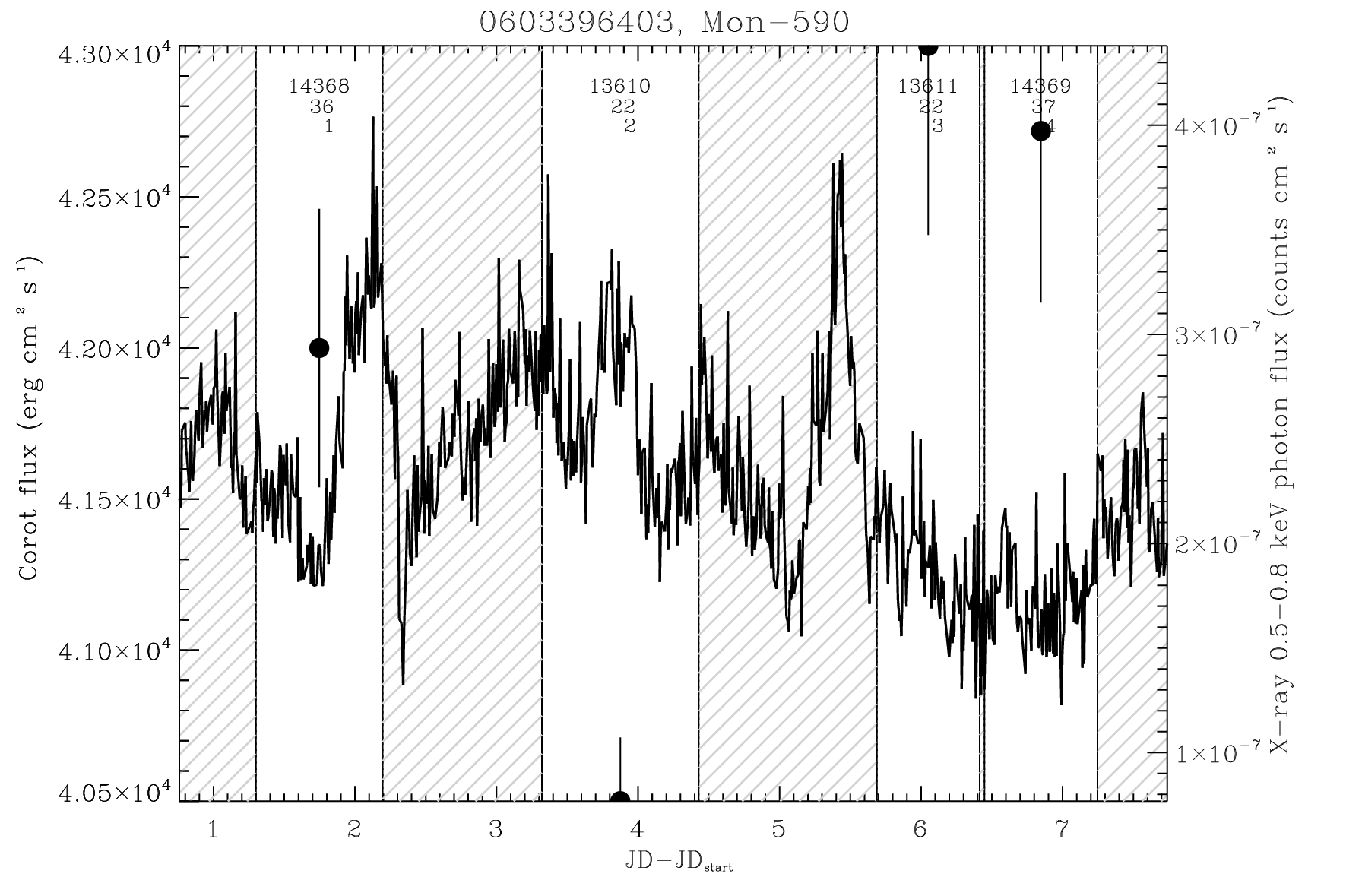}
	\includegraphics[width=9.5cm]{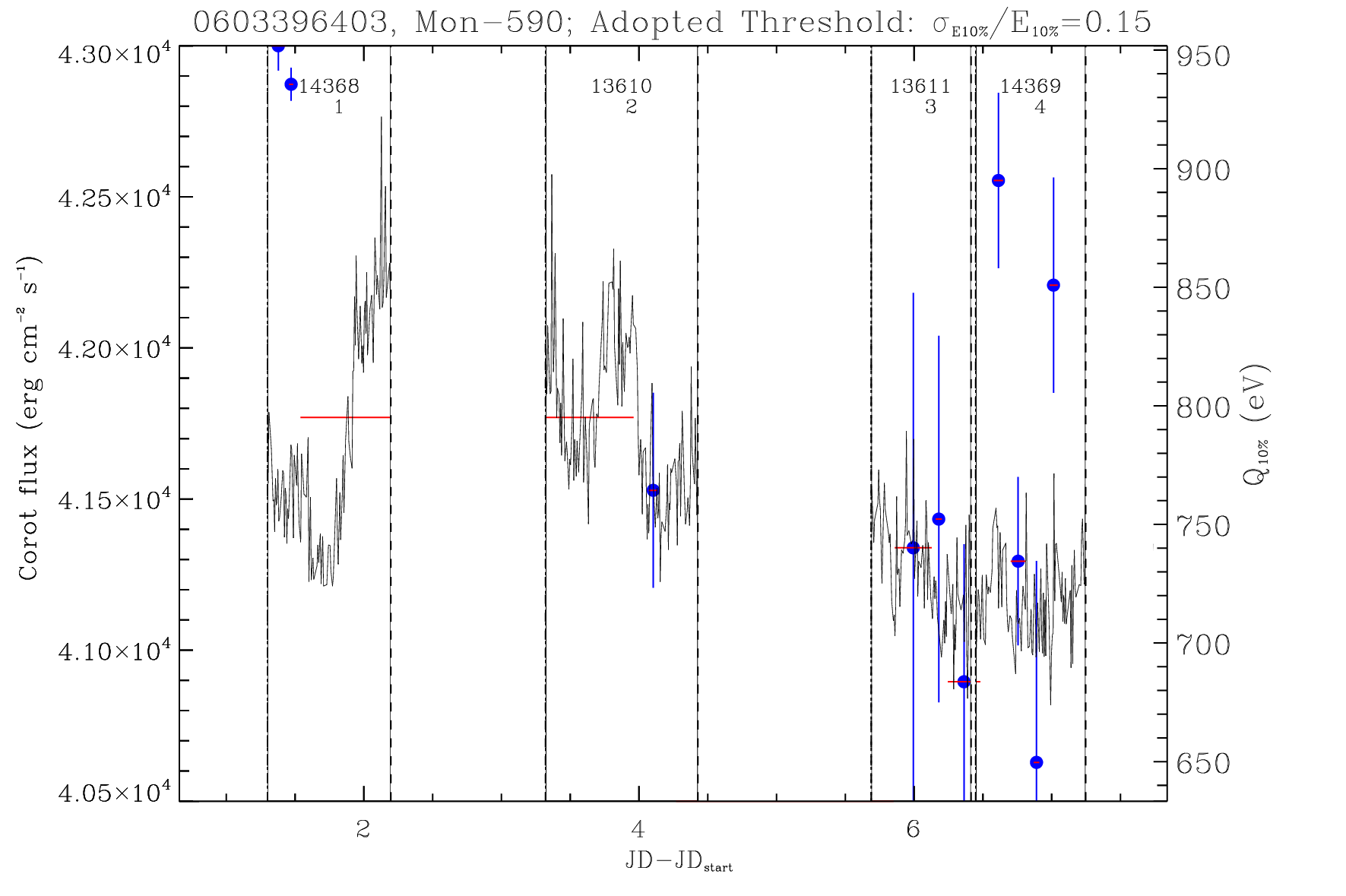}
	\includegraphics[width=8cm]{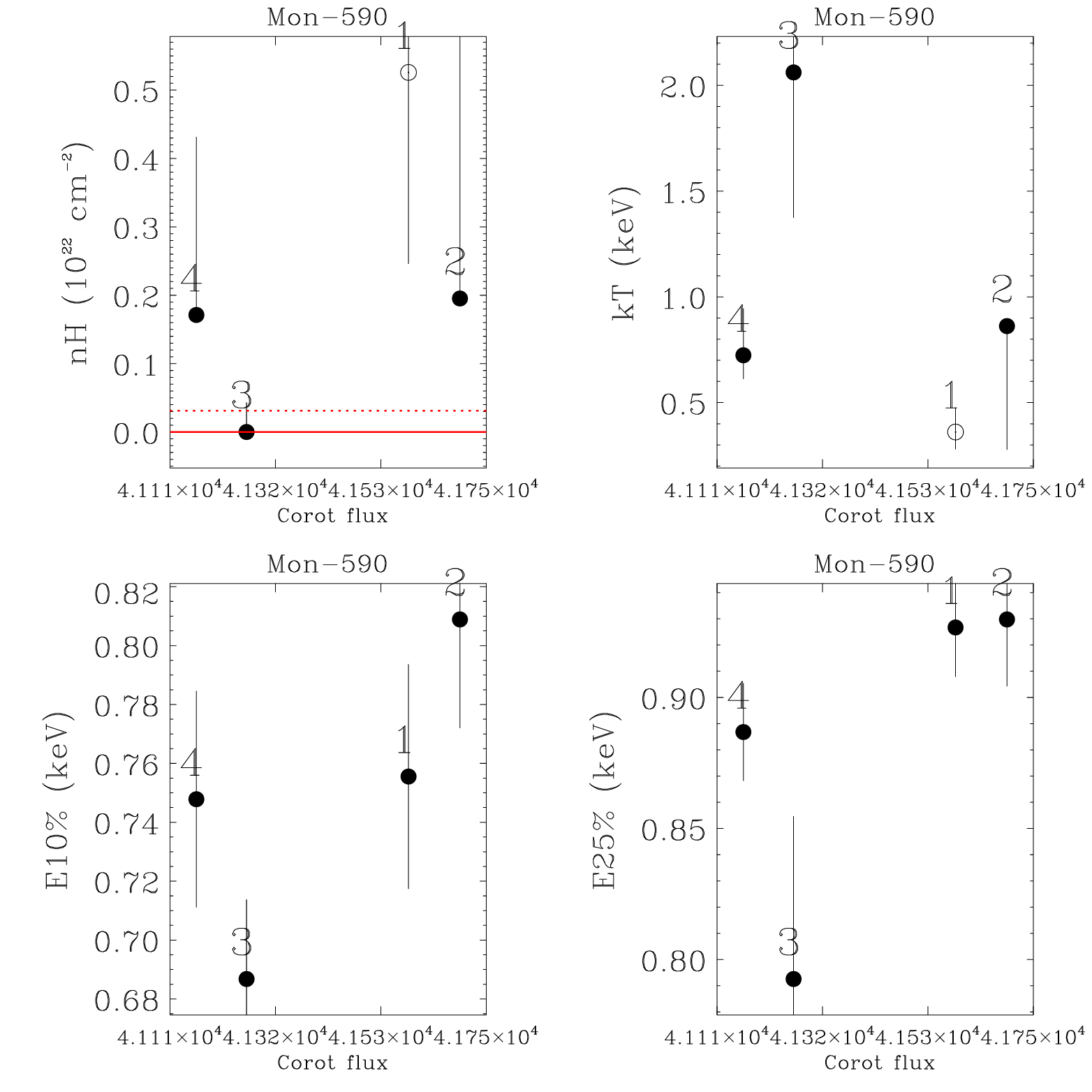}	
	\includegraphics[width=18cm]{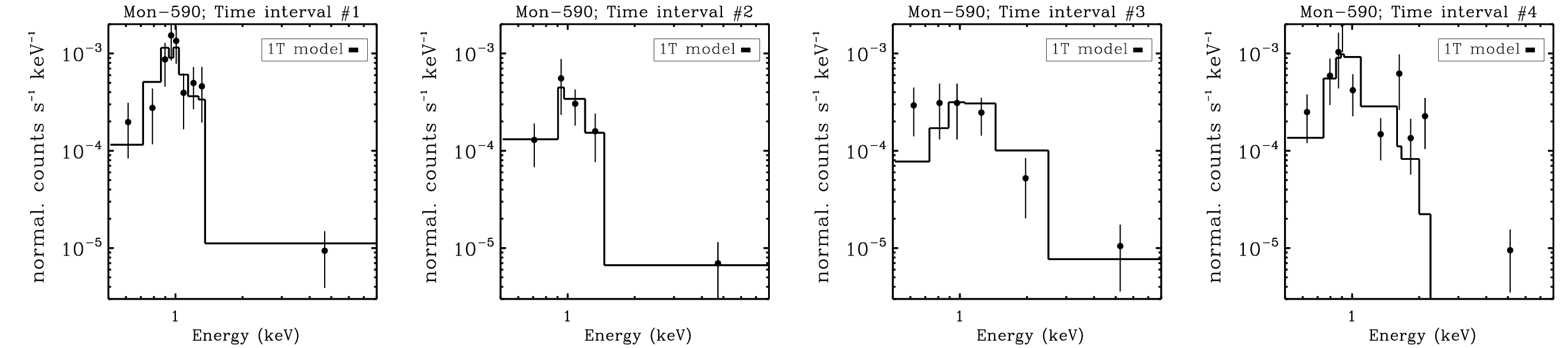}
	\caption{Variability and X-ray spectra of Mon-590, analyzed both as a dipper and burster, but with few X-ray photons detected.}
	\label{variab_others_14}
	\end{figure}

	\begin{figure}[]
	\centering	
	\includegraphics[width=9.5cm]{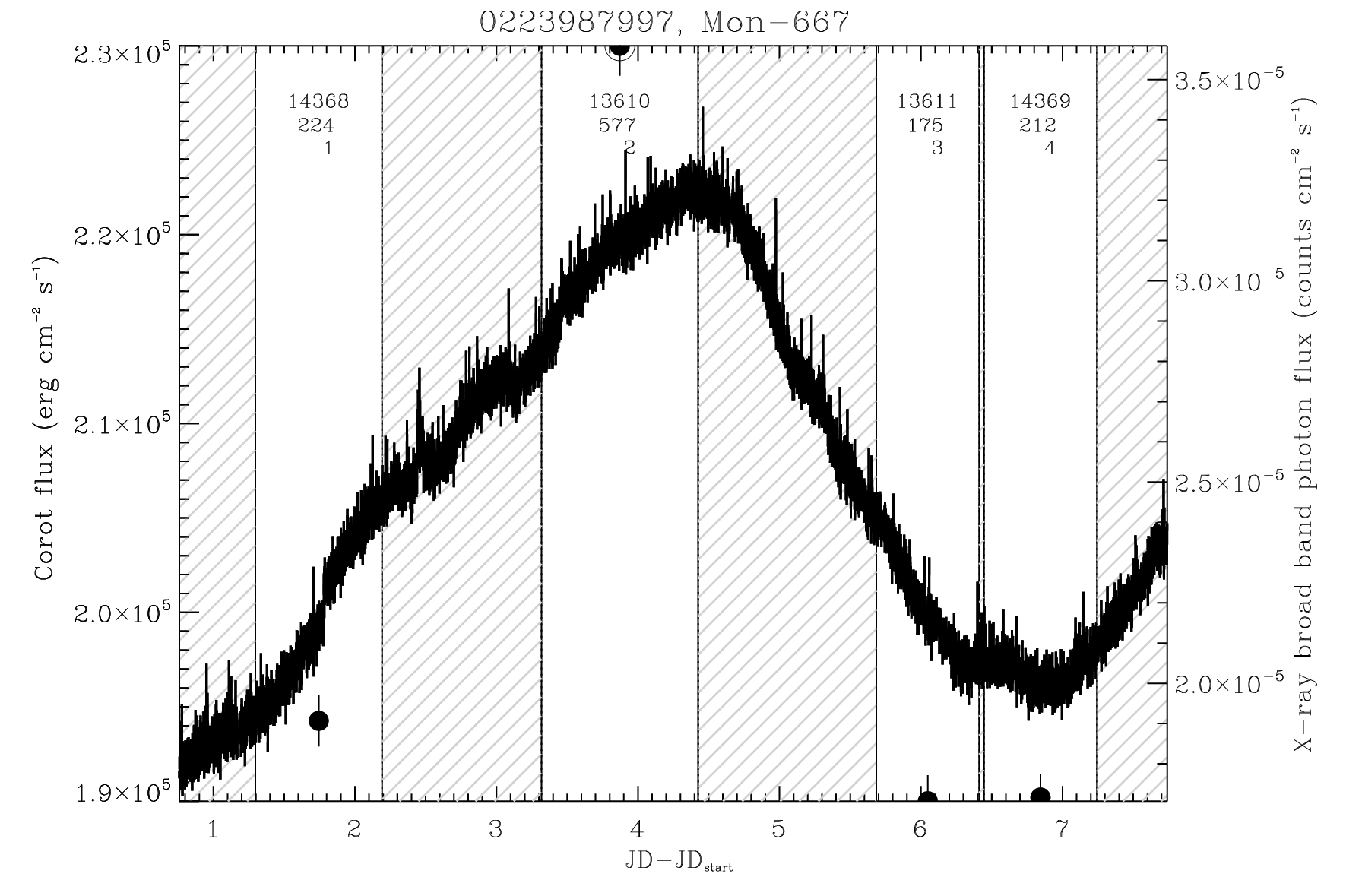}
	\includegraphics[width=8cm]{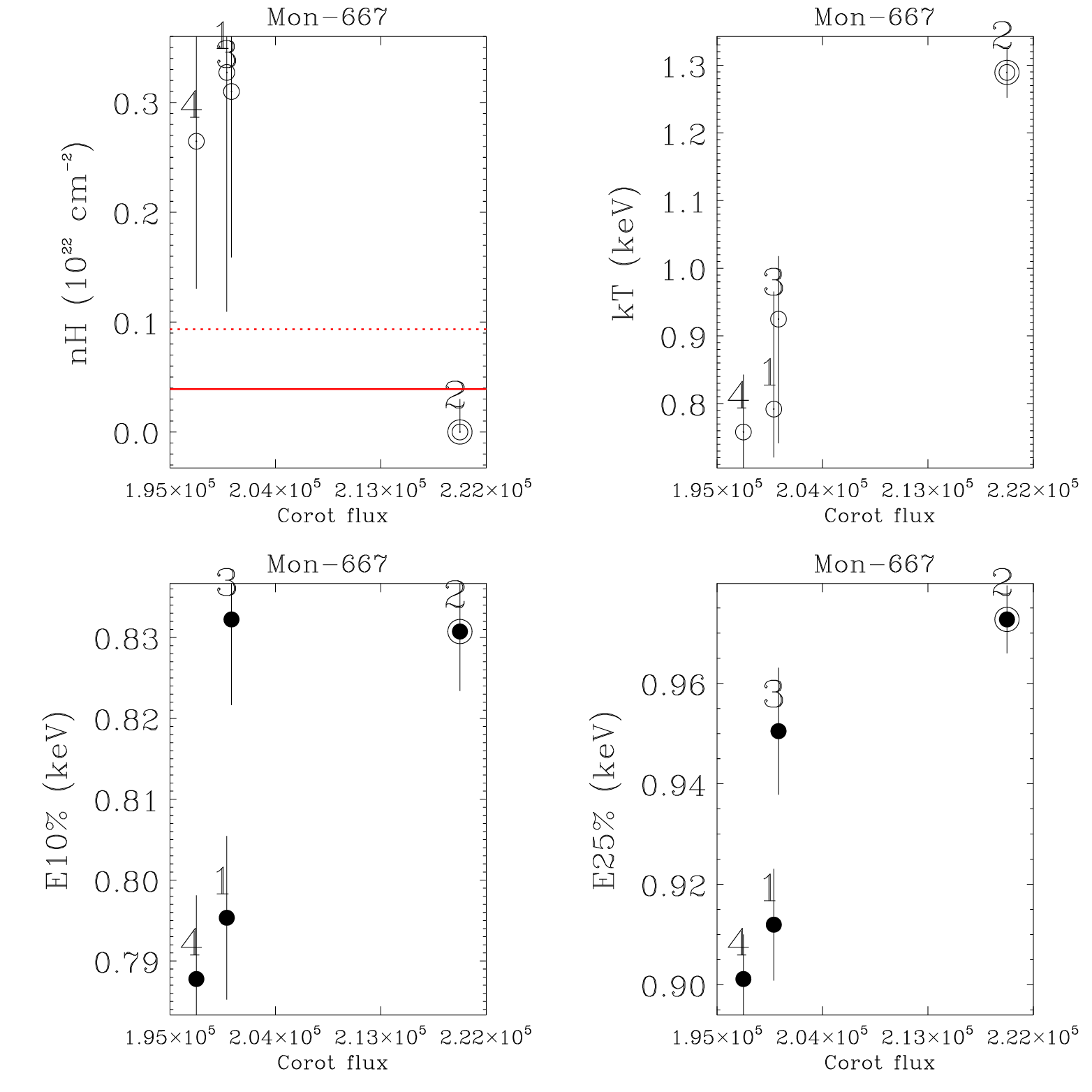}
	\includegraphics[width=18cm]{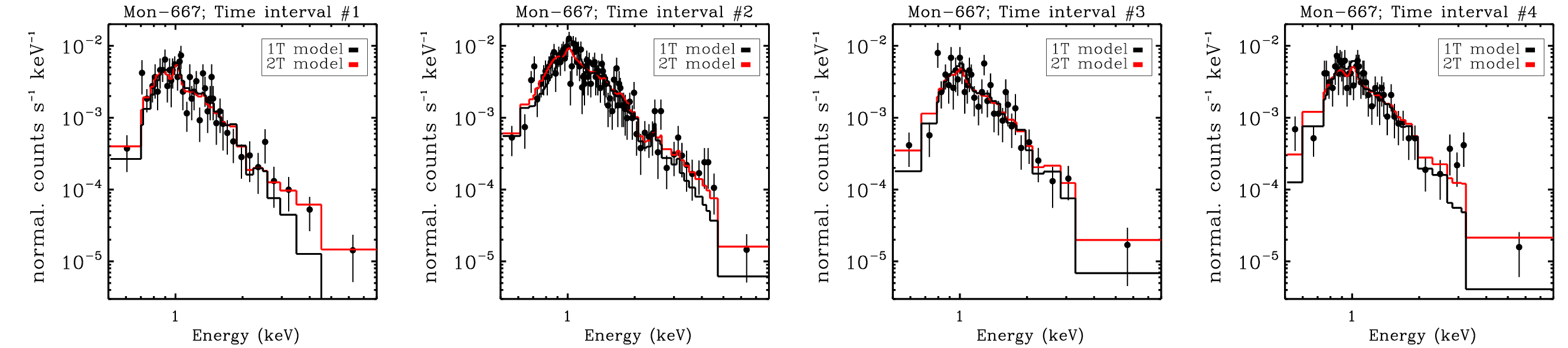}
	\caption{Variability and X-ray spectra of Mon-667. During the {\em Chandra} frames, Mon-667 mimics a quasi-periodic optical variability, but the CoRoT light curve shows evident dips after the {\em Chandra} observations. We analyzed it as a dipper, but the fit of the X-ray spectrum with 1T and 2T thermal plasma models are never well constrained.}
	\label{variab_others_15}
	\end{figure}

	\begin{figure}[]
	\centering	
	\includegraphics[width=9.5cm]{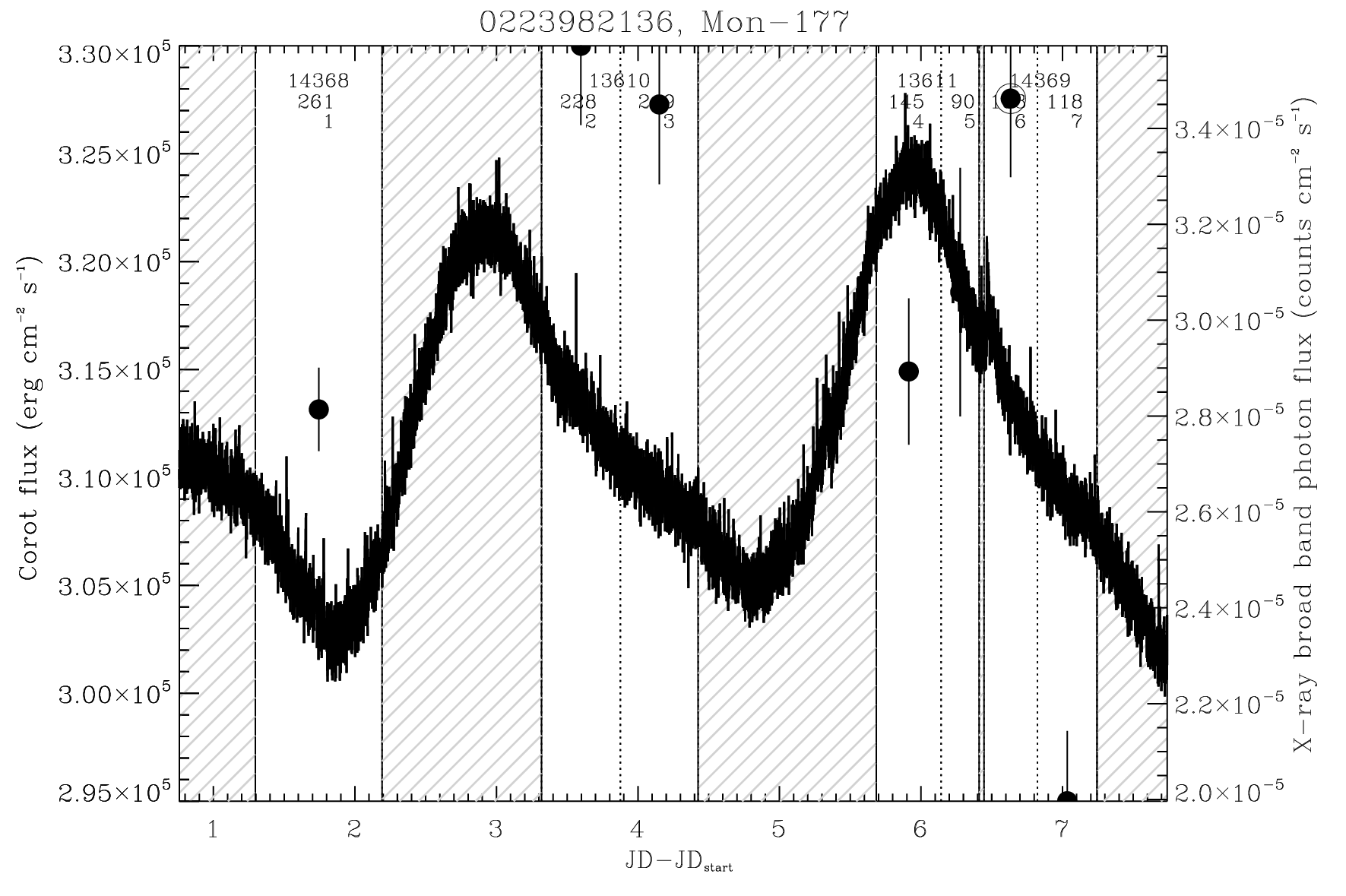}
	\includegraphics[width=8cm]{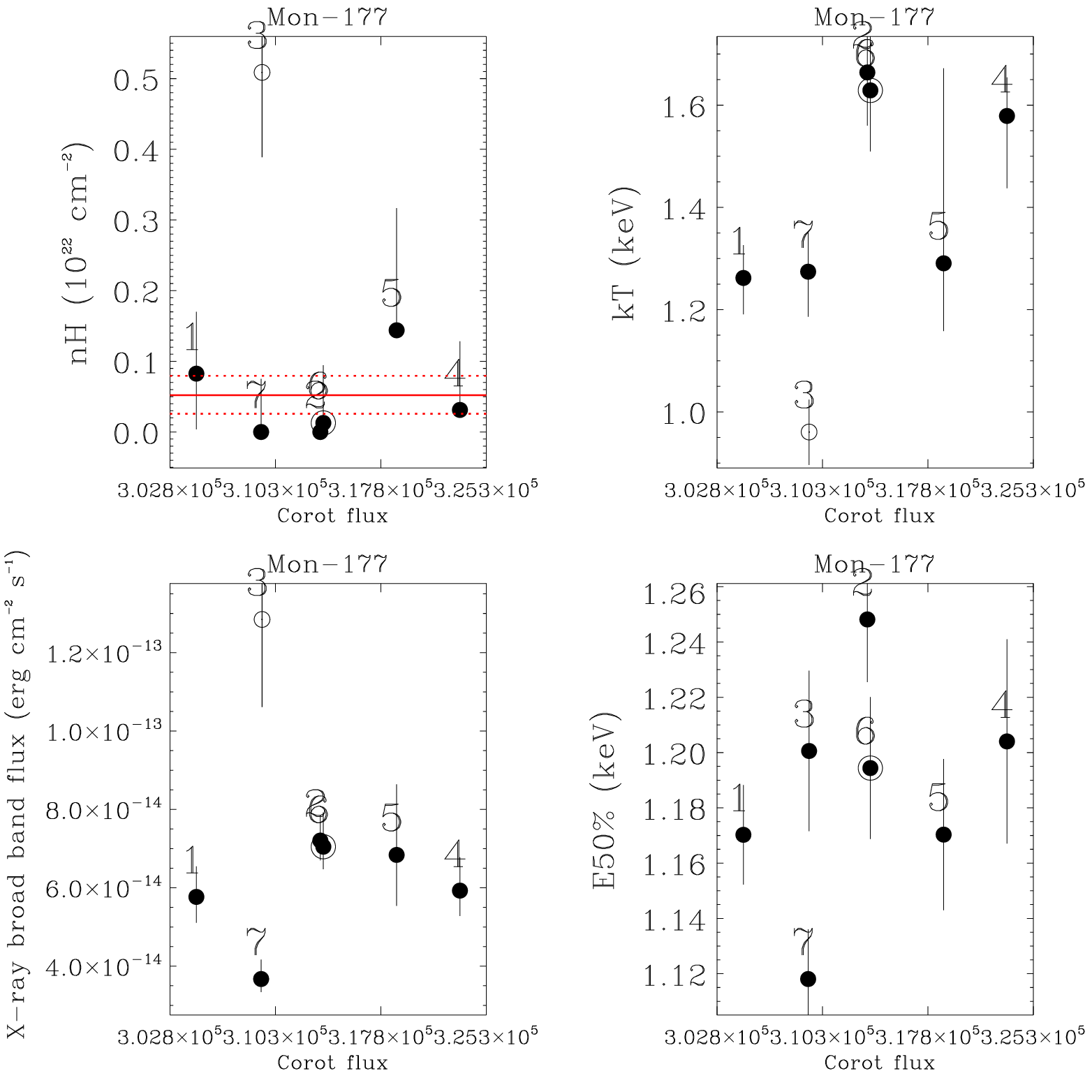}	
	\includegraphics[width=18cm]{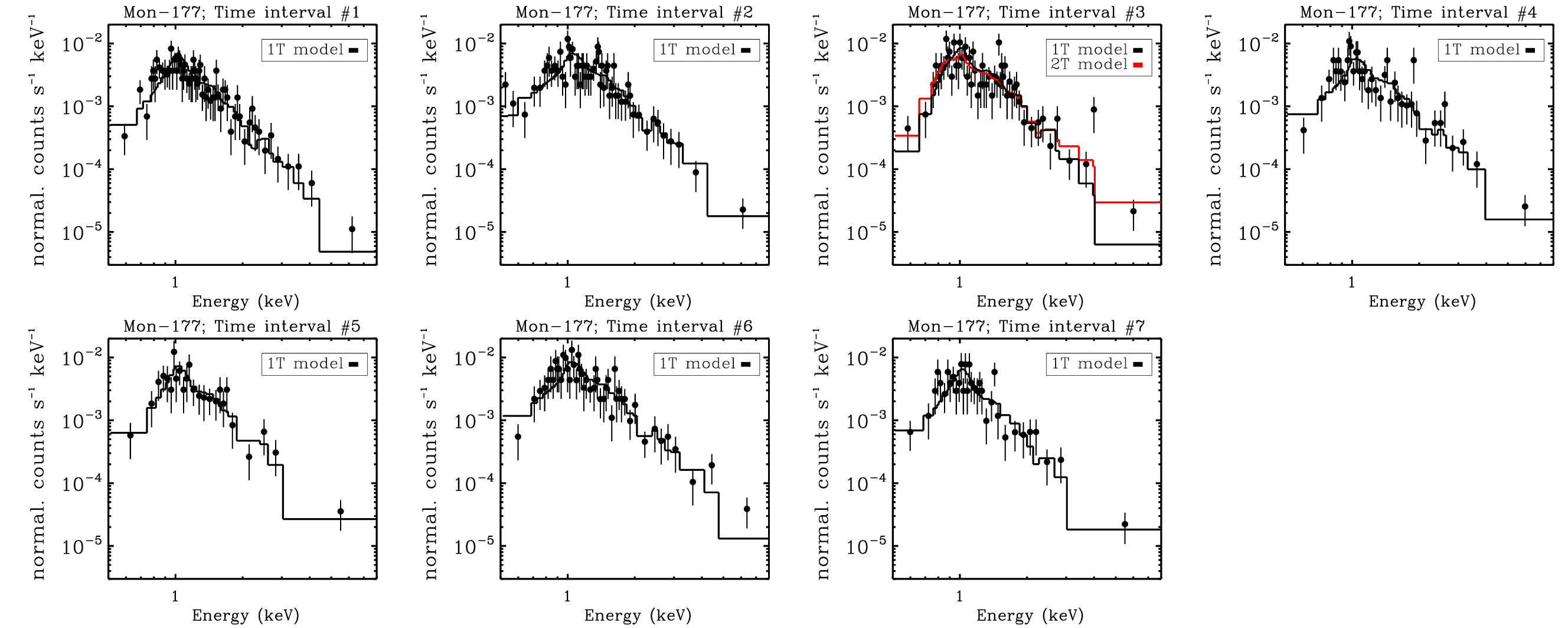}
	\caption{Variability and X-ray spectra of Mon-177, showing periodic optical variability but no interesting X-ray variability.}
	\label{variab_others_16}
	\end{figure}

	\begin{figure}[]
	\centering	
	\includegraphics[width=9.5cm]{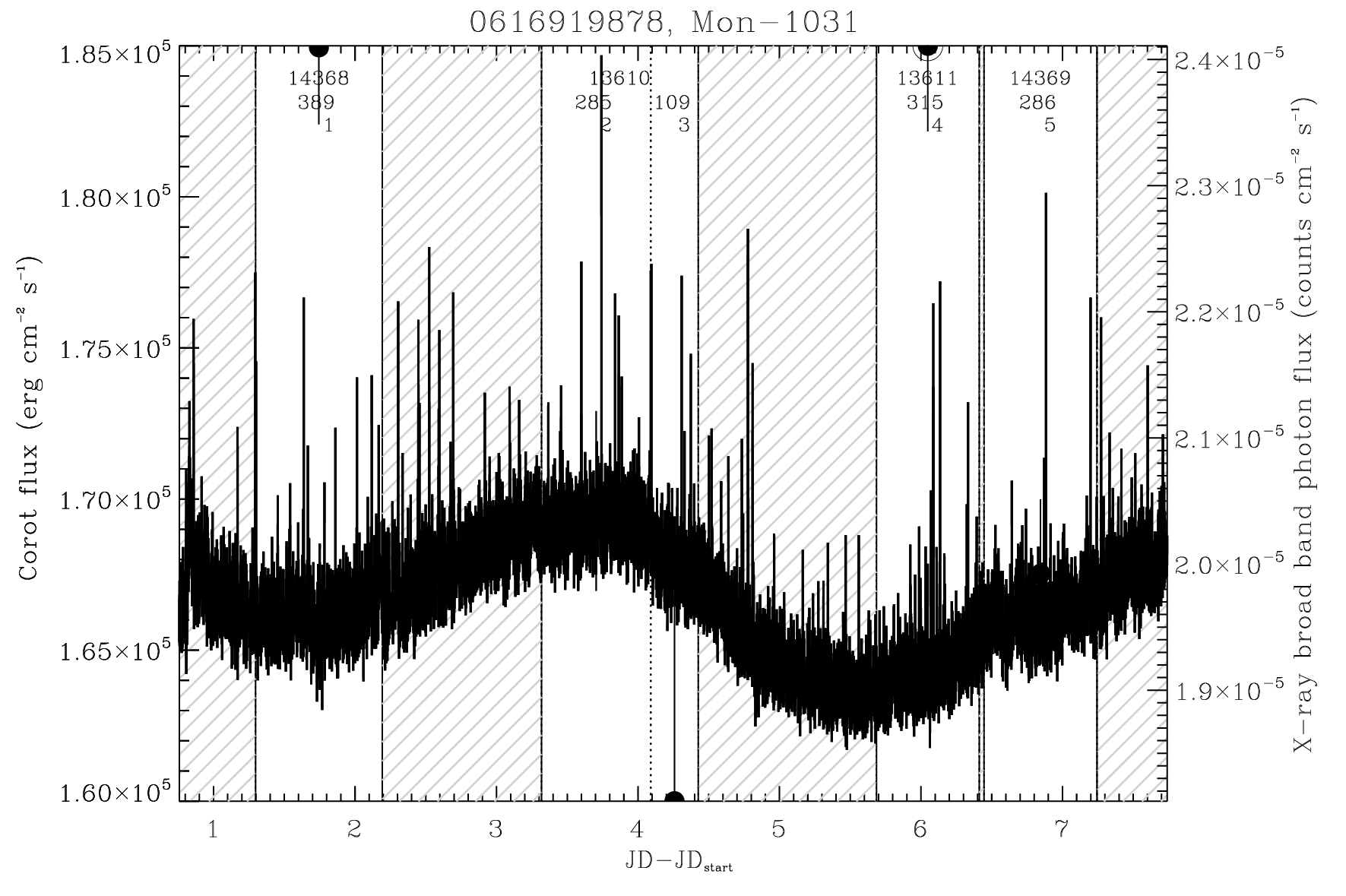}
	\includegraphics[width=8cm]{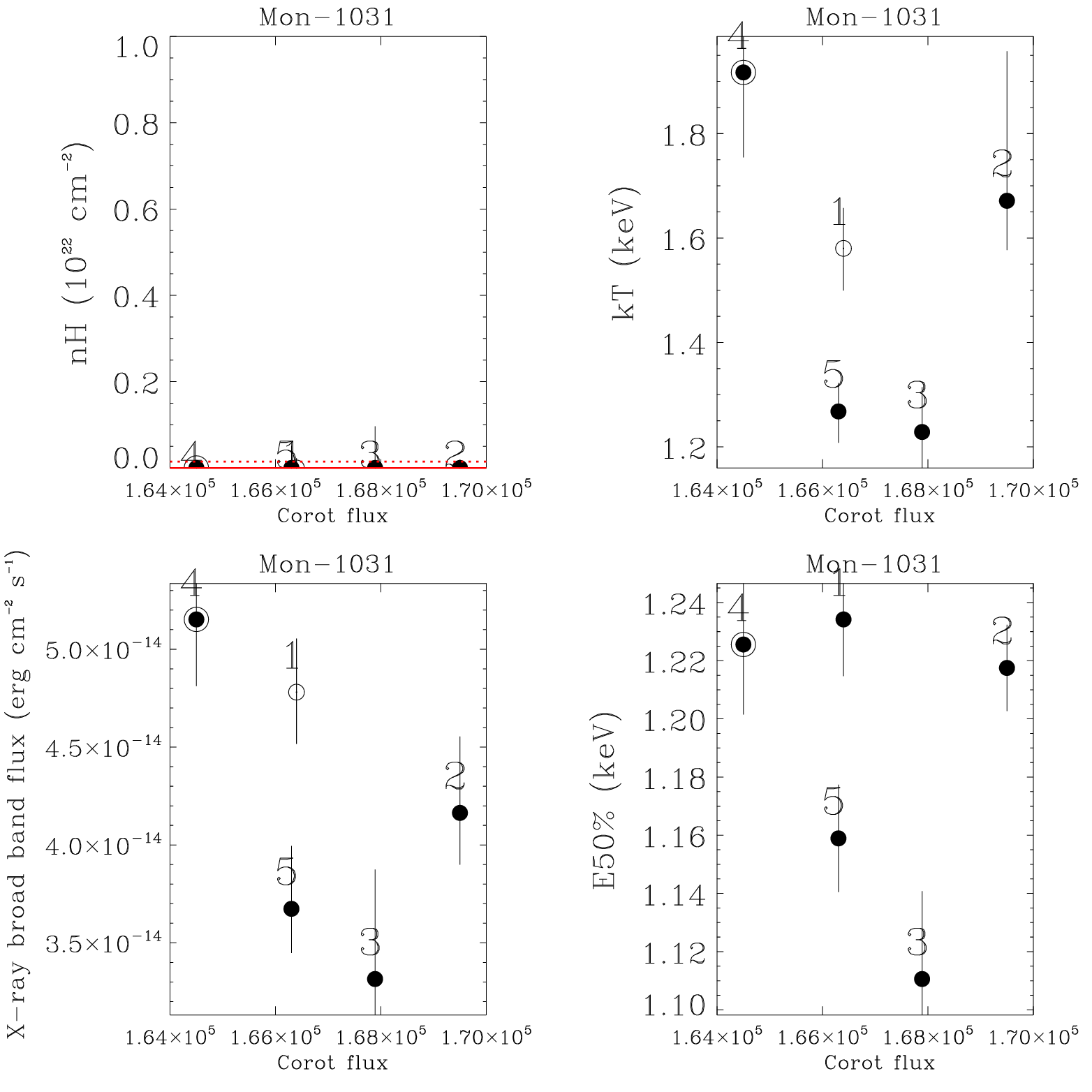}	
	\includegraphics[width=18cm]{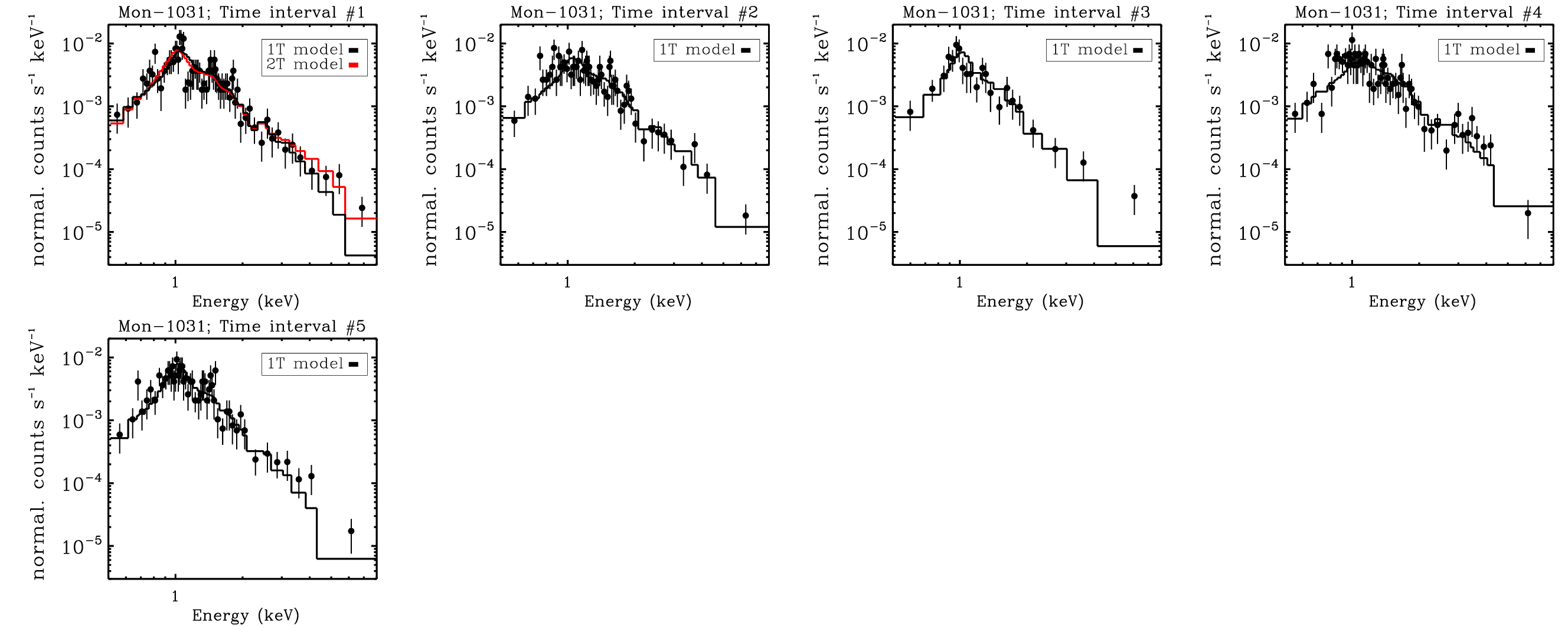}
	\caption{Variability and X-ray spectra of Mon-1031, showing periodic optical variability but no interesting X-ray variability.}
	\label{variab_others_17}
	\end{figure}

	\begin{figure}[]
	\centering	
	\includegraphics[width=9.5cm]{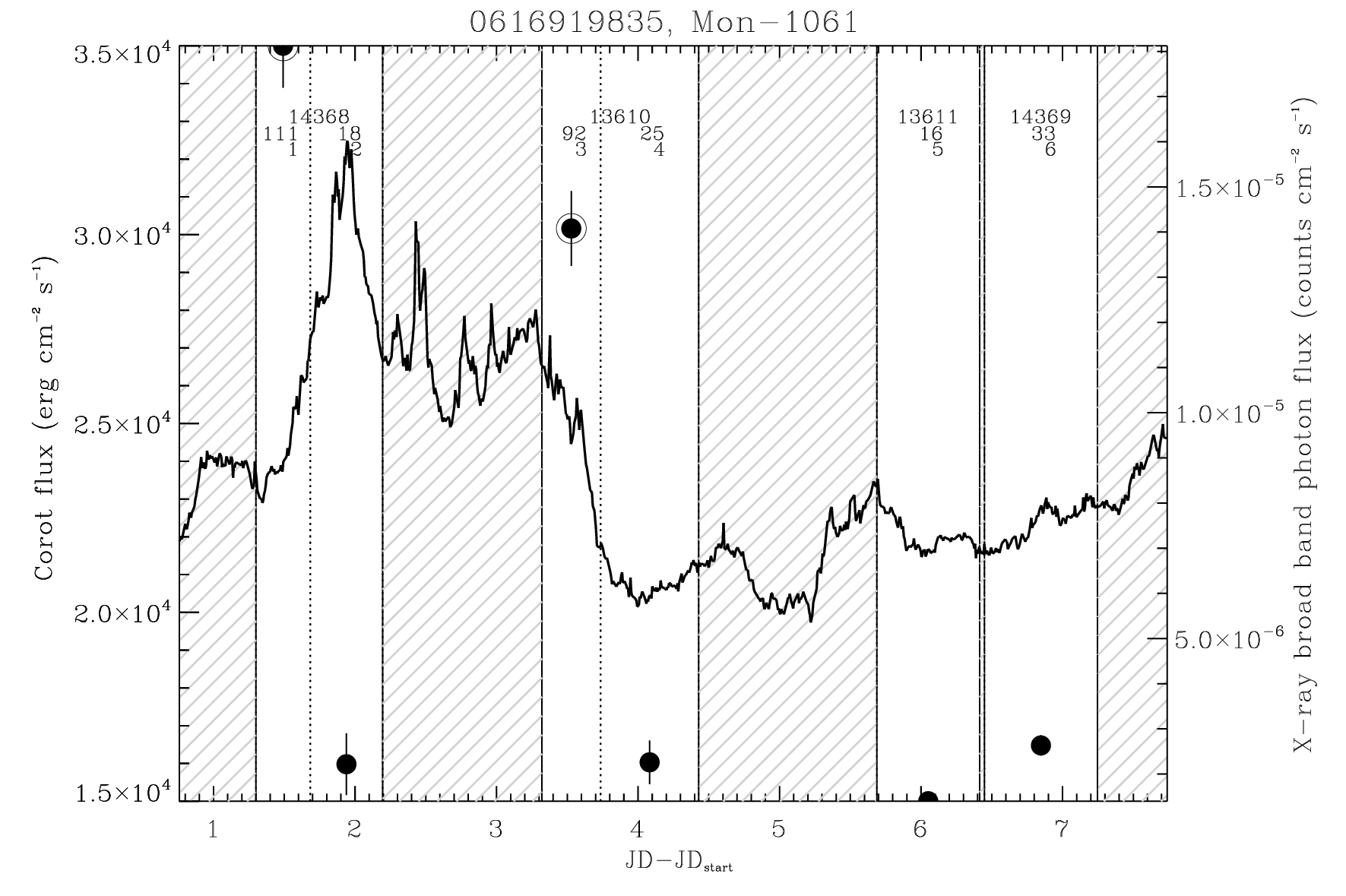}
	\includegraphics[width=9.5cm]{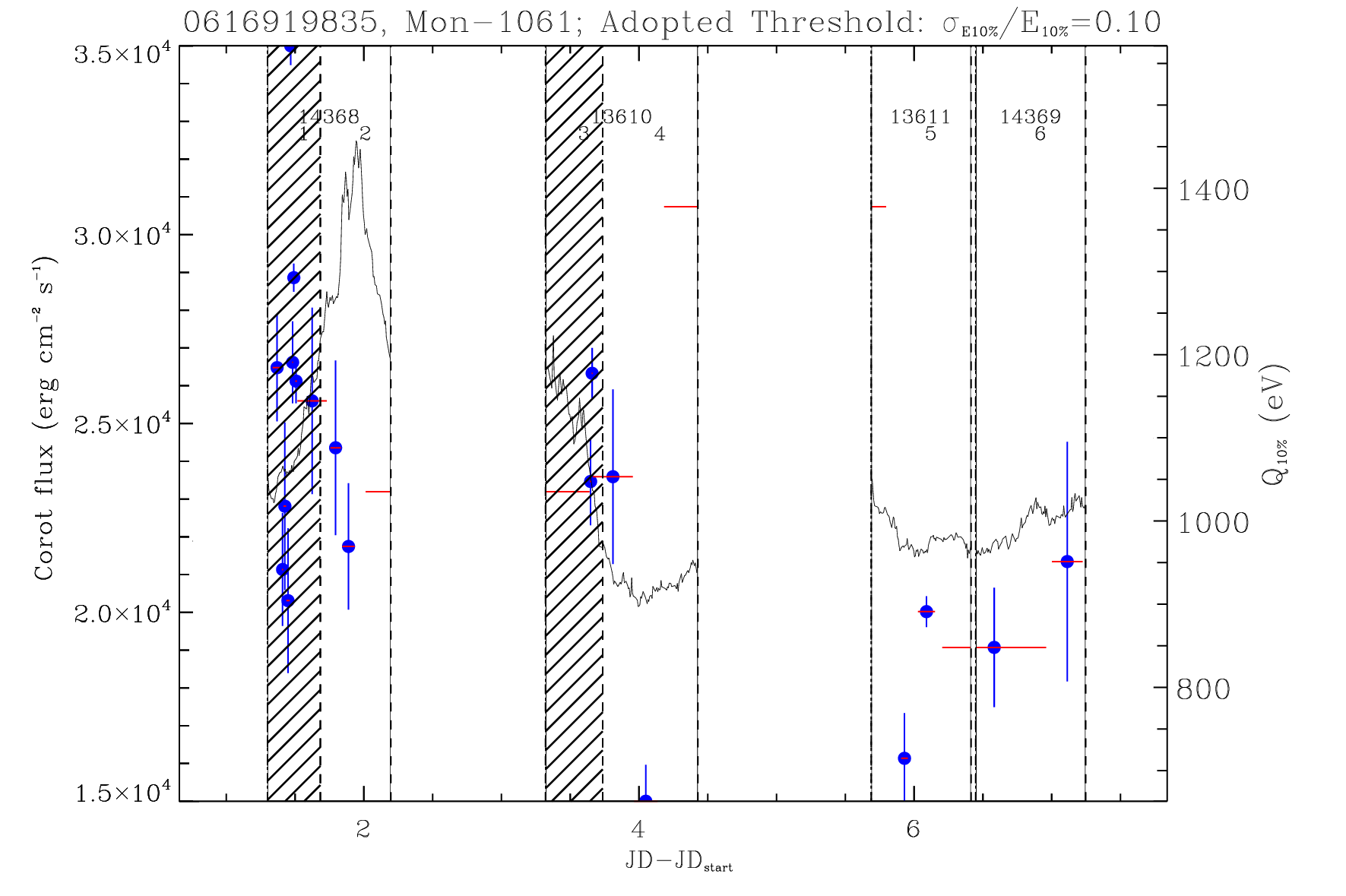}
	\includegraphics[width=8cm]{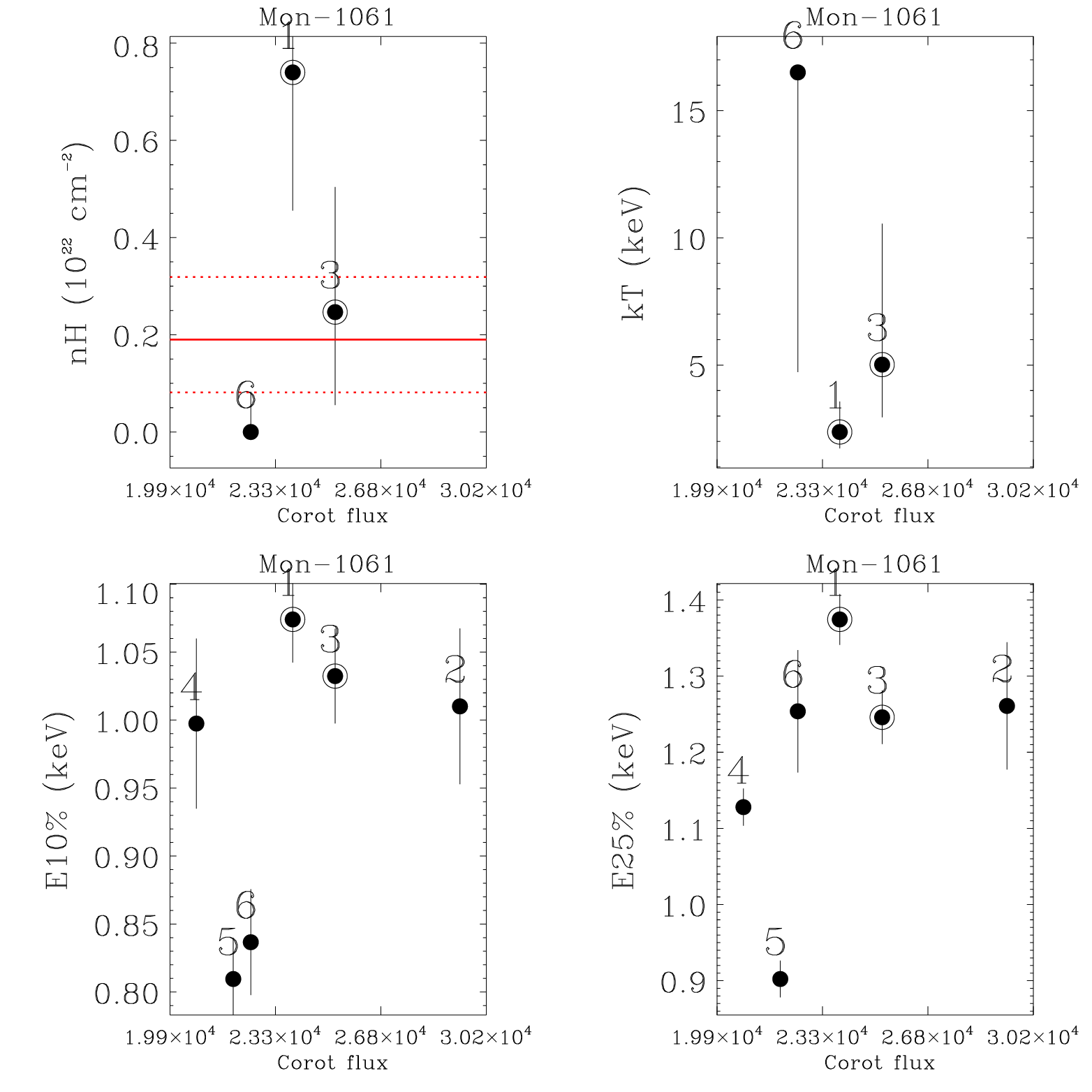}	
	\includegraphics[width=18cm]{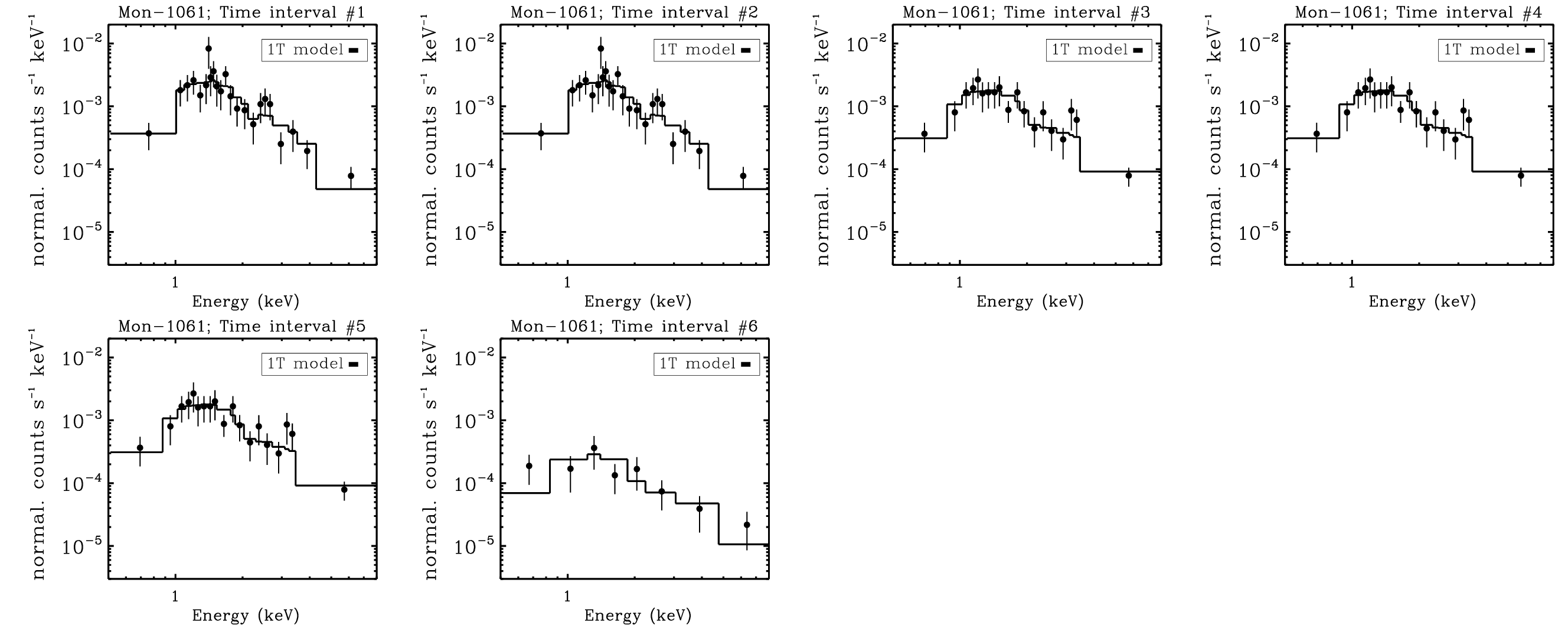}
	\caption{Variability and X-ray spectra of Mon-1061, which has not been analyzed since very few X-ray photons are detected during the relevant optical features.}
	\label{variab_others_18}
	\end{figure}

\clearpage
	\begin{figure}[]
	\centering	
	\includegraphics[width=9.5cm]{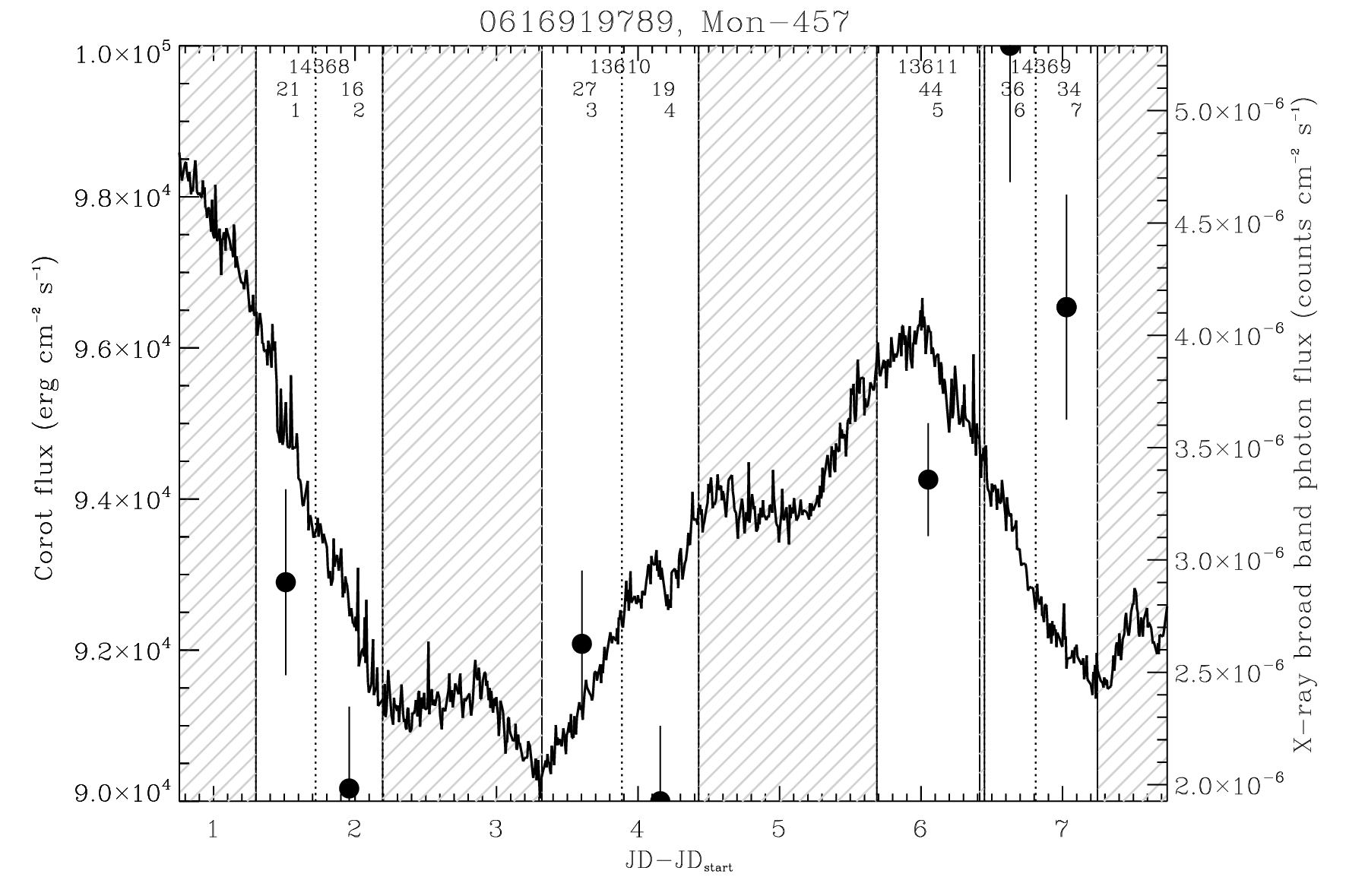}
	\includegraphics[width=9.5cm]{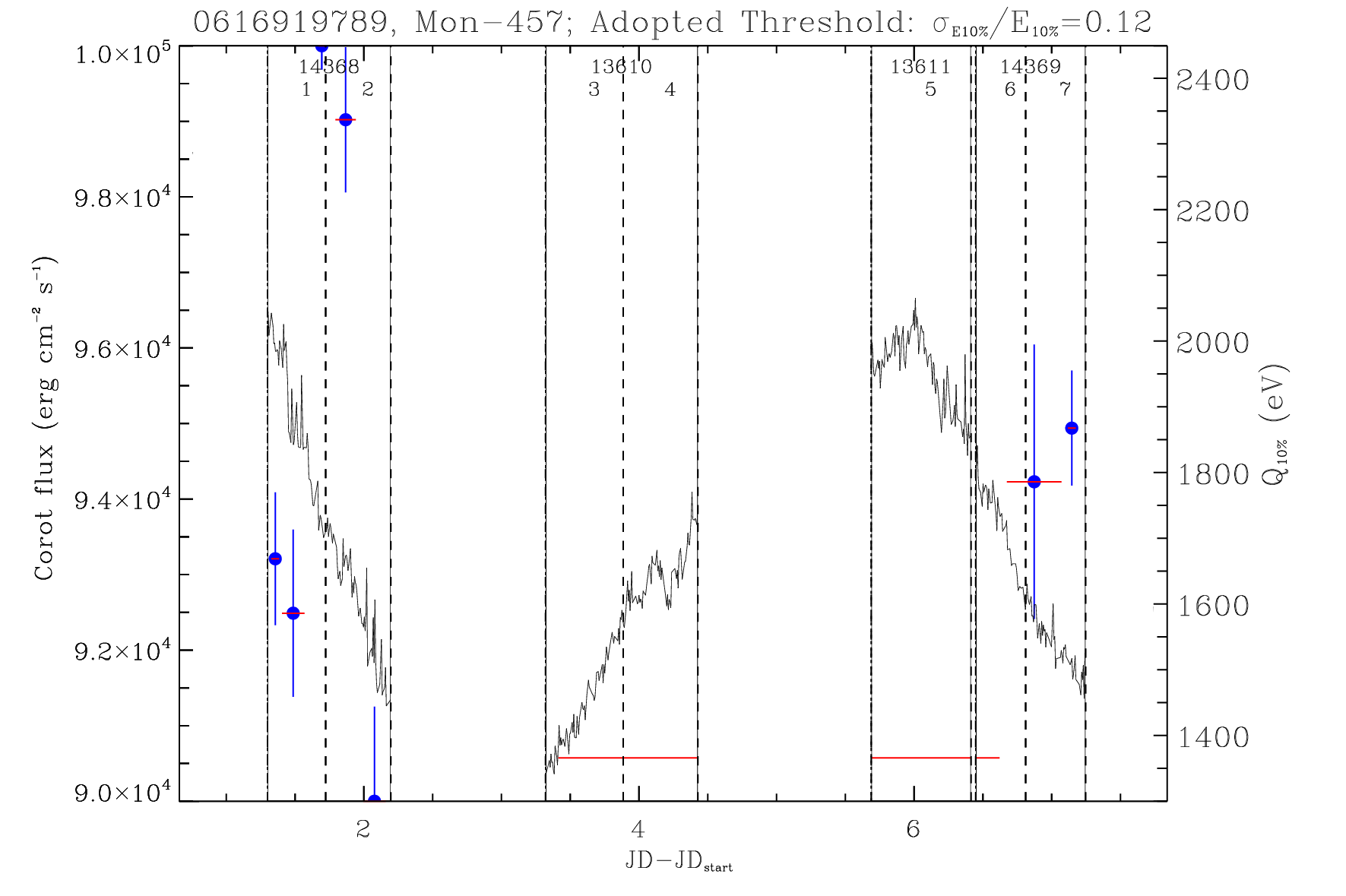}
	\includegraphics[width=8cm]{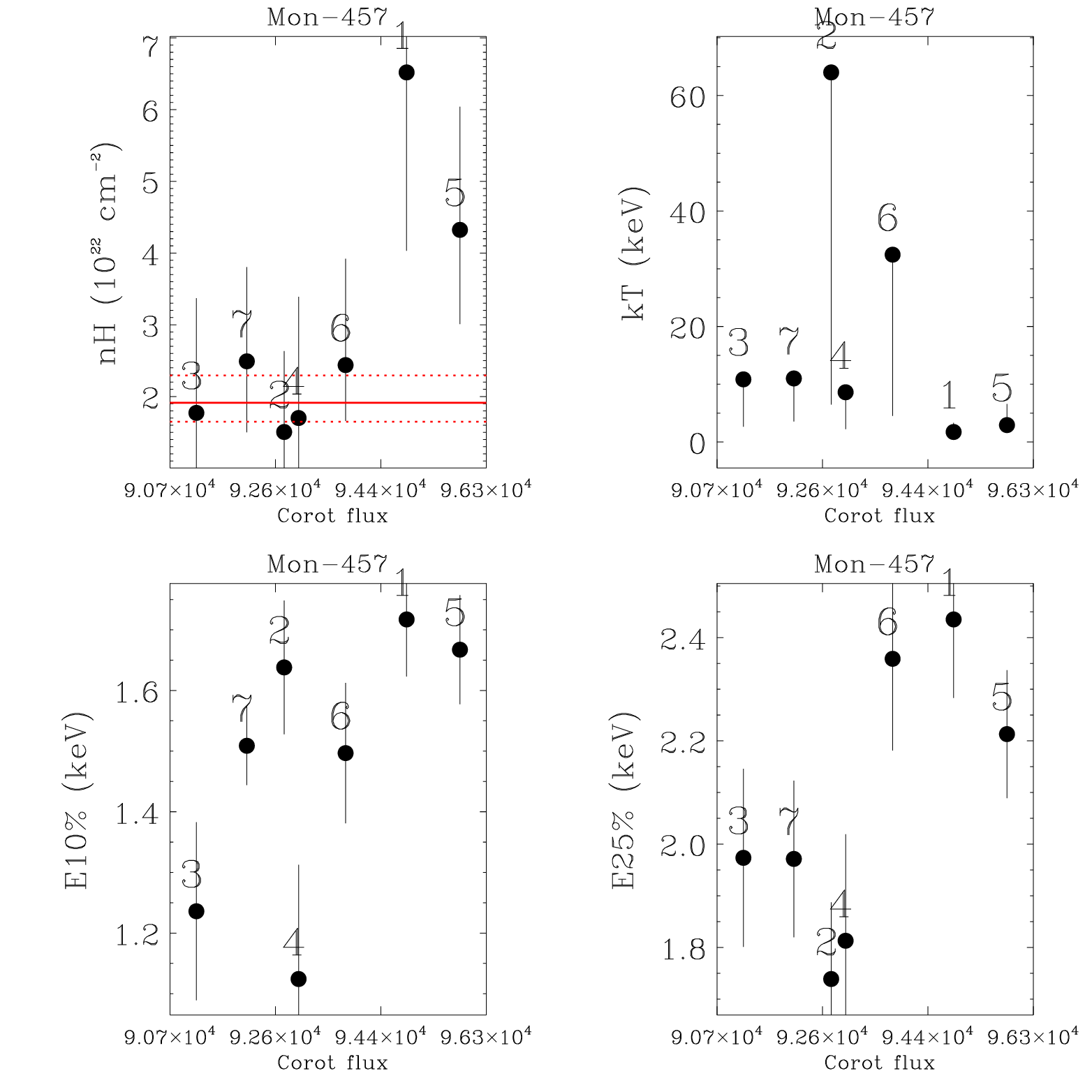}
	\includegraphics[width=18cm]{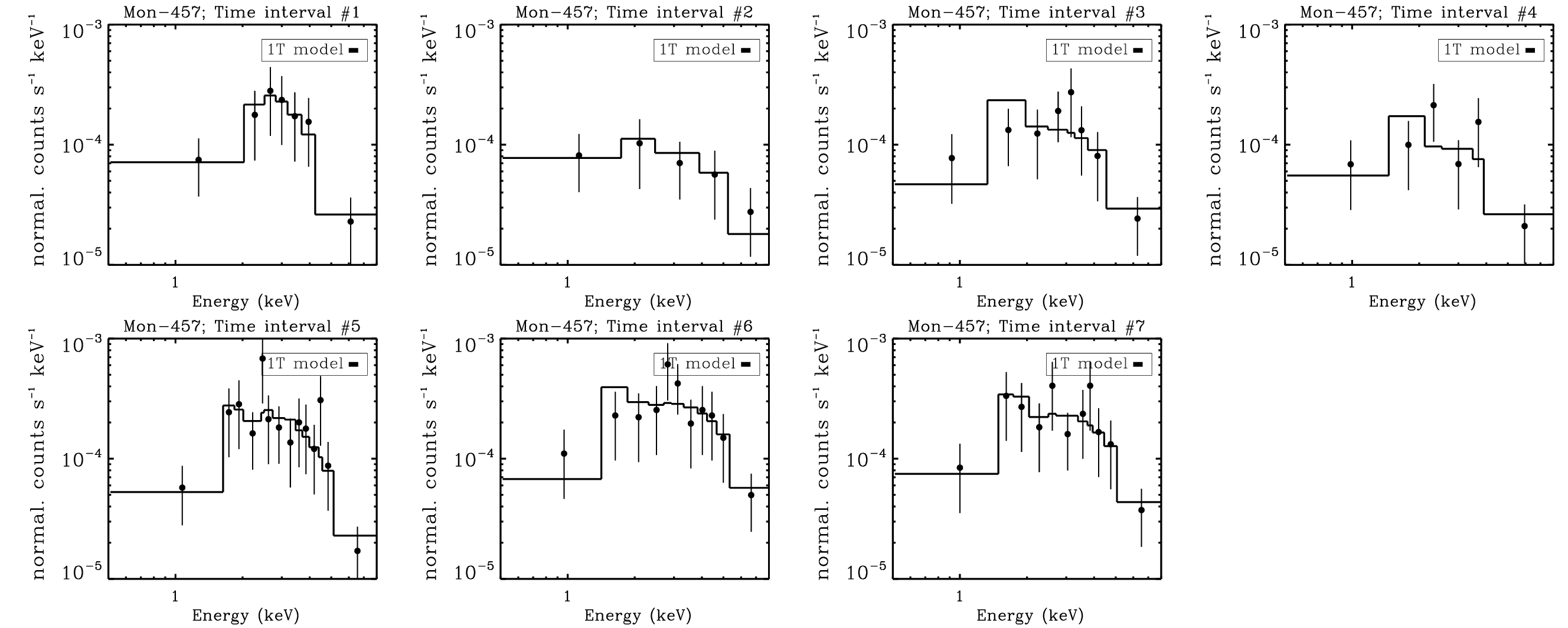}
	\caption{Variability and X-ray spectra of Mon-457, which has not been analyzed since very few X-ray photons are detected during the relevant optical features.}
	\label{variab_others_19}
	\end{figure}

	\begin{figure}[]
	\centering	
	\includegraphics[width=9.5cm]{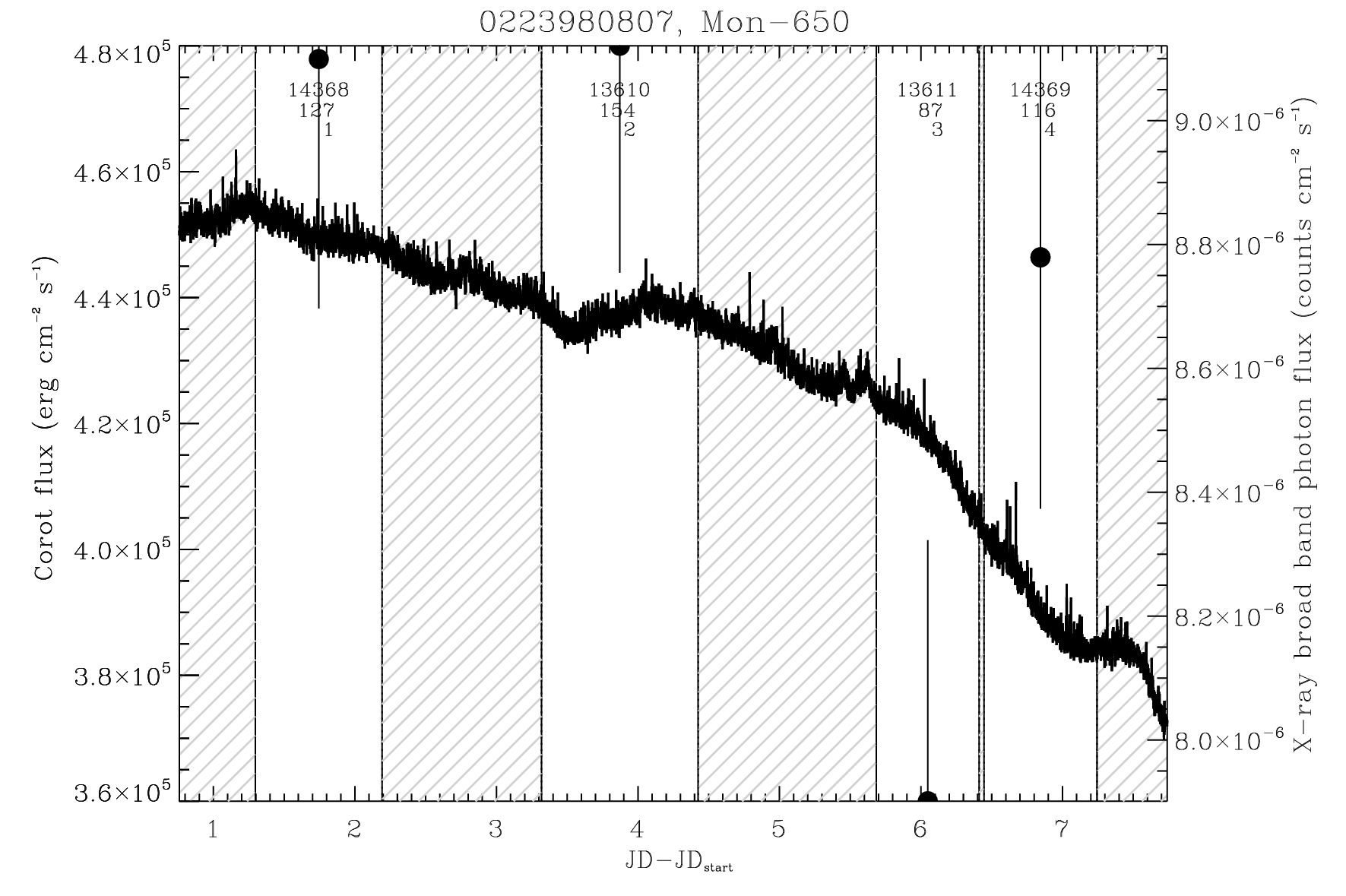}
	\includegraphics[width=9.5cm]{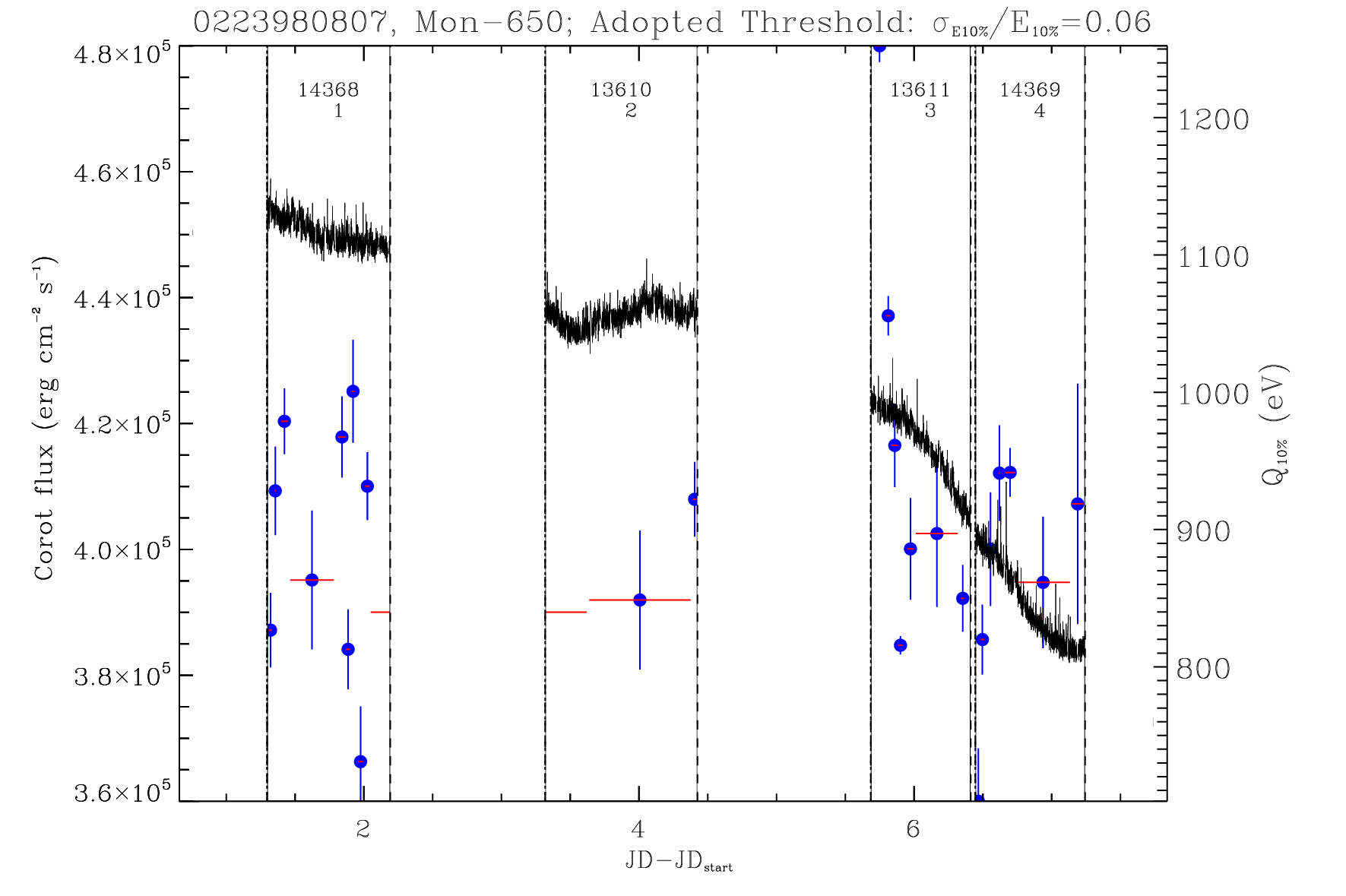}
	\includegraphics[width=8cm]{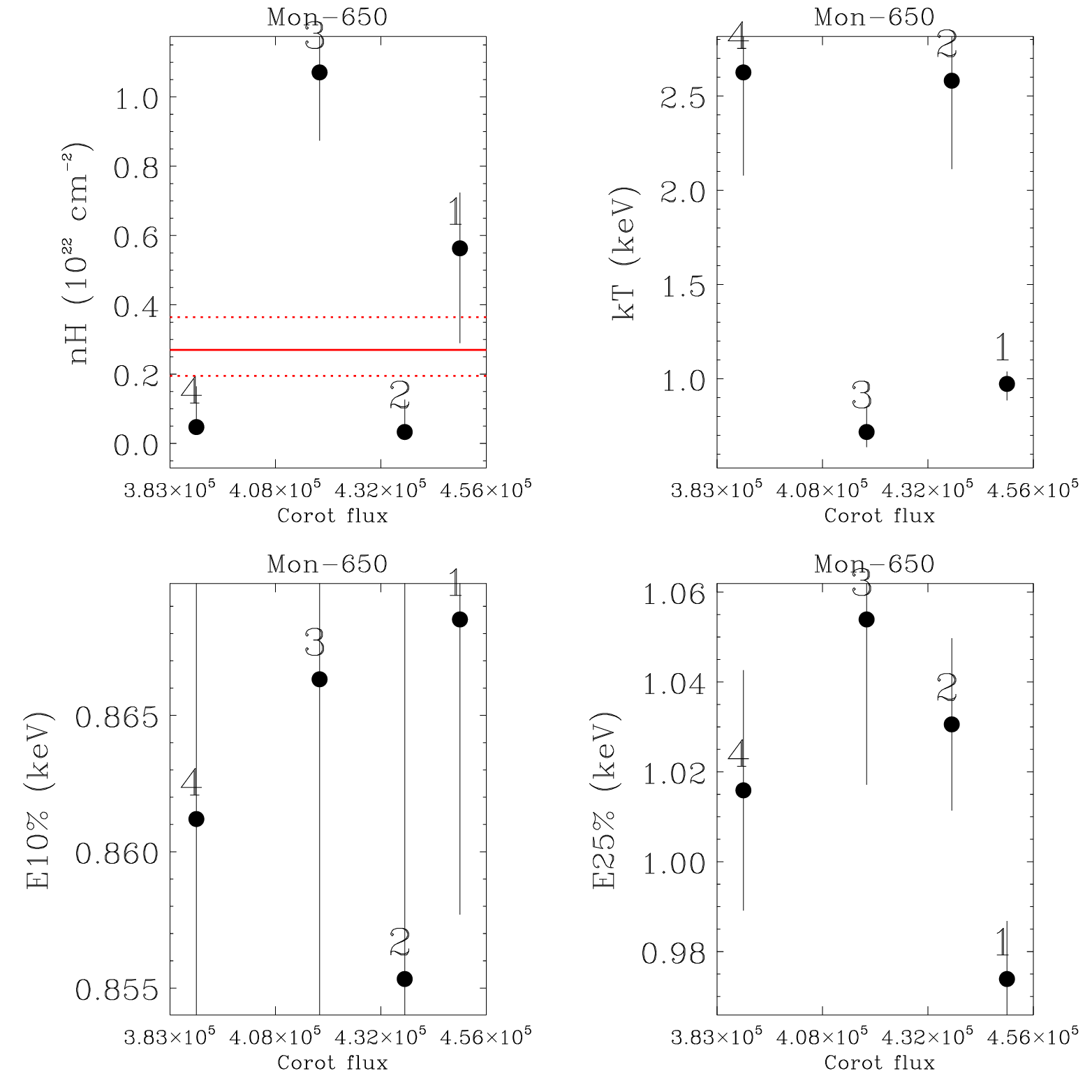}	
	\includegraphics[width=18cm]{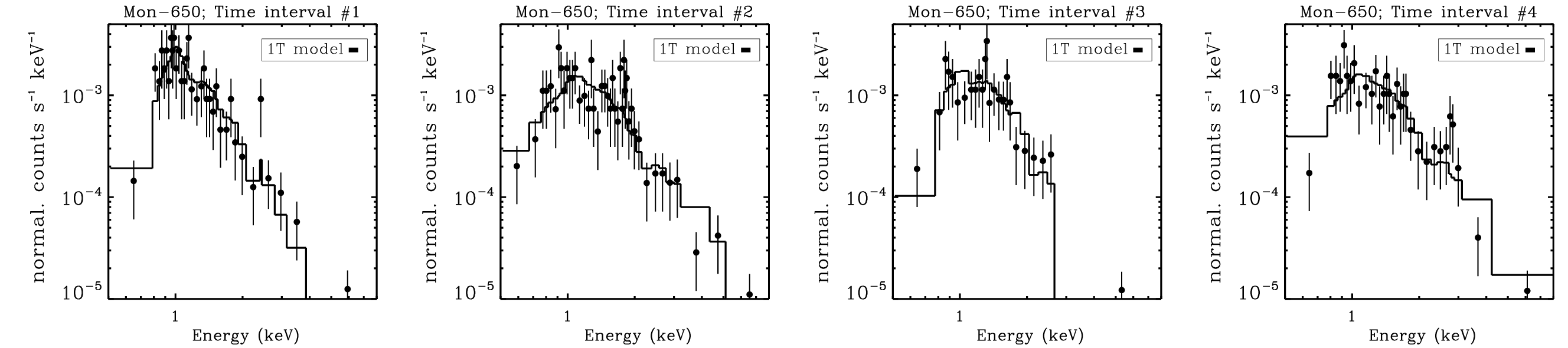}
	\caption{Variability and X-ray spectra of Mon-650, whose CoRoT light curve is dominated by large dips not occurring during the {\em Chandra} observations.}
	\label{variab_others_20}
	\end{figure}

	\begin{figure}[]
	\centering	
	\includegraphics[width=9.5cm]{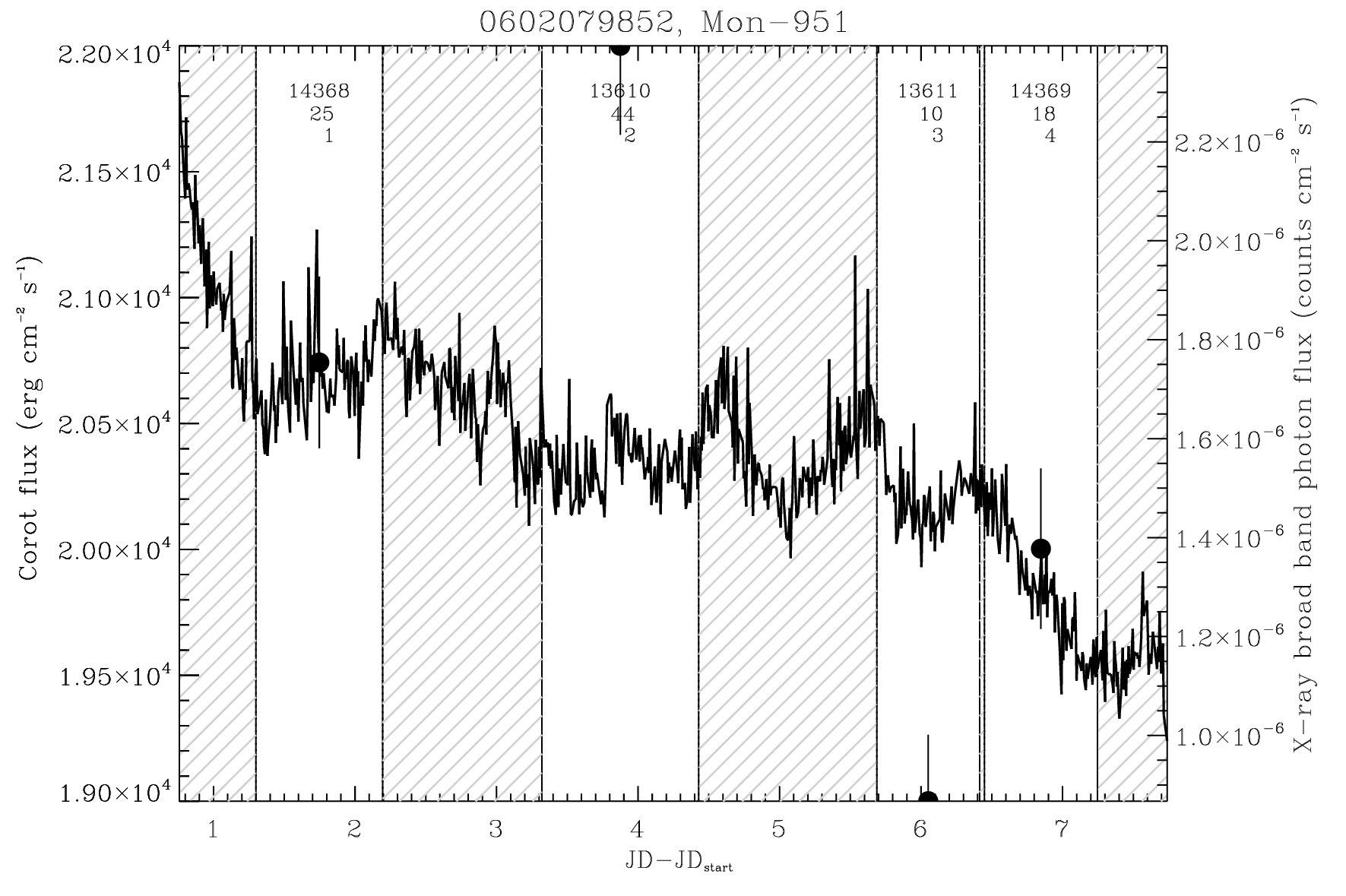}
	\includegraphics[width=9.5cm]{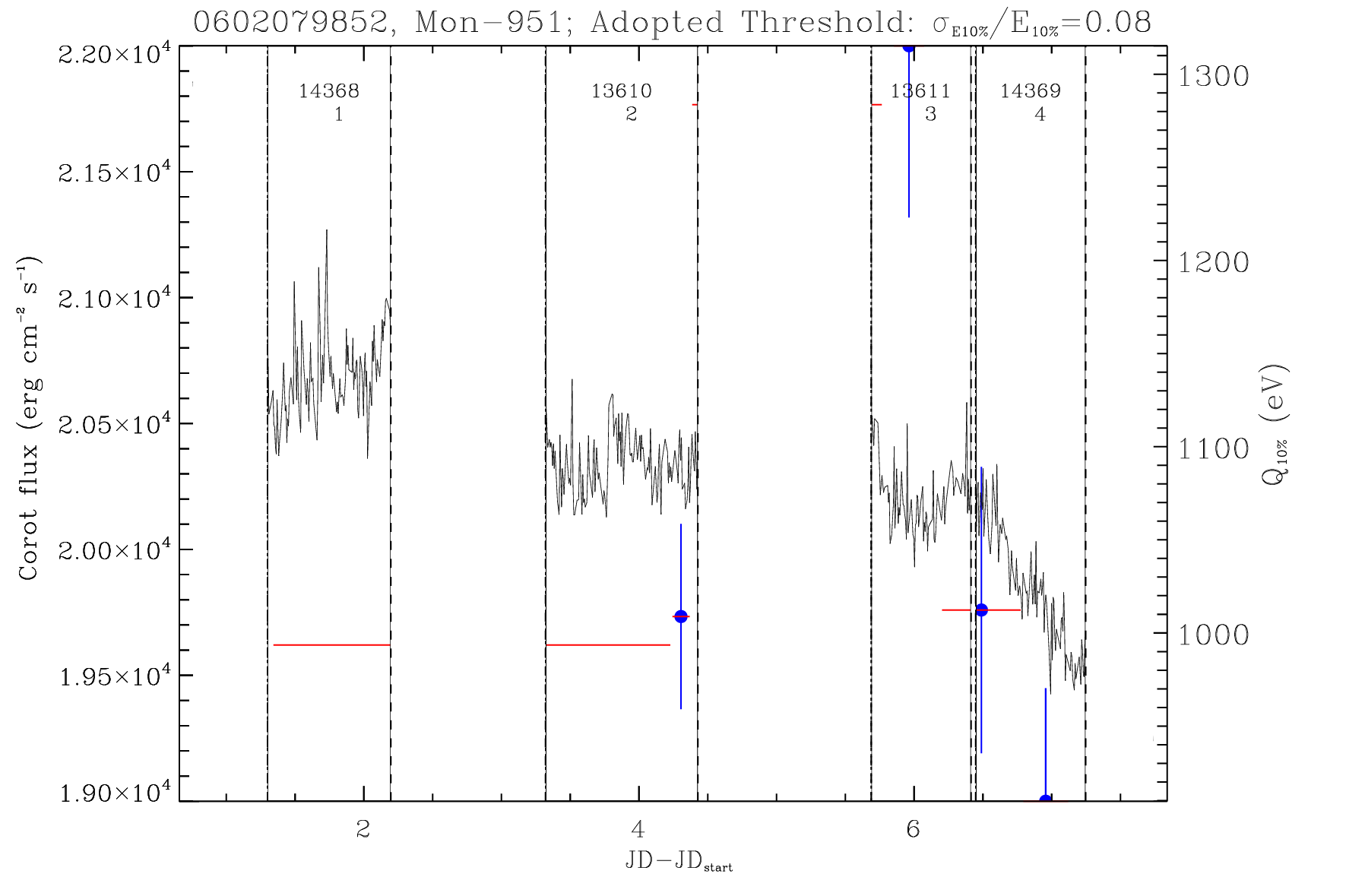}
	\includegraphics[width=8cm]{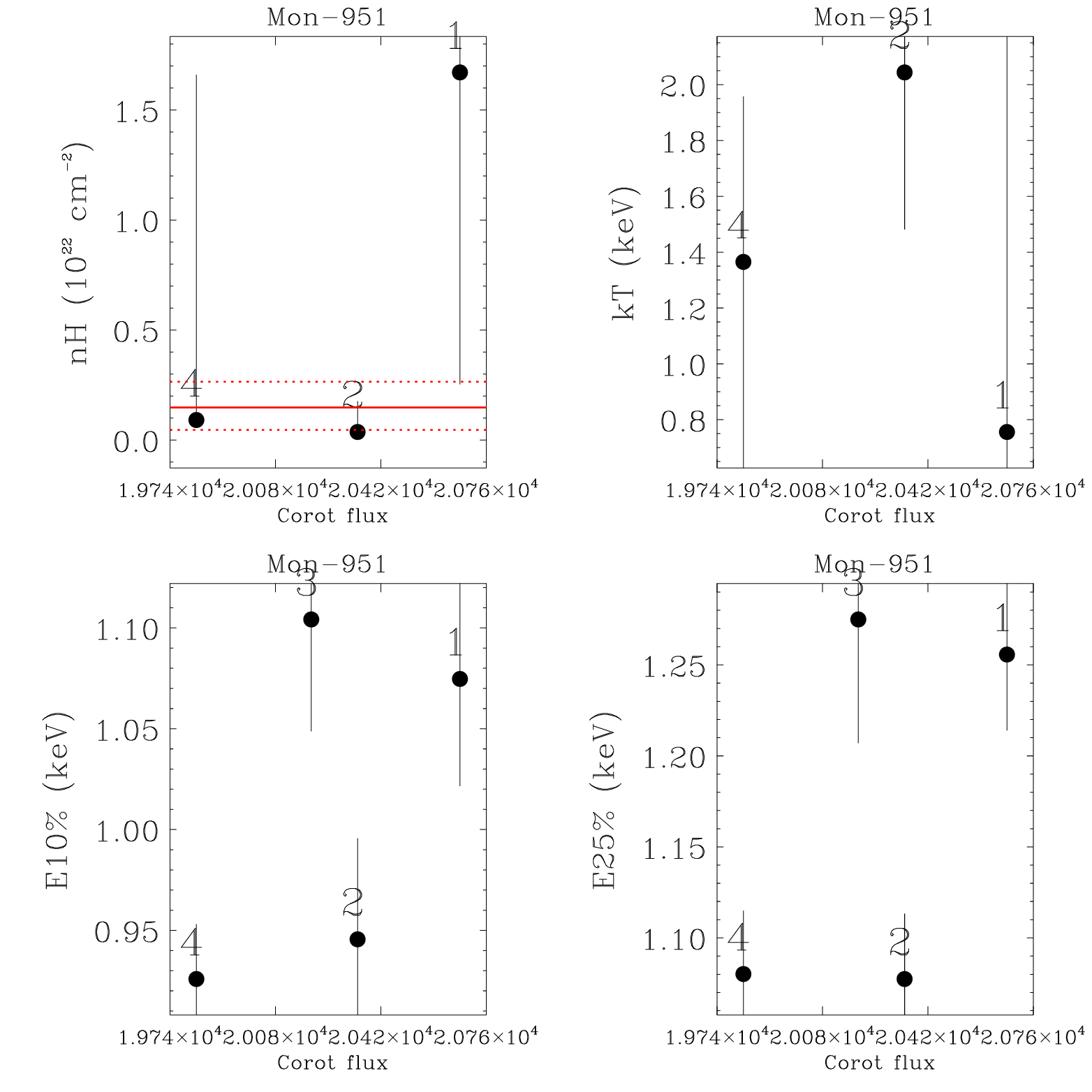}	
	\includegraphics[width=18cm]{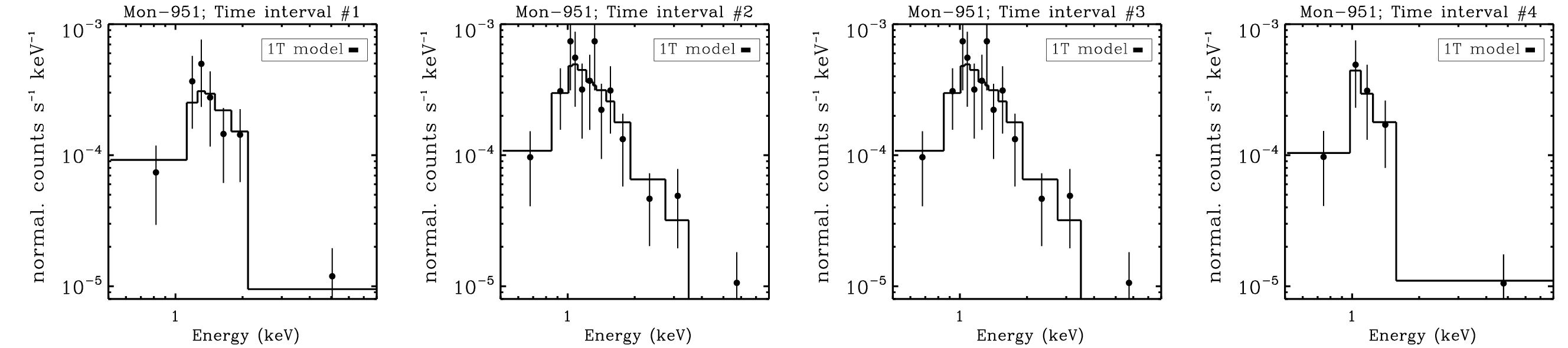}
	\caption{Variability and X-ray spectra of Mon-951, with no interesting variability of the X-ray properties observed and few photons detected.}
	\label{variab_others_21}
	\end{figure}

	\begin{figure}[]
	\centering	
	\includegraphics[width=9.5cm]{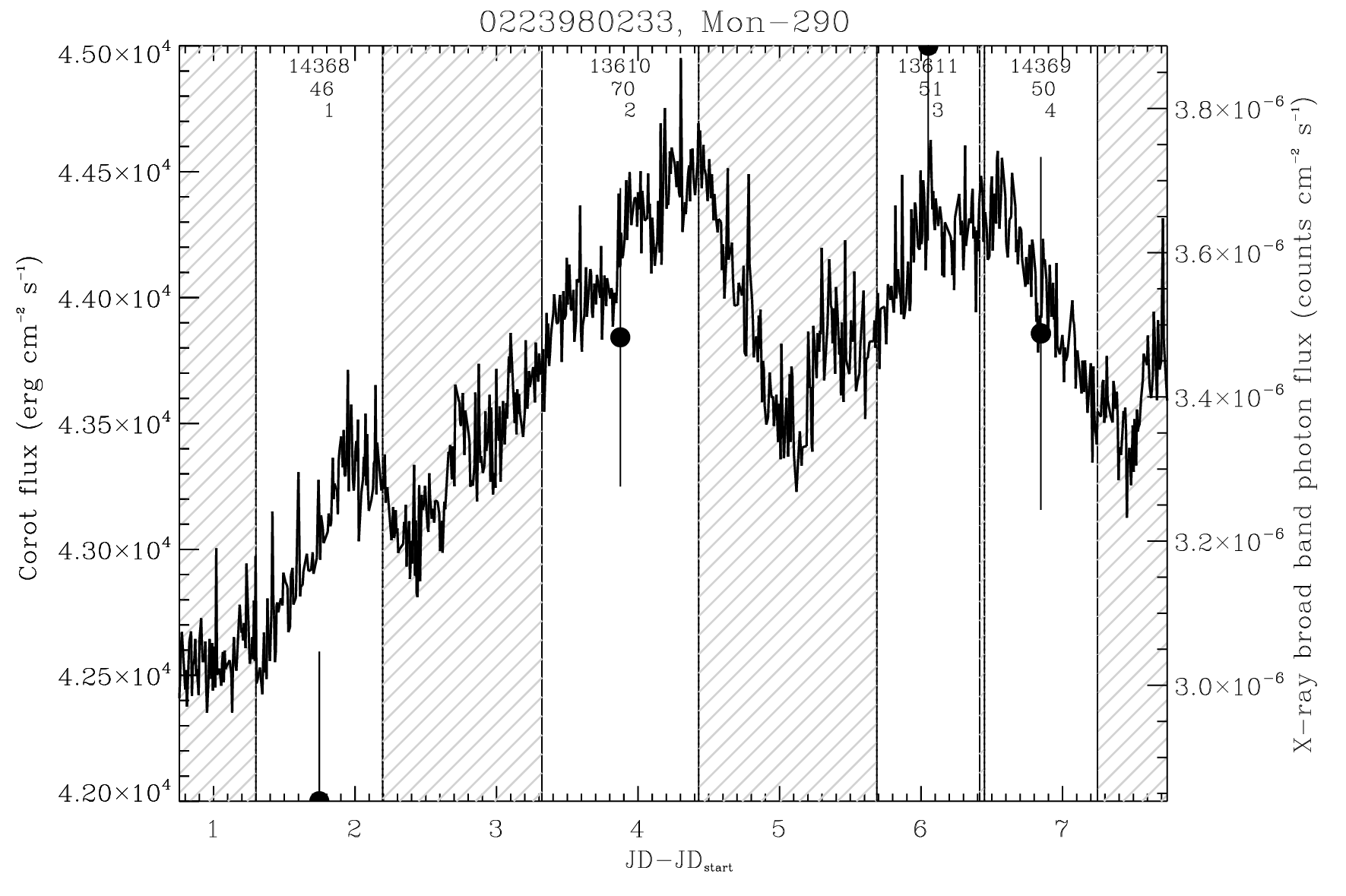}
	\includegraphics[width=9.5cm]{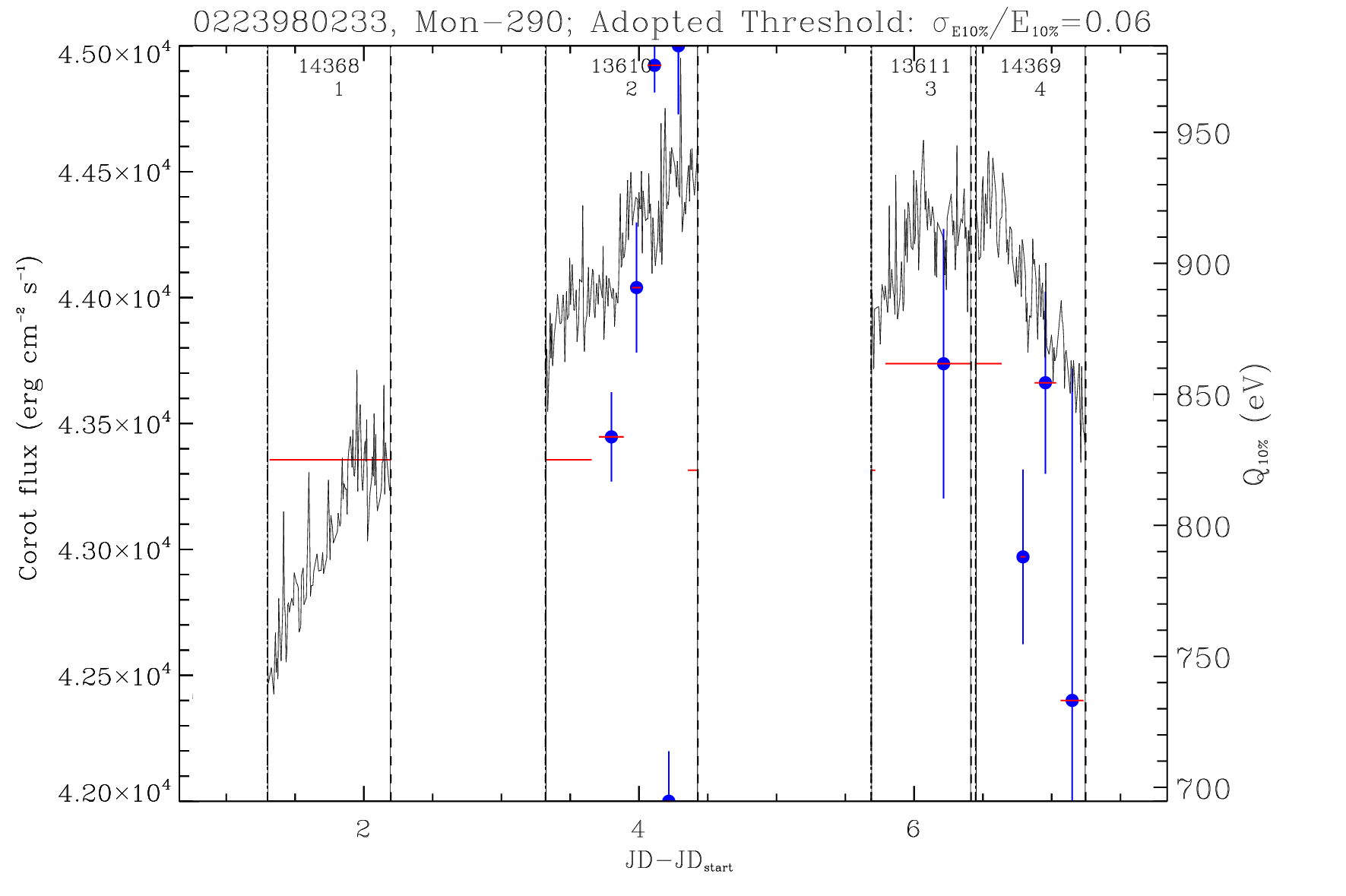}
	\includegraphics[width=8cm]{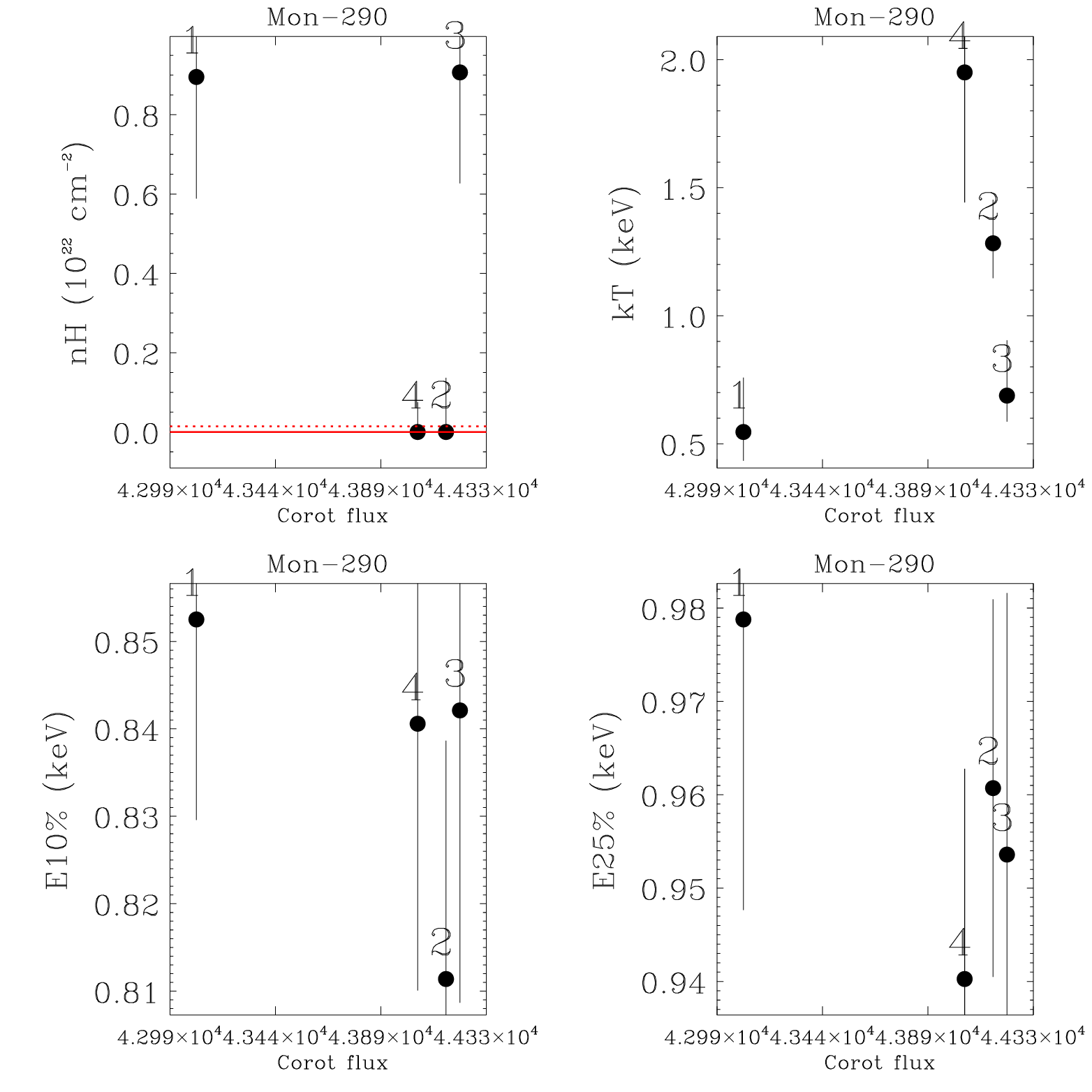}	
	\includegraphics[width=18cm]{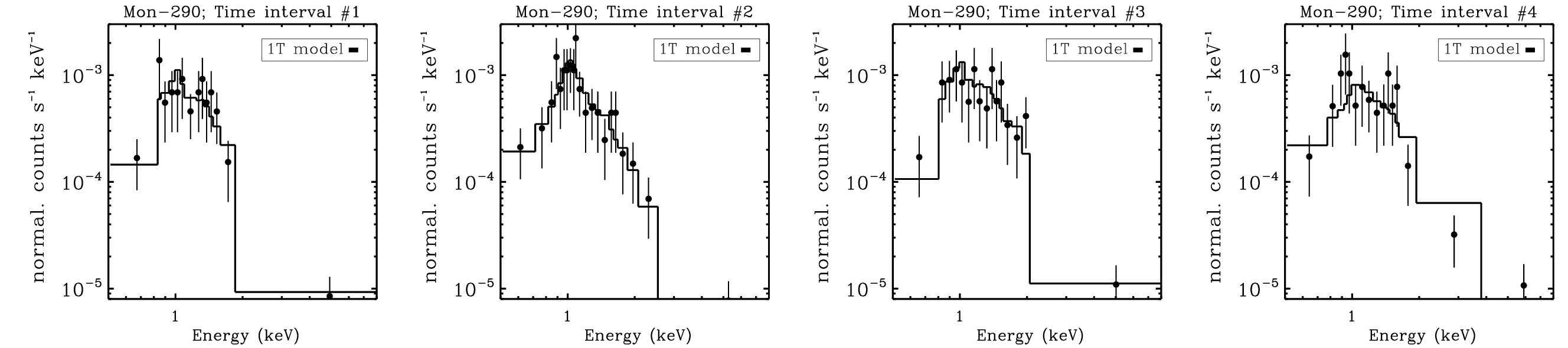}
	\caption{Variability and X-ray spectra of Mon-290, analyzed as a burster, but with no interesting variability of the X-ray properties observed.}
	\label{variab_others_22}
	\end{figure}

	\begin{figure}[]
	\centering	
	\includegraphics[width=9.5cm]{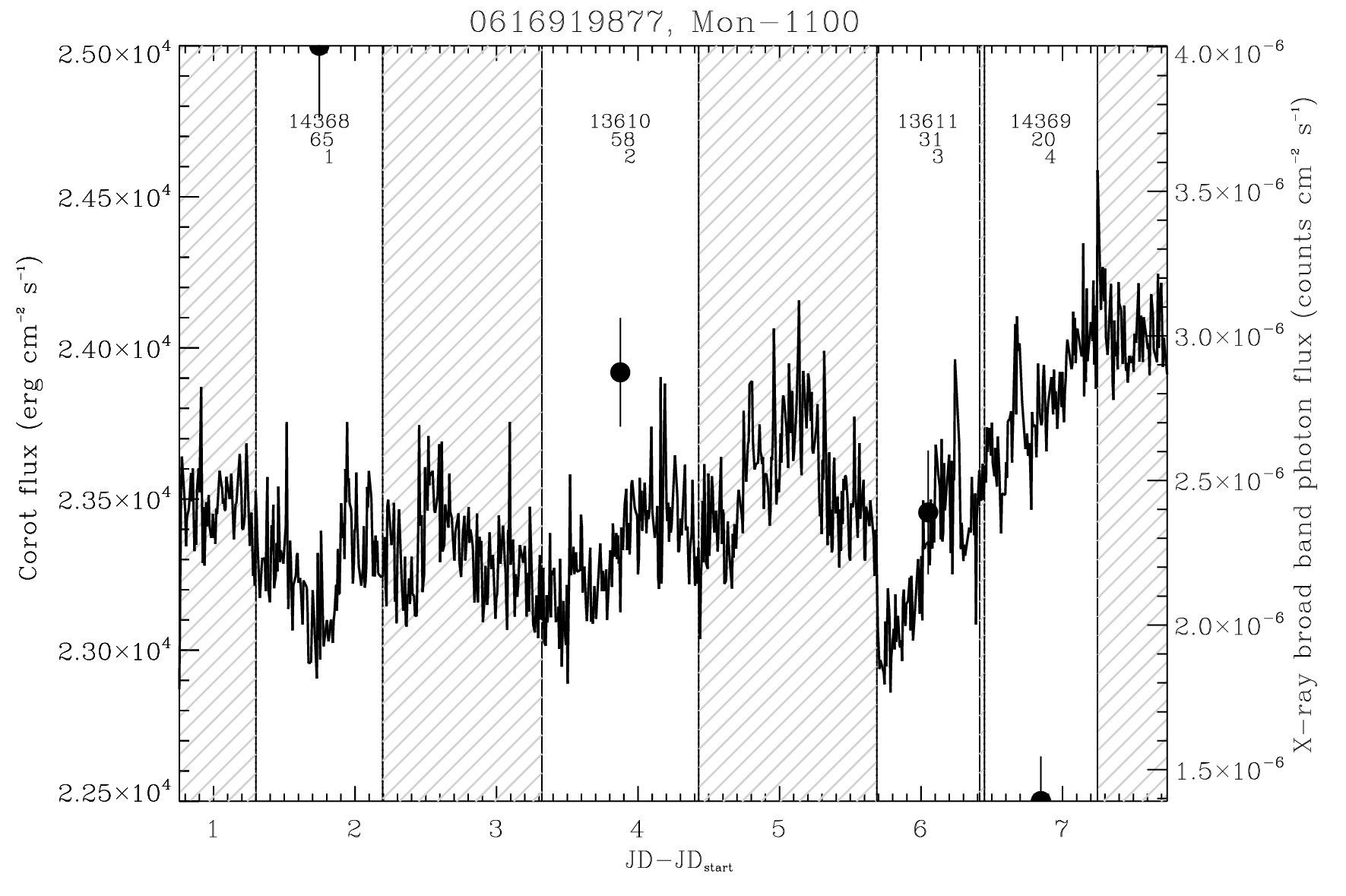}
	\includegraphics[width=9.5cm]{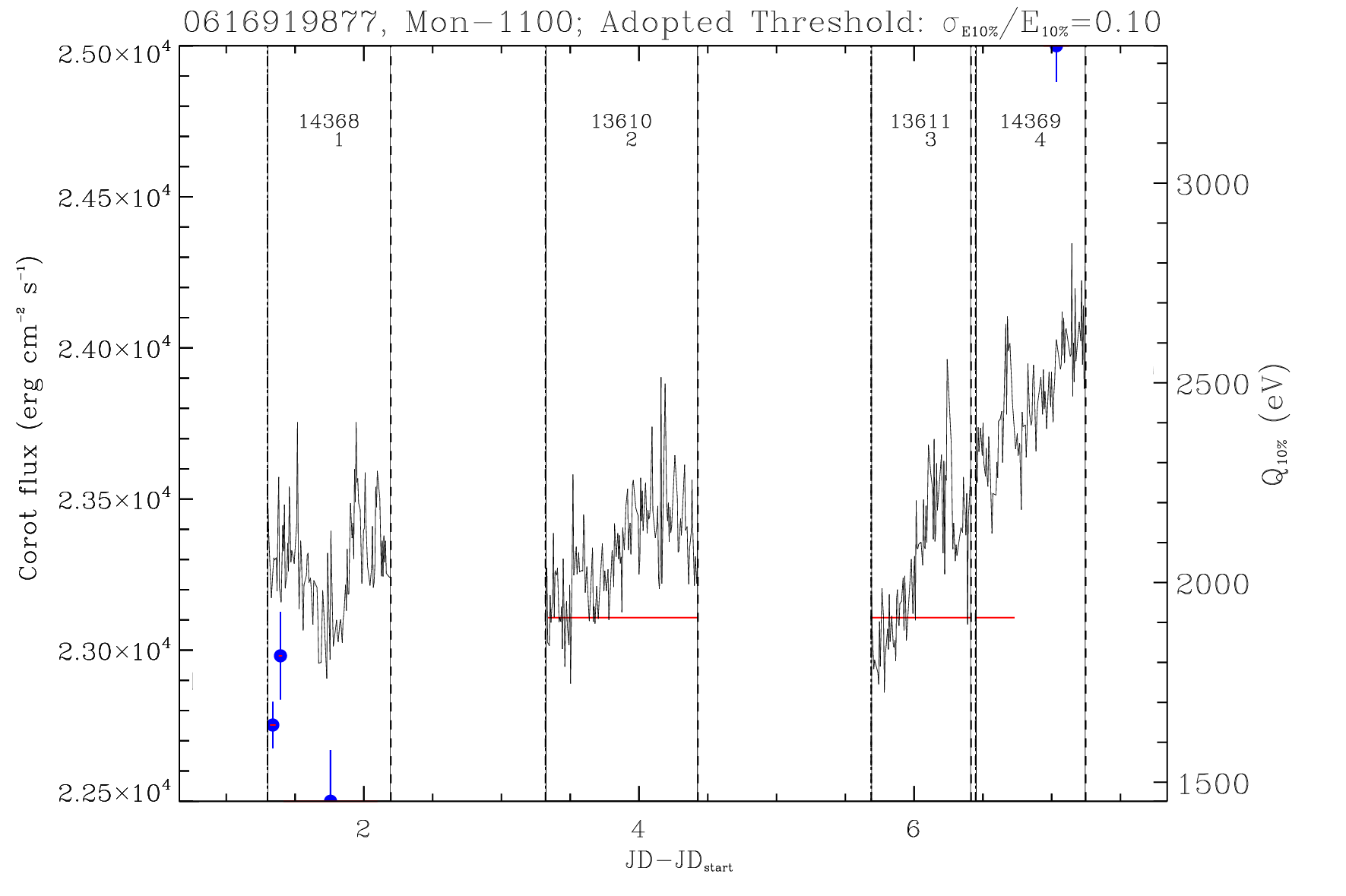}
	\includegraphics[width=8cm]{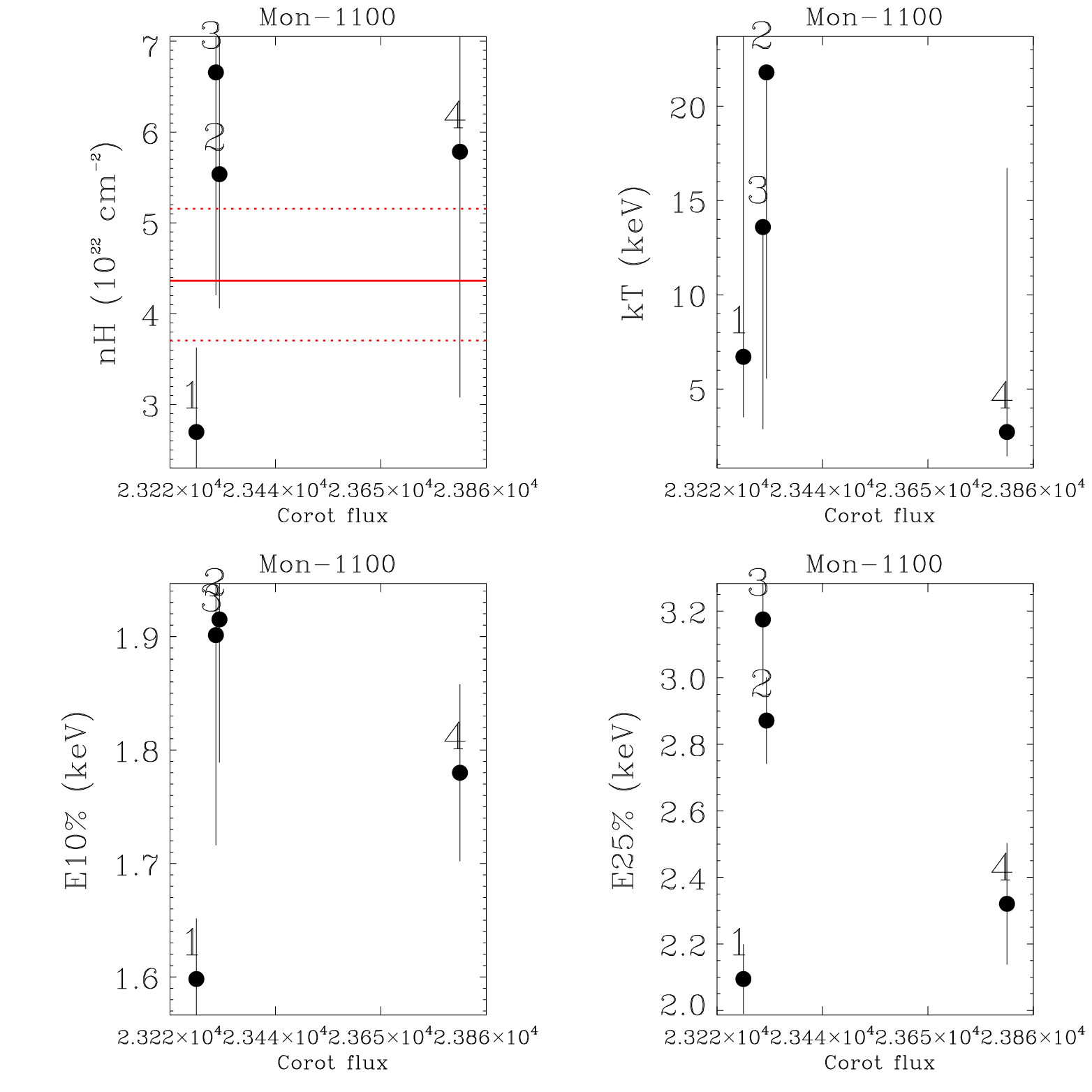}
	\includegraphics[width=18cm]{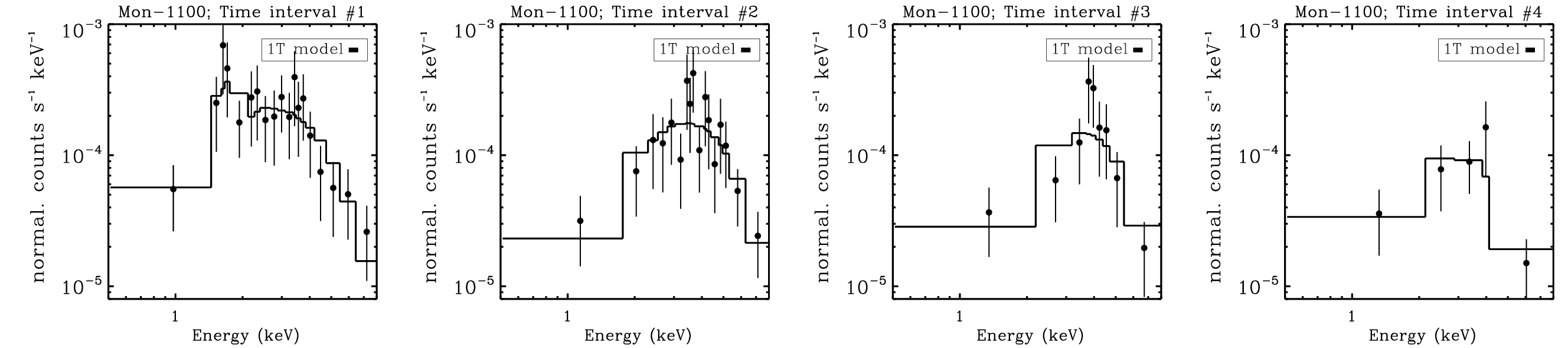}
	\caption{Variability and X-ray spectra of Mon-1100, with no interesting variability of the X-ray properties observed.}
	\label{variab_others_23}
	\end{figure}

	\begin{figure}[]
	\centering	
	\includegraphics[width=9.5cm]{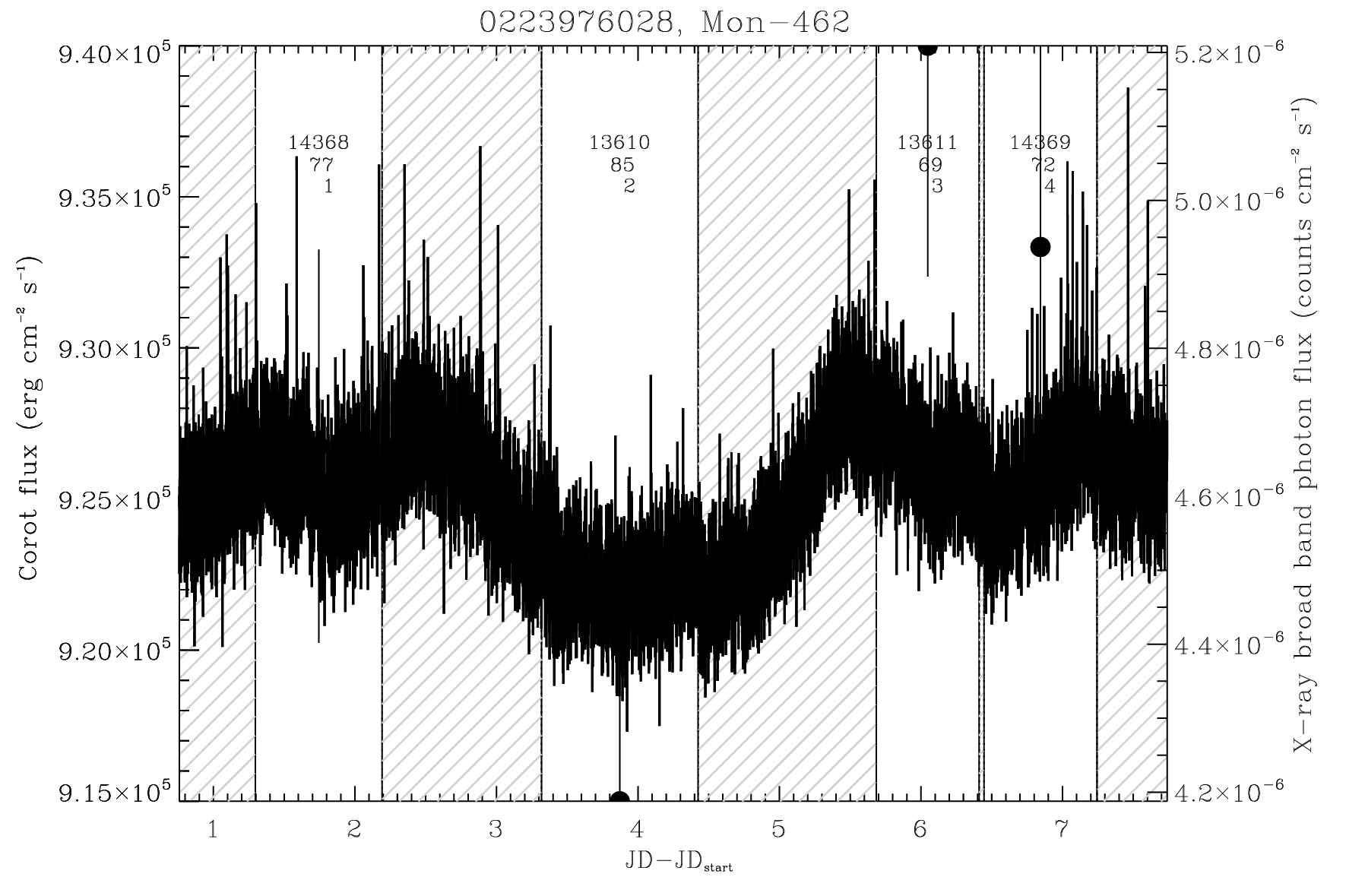}
	\includegraphics[width=9.5cm]{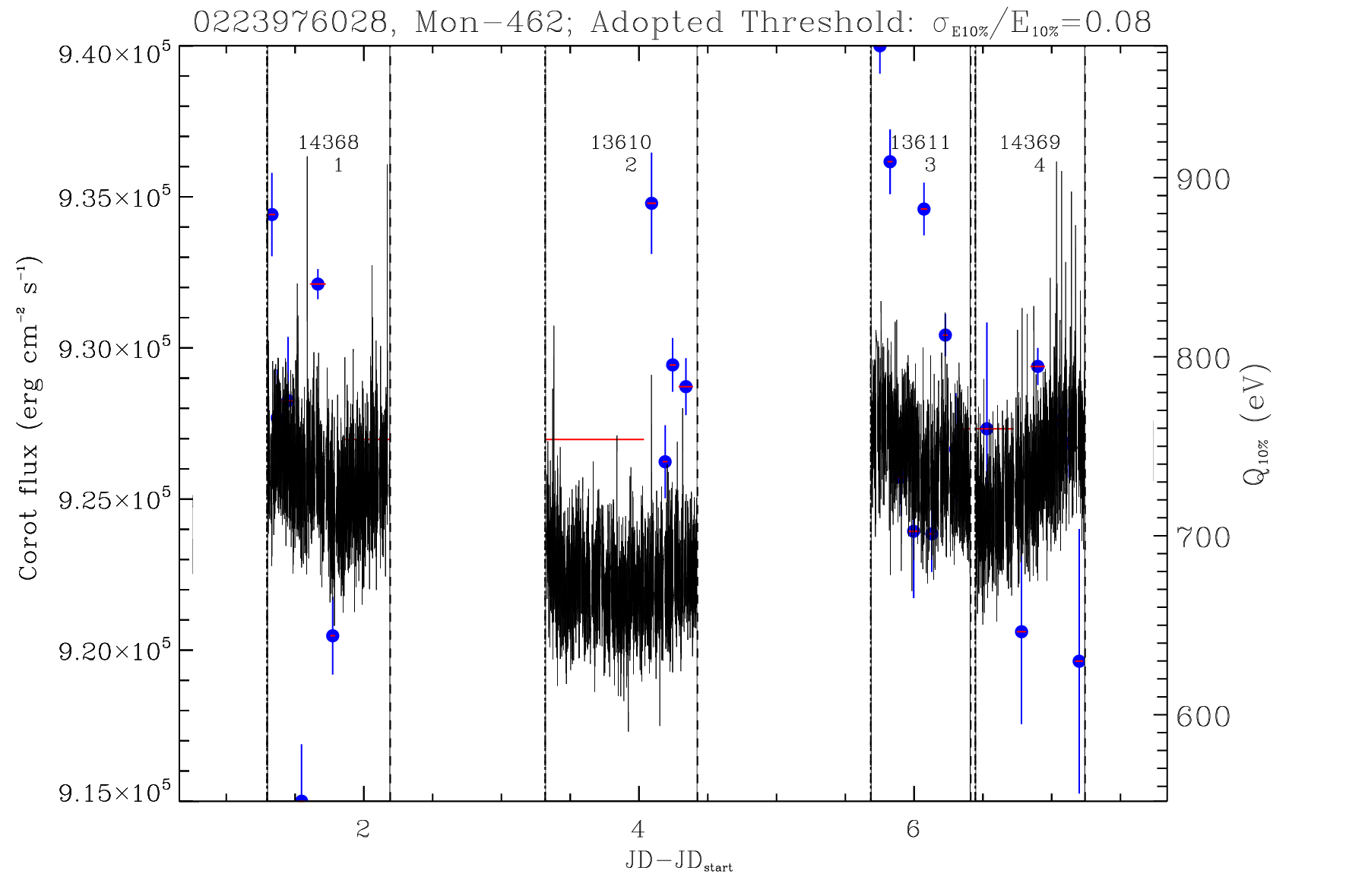}
	\includegraphics[width=8cm]{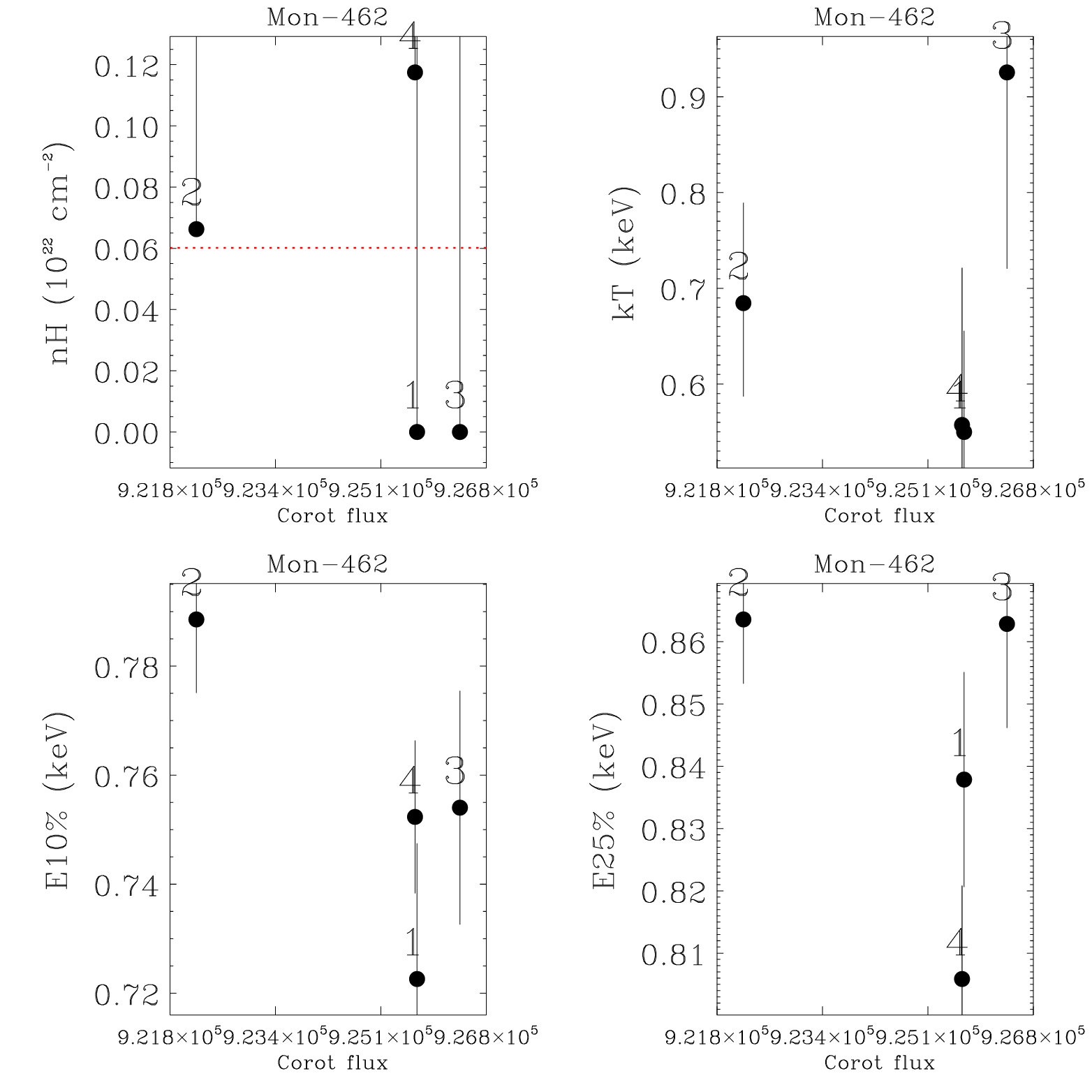}
	\includegraphics[width=18cm]{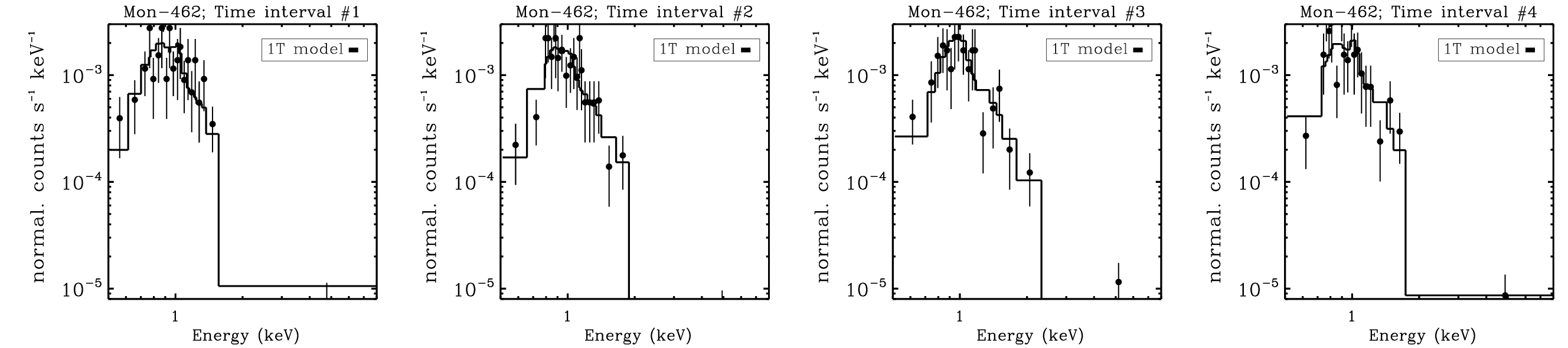}
	\caption{Variability and X-ray spectra of Mon-462, analyzed as a dipper, with a large optical dip observed, which however does not correspond to a significant increase of N$_H$.}
	\label{variab_others_24}
	\end{figure}

	\begin{figure}[]
	\centering	
	\includegraphics[width=9.5cm]{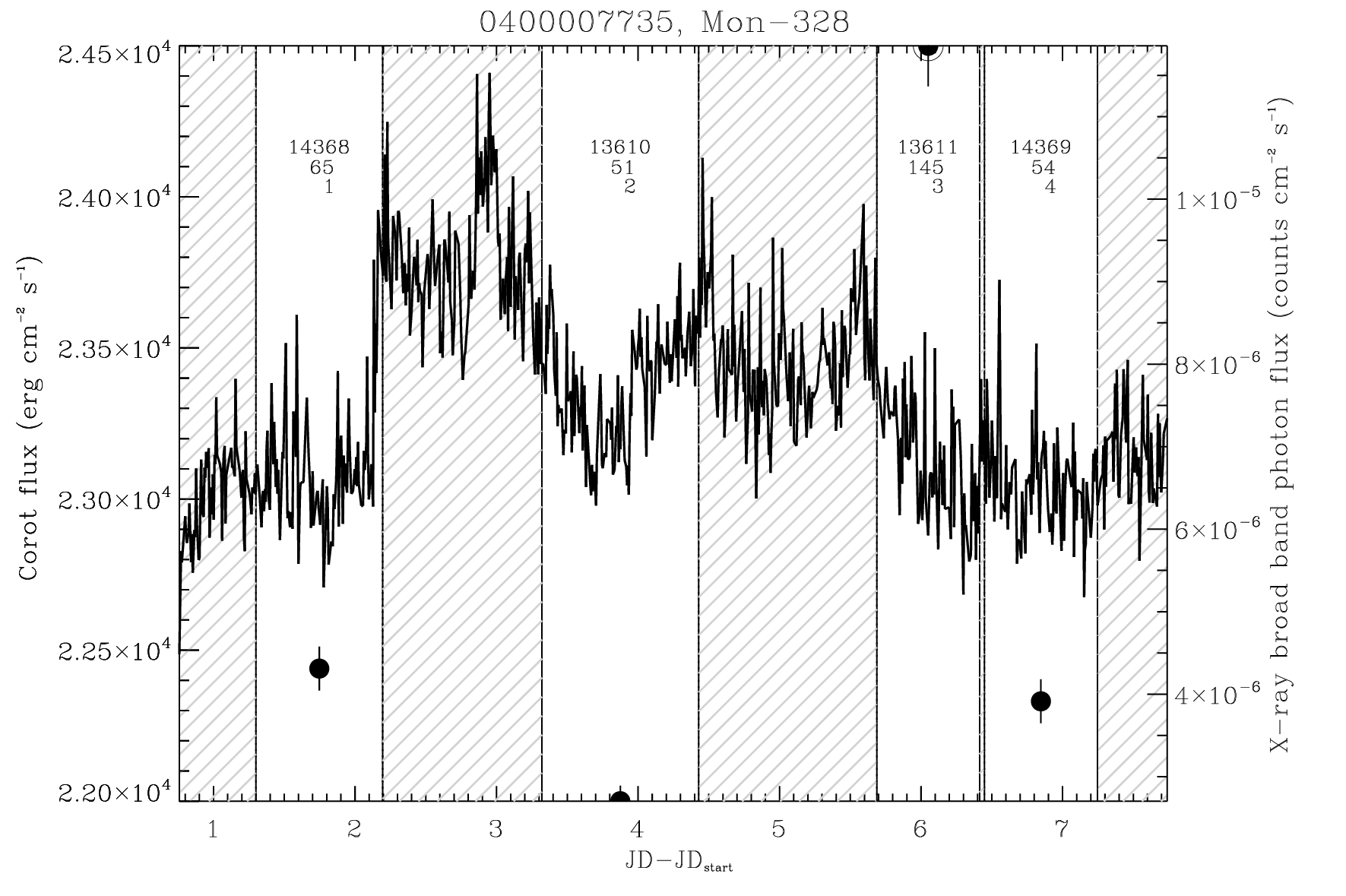}
	\includegraphics[width=9.5cm]{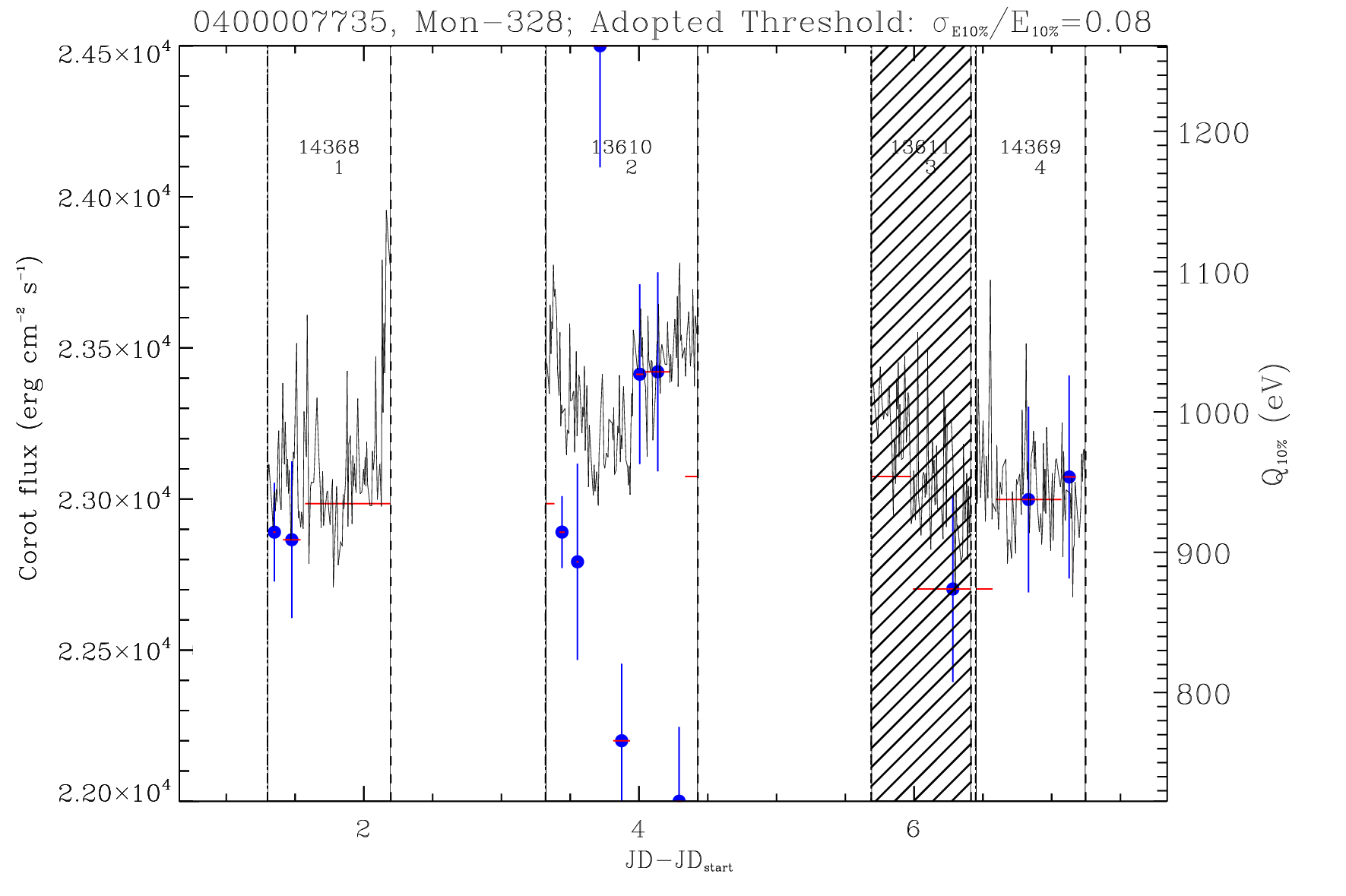}
	\includegraphics[width=8cm]{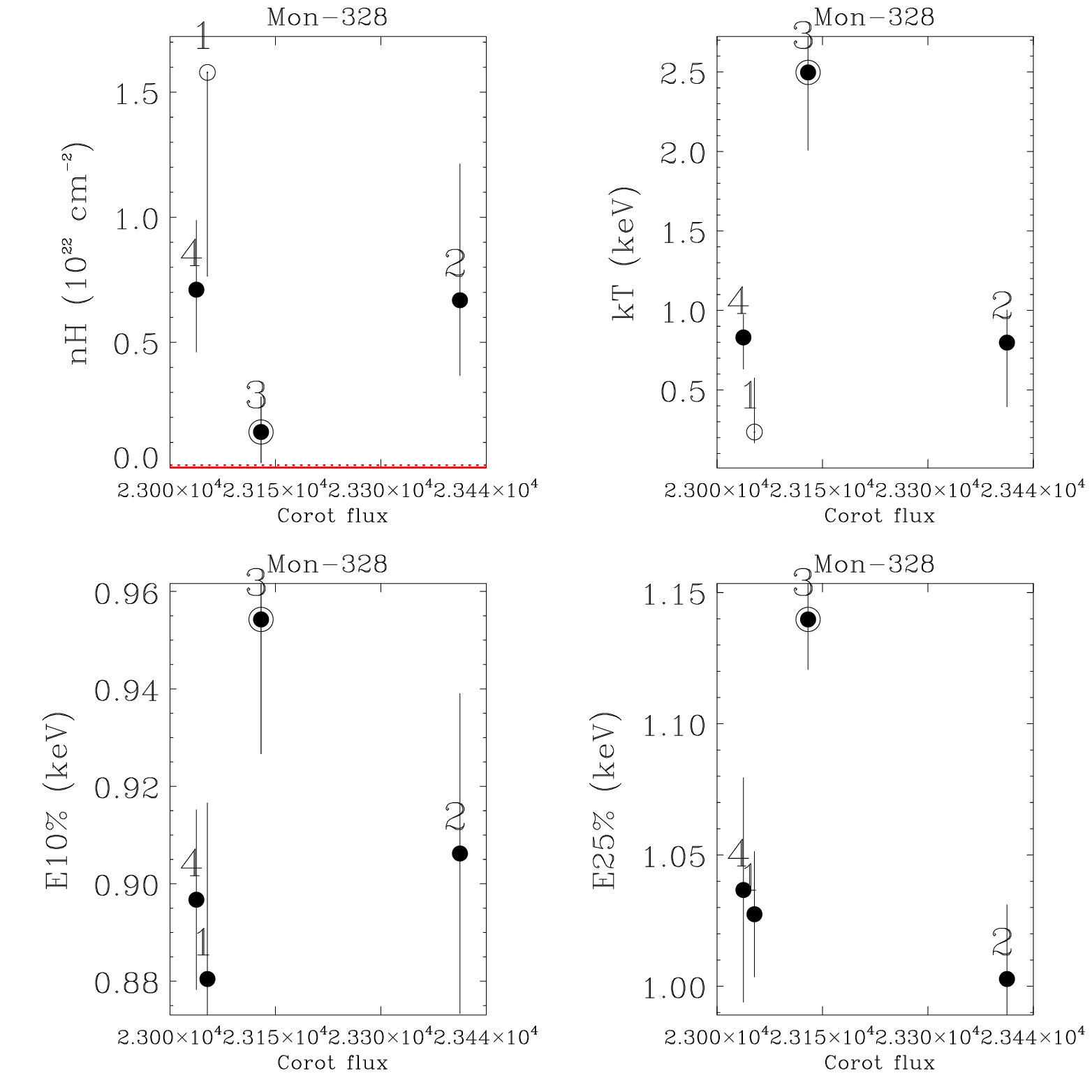}
	\includegraphics[width=18cm]{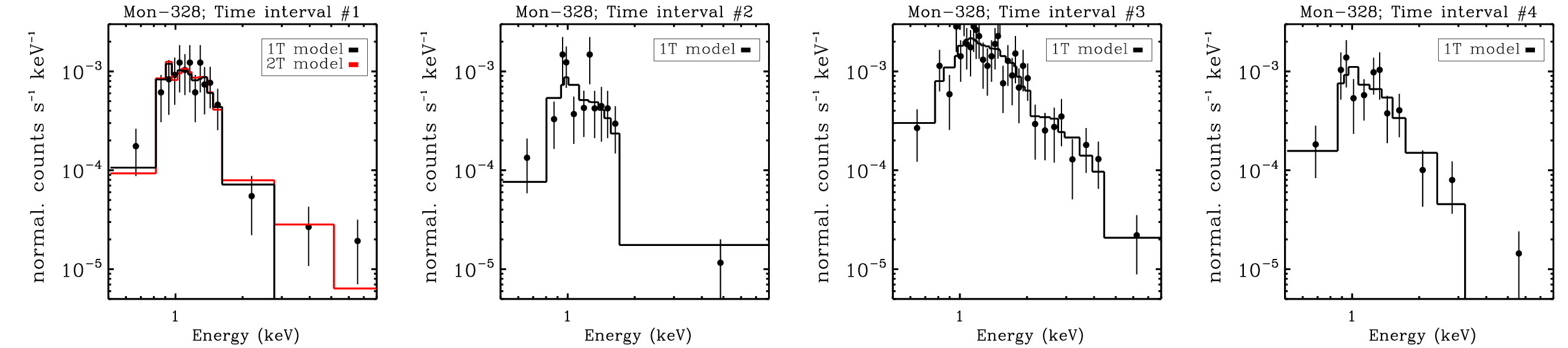}
	\caption{Variability and X-ray spectra of Mon-328, with no evident optical features observed during the {\em Chandra} frames.}
	\label{variab_others_25}
	\end{figure}

	\begin{figure}[]
	\centering	
	\includegraphics[width=9.5cm]{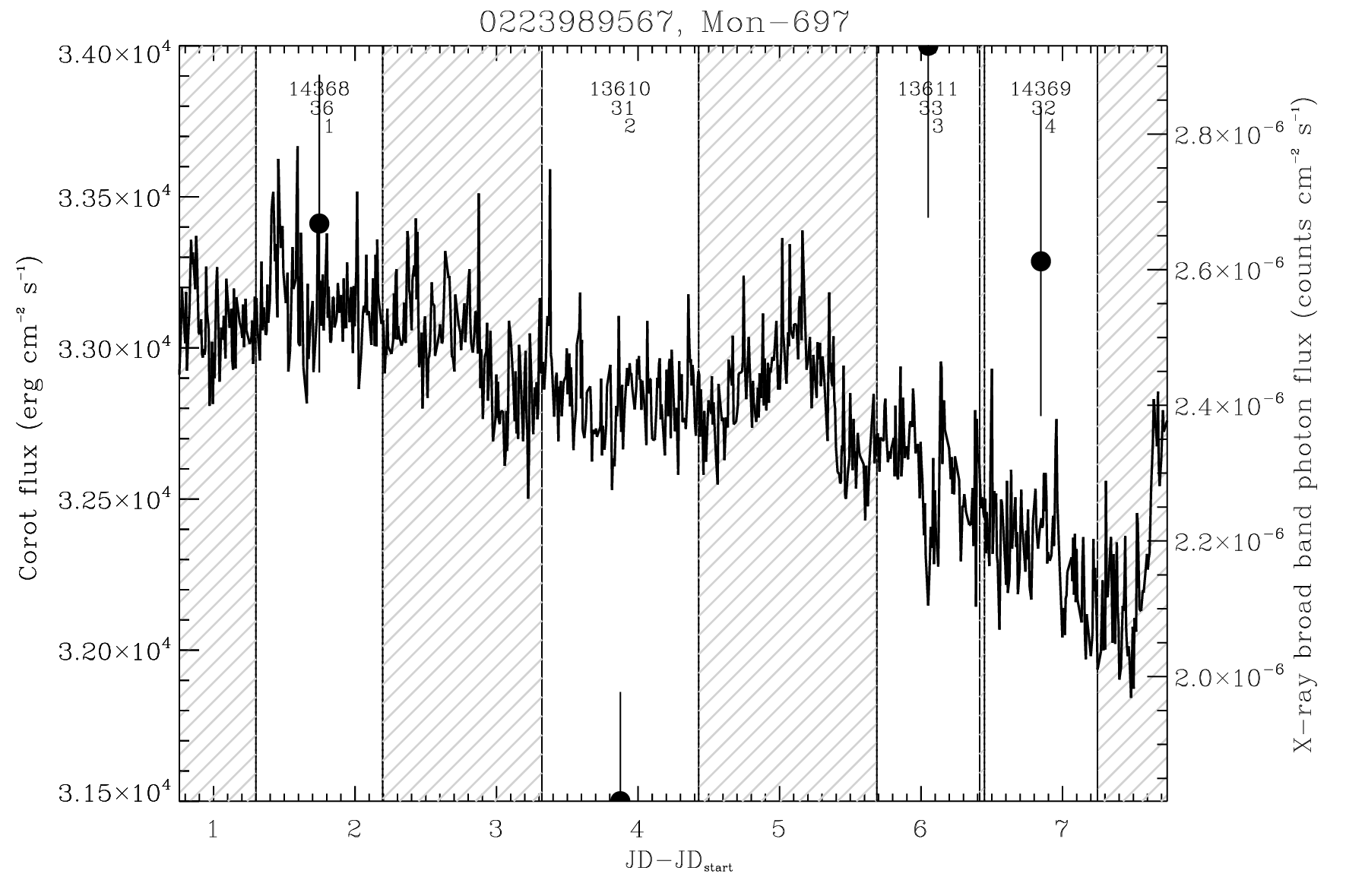}
	\includegraphics[width=9.5cm]{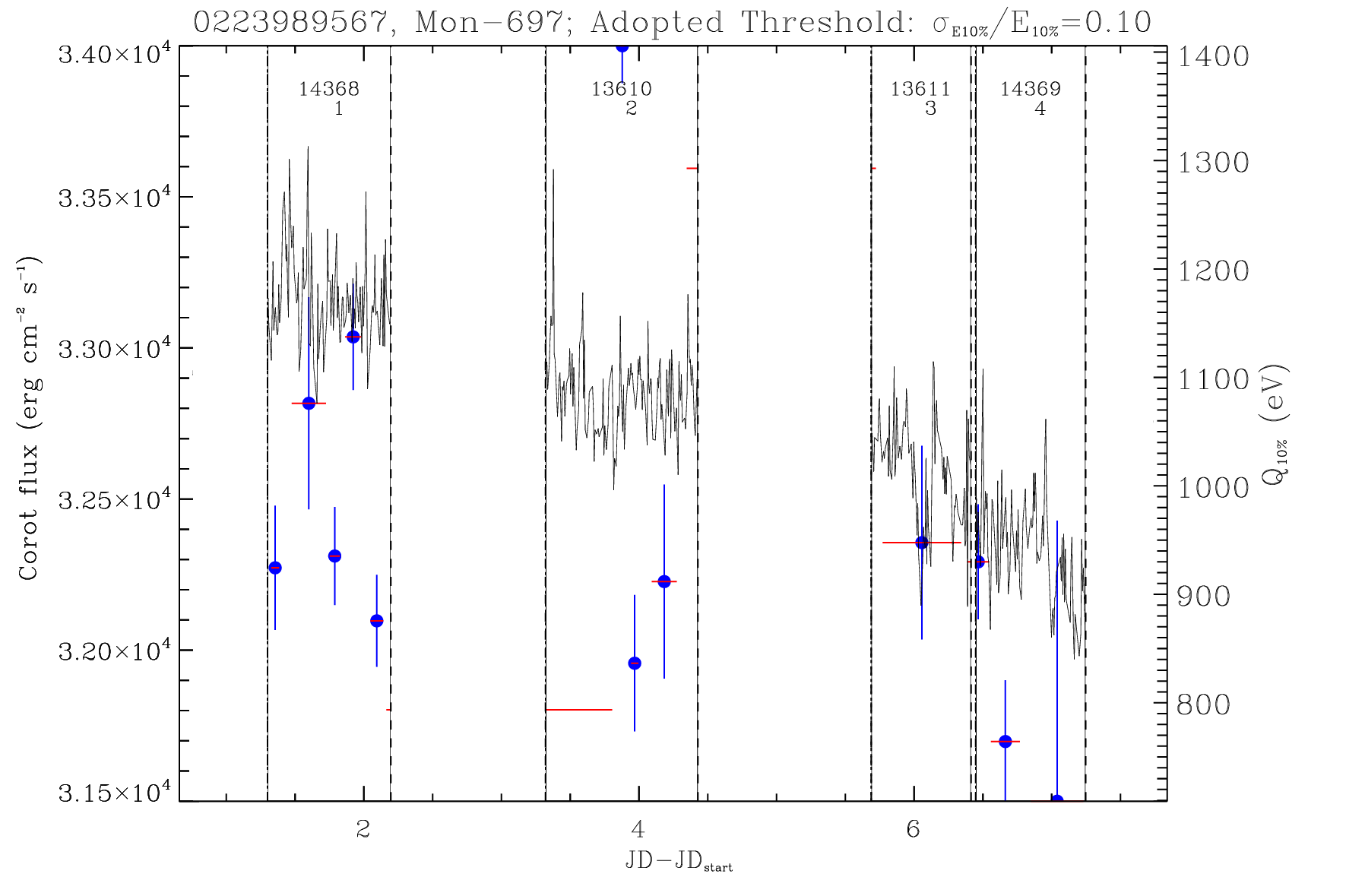}
	\includegraphics[width=8cm]{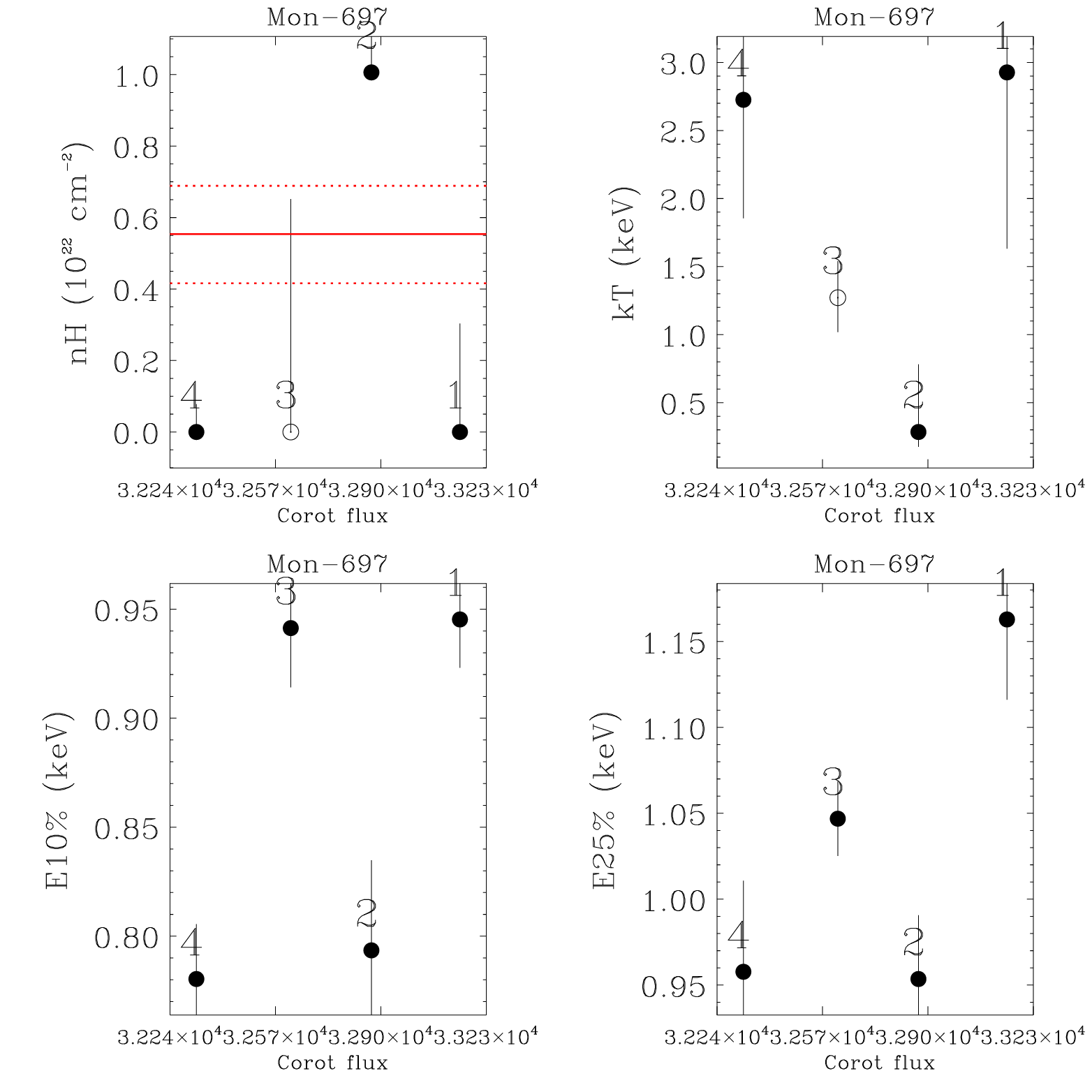}	
	\includegraphics[width=18cm]{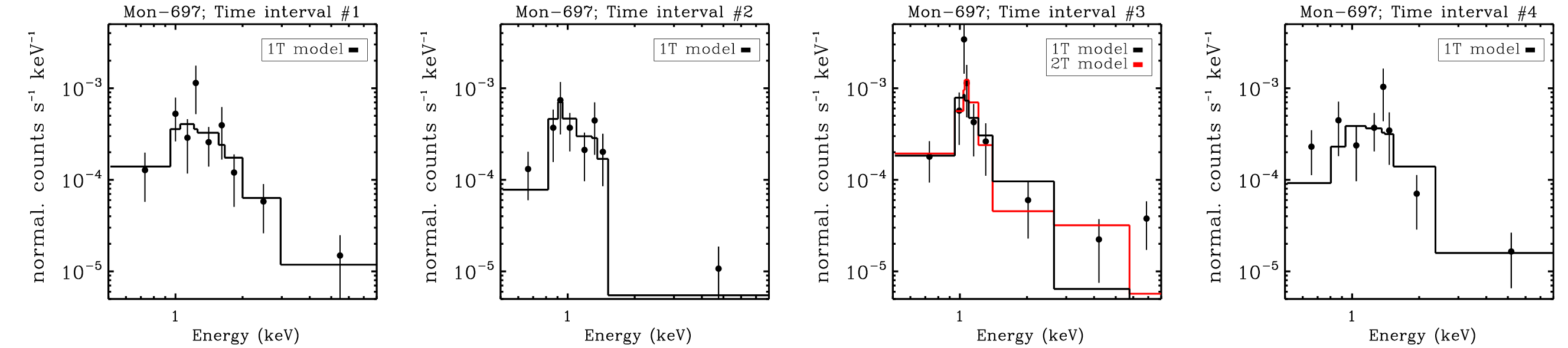}
	\caption{Variability and X-ray spectra of Mon-697, with no evident optical features observed during the {\em Chandra} frames and few X-ray photons detected.}
	\label{variab_others_26}
	\end{figure}

	\begin{figure}[]
	\centering	
	\includegraphics[width=9.5cm]{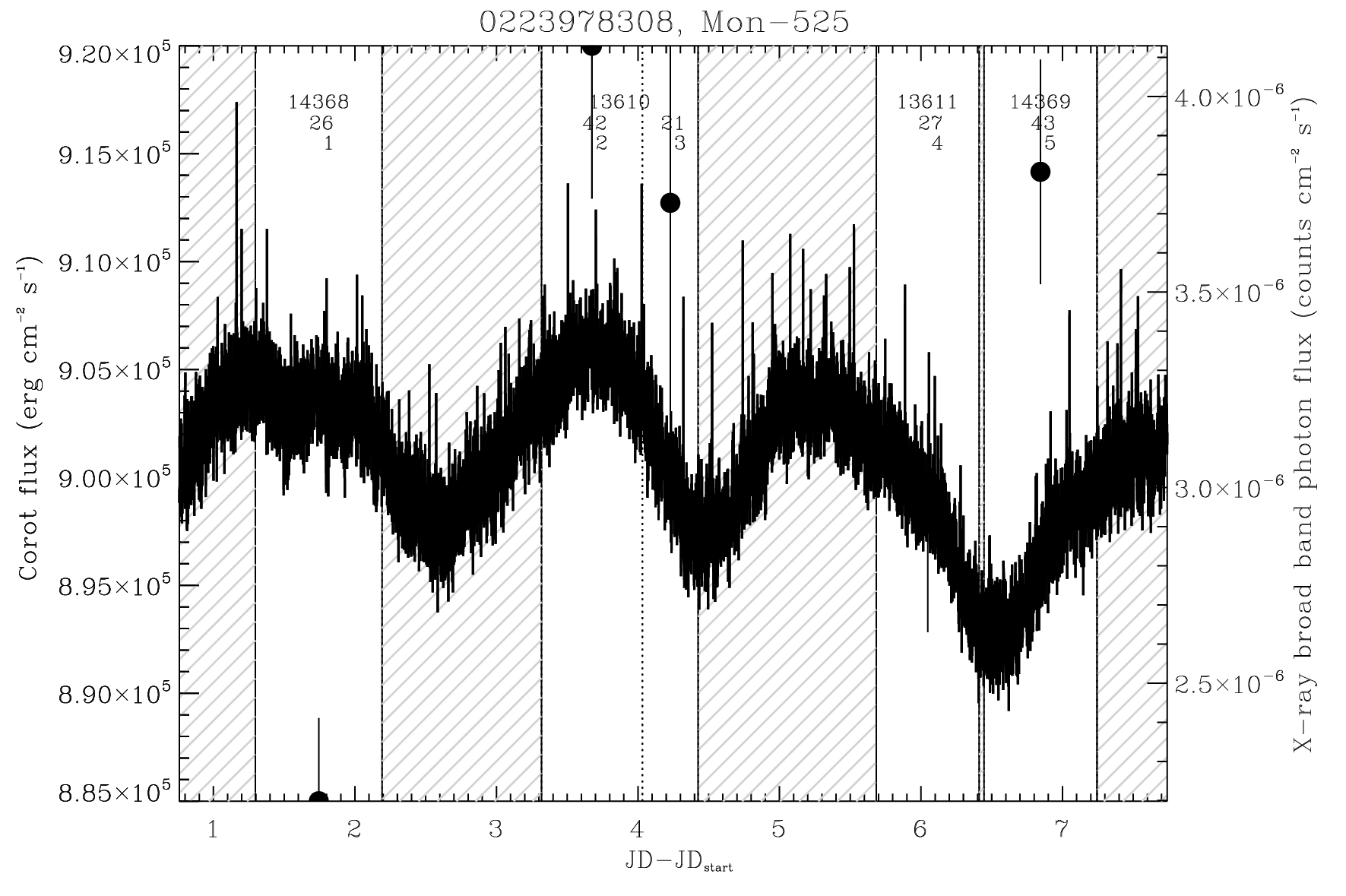}
	\includegraphics[width=9.5cm]{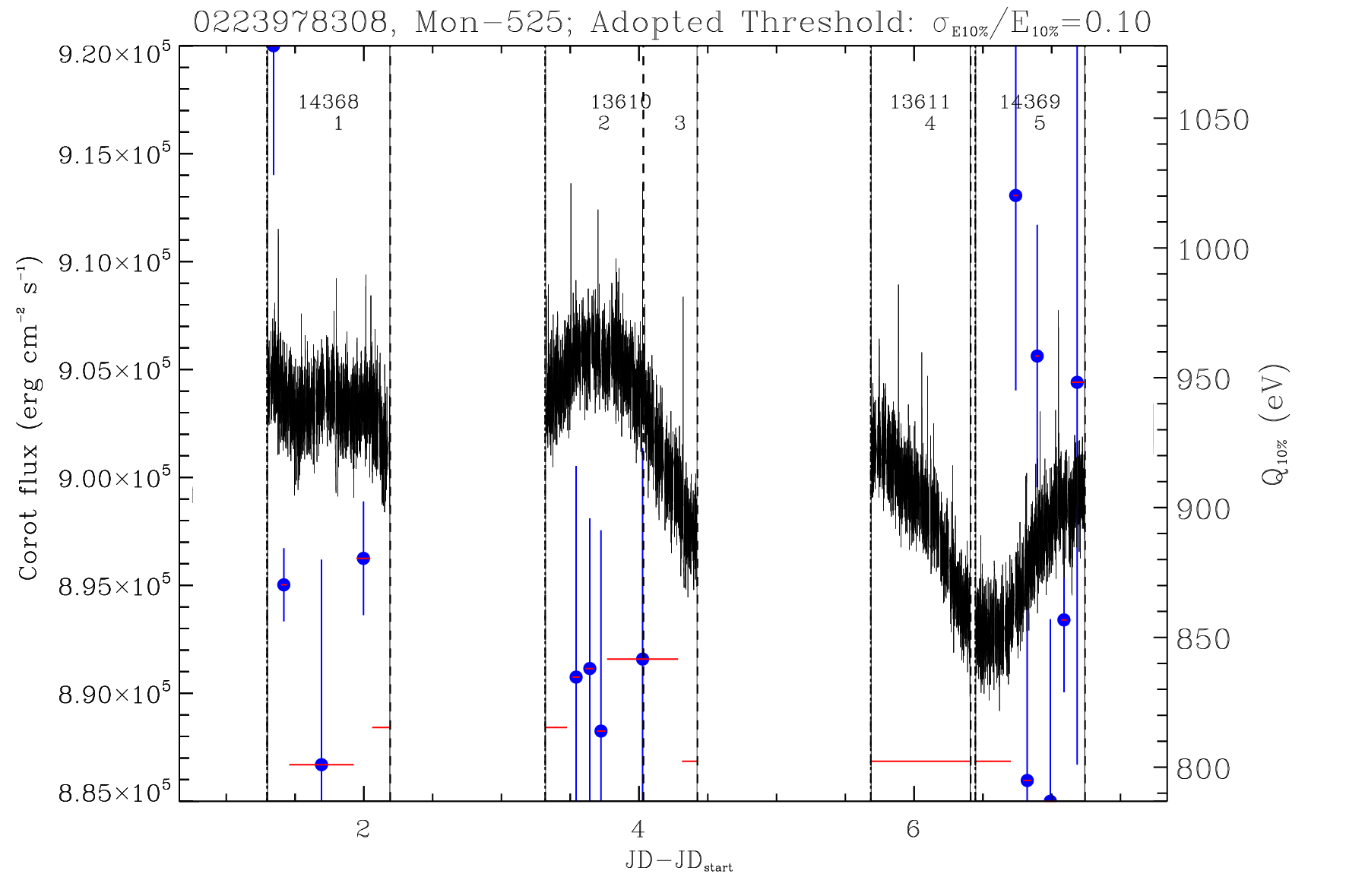}
	\includegraphics[width=8cm]{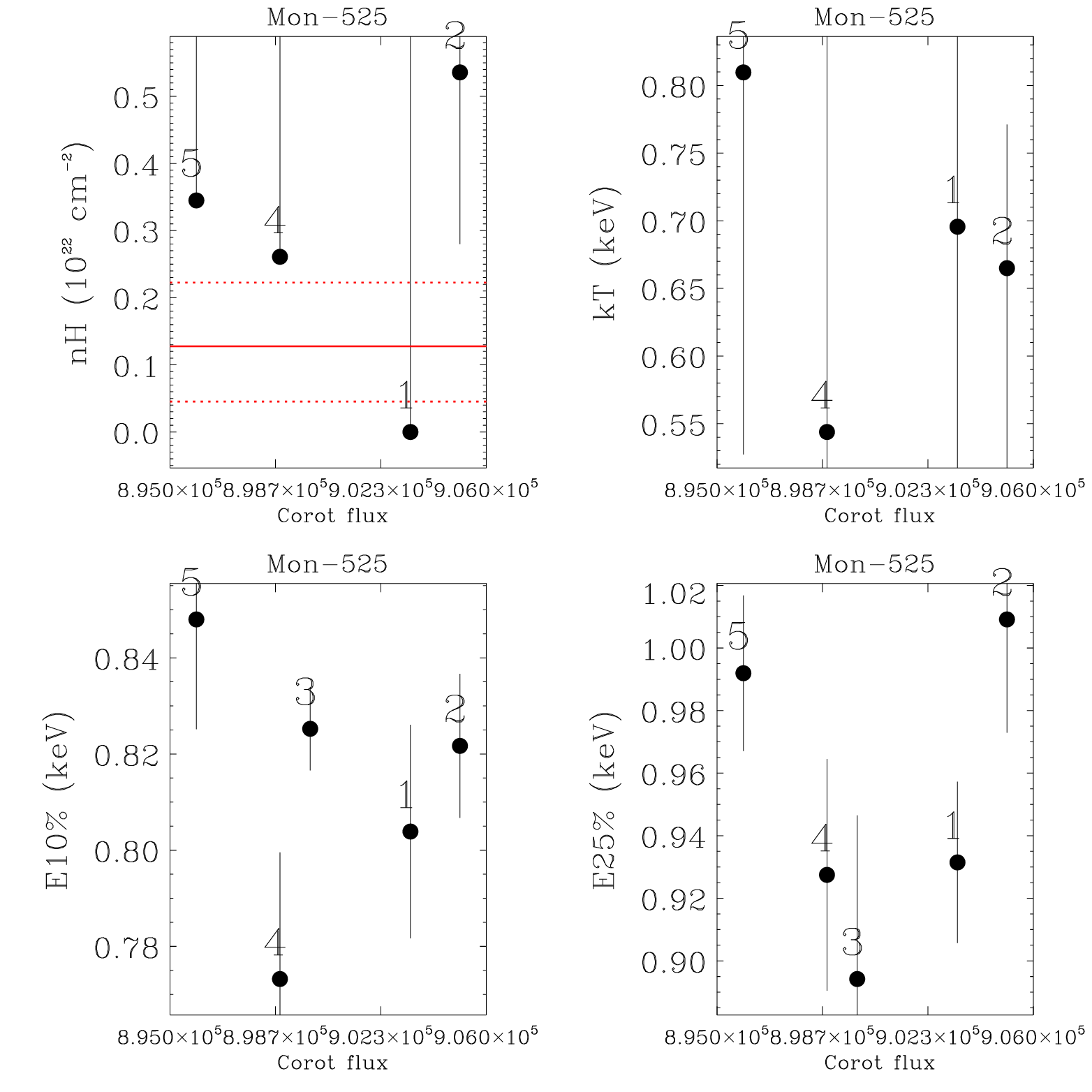}
	\includegraphics[width=18cm]{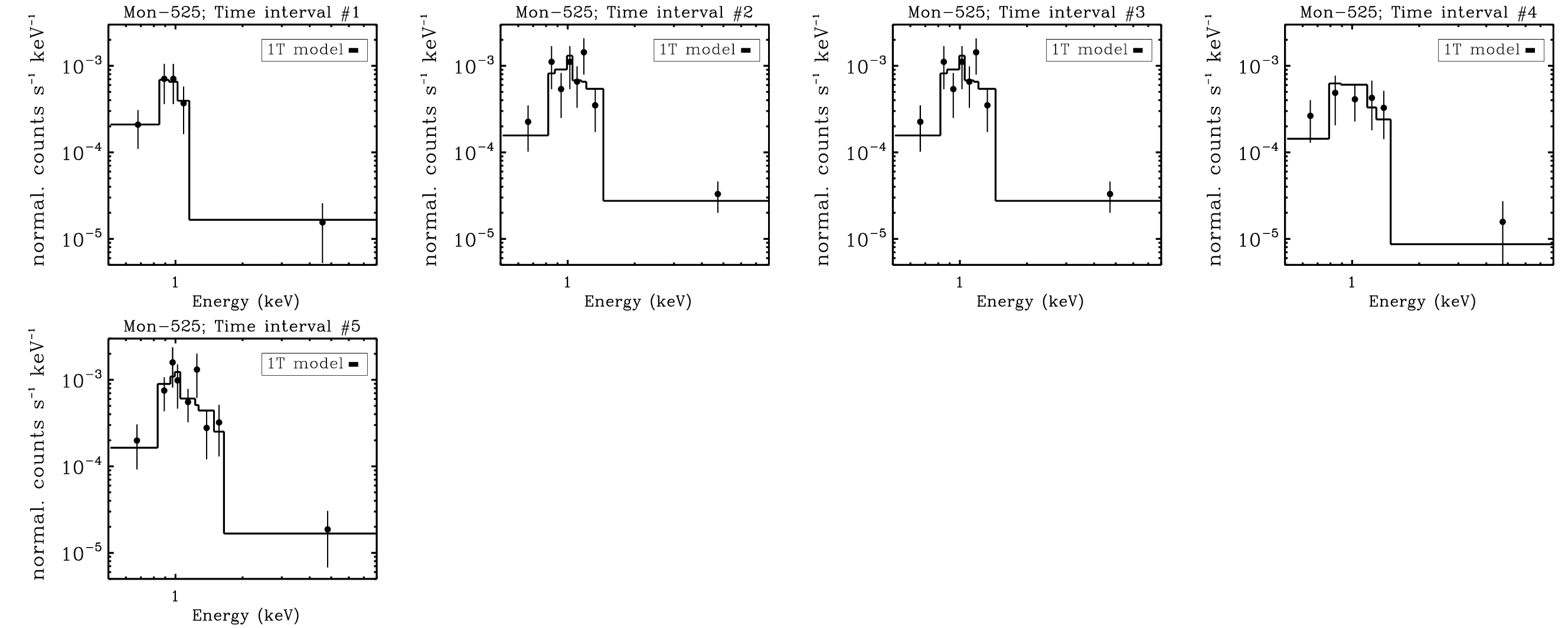}
	\caption{Variability and X-ray spectra of Mon-525, analyzed as a dipper. The CoRoT light curve shows dips that do not correspond to significant variations of N$_H$.}
	\label{variab_others_27}
	\end{figure}

	\begin{figure}[]
	\centering	
	\includegraphics[width=9.5cm]{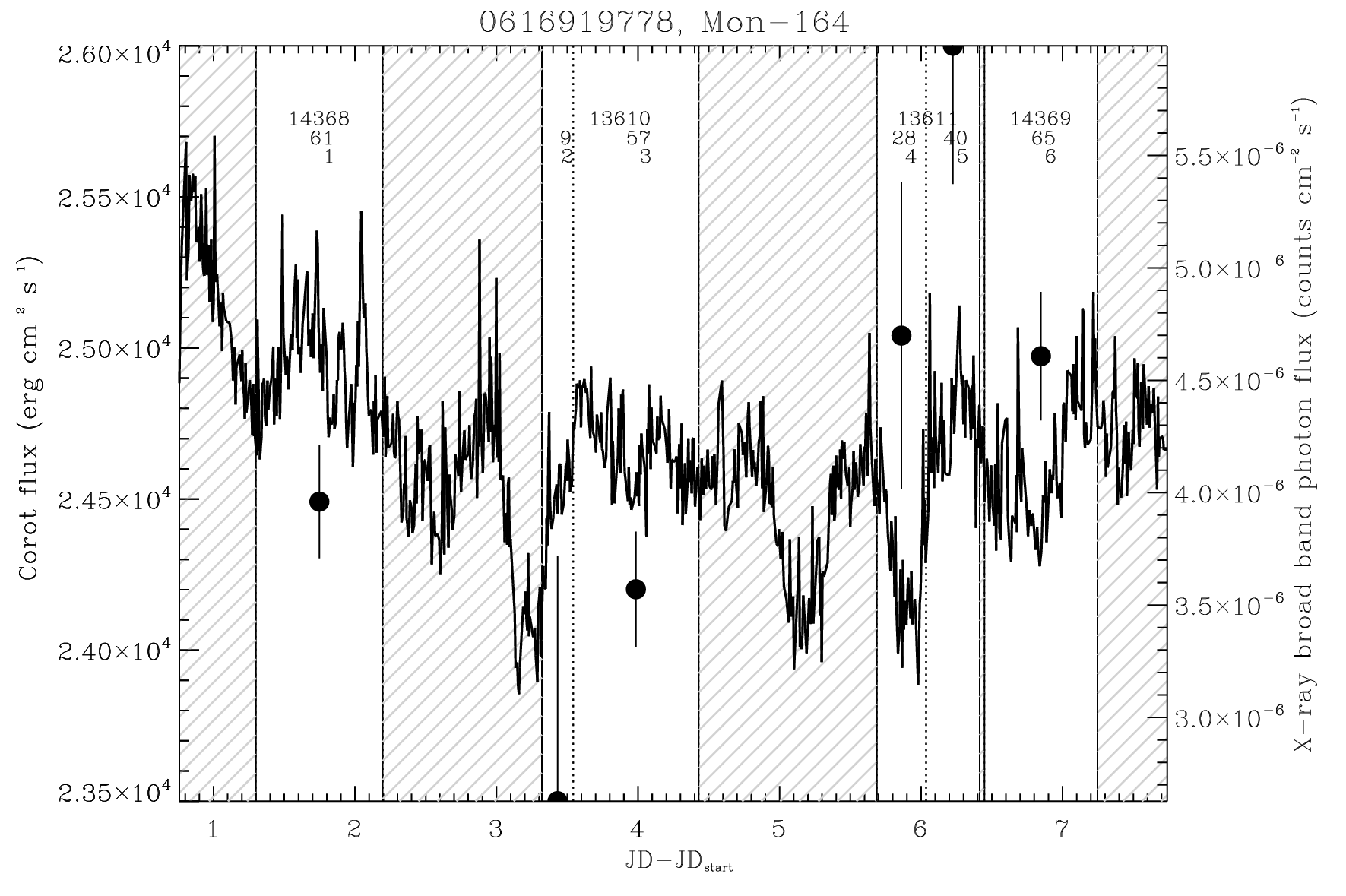}
	\includegraphics[width=9.5cm]{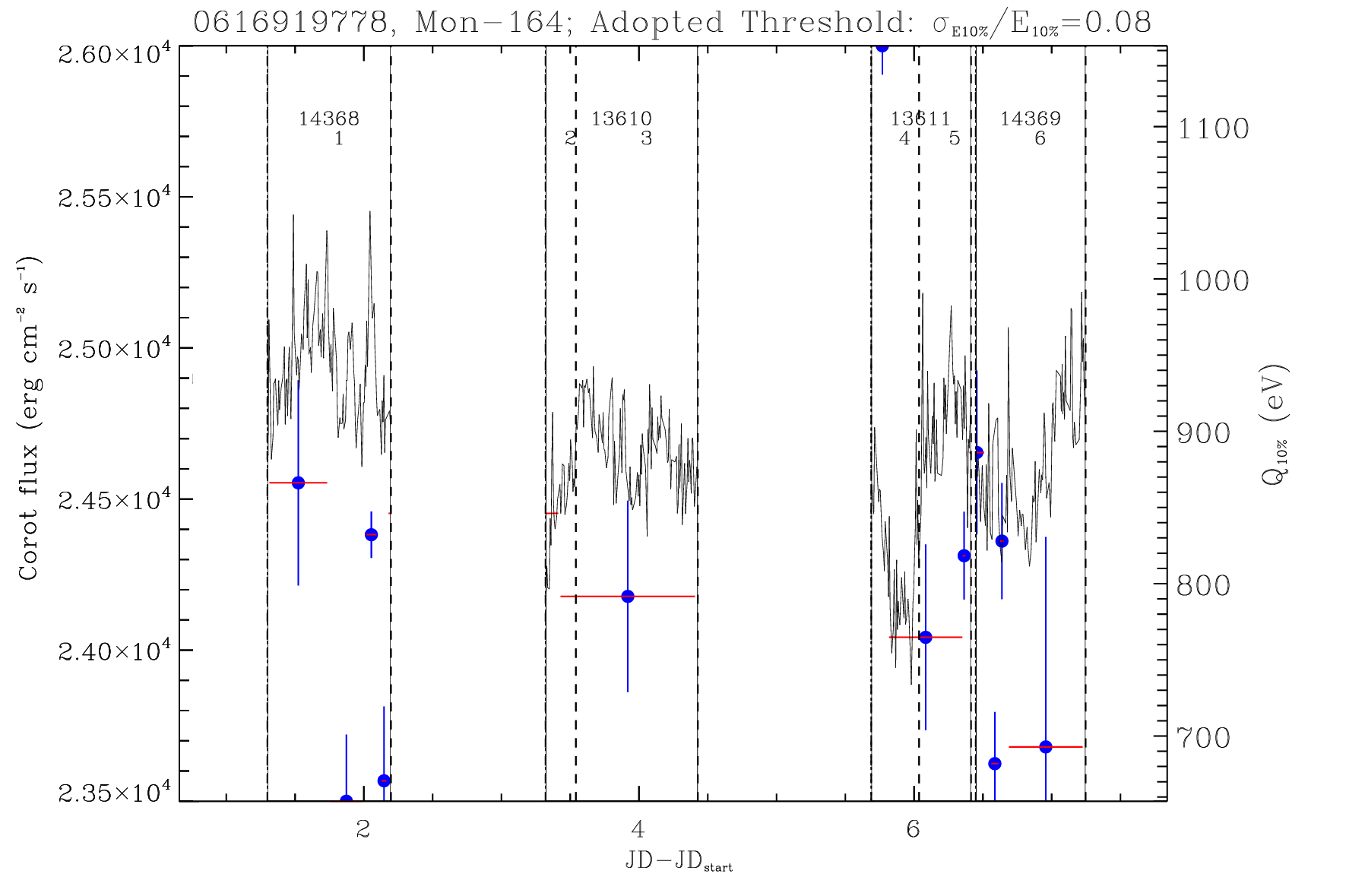}
	\includegraphics[width=8cm]{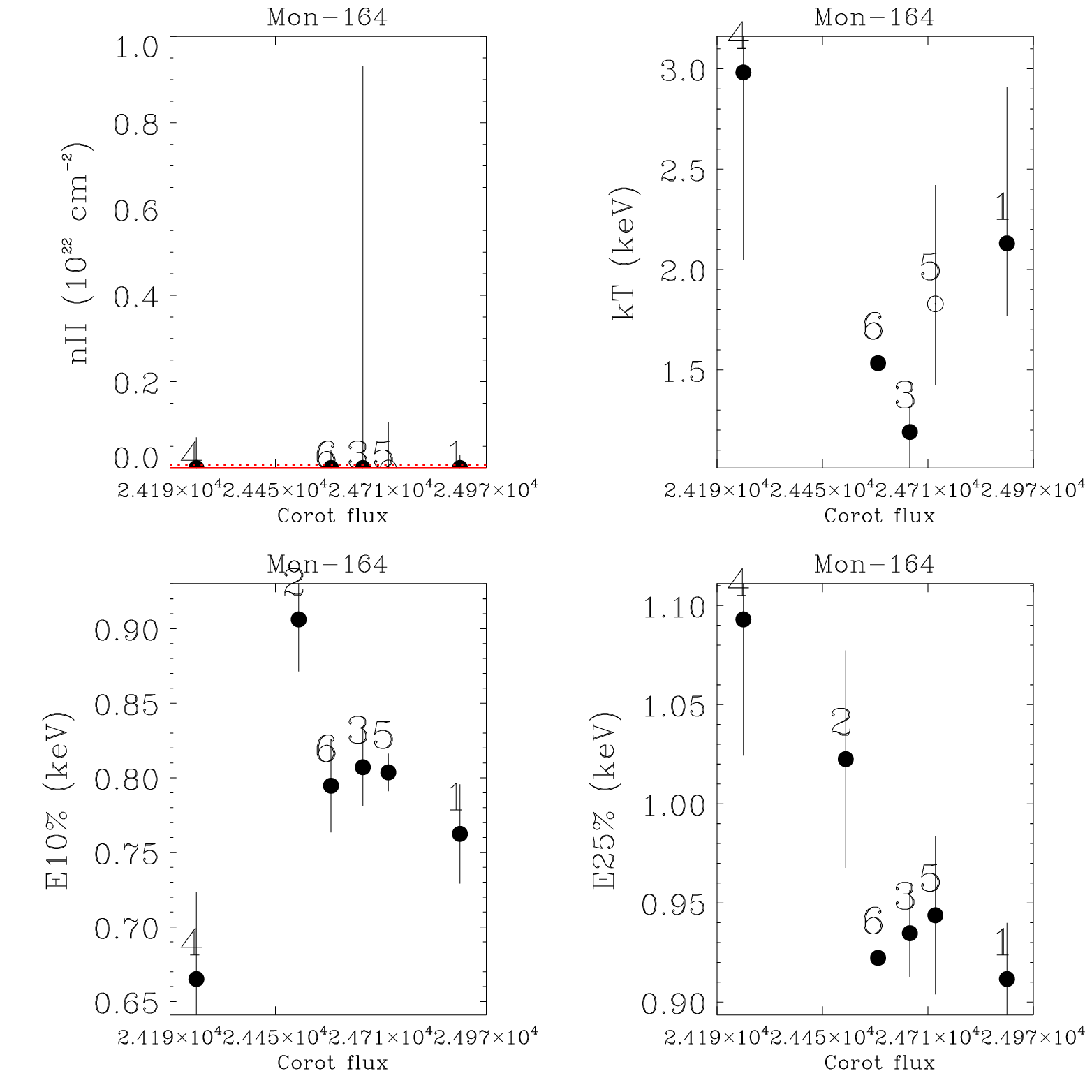}
	\includegraphics[width=18cm]{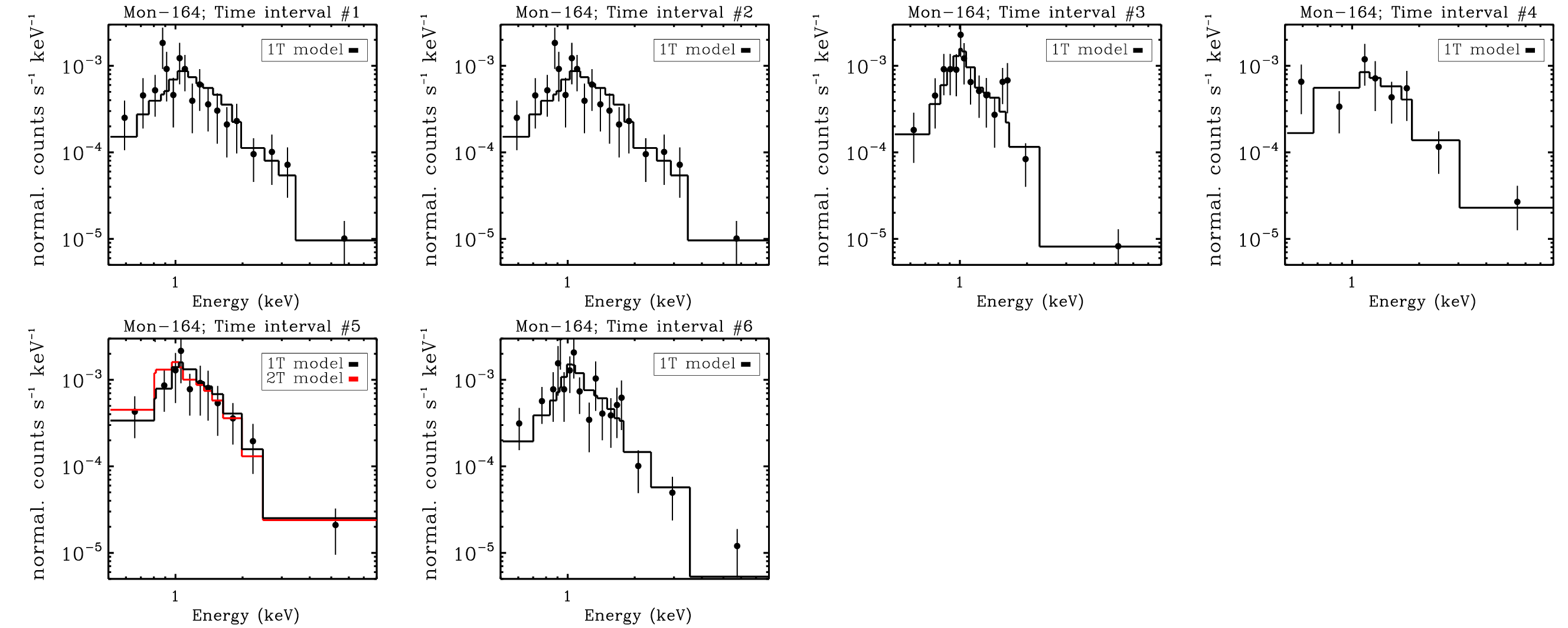}
	\caption{Variability and X-ray spectra of Mon-164, analyzed both as a dipper and burster, and not showing any relevant simultaneous optical and X-ray variability.}
	\label{variab_others_28}
	\end{figure}

	\begin{figure}[]
	\centering	
	\includegraphics[width=9.5cm]{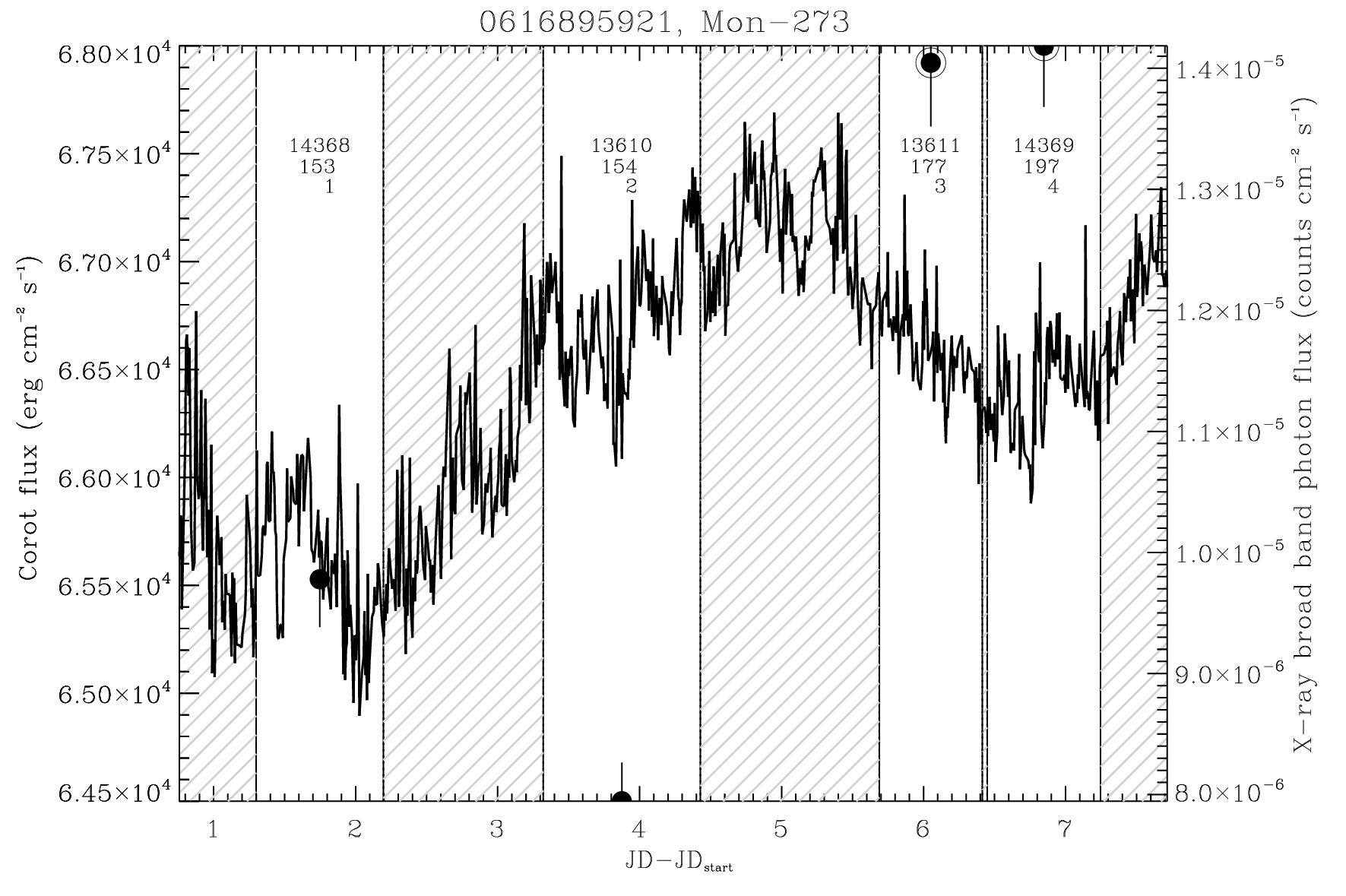}	
	\includegraphics[width=9.5cm]{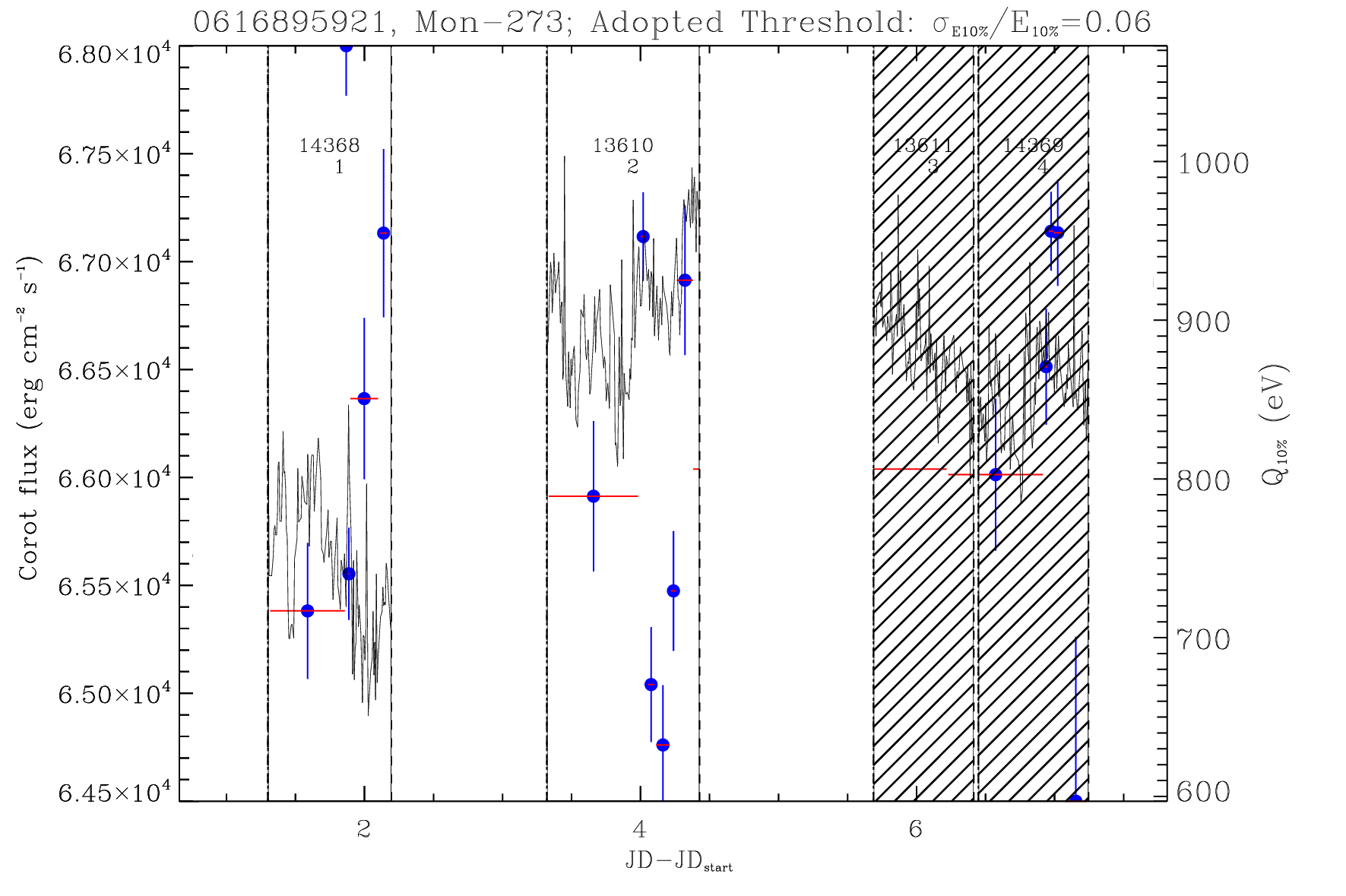}
	\includegraphics[width=8cm]{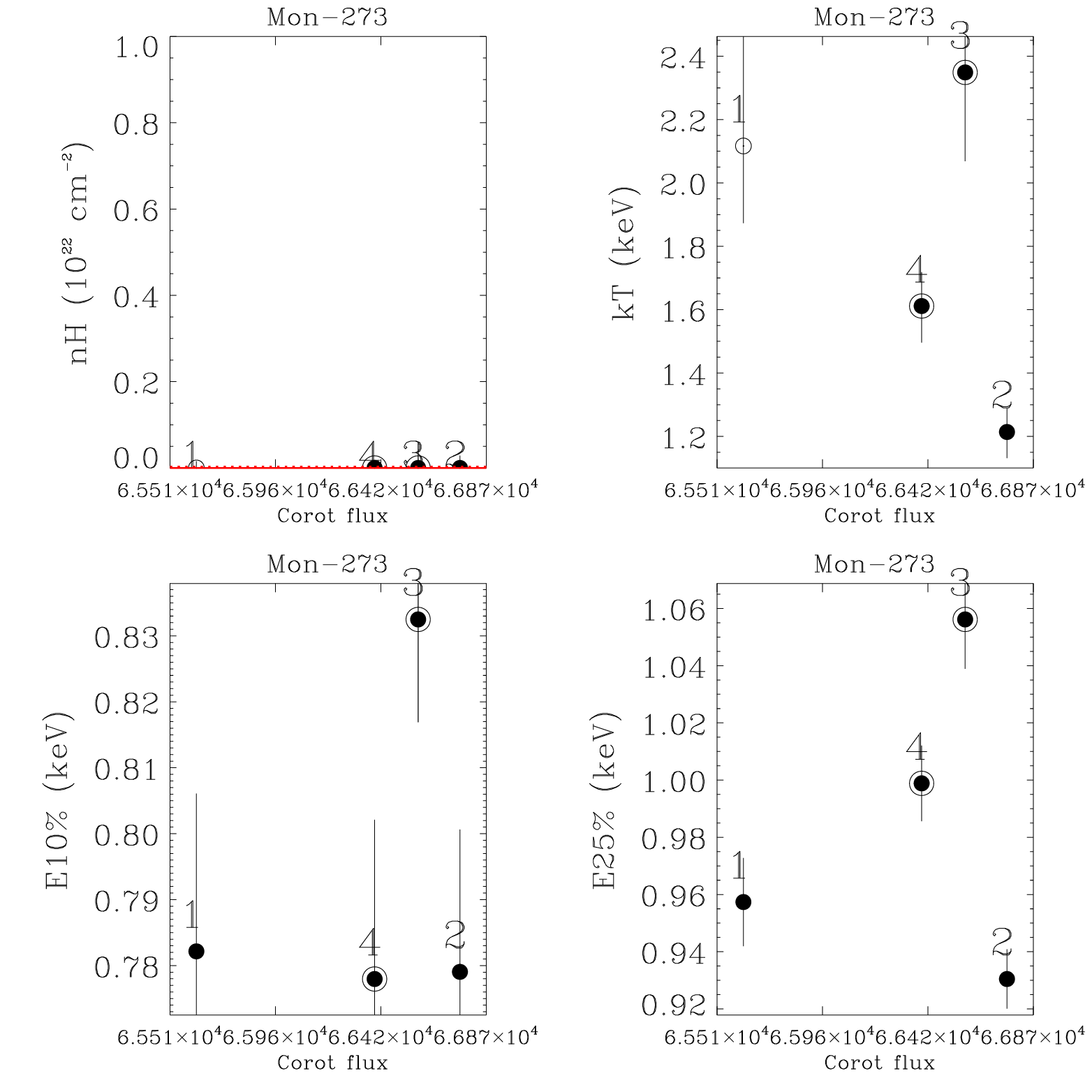}
	\includegraphics[width=18cm]{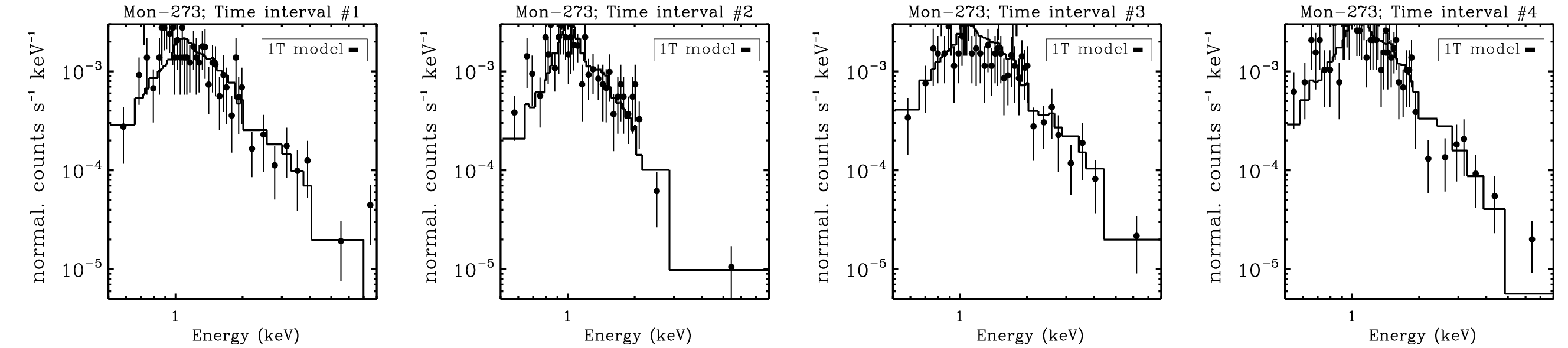}
	\caption{Variability and X-ray spectra of Mon-273, which have not been analyzed since the CoRoT mask around its the position is contaminated by a nearby bright source.}
	\label{variab_others_29}
	\end{figure}

	\begin{figure}[]
	\centering	
	\includegraphics[width=9.5cm]{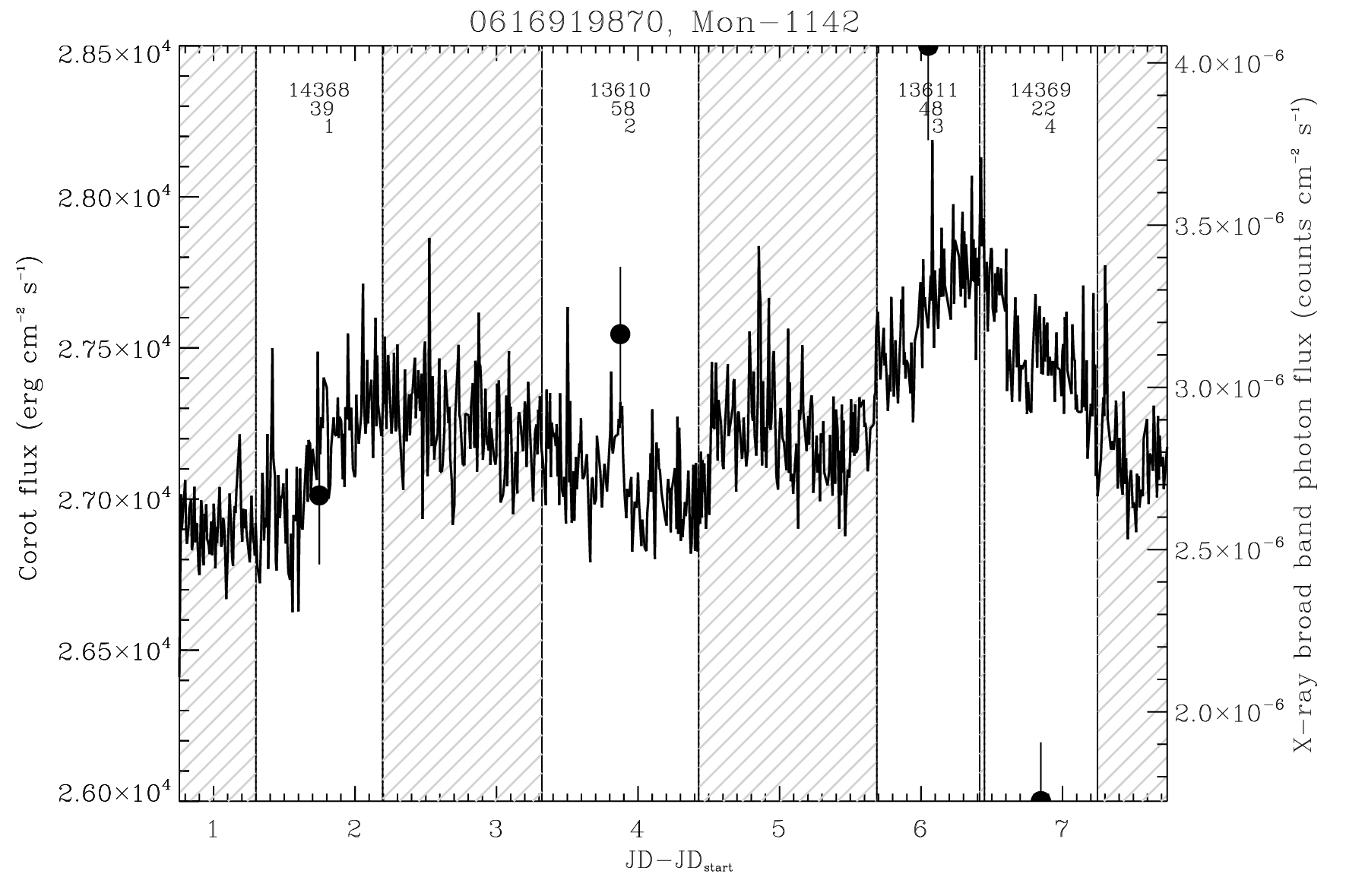}
	\includegraphics[width=9.5cm]{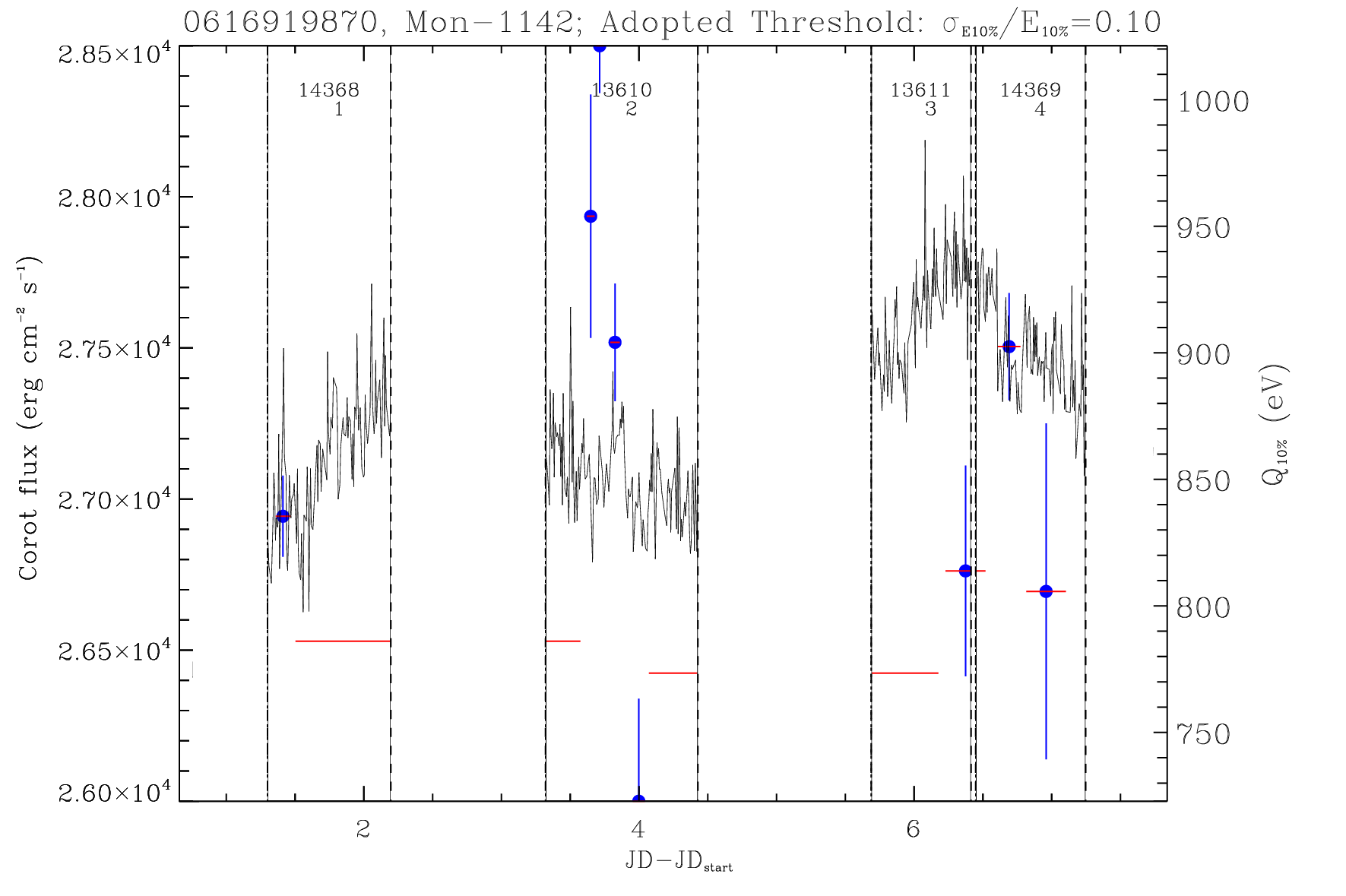}
	\includegraphics[width=8cm]{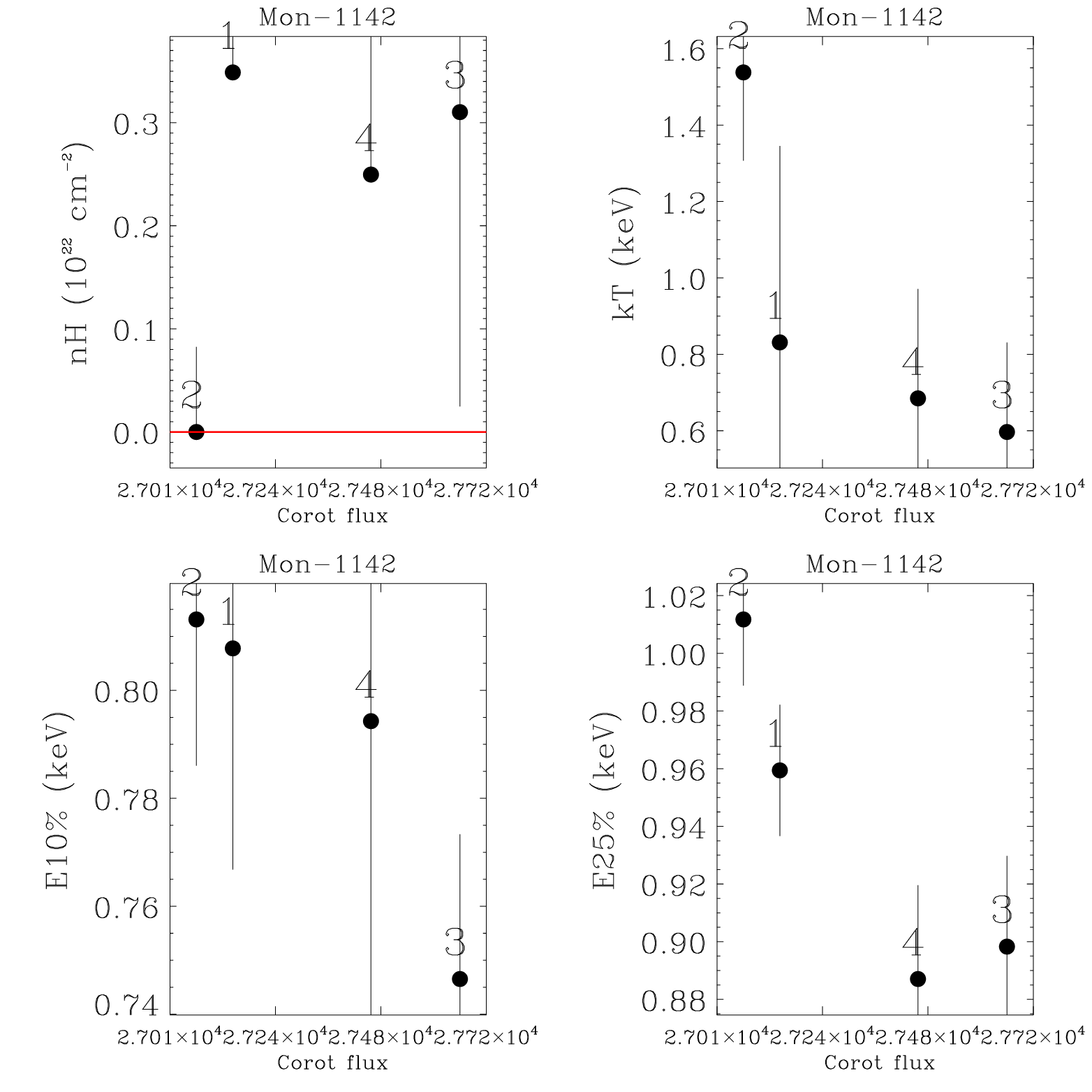}	
	\includegraphics[width=18cm]{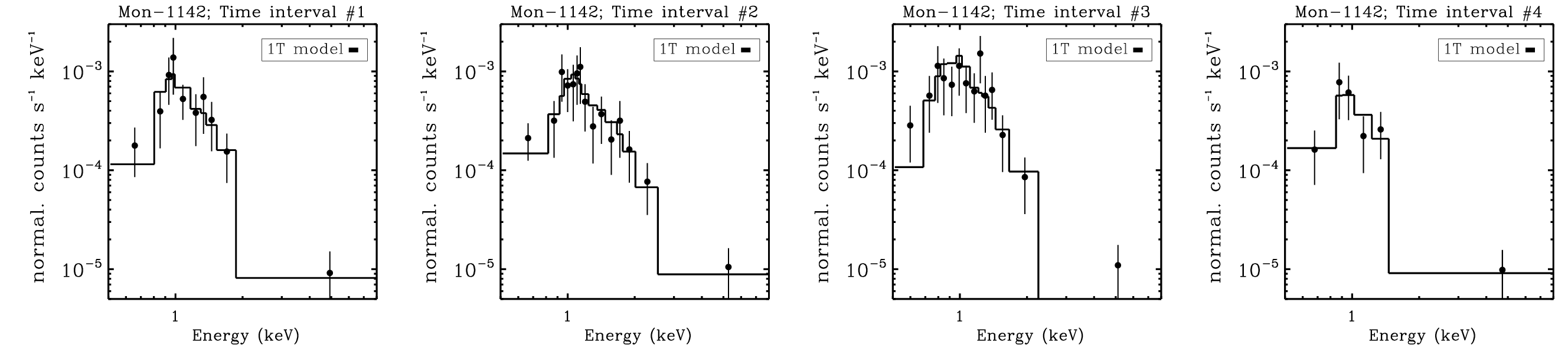}
	\caption{Variability and X-ray spectra of Mon-1142, analyzed both as a dipper and burster, and not showing any relevant variability.}
	\label{variab_others_30}
	\end{figure}

	\begin{figure}[]
	\centering	
	\includegraphics[width=9.5cm]{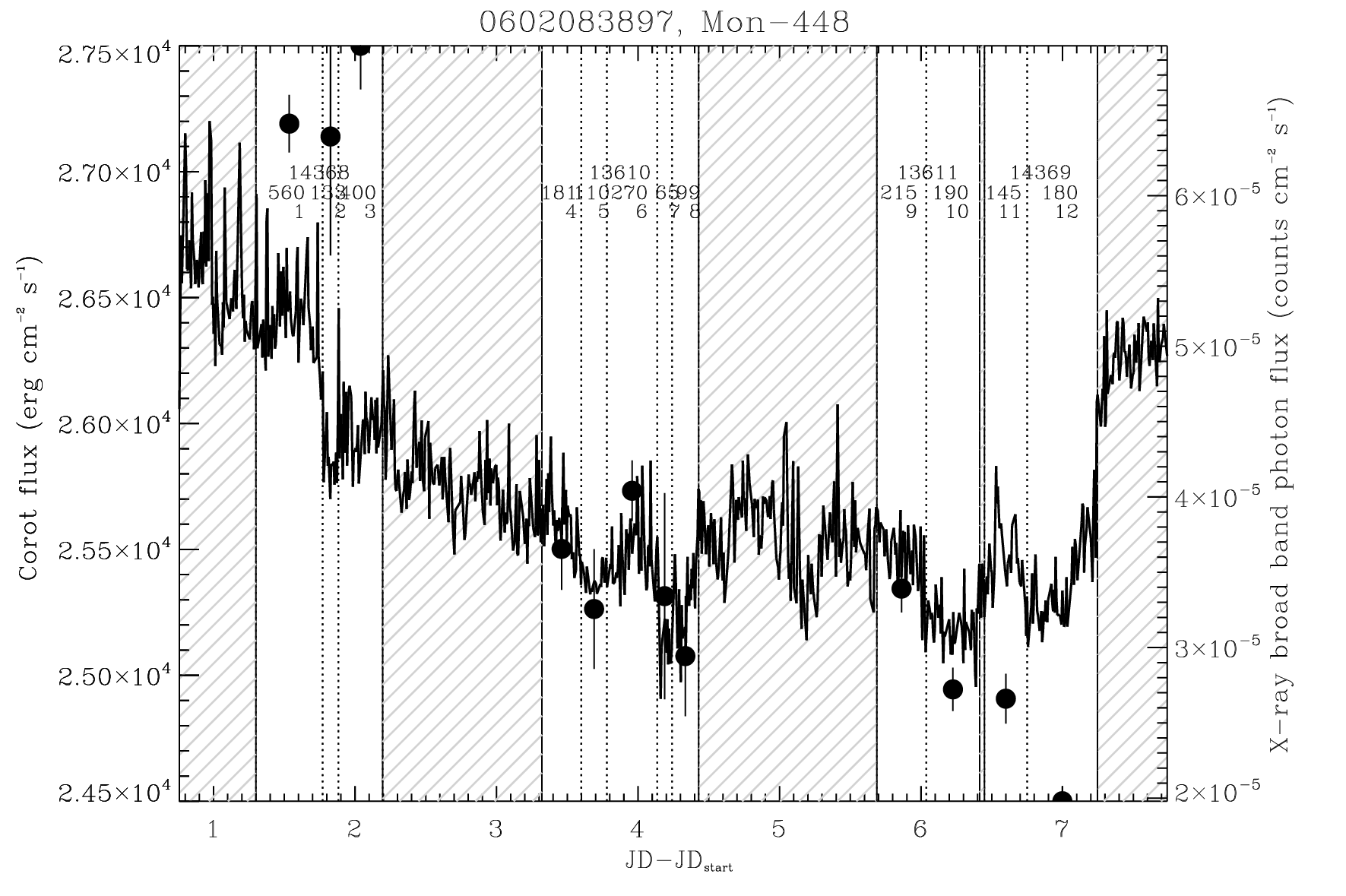}
	\includegraphics[width=9.5cm]{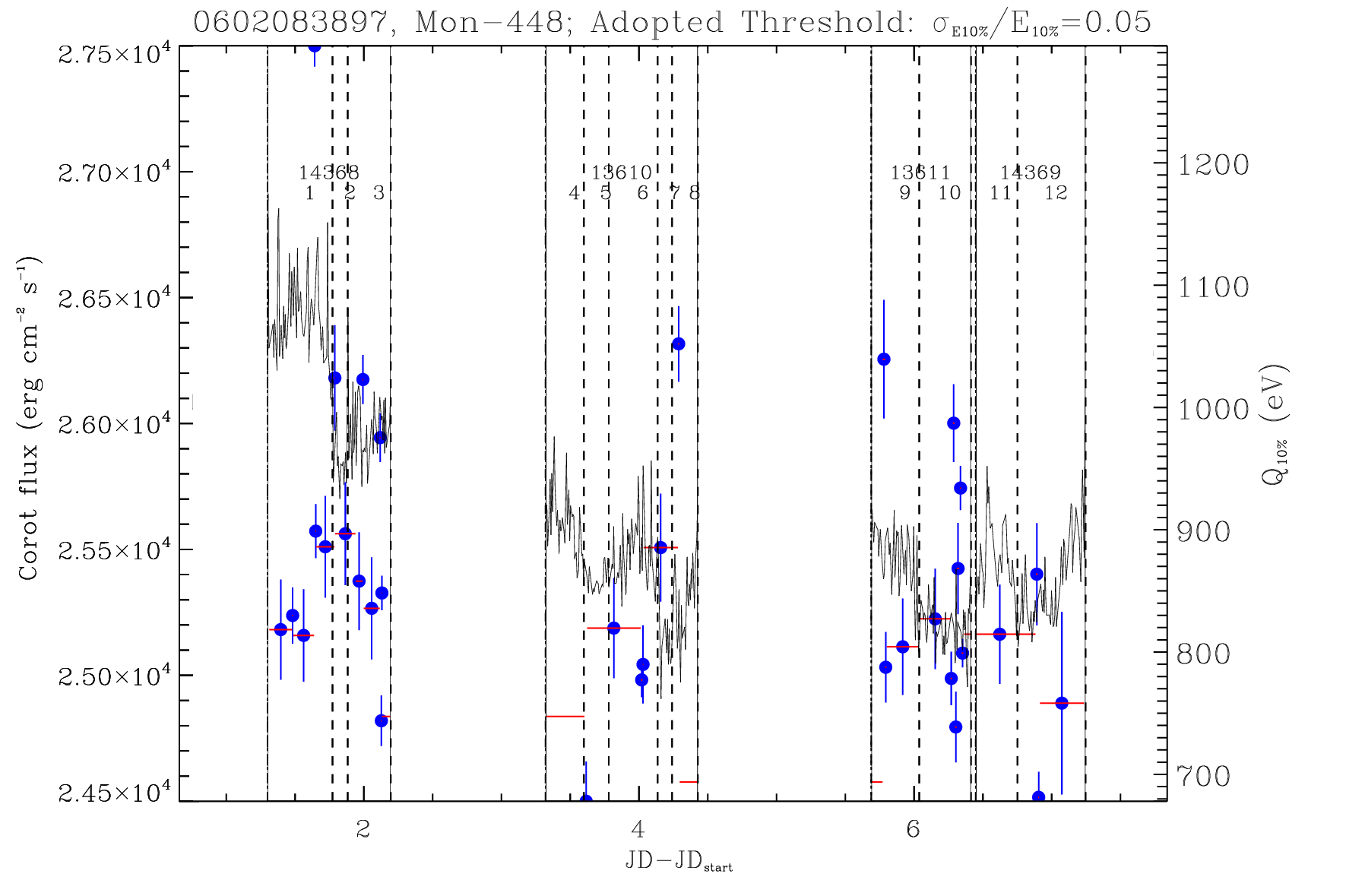}
	\includegraphics[width=8cm]{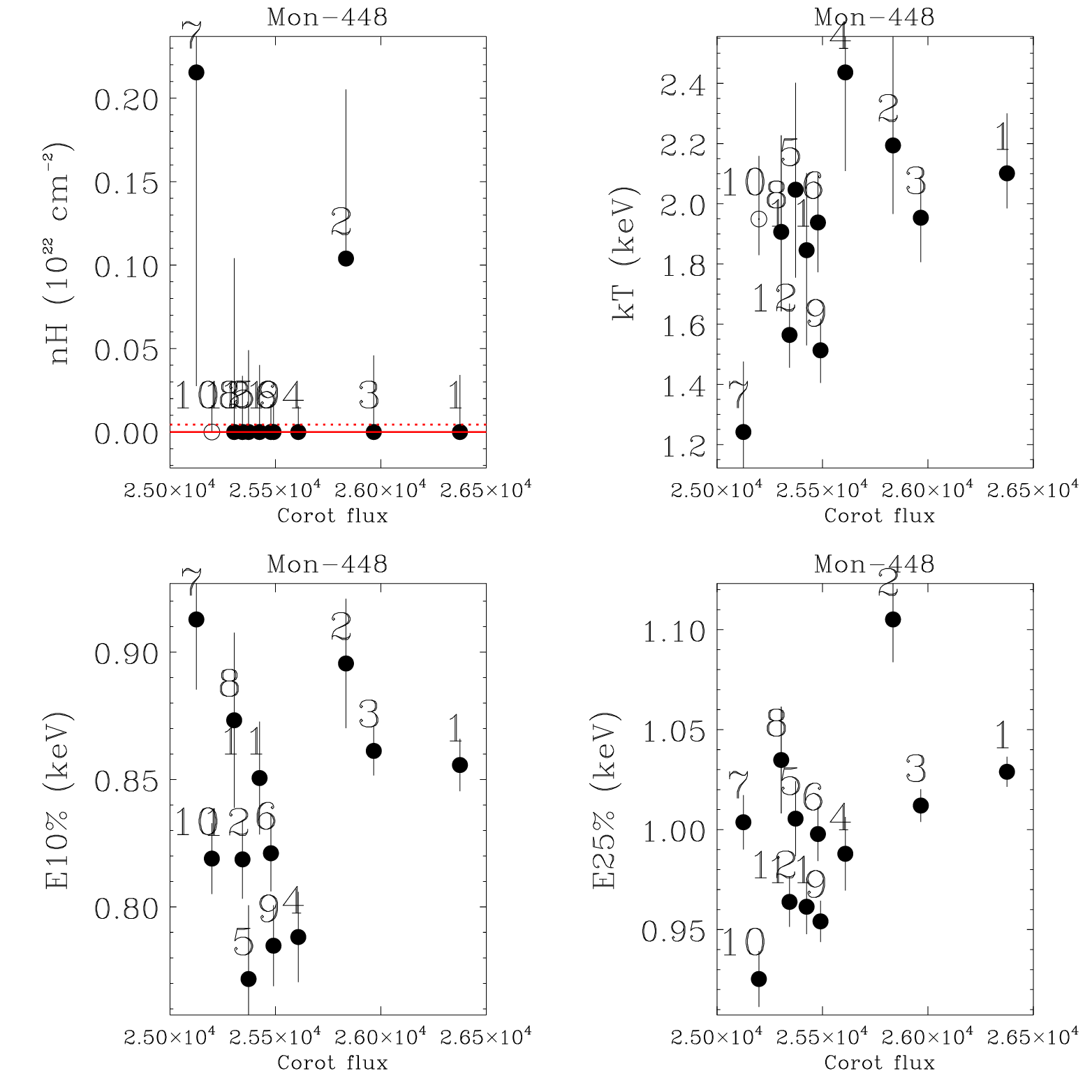}	
	\includegraphics[width=18cm]{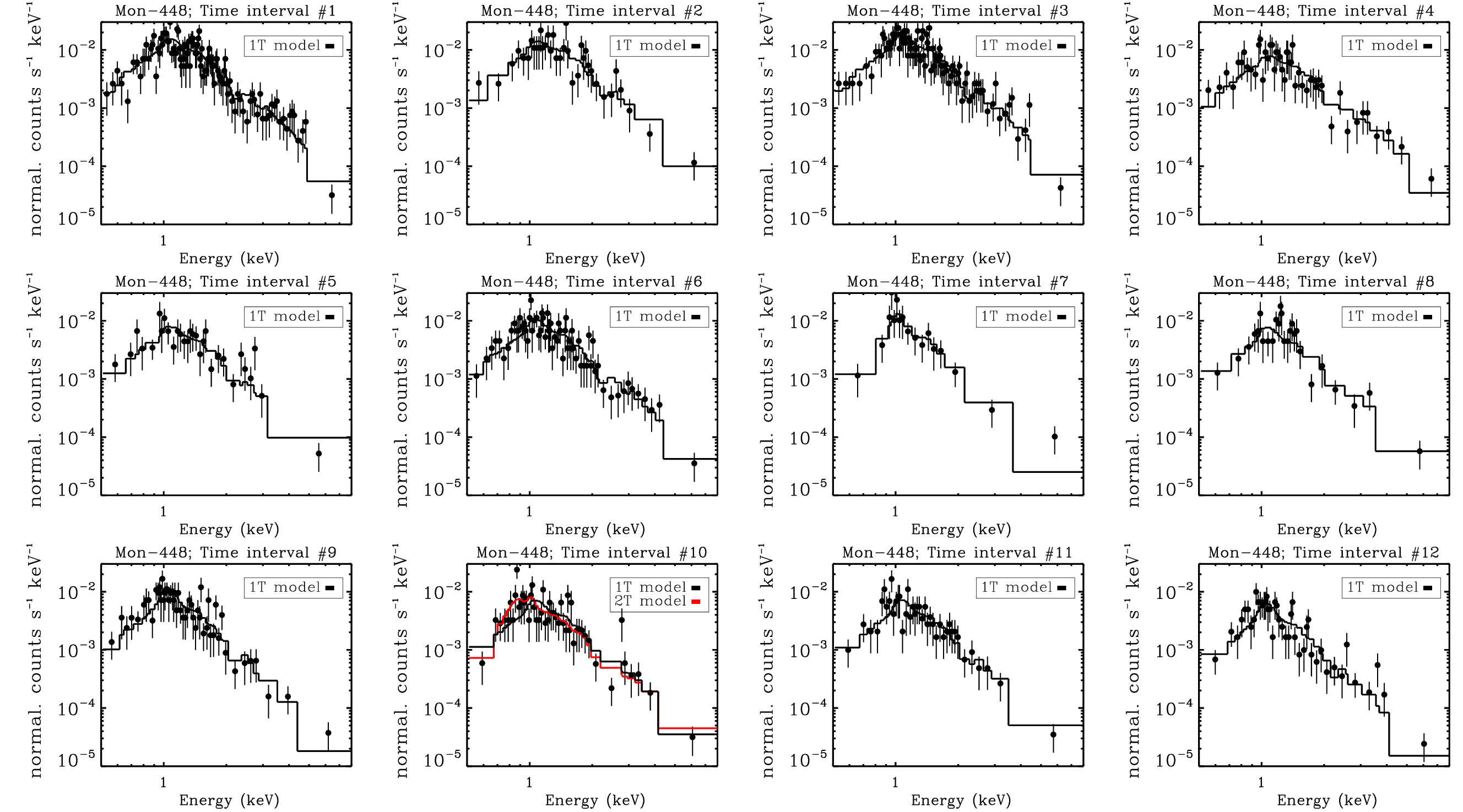}
	\caption{Variability and X-ray spectra of Mon-448, analyzed both as a dipper and burster. The features observed in the CoRoT light curve do not correspond to any significant variability of the X-ray properties.}
	\label{variab_others_31}
	\end{figure}

	\begin{figure}[]
	\centering	
	\includegraphics[width=9.5cm]{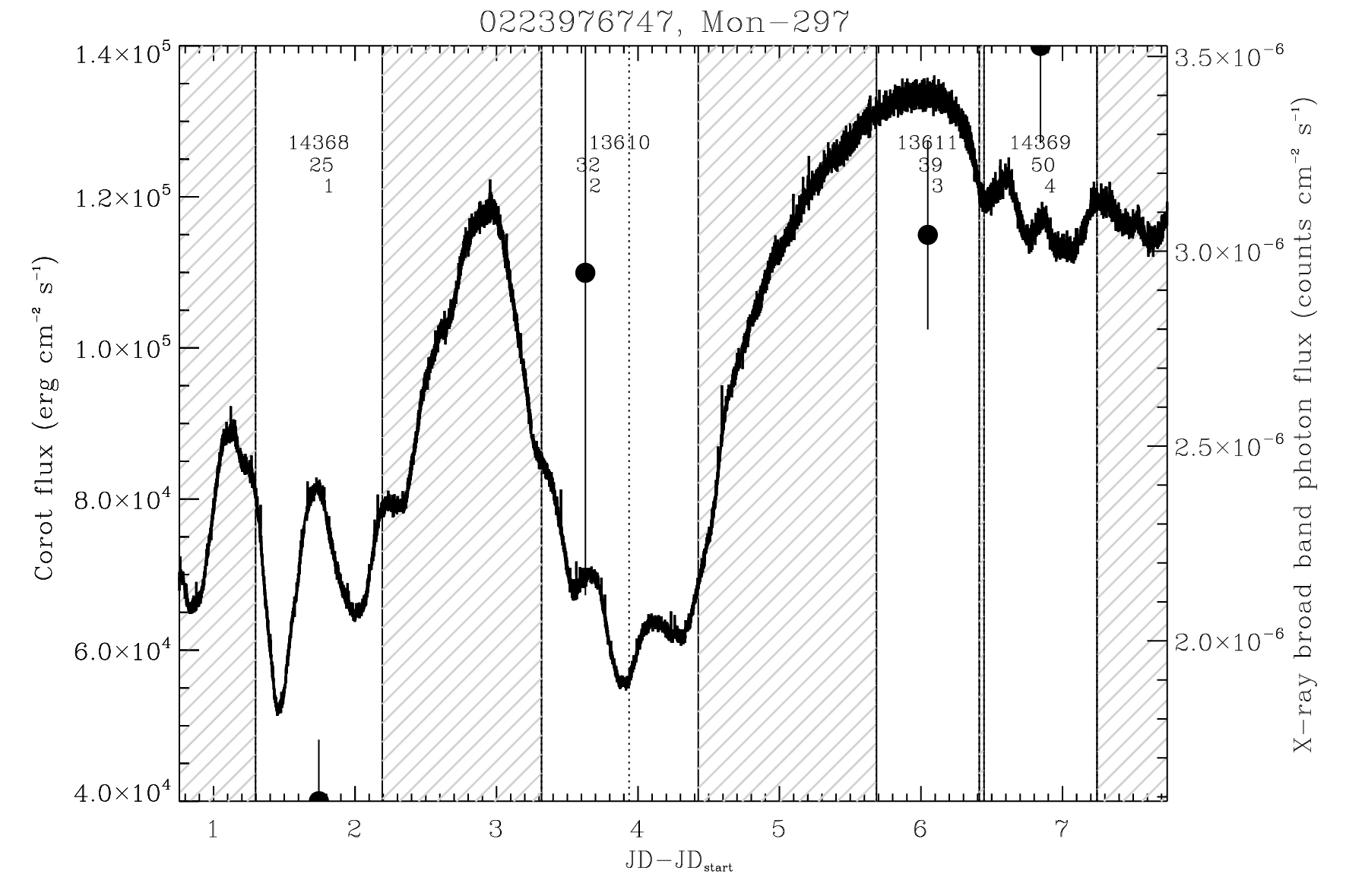}
	\includegraphics[width=9.5cm]{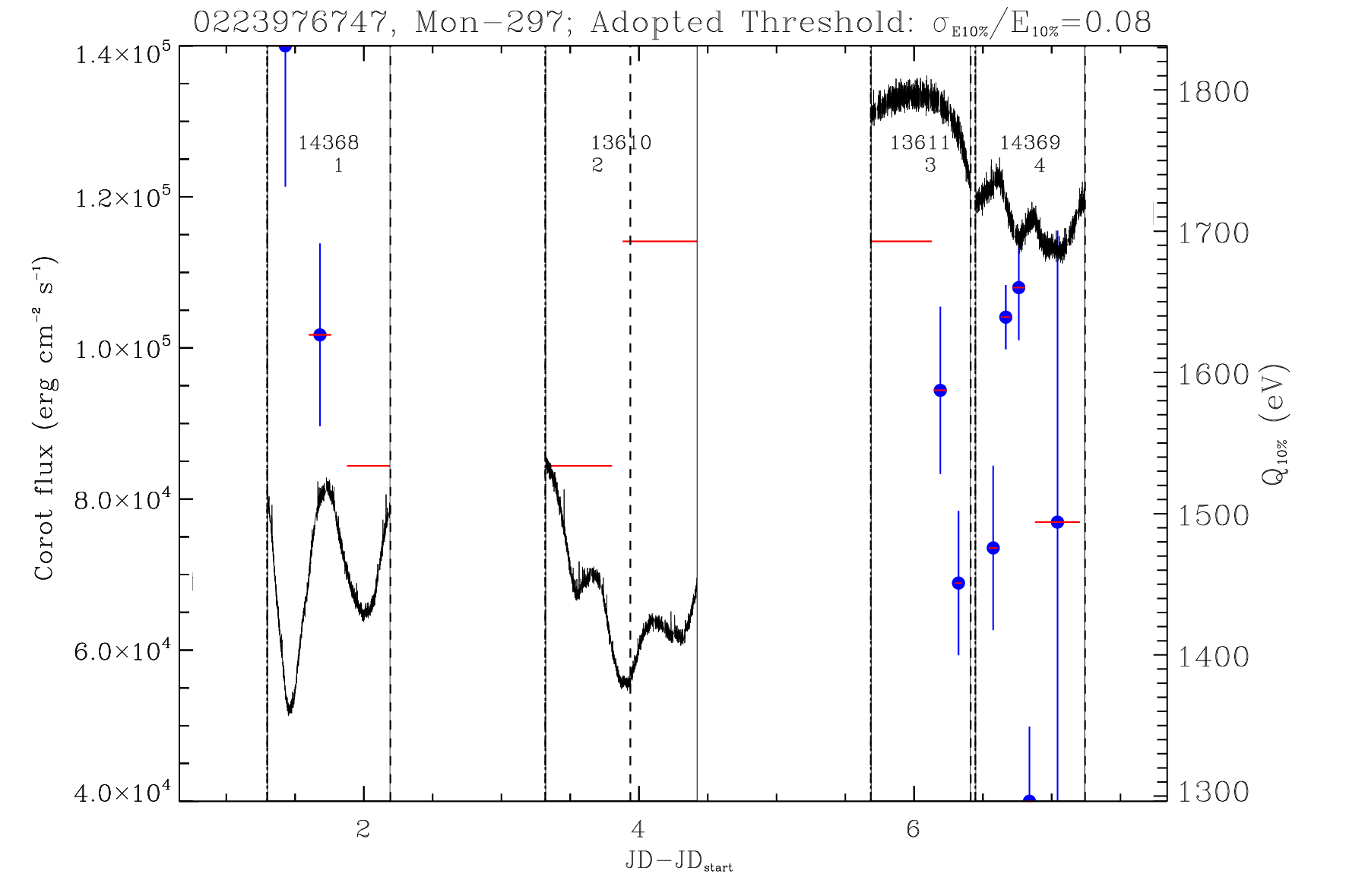}
	\includegraphics[width=8cm]{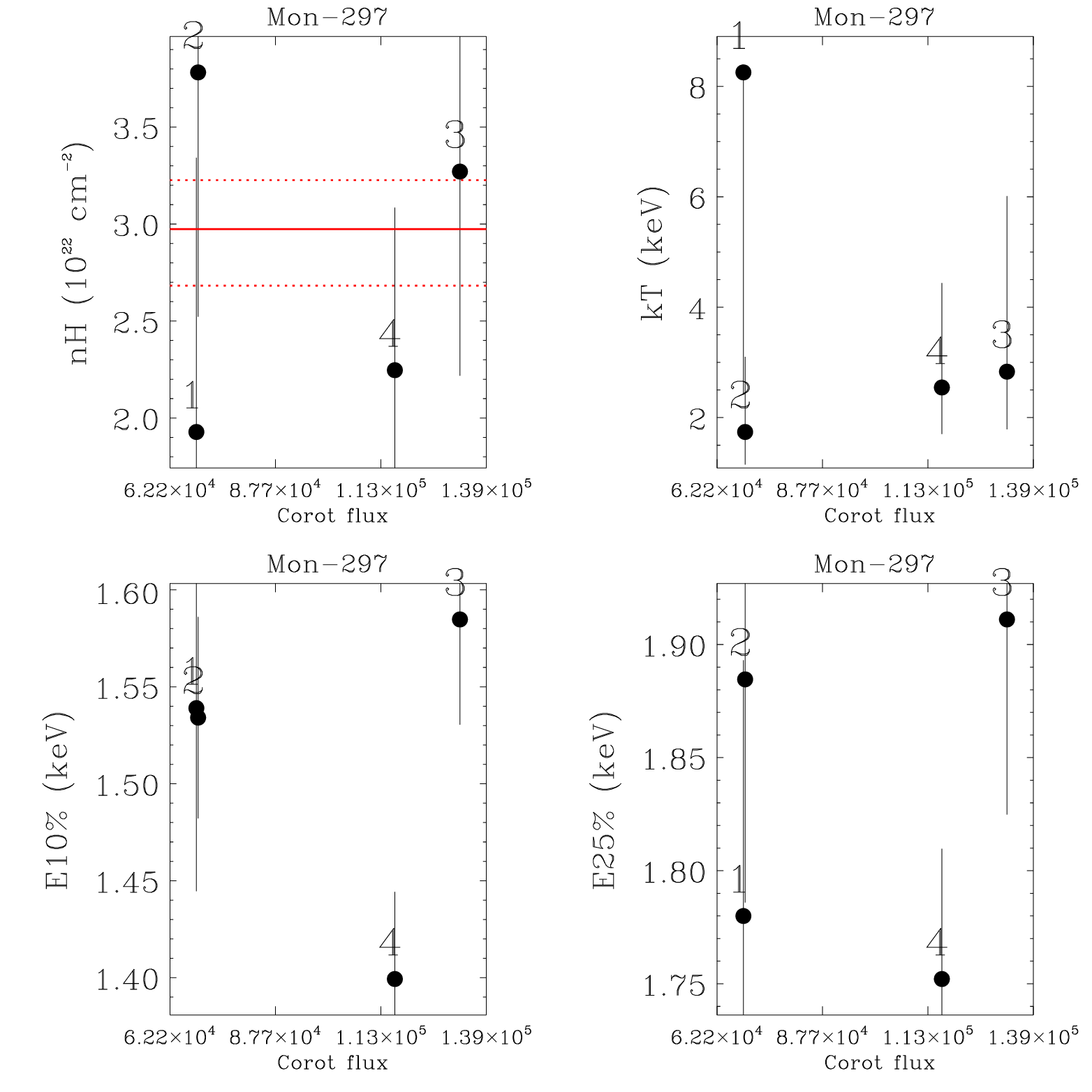}	
	\includegraphics[width=18cm]{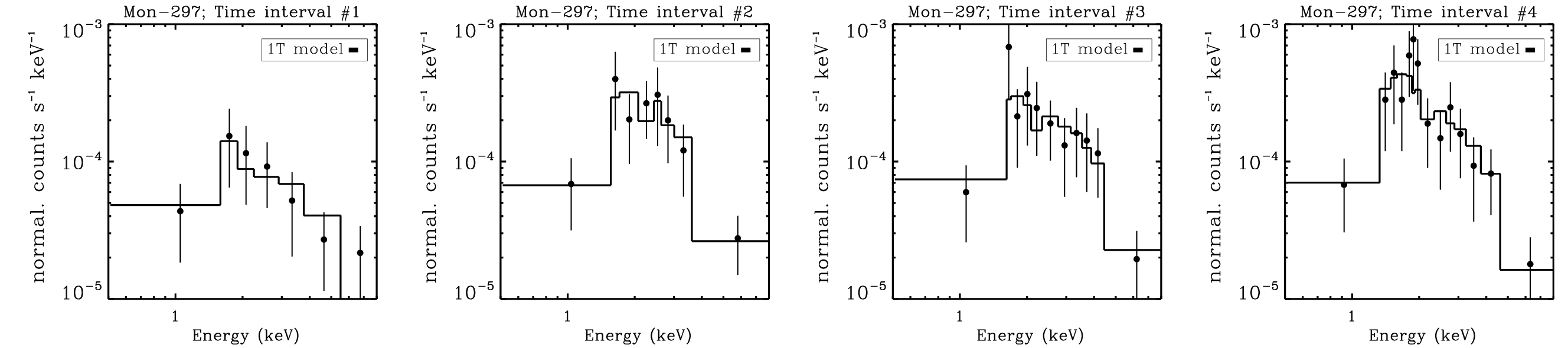}
	\caption{Variability and X-ray spectra of Mon-297, analyzed both as a dipper and burster. The features observed in the CoRoT light curve do not correspond to any significant variability of the X-ray properties given the few X-ray photons detected. The variability of E$_{10\%}$ during \#1 suggests that the X-ray spectrum may be harder during the first dip.}
	\label{variab_others_32}
	\end{figure}

	\begin{figure}[]
	\centering	
	\includegraphics[width=9.5cm]{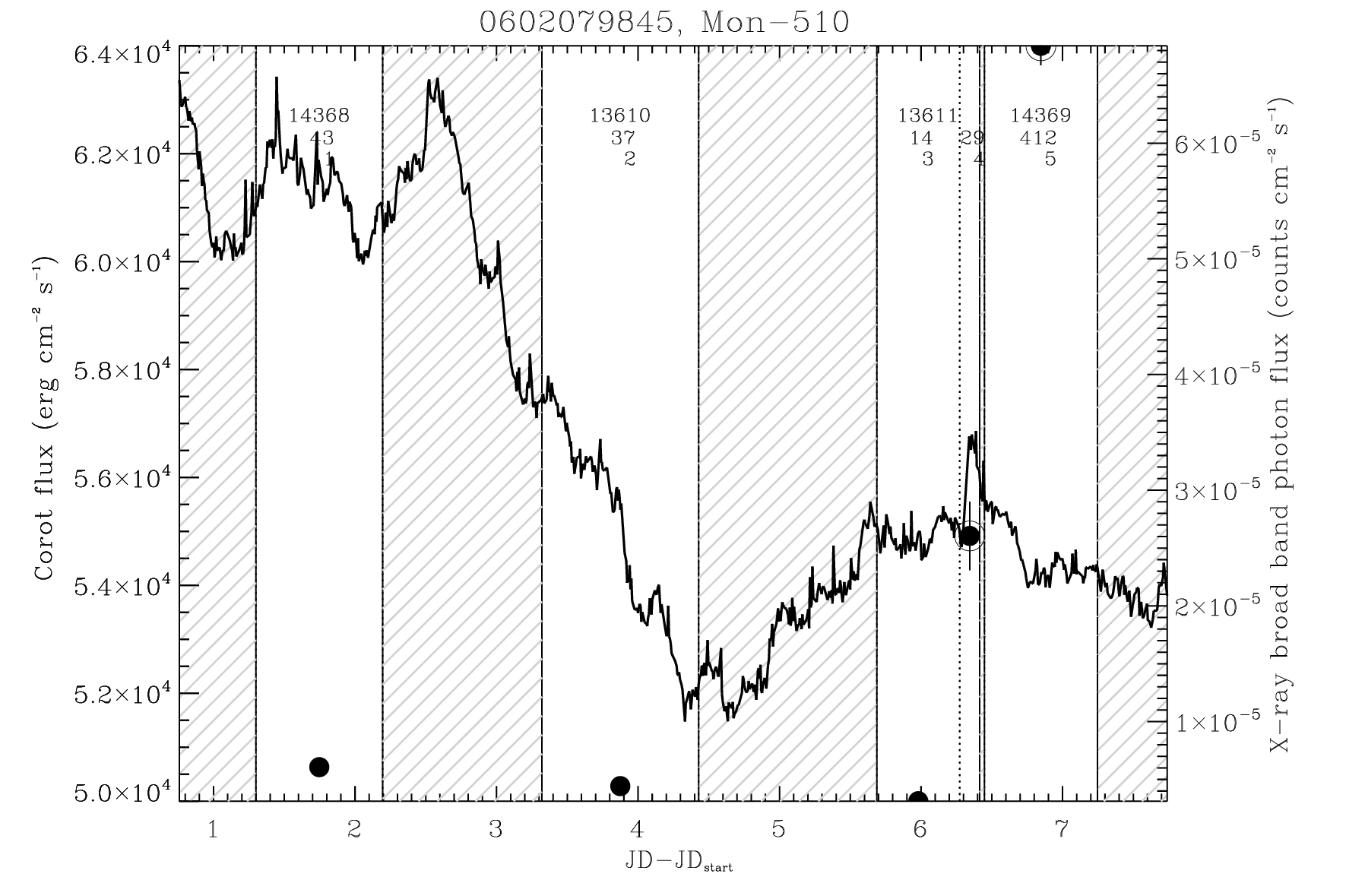}
	\includegraphics[width=9.5cm]{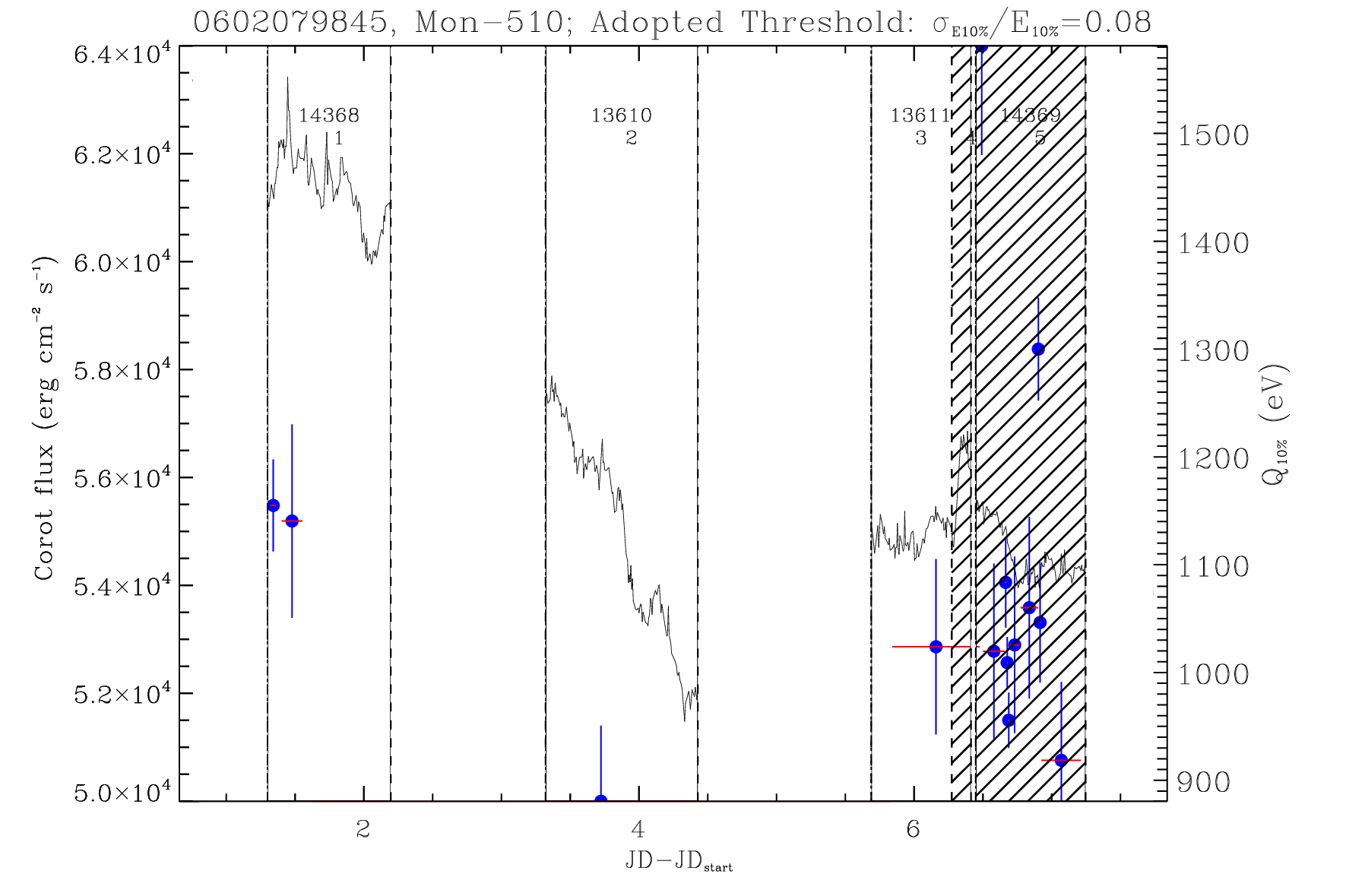}
	\includegraphics[width=8cm]{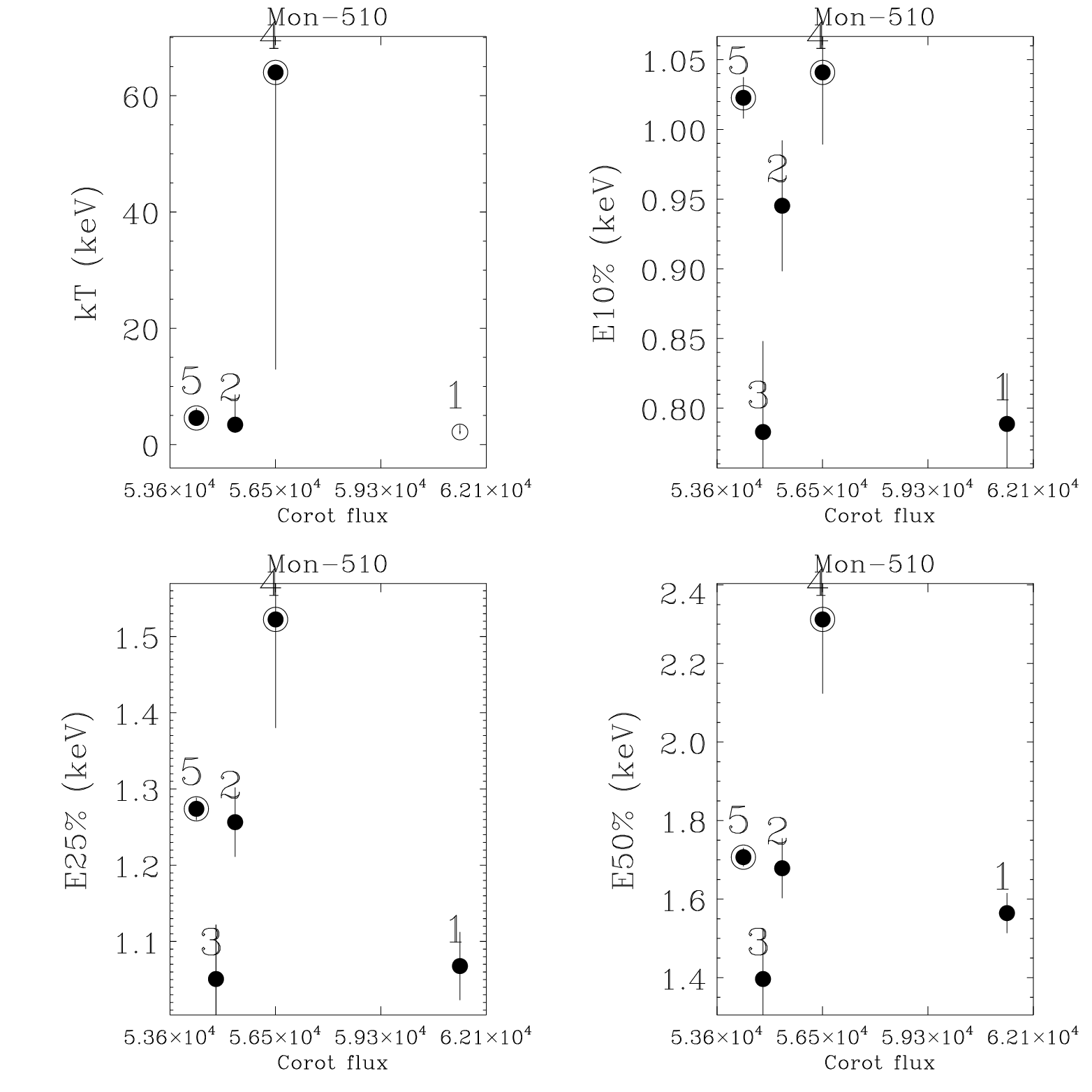}	
	\includegraphics[width=18cm]{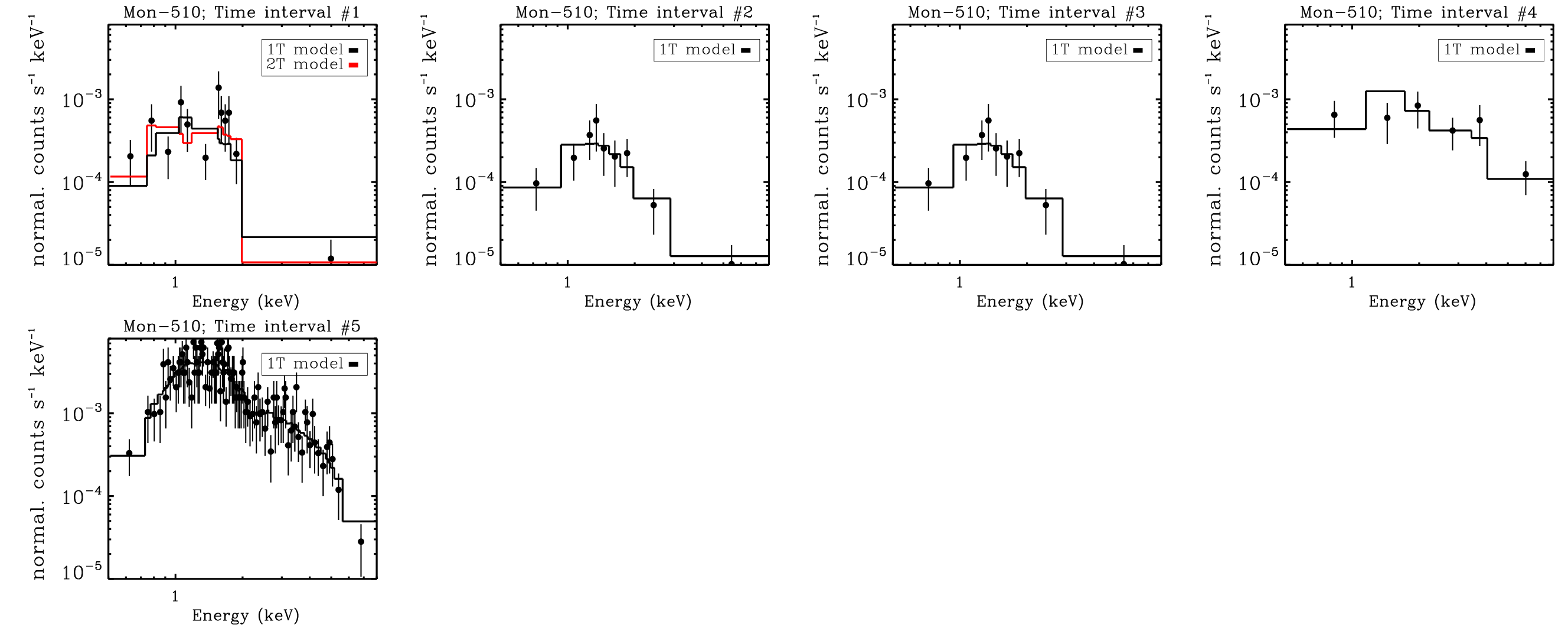}
	\caption{Variability and X-ray spectra of Mon-510, analyzed both as a burster. The features observed in the CoRoT light curve do not correspond to any significant variability of the X-ray properties.}
	\label{variab_others_33}
	\end{figure}

	\begin{figure}[]
	\centering	
	\includegraphics[width=9.5cm]{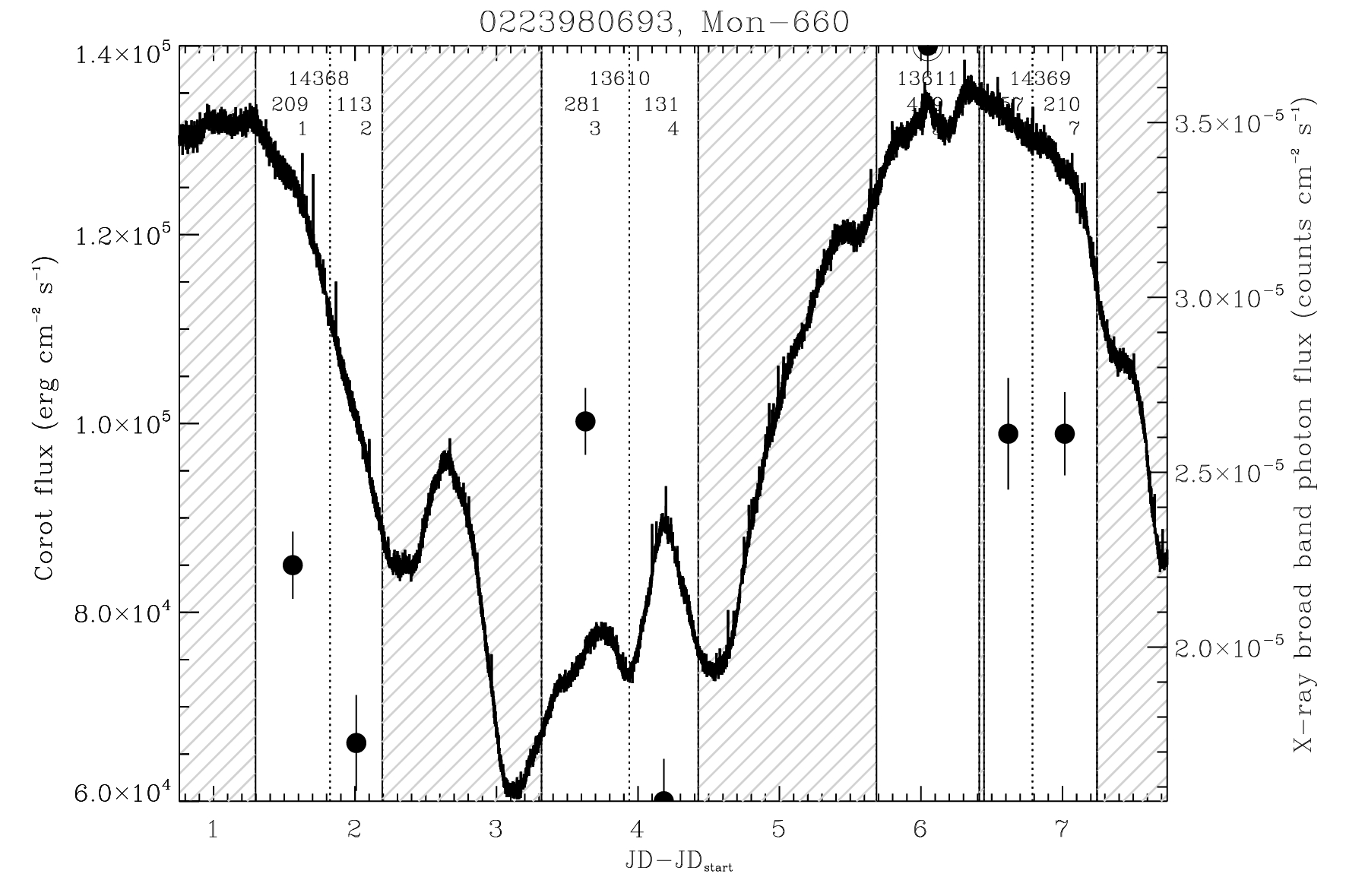}
	\includegraphics[width=9.5cm]{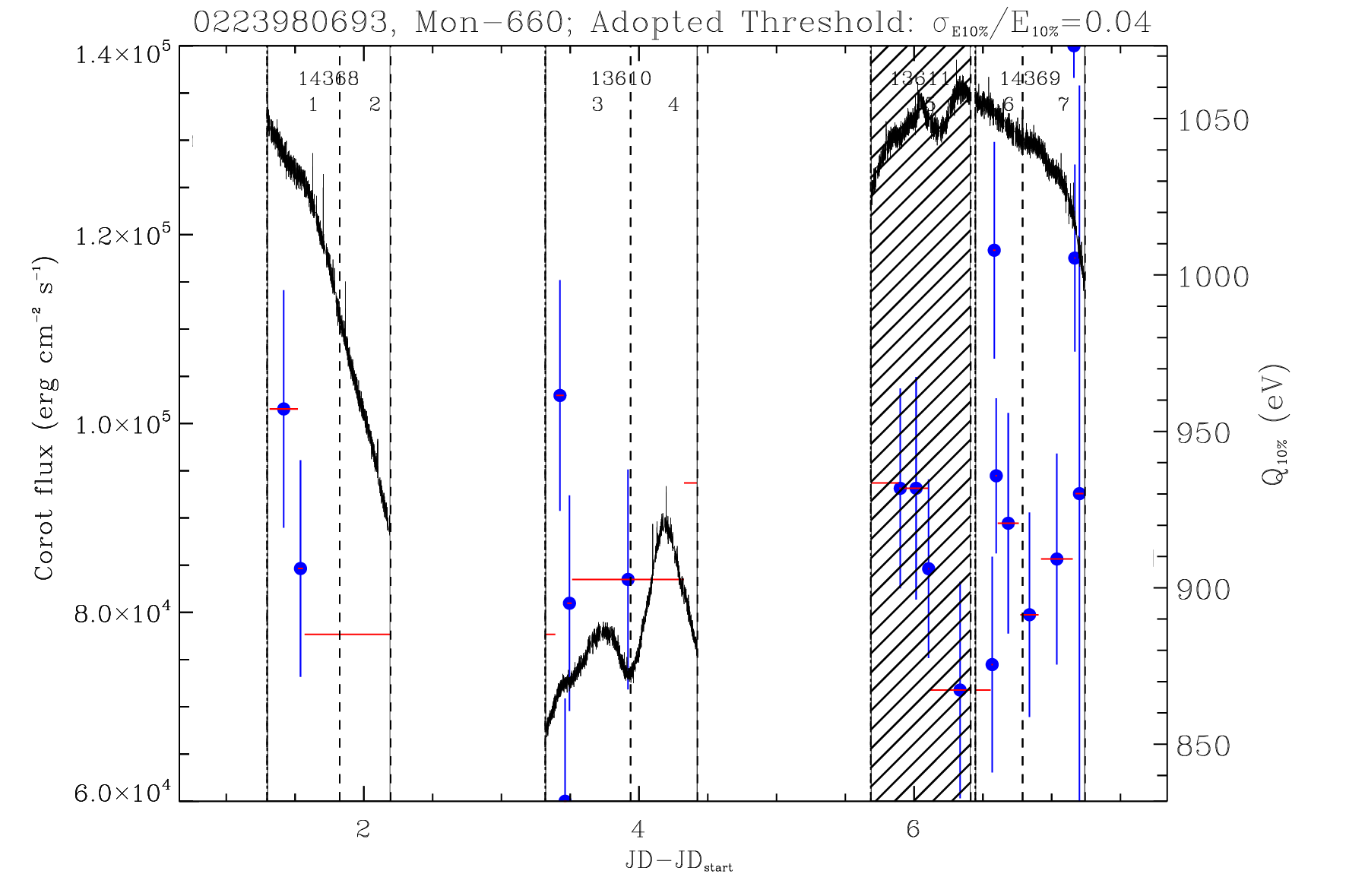}
	\includegraphics[width=8cm]{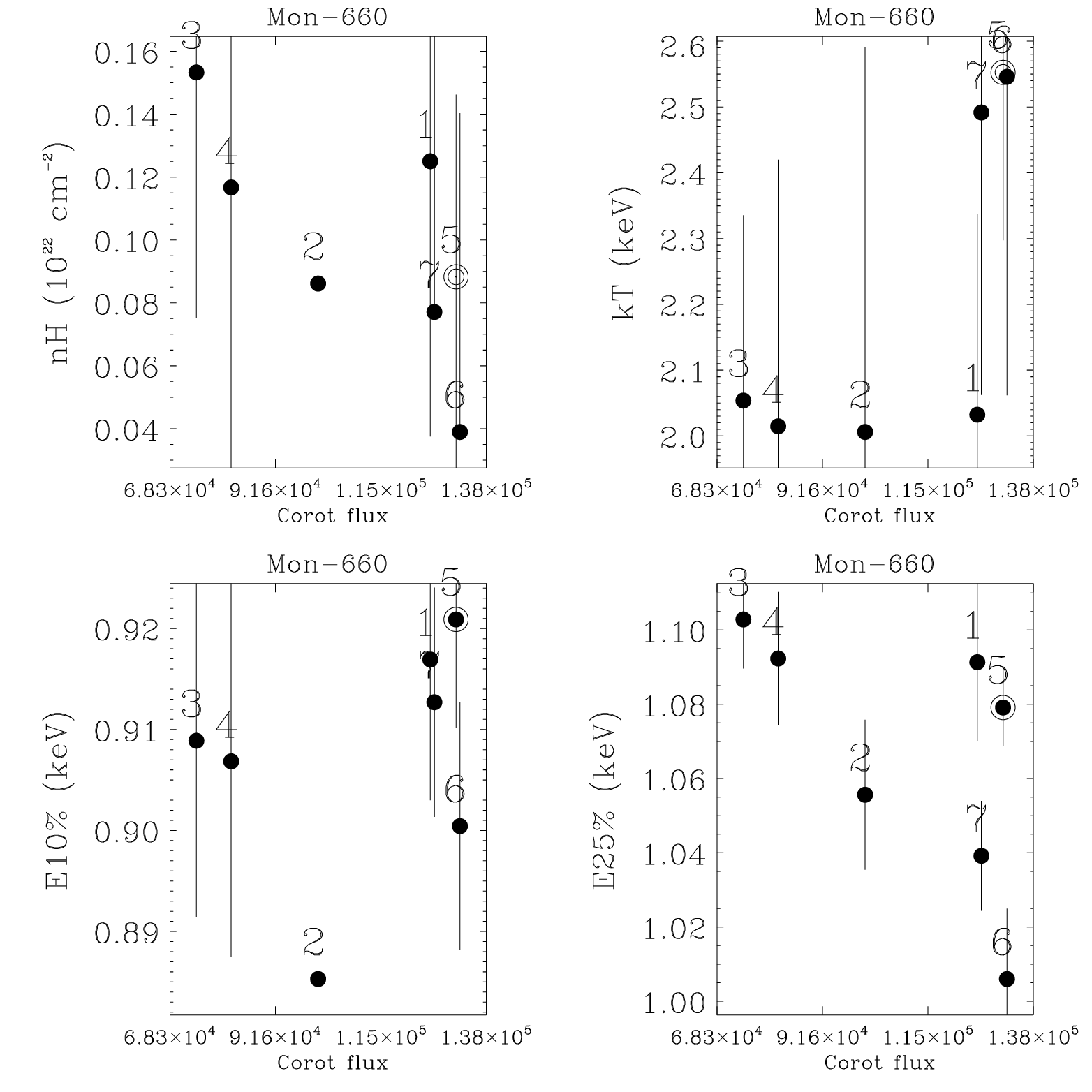}	
	\includegraphics[width=18cm]{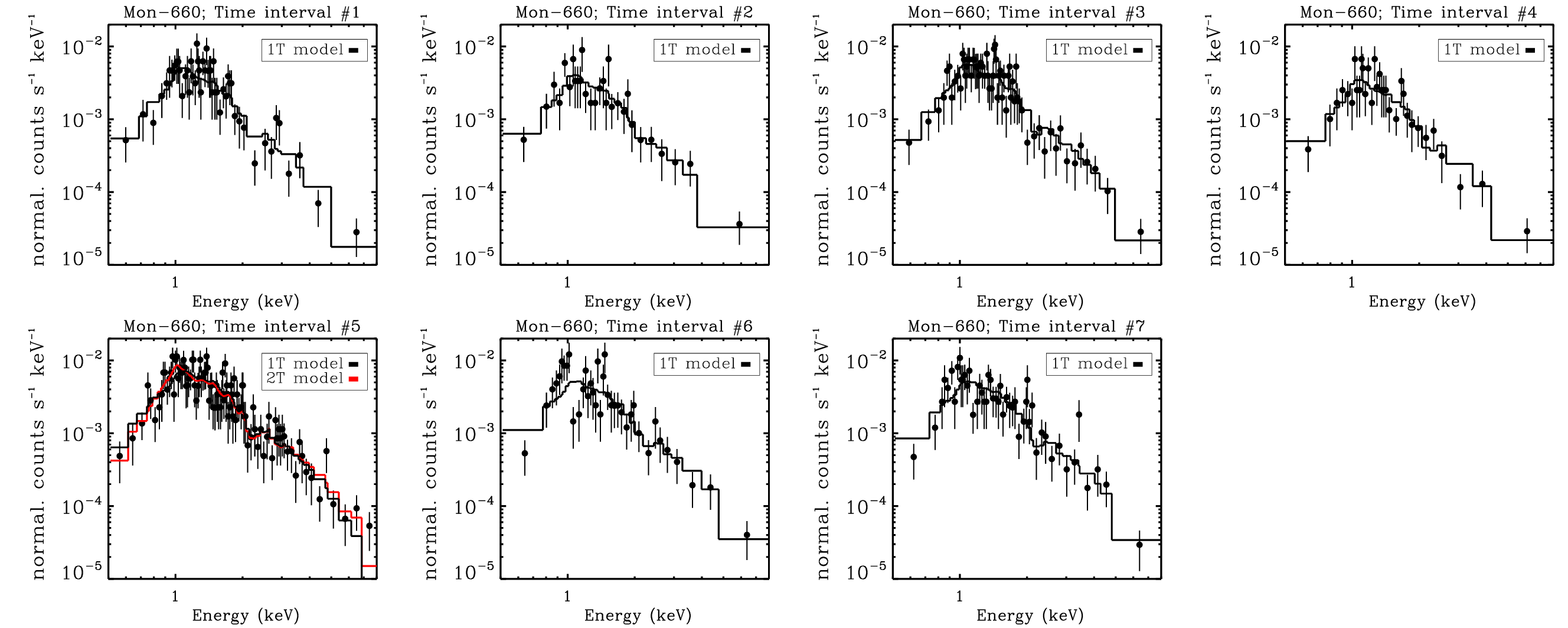}
	\caption{Variability and X-ray spectra of Mon-660, analyzed both as a dipper and burster. The features observed in the CoRoT light curve do not correspond to any significant variability of the X-ray properties.}
	\label{variab_others_34}
	\end{figure}

\newpage
\clearpage

%%%%%%%%%%%%%%%%%%%%%%%%%%%%%%%%%%%%%%%%%%%%%%%%%%%%%%%%%%%%%
\section{Entire CoRoT light curves of the stars discussed in this paper}
\label{all_LC}

In this appendix we show the whole CoRoT light curves of the stars with disks discussed in this paper. The $Chandra$ frames are the shaded time windows delimited by vertical lines.
	
\begin{figure}[]
\centering	
\includegraphics[width=9.0cm]{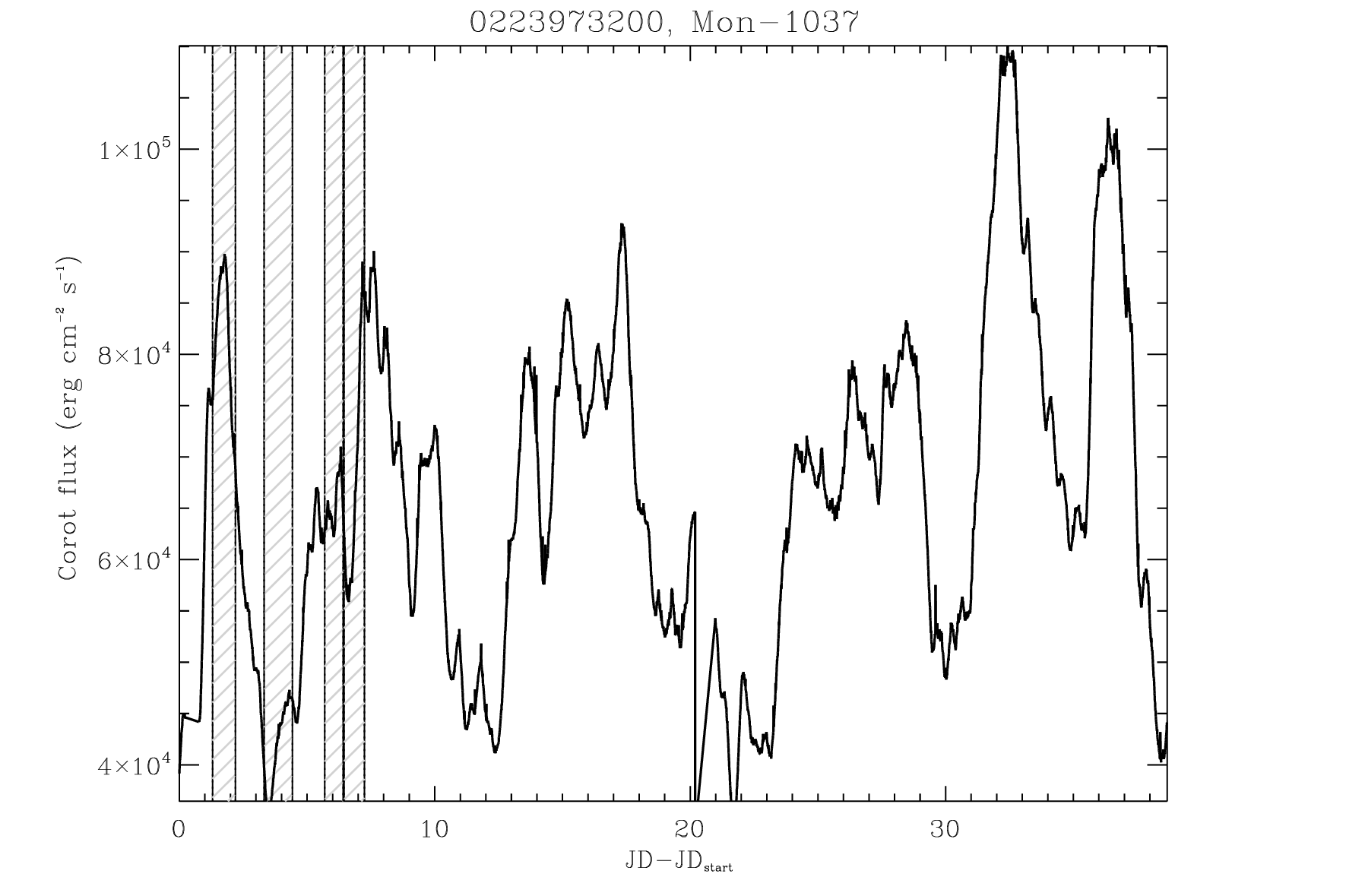}
\includegraphics[width=9.0cm]{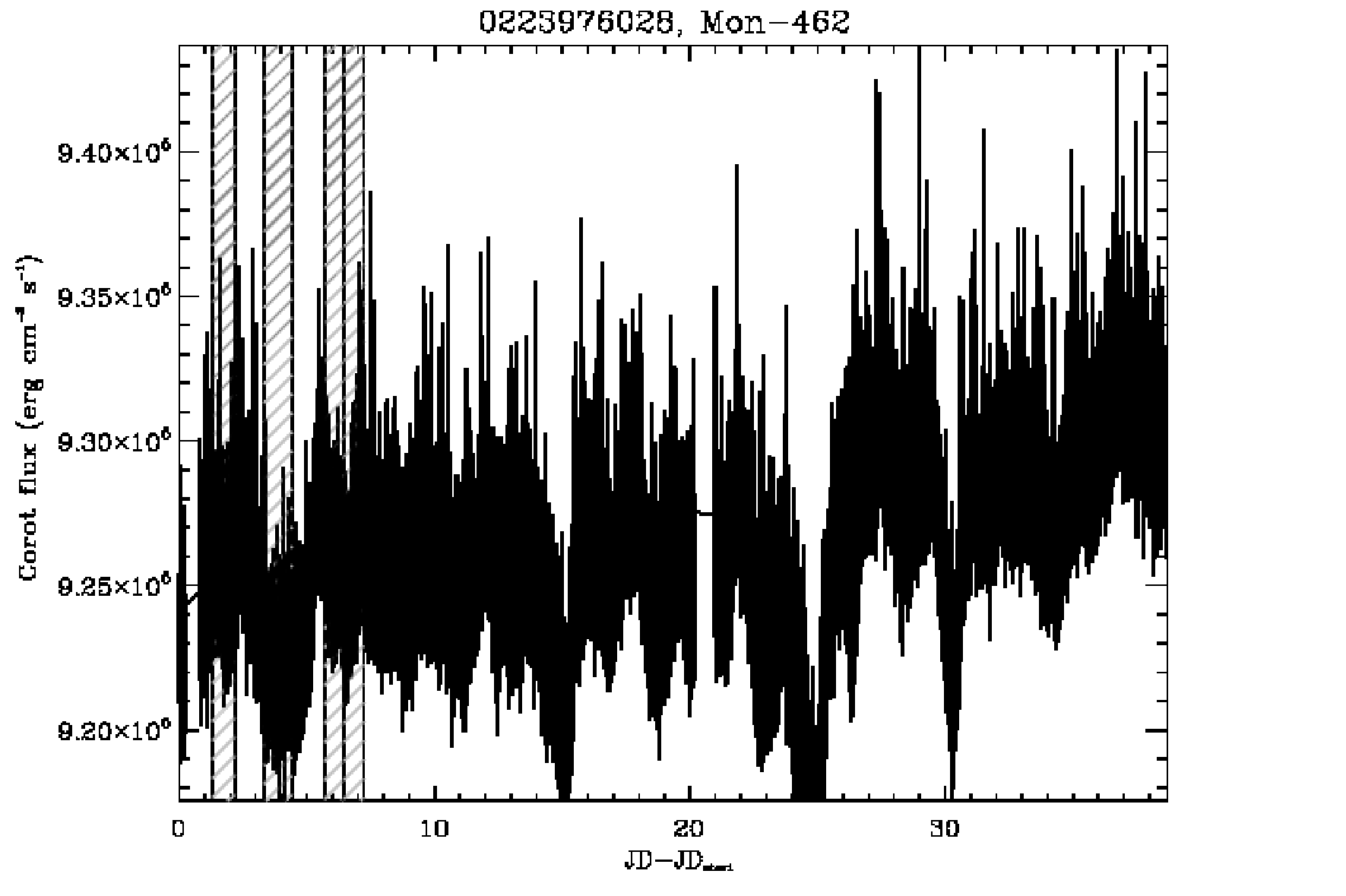}
\includegraphics[width=9.0cm]{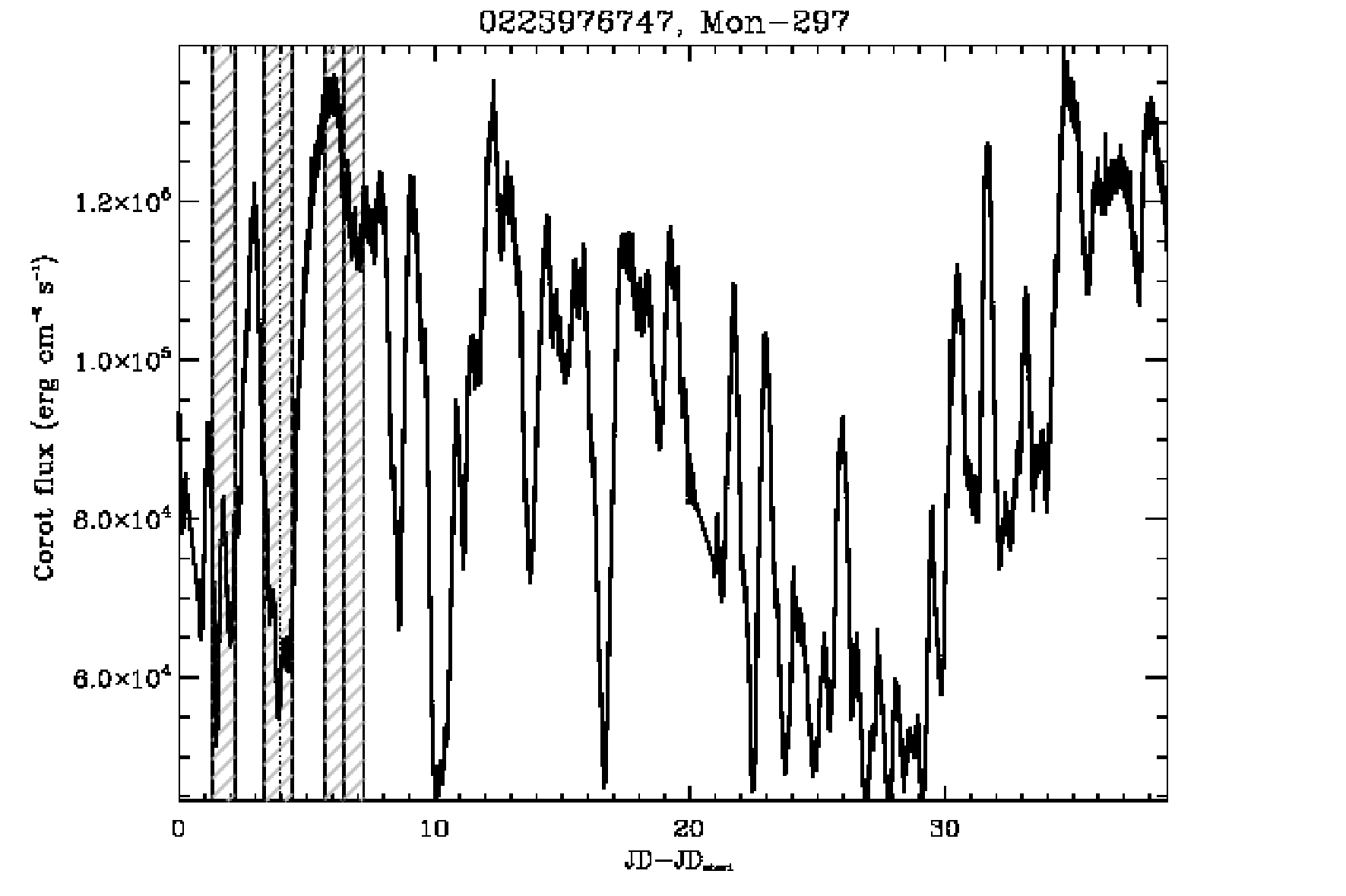}
\includegraphics[width=9.0cm]{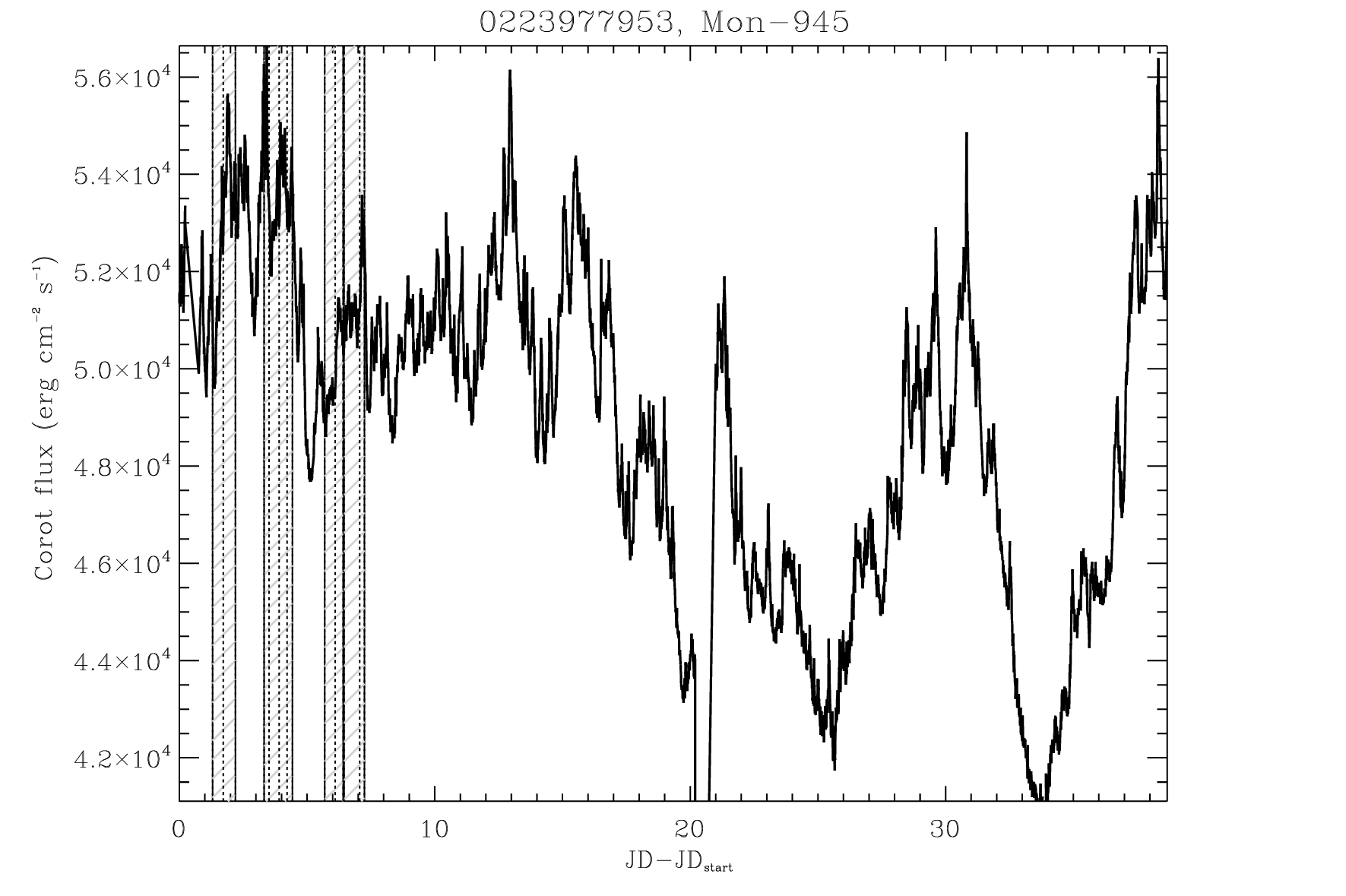}
\includegraphics[width=9.0cm]{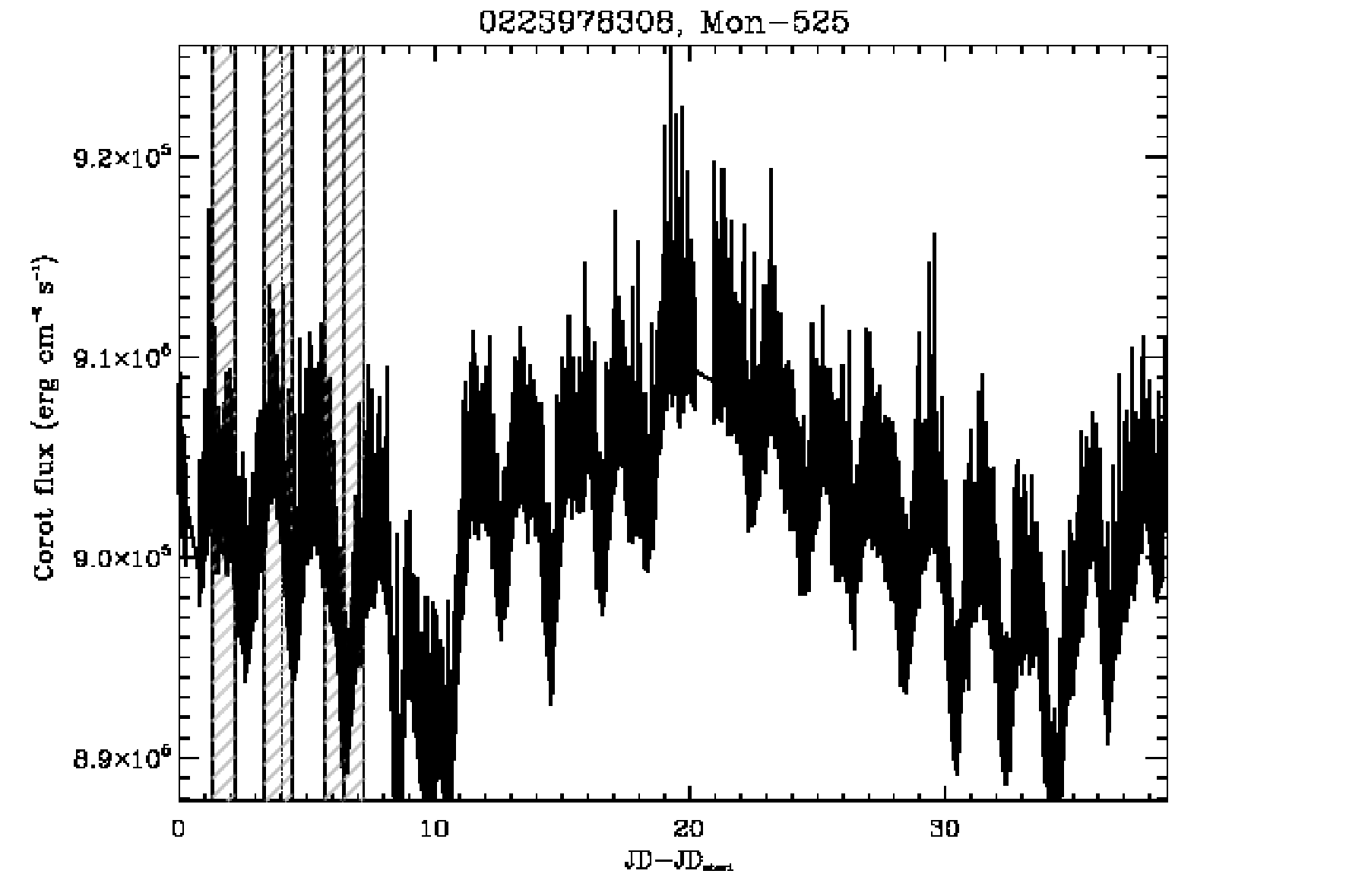}
\includegraphics[width=9.0cm]{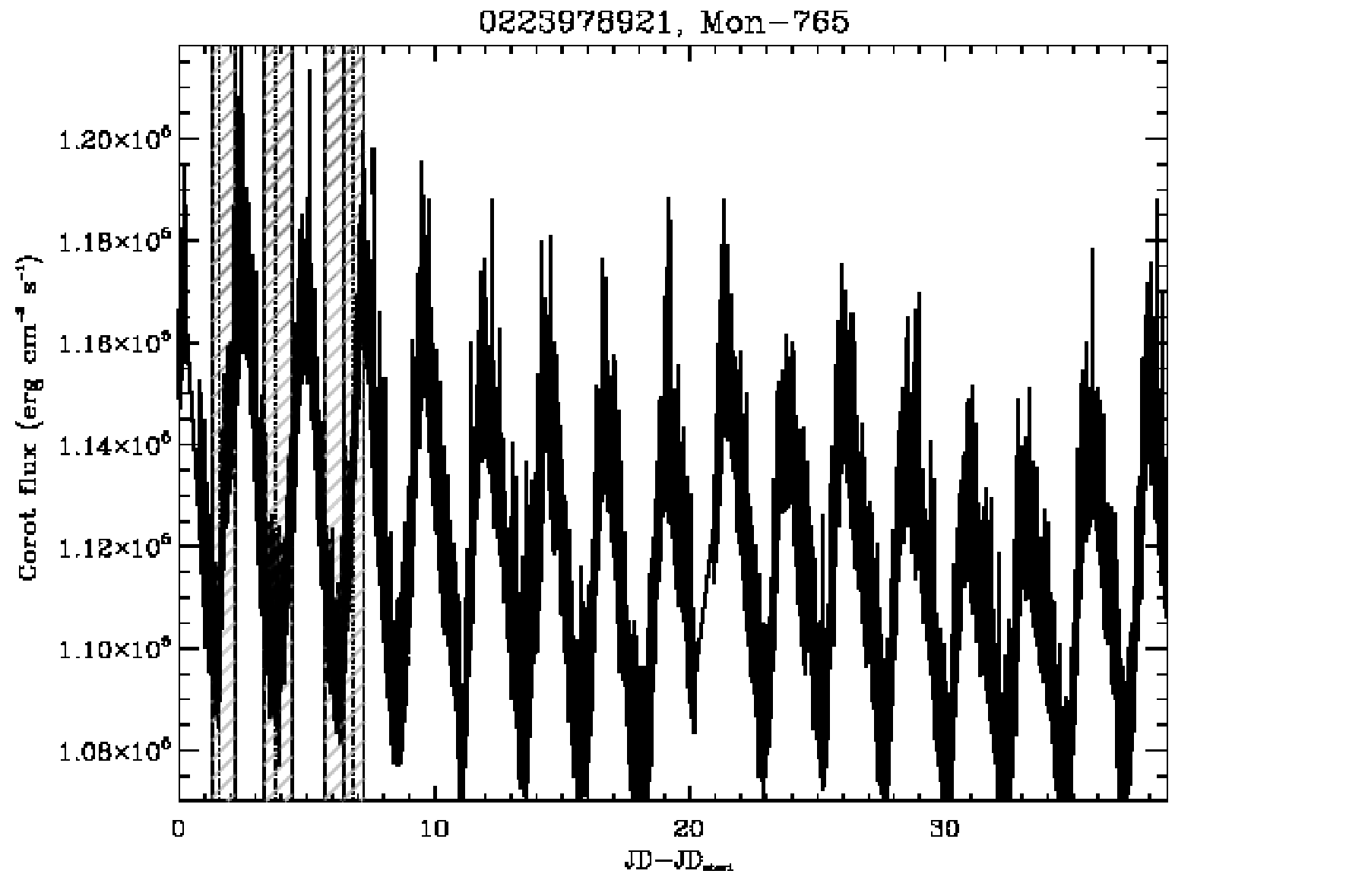}
\includegraphics[width=9.0cm]{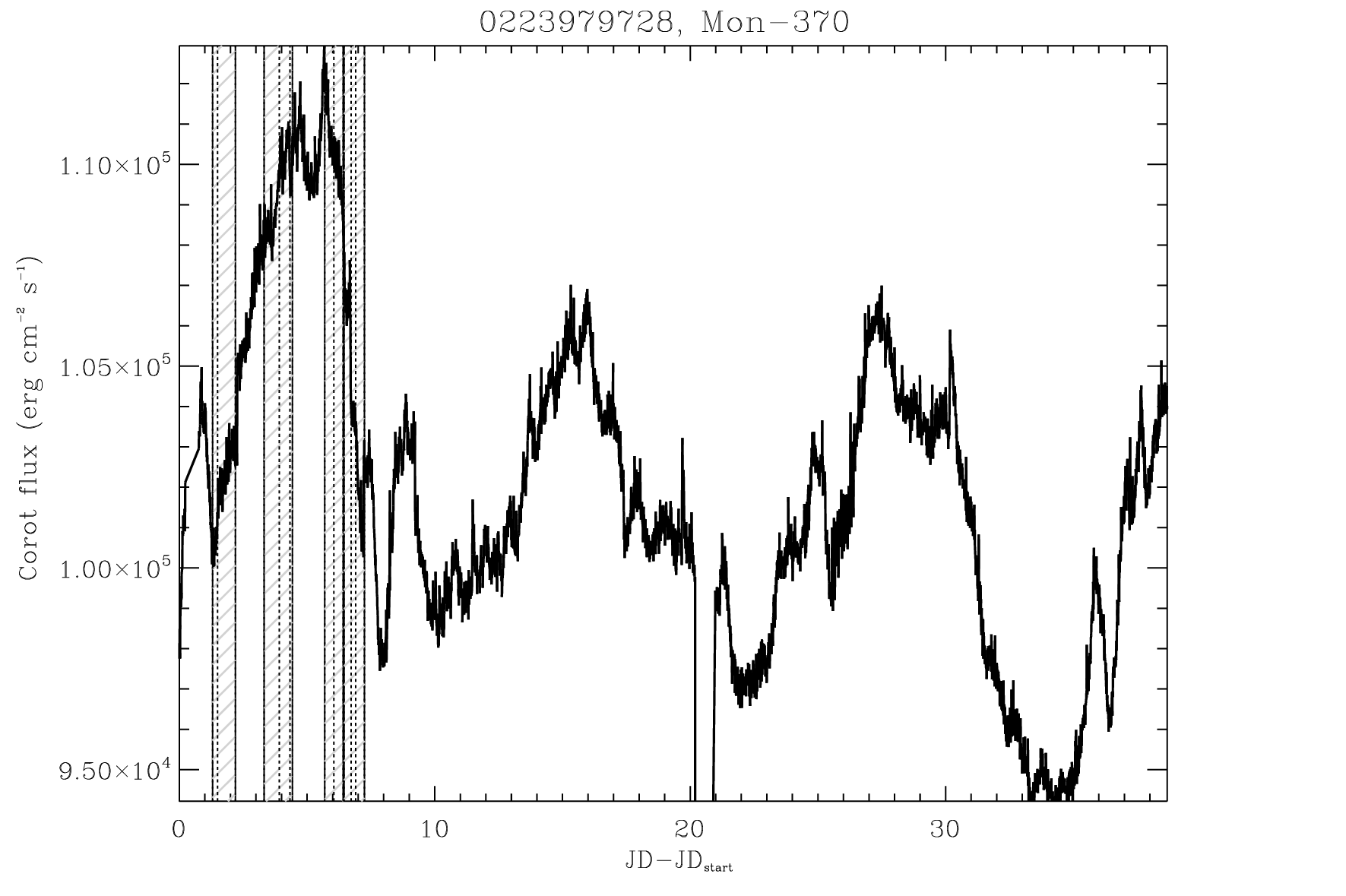}
\includegraphics[width=9.0cm]{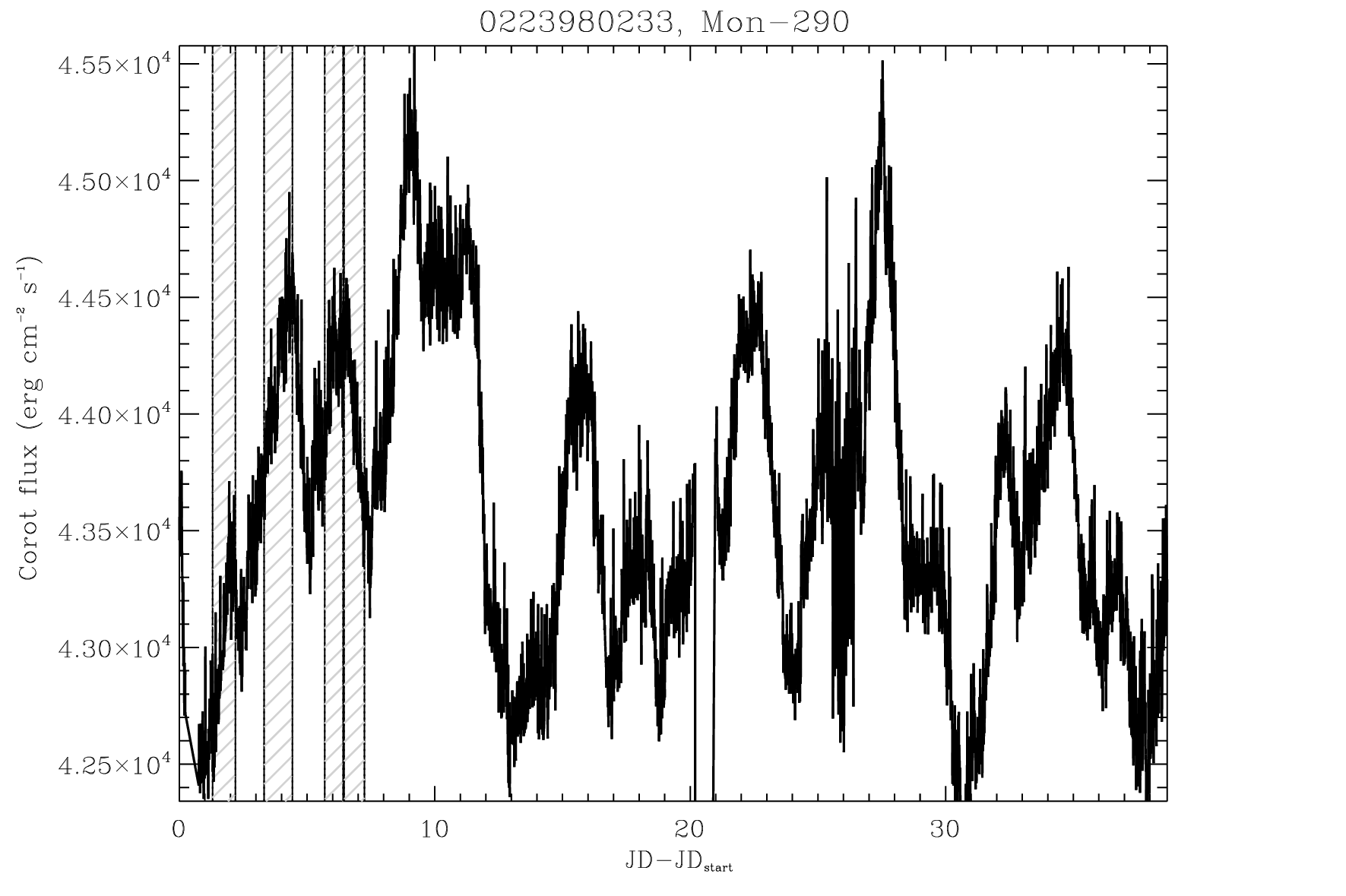}
\label{variab_others_41}
\end{figure}

\begin{figure}[]
\centering	
\includegraphics[width=9.0cm]{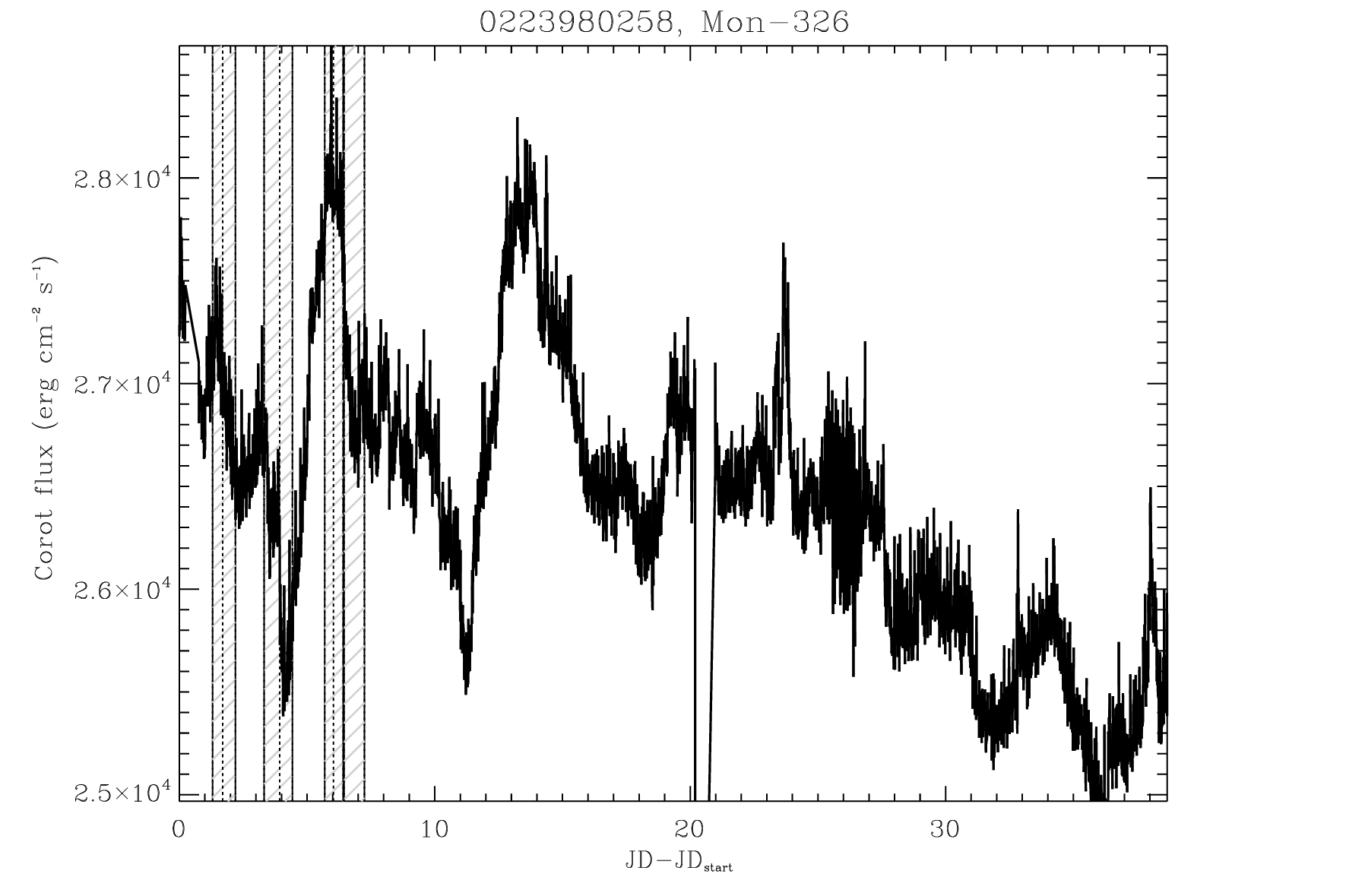}
\includegraphics[width=9.0cm]{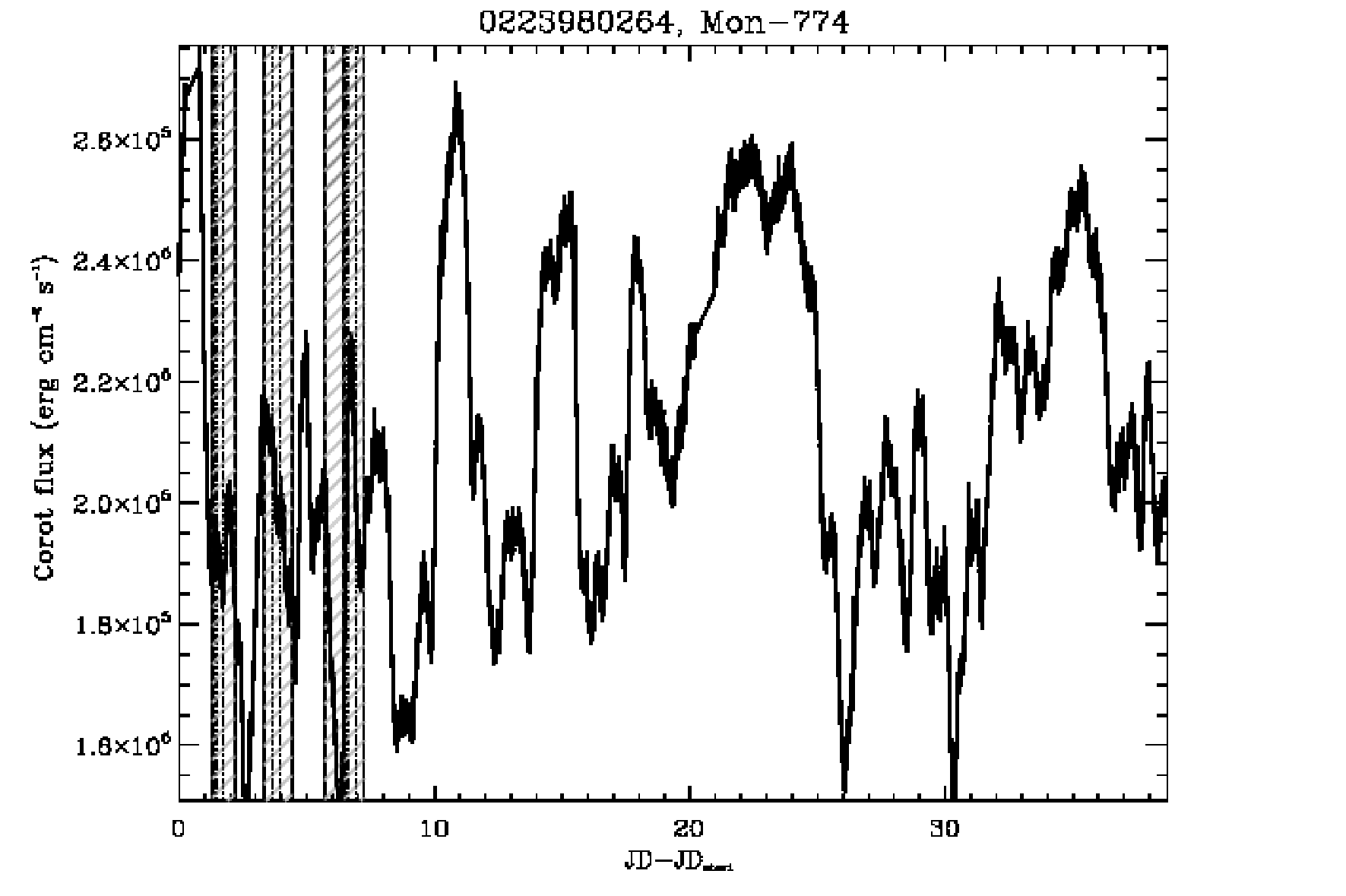}
\includegraphics[width=9.0cm]{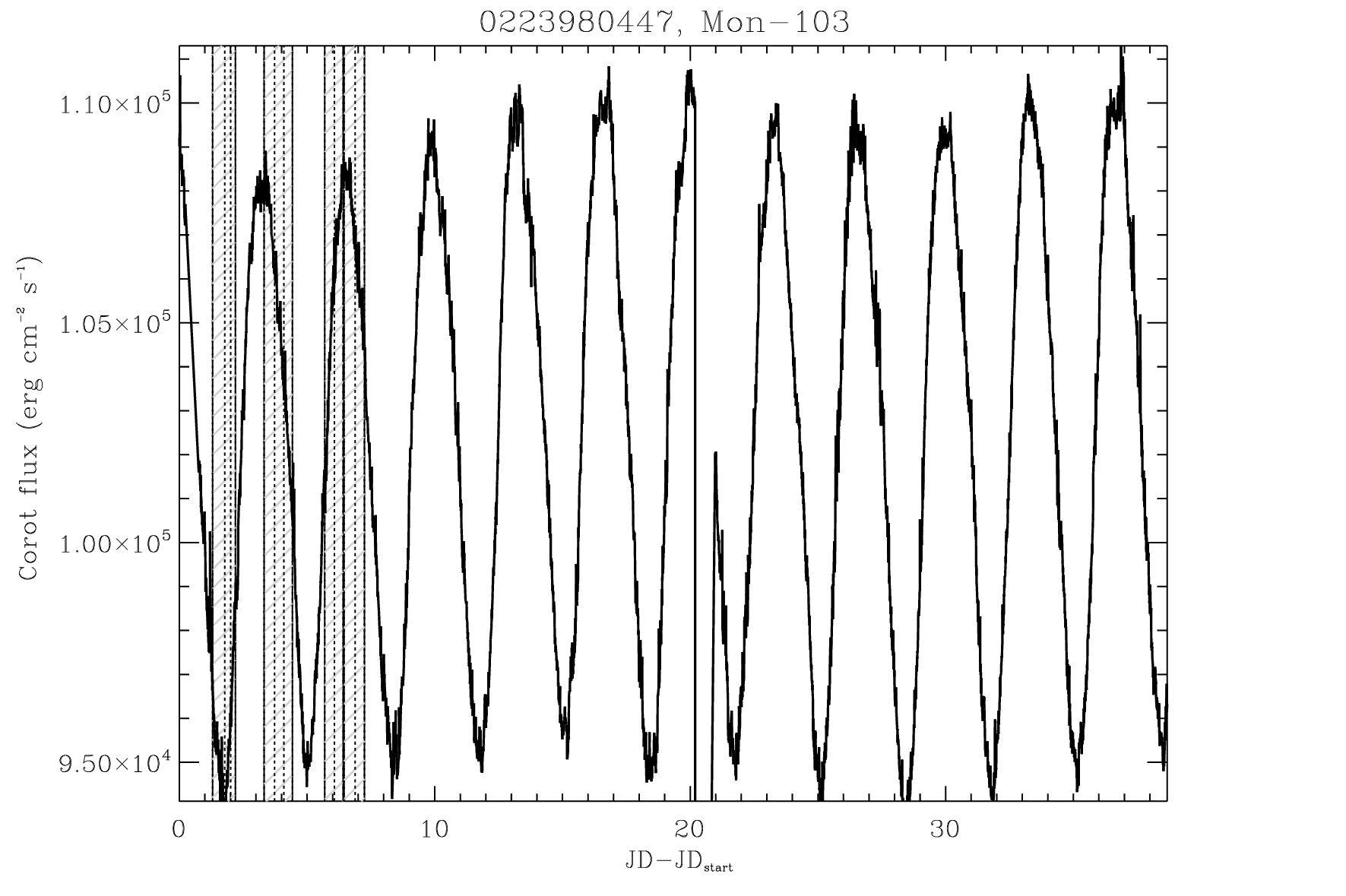}
\includegraphics[width=9.0cm]{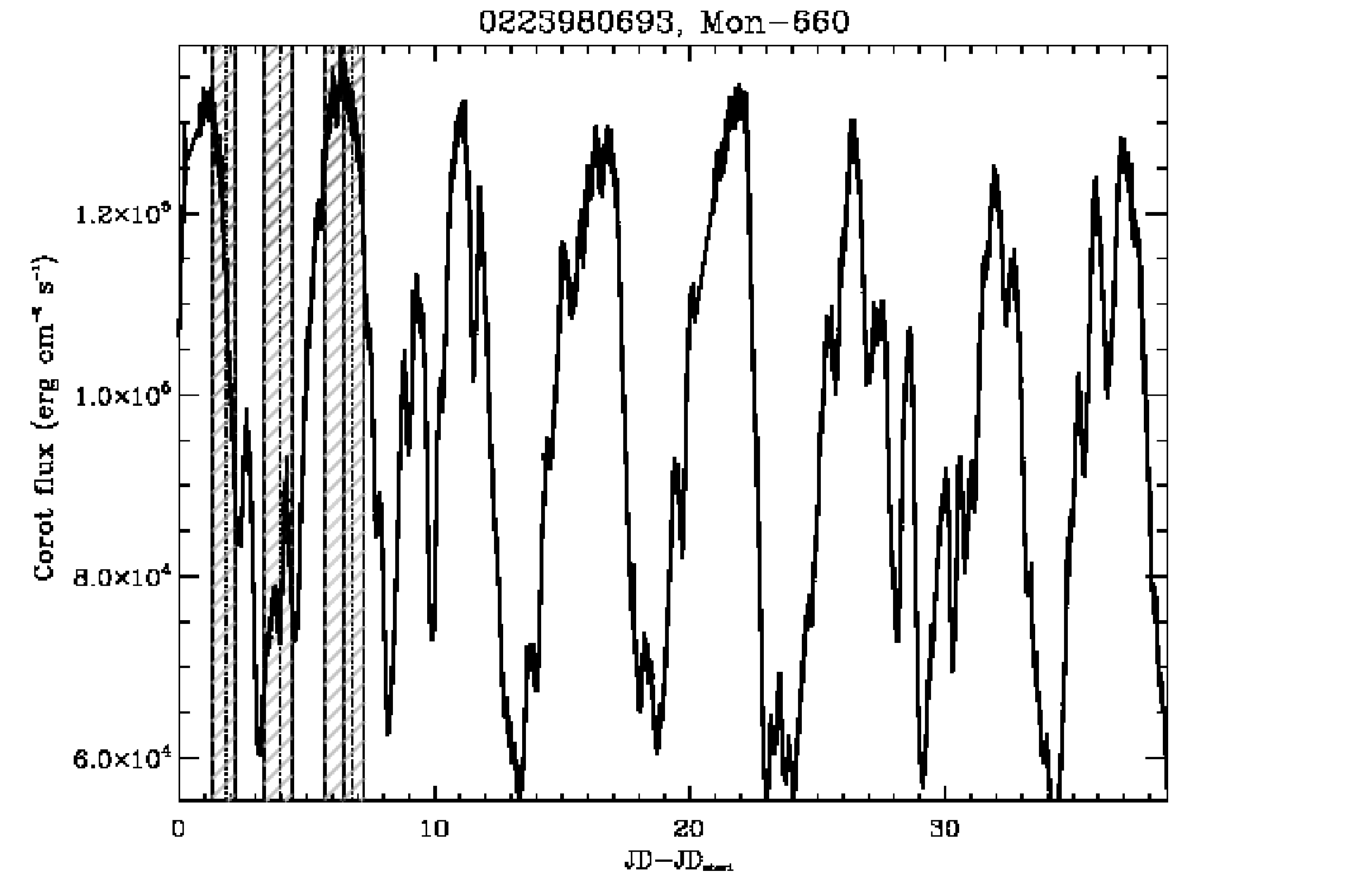}
\includegraphics[width=9.0cm]{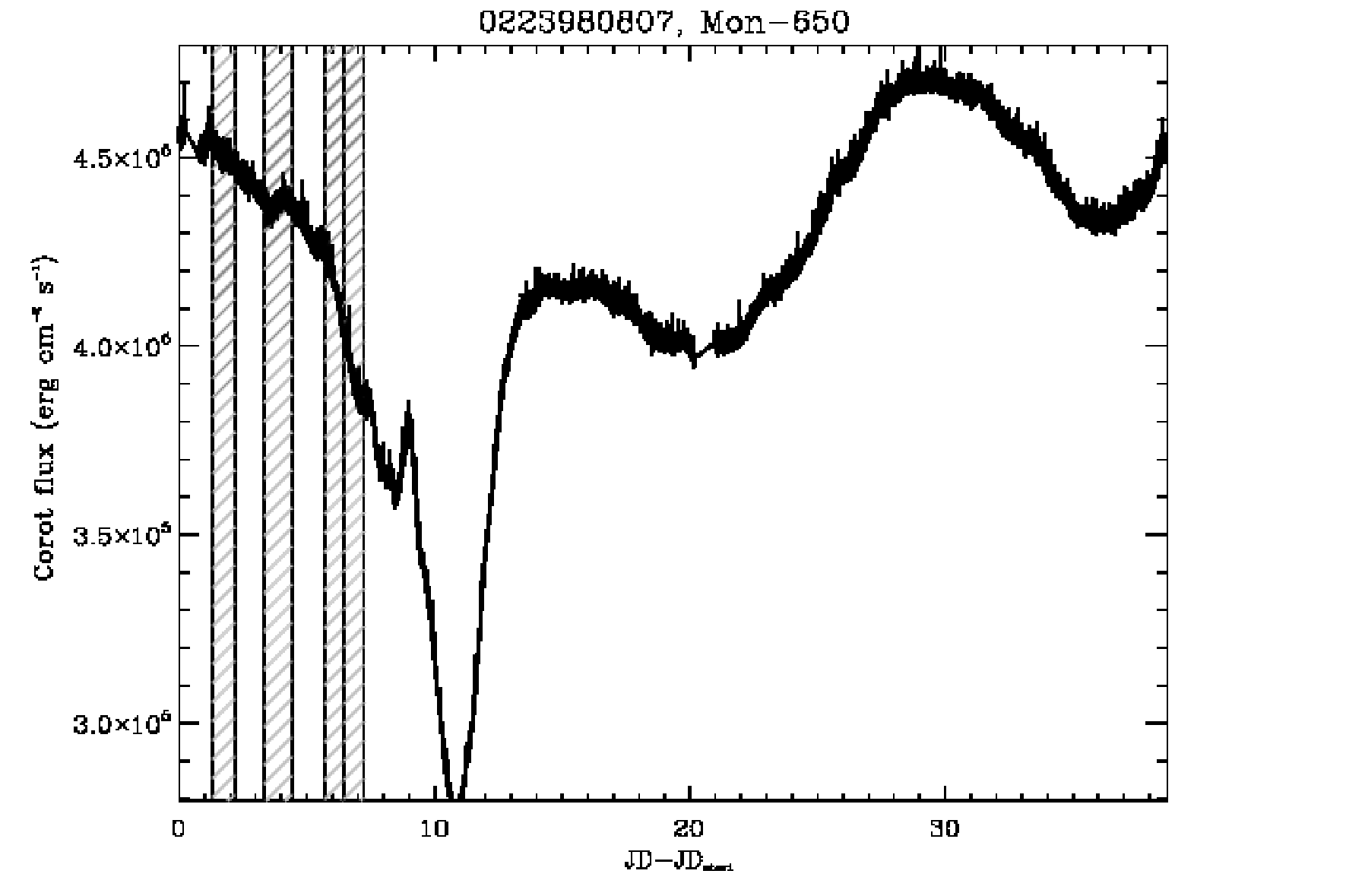}
\includegraphics[width=9.0cm]{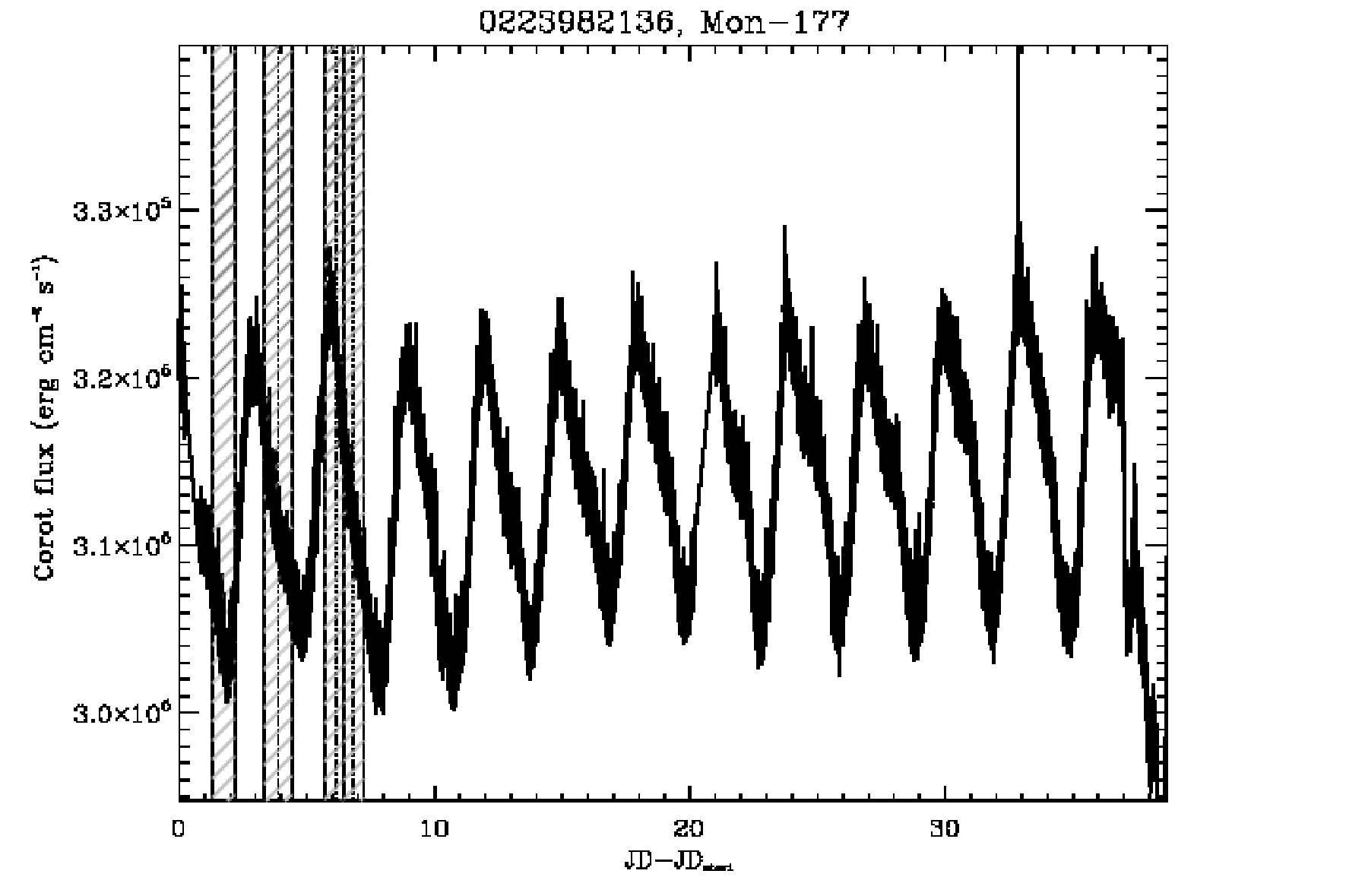}
\includegraphics[width=9.0cm]{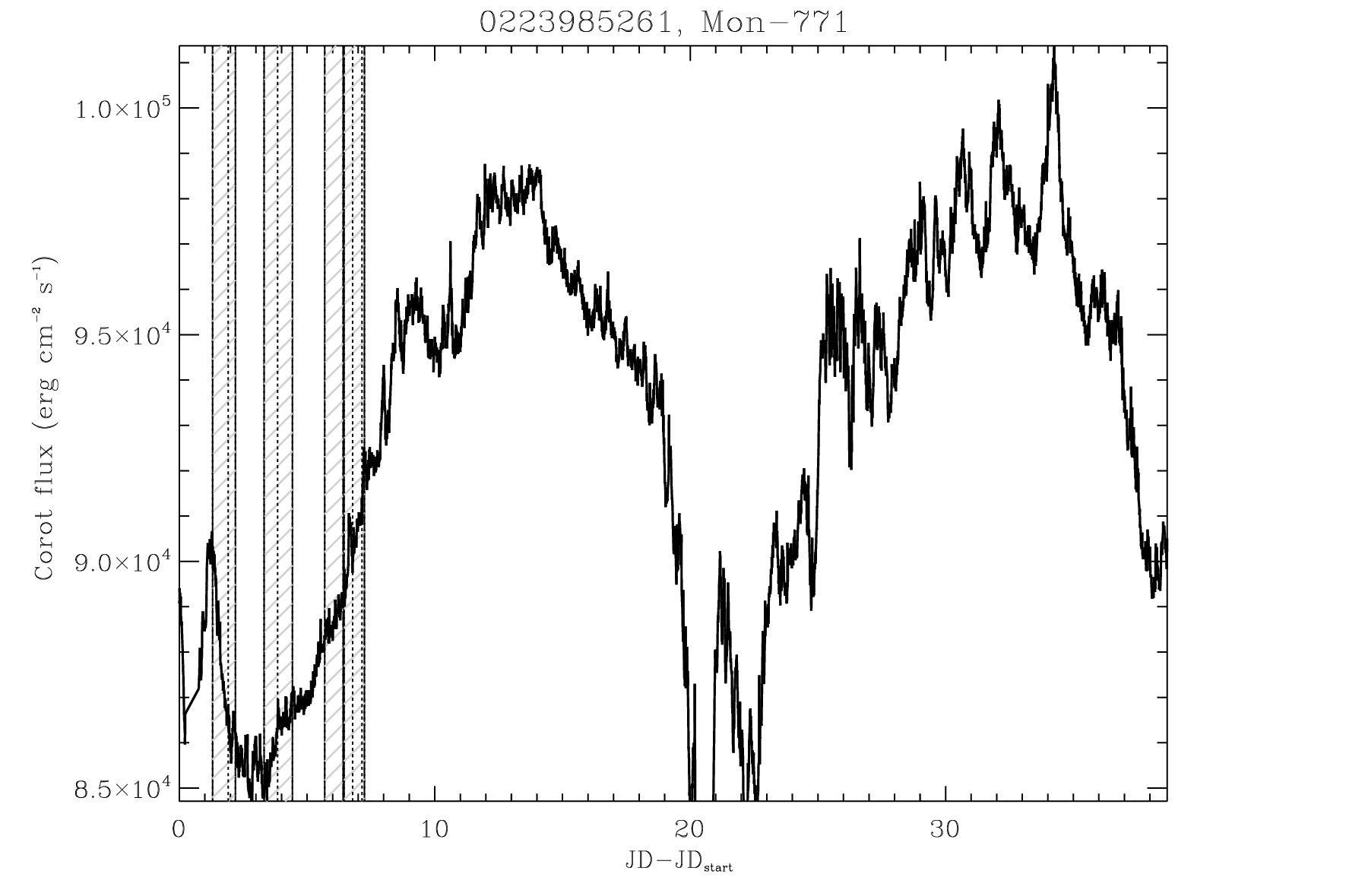}
\includegraphics[width=9.0cm]{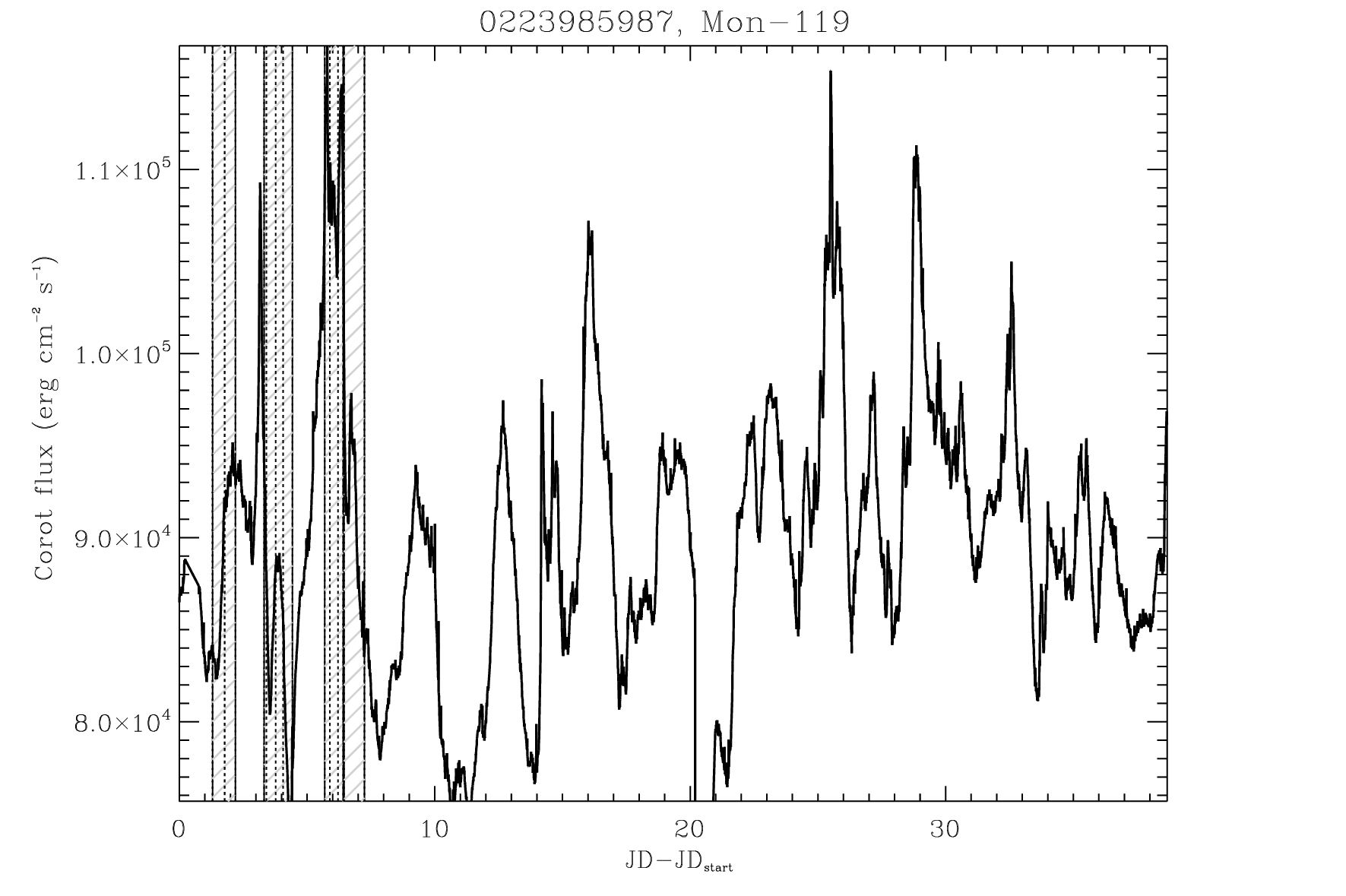}
\label{variab_others_42}
\end{figure}

\begin{figure}[]
\centering	
\includegraphics[width=9.0cm]{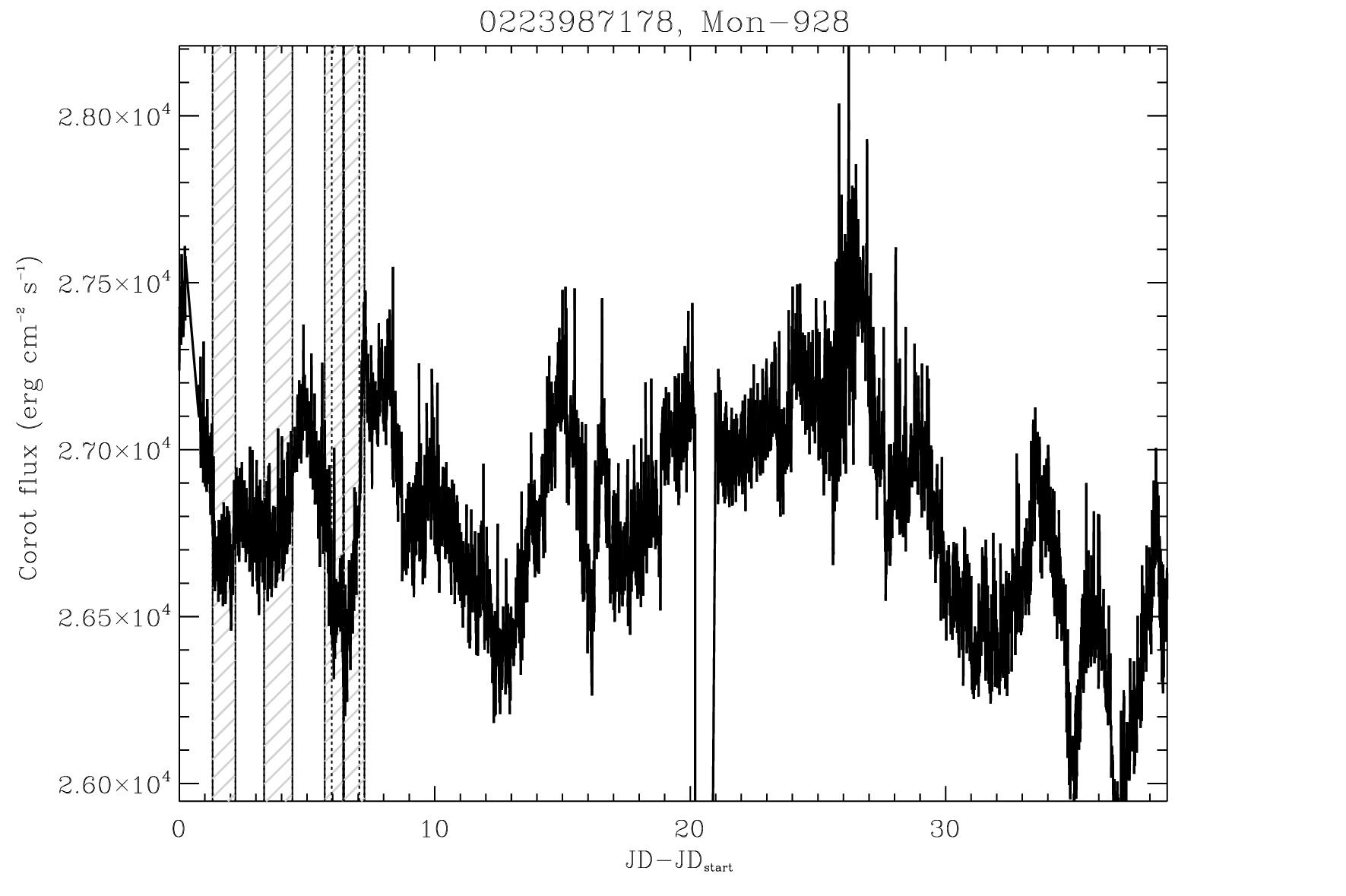}
\includegraphics[width=9.0cm]{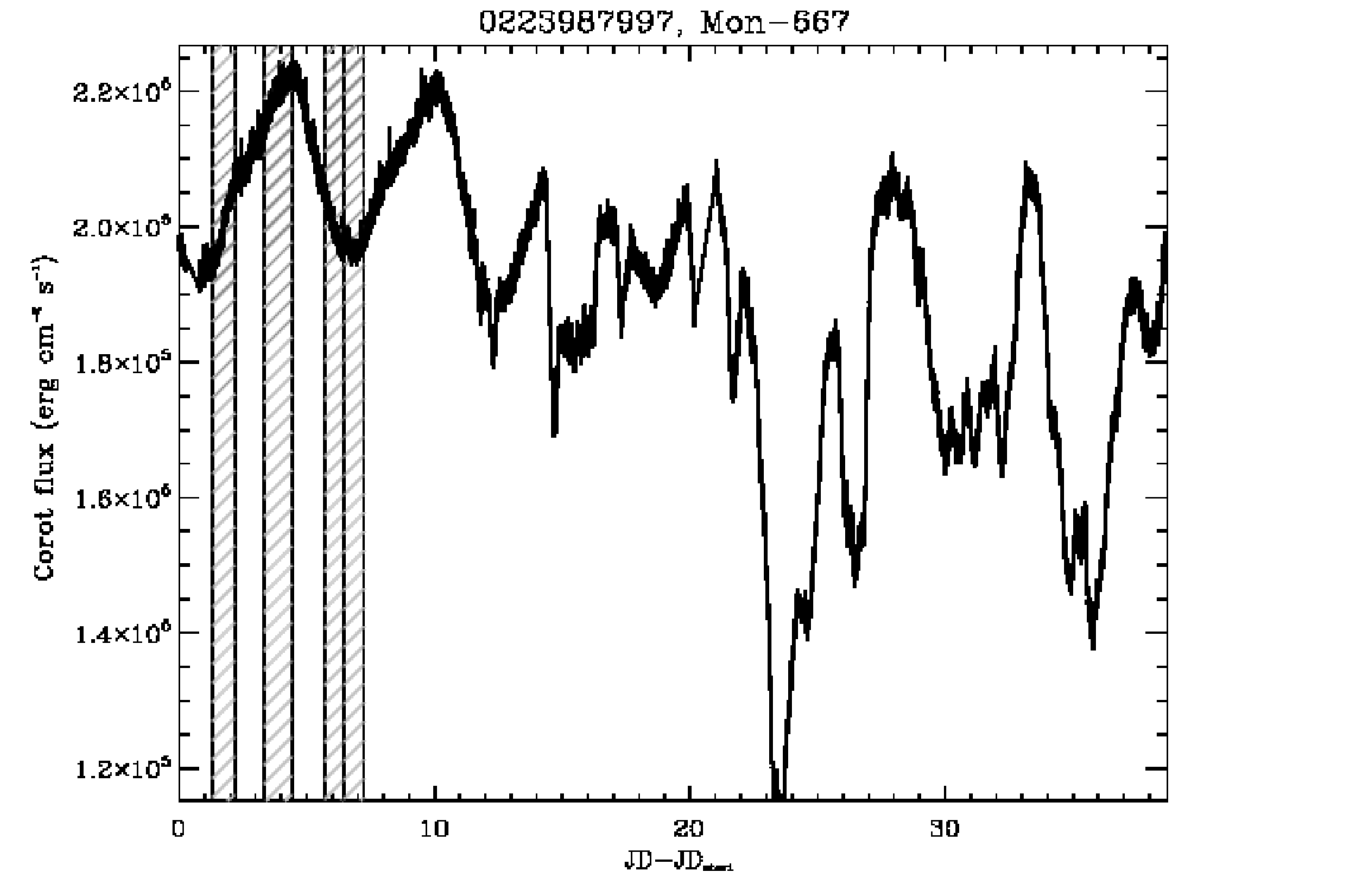}
\includegraphics[width=9.0cm]{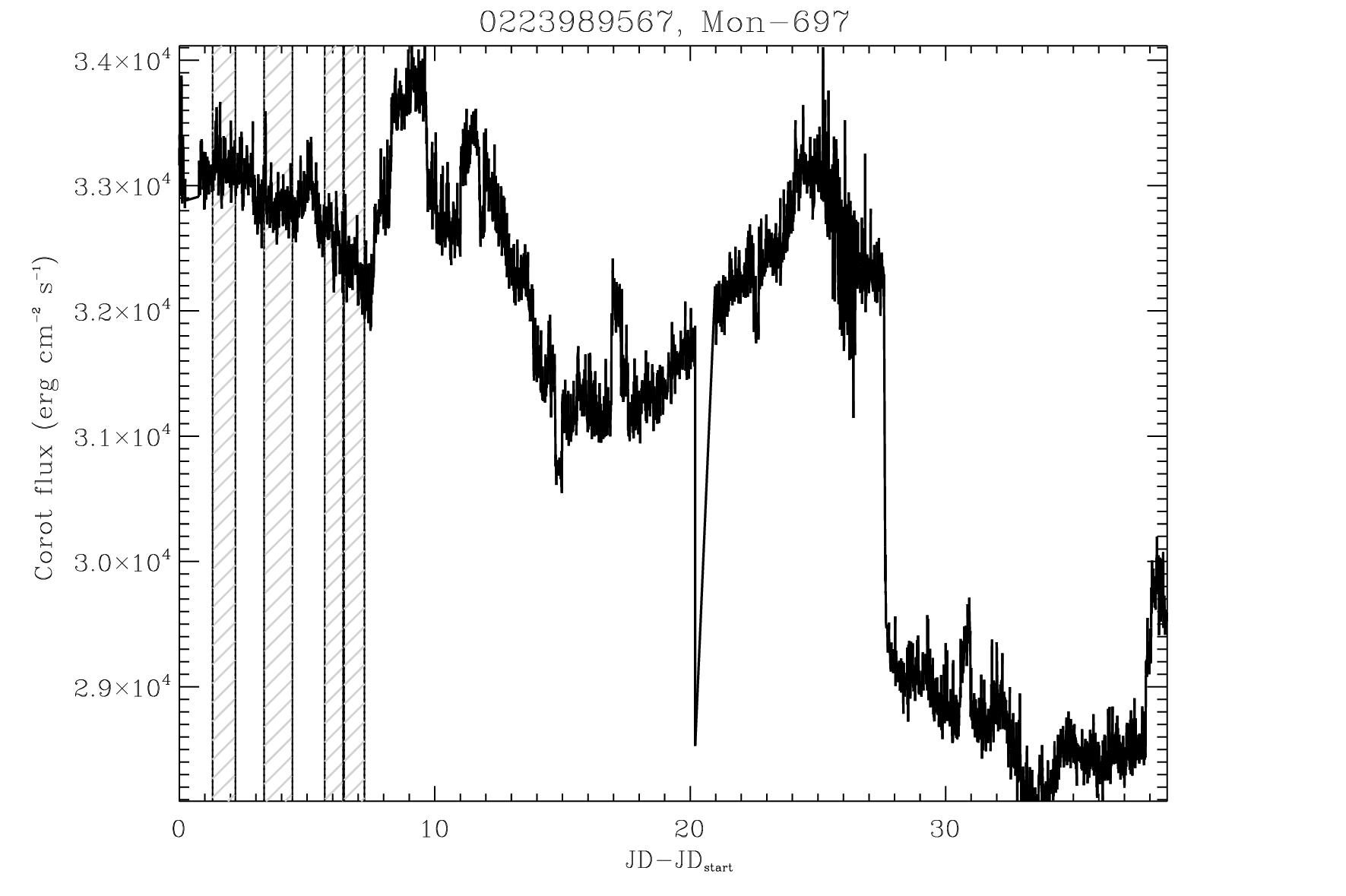}
\includegraphics[width=9.0cm]{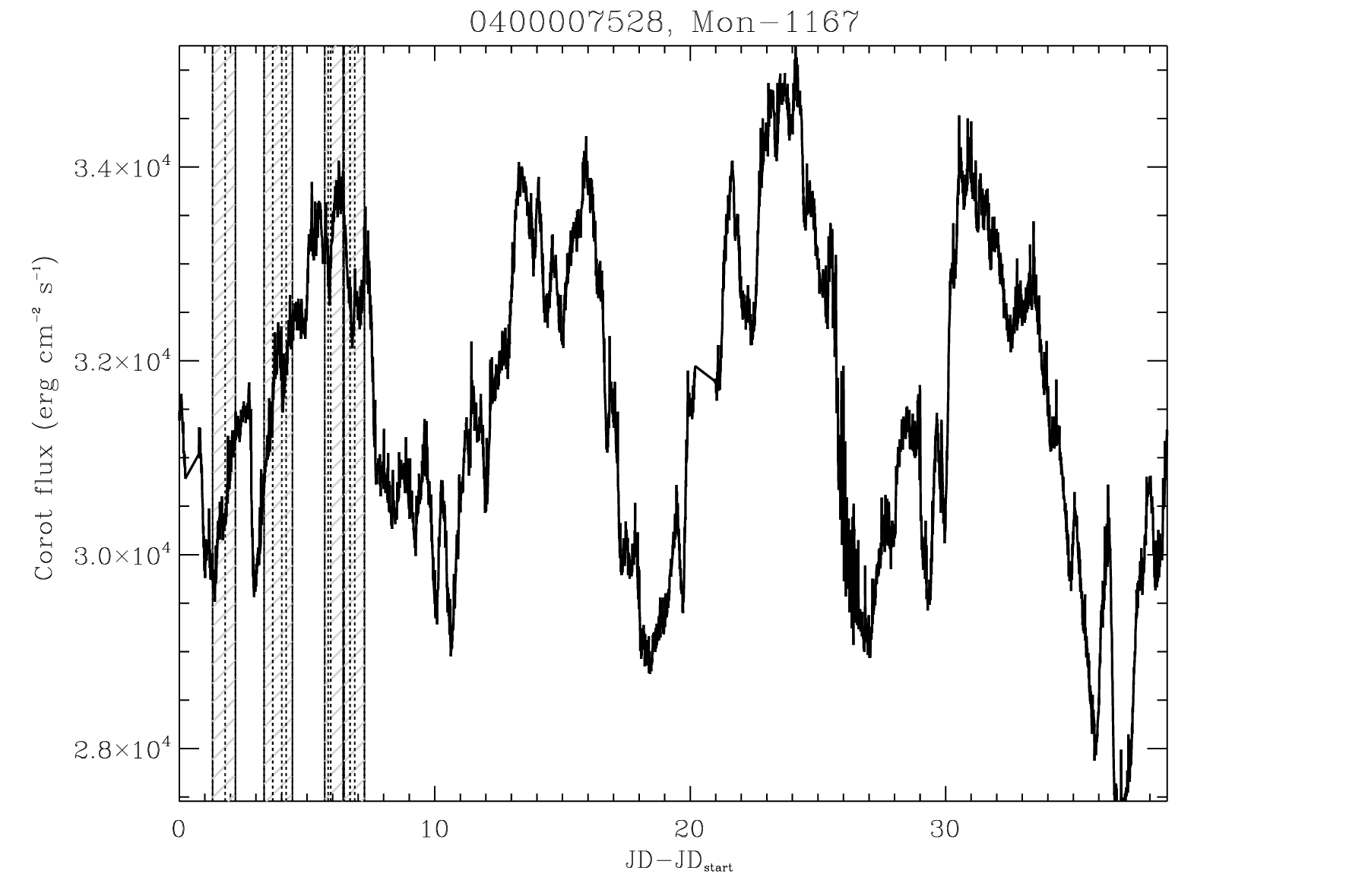}
\includegraphics[width=9.0cm]{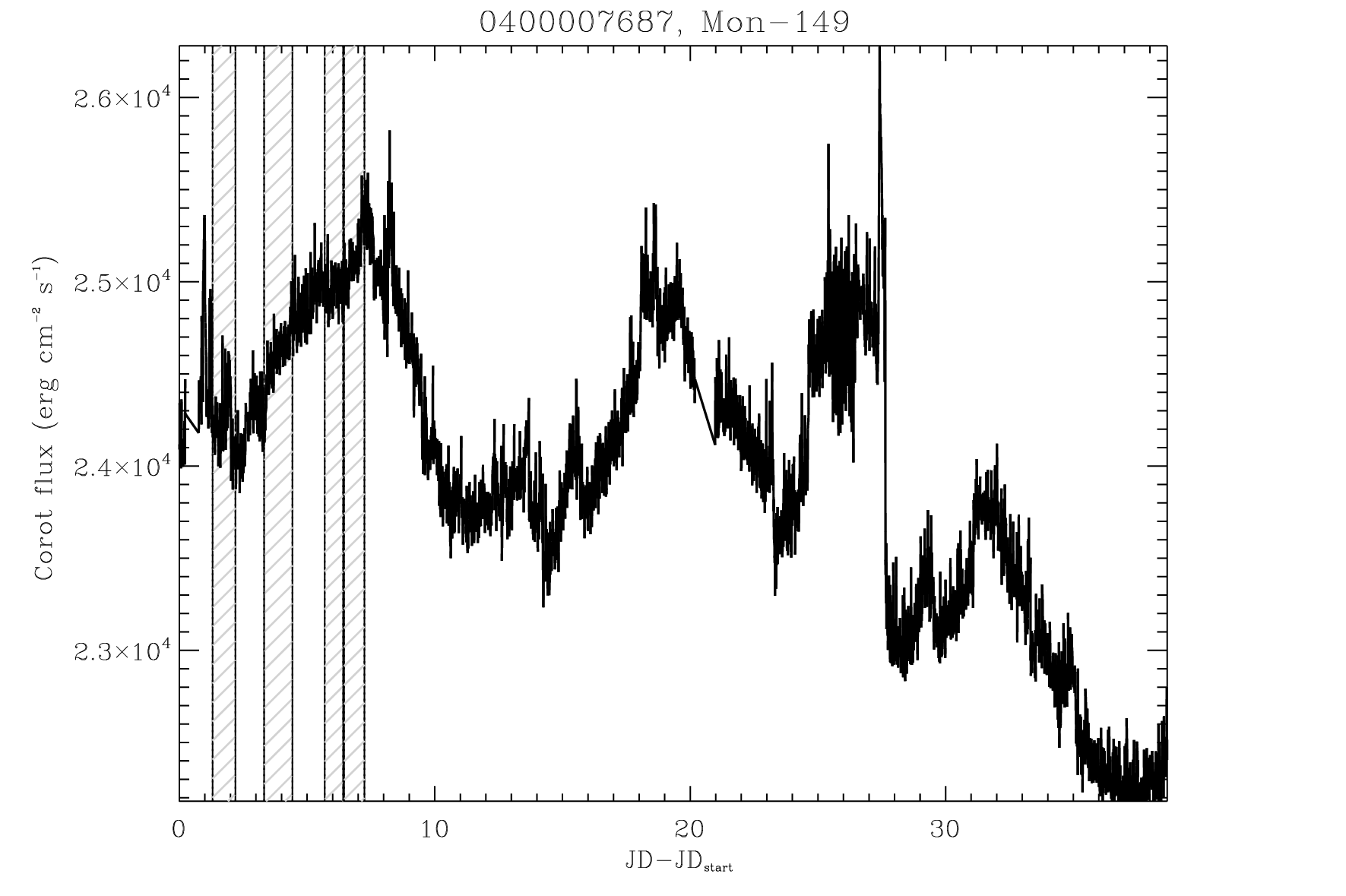}
\includegraphics[width=9.0cm]{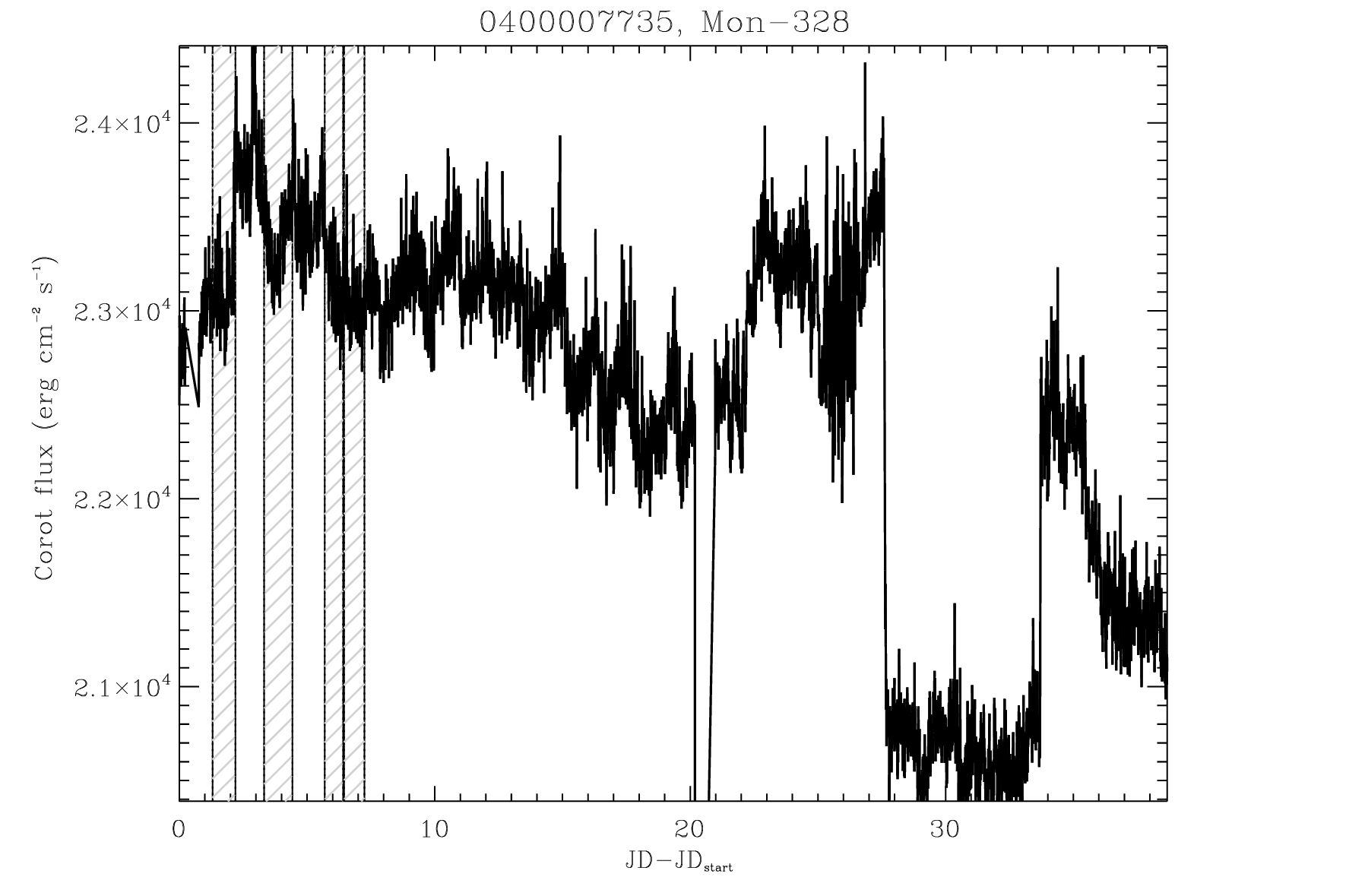}
\includegraphics[width=9.0cm]{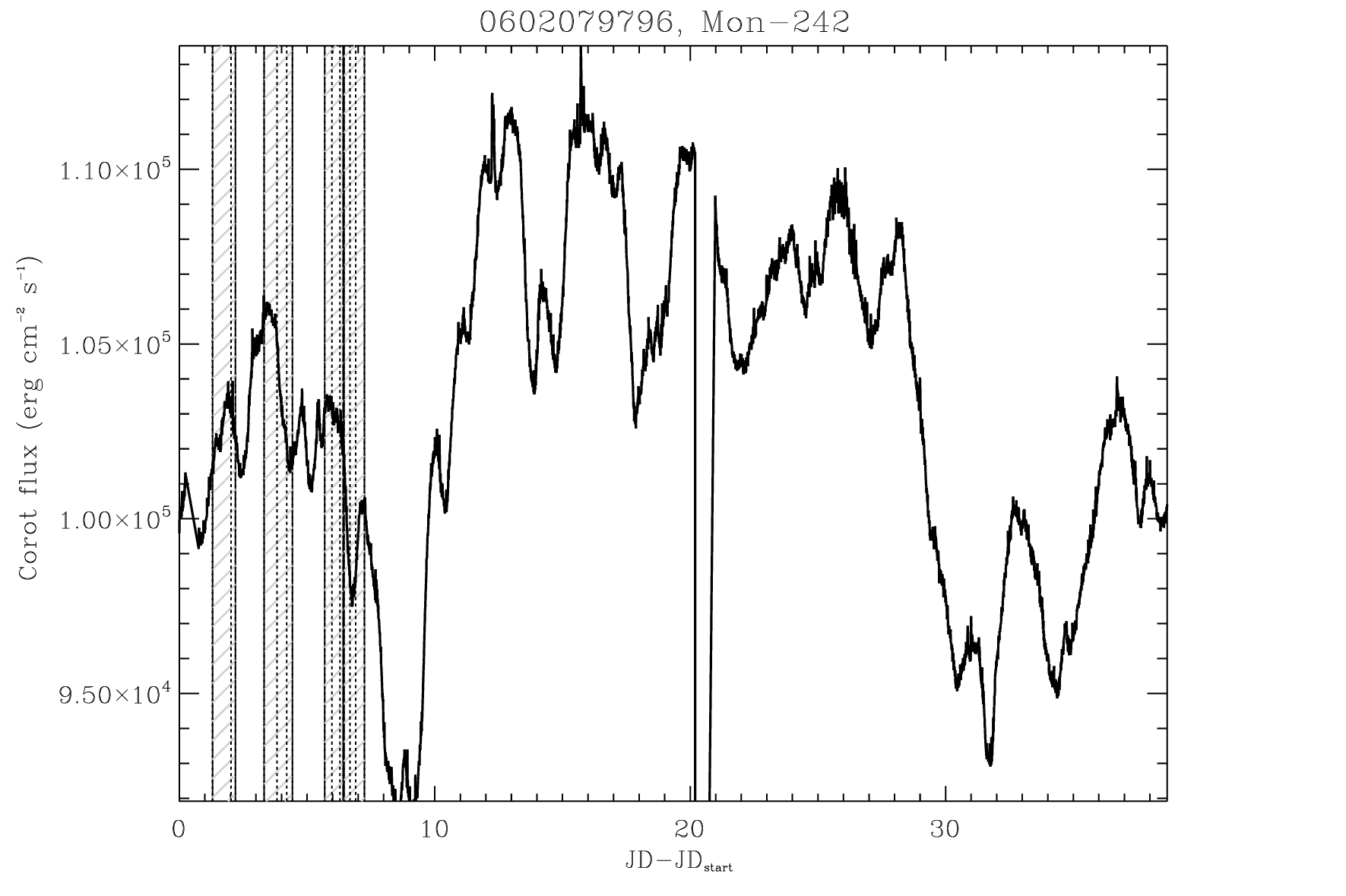}
\includegraphics[width=9.0cm]{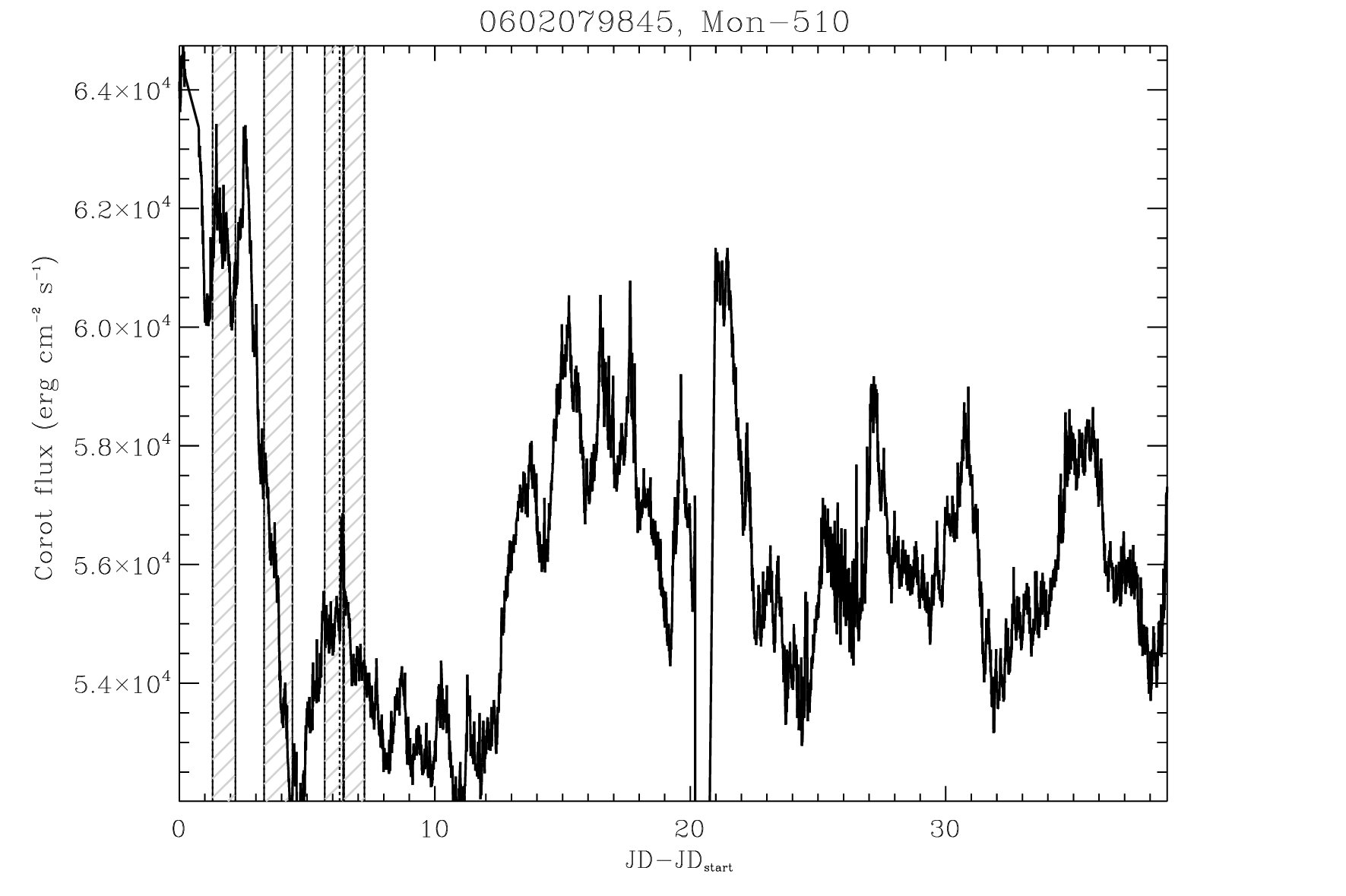}
\label{variab_others_43}
\end{figure}

\begin{figure}[]
\centering	
\includegraphics[width=9.0cm]{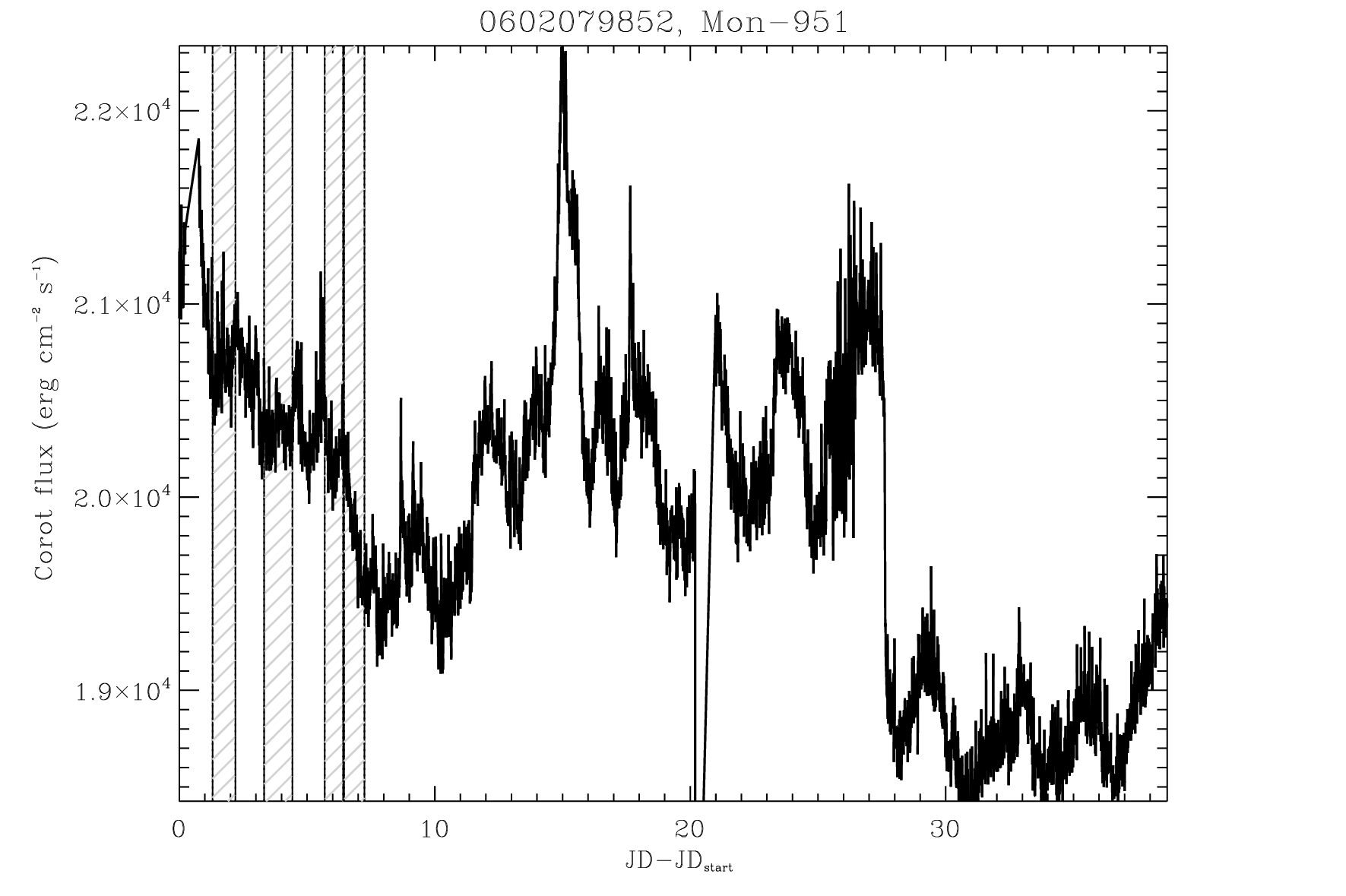}
\includegraphics[width=9.0cm]{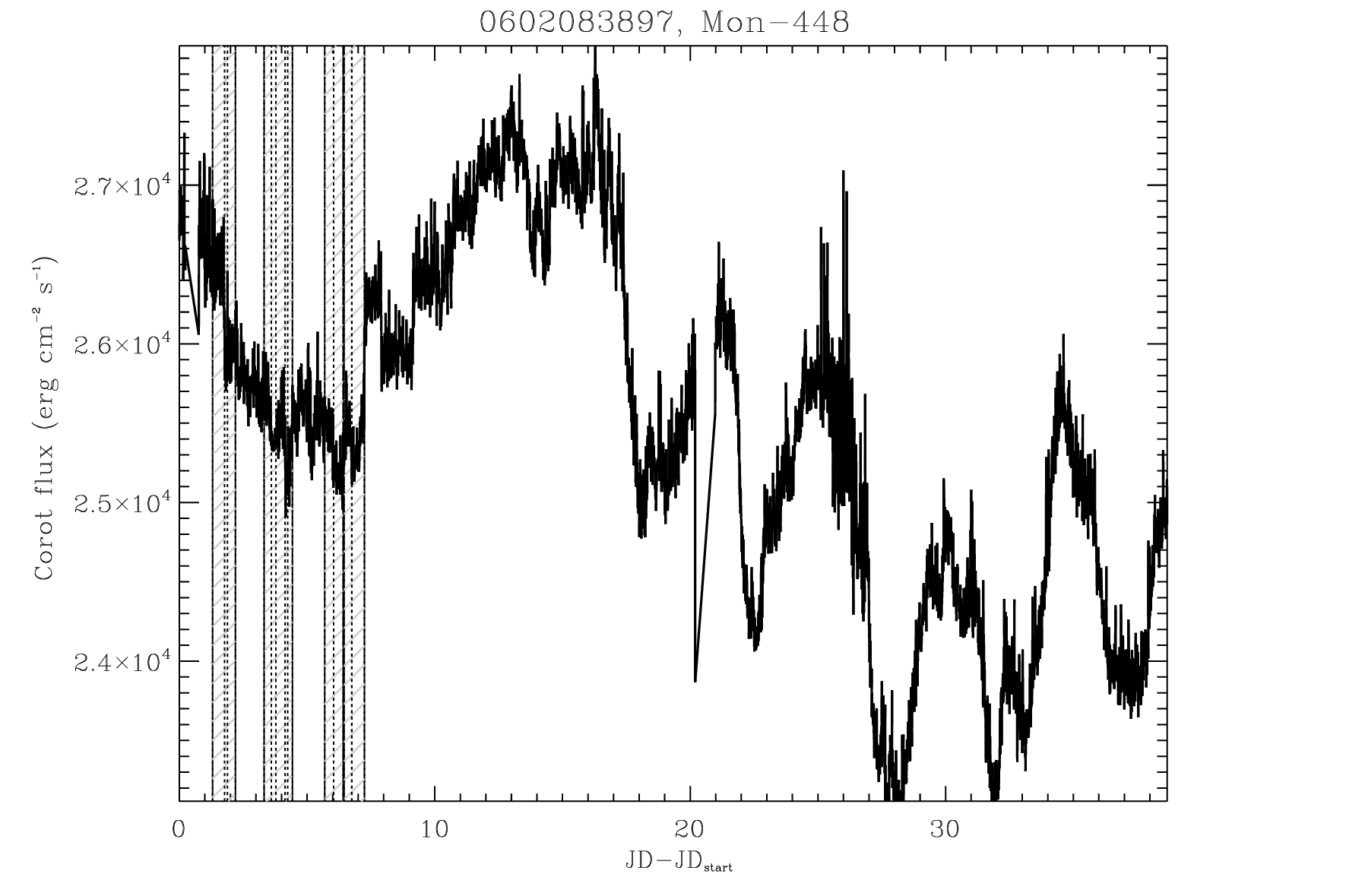}
\includegraphics[width=9.0cm]{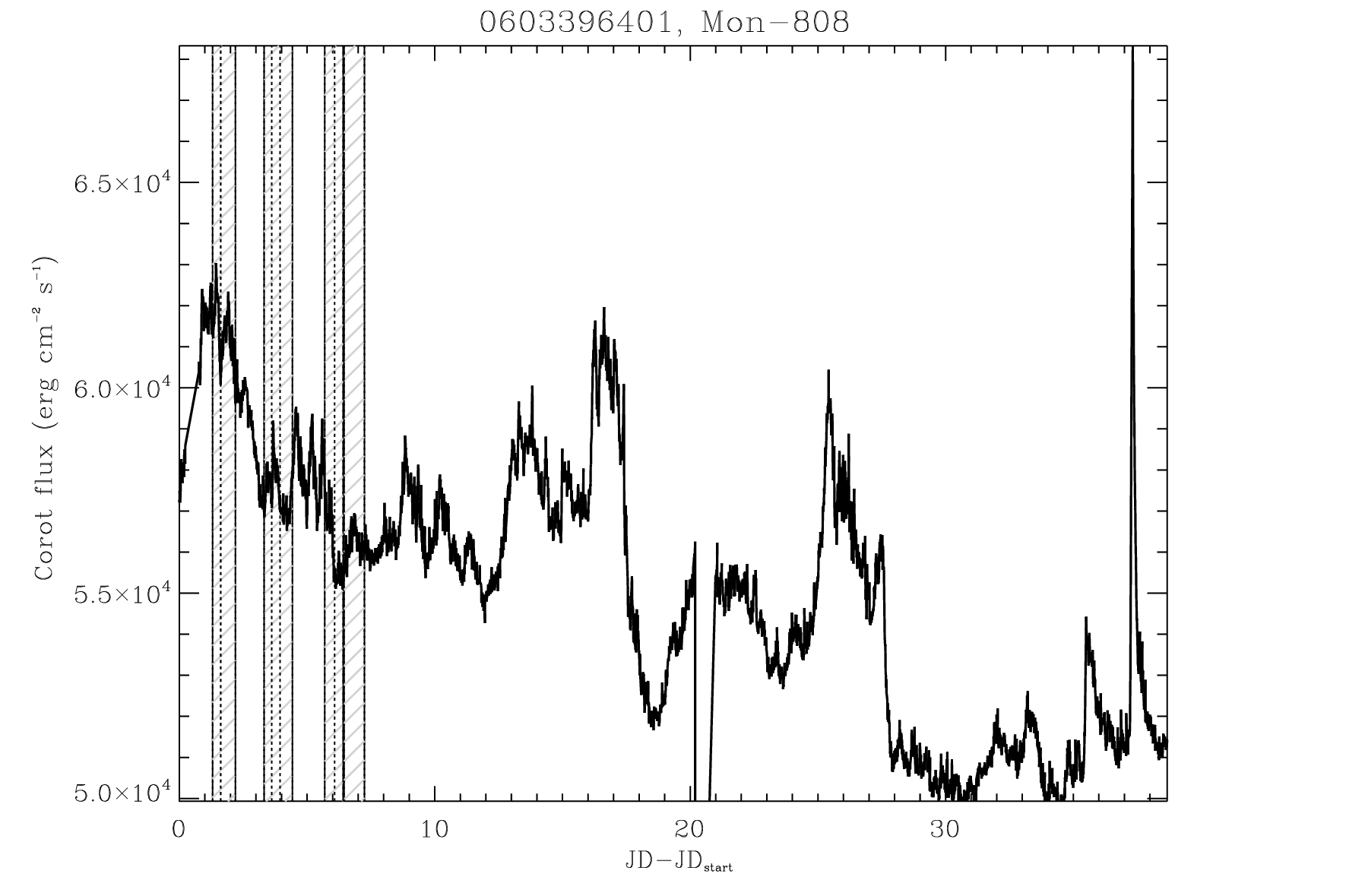}
\includegraphics[width=9.0cm]{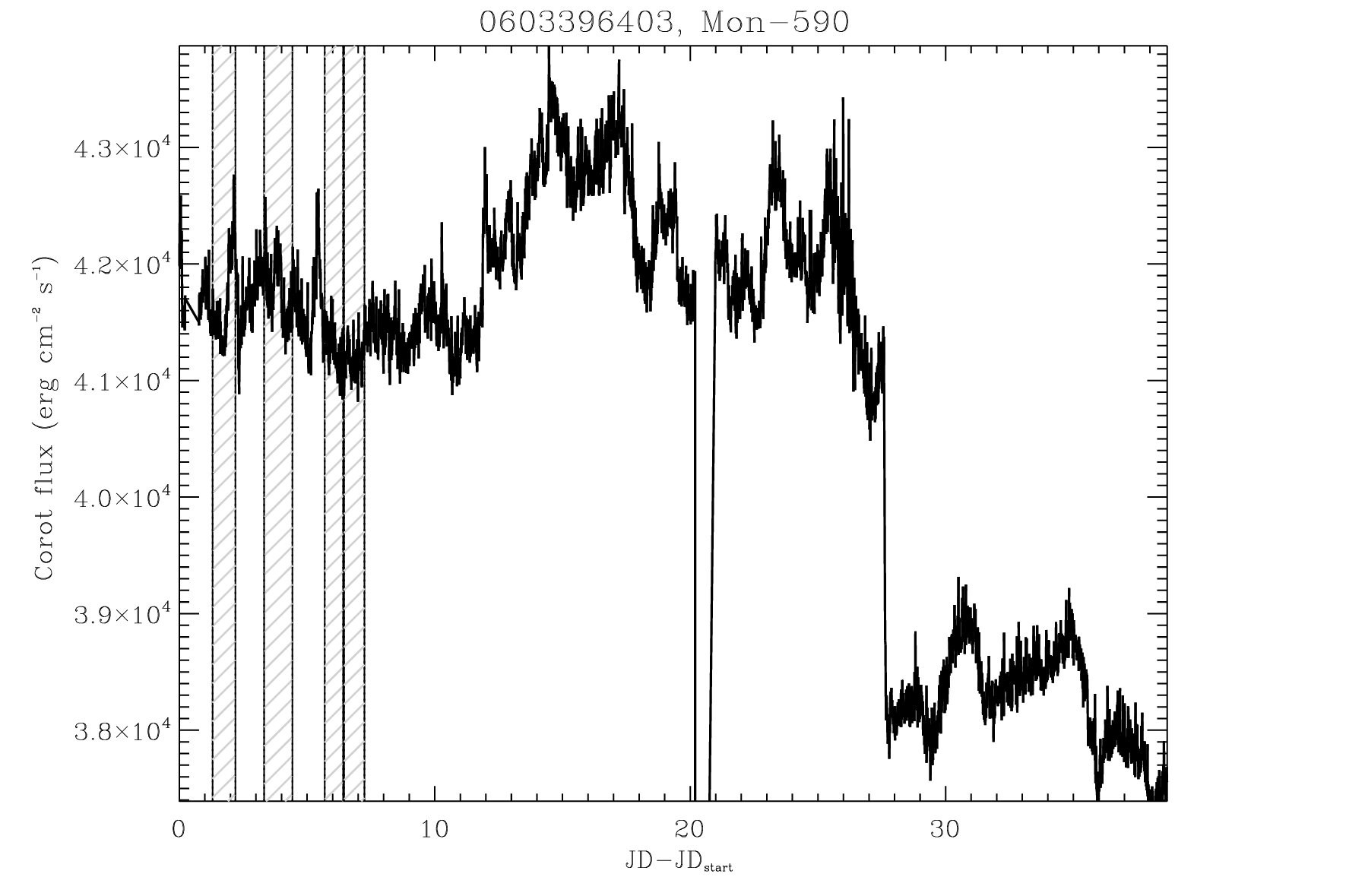}
\includegraphics[width=9.0cm]{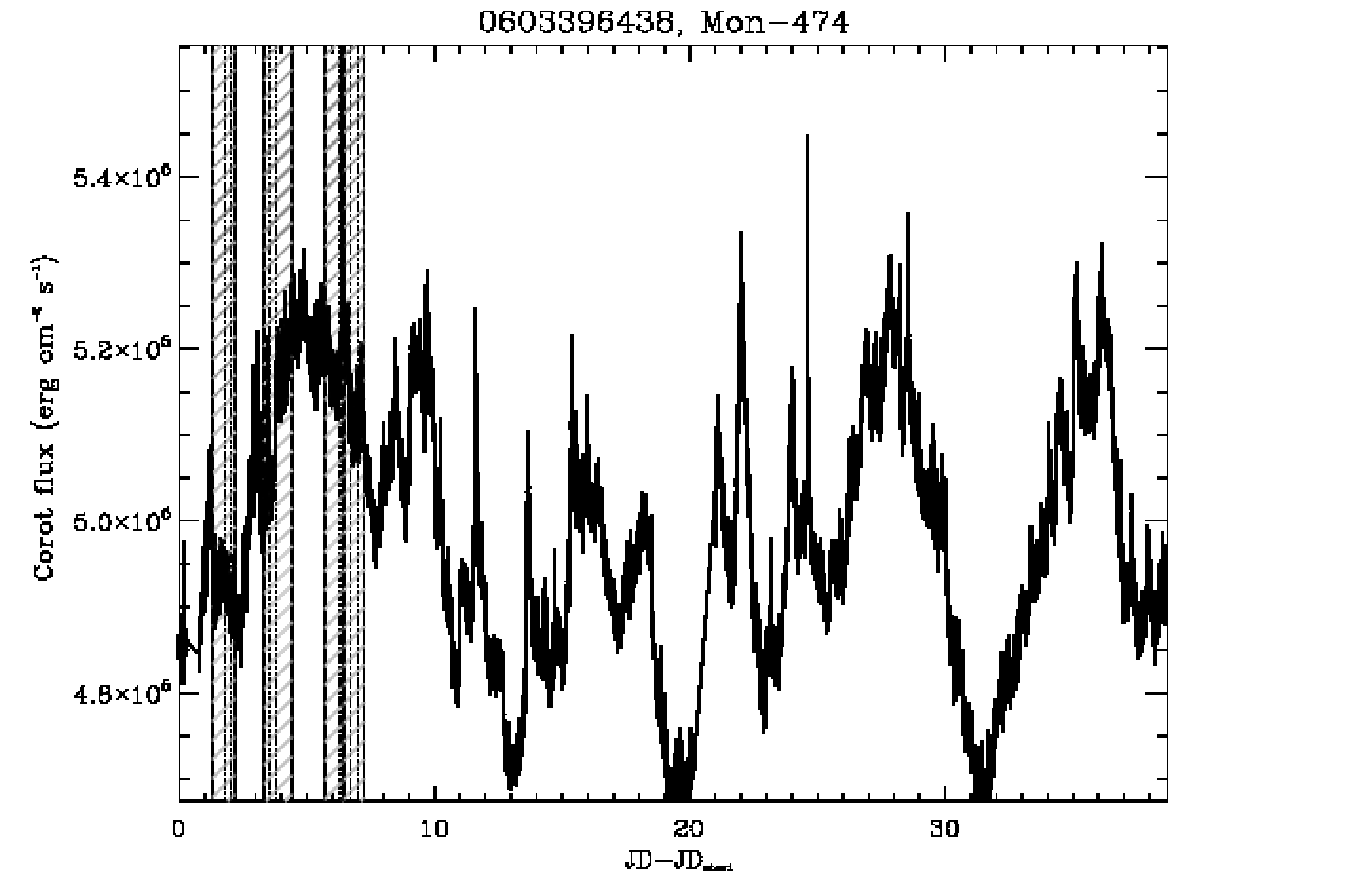}
\includegraphics[width=9.0cm]{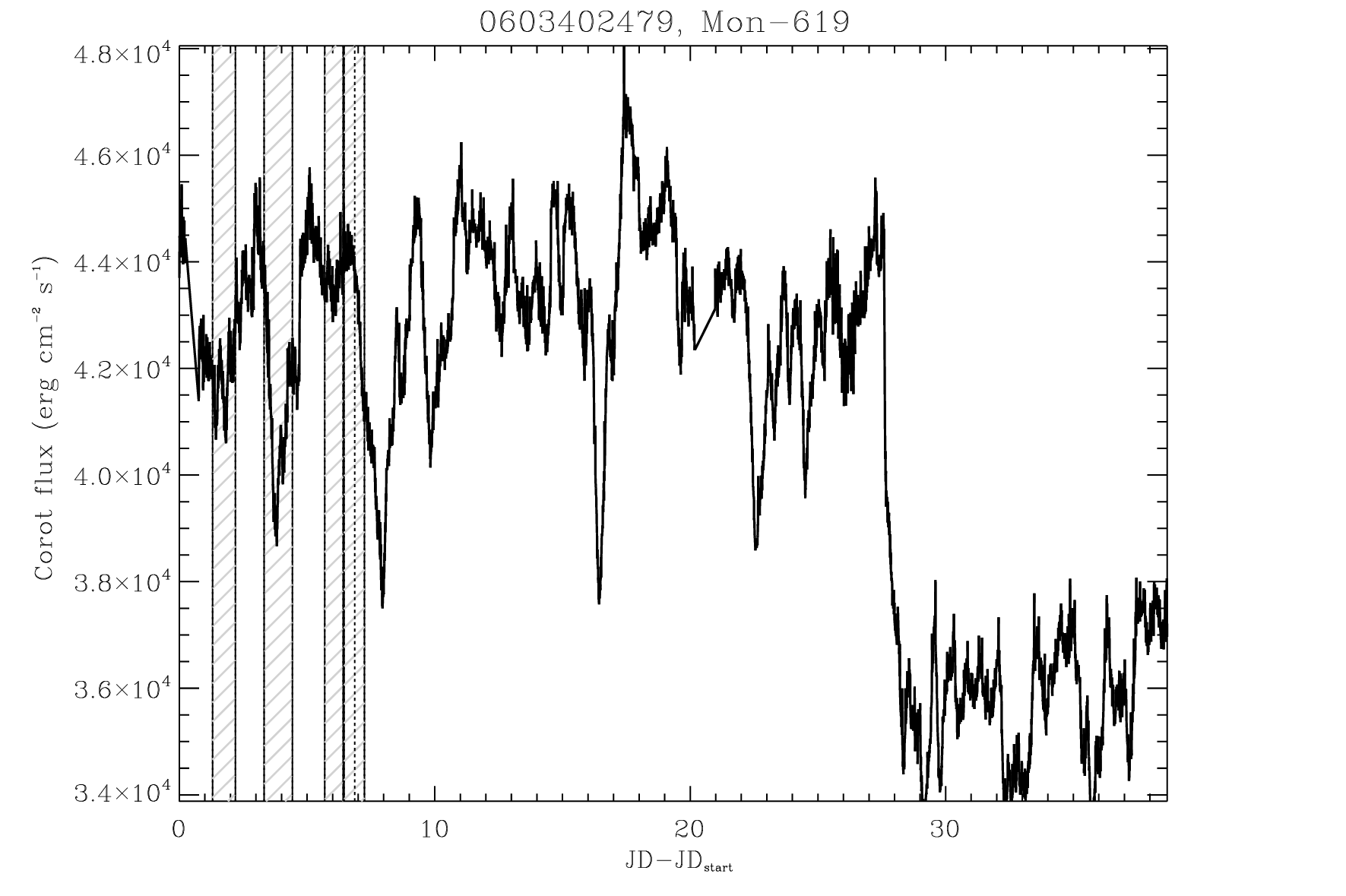}
\includegraphics[width=9.0cm]{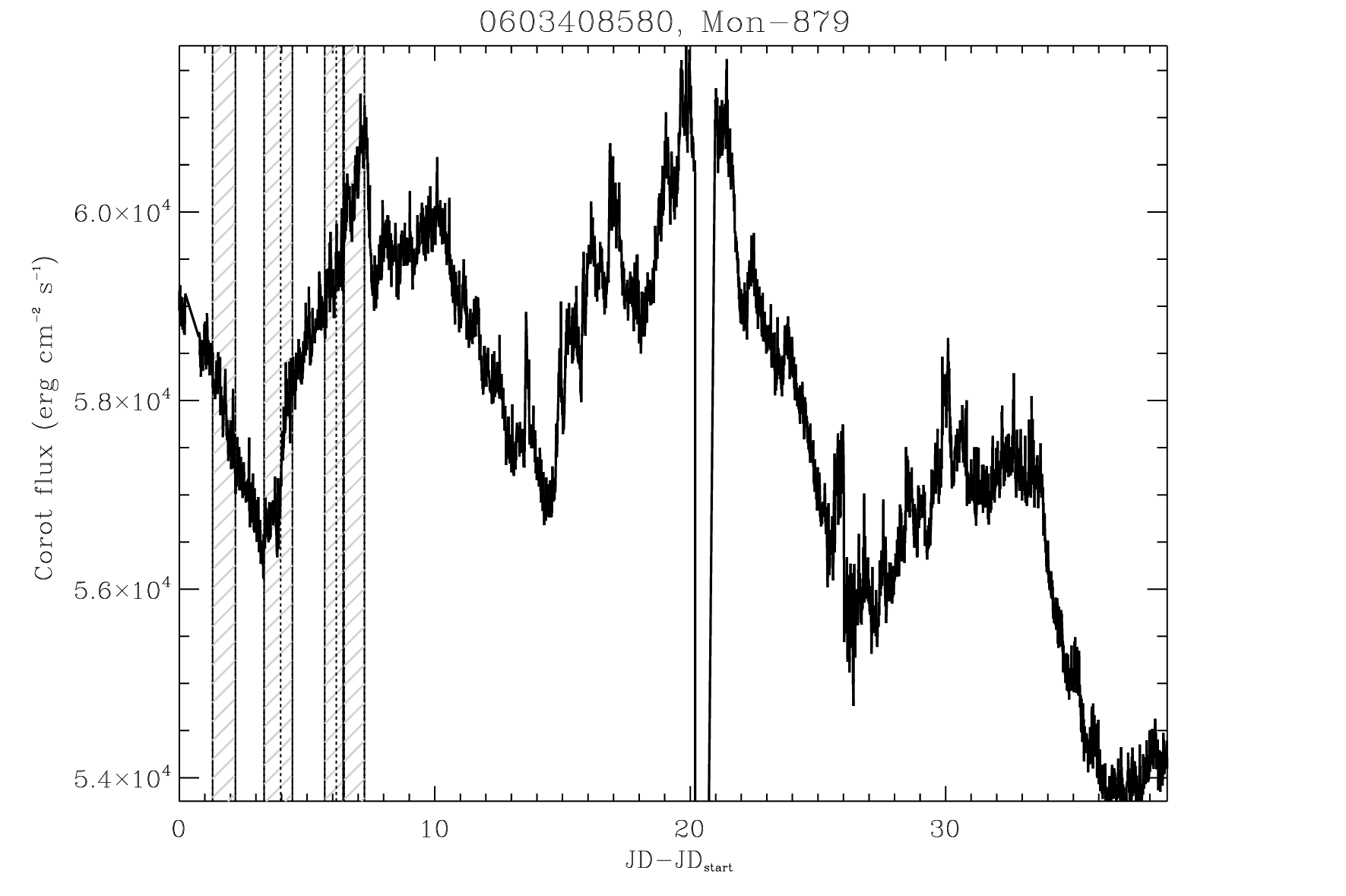}
\includegraphics[width=9.0cm]{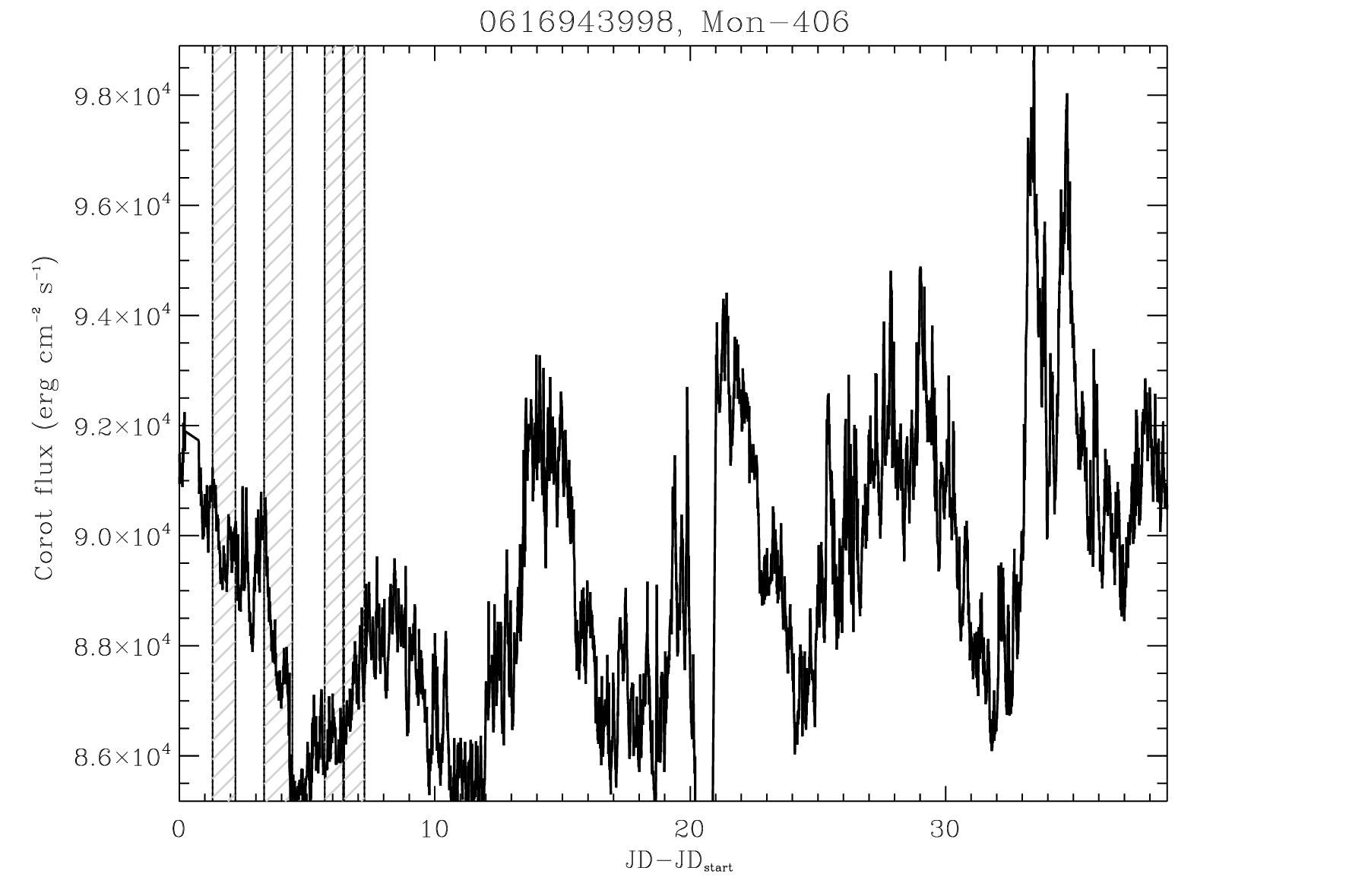}
\label{variab_others_44}
\end{figure}

\begin{figure}[]
\centering	
\includegraphics[width=9.0cm]{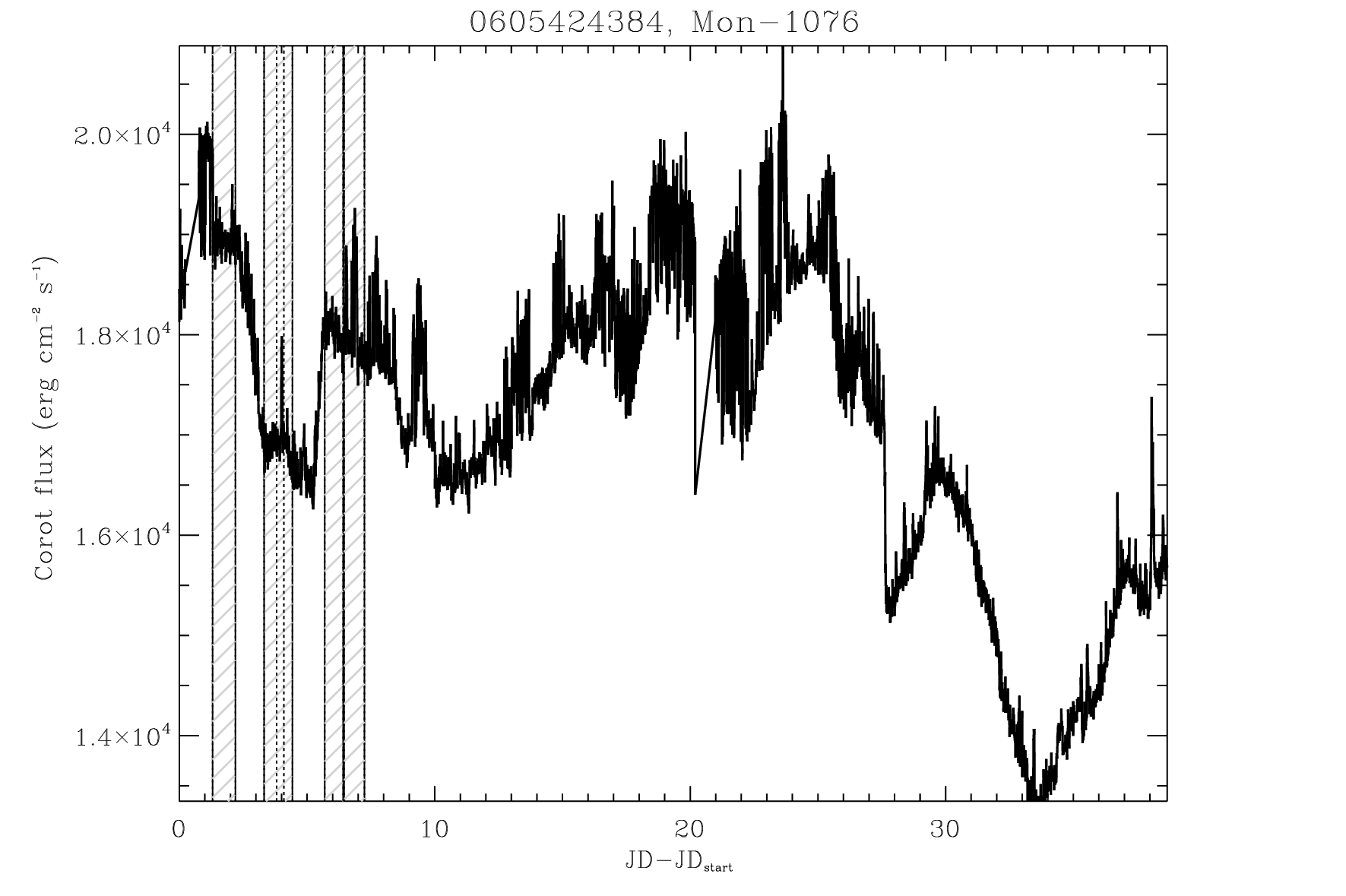}
\includegraphics[width=9.0cm]{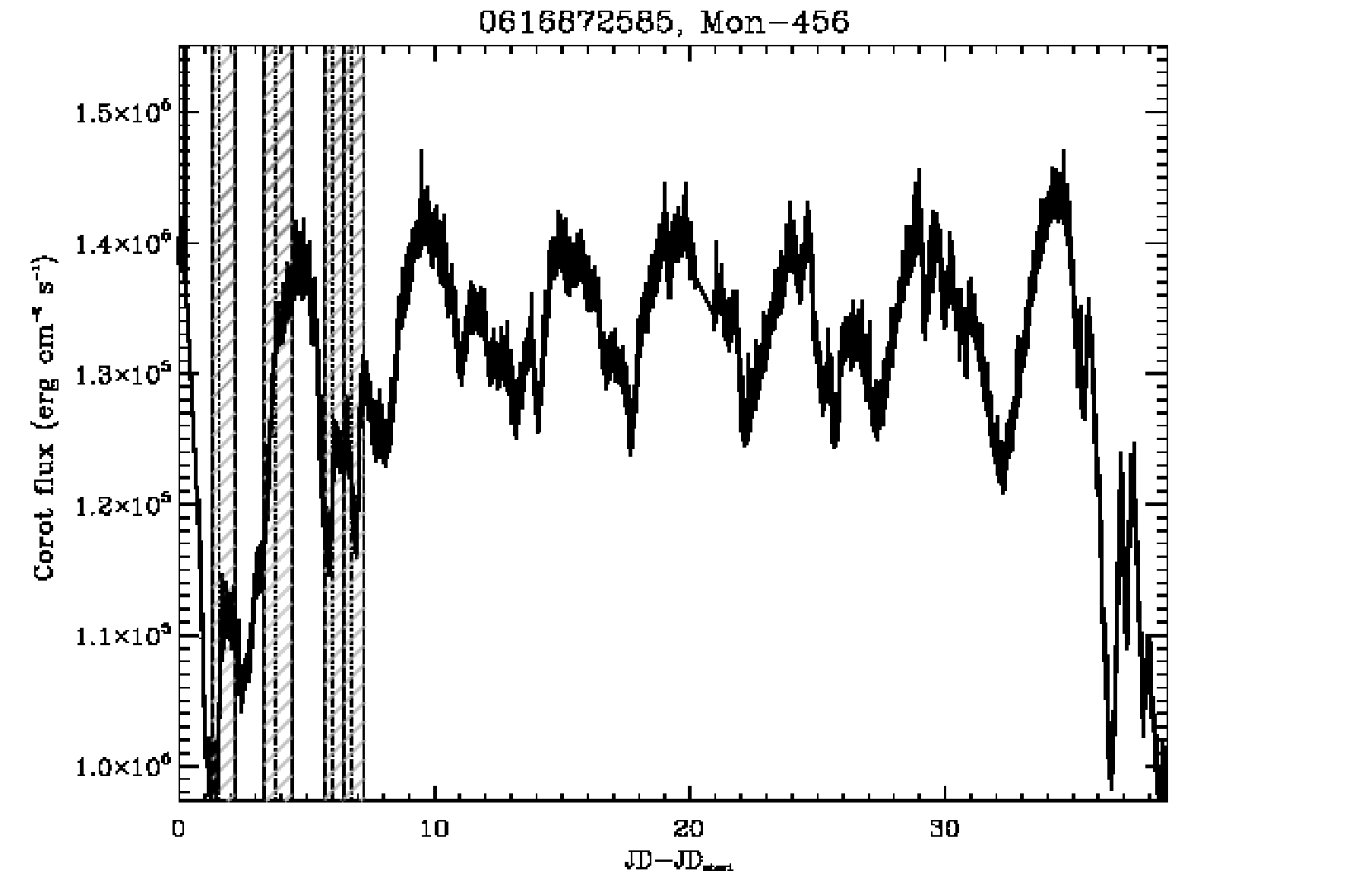}
\includegraphics[width=9.0cm]{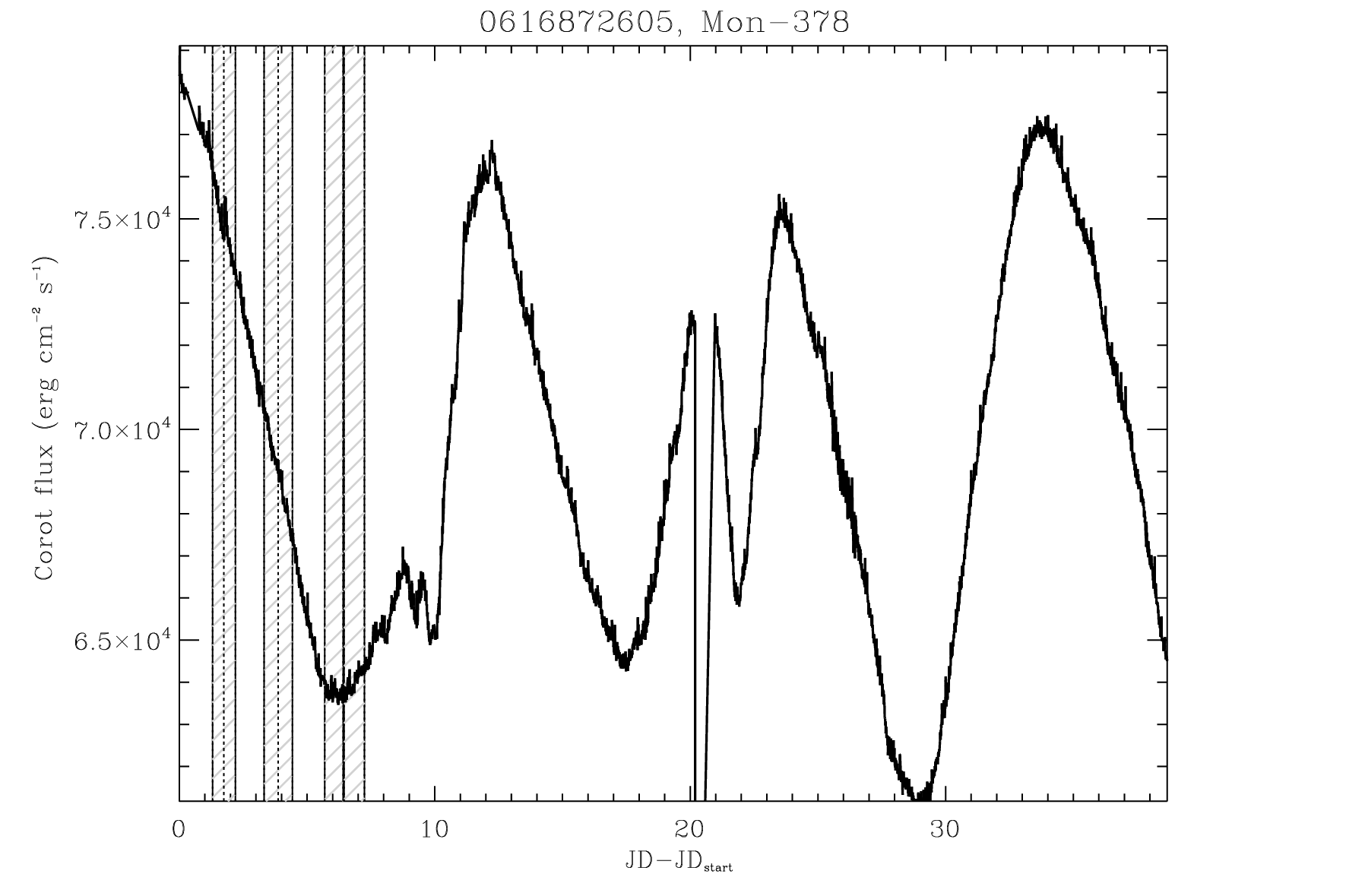}
\includegraphics[width=9.0cm]{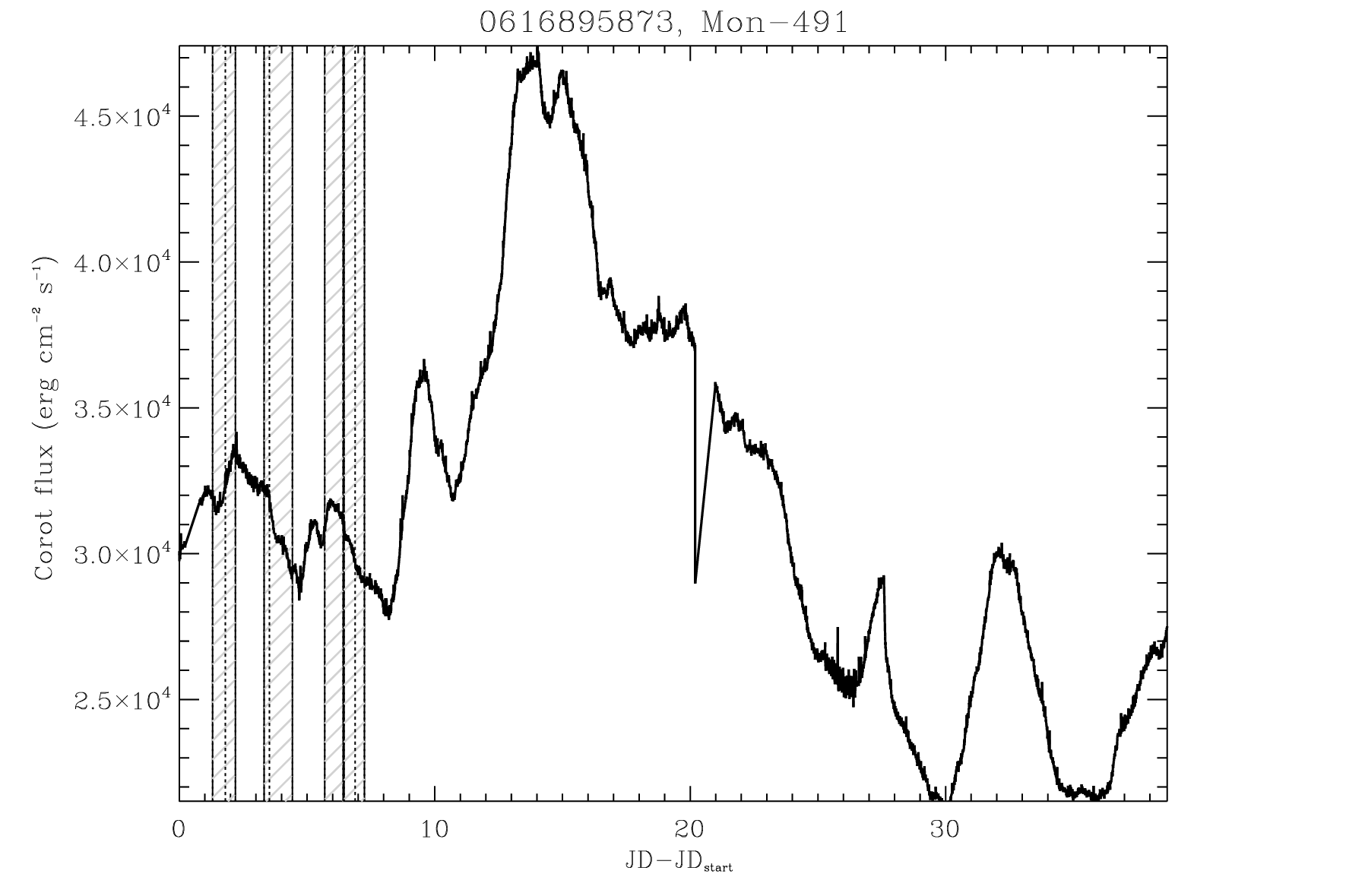}
\includegraphics[width=9.0cm]{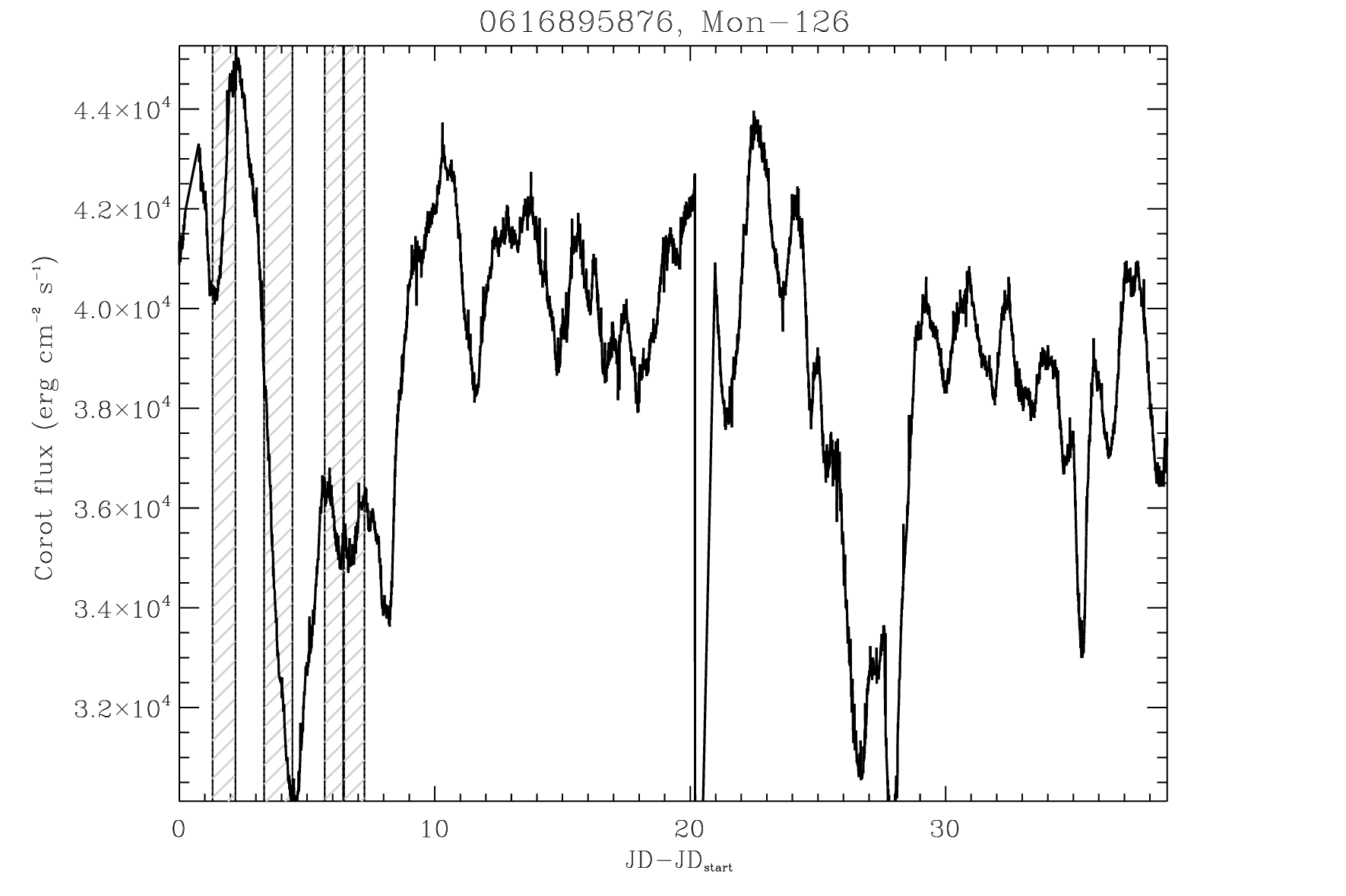}
\includegraphics[width=9.0cm]{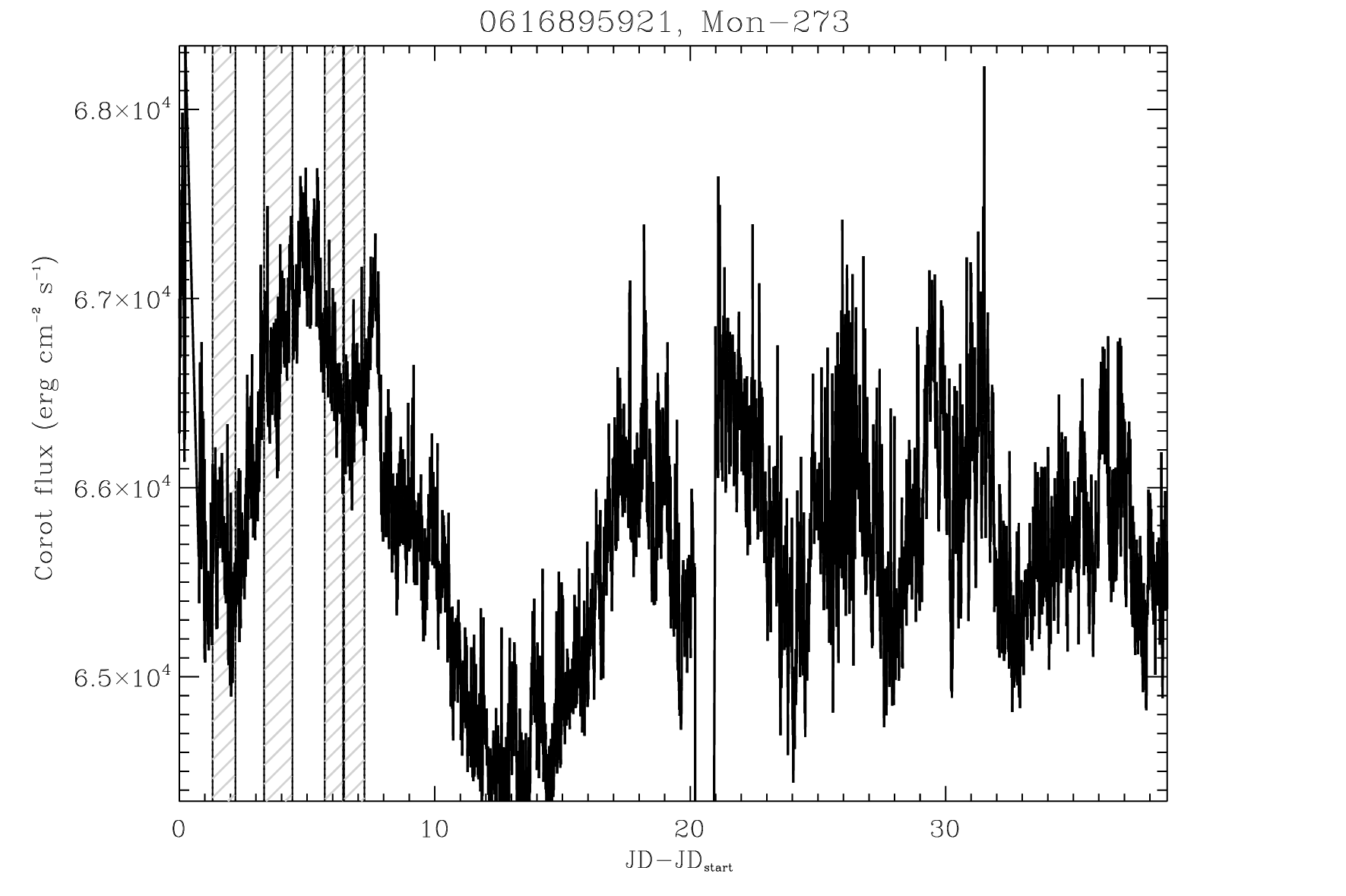}
\includegraphics[width=9.0cm]{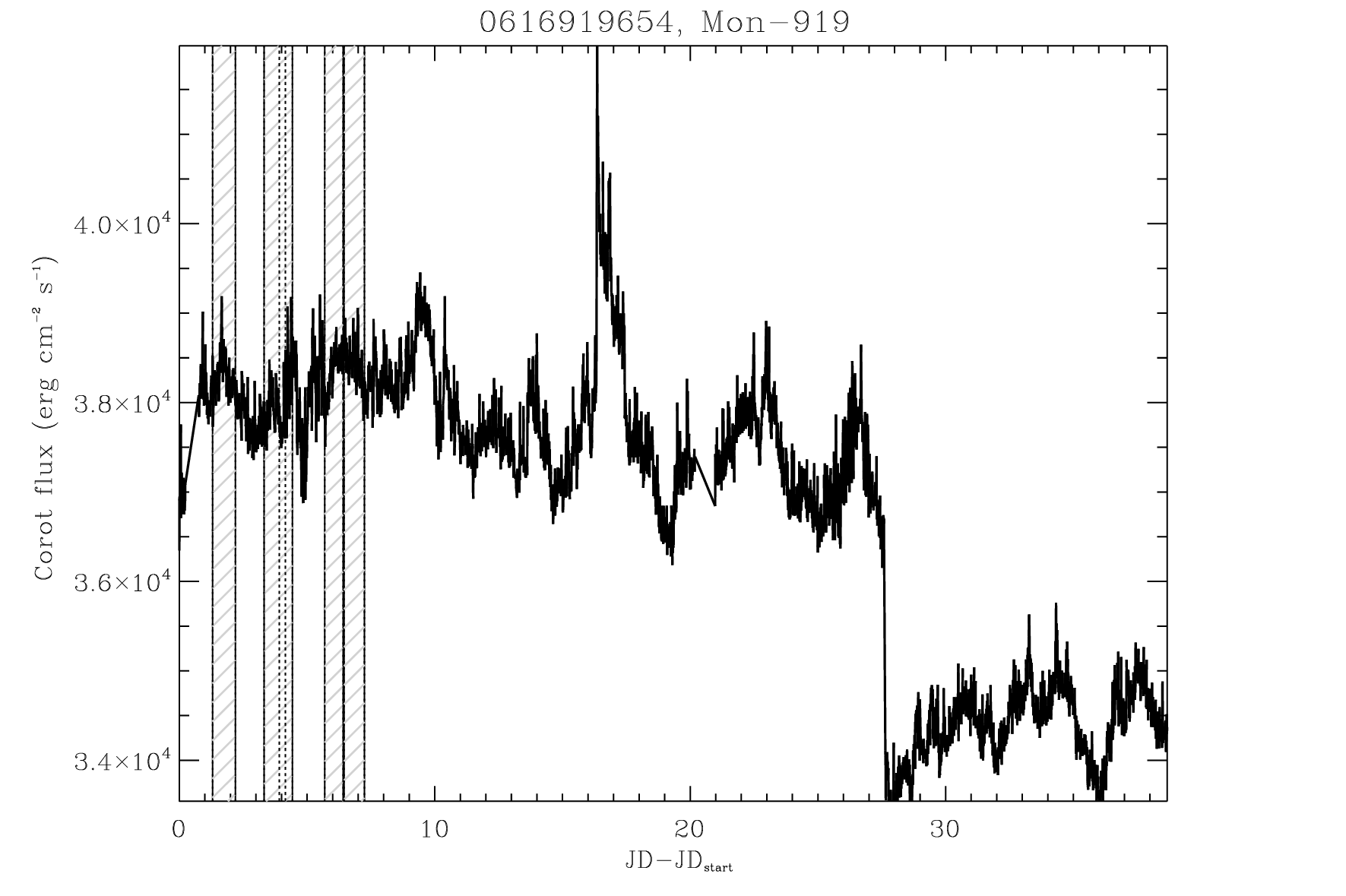}
\includegraphics[width=9.0cm]{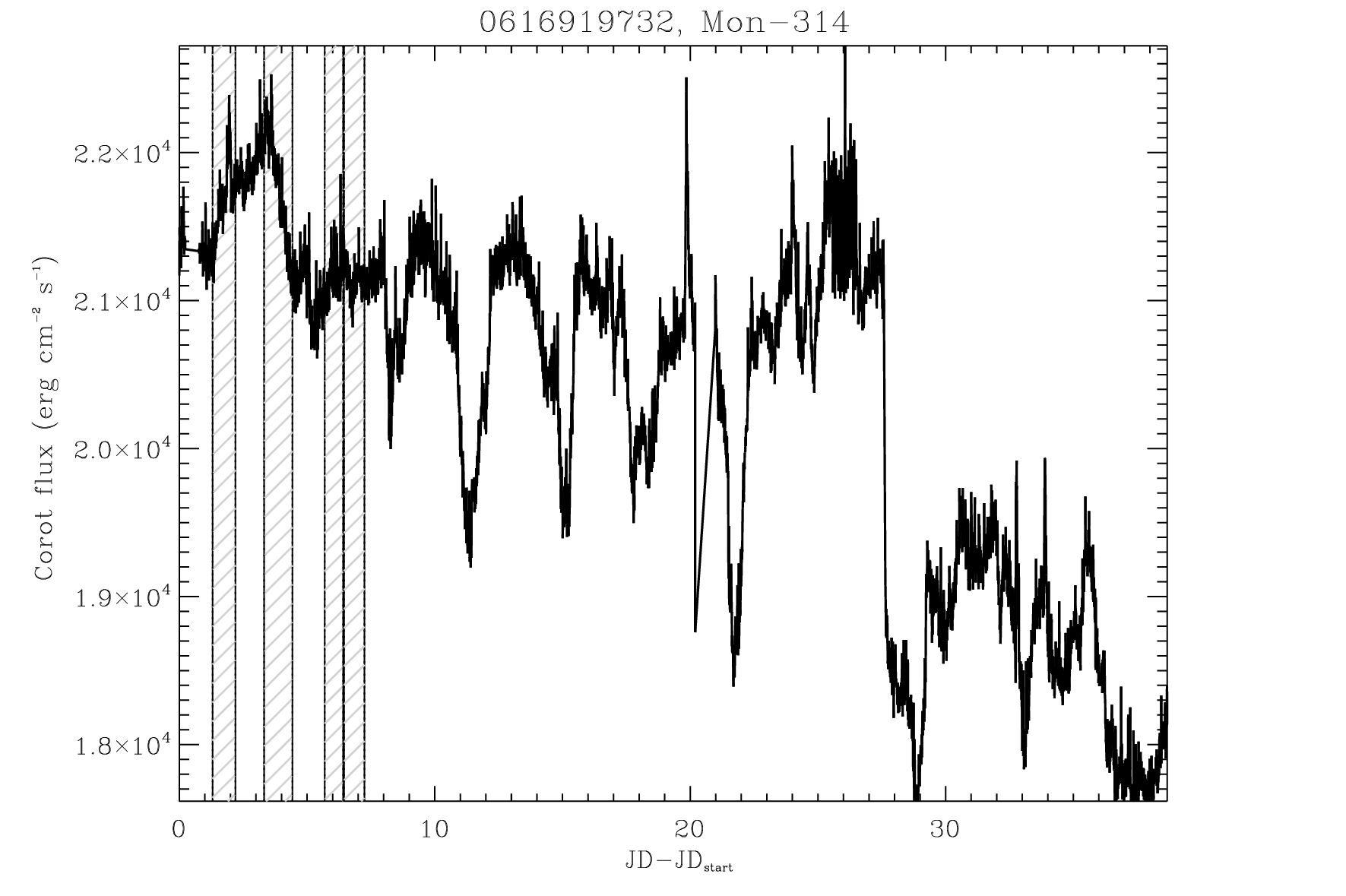}
\label{variab_others_45}
\end{figure}

\begin{figure}[]
\centering	
\includegraphics[width=9.0cm]{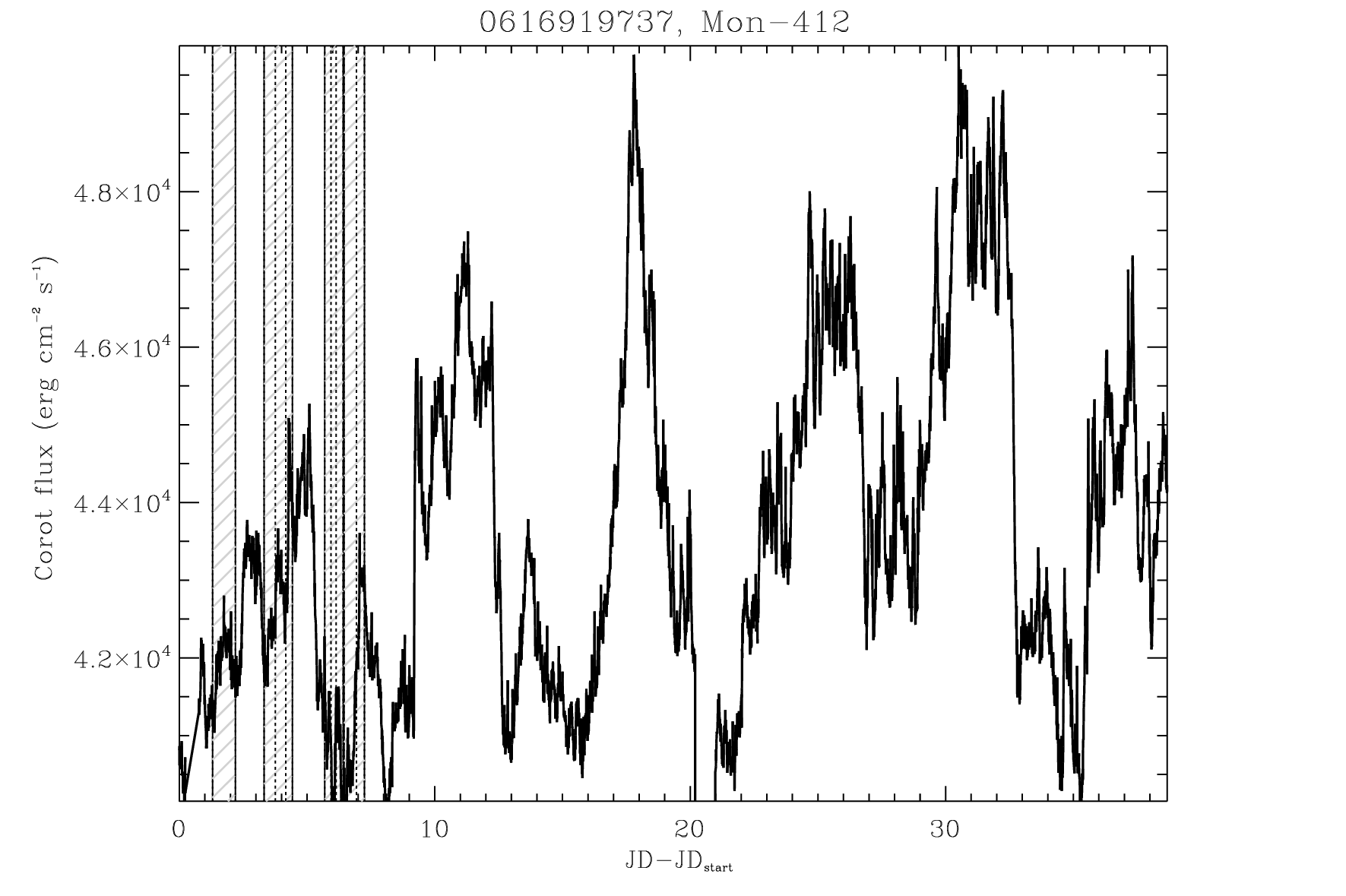}
\includegraphics[width=9.0cm]{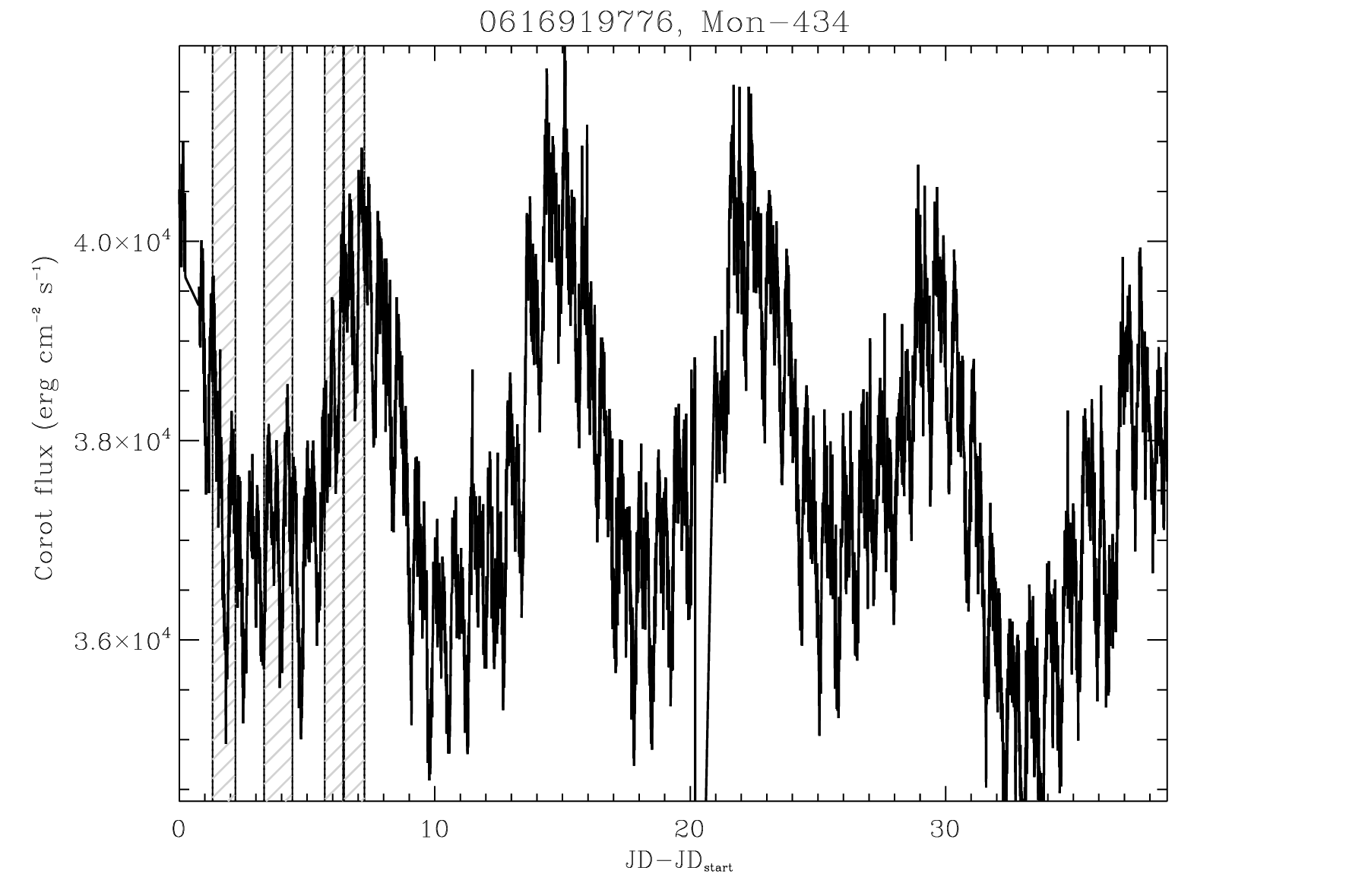}
\includegraphics[width=9.0cm]{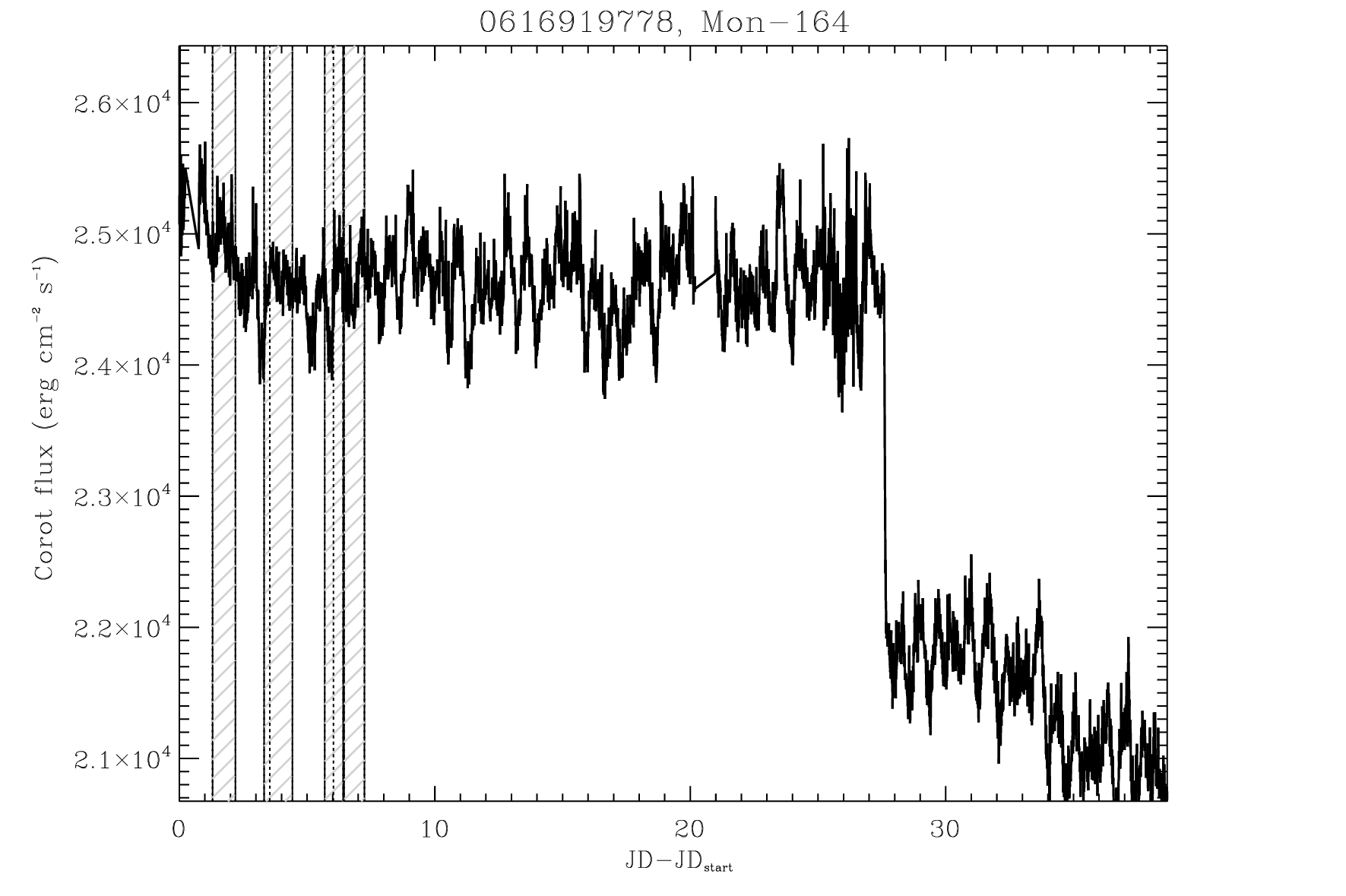}
\includegraphics[width=9.0cm]{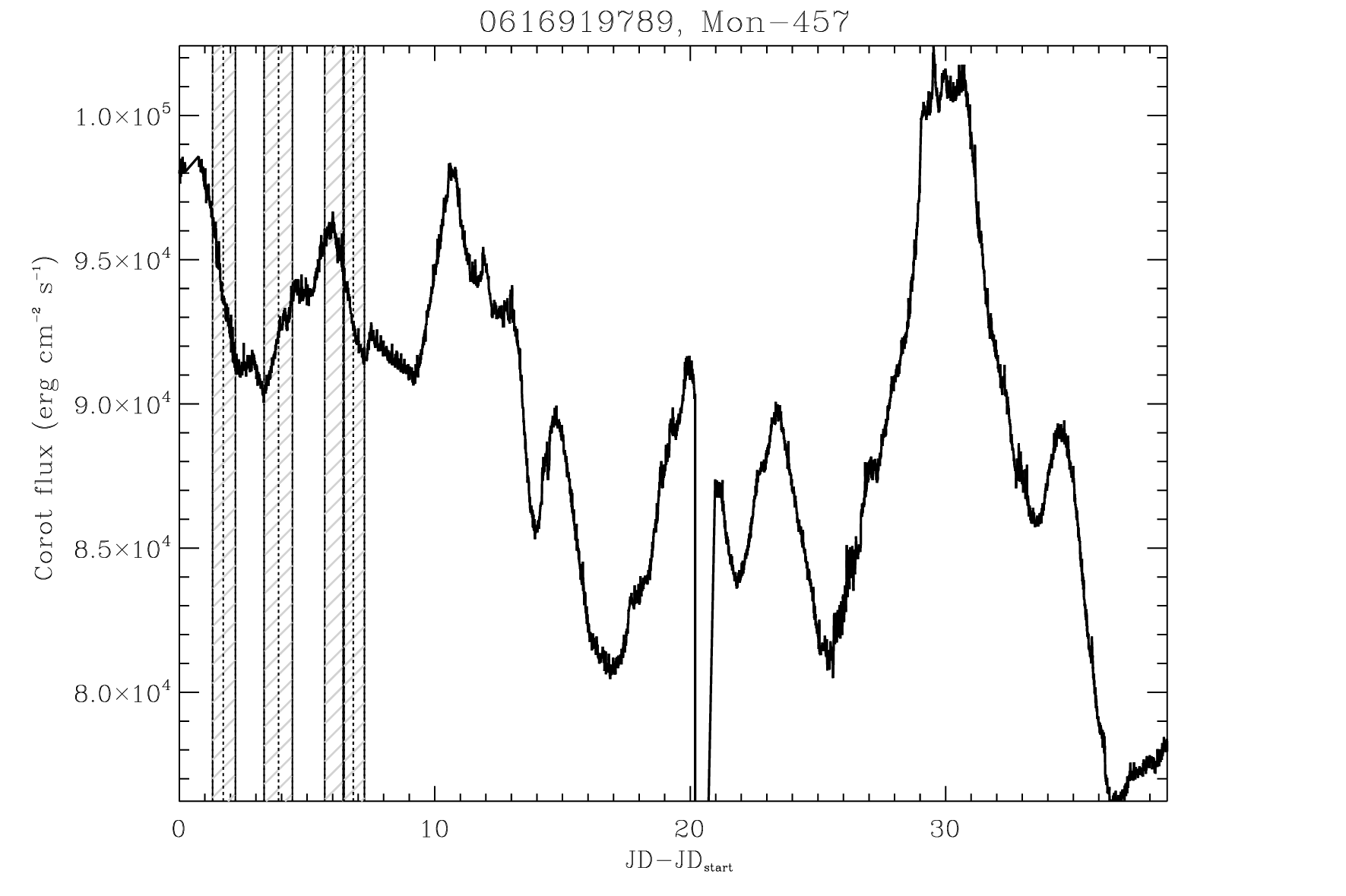}
\includegraphics[width=9.0cm]{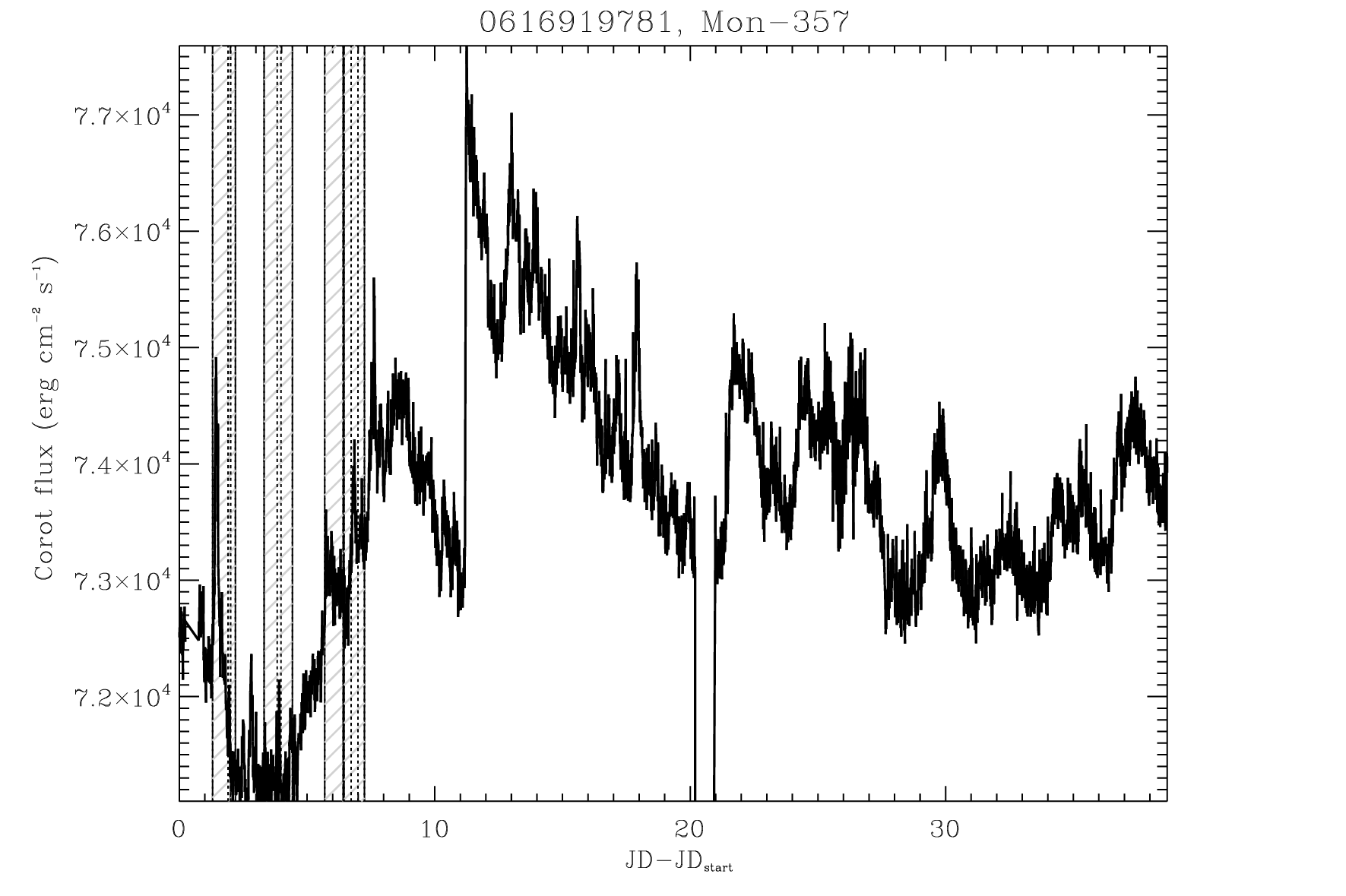}
\includegraphics[width=9.0cm]{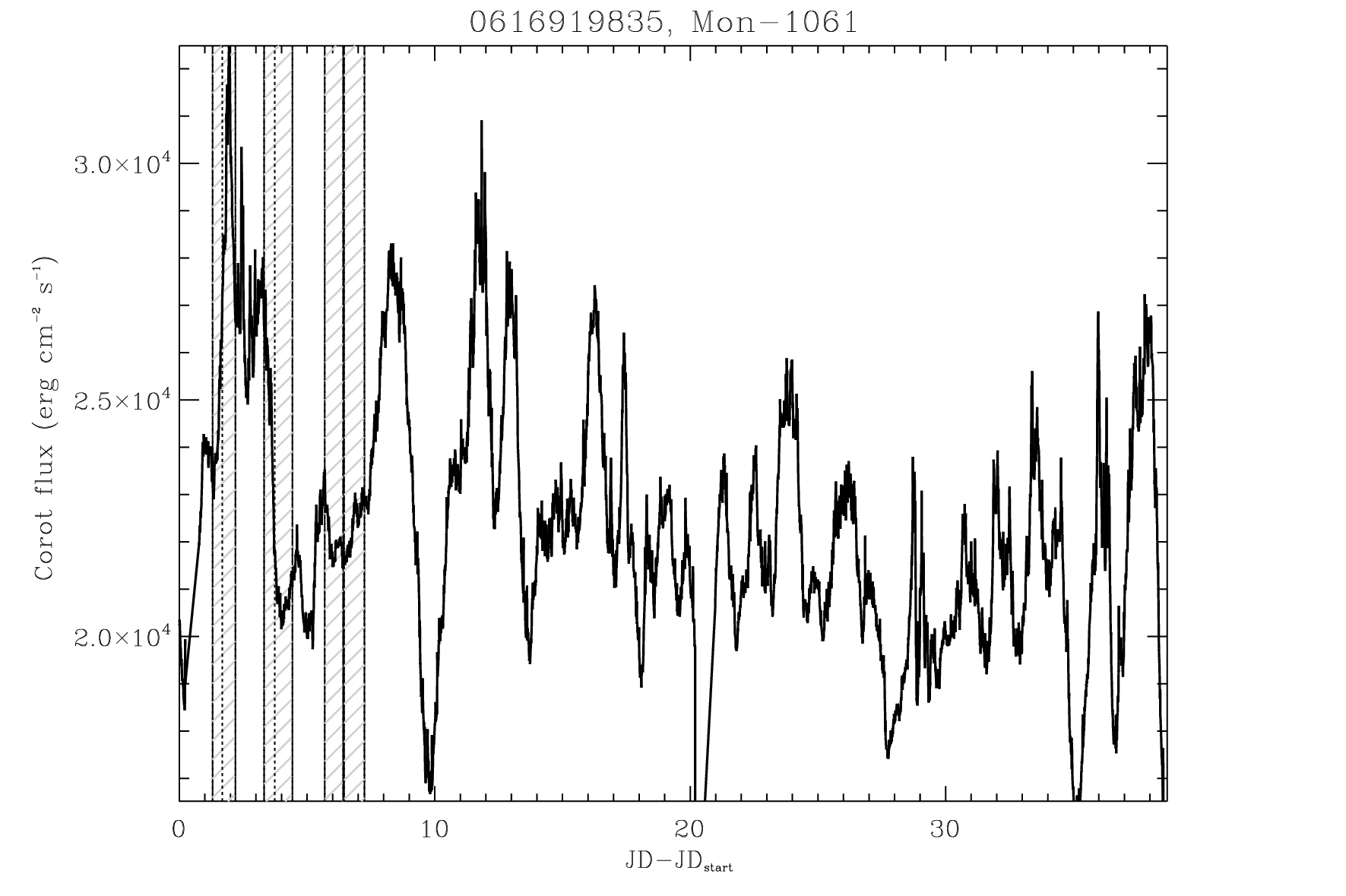}
\includegraphics[width=9.0cm]{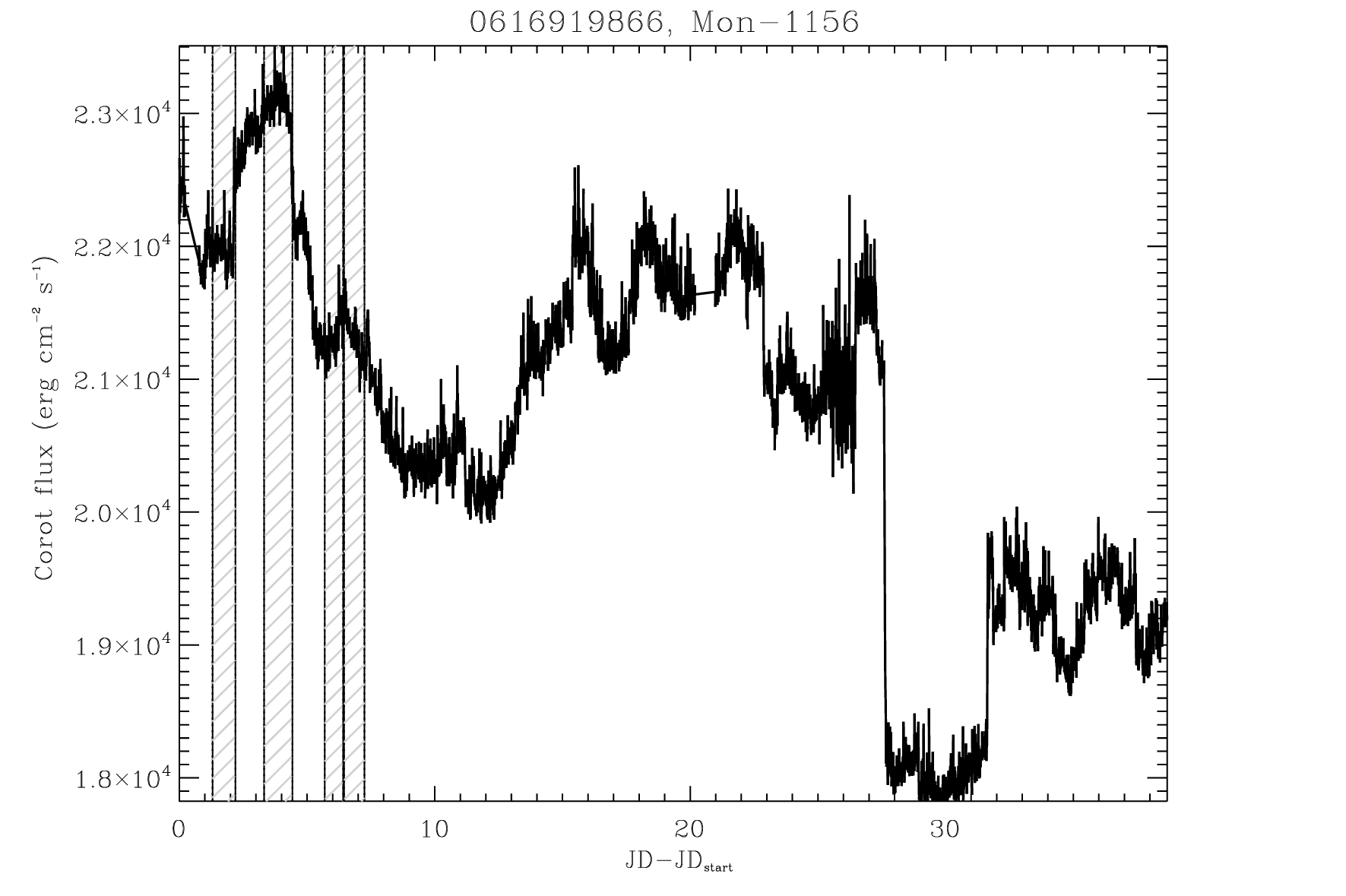}
\includegraphics[width=9.0cm]{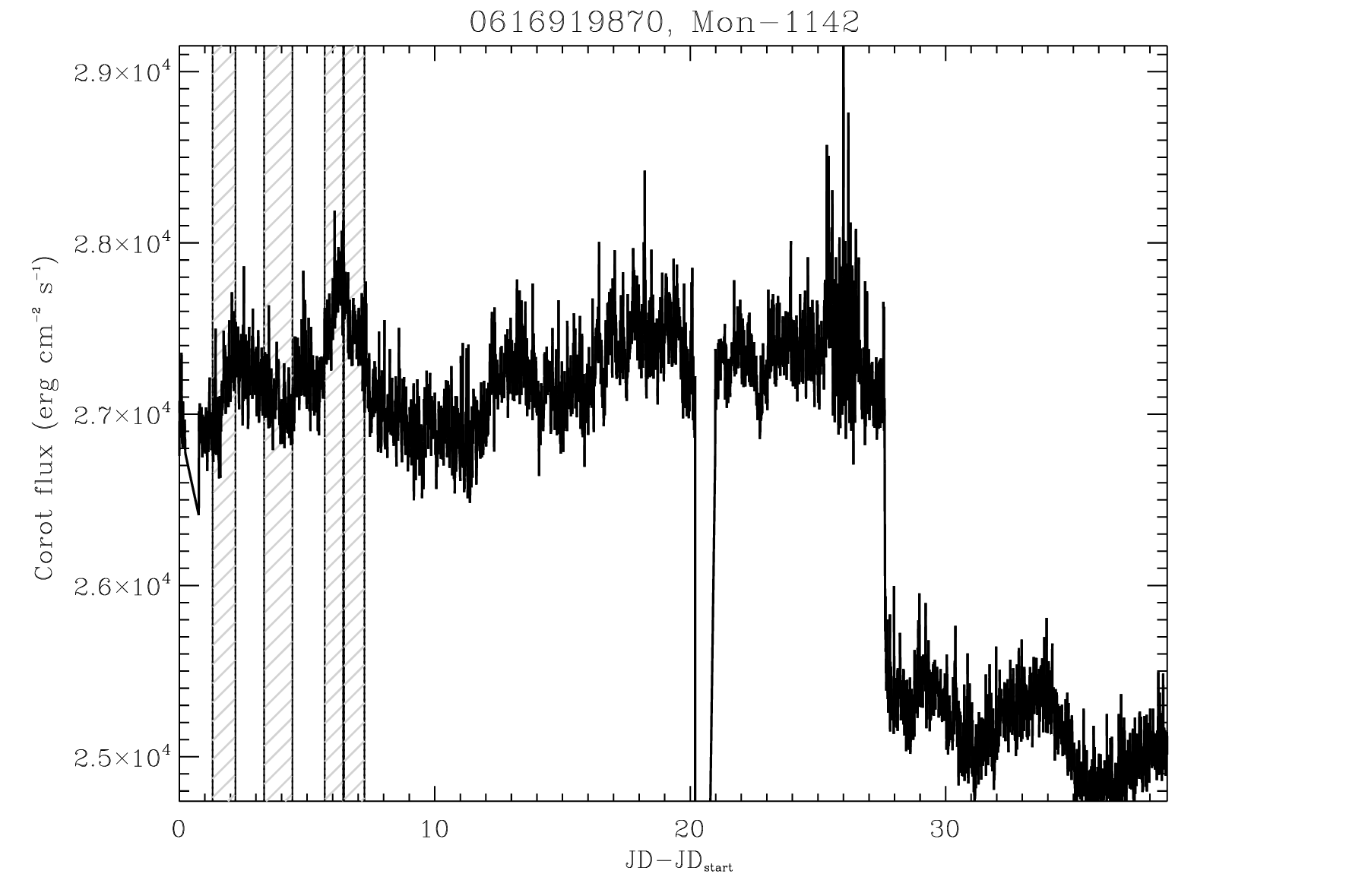}
\label{variab_others_46}
\end{figure}

\begin{figure}[]
\centering	
\includegraphics[width=9.0cm]{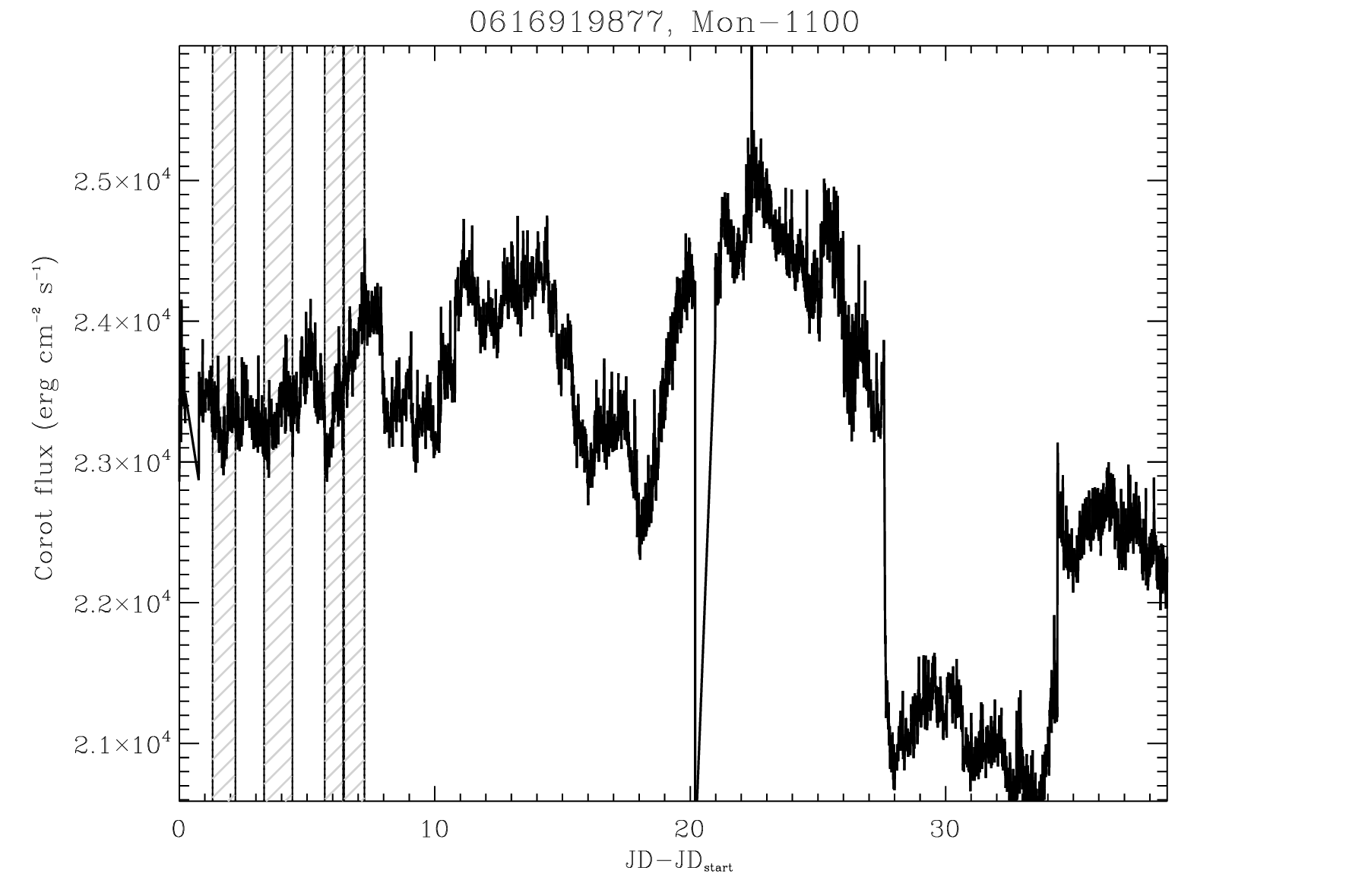}
\includegraphics[width=9.0cm]{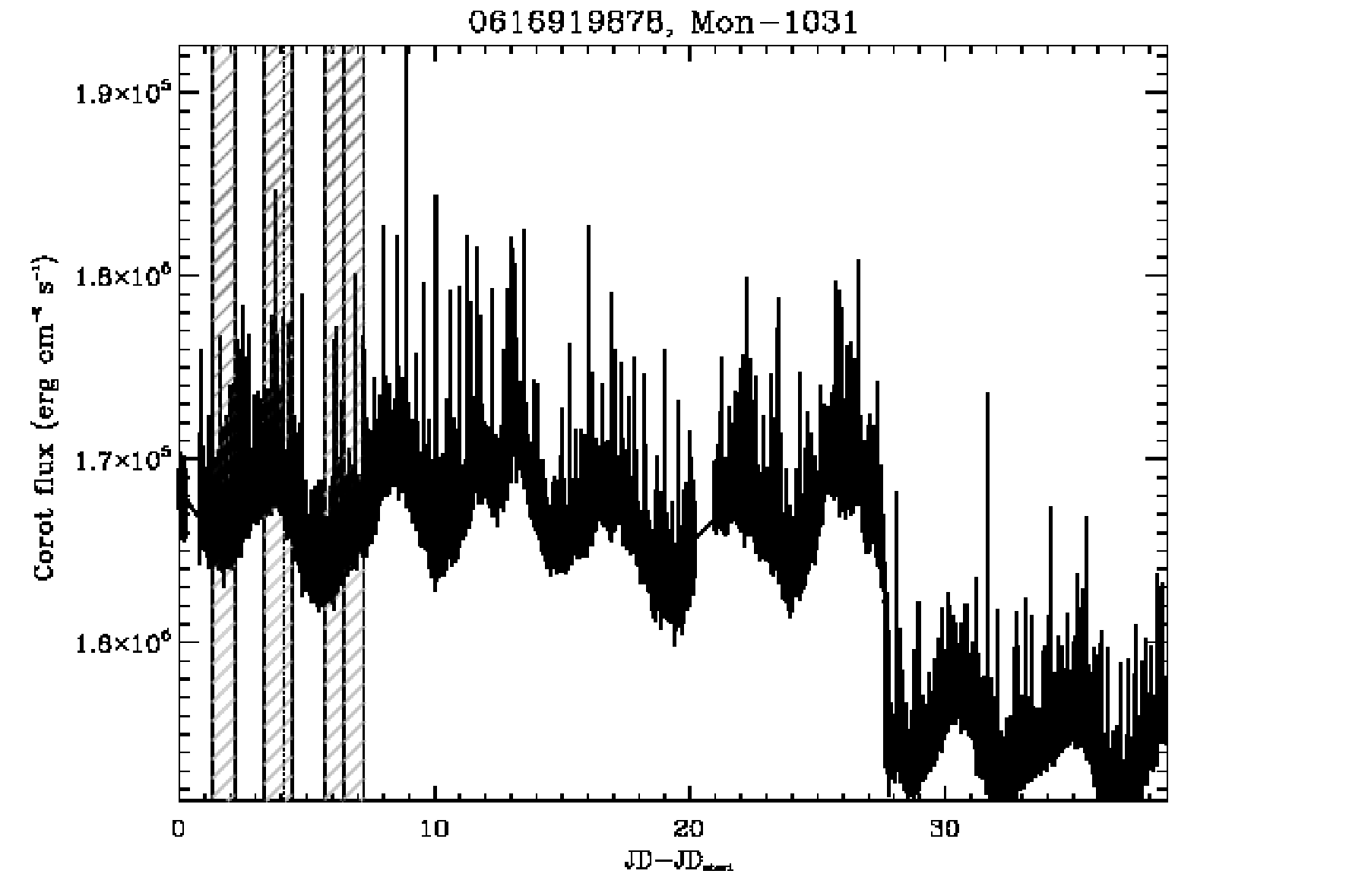}
\includegraphics[width=9.0cm]{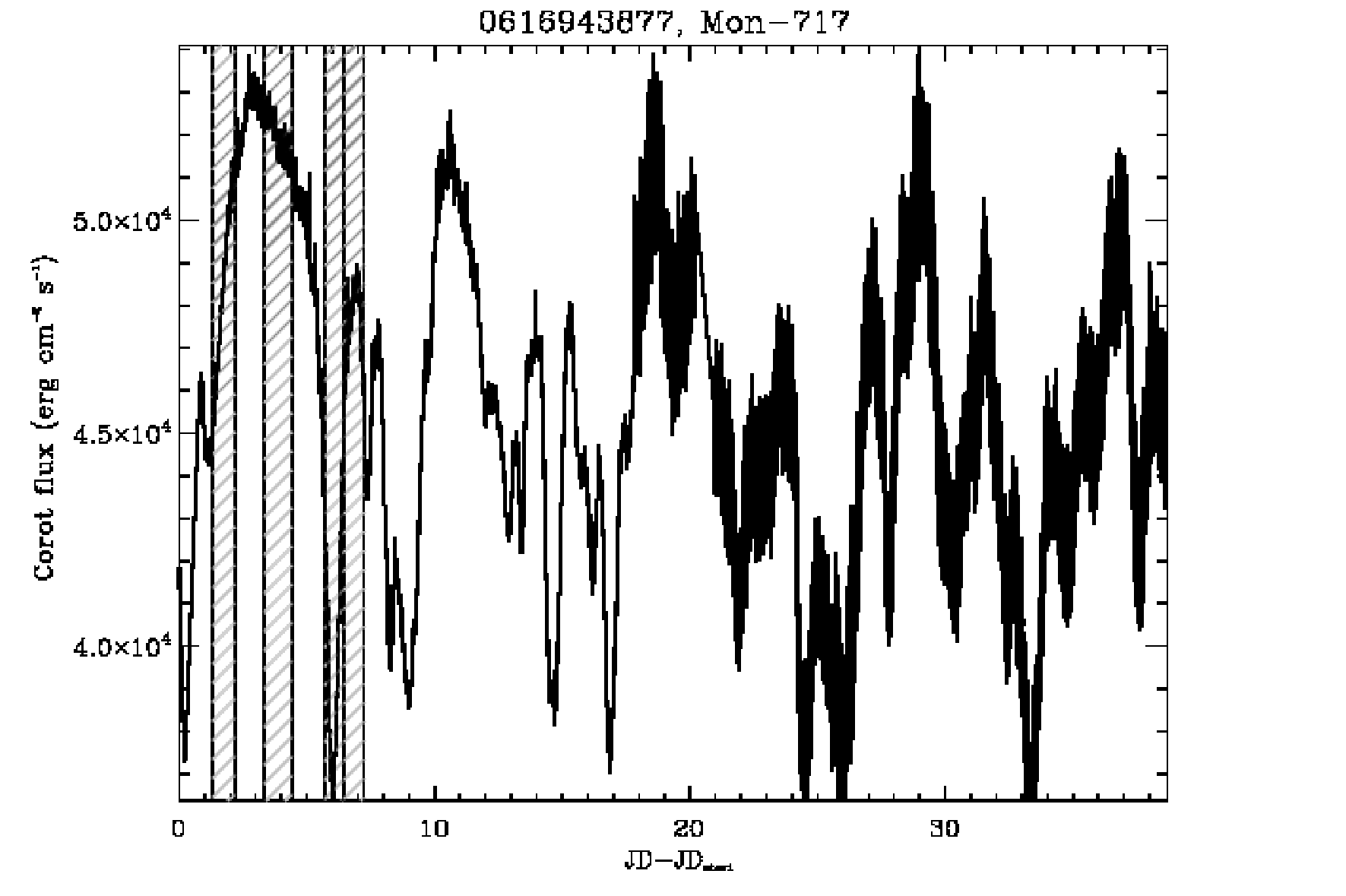}
\includegraphics[width=9.0cm]{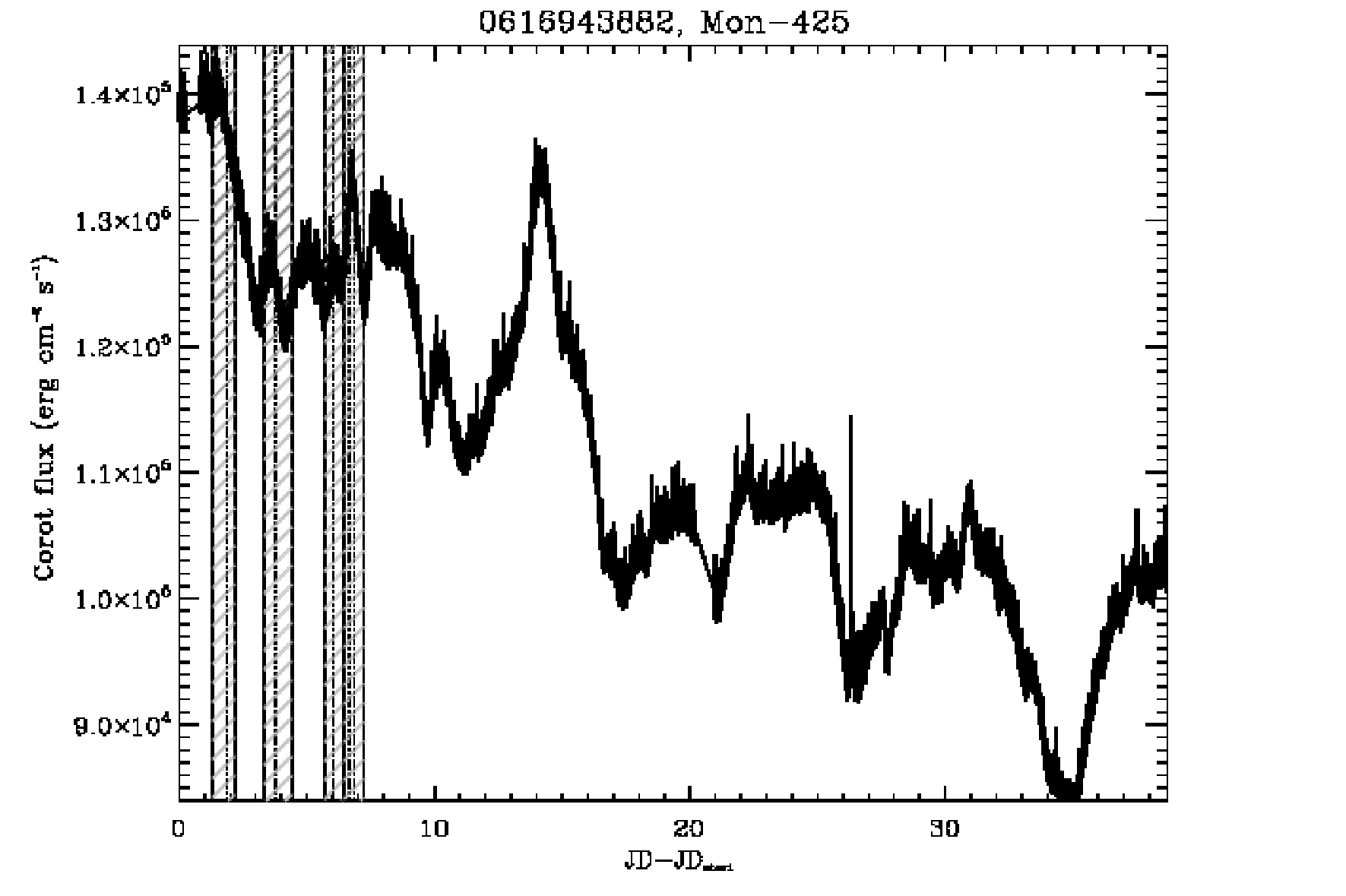}
\includegraphics[width=9.0cm]{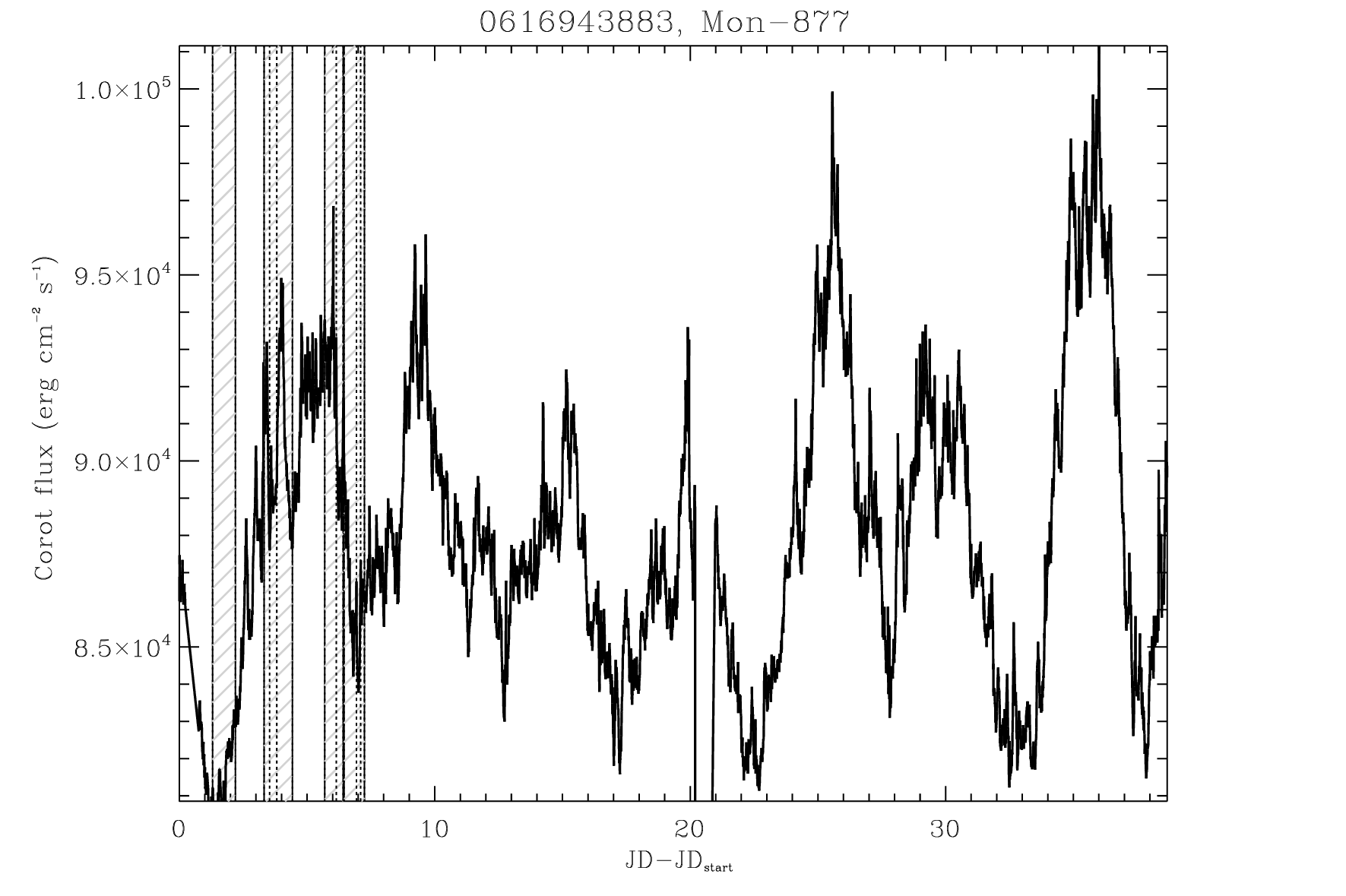}
\label{variab_others_47}
\end{figure}

\end{appendix}
\end{onecolumn}

\end{document}